\PassOptionsToPackage{table}{xcolor}
\documentclass[fleqn,review]{elsarticle}

\usepackage[utf8]{inputenc}
\usepackage{slashed}
\usepackage{tabularx}
\usepackage{braket}
\usepackage{slashed}
\usepackage{bm}
\usepackage{dcolumn}
\usepackage{array}
\usepackage{amsmath,mathtools}
\usepackage{tablefootnote}
\renewcommand{\Re}{\operatorname{Re}}
\renewcommand{\Im}{\operatorname{Im}}
\usepackage{amssymb}
\usepackage{multirow}
\usepackage{subcaption}
\usepackage{ragged2e} 
\usepackage{etoolbox}
\AtBeginEnvironment{equation}{%
  \setlength{\abovedisplayshortskip}{\abovedisplayskip}%
}
\newcommand{\massdevinline}[2]{\mbox{#1\,{\scriptsize #2}}}
\usepackage{tensor}
\usepackage{comment}
\usepackage{dashrule}
\usepackage{enumitem}
\graphicspath{{./fig/}}

\usepackage{kotex}
\usepackage{ytableau}
\usepackage{amsthm}
\usepackage{svg}
\usepackage{indentfirst}

\usepackage[font=normalsize,labelfont=bf]{caption}

\usepackage{multicol}

\newlist{notationlist}{description}{1}
\setlist[notationlist]{
  style=standard,
  leftmargin=0pt,
  labelwidth=0pt,
  labelsep=0.45em,
  itemindent=0pt,
  font=\normalfont,
  itemsep=0.05ex,
  parsep=0pt,
  topsep=0.20ex,
  partopsep=0pt
}
\newcommand{\notationgroup}[1]{\par\medskip\noindent\textbf{#1}\par\smallskip}

\makeatletter
\renewcommand\section{\@startsection{section}{1}{0pt}%
  {-3.5ex plus -1ex minus -.2ex}%
  {2.3ex plus.2ex}%
  {\normalfont\bfseries\raggedright}} 
\renewcommand\subsection{\@startsection{subsection}{2}{0pt}%
  {-3.25ex plus -1ex minus -.2ex}%
  {1.5ex plus .2ex}%
  {\normalfont\itshape\raggedright}} 
\makeatother

\ytableausetup{
  boxsize=1.28em,   
  centertableaux   
}
\usepackage{enumitem}
\usepackage{amsmath}
\usepackage{tikz}
\usetikzlibrary{decorations.pathreplacing}

\allowdisplaybreaks
\usepackage{booktabs}

\usepackage{threeparttable}

\usepackage[colorlinks,citecolor=blue,linktoc=all,linkcolor=cyan]{hyperref}
\usepackage{graphicx}

\usepackage[T1]{fontenc}
\usepackage{dsfont}   
\usepackage{mathrsfs} 
\usepackage{slashed}  
\usepackage{amsmath}
\usepackage{amssymb}
\usepackage{amsbsy}
\usepackage{amsfonts}

\usepackage{xcolor}
\usepackage{tabularray}
\UseTblrLibrary{booktabs}

\definecolor{headgray}{RGB}{236,239,242}

\definecolor{devred}{HTML}{D92D20}
\definecolor{devgreen}{HTML}{2F9E44}
\definecolor{deltaBlue}{HTML}{2F55D4}
\definecolor{yukawaGreen}{HTML}{2E8B57}
\definecolor{gaussOrange}{HTML}{B55A11}

\newcommand{\pdev}[1]{\textcolor{devred}{#1}}
\newcommand{\ndev}[1]{\textcolor{devgreen}{#1}}
\newcommand{\massdev}[2]{\shortstack[c]{#1\\[-1pt]\footnotesize #2}}
\newcommand{\mthreeq}[2]{%
  \makebox[\linewidth][c]{%
    \footnotesize\textcolor{#2}{\(\mathstrut #1\)}%
  }%
}

\usepackage[table]{xcolor}
\usepackage{colortbl}

\definecolor{headgray}{RGB}{236,239,242}

\usepackage{setspace}
\AtBeginEnvironment{thebibliography}{%
  \setstretch{0.90}
  \setlength{\bibsep}{1pt plus 0.3ex}
}

\numberwithin{equation}{section}
\numberwithin{table}{section}
\numberwithin{figure}{section}

\journal{Progress in Particle and Nuclear Physics}

\usepackage{titlesec}
\usepackage{sectsty}
\titleformat{\section}{\normalfont\Large\bfseries}{\thesection}{1em}{}
\titleformat{\subsection}{\normalfont\large\bfseries}{\thesubsection}{1em}{}
\titleformat{\subsubsection}{\normalfont\normalsize\bfseries}{\thesubsubsection}{1em}{}

\begin{document}
	
\begin{frontmatter}
		
\title{Three-body forces in the quark model}

\author[inst1]{Jongheon Baek}
\ead{bjh6941@yonsei.ac.kr}

\author[inst1]{Aaron Park\corref{cor1}}
\ead{aaron.park@yonsei.ac.kr}

\author[inst2,inst3]{Emiko Hiyama}
\ead{hiyama@riken.jp}

\author[inst4]{Sungsik Noh}
\ead{sungsiknoh@kangwon.ac.kr}

\author[inst1]{Hyeongock Yun}
\ead{mero0819@yonsei.ac.kr}

\author[inst5,inst6]{Kyong Chol Han}
\ead{kchan@pvamu.edu}

\author[inst1]{Su Houng Lee\corref{cor2}}
\ead{suhoung@yonsei.ac.kr}

\address[inst1]{Department of Physics and Institute of Physics and Applied Physics, Yonsei University, Seoul 03722, Republic of Korea}
\address[inst2]{RIKEN Nishina Center for Accelerator-Based Science, Wako, Saitama 351-0198, Japan}
\address[inst3]{Department of Physics, Graduate School of Science, Tohoku University, Sendai 980-8578, Japan}
\address[inst4]{Division of Science Education, Kangwon National University, Chuncheon 24341, Republic of Korea}
\address[inst5]{Department of Physics, Prairie View A\&M University, Prairie View, TX 77446, USA}
\address[inst6]{Institute for Quantum Science and Engineering, Texas A\&M University, College Station, TX 77843, USA}

\cortext[cor1]{Corresponding author: aaron.park@yonsei.ac.kr}
\cortext[cor2]{Corresponding author: suhoung@yonsei.ac.kr}
		
\begin{abstract}

The quark model provides an intuitive and effective framework for understanding hadron spectra. Its effective Hamiltonian can be shown to be rooted in quantum chromodynamics (QCD) through a combination of lattice gauge and effective theories. In this work, we begin with a comprehensive overview of how the quark model Hamiltonian is connected to the underlying theory of the strong interaction, namely QCD. We then discuss a long-standing difficulty that arises when attempting to simultaneously describe meson and baryon spectra with a common two-body Hamiltonian. In particular, a two-body interaction calibrated to mesons leaves systematic residual deviations in the baryon sector, with the largest discrepancies appearing in light-quark systems and decreasing as heavier quarks are introduced.

By introducing a short-range, color-spin-dependent three-quark interaction, we substantially reduce this long-standing discrepancy and achieve a simultaneous high-accuracy description of meson and baryon ground-state spectra. We further show that the spatial dependence of this interaction is important: mass-scaled finite-range profiles strongly improve the baryon spectrum, whereas a flavor-independent common physical range does not remove the systematic residual pattern. The resulting interaction is tested against additional ground-state baryons not used to determine its couplings and through meson--baryon compatibility analyses using several widely adopted quark-model Hamiltonians. The need to supplement a two-body Hamiltonian by a three-body interaction is methodologically analogous to nuclear physics, where for nucleon--nucleon interactions constructed within meson-exchange or chiral effective field theory, reproducing the binding properties of the triton and nuclei with $A>2$ requires three-body nuclear forces.

We also provide a detailed discussion of the Gaussian expansion method (GEM) and the infinitesimally shifted Gaussian technique (ISG), which are used for quantitative few-quark calculations across the light, strange, and heavy sectors. Coupled-channel and resonance formalisms are reviewed to clarify how continuum effects and non-valence configurations can modify the spectrum. We also examine radial and orbital excitations to identify the region in which a static compact valence quark Hamiltonian remains quantitatively reliable.

Subsequently, we present a detailed analysis and derivation of formulae for calculating the color and spin matrix elements of two- and three-body operators in baryons and multiquark exotic configurations. This section serves both as a review and as a source of new formulae that will be useful for future calculations.

The present work demonstrates that a properly formulated short-range three-body interaction is an essential ingredient for consistently describing meson and baryon ground-state spectra within the conventional valence quark-model framework. The same operator framework can be extended to compact multiquark configurations and provides a basis for investigating how many-quark interactions may influence exotic hadrons and systems in which several quarks participate simultaneously.

\end{abstract}

\end{frontmatter}

	
\thispagestyle{empty}
\tableofcontents
	


\section{Introduction}

Understanding the baryon spectrum with high precision remains a fundamental challenge in non-perturbative quantum chromodynamics (QCD)~\cite{Sil_Brac1996,Roberts2016}. \textit{Ab initio} approaches --- including lattice QCD and continuum functional methods --- have seen significant developments, yet a compact, QCD-motivated effective description remains essential for organizing the hadron resonance spectrum and diagnosing the leading interaction mechanisms. Constituent quark models, which treat baryons as bound states of three quarks, have historically achieved qualitative success using two-body quark interactions.  In 1964, Murray Gell-Mann (and independently George Zweig) introduced the quark model, postulating that hadrons are composed of a small number of constituent quarks bound by the strong interaction~\cite{Gell-Mann1964,Zweig1964}. The color degree of freedom, introduced in 1964--1965 to reconcile baryons such as the $\Delta^{++}$ with Fermi statistics~\cite{Greenberg1964,HanNambu1965}, became the gauge charge of QCD in the early 1970s. Quark-confining potentials and inter-quark forces inspired by QCD were later formulated to describe hadron spectroscopy. De R\'ujula, Georgi, and Glashow added an effective one-gluon-exchange (OGE) hyperfine interaction, which accounts for the $N$--$\Delta$ splitting and other mass relations~\cite{DeRujula:1975}. Around the same time, the MIT bag model was proposed as an alternative description of hadrons, treating quarks as quasi-free particles confined inside a finite ``bag'' by a vacuum pressure~\cite{ChodosJaffe1974,Chodos:1974pn}. By the 1980s, pairwise potential models such as the nonrelativistic Isgur--Karl model for baryons~\cite{Isgur1979} and the relativized Godfrey--Isgur model for mesons~\cite{Godfrey1985} reproduced much of the observed spectrum. Dyson--Schwinger studies developed a Poincar\'e-covariant continuum approach that incorporates dynamical chiral symmetry breaking and describes many meson observables, with increasing application to baryons~\cite{Roberts1994}. Additionally, chiral quark models were developed in the 1990s that replaced gluon-mediated spin-spin forces with Goldstone-boson exchange between constituent quarks, offering an alternative explanation for the hyperfine splittings in hadrons.

Pairwise models nevertheless share a systematic problem: a single parameter set has been difficult to fit to the meson and baryon spectra at the same time~\cite{CapstickRoberts2000,KlemptRichard2010,CredeRoberts2013}. This difficulty manifests as systematic, flavor-dependent deviations in the calculated baryon masses relative to experimental values, with light-quark baryons typically overestimated and heavy-quark baryons correspondingly less overestimated or even underestimated.  For over half a century, various refinements were attempted to address this ``flavor-dependency'' puzzle. 
Throughout this work, we use this term to denote the systematic difficulty of describing the meson and baryon spectra simultaneously with a single common two-body quark-model Hamiltonian. For a more detailed historical account of this issue, see  Section~\ref{sec:discuss}.
For example, some models included flavor-dependent short-range forces induced by instantons~\cite{Blask1990,Loring2001a,Loring2001} or incorporated meson-exchange (one-boson exchange) interactions between quarks to better account for the hyperfine splitting. While these ingredients (and others, such as adjusting quark masses or including phenomenological mixing terms) did improve fits in certain cases, they did not fully resolve all discrepancies within a single, unified framework. In particular, all widely used constituent quark models over the past decades have required separate fine-tuning for different sectors or introduced \emph{ad hoc} terms, undermining their predictive power~\cite{CapstickRoberts2000,KlemptRichard2010,CredeRoberts2013}. These persistent failures to eliminate flavor-dependent mass differences using only two-body interactions strongly suggested that an additional ingredient was missing from the conventional quark model dynamics.

At the same time, there is growing evidence that quark three-body forces play an indispensable role in baryon structure. An analogy is found in nuclear physics: three-nucleon forces are essential for reproducing the properties of light nuclei that cannot be explained by two-nucleon interactions alone. For example, the binding energies of few-nucleon systems (with $A \geq 3$) are underpredicted by models that include only two-body forces, and the inclusion of three-body terms is required for quantitative agreement~\cite{Fujita:1957zz,Brown:1968qla}.\footnote{This statement refers to nucleon--nucleon interactions constructed within meson-exchange or chiral effective-field-theory frameworks. Three-nucleon forces are not separately observable, so the size required depends on the representation of the pair interaction~\cite{PolyzouGlockle1990,Polyzou2010,Hammer2013}; for example, nonlocal nucleon--nucleon interactions that describe the two-nucleon data with comparable accuracy reproduce the ${}^3$H and ${}^3$He binding energies without an explicit three-nucleon force~\cite{Doleschall2000,Doleschall2003}. The corresponding statement for the quark model is discussed in  Section~\ref{subsec:three_quark_definition}.}  In chiral effective field theory (EFT), such forces arise naturally as higher-order contributions and have been found to influence both nuclear binding and scattering observables~\cite{Hammer2013,Epelbaum:2005pn,vanKolck1994}.

In the context of quark models, the possibility of quark three-body forces has been discussed for some time~\cite{Oka1980}. Lattice QCD studies of the static three-quark potential show that baryonic confinement is not strictly pairwise at the level of flux-tube geometry: the ground-state flux distribution is better described by a $Y$-shaped configuration than by a purely pairwise $\Delta$-type picture~\cite{Takahashi2001}. This interpretation is also consistent with earlier flux-tube modeling~\cite{IsgurPaton1985}. This observation establishes that the asymptotic three-source flux geometry is not exactly represented by a sum of independent pairwise strings.\footnote{It does not, however, imply that pairwise confinement is inadequate for compact baryons: because the $Y$- and $\Delta$-type lengths are numerically close over typical hadronic geometries, the latter remains a useful approximation and is the confinement prescription adopted in our baseline Hamiltonian~\cite{Alexandrou2002}. The long-distance $Y$-string geometry and the short-distance perturbative three-line contribution~\cite{BrambillaGhiglieriVairo2010} are reviewed as distinct examples of three-source QCD dynamics; neither is identified with the finite-range color--spin three-quark interaction fitted below.} 

In a quark model, one can naturally introduce simple three-body quark interactions, as the color SU(3) group admits two independent three-body operators proportional to the structure constants $f^{abc}$ and $d^{abc}$, where $a$, $b$ and $c$ are the color indices of the three quarks. The antisymmetric $f^{abc}$ type does not contribute directly to the energy of a color singlet baryon state. In contrast, the symmetric $d^{abc}$ operator does contribute, and early quark model studies have incorporated such a term as a phenomenological correction~\cite{Desplanques:1992rf}. Previous studies have examined how this interaction influences the nucleon mass as well as the stability of various exotic states~\cite{Pepin:2001is,wpark_dibaryon}. However, a flavor-independent quark three-body contact force simply adds a constant shift to all baryon masses, thereby failing to account for the observed flavor-dependent deviations within the baryon multiplets~\cite{Semay1999}. Moreover, the origin of the $d^{abc}$-type interaction remains obscure, which has forced practitioners to introduce \emph{ad hoc} prescriptions for any terms that involve antiquarks to ensure charge conjugation invariance. As a result, these extensions lack a unified theoretical basis and undermine the predictive power of conventional quark models.

Quark models that neglect three-body forces exhibit persistent shortcomings in reconciling meson and baryon spectra. Models fitted to the meson spectrum tend to overestimate baryon masses, and vice versa, suggesting that a missing ingredient is required to achieve a unified description of hadrons~\cite{Pepin:2001is,wpark_dibaryon,Karliner,Glozman1996,Semay1994}. For instance, the existence of a three-body term is mentioned to improve light-baryon mass predictions once the meson parameters are fixed~\cite{SilvestreBrac1996}, but it was never actually implemented. This observation highlights the need for a more general quark three-body interaction that can introduce the necessary flavor and spin dependencies to improve agreement with experimental data.

Ref.~\cite{noh2025inevitable} demonstrated that three-body forces improve the overall fit and help resolve long-standing discrepancies such as the $N-\Delta$ mass splitting (hereafter termed the ``spin splitting''). However, the flavor dependency remained. Although quark model studies have attempted for over fifty years to describe both meson and baryon spectra using a single parameter set, all widely used models have failed to reduce flavor-dependent deviations without resorting to separate refitting~\cite{DeRujula:1975,Semay1994,IsgurKarl:1978,CapstickIsgur:1986}. Whether fitting baryons and mesons separately or together, flavor-dependency always appears.

We introduce an explicit three-quark interaction into the constituent quark model and calculate baryon mass spectra.
First, we determine the two-body parameters from the ground-state meson spectrum.
We then calculate the baryon spectrum using the same parameter set and a converged three-Jacobi GEM basis containing 20 Gaussian ranges for each Jacobi coordinate together with the internal $SS$, $PP$, $DD$, and $FF$ components.
We use the resulting baryon residuals to fit the strengths and to test the proposed color--spin operator basis.
The spatially contact color--spin interaction substantially corrects the common baryon mass offset but leaves a strong flavor-dependent residual pattern.
We therefore introduce mass-scaled Yukawa and Gaussian profiles whose length scales are fixed by the reduced mass $m_{ij}$ already determined from the meson calibration.
The finite-range extension refits the same three couplings $A$, $B$, and $C$ but introduces no additional continuous range parameter.

We test the interaction in meson--baryon compatibility scans for the $u,d,s,c,b$ sectors, with alternative OGE-based two-body Hamiltonians, and on baryons outside the principal calibration set.
We also examine radial and orbital excitations to identify the domain in which a static compact valence Hamiltonian remains quantitatively reliable.
Across the fitting schemes and data sets examined here, the short-range three-quark interaction improves compact ground-state spectroscopy.
Many excited or threshold-sensitive states still require relativistic, chiral, long-distance, or coupled-channel dynamics.

This addition has important implications for the stability of exotic multiquark states and for the nature of flavor-dependent nuclear three-body forces---the latter being crucial to understanding the hyperon puzzle in neutron stars~\cite{nature1,science1}. At higher densities, many-quark interactions may also affect the properties of quark matter, although their role in the QCD phase structure remains to be clarified~\cite{AaronPark2021}.

The article is organized as follows.  Sections \ref{sec:model_and_meson}--\ref{sec:short_range} review the constituent-quark framework and develop the finite-range connected three-quark interaction.  Section~\ref{sec:udscb_benchmark} tests meson--baryon compatibility across flavor sectors and alternative OGE-based Hamiltonians,  Section~\ref{sec:excitation_spectra} benchmarks radial and orbital excitations, and  Section~\ref{sec:3bd_operators} extends the operator analysis to compact multiquark configurations. Technical conventions and numerical validations are collected in the notation guide and appendices.

\section{Quark models and QCD}\label{sec:model_and_meson}
This section reviews the QCD motivation for the main ingredients of constituent-quark models.
These models use effective constituent masses and interactions, often in a non-relativistic Hamiltonian.
We first review dynamical mass generation, from the NJL model to Dyson--Schwinger and lattice-QCD studies, and explain why a dressed-quark mass is useful in low-energy models.
We then discuss static and spin-dependent interactions, chiral interactions, coupled channels, relativistic formulations, multiquark applications, and the numerical methods used later.

\subsection{Quasiparticle picture of constituent (dressed) quarks in QCD}\label{quasiparticle_pic}

The constituent-quark picture has been useful for hadron spectra, magnetic moments, and form factors.
Lattice and continuum studies show that the quark mass function grows strongly at low momentum.
This supports a dressed-quark quasiparticle picture in low-energy models.
Constituent quarks are not physical asymptotic states, and their effective masses and interactions remain model dependent.

In Landau gauge, the gluon two-point function can be written as 
\begin{equation}\label{landau_gauge_gluon_prop}
D_{\mu\nu}(p)=\frac{d(p^2)}{p^2}\left(\delta_{\mu\nu}-\frac{p_\mu p_\nu}{p^2}\right),
\end{equation}
where $d(p^2)$ is the Landau-gauge gluon dressing function.
A convenient fit to lattice data takes the form~\cite{Iida2005}
\begin{equation}
d(p^2)=Z\frac{p^4+ap^2}{p^4+b^2}, \quad a=1.94\;\text{GeV}^2,\quad b=0.486\;\text{GeV}^2, \quad Z=1.80.
\end{equation}
This form has the following behavior: (i) As $p\to 0$, $d(p^2)\propto p^2$ indicates screening at long wavelengths (decoupling), so $D_{\mu\nu}(p)$ tends to a finite constant---IR suppression relative to $1/p^2$. (ii) It exhibits a broad enhancement around
$p\sim0.6$--$1~\mathrm{GeV}$, which amplifies the interaction strength at hadronic momenta before returning smoothly to perturbative behavior in the UV. 
In the deep infrared, this form defines an effective screening scale $m_g$ and the corresponding correlation length $m_g^{-1}$.
We use this correlation length to set the order of magnitude of the hyperfine range and take $r_0\sim1/m_g$.

The Euclidean quark two-point function is $S(p)={Z(p^2)}/[{i\slashed{p}+M(p^2)}]$, where $M(p^2)$ is the momentum-dependent mass function and $Z(p^2)$ the wave-function renormalization.\footnote{Equivalently, $S^{-1}(p)=i\slashed{p} A(p^2)+B(p^2)$ with $Z(p^2)=1/A(p^2)$ and $M(p^2)=B(p^2)/A(p^2)$.}  At high momenta, $M(p^2)$ reproduces the perturbative running current mass; toward the infrared ($p^2\to0$), non-perturbative effects dramatically enhance $M(p^2)$ to values of order a few hundred MeV. Lattice studies establish that an infrared plateau in $M(p)$ --- the ``constituent'' mass --- arises from dynamical chiral symmetry breaking. For example, in the chiral limit, Refs.~\cite{Iida2005,Bowman2002,Zhang2003,Bowman2002proc} report the empirical form
\begin{equation}
M(p^2)=\frac{M_0}{1+(p/\beta)^\gamma},
\end{equation}
with best-fit parameters $M_0 \simeq 260$ MeV, $\beta\simeq870~\text{MeV}$, and $\gamma \simeq 3$. Ref.~\cite{Skullerud2001} performed a systematic quenched study using ${\mathcal{O}}(a)$-improved Wilson fermions ($a$ denotes lattice spacing), finding $M(0)=298(8)(30)$ MeV in the chiral limit on a $16^3\times48$ lattice at $a\simeq0.1$ fm. Ref.~\cite{Bowman2005} presents the full momentum-dependent mass function $M(p^2)$ in both quenched and $2+1$-flavor (unquenched\footnote{Here ``unquenched'' has different meanings: in lattice QCD it denotes the inclusion of dynamical sea quarks, whereas in quark models it denotes explicit coupling of valence states to hadronic continua.}) simulations, showing a clear plateau in $M(p^2)$ at low momenta (indicative of a few-hundred-MeV quasiparticle mass). 
Ref.~\cite{Oliveira2019} confirmed on finer lattices ($a\approx0.06$--0.09~fm) with near-physical pion masses that $M(p)$ plateaus at 300--400~MeV for $p\lesssim500$ MeV, largely independent of the light-quark mass in the chiral regime. Removing low-lying Dirac modes or center vortices destroys this plateau, collapsing $M(p)$ back toward the small bare mass and thereby identifying the non-perturbative vacuum structure responsible for quasiparticle behavior~\cite{Kamleh2024}. As the near-zero Dirac eigenmodes (and their density) are responsible for chiral symmetry breaking, quark-mass generation is linked to chiral symmetry breaking~\cite{Banks1980}. The lattice mass function justifies modeling each light quark as a single entity with an effective mass of order $M(0)$, i.e.,~a few hundred MeV, comparable to the constituent masses used in quark models.

Similarly, the Dyson--Schwinger equation for the quark propagator contains the dressed gluon propagator together with the dressed quark--gluon vertex.
The combined interaction strength at intermediate momenta, rather than the gluon propagator alone, gives the quark a large momentum-dependent self-energy $\Sigma(p)=i\slashed{p}\Sigma_v(p^2)+\Sigma_s(p^2)$.
One then finds, with $\Sigma_v$ controlling the wave-function renormalization and $\Sigma_s$ generating the dynamical mass~\cite{Roberts1994,Langfeld1996,Bender2001},
\begin{equation}
M(p^2)=\frac{m+\Sigma_s(p^2)}{1+\Sigma_v(p^2)}
  =\frac{B(p^2)}{A(p^2)}
  \sim \text{hundreds of MeV at } p^2\to0,
\end{equation}
which defines a dominant mass scale characteristic of a massive quasiparticle.

\paragraph{Constituent masses in phenomenological models:} For this reason, constituent masses are treated as effective fit parameters.
For example, a typical parameter set used in non-relativistic AL1/Bhaduri-type potentials~\cite{SilvestreBrac1996,Bhaduri1981} is
\begin{align}
m_{u}=m_d\simeq0.33\;\text{GeV}; \quad m_s\simeq 0.60\;\text{GeV}; \quad m_c\simeq 1.87\;\text{GeV};\quad m_b\simeq 5.26\;\text{GeV},\label{params_mass_nonrel}
\end{align}
and that used in the relativized Godfrey-Isgur (GI) model~\cite{Godfrey1985} is
\begin{align}
m_{u}=m_d\simeq0.22\;\text{GeV}; \quad m_s\simeq 0.42\;\text{GeV}; \quad m_c\simeq 1.63\;\text{GeV};\quad m_b\simeq 4.98\;\text{GeV}.\label{params_mass_rel}
\end{align}
Light quarks are generally more model-dependent. In non-relativistic potential models, for example, one often introduces a constant constituent mass to absorb many unknown relativistic and vertex-structure effects. Depending on whether spin--orbit couplings, running coupling corrections, or smearing of the divergent term are included, the optimal $m_{u,d}$ can range from ${\sim}290$ MeV up to ${\sim}360$ MeV for non-relativistic quark models.
This tendency toward larger fitted constituent masses in non-relativistic models can be understood from the effective-mass description of relativistic kinetic energy~\cite{JaczkoDurand1998}, discussed below in  Section~\ref{sec:relativistic_quark_models}.
Hadronic loops can also shift the fitted constituent masses and short-range couplings.
In the present valence Hamiltonian, a smooth average part of these effects may be absorbed into the fitted parameters.

\subsection{Quark--quark interactions and low-energy QCD}\label{interquark_interaction}

We now turn to the static color-color potential extracted from rectangular Wilson loops. At short distances it is Coulombic with a running coupling; at hadronic scales it rises approximately linearly with separation distance $r$ --- i.e.,~the potential is proportional to $r$ --- signaling confinement. This behavior is well captured by the Cornell form, including an additive constant that effectively absorbs vacuum/self-energy contributions and is scheme dependent when compared to lattice determinations. Below we discuss the Coulomb term, the string tension, and the constant, together with scale setting and string breaking.

\subsubsection{Static potential}\label{VC_cornell} 
Ref.~\cite{Eichten1978} first postulated the Cornell potential in heavy quarkonium: a sum of a Coulomb term and a linear confining term. In their model of charmonium, the interquark potential was taken as
\begin{equation}\label{eq:pot_cornell}
V^C(r)=-\frac{3}{4}F^c_i F^c_j \Big(-\frac{\kappa}{r}+\sigma r-D\Big), \qquad (\kappa, \;\sigma,\;D>0)
\end{equation}
interpolating between one-gluon exchange (OGE) at small $r$ and a confining string at large $r$. Here the color factor $F_i^c F_j^c$ is included, where $F^c=\lambda^c/2$ is the SU$(3)_C$ generator, normalized by $\text{Tr}(F^aF^b)=\delta^{ab}/2$. Fitting to the observed $J/\psi,\psi'$ spectrum yielded $\kappa\approx0.48$ and a string-tension $\sigma\approx 0.18~{\rm GeV}^2$. This simple Cornell form successfully described the low-lying charmonium and bottomonium levels, explaining splittings and decay constants much better than a pure Coulomb or pure linear model alone. Indeed, most quark models include a short-range Coulomb term and a confining potential, whose strengths are determined by the color charges of the two interacting quarks. In lattice QCD, the static color interaction between heavy sources is extracted from Wilson loops.

The static quark--antiquark potential in QCD is defined via rectangular Wilson loops in the limit of large Euclidean time~\cite{Wilson1974}:
\begin{equation}\label{eq:wilson_loop}
V^C(r) = -\lim_{T\to\infty}\frac{d}{dT}\ln\langle W(r,T)\rangle,
\end{equation}
where $W(r,T)$  is the expectation value of a Wilson loop of spatial extent $r$ and temporal extent $T$. Because the Wilson loop is a closed contour, it is gauge invariant. In contrast, gauge-dependent quantities such as open Wilson lines or path-dependent static sources depend on the chosen path connecting the two heavy sources and therefore do not define a unique potential. In practice, lattice simulations extract $V^C(r)$ by computing smeared Wilson loops or temporal-link correlators (often in Coulomb gauge) to optimize ground-state overlap. The raw lattice static energy contains an additive UV-divergent self-energy, which is removed by an independent normalization convention such as $V(r_{\rm ref})=0$.
The Sommer parameters $r_0$ and $r_1$ instead set the length scale through the force conditions $r_0^2F(r_0)=1.65$~\cite{Sommer1994} and $r_1^2F(r_1)=1.0$, with $r_0\simeq0.5$ fm in QCD.
Choosing $r_{\rm ref}=r_0$ or $r_1$ is possible, but it is separate from the force-based definition of these scales. 
For example, Ref.~\cite{Bernard2000} sets $r_0\simeq 0.5$ fm and a variant $r_1$ via $r_1^2F(r_1)=1.0$ to match scales in full and quenched simulations (we will discuss further). 
Continuum extrapolations and improved lattice actions are used to control lattice-spacing effects.

In the short-distance (perturbative) regime, asymptotic freedom implies the potential is Coulombic $V(r)\simeq -\kappa/r$, with an effective coupling $\kappa\propto C_F \alpha_s(r)$ that decreases logarithmically as $r\to0$. Lattice studies confirm this behavior. For example, Ref.~\cite{Necco2002} computed the continuum static potential in pure SU(3) gauge theory for $0.05\;\text{fm}\lesssim r\lesssim 0.8\;\text{fm}$, and found excellent agreement with perturbative QCD down to $r\simeq0.15$ fm. Specifically, their continuum-extrapolated force and potential match the perturbative expansion (with the known running coupling) to within a few percent in this short-range regime. Similarly, Ref.~\cite{Bali2001} notes that lattice data at $r\lesssim0.2$ fm lie slightly above a pure Coulomb fit, indicating the expected weakening of the coupling (as also seen in Figure 4.2 of Ref.~\cite{Bali2001}). Thus, at short distance the lattice potential approaches the perturbative Coulomb form with a running coupling. The linear term is used to describe the intermediate-distance region.

In the intermediate regime (roughly 0.2--1.0~fm), lattice data exhibit a near-perfect match to the Cornell ansatz in Eq.~\eqref{eq:pot_cornell}.
For example, the quenched and 3-flavor QCD potentials are explicitly fitted to a ``Coulomb + linear'' form (with an additive constant, discussed below) and find excellent agreement (see Figure 4 in Ref.~\cite{Bernard2000}; three degenerate dynamical flavors). Likewise, Ref.~\cite{Bali2001} reviews that quenched lattice simulations consistently find a linear rise plus a Coulomb term at intermediate $r$, emphasizing that all phenomenological potentials reproducing quarkonium spectra are only slight variations of the Cornell form in the range $0.2\;\text{fm}\lesssim r\lesssim1\;\text{fm}$. Numerically, fits to lattice data give string tensions $\sigma$ of order $\text{(}430$--$ {450\;\text{MeV}\text{)}}^2$ and effective Coulomb coefficients $\kappa\sim0.3$--0.5 (in conventions where $\hbar=c=1$). For example, quenched potentials from several lattice spacings collapse onto a universal curve well described by a Cornell fit with $\kappa\simeq0.295$~\cite{Bali2001}. For the present meson-calibrated parameter set, the Coulomb coefficient is $\kappa=78.22998~\mathrm{MeV~fm}$, corresponding to $\kappa=0.3964$ in natural units.
The associated effective central coupling is $\alpha_s^{(C)}=3\kappa/4=0.2973$, which lies within the range commonly used in phenomenological Cornell-type potentials.

At large separations $r\gtrsim 1~\mathrm{fm}$, the non-perturbative confining dynamics of QCD dominates and the static potential is well described by a linear rise. In pure gauge (quenched) QCD this behavior persists indefinitely --- consistent with the area law for large Wilson loops --- and is characterized by the string tension $\sigma$. Ref.~\cite{Necco2002} shows that for $r\gtrsim 0.8\,r_0$ the continuum-extrapolated potential is described to high accuracy by the bosonic string prediction: in their Fig. 6 the lattice data closely follow the string curve over the quoted range, with only percent-level residuals. Accordingly, the long-distance slope extracted from such fits corresponds to the canonical scale $\sqrt{\sigma}\sim430~\mathrm{MeV}$, whereas the present meson calibration gives $\sigma=0.2259~\mathrm{GeV}^2$, or equivalently $\sqrt{\sigma}\simeq475~\mathrm{MeV}$.
This value is somewhat larger than the canonical Cornell scale but remains within the range encountered in phenomenological quark-model parametrizations. In full QCD (with light dynamical sea quarks), the potential is expected to eventually flatten at very large separations due to string breaking; nevertheless, lattice determinations of $V(r)$ up to distances of order 1~fm typically still exhibit a clear linear rise. Ref.~\cite{Bali2001} reviews that string-breaking effects set in only at larger separations, and that the potential extracted from standard Wilson-loop observables remains approximately linear over the $\mathcal{O}(1~\mathrm{fm})$ range commonly used in Cornell-type comparisons.

\subsubsection{The constant term and vacuum energy}\label{VC_constant}

The constant terms that appear in the color-color potential $V^C$ encode the energy difference between the true non-perturbative QCD vacuum and a perturbative-like interior. In the MIT bag model, the constant $B$ (the bag pressure) belongs to a different framework than the additive constant $-D$ in Cornell-type potentials, yet both effectively parametrize non-perturbative vacuum effects. Classically, massless QCD is scale invariant, but quantum effects break this via the trace anomaly. The QCD energy--momentum tensor satisfies~\cite{He2021}
\begin{equation}
  T^\mu{}_\mu \;=\; \frac{\beta(g)}{2g}\,G^{c}_{\mu\nu}G^{c,\mu\nu}
  \;+\; \sum_{f=1}^{n_q+n_Q} m_f\,(1+\gamma_m)\,\bar q_f q_f,
\end{equation}
where $n_q$ and $n_Q$ denote, respectively, the numbers of light and heavy flavors, $\beta(g)$ is the QCD beta function and $\gamma_m$ the mass anomalous dimension~\cite{Wang2024}. At leading order,
\begin{equation}
  \beta(g)= -\,\frac{11-(2/3)(n_q+n_Q)}{16\pi^2}\,g^3 \;+\; \mathcal{O}(g^5),
\end{equation}
and after dimensional regularization the heavy-quark condensate admits the following expansion~\cite{Generalis1984}
\begin{align}
  \big\langle \bar Q Q \big\rangle
  &= -\text{Tr}_{c,\gamma}\langle S_Q(0,0)\rangle=-\,\frac{1}{12\,m_Q}\left\langle \frac{\alpha_s}{\pi} G^2 \right\rangle
  + \mathcal{O}\Big(\frac{1}{m_Q^3}\Big).
\end{align}
Taking matrix elements, one sees that at leading order the heavy-flavor contribution $m_Q(1+\gamma_m)\langle\bar Q Q\rangle$ cancels against the per-flavor change in $(\beta/2g)\langle G^2\rangle$. Hence, to a good approximation one may drop the heavy-flavor term and evaluate the beta function with only the three light flavors, i.e.,~$n_f=3$ in $\beta(g)$. Keeping only the leading term of the beta function, the trace becomes
\begin{equation}
  \big\langle T^\mu{}_\mu \big\rangle
  \;=\; -\,\frac{9}{8}\left\langle \frac{\alpha_s}{\pi} G^2 \right\rangle
  \;+\; \sum_{f=1}^{n_q} m_f\,\big\langle \bar q_f q_f \big\rangle,
\end{equation}
so that in the light-quark chiral limit ($m_{u,d,s}\!\to\!0$) one has
$\langle T^{\mu}_{\mu}\rangle = -(9/8)\langle \alpha_s G^2/\pi \rangle$. The vacuum expectation values are nonzero due to the gluon condensate and spontaneous chiral symmetry breaking,
\begin{align}
  \left\langle \frac{\alpha_s}{\pi} G^2 \right\rangle &\;\approx\; (330~\text{MeV})^4, \qquad \langle \bar q q \rangle \;\approx\; -\,(250~\text{MeV})^3.
\end{align}
The existence of the trace anomaly together with a nonvanishing gluon condensate implies that the QCD vacuum carries a finite energy density. One may picture a hadron as a ``bubble'' of perturbative-like vacuum embedded in the true non-perturbative vacuum~\cite{He2021,Klinkhamer2009,Zhitnitsky2007}. The energy required to displace the vacuum scales with the volume, a feature captured by the MIT bag model~\cite{ChodosJaffe1974,Chodos:1974pn}:
\begin{equation}
  m_H \;=\; E_{\rm quarks} \;+\; E_{\rm fields} \;+\; B V \;-\; \frac{Z_0}{R}.
\end{equation}
Here $B$ is the bag pressure and $V$ and $R$ are the bag volume and radius, respectively. In the simplest version, the quark contribution is $E_{\rm quarks}=A/R$ (lowest spherical bag modes), $E_{\rm fields}=0$, and the last term is due to the Casimir energy. The volume term $BV$ encodes the work needed to create the bag and can be modeled as the difference between the external (true) vacuum and the interior vacuum. In terms of the trace,
\begin{equation}
  B \;\equiv\; \epsilon_{\rm in}-\epsilon_{\rm out}
  \;=\; \frac{1}{4}\Big(\big\langle T^\mu{}_\mu\big\rangle_{\rm in}
 - \big\langle T^\mu{}_\mu\big\rangle_{\rm out}\Big)
  \;\simeq\; \frac{9}{32}\left[\left\langle \frac{\alpha_s}{\pi}G^2\right\rangle_{\rm out}
  - \left\langle \frac{\alpha_s}{\pi}G^2\right\rangle_{\rm in}\right],
\end{equation}
up to light-quark mass terms.
Phenomenologically, typical values are $B^{1/4}\simeq 145$--170~MeV in hadron spectroscopy, with larger $B^{1/4}\sim 160$--200~MeV sometimes used in dense-matter EoS; precise numbers depend on scheme and active degrees of freedom. On the other hand, the gluon condensate scale is ${\langle \alpha_s G^2/\pi\rangle}^{1/4}\sim 0.33~\text{GeV}$, much larger than $B^{1/4}$, indicating that the effective interior gluon condensate is reduced by 20\%--50\% (depending on the value of $B^{1/4}=145$--200~MeV) relative to the outside (whereas the interior light-quark condensate is strongly suppressed, approaching zero)~\cite{Brown1987}. This is consistent with the general observation that even when chiral symmetry is restored (e.g.,~at high temperature), the gluon condensate remains non-vanishing~\cite{Lee1989}. 

Independently of its physical origin, in lattice QCD the constant term is intimately tied to the renormalization procedure. The expectation value of a Wilson loop $\langle W(r,T)\rangle\propto e^{-V(r)\,T}$ is computed on a spacetime lattice with UV cutoff $a^{-1}$. A static quark source then acquires a divergent self-energy $E_{\text{self}}(a)\propto 1/a$, so that the ``raw'' lattice potential takes the form $V_{\text{lat}}(r,a)=V_{\text{physical}}(r)+2E_{\text{self}}(a)$~\cite{Kawanai2016,Brambilla2022}. Because only energy differences are physical, one fixes the additive zero by choosing a reference distance $r_{\rm ref}$ and defining $V_{\rm ren}(r)=V_{\rm lat}(r,a)-V_{\rm lat}(r_{\rm ref},a)$, $V_{\rm ren}(r_{\rm ref})=0$.
Separately, the Sommer scale $r_0$ is determined from the force condition $r_0^2F(r_0)=1.65$, with $r_0\simeq0.5$ fm~\cite{Bernard2000,Sommer2014}.
One may choose $r_{\rm ref}=r_0$, but this is an additional normalization convention rather than part of the definition of $r_0$. When one fits this renormalized lattice data to the Cornell form in Eq.~\eqref{eq:pot_cornell}, the constant $-D$ no longer reflects purely vacuum energy. Instead, it combines that physics with the chosen renormalization scheme: if $V(r_0)=0$ is imposed, then by definition $-\kappa/r_0+\sigma r_0-D=0$, and changing to the alternate Sommer scale $r_1$ (defined by $r_1^2F(r_1)=1.0$) likewise shifts the fitted $D$. Thus, in lattice QCD the Cornell constant term is a scheme-dependent parameter rather than a direct measure of the vacuum energy~\cite{Bernard2000}.

Several benchmark studies illustrate this point. Ref.~\cite{Bali2001} presents a quenched QCD determination of the potential in the form given in Eq.~\eqref{eq:pot_cornell}; here $D$ depends on the lattice spacing and is fixed via the $r_0$ condition. Dynamical-quark (``unquenched'') computations, such as Ref.~\cite{Bernard2000} with three light flavors, find a deeper potential (larger $\kappa$), which in turn alters the renormalized constant $-D$. Studies of three-quark systems~\cite{Takahashi2001,Takahashi2002} fit a $Y$-shaped flux-tube potential plus a constant, finding a large 0.8--1.0 GeV offset required to form a baryon. Thus, the numerical value of $D$ depends strongly on simulation details (quenched vs unquenched, number of flavors), the system under study (quarkonium, meson, baryon), and the renormalization prescription. Representative values are summarized in Table~\ref{tab:cornell_lattice}. Phenomenologically, the spread of the fitted constant term, together with the variation of the other fitted quark model parameters, suggests that these parameters are effectively absorbing physics not explicitly encoded in a specific potential ansatz. As an illustration,  Section~\ref{sec:unquenching} shows how hadron loop effects can be incorporated effectively into the fitted parameters.
In non-relativistic models, the average energy shift from omitted relativistic kinetic effects may also be absorbed into $D$, while the remaining momentum-dependent effects can be partly accounted for by the fitted constituent masses, as shown in Table~\ref{tab:cornell_lattice} and Eqs.~\eqref{params_mass_nonrel}--\eqref{params_mass_rel}.

\begin{table}[!t]
  \centering
  \caption{\justifying Comparison of Cornell-type potential parameters in lattice QCD and phenomenological quark models.  We use the mesonic color singlet convention \(V(r)=-\kappa/r+\sigma r-D\) (natural units \(\hbar=c=1\)).  It should be noted that in lattice QCD the additive constant is renormalization/convention dependent (e.g.\ fixing \(V(r_{\rm ref})=0\)), whereas in quark models \(D\) effectively absorbs model-dependent self-energies (including average hadron-loop/unquenching effects) once constituent masses are fixed. See Section~\ref{sec:unquenching}.}
  \label{tab:cornell_lattice}
  \renewcommand{\arraystretch}{1.15}
\resizebox{\columnwidth}{!}{%
\begin{tblr}{
  colspec={
Q[l,wd=0.8cm]
Q[l,wd=3.2cm]
Q[l,wd=4.3cm]
Q[c,wd=1.6cm]
Q[c,wd=1.8cm]
Q[c,wd=3.8cm]
Q[l,wd=7.8cm]
  },
  row{1} = {bg=headgray, halign=c, valign=m},
  row{2,10} = {bg=headgray, halign=l, valign=m},
  cell{1}{1} = {}{halign=l},
  cell{1}{2} = {}{halign=l},
  cell{1}{3} = {}{halign=l},
  cell{1}{7} = {}{halign=l},
  hline{2} = {1-7}{0.3pt},
  hline{3} = {1-7}{0.3pt},
  hline{10} = {1-7}{0.3pt},
  hline{11} = {1-7}{0.3pt},
  rowsep = 1.5pt,
}
\toprule
Ref.
& $N_f$/Model
& Convention / scale / fit
& $\sigma$ (GeV$^2$)
& $\kappa$
& Const. term (GeV)
& Remarks \\

\SetCell[c=7]{l}{\textit{Lattice QCD (scheme-dependent constant)}} & & & & & & \\

\cite{Bernard2000}
& 3 (KS fermions)
& Scale via \(r_1^2F(r_1)=1.0\) (often fix potential at \(r_1\))
& \(\sim 0.18\)
& \(\sim 0.36\)
& ---
& Useful for comparing quenched vs dynamical effects in the force; constant usually not quoted as a physical number. \\

\cite{Necco2002}
& 0 (quenched, continuum)
& Typically presented up to an additive constant (often fixed at \(r_0\))
& \(\sim 0.21\)
& \(\sim 0.30\)
& (conv.-dep.)
& Good benchmark for the long-distance string regime; constant must be tied to the convention used. \\

\cite{Bali2001}
& 0 (quenched)
& Set \(V(r_0)=0\), with \(r_0\simeq0.5~\mathrm{fm}\)
& \(\sim 0.21\)
& \(\sim 0.295\)
& \(D=\sigma r_0-\kappa/r_0\) (report with chosen \(r_0\))
& When quoting \(D\), explicitly state the zero-point choice; otherwise \(D\) is not comparable across lattice papers. \\ 

\cite{Kawanai2012}
& \(2{+}1\) (PACS-CS; clover light quarks + RHQ charm)
& finite-\(m_c\) charmonium BS potential; fit to \(V(r)=-A/r+\sigma r+V_0\)
& \(0.155(6)\)
& \(\sim 0.81\text{--}0.86\)
& not directly quoted
& On-axis and full-set fits give \(A=0.861(17)\) and \(0.813(22)\), respectively; not a static Wilson-loop result. \\

\cite{Kawanai2014}
& 0 (quenched; RHQ charm)
& finite-\(m_c\) BS potential; several lattice spacings and volumes
& \(\sim 0.18\)
& \(\sim 0.35\text{--}0.44\)
& shifted const. only
& Representative charm-mass quenched BS values; the paper quotes \(V_0-E_{\rm ave}\sim -(0.32\text{--}0.38)\,\mathrm{GeV}\). \\

\cite{Takahashi2002}
& 0 (quenched)
& $3Q$ potential (not $Q\bar Q$)
& \(\sim 0.21\)
& \(\sim 0.14\)
& \(\sim 0.8\text{--}1.0\)
& Constant for the $3Q$ ($Y$-string) system; not directly comparable to the mesonic \(D\). \\

\cite{Kawanai2015}
& $2{+}1$ (PACS-CS; clover + RHQ)
& charmonium BS-potential (finite-$m_c$), Cornell fit
& \(0.162(12)\)
& \(0.713(83)\)
& const. not quoted
& Finite-mass lattice extractions (Bethe-Salpeter amplitude method; not a static Wilson-loop potential) \\

\SetCell[c=7]{l}{\textit{Quark models (constant fitted with spectrum)}} & & & & & & \\

\cite{Godfrey1985}
& GI (relativized)
& Meson fit
& \(\sim 0.18\)
& (running)
& (model-dep.)
& For a representative \(\kappa\), use an effective \(\kappa_{\rm eff}\) around \(r\sim 0.3\text{--}0.6~\mathrm{fm}\). \\

\cite{SilvestreBrac1996}
& AL1-type
& Meson + baryon fit
& \(0.1653\)
& \(0.5069\)
& \(0.8321\)
& Parameters quoted for the AL1 form; maps to \(V=-\kappa/r+\sigma r-D\) with \(\sigma=\lambda\), \(D=\Lambda\). \\

\cite{Eichten1978}
& Cornell ($Q\bar Q$)
& heavy quarkonium fit
& \(\sim 0.18\)
& \(\sim 0.48\text{--}0.52\)
& (model-dep.)
& Canonical Cornell scale: \(\sqrt{\sigma}\sim 430~\mathrm{MeV}\). \\

\cite{Barnes2005}
& GI (relativized)
& Charmonium fit
& \(0.1425\)
& \(0.7281\)
& 0.253
& Modern charmonium benchmark; the quoted central potential has no separate additive constant term. \\

\cite{BrauSemaySilvestreBrac2002}
& Semirelativistic + instanton
& Meson+baryon fit, but different constant offset
& \(0.210\)
& \(0.525\)
& \(0.691\) (mesons)
& Cornell core \( -\kappa/r + a r + C_M (+C_B \text{ in baryons})\); Baryon sector adds \(C_B=-0.033\) GeV and instanton-induced terms. \\

\cite{noh2025inevitable}
& Non-rel. OGE-based
& Meson fit
& \(0.221\)
& \(0.456\)
& \(1.084\)
& Phenomenological fit (variational method). \\

Present
& Non-rel. OGE-based
& Meson fit
& \(0.2259\)
& \(0.3964\)
& \(1.0982\)
& Phenomenological fit (variational method; GEM). \\
\bottomrule
\end{tblr}}
\end{table}

\subsubsection{Static three-quark potential}\label{sec:static_3bd}

Lattice QCD indicates that the long-distance ground-state static potential of three infinitely heavy quarks is well described by a ``Coulomb {\ensuremath{+}} $Y$-string'' ansatz~\cite{Takahashi2001,Takahashi2002,deForcrand2005}.
Using a gauge-invariant three-quark Wilson loop with ground-state enhancement, the potential is parametrized as
\begin{equation}\label{eq:Yansatz}
V_{3Q}(\{\mathbf{r}_i\})
= -\sum_{i<j}\frac{\kappa_{3Q}}{r_{ij}}
  + \sigma_{3Q}L_{\rm min}^{(Y)}
  - D_{3Q},
\end{equation}
where $r_{ij}=\vert \mathbf r_i-\mathbf r_j\vert $.
The quantity $L_{\rm min}^{(Y)}$ is the minimum total length of the flux-tube network joining the three quarks.
When all internal angles are smaller than $2\pi/3$, the three flux tubes meet at a Steiner point with angles $2\pi/3$.
The parameter $\sigma_{3Q}$ is the corresponding string tension, while $D_{3Q}$ is an additive convention-dependent vacuum constant.
In quenched SU(3) calculations, the $Y$-ansatz reproduces the static-potential data at the few-percent level.
The fitted parameters also satisfy $\sigma_{3Q}\simeq\sigma_{Q\bar Q}$ and $\kappa_{3Q}\simeq\kappa_{Q\bar Q}/2$ within the adopted conventions~\cite{Takahashi2001,Takahashi2002}.

Earlier lattice analyses found that a pairwise $\Delta$-ansatz also describes the available data rather well over the distance range considered~\cite{Alexandrou2002}.
Writing
\begin{equation}
L_{\Delta}\equiv\frac{1}{2}\left(r_{12}+r_{23}+r_{31}\right),
\end{equation}
the difference between $L_{\rm min}^{(Y)}$ and $L_{\Delta}$ is at most about 15\% for the relevant triangle geometries.
The pairwise $\Delta$ form is therefore a useful approximation at short and intermediate distances, although it does not represent the asymptotic ground-state flux-tube geometry.
Direct lattice studies of the flux distribution favor a $Y$-shaped configuration at large distance~\cite{deForcrand2005,Suganuma2004}.

In quark-model calculations, the $Y$-string contribution is often replaced by the simpler pairwise $\Delta$ form.
This replacement should be regarded as a numerical approximation to the static confinement energy rather than as an operator identity.
At short distance, perturbative QCD also contains a genuine three-line contribution to the static three-quark potential, which first appears at NNLO in addition to the pairwise terms~\cite{BrambillaGhiglieriVairo2010}.
The long-distance $Y$-string term and the perturbative NNLO three-line term have different origins, and neither is identified here with the finite-range color-spin three-quark interaction introduced later.

\subsection{The color-spin and other spin-dependent interactions}\label{VCS_intro}
The hyperfine (chromomagnetic, or color-spin) interaction splits hadrons that differ only in their spin coupling. For instance, if one ignores spin-dependent interactions in the quark model, the proton ($\mathcal{J}^P=1/2^+$) and the $\Delta$ ($\mathcal{J}^P=3/2^+$) will have a degenerate mass. In reality, their observed mass difference is about 300~MeV, which represents a substantial fraction of the baryon mass. The color-spin interaction can also be systematically derived starting from the heavy-quark limit. At leading order it arises from one-gluon exchange and is analogous to the Breit-Fermi magnetic interaction in QED.  In lattice QCD one computes it by inserting chromoelectric/magnetic field operators into Wilson loops, and in effective field theory it appears as a higher-order operator (the $\boldsymbol{\sigma}\cdot\mathbf{B}$ term) in NRQCD/pNRQCD.

\subsubsection{Spin-dependent interactions in the one-gluon exchange (OGE)} \label{sec:OGE_color_spin}

In quark models the potential is usually written as
\begin{equation}
  V(\mathbf{r})
  =
  V^C(\mathbf{r})
  +
V^{CS}(\mathbf{r})
+ V^T(\mathbf{r})
+ V^{LS}(\mathbf{r})
+ V^{ALS}(\mathbf{r})
  + \cdots ,
\end{equation}
where $V^C$ is the spin-independent static potential, $V^{CS}$ the color-spin (hyperfine) interaction, $V^T$ the tensor force, and $V^{LS}$, $V^{ALS}$ the symmetric and antisymmetric spin--orbit terms. These structures follow from a Pauli (Foldy-Wouthuysen) reduction of the OGE with a vector current $\gamma_\mu\otimes\gamma^\mu$ and a subsequent Fourier transform to coordinate space~\cite{Godfrey1985,Eichten1978,Eichten1981}. We write the full central potential as
\begin{equation}
  V_{ij}^C(r_{ij}) = -\frac{3}{4}F_i^c F_j^c\,[V_V(r_{ij}) + V_S(r_{ij})],
\end{equation}
with $V_V$ the vector piece and $V_S$ a phenomenological scalar contribution usually associated with confinement~\cite{Godfrey1985,Eichten1981}. For a central vector potential (the timelike vector piece of the interaction),
\begin{equation}
  V_V(r_{ij}) = -\frac{4}{3}\frac{\alpha_s(r_{ij})}{r_{ij}},
\end{equation}
which represents the central part of one gluon exchange entering the vector component of the Cornell potential.

Matching OGE to the Breit-Fermi Hamiltonian (Pauli reduction followed by a Fourier transform) yields a spin-spin interaction that splits into a contact and a tensor piece:
\begin{align}\label{V_CS_cont}
  V^{CS}_{\rm cont}(\mathbf r_{ij})
  = -\,\frac{8\pi\,\alpha_s}{3\,m_i m_j}\,
 F_i^cF_j^c\;\mathbf{S}_i\cdot\mathbf{S}_j\,
 \delta^{(3)}(\mathbf r_{ij}), \qquad V^{T}(\mathbf r_{ij})
  = -\frac{\alpha_s}{m_i m_j}\,
 F_i^c F_j^c\,
 \frac{1}{r_{ij}^3}\left[
   3(\mathbf{S}_i\cdot\hat{\mathbf r}_{ij})
(\mathbf{S}_j\cdot\hat{\mathbf r}_{ij})
   -\mathbf{S}_i\cdot\mathbf{S}_j
 \right],
\end{align}
with $\mathbf{S}_i=\boldsymbol{\sigma}_i/2$ the spin of the $i$th quark and $\hat{\mathbf r}_{ij}=\mathbf r_{ij}/r_{ij}$. The diagonal tensor matrix element vanishes in a pure $L=0$ state, but the tensor operator can couple $L=0$ and $L=2$ components~\cite{Godfrey1985,Eichten1981}.
The hyperfine splitting between vector and pseudoscalar quarkonia is therefore dominated by $V^{CS}_{\rm cont}$, while the tensor interaction contributes through components with nonzero orbital angular momentum.
This conventional tensor-induced $S$--$D$ mixing should be distinguished from the internal $SS\leftrightarrow DD$ correlation generated by the $\lambda=2$ multipole of a cross-pair central interaction in Eq.~\eqref{eq:cross_pair_multipole_expansion}.  For a color singlet $Q\bar Q$ state, $\langle F_i^c F_j^c\rangle=-4/3$ and one recovers the standard estimate
\begin{equation}
  \Delta_{\rm HF}({}^3S_1-{}^1S_0)
  \simeq \frac{32\pi\,\alpha_s}{9\,m_Q^2}\,|\psi(0)|^2,
\end{equation}
which reproduces the observed $J/\psi-\eta_c$ and $\Upsilon-\eta_b$ splittings for reasonable choices of $\alpha_s$ and $\vert \psi(0)\vert ^2$~\cite{Eichten1978}.  The form $\propto\delta^{(3)}(\mathbf r)$ is strictly justified in the heavy-quark regime, where the typical separation $r\sim 1/(m_Q\alpha_s)$ is small and long-range non-perturbative effects are suppressed; for light quarks one replaces the delta function by a smeared short-range interaction (often Gaussian) to mimic finite-size and relativistic effects.

The same Breit-Fermi reduction also generates spin--orbit terms. The spin--orbit interaction then splits into a symmetric and an antisymmetric piece,
\begin{align}
V_{ij}^{LS}(r_{ij})
= -\frac{3}{4}F_i^c F_j^c\,
   A_{ij}(r_{ij})\,
   \mathbf{L}_{ij}\cdot(\mathbf{S}_i+\mathbf{S}_j),\qquad V_{ij}^{ALS}(r_{ij}) = -\frac{3}{4}F_i^c F_j^c\,
   B_{ij}(r_{ij})\,
   \mathbf{L}_{ij}\cdot(\mathbf{S}_i-\mathbf{S}_j),
\end{align}
where $\mathbf{L}_{ij}=\mathbf r_{ij}\times(-i\nabla_{ij})$. In the Pauli reduction, the symmetric coefficient $A_{ij}(r)$ contains terms proportional to $1/m_i^2+1/m_j^2$ and $1/(m_i m_j)$. The antisymmetric coefficient $B_{ij}(r)$ is proportional to $1/m_i^2-1/m_j^2$.  In particular, for equal-mass constituents the antisymmetric term vanishes, and the net spin--orbit force is governed by the balance between the vector and scalar contributions.  This vector-scalar balance is phenomenologically nontrivial: in a pure short-range vector exchange picture, the spin--orbit force is typically too large, whereas the observed $P$-wave fine structure of quarkonia requires a substantially reduced net $\mathbf L\cdot\mathbf S$ interaction.  In potential models this reduction is achieved by an inverted spin--orbit contribution associated with the confining interaction, which is kinematically inevitable as a Thomas-precession term and is often modeled effectively by a Lorentz-scalar potential: it enters with the opposite sign to the OGE spin--orbit piece, leading to a partial cancellation in the coefficient $A_{ij}(r)\propto V_V'(r)-V_S'(r)$.  This cancellation helps keep the ${}^3P_J$ splittings at the observed ($\mathcal{O}(10)$-MeV) scale and reproduces the characteristic ordering ${}^3P_0<{}^3P_1<{}^3P_2$ in heavy quarkonia, while simultaneously explaining why the splittings are much smaller than naive OGE matrix elements would suggest in light $q\bar q$ mesons.  As one goes from $b\bar b$ to $c\bar c$ and further to light quarks, the $P$-wave radii increase and the relative weight of the long-range (Thomas/confining) piece grows, making the cancellation progressively more important~\cite{Godfrey1985,Eichten1981}. So, OGE alone cannot describe the fine structure; the Thomas-precession term from confinement must be included.

\paragraph{Link to the non-perturbative gluon propagator:} The contact form of the hyperfine interaction follows from the Breit-Fermi identity for vector exchange (color singlet normalized)~\cite{Godfrey1985,Eichten1981},
\begin{equation}\label{Breit_Fermi_idd}
V^{SS}(\mathbf{r})=\frac{2}{3m_im_j}\mathbf{S}_i\cdot \mathbf{S}_j \nabla^2 V_V(\mathbf{r}),\qquad V_V(\mathbf{r})=-\frac{16\pi}{3} \int\frac{d^3\mathbf{k}}{(2\pi)^3}e^{i\mathbf{k}\cdot \mathbf{r}}\alpha_s(\mathbf{k}^2)D(\mathbf{k}^2),
\end{equation}
where we define $V^{SS}$ from $V^{CS} = (-3/4)F_i^c F_j^c \, V^{SS}$ for notational convenience, and $D(\mathbf{k}^2)$ is the (static) gluon propagator. 
In perturbation theory $D(\mathbf{k}^2)=1/\mathbf{k}^2$, so $-\nabla^2(1/r)=4\pi \delta^{(3)}(\mathbf{r})$ gives the usual contact term. In Eq.~\eqref{Breit_Fermi_idd} the function $V_V(r)$ denotes the central color-Coulomb potential: since $V_V(r) =-(4/3)\alpha_s/r$, $\nabla^2 V_V(r)$ reproduces the contact term. If instead we use the IR-dressed propagator, the $\delta$-function is automatically resolved into a finite-range interaction. For the following decoupling-type propagator~\cite{Aguilar2008,Boucaud2008}, one obtains
\begin{equation}
D(\mathbf{k}^2)=\frac{Z(0)}{\mathbf{k}^2+m_g^2}, \qquad\Rightarrow \qquad V^{SS}(\mathbf{r})=\frac{32\pi \alpha_s Z(0)}{9m_im_j}\,\mathbf{S}_i\cdot \mathbf{S}_j\,\Big[\delta^{(3)}(\mathbf{r})-\frac{m_g^2}{4\pi}\frac{e^{-m_g r}}{r}\Big], 
\end{equation}
since $\nabla^2(e^{-m_g r}/r)=m_g^2e^{-m_g r}/r\,-4\pi\delta^{(3)}(\mathbf{r})$. The Yukawa-like piece is a normalized short-range smearing function,
\begin{equation}
\delta^{(3)}_{m_g}(\mathbf{r})\equiv \frac{m_g^2}{4\pi}\frac{e^{-m_g r}}{r},\qquad \int d^3\mathbf{r}\,\delta^{(3)}_{m_g}(\mathbf{r})=1,
\end{equation}
with width $r_0\sim1/m_g$ ($r_0\simeq0.3$--0.4~fm for $m_g\simeq0.5$--0.7 GeV). Thus, replacing $\delta^{(3)}(\mathbf{r})$ by a finite-range smear is nothing but using the non-perturbative propagator in the OGE. Likewise, if one employs a Gaussian-IR interaction (e.g.,~$\alpha_s D\propto e^{-\mathbf{k}^2/w^2}$ as in Maris-Tandy-type ansatz~\cite{Maris1999}), the Fourier transform yields a Gaussian smear $\delta_{w}^{(3)}(\mathbf{r})\propto w^3 e^{-(wr)^2/4}$, again fixing the smearing length $r_0\sim 1/w$.

In the gauge-invariant pNRQCD/field-correlator formulation, the same physics appears via the chromomagnetic correlator $\langle B_i^c(x)B_i^c(0)\rangle$~\cite{Brambilla2000}, which decays with a correlation length $\lambda_B\sim 0.2$--0.3~fm~\cite{Giacomo2002}; inserting this into the Eichten-Feinberg expressions reproduces a Yukawa- or Gaussian-type radial dependence for $V^{SS}$. In all cases the smearing scale is set by the IR structure of the gluon two-point function, or equivalently the $BB$ correlator, rather than by an \emph{ad hoc} regulator. The same finite-range logic motivates the Yukawa- and Gaussian-type profiles introduced for the connected three-quark interaction in  Section~\ref{sec:short_range}.

\subsubsection{Spin-dependent potentials from lattice QCD and quark-model phenomenology}

Lattice QCD provides a first-principles determination of the spin-dependent corrections to the static quark--antiquark potential. In practice one computes correlators of the quark--antiquark Wilson loop with insertions of chromomagnetic/electric field operators. For example, one can attach a pair of field-strength insertions (at relative time separation) to the static quark--antiquark source; the correlators yield the spin--orbit, tensor and spin-spin potentials in integral form~\cite{Koma2007,Koike1987}. Early lattice studies (1980s--90s) qualitatively confirmed that the spin--orbit potential contains a long-range confining piece (linked to the string tension) while the tensor and spin-spin parts are short-ranged, as expected from one-gluon exchange. However, quantitatively extracting these small potentials was challenging. Ref.~\cite{Bali2001} reviews that initial results for the difference $V'_2-V'_1$ (related to spin--orbit; to be defined below) were a factor 3--4 below perturbative expectation, due to lattice renormalization effects. This is remedied by renormalizing the field insertions (e.g.,~Huntley-Michael improvement; see Refs.~\cite{Campbell1988,Huntly1987}): then the Gromes relation (a consequence of Lorentz invariance linking spin--orbit forces to the static potential; for detail, see Ref.~\cite{Gromes1984}) is satisfied to a few percent. With these improvements, Ref.~\cite{Koma2007} obtained high-precision lattice data for all spin-dependent potentials up to intermediate distances $r\sim0.6$ fm. The spin-dependent potential is summarized in the form:
\begin{align}\label{LQCD_SD_pot}
V_{12}(r)=&\bigg(\frac{\mathbf{L}_1\cdot \mathbf{S}_1}{m_1^2}-\frac{\mathbf{L}_2\cdot \mathbf{S}_2}{m_2^2}\bigg)\frac{(2c_F^{(+)}-1)V_0'(r)+2c_F^{(+)}V_1'(r)}{2r} + \left(\frac{\mathbf{L}_1\cdot \mathbf{S}_2}{m_1m_2}-\frac{\mathbf{L}_2\cdot \mathbf{S}_1}{m_1m_2}\right)\frac{c_F^{(+)}V_2'(r)}{r}\nonumber\\
&+\bigg(\frac{\mathbf{L}_1\cdot \mathbf{S}_1}{m_1^2}+\frac{\mathbf{L}_2\cdot \mathbf{S}_2}{m_2^2}\bigg)\frac{c_F^{(-)}(V_0'(r)+V_1'(r))}{r} + \left(\frac{\mathbf{L}_1\cdot \mathbf{S}_2}{m_1m_2}+\frac{\mathbf{L}_2\cdot \mathbf{S}_1}{m_1m_2}\right)\frac{c_F^{(-)}V_2'(r)}{r} \nonumber\\
&+\frac{c_F^{(1)}c_F^{(2)}V_3(r)}{m_1m_2}\bigg(\frac{(\mathbf{S}_1\cdot \mathbf{r})(\mathbf{S}_2\cdot \mathbf{r})}{r^2}-\frac{\mathbf{S}_1\cdot\mathbf{S}_2}{3}\bigg)+\frac{\mathbf{S}_1\cdot\mathbf{S}_2}{3m_1m_2}\left(c_F^{(1)}c_F^{(2)}V_4(r)-48\pi \alpha_s d_vC_F\delta^3(\mathbf{r})\right),
\end{align}
where $\mathbf{r}$ denotes the interquark distance, $c_F^{(i)}(\mu,m_i)\;(i=1,2)$ is the matching coefficient in the (p)NRQCD Lagrangian which multiplies the term $\boldsymbol{\sigma}\cdot\mathbf{B}/2m_i$, whose scale is schematically in the form of $c_F(\mu)=1+\mathcal{O}(\alpha_s(\mu))$. Here $c_F^{(\pm)}=(c_F^{(1)}\pm c_F^{(2)})/2$. For equal masses, $c_F^{(-)}$ vanishes. $C_F=4/3$ is the Casimir charge of the fundamental representation. $V_0$ denotes a spin-independent static potential (e.g.,~modeled by Cornell form). Here $d_v$ denotes the Wilson coefficient of the local four-quark operator in the (p)NRQCD description. All four of the spin-dependent corrections up to $\mathcal{O}(1/m^2)$ in the heavy-quark potential --- namely the two spin--orbit terms $V_1'(r)$, $V_2'(r)$, the tensor term $V_3(r)$, and the hyperfine term $V_4(r)$ --- have been determined and are found to exhibit the following characteristic behaviors: 
\begin{itemize}
\item $V'_1(r)$: Negative for all $r\gtrsim0.25$ fm. Approximately constant (plateau) beyond about 0.25~fm, indicating a long-range, Lorentz-scalar-like confining contribution. A constant fit gives $V_1'(r)\simeq-\sigma_{v1}$, $\sigma_{v1}a^2=0.0362(4)$ with $\chi^2/N_{\text{d.o.f.}}=0.13$.
\item $V_2'(r)$: Positive, monotonically decreasing with $r$, with a finite tail persisting to $r\sim 0.6$ fm. Well described by a Coulomb-plus-string tail: $V_2'(r)\simeq \sigma_{v2}+c/r^2$, $\sigma_{v2}a^2=0.0070(7)$, $c=0.288(7)$ with $\chi^2/N_{\text{d.o.f.}}=0.22$. This long-range piece supplies the remainder of the static force when combined with $V_1'$ via Gromes relation $V_2'-V_1'=V_0'$.
\item $V_3(r)$: Positive at short distances, falling off rapidly. Best fit by a power law slightly shallower than $1/r^3$ as $V_3(r)\simeq3c'/r^p$, $p=2.80(6)$, $c'a^{p-3}=0.171(10)$ with $\chi^2/N_{\text{d.o.f.}}=0.79$, rather than the pure one-gluon-exchange form $3/r^3$.
\item $V_4(r)$: Negative at very short distances and close to zero for $r\gtrsim0.2$ fm.
The authors tested the exploratory form $V_4(r)\simeq-g'm_g^2e^{-m_gr}/r+4\sigma_{v4}/r$, consisting of a negative pseudoscalar-like Yukawa contribution and a small $1/r$ term~\cite{Koma2007}.
In the quenched calculation, $m_g=2.47~\mathrm{GeV}$ was fixed to a pseudoscalar glueball mass rather than to a pion or a gluelump scale.
Here $m_g$ follows the notation of Ref.~\cite{Koma2007} and is unrelated to the effective infrared screening mass discussed above.
The fit gives $g'=0.292(12)$ and $\sigma_{v4}a^2=0.0015(3)$ with $\chi_{\rm min}^2/N_{\rm d.o.f.}=5.1$.
The fit improves substantially when $m_g$ is treated as a free parameter, so this form should be regarded as a possible parametrization of the short-distance behavior rather than as a determination of a gluonic mass scale.
\end{itemize}
Together, these potentials satisfy the Gromes relation to about 8\% accuracy at intermediate distances in this quenched calculation.
The lattice results support a long-range contribution in the spin--orbit sector and mainly short-range behavior for the tensor and hyperfine potentials.
The clearest non-perturbative long-range contribution appears in the spin--orbit potentials, while the fitted form of $V_4$ allows a small nonzero tail.
These lattice potentials can be used as inputs to phenomenological quark models.

In non-relativistic quark models the spin-spin (hyperfine) interaction governed by $V_4$ and its associated contact term naturally emerge as the dominant spin-dependent interaction for $S$-wave ground states, while the spin--orbit terms $V_1,\; V_2$ and the tensor term $V_3$ contribute to the masses of excited states. As a result, quark model practitioners often treat $V_4$ as the dominant spin correction and include $V_1, \;V_2$, and $V_3$ only as smaller perturbations needed to reproduce fine and tensor splittings involving nonzero orbital modes. In addition to the Cornell (static) part, often-used phenomenological $V^{CS}$ potentials (instead of $V_4$) are
\begin{equation}\label{VCS_form_two_types}
V^{CS}_{ij}(r)=-\frac{3}{4}\frac{\mu_{ij}\mu_{ij}'}{m_i m_j c^4}\frac{\hbar c}{r}e^{-\mu_{ij}r/\hbar c}F_i^cF_j^c \boldsymbol{\sigma}_i\cdot \boldsymbol{\sigma}_j\quad \text{or} \quad  -\frac{3}{4}\frac{\mu_{ij}\mu_{ij}'}{m_i m_j c^4 }\frac{\hbar c}{r}e^{-(\mu_{ij}r/\hbar c)^2} F_i^cF_j^c \boldsymbol{\sigma}_i\cdot \boldsymbol{\sigma}_j,
\end{equation}
where $r$ is the interquark distance and the parameters $\mu_{ij}$ and $\mu_{ij}'$ are parametrized as
\begin{equation}\label{eq:VCS_range_parameters}
\mu_{ij}=\left(\alpha+\frac{\beta\; m_i\,m_j}{m_i+m_j}\right),\quad \mu'_{ij}=\left(\gamma+\frac{\delta\;m_i\,m_j}{m_i+m_j}\right).
\end{equation}
The parameters $\alpha$, $\beta$, $\gamma$, and $\delta$ are phenomenological fit constants.
They include unresolved higher-order, relativistic, and non-perturbative effects in the color-spin interaction.
Their mass dependence allows the same functional form to describe hyperfine splittings in both light and heavy mesons.

In the heavy-quark limit, the regulators in Eq.~\eqref{VCS_form_two_types} reduce the spatial part toward a contact interaction.
We define the reduced mass as $m_{ij}\equiv {2m_i m_j}/({m_i+m_j})$, with the factor of 2 from the Jacobi-coordinate convention in Eq.~\eqref{Jacobi_coord}. In the limit $m_{ij}\to\infty$ (i.e.,~$m_i,m_j\to\infty$ at fixed mass ratio), the smearing radius shrinks as $(1/\mu_{ij})\to0$ provided $\beta>0$. Using the standard distributional identities, one finds

\begin{align}\label{approaches_to_delta_general}
  &\frac{\mu_{ij} \mu_{ij}'}{m_i m_j c^4}\frac{\hbar c}{r}e^{-\mu_{ij}r/\hbar c}
  \xrightarrow[m_{ij}\to\infty]{}
  \frac{4\pi(\hbar c)^3}{m_i m_j c^4}\bigg(\frac{\mu_{ij}'}{\mu_{ij}}\bigg)\delta^{(3)}(\mathbf{r}), \nonumber\\
  &\frac{\mu_{ij} \mu_{ij}'}{m_i m_j c^4}\frac{\hbar c}{r}e^{-(\mu_{ij}r/\hbar c)^2}
  \xrightarrow[m_{ij}\to\infty]{}
  \frac{2\pi(\hbar c)^3}{m_i m_j c^4}\bigg(\frac{\mu_{ij}'}{\mu_{ij}}\bigg)\delta^{(3)}(\mathbf{r}).
\end{align}
The delta-function coefficient is fixed by the heavy-mass limit of $(\mu_{ij}'/\mu_{ij})\to\delta/\beta$. 
Accordingly, $V^{CS}_{ij}$ approaches a contact color-spin operator with the usual $1/(m_i m_j)$ coefficient, while the increasing reduced mass makes the interaction progressively shorter-ranged through $1/\mu_{ij}\sim 2/(\beta m_{ij})$ for heavy pairs. The net hyperfine splitting will then be determined by the short-distance coefficient and the wave function near the origin (short-range nature). Comparing Eq.~\eqref{approaches_to_delta_general} with the standard OGE contact term in the convention of Eq.~\eqref{V_CS_cont}, one finds $\delta/\beta \rightarrow 2\alpha_s/9$ for the Yukawa regulator, $\delta/\beta\rightarrow 4\alpha_s/9$ for the Gaussian regulator.

\subsubsection{Static and spin-dependent potentials from (p)NRQCD}\label{sec:pNRQCD_spin_dep}

Although a quark-model potential cannot be derived from QCD in a closed form, the heavy-quark limit admits a precise EFT formulation in which a potential description becomes meaningful.  In this regime pNRQCD clarifies (i) where a ``potential'' enters as a Wilson coefficient after integrating out hard and soft modes, and (ii) which operator structures are guaranteed to reproduce the familiar spin-dependent interactions used in spectroscopy.  This provides an EFT interpretation of the dominant two-body interaction employed in phenomenological Hamiltonians, while making explicit which pieces are short-distance, which are ultrasoft, and which are genuinely non-perturbative.

For a heavy $Q\bar Q$ system the standard scale hierarchy reads $m \gg mv \gg mv^2$, where $m$ is the hard scale, $mv$ the soft relative momentum scale, and $mv^2$ the ultrasoft (US) binding-energy scale.  Matching QCD $\to$ NRQCD integrates out hard modes $\sim m$ and yields a non-relativistic operator expansion with short-distance Wilson coefficients.  Matching NRQCD $\to$ pNRQCD then integrates out soft modes $\sim mv$; at this step the color singlet and color-octet ``potentials'' appear as matching coefficients multiplying bilinears of the corresponding $Q\bar Q$ fields.  Ultrasoft modes at $\sim mv^2$ remain dynamical and contribute through explicit US interactions organized in a multipole expansion rather than being absorbed into the static potential~\cite{Brambilla2005}.

At leading order in the multipole expansion, the pNRQCD Lagrangian takes the schematic form
\begin{equation}
\label{eq:pNRQCD_LO}
\mathcal{L}_{\text{pNRQCD}}^{\text{LO}} = \int d^3\mathbf{r}\; \bigg[ S^\dagger \bigg( i\partial_0 - \frac{\mathbf{p}^2}{m} - V_\mathbf{1}(r;\mu) \bigg) S + O^\dagger \bigg( iD_0 - \frac{\mathbf{p}^2}{m} - V_\mathbf{8}(r;\mu) \bigg) O \bigg] +\mathcal{L}_{\text{US}},
\end{equation}
where $S$ and $O$ denote the singlet and octet $Q\bar Q$ fields (with relative coordinate $\mathbf{r}$), $\mathbf{p}\equiv -i\nabla_{\mathbf{r}}$, and $D_0$ is the ultrasoft covariant derivative acting on the octet. Beyond Eq.~\eqref{eq:pNRQCD_LO}, the multipole expansion introduces chromoelectric dipole couplings ($\propto \mathbf r\cdot \mathbf E$) that generate the US effects~\cite{Brambilla2005}.

From this viewpoint, the singlet static potential follows from QCD itself. The large-time behavior of a rectangular Wilson loop defines the static energy, which in pNRQCD is organized into a matching coefficient (the singlet potential) plus US contributions,
\begin{equation}
E_{\rm stat}(r)=V_\mathbf{1}(r;\mu)+\delta E_{\rm us}(r;\mu),
\end{equation}
with the $\mu$ dependence canceling in the sum~\cite{Brambilla2005}.  Accordingly, phenomenological potentials may be viewed as parametrizations of this well-defined QCD observable (or of a correspondingly defined static potential in a specified scheme) in the distance range relevant to spectroscopy.

At short distances, pNRQCD gives a more precise identification of the coupling entering the vector potential.
In the potential scheme, the static singlet potential may be written as
\begin{equation}
\label{eq:alphaV_short}
V_\mathbf{1}(r;\mu) = -\frac{C_F\,\alpha_V(1/r;\mu)}{r},
\end{equation}
where $\alpha_V$ has a perturbative expansion in $\alpha_s$.
The Coulomb term $-\kappa/r$ of a quark model can therefore be identified with this short-distance behavior, with $\kappa\simeq C_F\alpha_V$ in the weak-coupling regime.
The perturbative relation between $\alpha_V$ and $\alpha_s$ is known to high loop order~\cite{Anzai2010,Smirnov2010}.

Spin-dependent potentials give a connection to standard quark-model practice.  In addition to the scale-separation hierarchy $m\gg mv\gg mv^2$ that underlies the construction of pNRQCD, there is an independent relativistic (Breit-Fermi reduced) expansion in inverse powers of the heavy-quark mass.  Accordingly, the pNRQCD Hamiltonian (or, equivalently, the potential sector of the EFT) is organized as
\begin{equation}
V(r)=V^{(0)}(r)+\frac{V^{(1)}(r)}{m}+\frac{V^{(2)}(r)}{m^2}+\cdots,
\end{equation}
where the heavy-mass limit defines the leading static interaction: the singlet static potential $V_{\mathbf{1}}(r;\mu)$ belongs to $V^{(0)}(r)$.  The familiar fine- and hyperfine-structure operators used in spectroscopy arise at ${\mathcal{O}}(1/m^2)$, encoded in spin-dependent potentials (spin-spin, spin--orbit, tensor, \emph{etc.}). In pNRQCD these terms are determined by NRQCD Wilson coefficients (such as $c_F$) and by field-strength insertions in Wilson loops~\cite{Brambilla2005}. The complete ${\mathcal{O}}(1/m^2)$ spin-dependent sector is therefore more general than the compact forms commonly used in spectroscopy. For contact with standard quark model Hamiltonians, it is sufficient to display the short-distance pure-vector limit, in which the familiar Breit-Fermi structures are recovered. In that limit, the spin-spin (hyperfine), spin--orbit, and tensor interactions can be written schematically as (for equal mass $m_1=m_2=m$)
\begin{align}\label{VCSgen}
V^{SS}(r)
&= \frac{2\,c_F^{(1)}(\mu)\,c_F^{(2)}(\mu)}{3m^2}\,
\nabla^2 V_V(r)\;\mathbf{S}_1\cdot\mathbf{S}_2,
\qquad
\nabla^2\left(-\frac{C_F\alpha_s}{r}\right)
= 4\pi C_F\alpha_s\,\delta^{(3)}(\mathbf{r}),
\\
V^{LS}(r)
&= \frac{3}{2m^2}\,
\frac{1}{r}\frac{dV_V(r)}{dr}\;\mathbf{L}\cdot\mathbf{S},
\qquad
\frac{1}{r}\frac{d}{dr}\left(-\frac{C_F\alpha_s}{r}\right)
= \frac{C_F\alpha_s}{r^3},
\nonumber\\
V^{T}(r)
&= \frac{c_F^{(1)}(\mu)\,c_F^{(2)}(\mu)}{3m^2}\,
\left(\frac{1}{r}\frac{dV_V(r)}{dr}-\frac{d^2V_V(r)}{dr^2}\right)\;S_{12}^{(S)},
\qquad
\left(\frac{1}{r}\frac{d}{dr}-\frac{d^2}{dr^2}\right)\left(-\frac{C_F\alpha_s}{r}\right)
= \frac{3C_F\alpha_s}{r^3}. \nonumber
\end{align}
where $\mathbf{S}\equiv \mathbf{S}_1+\mathbf{S}_2$ and $S_{12}^{(S)}\equiv 3(\mathbf{S}_1\cdot\hat{\mathbf r})(\mathbf{S}_2\cdot\hat{\mathbf r})-\mathbf{S}_1\cdot\mathbf{S}_2$ denotes the tensor operator in the spin-operator convention. In the Pauli-matrix convention $S_{12}=4S_{12}^{(S)}$.
This reproduces the familiar contact term in the Coulombic limit.\footnote{In phenomenological Hamiltonians the $\delta^{(3)}(\mathbf r)$ structure is usually regulated by a finite-range function as in Eq.~\eqref{VCS_form_two_types}. This is a model regularization of unresolved short-distance structure and should not be identified directly with the pNRQCD factorization scale $\mu$.}   For unequal quark masses $m_1\neq m_2$, the spin--orbit sector generically splits into a symmetric and an antisymmetric component, $V^{LS}(r)=V^{LS}_{\rm S}(r)\,\mathbf L\cdot\mathbf S+V^{LS}_{\rm A}(r)\,\mathbf L\cdot(\mathbf S_1-\mathbf S_2)$, with $V^{LS}_{\rm A}\propto (1/m_1^2-1/m_2^2)$; the latter (ALS) term vanishes for $m_1=m_2$ but can be phenomenologically important in heavy-light systems, where it induces mixing between states with the same $\mathcal{J}^P$, e.g.,~${}^1P_1$--${}^3P_1$.  In many quark-model realizations the short-distance regularization is most cleanly implemented at the level of the underlying vector potential, e.g.,~by replacing $1/r$ with a regulated smearing function. Then the regulated spin-spin, spin--orbit, and tensor interactions follow automatically by applying the corresponding derivative operators, $V^{SS}\propto \nabla^2 V_V$, $V^{LS}\propto (1/r)\,dV_V/dr$, and $V^{T}\propto \bigl[(1/r)\,d/dr-d^2/dr^2\bigr]V_V$.  In this way, we obtain the regulated spin-spin, spin--orbit, and tensor terms used in the phenomenological Hamiltonian.

Returning to the scale ordering $m \gg mv \gg mv^2$, pNRQCD provides a controlled framework for assessing how higher-order perturbative and ultrasoft effects reshape the static interaction in the intermediate region.  Increasing the perturbative accuracy and resumming the ultrasoft logarithms improves the description of lattice determinations of the static energy over a progressively wider range of $r$; this ``progressive tracking'' is explicitly demonstrated, for example, by the sequence of fixed-order and log-resummed predictions up to N\textsuperscript{3}LL accuracy~\cite{Brambilla2009}.  The ultrasoft logarithmic contribution entering at N\textsuperscript{4}LO is also known~\cite{Brambilla2007}, further sharpening the EFT control over the transition region where the interaction begins to depart from a purely Coulombic shape.

\subsubsection{Relativistic effects in quark models}
\label{sec:relativistic_quark_models}

\paragraph{Relativistic kinematics and higher-order interactions:}

Relativistic effects become important when the typical quark momentum is not small compared with the constituent mass. This is most evident in light mesons, in heavy-light systems, and in radially or orbitally excited states. In a purely non-relativistic model, many of these effects are hidden in effective parameters. One then tends to use larger constituent masses and stronger short-range smearing to mimic missing relativistic kinematics and momentum-dependent vertex effects~\cite{Godfrey1985,Bhaduri1981}.

In relativized quark models, part of this burden is treated explicitly. The kinetic energy is replaced by $E_i=\sqrt{\mathbf p^2+m_i^2}$, the spin-dependent factors are promoted schematically from $1/m_i$ and $1/(m_i m_j)$ to $1/E_i$ and $1/(E_iE_j)$, and the short-range interaction acquires momentum-dependent and smeared structures~\cite{Godfrey1985,CapstickIsgur:1986}. For this reason, relativized fits usually prefer smaller constituent masses than non-relativistic ones. A typical example is $m_{u,d}\simeq0.22$ GeV in the Godfrey--Isgur model, compared with $m_{u,d}\sim0.30$--0.35 GeV in standard non-relativistic fits~\cite{Godfrey1985,Bhaduri1981,JaczkoDurand1998}.\footnote{For an equal-mass quark--antiquark system, expanding the relativistic kinetic energy about a representative squared momentum yields a non-relativistic form with an enlarged effective mass and a residual correction whose average can be represented by an additive energy shift; see Eq. (6) and the subsequent discussion in Ref.~\cite{JaczkoDurand1998}. This motivates absorbing the average spectral effects of omitted relativistic kinetic terms into the fitted constituent masses and the constant $D$. Agreement in the spectrum does not by itself ensure accurate wave functions or transition matrix elements.} 

These corrections are needed once one goes beyond the gross structure of the lowest states. They improve radial and orbital level spacings, the $P$-wave fine structure, and the spin-dependent mixing in unequal-mass systems~\cite{Eichten1994,Godfrey2016}. They are also important for radiative transitions, especially for hindered M1 processes and for E1 amplitudes that are sensitive to the detailed wave functions~\cite{Ganhold2021}. The same trend appears in open-flavor strong decays, where relativized wave functions and the proper spin--orbit/tensor structure are needed for realistic ${}^3P_0$ amplitudes~\cite{Barnes2005,Eichten2006}. In the baryon sector, relativized three-quark models similarly improve the excited $N$ and $\Delta$ spectra and their decay properties~\cite{CapstickRoberts2000,CapstickIsgur:1986}. 
Therefore, a non-relativistic Hamiltonian well describes the gross pattern of ground states, but it is not sufficient for precision spectroscopy, decay widths, or higher excitations. Relativistic kinematics and momentum-dressed interactions are tested in the ground-state meson--baryon compatibility analysis of  Section~\ref{sec:udscb_benchmark}, while the role of noncentral and relativistic terms in radial and orbital spectra is examined in  Section~\ref{sec:excitation_spectra}.

\paragraph{Poincar\'e-invariant constituent-quark models:}
Constituent-quark models can be formulated within relativistic Hamiltonian dynamics, in which the states carry a unitary representation of the Poincar\'e group.
Dirac's instant, front, and point forms differ in the generators chosen to be interaction free~\cite{Dirac1949,KeisterPolyzou1991}.
In the Bakamjian--Thomas construction, interactions enter an internal mass operator that commutes with the noninteracting spin and the chosen kinematic variables; the remaining generators are constructed to satisfy the Poincar\'e algebra~\cite{KeisterPolyzou1991,BakamjianThomas1953}.
Embedding a given mass operator in this way preserves its rest-frame spectrum, while structure observables also depend on the boosts and current operators.
Applications to nucleon electromagnetic form factors illustrate the importance of relativistic boosts and the sensitivity to the current prescription~\cite{WagenbrunnEtAl2001,MeldeEtAl2007}.
In particular, Ref.~\cite{BergerWagenbrunnPlessas2004} found sizable effects from relativistic currents and boosts on baryon charge radii and magnetic moments, even in the $Q^2\to0$ limit.
Small external momentum transfer therefore does not by itself justify a non-relativistic treatment of electromagnetic observables.
For more than two particles, the Bakamjian--Thomas construction does not generally satisfy cluster separability, which requires independent subsystems to transform independently when their mutual interactions are removed.
Restoring this property requires additional unitary transformations that generally induce many-body operators~\cite{KeisterPolyzou1991,CoesterPolyzou1982}.
These induced contributions should be distinguished from a separately parametrized dynamical three-quark interaction; their relation must be established within a common representation.

\subsubsection{The color-spin interaction and heavy-heavy Coulomb effects in multiquark systems}\label{sec_V_CS_dom} 

When assessing whether a compact multiquark state is energetically favored over its well-separated hadrons, the following two cases should be distinguished: (i) states composed of light quarks with similar masses; and (ii) systems that contain heavy quarks. In case (i), the difference between the color-color interaction of the compact multiquark configuration and that of the two separated hadrons cancels if the compact multiquark is also a color singlet. In this case, the color-spin interaction  typically governs the stability of the multiquark configuration against decay into separate hadrons or the formation of a loosely bound molecular state. In case (ii), however, in addition to the color-spin effect, the presence of a compact $QQ$ pair introduces an enhanced color-Coulomb attraction relative to the lowest threshold, which consists of two heavy-light clusters.

We compare a color singlet multiquark with $n$ constituents against its lowest color singlet threshold composed of two clusters with $n_1$ and $n_2$ constituents ($n=n_1+n_2$, no coupled-channel effect considered). Using standard generators $F_i^c=\lambda_i^c/2$ for (anti)fundamental quarks (with $F_{\bar q}^c=-(F_q^c)^{*}$), we write a pairwise Hamiltonian with a color-spin term:
\begin{align}
H &= K + V^C + V^{CS},\qquad 
V^C=-\frac{3}{4}\sum_{i<j} \Big[V_{\rm lin}(r_{ij})+V_{\rm C}(r_{ij})-D\Big]\,F_i^cF_j^c,\qquad
V^{CS} = -A \sum_{i<j}\frac{F_i^cF_j^c\,\boldsymbol{\sigma}_i\cdot\boldsymbol{\sigma}_j}{m_i m_j},
\end{align}
where $V_{\rm lin}= \sigma r$ and $V_{\rm C}(r)= -\kappa/r\propto-\alpha_s/r$. We define the energy difference between the multiquark system and its
separated two-hadron threshold.
\begin{equation}
\Delta E \equiv \big\langle H\big\rangle_{n}-\sum_{h=1,2}\big\langle H\big\rangle_{n_h}=\Delta E^K+\Delta E^C+\Delta E^{CS}.
\end{equation}

\paragraph{Kinetic energy:}

It is convenient to use Jacobi coordinates: the $n$-body internal kinetic energy decomposes into $n-1$ relative modes. A compact configuration thus has one additional localized mode compared to its two separated hadrons, which only carry $(n_1-1)$ and $(n_2-1)$ internal modes in each cluster ($n=n_1+n_2$), while the relative motion between the two hadrons is unbound at threshold. This extra mode, associated with the relative distance between the two clusters and reduced mass $M_{12}=2M_1M_2/(M_1{+}M_2)$ (cluster masses $M_{1,2}$), implies that the additional attraction of the compact state must be large enough to overcome the extra kinetic energy.

\paragraph{Color-electric part (linear vs Coulomb):}
For clarity, we again consider the Cornell-type color-electric potential explicitly as
\begin{equation}\label{temp_cornell}
V^C=\sum_{i<j}V^C_{ij},\qquad 
V^C_{ij}=-\frac{3}{4}F_i^cF_j^c\bigg(-\frac{\kappa}{r_{ij}}+\sigma r_{ij}-D\bigg),
\end{equation}
with the standard generators $F_i^c$ for quark $i$. This is a sum of two-body interactions, for which the following simplified rules can be applied.

\begin{itemize}
\item[(1)] \underline{All quarks have the same mass}: Consider comparing the color-color interaction between a compact configuration of {$n$} quarks to that of two hadrons with $n_1$ and $n_2$ quarks, respectively.  Furthermore, we assume that both the multiquark state and the hadrons have the same spatial size.  Then, the matrix element of the spatial dependence $f(r_{ij})$, e.g.,~Eq.~\eqref{temp_cornell}, will be the same for all pairs.  Then, it is enough to compare the color factors only, which has the following value.
\begin{equation}
\sum_{i<j}^n F_i^c F_j^c=\frac{1}{2}C_2(R_\text{tot})-\frac{1}{2}\sum_{i=1}^n C_2(R_i) = \frac{1}{2}C_2(R_\text{tot})-\frac{2}{3}n.
\end{equation}
Here, $C_2(R_{\text{tot}})$ denotes the quadratic Casimir of the total color irreducible representation $R_{\text{tot}}$, and $C_2(R_i)$  
is equal to $4/3$ for  individual quarks. 
$C_2(R)=0$ for a color singlet configuration  so that the color factor is linearly proportional to the number of quarks $n$.  
Therefore, subtracting the color factors of the individual clusters composed of $n_1$ and $n_2$ quarks from that of the total multiquark states composed of $n=n_1+n_2$ quarks, one finds that the color factors cancel if all the clusters are in the color singlet state.
\item[(2)] \underline{Two (or more) heavy quarks separated across clusters}: The cancellation discussed in the previous equal-mass case no longer holds when the quark masses differ, because the spatial part of {$V_{ij}$} cannot be factored out as a common factor. In a compact configuration containing heavy quarks $Q$, the $QQ$ pair can occupy in the attractive $\overline{\mathbf 3}$ channel ($\mathbf{1}$ for $Q\bar Q$) at distances of order the Bohr scale,
\begin{equation}\label{color-electric_dom}
r_B^{(QQ)} \sim \frac{2}{m_{QQ}\,|C_{QQ}|\,\alpha_s},\qquad 
E_{\rm Coul}^{(QQ)} \sim -\frac{m_{QQ}}{4}\,\big(C_{QQ}\alpha_s\big)^2,
\qquad (C_{QQ}=F_i^cF_j^c),
\end{equation}
where $m_{QQ}$ denotes twice the standard two-body reduced mass, consistent with the convention for $m_{ij}$ above. At thresholds such as $Q\bar q+Q\bar q$ or $Qqq+Q\bar q$, the heavy quarks reside in separate hadrons and do not experience this short-distance $QQ$ attraction. Therefore a compact state that localizes the heavy pair gains a dominant (attractive) Coulomb excess~\cite{Karliner2017_2}.
\end{itemize}

\paragraph{Color-spin interaction:}
After separating the color-electric contribution, the remaining spin-dependent contribution is
\begin{align}
\Delta E^{CS}
&= \big\langle V^{CS}\big\rangle_{n}-\sum_h \big\langle V^{CS}\big\rangle_{n_h}
= -A \sum_{i<j}\frac{\big\langle F_i^cF_j^c\,\boldsymbol{\sigma}_i\cdot\boldsymbol{\sigma}_j\big\rangle_{n}}{m_i m_j}
+ A \sum_h\sum_{i<j\in n_h}\frac{\big\langle F_i^cF_j^c\,\boldsymbol{\sigma}_i\cdot\boldsymbol{\sigma}_j\big\rangle_{n_h}}{m_i m_j}.
\end{align}
In the equal-mass (or light-dominated) limit where the spatial matrix elements can be treated as common weights, the channel dependence of the chromomagnetic expectation value can be assessed algebraically using the following $\mathrm{SU}(6)_{CS}$ identity:
\begin{align}\label{eq:27Cas}
-\Big\langle \sum_{i<j}\lambda_i\cdot\lambda_j\;\boldsymbol\sigma_i\cdot\boldsymbol\sigma_j\Big\rangle 
= -4\,C_2^{(CS)} + 2\,C_2^{(C)} + \frac{4}{3}\,S(S+1) + 8n,
\end{align}
where $C_2^{(CS)}$ and $C_2^{(C)}$ are the quadratic Casimir operators of the total SU$(6)_{CS}$ and SU$(3)_C$ irreps of the $n$-constituent state, and $S$ is the total spin.  Since $C_2^{(C)}=0$ for an overall color singlet, Eq.~\eqref{eq:27Cas} reduces the evaluation of the dominant color-spin factor to a small set of discrete quantum numbers $\{C_2^{(CS)},S,n\}$.  To assess whether a multiquark state has a strong attraction one has to compare the value of Eq.~\eqref{eq:27Cas} in the multiquark state with that in the two-hadron threshold.  This yields a fast ordering of multiquark channel candidates before any explicit spatial calculation.
For ground states where the spatial wave function is symmetric, one can use the total antisymmetry property of the color-spin-flavor quantum number and re-express Eq.~\eqref{eq:27Cas} in terms of the total number of quarks and flavor quantum numbers.
For SU(2) and SU(3), the corresponding relations are given below.
For SU(2),
\begin{align}\label{eq:27Cas-2}
-\Big\langle \sum_{i<j}\lambda_i\cdot\lambda_j\;\boldsymbol\sigma_i\cdot\boldsymbol\sigma_j\Big\rangle 
= \frac{4}{3}n(n-6) +4I(I+1) + 2\,C_2^{(C)} + \frac{4}{3}\,S(S+1), 
\end{align}
where $I$ is the isospin.
For flavor SU(3),
\begin{align}\label{eq:27Cas-3}
-\Big\langle \sum_{i<j}\lambda_i\cdot\lambda_j\;\boldsymbol\sigma_i\cdot\boldsymbol\sigma_j\Big\rangle 
= n(n-10) +4C_2^{(F)} + 2\,C_2^{(C)} + \frac{4}{3}\,S(S+1), 
\end{align}
where $C_2^{(F)}$ is the quadratic Casimir operator of the flavor SU(3). For flavor SU(4),
\begin{align}\label{eq:27Cas-4}
&-\Big\langle \sum_{i<j}\lambda_i\cdot\lambda_j\;\boldsymbol\sigma_i\cdot\boldsymbol\sigma_j\Big\rangle 
= \frac{5}{6}n(n-\frac{72}{5}) +4C_2^{\rm SU(4)} + 2\,C_2^{(C)} + \frac{4}{3}\,S(S+1),
\end{align}
where $C_2^{(SU(4))}$ is the quadratic Casimir operator of the flavor SU(4). 
One can go from the SU(2) relation to the SU(3) relation by noting $n=p+2q$ and $I=p/2$,\footnote{For an SU(3) Young tableau without a complete column of three, $n=p+2q$ and $I=p/2$ reproduce the corresponding relation between the SU(2) and SU(3) expressions.}  which leads to $C_2^{(F)}=I(I+1)+n(n+6)/{12}$. One notes that the first term in Eq.~\eqref{eq:27Cas-3} is negative if $n=6$, which is the origin of the strong attraction in the $H$-dibaryon. 
Compared to the color factors given in Eq.~\eqref{temp_cornell}, the color-spin factors are not linear in the number of quarks and could produce attraction when the number of quarks is less than 6 or 10 for 2 and 3 flavors, respectively. 

In heavy-rich channels, however, a compact $QQ$ subsystem can produce a sizable attractive Coulomb contribution,  whereas the color spin interaction is suppressed by mass factor $1/(m_i m_j)$.  As a result, for configurations such as the $T_{cc}$, which is dominated by the color anti-triplet for the $cc$ and triplet for the light $\bar{u}\bar{d}$, the Coulomb attraction in the $cc$ pair and the color-spin attraction in the $\bar u\bar d$ pair overwhelm the other repulsive contributions, leading to a possible compact configuration.

\paragraph{Color basis for tetraquarks, pentaquarks, and dibaryons:}

The preceding discussion provides a simple formula for the color-spin matrix element when the wavefunction is totally antisymmetric. To evaluate these quantities in practice when quarks with different masses are included, one must first specify an explicit color basis for each multiquark sector. We therefore summarize convenient color-coupling bases for tetraquarks, pentaquarks, and dibaryons, together with the recoupling relations between diquark and meson--meson (or baryon-meson/baryon-baryon) schemes.

For the lowest orbital configuration in which the relevant identical quarks occupy a symmetric spatial state, the Pauli principle constrains the internal $\mathrm{Color}\otimes\mathrm{Spin}\otimes\mathrm{Flavor}$ structure.\footnote{A general state with total orbital angular momentum $J=0$ need not be totally symmetric under all particle exchanges. See \ref{bases}. Here we assume totally symmetric $S$-wave orbital states.} 
For a tetraquark configuration $q_1 q_2 \bar q_3 \bar q_4$ the color space is $\mathbf{3}_C \otimes \mathbf{3}_C \otimes \bar{\mathbf{3}}_C \otimes \bar{\mathbf{3}}_C$.
A convenient starting point is the diquark-antidiquark coupling scheme. The two-quark and two-antiquark subsystems decompose as $\mathbf{3}_C \otimes \mathbf{3}_C = \bar{\mathbf{3}}_C \oplus \mathbf{6}_C$, $\bar{\mathbf{3}}_C \otimes \bar{\mathbf{3}}_C = \mathbf{3}_C \oplus \bar{\mathbf{6}}_C$, and color singlet states are obtained by combining conjugate irreps, $(\bar{\mathbf{3}}_C\otimes \mathbf{3}_C) \supset \mathbf{1}_C$, $(\mathbf{6}_C \otimes \bar{\mathbf{6}}_C) \supset \mathbf{1}_C$.
Thus there are two independent color singlet structures, which one may take as the basis vectors $\lvert C_{12;34}^{\bar{\mathbf{3}}-\mathbf{3}}\rangle  = [(q_1 q_2)_{\bar{\mathbf{3}}}\,(\bar q_3 \bar q_4)_{\mathbf{3}}]_{\mathbf{1}}$, $\lvert C_{12;34}^{\mathbf{6}-\bar{\mathbf{6}}}\rangle  = [(q_1 q_2)_{\mathbf{6}}\,(\bar q_3 \bar q_4)_{\bar{\mathbf{6}}}]_{\mathbf{1}}$, where the subscripts identify the quark pairs and the superscripts identify their color states. Alternatively, in a meson--meson coupling scheme one couples quark--antiquark pairs to color singlets or octets, $\lvert C_{13;24}^{(\mathbf{1}-\mathbf{1})}\rangle  = [(q_1 \bar q_3)_{\mathbf{1}}\,(q_2 \bar q_4)_{\mathbf{1}}]_{\mathbf{1}}$, $\lvert C_{13;24}^{(\mathbf{8}-\mathbf{8})}\rangle  = [(q_1 \bar q_3)_{\mathbf{8}}\,(q_2 \bar q_4)_{\mathbf{8}}]_{\mathbf{1}}$, and similarly for the pairing $(14)(23)$. The diquark and meson--meson color bases are related by an orthogonal transformation:
\begin{equation}
\begin{pmatrix}
|C_{12;34}^{\bar{\mathbf{3}}-\mathbf{3}}\rangle \\ |C_{12;34}^{\mathbf{6}-\bar{\mathbf{6}}}\rangle
\end{pmatrix} = \begin{pmatrix}
\sqrt{1/3} & -\sqrt{2/3} \\ \sqrt{2/3} & \sqrt{1/3}
\end{pmatrix} \begin{pmatrix}
|C_{13;24}^{(\mathbf{1}-\mathbf{1})}\rangle \\ |C_{13;24}^{(\mathbf{8}-\mathbf{8})}\rangle
\end{pmatrix}.
\end{equation}

For pentaquarks with four quarks and one antiquark, $q_1 q_2 q_3 q_4 \bar q_5$, the color configuration is  $\mathbf{3}_C \otimes \mathbf{3}_C \otimes \mathbf{3}_C \otimes \mathbf{3}_C \otimes \bar{\mathbf{3}}_C$,
and again the physical states must be overall singlets. To form a color singlet by combining with an antitriplet, the four-quark system must be in a color triplet representation. In this case, the color basis that yields a color singlet can be represented using Young tableaux as follows:

\begin{equation}
|C_1\rangle = \left(
\begin{ytableau}
1 & 2 \\
3 \\
4 \\
\end{ytableau},
\begin{ytableau}
\overline{5}
\end{ytableau} \right), \qquad 
|C_2\rangle = \left(
\begin{ytableau}
1 & 3 \\
2 \\
4 \\
\end{ytableau},
\begin{ytableau}
\overline{5}
\end{ytableau} \right), \qquad
|C_3\rangle = \left(
\begin{ytableau}
1 & 4 \\
2 \\
3 \\
\end{ytableau},
\begin{ytableau}
\overline{5}
\end{ytableau} \right)
\end{equation}
Among these basis states, $\lvert C_3\rangle$ corresponds to ``baryon {\ensuremath{+}} meson'' clustering, where three of the quarks form a color singlet and the remaining quark--antiquark pair also forms a color singlet meson $\lvert C_{123;45}^{\mathbf{1}-\mathbf{1}}\rangle  = [(q_1 q_2 q_3)_{\mathbf{1}}\,(q_4 \bar q_5)_{\mathbf{1}}]_{\mathbf{1}}$.
In addition, there exist hidden-color configurations where the three-quark subsystem is in an octet and the $q \bar q$ pair is also in an octet, $\lvert C^{\mathbf{8}-\mathbf{8}}_{123;45}\rangle  = [(q_1 q_2 q_3)_{\mathbf{8}}\,(q_4 \bar q_5)_{\mathbf{8}}]_{\mathbf{1}}$, corresponding to $\lvert C_1\rangle$ and $\lvert C_2\rangle$. To display the internal color-coupling structure explicitly, the basis states can be written as follows:
\begin{eqnarray}
|C_1\rangle = [[[(q_1 q_2)_{\mathbf{6}}q_3]_{\mathbf{8}}q_4]_{\mathbf{3}}\bar{q}_5]_{\mathbf{1}},\qquad |C_2\rangle = [[[(q_1 q_2)_{\bar{\mathbf{3}}}q_3]_{\mathbf{8}}q_4]_{\mathbf{3}}\bar{q}_5]_{\mathbf{1}}, \qquad |C_3\rangle = [[[(q_1 q_2)_{\bar{\mathbf{3}}}q_3]_{\mathbf{1}}q_4]_{\mathbf{3}}\bar{q}_5]_{\mathbf{1}}.
\end{eqnarray}
A complementary coupling scheme is the diquark-diquark--antiquark picture, where two quarks form a diquark in either $\bar{\mathbf{3}}_C$ or $\mathbf{6}_C$:
$(q_1 q_2)_{\bar{\mathbf{3}},\mathbf{6}}$, $(q_3 q_4)_{\bar{\mathbf{3}},\mathbf{6}}$, the two diquarks are combined to form a color triplet $\mathbf{3}_C$: $\bigl[(q_1 q_2)_{R_1}\,(q_3 q_4)_{R_2}\bigr]_{\mathbf{3}}$, and then this triplet is coupled with the antiquark $\bar{\mathbf{3}}_C$ to a color singlet. In this coupling scheme, the basis states can be expressed as follows
in the Young-tableau basis introduced above.
\begin{align}
&[[(q_1 q_2)_{\mathbf{6}}(q_3 q_4)_{\mathbf{\bar{3}}}]_{\mathbf{3}}\,\bar q_5]_{\mathbf{1}}=|C_1\rangle, \nonumber\\
&[[(q_1 q_2)_{\bar{\mathbf{3}}}(q_3 q_4)_{\mathbf{6}}]_{\mathbf{3}}\,\bar q_5]_{\mathbf{1}}=\frac{1}{\sqrt{3}}|C_2\rangle + \frac{\sqrt{2}}{\sqrt{3}}|C_3 \rangle, 
\qquad [[(q_1 q_2)_{\bar{\mathbf{3}}}(q_3 q_4)_{\mathbf{\bar{3}}}]_{\mathbf{3}}\,\bar q_5]_{\mathbf{1}}=\frac{\sqrt{2}}{\sqrt{3}}|C_2\rangle - \frac{1}{\sqrt{3}}|C_3 \rangle. 
\end{align}
Using these color basis sets, we construct the wave function of the multiquark system by applying the Pauli exclusion to the four-quark subsystem $q_1 q_2 q_3 q_4$: for $S$-wave states the spatial part is totally symmetric, so the product of color, spin, and flavor Young diagrams of the four-quark cluster must be a totally antisymmetric diagram $[1^4]$ under permutations. 

Next, we consider dibaryons, which are six-quark color-singlet systems.
In terms of Young diagrams, the color singlet of a dibaryon corresponds to [222], and in this case the color basis can be represented using Young tableaux as follows:

\begin{equation}
|C_1\rangle = \begin{ytableau}
1 & 2 \\
3 & 4 \\
5 & 6 \\
\end{ytableau}, \qquad 
|C_2\rangle = \begin{ytableau}
1 & 3 \\
2 & 4 \\
5 & 6 \\
\end{ytableau}, \qquad
|C_3\rangle = \begin{ytableau}
1 & 2 \\
3 & 5 \\
4 & 6 \\
\end{ytableau}, \qquad
|C_4\rangle = \begin{ytableau}
1 & 3 \\
2 & 5 \\
4 & 6 \\
\end{ytableau}, \qquad
|C_5\rangle = \begin{ytableau}
1 & 4 \\
2 & 5 \\
3 & 6 \\
\end{ytableau}.
\end{equation}
The simplest color singlet configuration is the baryon-baryon cluster, where two sets of three quarks are both singlets: $\lvert C_5\rangle  = [(q_1 q_2 q_3)_{\mathbf{1}}\,(q_4 q_5 q_6)_{\mathbf{1}}]_{\mathbf{1}}$.
The four hidden-color states $\lvert C_1\rangle,\ldots,\lvert C_4\rangle$ all have the color coupling $[(q_1 q_2 q_3)_{\mathbf{8}}\,(q_4 q_5 q_6)_{\mathbf{8}}]_{\mathbf{1}}$. The explicit internal color-coupling structures of color bases can be represented as follows:
\begin{align}
&|C_1 \rangle = [[(q_1 q_2)_{\mathbf{6}}q_3]_{\mathbf{8}}[(q_5 q_6)_{\mathbf{6}}q_4]_{\mathbf{8}}]_{\mathbf{1}}, \qquad |C_2 \rangle =[[ (q_1 q_2)_{\bar{\mathbf{3}}}q_3]_{\mathbf{8}}[(q_5 q_6)_{\mathbf{6}}q_4]_{\mathbf{8}}]_{\mathbf{1}},\nonumber\\
&|C_3 \rangle = [[(q_1 q_2)_{\mathbf{6}}q_3]_{\mathbf{8}}[(q_5 q_6)_{\bar{\mathbf{3}}}q_4]_{\mathbf{8}}]_{\mathbf{1}},\qquad |C_4 \rangle = [[(q_1 q_2)_{\bar{\mathbf{3}}}q_3]_{\mathbf{8}}[(q_5 q_6)_{\bar{\mathbf{3}}}q_4]_{\mathbf{8}}]_{\mathbf{1}},\nonumber\\
&|C_5 \rangle = [[(q_1 q_2)_{\bar{\mathbf{3}}}q_3]_{\mathbf{1}}[(q_5 q_6)_{\bar{\mathbf{3}}}q_4]_{\mathbf{1}}]_{\mathbf{1}}.
\end{align}
An alternative coupling scheme is the diquark-diquark-diquark configuration. Color singlets can then be formed by combining three diquarks into a singlet. In this case as well, it can be expressed as follows using the Young tableau basis introduced above.
\begin{align}
&[[(q_1 q_2)_{\mathbf{6}}(q_3 q_4)_{\mathbf{6}}]_{\bar{\mathbf{6}}}(q_5 q_6)_{\mathbf{6}}]_{\mathbf{1}} = |C_1 \rangle, \qquad [[(q_1 q_2)_{\bar{\mathbf{3}}}(q_3 q_4)_{\bar{\mathbf{3}}}]_{\bar{\mathbf{6}}}(q_5 q_6)_{\mathbf{6}}]_{\mathbf{1}} = |C_2 \rangle, \qquad [[(q_1 q_2)_{\mathbf{6}}(q_3 q_4)_{\bar{\mathbf{3}}}]_{\mathbf{3}}(q_5 q_6)_{\bar{\mathbf{3}}}]_{\mathbf{1}} = |C_3 \rangle,\nonumber\\
&[[(q_1 q_2)_{\bar{\mathbf{3}}}(q_3 q_4)_{\bar{\mathbf{3}}}]_{\mathbf{3}}(q_5 q_6)_{\bar{\mathbf{3}}}]_{\mathbf{1}} = \frac{\sqrt{2}}{\sqrt{3}}|C_4 \rangle -\frac{1}{\sqrt{3}}|C_5 \rangle, \qquad [[(q_1 q_2)_{\bar{\mathbf{3}}}(q_3 q_4)_{\mathbf{6}}]_{\mathbf{3}}(q_5 q_6)_{\bar{\mathbf{3}}}]_{\mathbf{1}} = \frac{1}{\sqrt{3}}|C_4 \rangle +\frac{\sqrt{2}}{\sqrt{3}}|C_5 \rangle.
\end{align}
This three-diquark configuration can be useful for discussing diquark correlations and for organizing short-range dynamics in multibaryon states. For $S$-wave dibaryons the overall permutation symmetry of color $\otimes$ spin $\otimes$ flavor must be totally antisymmetric $[1^6]$, which strongly constrains the allowed spin-flavor multiplets. In practice, one constructs a basis of color singlet states made of $\ket{C^{\mathbf{1}-\mathbf{1}}_{qqq;qqq}}$ and several linearly independent hidden-color states $\ket{C^{\mathbf{8}-\mathbf{8}}_{qqq;qqq}}$, or equivalently diquark-coupled singlets in a $2+2+2$ scheme, and then imposes the Pauli principle by combining them with appropriate spin-flavor states. These color bases and recoupling conventions are used explicitly in  Section~\ref{sec:3bd_operators} to evaluate the two- and three-body matrix elements of tetraquark, pentaquark, and dibaryon configurations.

\subsection{Unquenched quark models and coupled-channel effects}\label{sec:unquenching}

While traditional quark models treated hadrons as isolated few-quark systems, it became clear in the 1980s that coupling to meson--meson or meson--baryon channels is often crucial~\cite{Eichten1978,Eichten1980}. This realization followed the spirit of ``unquenching'' in lattice QCD (including sea quark effects) and led to unquenched quark models. These models explicitly include quark--antiquark pair creation effects, usually implemented via a $^3P_0$ pair-creation mechanism, to mix the original $qqq$ or $q\bar{q}$ states with higher Fock components~\cite{Santopinto2013}. In the baryon sector, mixing between valence $qqq$ states and $(qqq)(q\bar q)$ components can affect the nucleon strangeness content and electromagnetic observables. Coupled-channel dynamics has long been invoked to address the missing-resonance problem and the anomalously low Roper $N(1440)$: mixing with $N\pi$- and $N\sigma$-like continua can generate sizeable downward self-energies and redistribute strength from ``bare'' valence levels into broad, weakly excited structures~\cite{Krehl2000,SuzukiEtAl2010}.

\subsubsection{Formulation of unquenching and channel coupling}

In an unquenched constituent-quark model, we treat the bare valence configurations and the hadronic scattering channels in one Hamiltonian~\cite{Santopinto2013,Swanson2006,Barnes2008}. 
In the chosen truncated model space, we denote by $\mathsf P$ the space of all retained valence eigenstates with the chosen conserved quantum numbers and by $\mathsf Q$ the explicitly retained meson--meson or meson--baryon channels:
\begin{equation}
H_0\ket{A_i}=M_i^{(0)}\ket{A_i},\qquad \mathsf P=\sum_i\ket{A_i}\bra{A_i},\qquad \mathsf Q=1-\mathsf P.
\label{eq:unquenched_PQ_spaces}
\end{equation}
Here, 1 is the identity in the adopted model space. Therefore, $\mathsf P+\mathsf Q=1$ and $\mathsf P\mathsf Q=0$ within this truncation. We choose the bare-state basis so that it diagonalizes $H_{\mathsf P\mathsf P}$,
\begin{equation}
H_{\mathsf P\mathsf P}=\mathsf P H_0\mathsf P=\sum_i M_i^{(0)}\ket{A_i}\bra{A_i}.
\label{eq:unquenched_HPP_bare_basis}
\end{equation}
The full Hamiltonian can be decomposed as
\begin{equation}
H=H_{\mathsf P\mathsf P}+H_{\mathsf Q\mathsf Q}+T_{\mathsf P\mathsf Q}+T_{\mathsf Q\mathsf P},\qquad H_{\mathsf P\mathsf P}=\mathsf P H\mathsf P,\qquad T_{\mathsf P\mathsf Q}=\mathsf P H\mathsf Q,
\label{eq:unquenched_block_H}
\end{equation}
The remaining blocks are defined in the same way. 
The continuum Hamiltonian $H_{\mathsf Q\mathsf Q}$ contains the free two-hadron propagation and may also include direct interactions and rescattering among the continuum channels. 
The two projected Schr\"odinger equations are
\begin{align}
(E-H_{\mathsf P\mathsf P})\mathsf P\ket{\Psi}=T_{\mathsf P\mathsf Q}\mathsf Q\ket{\Psi},\qquad (E-H_{\mathsf Q\mathsf Q})\mathsf Q\ket{\Psi}=T_{\mathsf Q\mathsf P}\mathsf P\ket{\Psi}.
\label{eq:unquenched_block_equations}
\end{align}
Eliminating the continuum component gives ($R_{\mathsf Q}(E)=({E-H_{\mathsf Q\mathsf Q}+i0})^{-1}$)
\begin{align}
\mathsf Q\ket{\Psi}=R_{\mathsf Q}(E)T_{\mathsf Q\mathsf P}\mathsf P\ket{\Psi},\qquad\left[E-H_{\mathsf P\mathsf P}-\Sigma_{\mathsf Q}(E)\right]\mathsf P\ket{\Psi}=0,\qquad \Sigma_{\mathsf Q}(E)=T_{\mathsf P\mathsf Q}R_{\mathsf Q}(E)T_{\mathsf Q\mathsf P}.
\label{eq:unquenched_feshbach_equation}
\end{align}
\eqref{eq:unquenched_feshbach_equation} is the Feshbach--Bloch--Horowitz equation and is exact within the chosen $\mathsf P\oplus\mathsf Q$ model space~\cite{Feshbach1958,BlochHorowitz1958,Hammer2016}. 
We use the same projection in  Section~\ref{subsec:three_quark_definition}, where the non-valence $qqqg$ sector is eliminated and its low-energy contribution is separated into one-body, pairwise, and connected three-quark terms. When several bare states are retained, $\Sigma_{\mathsf Q}(E)$ becomes a matrix and produces both diagonal self-energies and continuum-induced mixing. 
Therefore, the physical bound-state or resonance energies are determined by
\begin{equation}
\det\!\left[E\mathbf 1_{\mathsf P}-H_{\mathsf P\mathsf P}-\Sigma_{\mathsf Q}(E)\right]=0,
\label{eq:unquenched_pole_condition}
\end{equation}
and cannot in general be obtained by adding the same mass correction to every valence eigenvalue.

In most unquenched quark-model applications, $T$ is represented by the ${}^3P_0$ pair-creation operator, which creates a color-singlet and flavor-singlet $q\bar q$ pair in a spin--triplet relative $P$ wave coupled to total $\mathcal J^{PC}=0^{++}$~\cite{CapstickRoberts2000,Santopinto2013,Yaouanc1974}.
Up to convention-dependent overall normalization factors, its momentum-space form may be written as (we use the relative momentum $\mathbf q=({\mathbf p-\mathbf p'})/{2}$ and standard solid harmonic notation $\mathcal Y_{1m}(\mathbf q)=qY_{1m}(\hat{\mathbf q})$)
\begin{align}
T_{{}^3P_0}=\gamma\sum_{m=-1}^{1}\mathcal C^{11;0}_{m,-m;0}\int\frac{d^3\mathbf p\,d^3\mathbf p'}{(2\pi)^3}\,\delta^{(3)}(\mathbf p+\mathbf p')\,F(\mathbf q)\,\mathcal Y_{1m}(\mathbf q)\left[\chi_{1,-m}\phi_0\omega_0\right]_{\alpha\beta}\,b_\alpha^\dagger(\mathbf p)d_\beta^\dagger(\mathbf p'),
\label{eq:3P0_transition_operator}
\end{align}
where $\gamma$ is the effective pair-creation strength, while $\chi_{1m}$, $\phi_0$, and $\omega_0$ denote the spin--triplet, flavor-singlet, and color-singlet wave functions of the created pair, respectively. 
The standard pointlike ${}^3P_0$ vertex corresponds to $F(\mathbf q)=1$, while a nontrivial $F(\mathbf q)$ defines a regulated or otherwise modified pair-creation vertex.
The partial-wave transition amplitude is
\begin{equation}
\mathcal M_{A_i\to BC}^{l,\mathcal J}(k)\equiv\bra{BC;k,l,\mathcal J}T\ket{A_i},
\label{eq:transition_amp}
\end{equation}
and, for $T=T_{\mbox{}^3P_0}$, it reduces to spin, flavor, and color recoupling coefficients multiplying spatial overlap integrals of the valence-hadron wave functions.
A mass-dependent $\gamma^{\rm eff}$ or an equivalent suppression factor is sometimes introduced to reduce the creation of heavier quark pairs~\cite{CapstickRoberts2000,Santopinto2010}.

If $H_{\mathsf Q\mathsf Q}$ is approximated by noninteracting two-hadron channels, the operator self-energy in Eq.~\eqref{eq:unquenched_feshbach_equation} reduces to
\begin{equation}
\Sigma_{ij}(E)=\sum_{BC l\mathcal J}\int_0^\infty\frac{dk\,k^2}{(2\pi)^3}\,\frac{\mathcal M_{A_i\to BC}^{l,\mathcal J}(k)\,\mathcal M_{BC\to A_j}^{l,\mathcal J}(k)}{E-E_B(k)-E_C(k)+i0},
\label{self_e}
\end{equation}
where $E_{B,C}(k)=\sqrt{M_{B,C}^2+k^2}$ for relativistic kinematics.
The phase-space prefactor in Eq.~\eqref{self_e} follows the stated continuum normalization and may equivalently be absorbed into the definition of $\mathcal M$ under another convention.
On the real axis, Hermiticity relates the two transition amplitudes, and the diagonal numerator becomes $\vert \mathcal M_{A_i\to BC}^{l,\mathcal J}(k)\vert ^2$.
For a single retained valence state, we write $\Sigma_A(E)\equiv\Sigma_{AA}(E)$.
Nothing in Eqs.~\eqref{eq:unquenched_feshbach_equation}--\eqref{self_e} requires the ${}^3P_0$ ansatz specifically.
The same formalism applies when $T_{\mathsf P\mathsf Q}$ is obtained from a microscopic quark-rearrangement interaction, a flux-tube-breaking operator, a chiral hadronic vertex, or another transition mechanism, provided that $T_{\mathsf P\mathsf Q}$ and $H_{\mathsf Q\mathsf Q}$ are specified consistently.

When the continuum channels interact, we write $H_{\mathsf Q\mathsf Q}=H_{\mathsf Q\mathsf Q}^{(0)}+V_{\mathsf Q\mathsf Q}$, with $R_{\mathsf Q}^{(0)}(E)=({E-H_{\mathsf Q\mathsf Q}^{(0)}+i0})^{-1}$, where $V_{\mathsf Q\mathsf Q}$ contains elastic and channel-changing interactions among the retained hadrons. In this case, the compact--continuum vertices are dressed by continuum rescattering and satisfy the following Lippmann--Schwinger-type equations, where the integrations over intermediate momenta and the sums over continuum-channel indices are understood:
\begin{align}
\Gamma_{\mathsf Q\mathsf P}(E)=T_{\mathsf Q\mathsf P}+V_{\mathsf Q\mathsf Q}R_{\mathsf Q}^{(0)}(E)\Gamma_{\mathsf Q\mathsf P}(E),\qquad \Gamma_{\mathsf P\mathsf Q}(E)=T_{\mathsf P\mathsf Q}+\Gamma_{\mathsf P\mathsf Q}(E)R_{\mathsf Q}^{(0)}(E)V_{\mathsf Q\mathsf Q}.
\label{eq:unquenched_dressed_vertices}
\end{align}
Their partial-wave matrix elements are denoted by $\Gamma_{\alpha i}^{\mathsf Q\mathsf P}(E;k)=\bra{\alpha;k}\Gamma_{\mathsf Q\mathsf P}(E)\ket{A_i}$ and $\Gamma_{i\alpha}^{\mathsf P\mathsf Q}(E;k)=\bra{A_i}\Gamma_{\mathsf P\mathsf Q}(E)\ket{\alpha;k}$.
Thus the free denominator in Eq.~\eqref{self_e} cannot in general be retained together with an undressed transition amplitude when continuum rescattering is important.
The full resolvent $R_{\mathsf Q}(E)$ retains both the rescattering series and any poles dynamically generated within $H_{\mathsf Q\mathsf Q}$, while the dressed vertices in Eq.~\eqref{eq:unquenched_dressed_vertices} provide the corresponding left and right couplings to an asymptotic on-shell channel.

Up to this point, we have assumed that the $\mathsf P$ and $\mathsf Q$ sectors are orthogonal. However, the antisymmetrized cluster basis used in a microscopic resonating-group method (RGM) is generally nonorthogonal~\cite{Saito1977,FujiwaraEtAl2000LSRGM}. 
In this case, the equation in coefficient space takes the generalized form
\begin{equation}
\begin{pmatrix}
H_{\mathsf P\mathsf P} & T_{\mathsf P\mathsf Q}\\
T_{\mathsf Q\mathsf P} & H_{\mathsf Q\mathsf Q}
\end{pmatrix}
\begin{pmatrix}
c_{\mathsf P}\\
c_{\mathsf Q}
\end{pmatrix}
=
E
\begin{pmatrix}
\mathbf 1_{\mathsf P} & \mathcal O_{\mathsf P\mathsf Q}\\
\mathcal O_{\mathsf Q\mathsf P} & \mathcal O_{\mathsf Q\mathsf Q}
\end{pmatrix}
\begin{pmatrix}
c_{\mathsf P}\\
c_{\mathsf Q}
\end{pmatrix}.
\label{eq:unquenched_RGM_generalized_equation}
\end{equation}
In Eq.~\eqref{eq:unquenched_RGM_generalized_equation}, the $\mathsf P$ and $\mathsf Q$ labels identify the chosen compact and cluster basis blocks before norm-kernel orthogonalization rather than two mutually orthogonal coefficient subspaces.
The off-diagonal quantity entering the eliminated equation is therefore the generalized transition kernel
\begin{equation}
\mathcal W_{\mathsf P\mathsf Q}(E)=T_{\mathsf P\mathsf Q}-E\mathcal O_{\mathsf P\mathsf Q},\qquad \mathcal W_{\mathsf Q\mathsf P}(E)=T_{\mathsf Q\mathsf P}-E\mathcal O_{\mathsf Q\mathsf P},
\label{eq:unquenched_generalized_vertex}
\end{equation}
and the continuum resolvent becomes $[E\mathcal O_{\mathsf Q\mathsf Q}-H_{\mathsf Q\mathsf Q}]^{-1}$.
The inverse is understood on the Pauli-allowed subspace after zero-norm components of $\mathcal O_{\mathsf Q\mathsf Q}$ have been removed.
The corresponding Schur-complement self-energy is
\begin{equation}
\Sigma_{\mathsf Q}^{\rm RGM}(E)
=
\mathcal W_{\mathsf P\mathsf Q}(E)
\left[E\mathcal O_{\mathsf Q\mathsf Q}-H_{\mathsf Q\mathsf Q}\right]^{-1}
\mathcal W_{\mathsf Q\mathsf P}(E).
\label{eq:unquenched_RGM_self_energy}
\end{equation}
Equivalently, relative to a free orthonormal channel Hamiltonian $H_{\mathsf Q\mathsf Q}^{(0)}$, one may define the energy-dependent RGM interaction $V_{\mathsf Q\mathsf Q}^{\rm RGM}(E)
=
H_{\mathsf Q\mathsf Q}
-
H_{\mathsf Q\mathsf Q}^{(0)}
-
E\left(\mathcal O_{\mathsf Q\mathsf Q}-\mathbf 1_{\mathsf Q}\right)$, so that $E\mathcal O_{\mathsf Q\mathsf Q}-H_{\mathsf Q\mathsf Q}
=
E-H_{\mathsf Q\mathsf Q}^{(0)}-V_{\mathsf Q\mathsf Q}^{\rm RGM}(E)$.
The RGM-dressed compact--continuum vertices therefore obey
\begin{align}
\Gamma_{\mathsf Q\mathsf P}^{\rm RGM}(E)
=
\mathcal W_{\mathsf Q\mathsf P}(E)
+
V_{\mathsf Q\mathsf Q}^{\rm RGM}(E)
R_{\mathsf Q}^{(0)}(E)
\Gamma_{\mathsf Q\mathsf P}^{\rm RGM}(E), \qquad
\Gamma_{\mathsf P\mathsf Q}^{\rm RGM}(E)
=
\mathcal W_{\mathsf P\mathsf Q}(E)
+
\Gamma_{\mathsf P\mathsf Q}^{\rm RGM}(E)
R_{\mathsf Q}^{(0)}(E)
V_{\mathsf Q\mathsf Q}^{\rm RGM}(E).
\label{eq:unquenched_RGM_dressed_vertices}
\end{align}
Eqs.~\eqref{eq:unquenched_RGM_self_energy} and~\eqref{eq:unquenched_RGM_dressed_vertices} give the momentum-space Lippmann--Schwinger RGM construction~\cite{FujiwaraEtAl2000LSRGM}. If the norm kernel is explicitly orthogonalized, Eq.~\eqref{eq:unquenched_RGM_generalized_equation} is transformed back to an ordinary Feshbach equation. In this case, the same exchange and normalization effects appear in the transformed energy-dependent or nonlocal interactions.

For a one-dimensional $\mathsf P$ space containing a single stable bound state below all included thresholds, and for orthogonal sectors with energy-independent vertices and noninteracting continuum channels, the normalized state can be written as
\begin{equation}
\ket{\Psi_A}
=
\sqrt{Z_A}
\left[
\ket{A}
+
\sum_{BC l\mathcal J}
\int_0^\infty
\frac{dk\,k^2}{(2\pi)^3}
\frac{
\ket{BC;k,l,\mathcal J}
\mathcal M_{A\to BC}^{l,\mathcal J}(k)
}{
M_A-E_B(k)-E_C(k)
}
\right].
\label{eq:unquenched_bound_state}
\end{equation}
Writing $\Sigma_A(E)=\sum_{BC}\Sigma_{A,BC}(E)$, its normalization gives
\begin{align}
Z_A =
\left[ 1- \left. \frac{\partial\Sigma_A}{\partial E} \right|_{E=M_A} \right]^{-1},
\qquad
X_{A,BC} = - Z_A \left. \frac{\partial\Sigma_{A,BC}}{\partial E} \right|_{E=M_A},
\qquad
1=Z_A+\sum_{BC}X_{A,BC}.
\label{eq:bound_compositeness_sum_rule}
\end{align}
Under these conditions, $Z_A$ is the probability of the valence component in the specified $\mathsf P$ space, and $X_{A,BC}$ is the compositeness probability of channel $BC$~\cite{HyodoJidoHosaka2012,HyodoCompositeness2013,SekiharaHyodoJido2015}. 
When the continuum channels interact, we introduce the projector $\mathsf Q_\alpha$ for the asymptotic channel $\alpha$, with $\sum_\alpha\mathsf Q_\alpha=\mathsf Q$. 
In an energy-independent orthogonal representation, the corresponding bound-state component is
\begin{equation}
X_{A,\alpha}
=
Z_A
\bra{A}
T_{\mathsf P\mathsf Q}
R_{\mathsf Q}(M_A)
\mathsf Q_\alpha
R_{\mathsf Q}(M_A)
T_{\mathsf Q\mathsf P}
\ket{A},
\label{eq:bound_compositeness_interacting}
\end{equation}
Only when the channel resolvent is block diagonal does Eq.~\eqref{eq:bound_compositeness_interacting} reduce channel by channel to $-Z_A\Sigma_{A,\alpha}'(M_A)$. 
The same algebraic expressions can be analytically continued to an unstable state, but the resulting quantities are then no longer probabilities.

Let us now consider the case where at least one retained channel is open. 
For a real energy below all retained thresholds and away from a discrete $\mathsf Q$-space pole, the physical-sheet resolvent has no on-shell singularity, and the effective Hamiltonian can be chosen Hermitian. 
Above an open threshold, the outgoing-wave prescription produces a unitarity cut,
\begin{equation}
\Im\Sigma_{\mathsf Q}(E+i0)
=
-\pi
T_{\mathsf P\mathsf Q}
\delta(E-H_{\mathsf Q\mathsf Q})
T_{\mathsf Q\mathsf P},
\label{eq:unquenched_imaginary_self_energy}
\end{equation}
so the reduced Hamiltonian becomes an energy-dependent non-Hermitian optical Hamiltonian.
The physical resonance is then defined by a pole on an appropriate unphysical Riemann sheet,
\begin{equation}
E_{\rm pole}
=
M_R-\frac{i}{2}\Gamma_R,
\qquad
\det\!\left[
E_{\rm pole}\mathbf 1_{\mathsf P}
-
H_{\mathsf P\mathsf P}
-
\Sigma_{\mathsf Q}^{(\boldsymbol\eta)}(E_{\rm pole})
\right]
=
0.
\label{eq:unquenched_complex_pole}
\end{equation}
For each channel $\alpha$, define the on-shell momentum by
\begin{equation}
E
=
E_{B_\alpha}\!\left(k_\alpha\right)
+
E_{C_\alpha}\!\left(k_\alpha\right),
\qquad
k_\alpha^{(\boldsymbol\eta)}(E)
=
\eta_\alpha k_\alpha^{(+)}(E),
\qquad
\eta_\alpha=\pm1,
\label{eq:unquenched_sheet_momentum}
\end{equation}
where the physical branch $k_\alpha^{(+)}$ is chosen with $\Im k_\alpha^{(+)}>0$ below threshold.
With non-relativistic two-body kinematics in channel $\alpha$, this convention reduces to $k_\alpha^{(+)}(E) = \sqrt{2\mu_\alpha(E-E_\alpha^{\rm th})}$.
The sheet vector $\boldsymbol\eta=(\eta_1,\eta_2,\ldots)$ shows which channel cuts are crossed when we continue from the physical sheet. Between two successive thresholds, the sheet adjacent to the physical axis has $\eta_\alpha=-1$ for the channels that are already open and $\eta_\alpha=+1$ for the channels that are still closed. At a complex pole, the meaning of ``open'' and ``closed'' is determined by this continuation path and not only by $\Re E_{\rm pole}$. A closed channel kept on its physical branch still gives a dispersive virtual correction. Its on-shell discontinuity is added only when its cut is crossed, as may be needed when searching for virtual or shadow poles.

If the self-energy integral is evaluated at complex $E$ along the original contour, it remains on the reference sheet. 
To cross an open-channel cut, the contour must be deformed through the on-shell singularity. The residue from this singularity gives the Riemann-sheet jump~\cite{SuzukiSatoLee2009}.
We denote the full interacting continuum resolvent on the chosen reference sheet by $R_{\mathsf Q}^{\rm ref}(E)$ and write
\begin{equation}
\Sigma_{\mathsf Q}^{\rm ref}(E)
=
T_{\mathsf P\mathsf Q}
R_{\mathsf Q}^{\rm ref}(E)
T_{\mathsf Q\mathsf P}
\end{equation}
in an orthogonal representation.
For the normalization of Eq.~\eqref{self_e}, define the analytically continued two-body phase-space factor by
\begin{equation}
\rho_\alpha(E)
=
\left.
\frac{k^2}{(2\pi)^3}
\left[
\frac{d}{dk}
\left(
E_{B_\alpha}(k)+E_{C_\alpha}(k)
\right)
\right]^{-1}
\right|_{k=k_\alpha(E)}.
\label{eq:unquenched_phase_space}
\end{equation}
With the present reference-sheet convention, the jump associated with crossing channel $\alpha$ is
\begin{equation}
J_{ij}^{\alpha}(E)
=
-2\pi i\,
\rho_\alpha(E)\,
\Gamma_{i\alpha}^{\mathsf P\mathsf Q}
\!\left(E;k_\alpha(E)\right)
\Gamma_{\alpha j}^{\mathsf Q\mathsf P}
\!\left(E;k_\alpha(E)\right),
\label{eq:unquenched_sheet_jump}
\end{equation}
where the left and right on-shell vertices are the solutions of Eq.~\eqref{eq:unquenched_dressed_vertices}, or of its RGM generalization when norm and exchange kernels are retained.
For the non-relativistic radial normalization $k^2dk/[2\pi^2(\hbar c)^3]$, Eq.~\eqref{eq:unquenched_sheet_jump} becomes
\begin{equation}
J_{ij}^{\alpha}(E)
=
-i
\frac{
\mu_\alpha k_\alpha(E)
}{
\pi(\hbar c)^3
}
\Gamma_{i\alpha}^{\mathsf P\mathsf Q}
\!\left(E;k_\alpha(E)\right)
\Gamma_{\alpha j}^{\mathsf Q\mathsf P}
\!\left(E;k_\alpha(E)\right).
\end{equation}
The overall sign of $J^\alpha$ is tied to the stated reference-sheet and momentum conventions, and the analytically continued self-energy is written consistently as
\begin{equation}
\Sigma_{ij}^{(\boldsymbol\eta)}(E)
=
\Sigma_{ij}^{\rm ref}(E)
+
\sum_{\alpha:\,\eta_\alpha=-1}
J_{ij}^{\alpha}(E).
\label{eq:unquenched_sheet_self_energy}
\end{equation}
If continuum rescattering is neglected, $\Gamma_{\mathsf Q\mathsf P}\to T_{\mathsf Q\mathsf P}$ and $\Gamma_{\mathsf P\mathsf Q}\to T_{\mathsf P\mathsf Q}$. In a coupled-channel or RGM calculation, however, the dressed vertices must be used because the cut is taken between asymptotic scattering states and not between bare channel basis states. 
After analytic continuation, $\Gamma_{\mathsf P\mathsf Q}$ and $\Gamma_{\mathsf Q\mathsf P}$ are independent left and right functions of the same complex energy and are not related by complex conjugation.

For an isolated narrow resonance dominated by one bare state, one often uses the relations
\begin{equation}
M_R-M_A^{(0)}
\simeq
\Re\Sigma_A(M_R),
\qquad
\Gamma_R
\simeq
-\frac{
2\Im\Sigma_A(M_R)
}{
1-\Re\Sigma_A'(M_R)
}.
\label{eq:narrow_resonance_mass_width}
\end{equation}
These relations are narrow-width approximations. The second one reduces to $\Gamma_R\simeq-2\Im\Sigma_A(M_R)$ only when the energy dependence of the self-energy is weak. For a broad or threshold-adjacent state, one should instead solve the complex pole equation.

For a general multistate resonance, we define the reduced inverse propagator by
\begin{equation}
\mathsf D^{(\boldsymbol\eta)}(E)
=
E\mathbf 1_{\mathsf P}
-
H_{\mathsf P\mathsf P}
-
\Sigma_{\mathsf Q}^{\rm ref}(E)
-
\sum_{\alpha:\,\eta_\alpha=-1}
J^\alpha(E).
\label{eq:unquenched_inverse_propagator}
\end{equation}
At a simple pole, the right and left pole vectors satisfy $\mathsf D^{(\boldsymbol\eta)}(E_{\rm pole})\Psi_R=0$ and $\Psi_L^T\mathsf D^{(\boldsymbol\eta)}(E_{\rm pole})=0$. 
For a stable bound-state pole on the real axis, $\mathsf D(M_A)$ is Hermitian under the assumptions given above, and we may choose $\Psi_L=\Psi_R^{*}$ and $\Psi_L^T=\Psi_R^\dagger$. 
For a resonance pole on an unphysical sheet, $\mathsf D^{(\boldsymbol\eta)}(E_{\rm pole})$ is generally non-Hermitian, and $\Psi_L$ and $\Psi_R$ are independent biorthogonal pole vectors. 
Here, the superscript $T$ only changes the independently determined left column vector into a row vector and does not mean complex conjugation. 
If the analytically continued representation is complex symmetric, $[\mathsf D^{(\boldsymbol\eta)}]^T=\mathsf D^{(\boldsymbol\eta)}$, we may choose $\Psi_L=\Psi_R$. 
In this special case, the bilinear product becomes the transpose, or $c$-product. The pole part of the reduced propagator is normalized by
\begin{align}
\left[\mathsf D^{(\boldsymbol\eta)}(E)\right]^{-1}\simeq \frac{ \Psi_R\Psi_L^T }{ (E-E_{\rm pole})\mathcal N }, \qquad  \mathcal N = \Psi_L^T \mathsf D^{(\boldsymbol\eta)\prime}(E_{\rm pole}) \Psi_R.
\label{eq:unquenched_pole_normalization}
\end{align}

To separate the bare-state and continuum-channel contributions, we define $\mathsf P_i=\ket{A_i}\bra{A_i}$ for bare state $i$ and introduce the asymptotic channel projectors satisfying
\begin{equation}
\mathsf Q_\alpha\mathsf Q_\beta = \delta_{\alpha\beta}\mathsf Q_\alpha,
\qquad \sum_\alpha\mathsf Q_\alpha = \mathsf Q.
\label{eq:unquenched_channel_projectors}
\end{equation}
For orthogonal sectors with energy-independent transition operators and continuum interactions, the propagation part of the channel decomposition is
\begin{equation}
X_\alpha
=
\mathcal N^{-1}
\Psi_L^T
T_{\mathsf P\mathsf Q}
R_{\mathsf Q}^{\rm ref}(E_{\rm pole})
\mathsf Q_\alpha
R_{\mathsf Q}^{\rm ref}(E_{\rm pole})
T_{\mathsf Q\mathsf P}
\Psi_R.
\label{eq:unquenched_channel_propagation}
\end{equation}
Here, the full reference-sheet resolvent appears on both sides of $\mathsf Q_\alpha$. This definition can also be used when $H_{\mathsf Q\mathsf Q}$ mixes the channels. 
In that case, however, a unique decomposition $\Sigma_{\mathsf Q}=\sum_\alpha\Sigma_\alpha$ need not exist. 
A crossed cut gives the additional contribution $C_\alpha^{\rm jump}=-\mathcal N^{-1}\Psi_L^TJ^{\alpha\prime}(E_{\rm pole})\Psi_R$, while $C_\alpha^{\rm jump}=0$ when the cut of channel $\alpha$ is not crossed on the selected sheet. 
The compact contribution is $Z_{P,i}=\mathcal N^{-1}\Psi_L^T\mathsf P_i\Psi_R$. 
If the continuum resolvent is block diagonal in $\alpha$, Eq.~\eqref{eq:unquenched_channel_propagation} reduces to $X_\alpha=-\mathcal N^{-1}\Psi_L^T\Sigma_\alpha^{{\rm ref}\,\prime}(E_{\rm pole})\Psi_R$.

In the common RGM representation in which the asymptotic $\mathsf Q$ basis has been orthonormalized, Eq.~\eqref{eq:unquenched_channel_propagation} is used with $T\to\mathcal W(E)$ and $R_{\mathsf Q}^{\rm ref}\to[E\mathbf 1_{\mathsf Q}-H_{\mathsf Q\mathsf Q}]_{\rm ref}^{-1}$.
If $\mathcal O_{\mathsf Q\mathsf Q}=\mathbf 1_{\mathsf Q}$ and only the off-diagonal overlap kernels generate explicit energy dependence, their contribution is
\begin{align}
Y_{\rm metric}
=
\mathcal N^{-1}
\Psi_L^T
\big[
&
\mathcal O_{\mathsf P\mathsf Q}
R_{\mathsf Q}^{\rm ref}
\mathcal W_{\mathsf Q\mathsf P}
+
\mathcal W_{\mathsf P\mathsf Q}
R_{\mathsf Q}^{\rm ref}
\mathcal O_{\mathsf Q\mathsf P}
\big]_{E=E_{\rm pole}}
\Psi_R.
\label{eq:unquenched_metric_contribution}
\end{align}
The exact pole-normalization identity is then
\begin{equation}
1
=
\sum_i Z_{P,i}
+
\sum_\alpha X_\alpha
+
\sum_{\alpha:\,\eta_\alpha=-1}
C_\alpha^{\rm jump}
+
Y_{\rm metric}.
\label{eq:resonance_residue_sum_rule}
\end{equation}
If $\mathcal O_{\mathsf Q\mathsf Q}\neq\mathbf 1_{\mathsf Q}$, the derivative of the propagation term contains the metric insertion $R_{\mathsf Q}^{\rm ref}\mathcal O_{\mathsf Q\mathsf Q}R_{\mathsf Q}^{\rm ref}$. 
In this case, a channel-by-channel interpretation requires either an explicit orthogonalization of the $\mathsf Q$-space basis with respect to $\mathcal O_{\mathsf Q\mathsf Q}$ or channel projectors defined with the same overlap metric. If any other overlap, interaction, or transition operator has an explicit energy dependence, one should differentiate the full generalized inverse propagator. All the resulting terms must then be included in Eq.~\eqref{eq:resonance_residue_sum_rule}. For a resonance, $Z_{P,i}$, $X_\alpha$, $C_\alpha^{\rm jump}$, and $Y_{\rm metric}$ are generally complex and are not ordinary probabilities~\cite{HyodoCompositeness2013,SekiharaHyodoJido2015}. Their full complex sum is fixed by Eq.~\eqref{eq:resonance_residue_sum_rule}. If one component is close to unity, has a small imaginary part, and is much larger than the other components, one may only say that this component is dominant. The separate values of $X_\alpha$ and $C_\alpha^{\rm jump}$ depend on the reference-sheet convention, but their combined contribution is fixed when the same convention is used throughout. The term $Y_{\rm metric}$ shows the norm-kernel contribution in the nonorthogonal representation. After exact orthogonalization, this term is moved into the transformed effective interactions and is not an additional observable probability. A pole may remain continuously connected to a valence seed even when its residue is dominated by a continuum channel. Therefore, the origin of the pole and the composition of its residue should be discussed separately.

The form factor $F(\mathbf q)$ in Eq.~\eqref{eq:3P0_transition_operator} is a phenomenological ingredient rather than an observable vertex function.
It regulates short-distance momentum components and may mimic a finite pair-creation range, but inside a loop it simultaneously specifies an off-shell continuation of the transition amplitude.
On-shell decay widths constrain the vertex only over the physical momentum region and do not uniquely determine its behavior at virtual momenta, especially for a state far below threshold or close to a rapidly varying threshold.
A form factor depending only on three-momentum also does not encode the full energy dependence, relativistic covariance, left-hand singularities, or short-distance effects of channels omitted from the model space.
Consequently, the dispersive mass shift, the fitted bare masses, and the extracted $Z$ and $X_\alpha$ can depend appreciably on the functional form and cutoff of $F$, except in special weak-binding limits where compositeness becomes less model dependent~\cite{Hammer2016,HyodoCompositeness2013,SekiharaHyodoJido2015}.
A controlled application should therefore vary the regulator, include the channels relevant at the energy of interest, and refit the parameters of $H_0$ and $T$ together, because a valence Hamiltonian fitted directly to physical masses may already absorb a smooth average part of the continuum self-energy.
Existing unquenched calculations also differ in whether the pole equation is solved perturbatively or self-consistently, whether rescattering is retained in $H_{\mathsf Q\mathsf Q}$, how nonorthogonal cluster kernels and unstable hadrons are treated, which Riemann sheets are searched, and whether the valence parameters are refitted after unquenching.
Quoted mass shifts or compositeness values from different implementations are therefore not directly comparable unless these conventions are specified.
The relation between this explicit $\mathsf P$--$\mathsf Q$, sheet-by-sheet construction and full-Hamiltonian complex- or real-scaling calculations is summarized in  Section~\ref{subsubsec:complex_real_scaling}.

\subsubsection{Mass shifts in hadron spectroscopy and loop-induced parametrization}

Within the self-energy formalism of Eq.~\eqref{self_e}, the self-energy shifts the pole position and acquires an imaginary part above an open threshold.
For a sufficiently narrow state evaluated near the real axis, its real and imaginary parts give the conventional mass-shift and width approximations in Eq.~\eqref{eq:narrow_resonance_mass_width}.
Explicit calculations in unquenched quark models have quantified these effects in several sectors of the spectrum.

For charmonium, Ref.~\cite{Barnes2008} computed hadron loop corrections generated by a ${}^3P_0$ pair-creation operator for the low-lying $1S$, $1P$, and $2S$ states.  The resulting self-energies are negative and typically of order 50--150~MeV, while the differences within a given $nL$ multiplet are considerably smaller than the common shift.  Light-meson masses in an unquenched chiral quark model were studied in Ref.~\cite{Chen2018}.  It was shown there that a naive ${}^3P_0$ vertex leads to negative mass shifts, which seems to be incompatible with the phenomenological success of conventional constituent-quark models, and that introducing suitable suppression factors in the pair-creation operator reduces the self-energies to the level of about 10\%--20\% of the corresponding bare masses while preserving the qualitative pattern of splittings within a multiplet~\cite{Chen2018}.

Coupled-channel treatments of charmonium based on Cornell-type potentials have emphasized the role of nearby $S$-wave thresholds.
The $\chi_{c1}(3872)$ provides a representative example because its proximity to the $D\bar D^{*}$ threshold and the tendency of quenched $c\bar c$ calculations to place the $\chi_{c1}(2P)$ seed above the observed region motivated coupled-channel and mixed-state interpretations.
Starting from such a quenched $c\bar c$ seed, inclusion of the $D\bar D^{*}$ hadronic loop can lower the dressed level toward the observed mass~\cite{Li2009}.
This supports a description in which a short-distance $c\bar c$ component coexists with a substantial continuum component, although the resulting mass and composition remain sensitive to the transition, regulator, and channel content~\cite{Huang2022-1,Ferretti2013}.
Similarly, in the bottomonium sector, self-energies due to coupling to open-bottom meson--meson channels have been computed using relativized quark model wave functions~\cite{Ferretti2012,Ferretti2014}.  For the low-lying $1S$, $2S$, and $1P$ bottomonium states the resulting mass shifts are negative and of the order of a few tens of MeV, while higher excitations near $B^{(*)}\bar B^{(*)}$ thresholds receive larger corrections, and the level spacings among the lowest states remain close to those of the underlying quenched spectrum~\cite{Ferretti2012,Ferretti2014}.  

Baryon self-energies from baryon-meson loops have been evaluated in a similar framework.  Ref.~\cite{Tecocoatzi2017} considered ground-state octet and decuplet baryons coupled to the baryon-meson continuum through a microscopic ${}^3P_0$ operator, including all ground-state baryons and pseudoscalar mesons in the intermediate states, and obtained negative mass shifts of several hundred MeV for all octet and decuplet baryons, with typical values such as \ensuremath{-}377 MeV for the nucleon and \ensuremath{-}410 MeV for the $\Delta$.  Channels containing $\pi$ and $K$ mesons were found to provide the dominant contributions whenever phase space allows~\cite{Tecocoatzi2017}.

A common procedure is to determine the valence Hamiltonian $H_0$ and the pair-creation vertex in a loop-aware fit.
The physical masses are then written as
\begin{equation}
M_i^{\rm phys} = M_i^{(0)}
+ \Re\Sigma_i(M_i^{\rm phys}),
\qquad
H_0\ket{i}= M_i^{(0)}\ket{i}.
\end{equation}
Here $\Sigma_i(E)$ is generated by the included meson--meson or meson--baryon channels~\cite{Swanson2006,Barnes2008,Geiger1991}.
Low-lying and spin-averaged states can be used to constrain the baseline parameters $(\kappa,\sigma,D,m_q)$.
Near-threshold and excited states give stronger constraints on the transition strength, regulator, and channel dependence~\cite{Swanson2006,Barnes2008}.

In several unquenched quark-model calculations, the leading loop correction is approximately common within a low-lying valence multiplet.
This behavior is most accurate when the relevant intermediate channels form nearly complete and approximately degenerate multiplets.
Threshold splittings, incomplete channel spaces, and channel-dependent phase space generate residual state-dependent shifts and configuration mixing~\cite{Barnes2008,Chen2018,Geiger1991,Kanwal2022}.
The smooth part of the correction can be absorbed into constituent masses and additive energy parameters.
The remaining part must be treated explicitly when it is comparable to the required spectroscopic accuracy.
Coupled-channel charmonium calculations show that a refit can preserve much of the low-lying level-spacing pattern even when the common mass shift is sizable~\cite{Li2009,Ni2024}.
The fitted constant $-D$ can absorb part of this smooth continuum dressing.
However, $D$ is also correlated with the constituent masses and the other potential parameters.
It should therefore not be interpreted as a direct measure of the hadronic-loop self-energy.

Taken together, these studies indicate that a smooth part of the continuum correction can often be absorbed into a refit of low-lying spectra~\cite{Santopinto2013,Barnes2008,Hammer2016}.
This approximation is less reliable for a state controlled by a nearby strongly coupled threshold.
In such cases, explicit coupled-channel dynamics is required.
From  Section~\ref{sec:baryon_GEM} onward, we use established ground-state mesons and baryons as the calibration set.
These states have relatively clear spectroscopic assignments and are generally less sensitive to orbital excitation and nearby-channel effects than highly excited states.
This choice does not imply that their individual hadronic-loop corrections vanish or that all of them are stable. 
It only reduces the expected sensitivity to omitted continuum dynamics. 
The distinction between a compact valence eigenvalue and a physical threshold pole is used explicitly in  Section~\ref{sec:excitation_spectra} when assessing threshold-sensitive excited states. 
Representative model calculations of these shifts are summarized in Table~\ref{tab:unquenching_shift}.

\begin{table}[!t]
  \centering
  \caption{\justifying Recent values of mass shifts from coupled-channel effects. Negative values indicate downward shifts. Each channel includes its spin-excited counterparts; for example, $K\bar K$ represents $K\bar K$, $\bar K K^*$, $K^* \bar K$, and $\bar{K}^* K^*$. We do not explicitly list all the channels for mesons. In these calculations, the loop-induced mass shifts are typically of order 10--100~MeV for deeply bound heavy quarkonia, but can reach several hundred MeV for light and heavy-light hadrons and for baryons strongly coupled to $S$-wave thresholds. Mass shift calculations in unquenched quark models are still not based on a universally agreed fitting prescription; for instance, Ref. \cite{Tecocoatzi2017} treats the bare energies $M_A^{(0)}$ as free parameters and determines them iteratively so that the physical masses satisfy $M_A^{\rm phys}=M_A^{(0)}+\Re\,\Sigma_A(M_A^{\rm phys})$, i.e.,~reproducing the observed masses after including the self-energy correction.}
  \label{tab:unquenching_shift}
  \resizebox{\columnwidth}{!}{%
\begin{tblr}{
  colspec={
Q[l,wd=4.2cm]
Q[l,wd=10.5cm]
Q[l,wd=4cm]
Q[l,wd=5.2cm]
  },
  row{1} = {bg=headgray, valign=m},
  hline{2} = {1-4}{0.3pt},
  rowsep = 1.5pt,
}
\toprule
\textbf{Meson} (type)
& \textbf{Shifts} (state or $n^{2S+1}L_\mathcal{J}$; MeV)
& \textbf{Valence scheme}
& \textbf{Coupled channels} \\

Light mesons ($\pi,\rho,\omega,\eta$)
& $\approx -37(\pi)$, $-100(\rho)$, $-66(\omega)$, $-65(\eta)$
& Chiral quark model
& $\pi\pi$, $\pi\rho$, $\rho\rho$, $K\bar K$, $\cdots$~\cite{Chen2018} \\

$D(c\bar{q})$ and excitations
& $-360(D)$, $-434(D^*)$, $-395(2^1S_0)$, $-393(2^3S_1)$
& OGE quark model
& $D\pi$, $D\eta$, $D\rho$, $D\omega$, $\cdots$~\cite{Hao2025} \\

$D_s(c\bar{s})$ and excitations
& $-304(1^1S_0)$, $-359(1^3S_1)$, $-373(2^1S_0)$, $-334(2^3S_1)$
& OGE quark model
& $DK$, $D_s \eta$, $D_s\phi$, $\cdots$~\cite{Hao2022} \\

Charmonium $(c\bar c)$ series
& $-57.8(\eta_c(1S))$, $-100.3(\eta_c(2S))$, $-66.8(J/\psi (1S))$, $-111.6(\psi(2S))$, $-86(\chi_{c0}(1P))$, $-91(\chi_{c1}(1P))$, $-125(\chi_{c0}(2P))$, $-135(\chi_{c1}(2P))$
& Chiral quark model
& $D\bar D$, $D_s \bar D_s$, $\cdots$~\cite{Chen2024} \\

Bottomonium $(b\bar b)$ series
& $-52(\eta_b)$, $-49(\Upsilon)$, $-53(\eta_b(2S))$, $-53(\Upsilon(2S))$, $-65(h_b)$
& ${}^3P_0$; bare $M_A^{(0)}$ fitted
& $BB$, $B_{s,c}B_{s,c}$, $(\eta_b\text{ or }\Upsilon)^2$~\cite{Ferretti2012} \\

$B(b\bar q)$ and excitations
& $-88.4(2^1S_0)$, $-45.5(2^3S_1)$, $-93.0(1P_1)$, $-41.3(1P'_1)$
& OGE quark model (rel)
& $B\pi$, $B\eta$, $B_sK$, $\cdots$~\cite{Ni2024} \\

Charmonium $(c\bar c)$ $2P$ series
& $-61.8(\chi_{c0}(2P))$, $-93.2(\chi_{c1}(2P))$, $-71.2(\chi_{c2}(2P))$, $-89.0(h_c(2P))$
& OGE quark model
& $D\bar D$, $D_s\bar D_s$~\cite{Man2024} \\

$B_s(b\bar s)$ and excitations
& $-114(1^1S_0)$, $-122(1^3S_1)$, $-146(2^1S_0)$, $-130(2^3S_1)$
& OGE quark model
& $BK$, $B_s\eta$, $B_s\phi$, $\cdots$~\cite{Hao2025Bs} \\
\bottomrule
\end{tblr}}
  \resizebox{\columnwidth}{!}{%
\begin{tblr}{
  colspec={
Q[l,wd=1.3cm]
Q[l,wd=1cm]
Q[l,wd=3.8cm]
Q[l,wd=18.5cm]
  },
  row{1} = {bg=headgray, halign=l, valign=m},
  hline{2} = {1-4}{0.3pt},
  rowsep = 1.5pt,
}
\toprule
\textbf{Baryon}
& \textbf{Shift}
& \textbf{Model}
& \textbf{Coupled channels} (their contributions to mass shift; MeV)~\cite{Tecocoatzi2017,GarciaTecocoatzi2018UQM} \\

$N$ & $-377$ & ${}^3P_0$ + flux-tube overlap & $N\pi (-178)$, $\Sigma K(-2)$, $\Delta \pi (-158)$, $N\eta(-8)$, $N\eta'(-5)$, $\Sigma^*K(-11)$, $\Lambda K(-15)$ \\

$\Delta$ & $-410$ & ${}^3P_0$ + flux-tube overlap & $N\pi (-120)$, $\Sigma K(-16)$, $\Delta \pi (-186)$, $\Delta \eta (-46)$, $\Delta\eta'(-25)$, $\Sigma^*K(-17)$ \\

$\Lambda$ & $-359$ & ${}^3P_0$ + flux-tube overlap & $N\bar K (-123)$, $\Sigma \pi (-78)$, $\Xi K(-3)$, $\Lambda \eta(-4)$, $\Lambda \eta'(-5)$, $\Sigma^*\pi(-124)$, $\Xi^*K(-22)$ \\

$\Sigma$ & $-345$ & ${}^3P_0$ + flux-tube overlap &
$N\bar K(-6)$, $\Sigma\pi(-79)$, $\Lambda\pi(-34)$, $\Sigma\eta(-22)$, $\Sigma\eta'(-5)$, $\Xi K(-29)$,
$\Delta\bar K(-113)$, $\Sigma^*\pi(-30)$, $\Sigma^*\eta(-20)$, $\Sigma^*\eta'(0)$, $\Xi^*K(-8)$ \\

$\Sigma^*$ & $-394$ & ${}^3P_0$ + flux-tube overlap &
$N\bar K(-33)$, $\Sigma\pi(-33)$, $\Lambda\pi(-60)$, $\Sigma\eta(-15)$, $\Xi K(-6)$,
$\Delta\bar K(-91)$, $\Sigma^*\pi(-101)$, $\Sigma^*\eta(-8)$, $\Sigma^*\eta'(-24)$, $\Xi^*K(-23)$ \\

$\Xi$ & $-328$ & ${}^3P_0$ + flux-tube overlap &
$\Sigma\bar K(-163)$, $\Lambda\bar K(-7)$, $\Xi\pi(-8)$, $\Xi\eta'(-16)$, $\Xi\eta(-4)$,
$\Sigma^*\bar K(-42)$, $\Xi^*\pi(-44)$, $\Xi^*\eta(-20)$ \\

$\Xi^*$ & $-376$ & ${}^3P_0$ + flux-tube overlap &
$\Sigma\bar K(-38)$, $\Lambda\bar K(-47)$, $\Xi\pi(-54)$, $\Xi\eta'(-16)$, $\Xi\eta(0)$,
$\Sigma^*\bar K(-138)$, $\Xi^*\pi(-39)$, $\Xi^*\eta(-2)$, $\Xi^*\eta'(-24)$, $\Omega K(-18)$ \\

$\Omega$ & $-374$ & ${}^3P_0$ + flux-tube overlap &
$\Xi \bar K(-163)$, $\Xi^* \bar K(-138)$, $\Omega \eta(-19)$, $\Omega \eta'(-54)$ \\
\bottomrule
\end{tblr}}

\resizebox{\columnwidth}{!}{%
\begin{tblr}{
  colspec={
Q[l,wd=3cm]
Q[l,wd=3cm]
Q[l,wd=2.5cm]
Q[l,wd=15cm]
  },
  row{1} = {bg=headgray, halign=l, valign=m},
  hline{2} = {1-4}{0.3pt},
  rowsep = 1.5pt,
}
\toprule
\textbf{Baryon} (type)
& \textbf{Quenched} (MeV)
& \textbf{Shift} (MeV)
& \textbf{Valence (unperturbed) model and scheme} \\

$\Lambda_c(2P,3/2^-)$ & $3012$ & $-75$ & OGE quark model and ${}^3P_0$ ($D^*N$ continuum)~\cite{Luo2020} \\

& & $-92$ & OGE quark model, ${}^3P_0$ ($D^*N$ continuum), $D^* N-D^*N$ interaction (unitarized)~\cite{Zhang2023} \\

$\Lambda_c(2P,1/2^-)$ & $2994$ & $-16$ & OGE quark model and ${}^3P_0$ ($D^*N$ continuum)~\cite{Luo2020} \\

& & $-53$ & OGE quark model, ${}^3P_0$ ($D^*N$ continuum), $D^* N-D^*N$ interaction (unitarized)~\cite{Zhang2023} \\

$\Omega_c(1P,1/2^-)$ & $-$ & $-97$ & OGE quark model, ${}^3P_0$; coupled to $\Xi_c \bar K$ and $\Xi_c' \bar K$ ($S$-wave); predicts $M_{\rm phys}=2945$ MeV~\cite{Luo2021} \\
\bottomrule
\end{tblr}}
  
\end{table}

\subsection{The chiral quark model}\label{chi_QM}

Constituent-quark models represent non-perturbative QCD effects, including dynamical chiral symmetry breaking (DCSB) and confinement, through effective quark masses and interactions. Beyond the Cornell potential with OGE, the chiral quark model implements the consequences of spontaneous chiral symmetry breaking at the quark level: Goldstone-boson exchange (GBE) among light quarks generates a characteristic flavor-spin force that reshuffles level orderings (e.g.\ the Roper pattern), and instanton-induced interactions (III; the 't~Hooft/KMT vertex) supply additional short-range, flavor-selective multi-quark forces not captured by OGE or GBE alone.

\subsubsection{The quark model from Goldstone-boson exchange (GBE)}

For $N_f=3$, the light-quark sector of QCD is described by
\begin{equation}
\mathcal L_{\rm QCD} = \bar q \left( i\slashed D-\hat m \right) q -
\frac14 G_{\mu\nu}^cG^{c,\mu\nu}, \qquad \hat m = \operatorname{diag}(m_u,m_d,m_s).
\end{equation}
The $N_f=2$ theory follows by restricting the flavor space to $q=(u,d)$.
In the chiral limit, the theory is invariant under
$\mathrm{SU}(N_f)_L\times\mathrm{SU}(N_f)_R$.
The QCD vacuum breaks this symmetry spontaneously to $\mathrm{SU}(N_f)_V$.
The breaking is characterized by a nonzero order parameter such as
$\langle\bar q q\rangle$ and by the appearance of pseudoscalar Goldstone modes.
The quark masses generate explicit breaking and lead to the Gell-Mann--Oakes--Renner (GMOR) relation~\cite{Roberts1994,Gasser1984,Gell-Mann1968}.

Dynamical chiral symmetry breaking also dresses the quark propagator,
\begin{equation}
S^{-1}(p) = Z^{-1}(p^2)
\left[ i\slashed p+M(p^2) \right].
\end{equation}
Lattice and Dyson--Schwinger calculations find an infrared constituent-like mass scale of several hundred MeV for light quarks~\cite{Bowman2005,Valcarce2005}.
This provides the basis for the constituent-quark description at hadronic scales.
In the following chiral constituent-quark description, the constant mass $M$ represents an effective low-momentum constituent mass associated with the dressed-quark mass function $M(p^2)$.
It is not the current-quark mass appearing in $\hat m$.
The flavor-dependent masses $m_i$ used in the exchange potential have the same effective constituent-mass interpretation, although their numerical values remain model dependent.

In one commonly used nonlinear chiral constituent-quark formulation, the Goldstone fields are introduced through
\begin{equation}
\mathcal L_{\chi{\rm QM}} = \bar Q \left( i\slashed\partial - M U^{\gamma_5}
\right) Q + \frac{f_\pi^2}{4} \operatorname{Tr} \left( \partial_\mu U\, \partial^\mu U^\dagger \right) +\cdots, \qquad U =\exp\left[
\frac{i\lambda^a\phi^a}{f_\pi}
\right],
\qquad
U^{\gamma_5} = \exp\left[
\frac{i\gamma_5\lambda^a\phi^a}{f_\pi}
\right].
\end{equation}
Expansion of $U^{\gamma_5}$ generates the constituent mass term and the coupling of the Goldstone fields to light constituent quarks~\cite{Valcarce2005,Manohar1984,Vijande2005}.

At leading order, pseudoscalar exchange $\phi\in\{\pi,K,\eta\}$ generates flavor-dependent spin-spin and tensor interactions between constituent quarks~\cite{Glozman1996,Valcarce2005,Vijande2005,Glozman1998}. A commonly used monopole-regulated coordinate-space form is~\cite{Valcarce2005,Vijande2005} \begin{align} 
V^{(\phi)}_{ij}(\mathbf r) ={}& \frac{g_{ch}^2}{4\pi} \frac{m_\phi^2}{12m_im_j} \frac{\Lambda_\phi^2}{\Lambda_\phi^2-m_\phi^2} m_\phi \mathcal F_{ij}^{(\phi)} \Bigg\{ \bigg[ Y(m_\phi r) - \frac{\Lambda_\phi^3}{m_\phi^3} Y(\Lambda_\phi r) \bigg] \boldsymbol{\sigma}_i\cdot\boldsymbol{\sigma}_j+ \bigg[ H(m_\phi r) - \frac{\Lambda_\phi^3}{m_\phi^3} H(\Lambda_\phi r) \bigg] S_{ij} \Bigg\}, \label{eq:GBE_general} 
\end{align} where $m_i$ denotes the constituent-quark mass and the flavor factors for the three exchanged pseudoscalar mesons are 
\begin{align} 
\mathcal F_{ij}^{(\pi)} &= \sum_{a=1}^{3}\lambda_i^a\lambda_j^a, \qquad \mathcal F_{ij}^{(K)} = \sum_{a=4}^{7}\lambda_i^a\lambda_j^a, \qquad \mathcal F_{ij}^{(\eta)} = \cos\theta_P\,\lambda_i^8\lambda_j^8-\sin\theta_P\mathbf{1}. \label{eq:GBE_flavor_factors} 
\end{align} Here $\lambda^a$ are the SU(3) flavor Gell-Mann matrices and $\theta_P$ is the pseudoscalar octet--singlet mixing angle associated with the use of the physical $\eta$ rather than the pure octet $\eta_8$ state. The last term in $\mathcal F_{ij}^{(\eta)}$ is proportional to the identity in flavor space. The radial functions are 
\begin{equation} 
Y(x)=\frac{e^{-x}}{x}, \qquad H(x)= \left( 1+\frac{3}{x}+\frac{3}{x^2} \right) Y(x). \end{equation}
The tensor operator is $S_{ij} = 3 \left( \hat{\mathbf r}\cdot\boldsymbol{\sigma}_i \right) \left( \hat{\mathbf r}\cdot\boldsymbol{\sigma}_j \right) - \boldsymbol{\sigma}_i\cdot\boldsymbol{\sigma}_j$. The precise short-distance form remains dependent on the regulator convention.

\paragraph{Phenomenology and practical implementation:}
In phenomenological applications, the GBE quark model covers two related but distinct strategies. In the ``pure'' GBE model, the flavor-spin force is taken as the dominant short-range hyperfine interaction~\cite{Glozman1996,Glozman1998}. In broader chiral constituent quark models, confinement is supplemented by regulated GBE together with a residual OGE core at very short distances and, in many cases, an intermediate-range scalar attraction~\cite{Valcarce2005,Vijande2005,Garcilazo2003}. The reason for these two alternative approaches is that spectroscopy constrains the observed hyperfine splittings, but does not by itself specify how they are decomposed into OGE and GBE contributions.

In the chiral quark model, GBE provides an important interaction for light quarks but can overestimate short-range and tensor forces if used alone. Its effects are primarily restricted to light-light quark pairs, while in heavy hadrons it influences the system indirectly through the light-quark sector, rather than acting as a dominant interaction between heavy quarks~\cite{Neubert1994}.

\paragraph{Spectroscopy, successes, and limitations:}
The distinctive operator of the GBE picture is the flavor-spin interaction
\begin{equation}\label{eq:GBE_flav_spin}
\mathcal{V}^{FS}\propto \sum_{i<j}\frac{1}{m_im_j}\lambda_i^F\lambda_j^F\,
\boldsymbol{\sigma}_i\cdot\boldsymbol{\sigma}_j,
\end{equation}
whose matrix elements are controlled by the flavor and spin symmetry of the state~\cite{Glozman1996,Glozman1998}. In GBE-based models, enhanced attraction in the symmetric flavor-spin configuration can lower the first positive-parity excitation relative to the negative-parity states.
This mechanism has been used to reproduce the qualitative Roper ordering in the light-baryon spectrum~\cite{Glozman1996,Glozman1998}.
In heavy baryons, the same logic relates the $\Lambda_Q-\Sigma_Q$ mass difference to the differing magnitude of Eq.~\eqref{eq:GBE_flav_spin} for the light diquark, although the two contributions have the same sign. This contrasts with the OGE color-spin interaction, for which the relevant terms differ not only in magnitude but also in sign.
Relativistic formulations improve parts of the low-lying nonstrange spectrum~\cite{Garcilazo2003,Garcilazo2005}.
GBE-based valence models can lower the calculated $\Lambda(1405)$ relative to simple OGE models.
They do not, however, reproduce its observed structure as a pure $qqq$ state.
Coupled-channel analyses find strong $\bar KN$ and $\pi\Sigma$ dynamics and two nearby poles~\cite{Hyodo2012,Jido2003}.

\subsubsection{Instanton-induced interactions}
\label{subsec:instanton_3body}
Instanton-induced interactions (III) provide the constituent-quark realization of the $\mathrm{U}_{A}(1)$ anomaly. For three light flavors, the underlying local interaction is the Kobayashi-Maskawa-'t~Hooft (KMT) determinant~\cite{tHooft1976,KobayashiMaskawa1970,Schafer1998,Diakonov2003},
\begin{equation}
  \mathcal L_{\mathrm{KMT}}
  =
  G_D \big[
\det_{f,f'=u,d,s}\!\bigl(\bar q_f (1+\gamma_5) q_{f'}\bigr)
+
\det_{f,f'=u,d,s}\!\bigl(\bar q_f (1-\gamma_5) q_{f'}\bigr)
  \big],
  \label{eq:L_KMT}
\end{equation}
which is totally antisymmetric in flavor and generates a local six-quark vertex. The six-quark KMT vertex gives the effective three-body contact interactions. Contracting two quarks into a condensate gives the two-body interactions. The potential forms used below follow from the leading non-relativistic reduction in the Weyl basis, with separate $qqq$, $qq$, and $q\bar q$ operators.

The three-body piece is the local $uds$ contact term before contraction,
\begin{equation}
  H_{\mathrm{III}}^{(3)}
  =
  V_0 \int d^3x\;
  \bar q_{Ri}(x)\bar q_{Rj}(x)\bar q_{Rk}(x)\,
  \mathcal O^{(3)}_{ijk}\,
   q_{Lk}(x) q_{Lj}(x) q_{Li}(x)
  + \mathrm{h.c.},
  \qquad (i,j,k)\sim (u,d,s),
  \label{eq:HIII_3body}
\end{equation}
with the explicit color-spin operator written in orbital $S$-wave,\footnote{We can expand the flavor antisymmetrizer $\mathcal{A}^F_{ijk}=[1-(12)_F-(23)_F-(31)_F+(123)_F+(132)_F]/6$ obtained from Eq.~\eqref{eq:HIII_3body} using the Pauli principle $(ij)_C(ij)_F(ij)_S=-1$. For example, $(12)_F$ term can be written as
\begin{equation*}
(12)_F=-(12)_C(12)_S=-\Big(\frac{1}{3}+\frac{1}{2}\lambda^c_1\lambda^c_2\Big)\Big(\frac{1}{2}+\frac{1}{2}\boldsymbol{\sigma}_1\cdot\boldsymbol{\sigma}_2\Big)=\cdots.
\end{equation*}
When antiquarks are included, the potential form of Eq.~\eqref{eq:VIII3_explicit} is obtained by an appropriate reordering of the fermion fields in Eq.~\eqref{eq:HIII_3body}.}  in the Takeuchi-Oka convention~\cite{TakeuchiOka1991}, as
\begin{align}
  \mathcal O^{(3)}_{ijk}
  =&\;
1
+ \frac{3}{32}
  (\lambda_i^c\lambda_j^c+\text{perms})
- \frac{9}{320}\, d_{abc}\,\lambda_i^a\lambda_j^b\lambda_k^c + \frac{9}{32}( \lambda_i^c\lambda_j^c\,\boldsymbol{\sigma}_i\cdot\boldsymbol{\sigma}_j+\text{perms})
  \nonumber\\
  &
+ \frac{27}{320}\, d_{abc}\,\lambda_i^a\lambda_j^b\lambda_k^c
\Bigl(
  \boldsymbol{\sigma}_i\cdot\boldsymbol{\sigma}_j
  + \boldsymbol{\sigma}_j\cdot\boldsymbol{\sigma}_k
  + \boldsymbol{\sigma}_k\cdot\boldsymbol{\sigma}_i
\Bigr)
- \frac{9}{64}\, f_{abc}\,\lambda_i^a\lambda_j^b\lambda_k^c\,
  (\boldsymbol{\sigma}_i\times\boldsymbol{\sigma}_j)\cdot\boldsymbol{\sigma}_k.
  \label{eq:OIII3_explicit}
\end{align}
At leading non-relativistic order this gives the local three-quark potential
\begin{equation}
  V_{\mathrm{III},ijk}^{(3)}
  =
  V_0\,
  \mathcal O^{(3)}_{ijk}\,
  \delta^{(3)}(\mathbf r_{ij})\,\delta^{(3)}(\mathbf r_{jk}),
  \label{eq:VIII3_explicit}
\end{equation}
which survives only for compact $uds$ flavor singlet configurations. For ordinary $qqq$ baryons the genuine three-body term vanishes because one cannot have a color singlet and flavor singlet $S$-wave three-quark configuration, whereas it survives for compact flavor singlet $uds$ subclusters in multiquark configurations.

After contracting one bilinear, the effective two-body Hamiltonian becomes
\begin{equation}
  H_{\mathrm{III}}^{(2)} = \sum_{i<j} V_0^{(2)}(i,j)\int d^3x\; \bar q_{Ri}(x)\bar q_{Rj}(x) \Bigl( 1+\frac{3}{32}\lambda_i^c\lambda_j^c +\frac{9}{32}\lambda_i^c\lambda_j^c\,\boldsymbol{\sigma}_i\cdot\boldsymbol{\sigma}_j \Bigr) q_{Lj}(x) q_{Li}(x) + \mathrm{h.c.}, \label{eq:HIII_2body}
\end{equation}
which in the $qq$ channel reduces to
\begin{equation}
  V_{\mathrm{III},ij}^{(2;qq)}
  =
  V_0^{(2)}(i,j)
  \Bigl(
1+\frac{3}{32}\lambda_i^c\lambda_j^c
+\frac{9}{32}\lambda_i^c\lambda_j^c\,\boldsymbol{\sigma}_i\cdot\boldsymbol{\sigma}_j
  \Bigr)
  \delta^{(3)}(\mathbf r_{ij}), \qquad (i\neq j)
  \label{eq:VIII2_qq_explicit}
\end{equation}
namely the familiar flavor-antisymmetric short-range instanton attraction. For a color-$\bar{\mathbf 3}_C$, spin--singlet $qq$ pair, one finds
$\langle\lambda_i^c\lambda_j^c\rangle=-8/3$ and
$\langle\boldsymbol{\sigma}_i\cdot\boldsymbol{\sigma}_j\rangle=-3$.
The color-spin operator in Eq.~\eqref{eq:VIII2_qq_explicit} then has the positive matrix element 3.
It also has the non-negative value $3/2$ in the color-$\mathbf 6_C$, spin--triplet channel and vanishes in the other two color-spin pair channels.
Thus, the sign of the two-body potential is fixed by the coupling $V_0^{(2)}(i,j)$.

The effective two-body coupling is related to the three-body coupling $V_0$ by~\cite{TakeuchiOka1991}
\begin{equation}\label{eq:iii_2_3_coup_rel}
V_0^{(2)}(i,j) = \frac12V_0 \left( \langle\bar q q\rangle - K_{\rm III}m_k^{\rm cur} \right)
= -\frac12V_0K_{\rm III}m_k, \qquad (i,j,k)\sim(u,d,s).
\end{equation}
Here $K_{\rm III}>0$ relates the constituent mass, the current mass, and the chiral condensate through $m_k =m_k^{\rm cur}-{\langle\bar q q\rangle}/{K_{\rm III}}$.
In the Takeuchi--Oka convention, $V_0>0$ and therefore $V_0^{(2)}(i,j)<0$.
The two-body instanton-induced interaction is consequently attractive in the flavor-antisymmetric $qq$ channels on which it acts. Crossing one leg to the antiquark channel and keeping the leading $q\bar q \to q\bar q$ term gives
\begin{equation}
  V_{\mathrm{III},i\bar j}^{(2;q\bar q)}
  =
  V_0^{(2)}(i,j)
  \Bigl(
1-\frac{3}{32}\lambda_i^c \bar\lambda_j^c
+\frac{9}{32}\lambda_i^c \bar\lambda_j^c\boldsymbol{\sigma}_i \cdot  {{\boldsymbol{\sigma}}}_j
  \Bigr)
  \delta^{(3)}(\mathbf r_{ij}),\qquad (i\neq j)
  \label{eq:VIII2_qbarq_explicit}
\end{equation}
where, for an antiquark, $\bar\lambda_j=-\lambda_j^{T}$. For a color singlet, spin singlet $q\bar q$ pair one then finds $\langle \lambda_i^c\bar{\lambda}_{j}^c\rangle=-16/3$ and $\langle \boldsymbol{\sigma}_i\cdot{\boldsymbol{\sigma}}_{j}\rangle=-3$, so that $\langle V_{\mathrm{III},i\bar j}^{(2;q\bar q)}\rangle = 6\,V_0^{(2)}(i,j)\delta^{(3)}(\mathbf r)$. Similarly, for the three-body operator in Eq.~\eqref{eq:OIII3_explicit}, even when an antiquark is included, one can obtain the color-spin structure of $V_{\text{III}}^{(3)}$. The contact interaction is typically smeared over a finite range, often associated with an effective instanton size (e.g.,~$\sim 0.3$ fm); however, this should be understood as a model parameter encoding the underlying nonlocality rather than a direct identification with the QCD instanton size. Higher order III tensor and (anti) spin--orbit terms can be found in Ref.~\cite{Takeuchi1998}.

\paragraph{Phenomenology and multiquark implications:}

The two-body instanton term $V^{(2)}_{\mathrm{III}}$ has nearly the same short-range spin-spin structure as the OGE color-spin interaction $V^{CS}$ (e.g.,~Eq.~\eqref{VCS_form_two_types}). For this reason, low-lying baryon spectroscopy constrains mainly the sum of the two contributions rather than each piece separately. A useful parametrization is
\begin{equation}
H = \sum_i m_i + K + V^{C}
+ \sum_{(ij)\in LL}\!\Big[(1-p_{\mathrm{III}})\,V^{CS}_{ij}
+ p_{\mathrm{III}}\,V^{(2)}_{\mathrm{III},ij}\Big]
+ \sum_{(ij)\in LH,HH}\! V^{CS}_{ij}
+ \sum_{(i,j,k)\sim(u,d,s)}p_{\rm III}V^{(3)}_{\mathrm{III},ijk} ,
\label{eq:Hmix_III}
\end{equation}
where $LL$, $LH$, and $HH$ denote light-light, light-heavy, and heavy-heavy pairs, respectively; the two-body instanton-induced interaction $V^{(2)}_{\mathrm{III}}$ acts only on $LL$ pairs. Whenever compact $uds$-singlet clusters are important, the genuine three-body term $V^{(3)}_{\mathrm{III}}$ is added explicitly. The total hyperfine strength is fixed from the $\Delta-N$ splitting, while the sharing parameter $p_{\mathrm{III}}$ is constrained from the $\eta-\eta'$ sector. With this treatment and Eq.~\eqref{eq:VIII2_qbarq_explicit}, the model can also determine the $\eta-\eta'$ mixing angle. For the underlying role of the KMT interaction in the $\eta-\eta'$ problem, see also Refs.~\cite{KobayashiMaskawa1970,Kunihiro2009}. One typically finds $p_{\mathrm{III}}\sim 0.3$--0.4, with a broader phenomenological range of roughly 0.25--0.6 depending on wave functions and operator conventions
\cite{TakeuchiOka1991,Takeuchi1998,TakeuchiOka1992,IrieOkaYasui2018}. 

The two- and three-body parts of III generally act in opposite directions in the compact multiquark channels studied in these models.
The two-body coupling $V_0^{(2)}$ is negative, while its color-spin operator is non-negative in the relevant flavor-antisymmetric $qq$ channels.
The two-body contribution is therefore attractive and becomes stronger when the configuration contains several such light-quark pairs.
The three-body coupling has the opposite sign.
Its contribution is usually repulsive in compact flavor-singlet $uds$ configurations, although its matrix element remains dependent on the color-spin-flavor channel~\cite{TakeuchiOka1991,TakeuchiOka1992,IrieOkaYasui2018}.
In many of the multiquark channels considered so far, the two-body attraction is larger than the three-body repulsion.
The net instanton-induced correction then remains attractive, while the three-body term reduces its magnitude.
This pattern is found, for example, in pentaquark calculations in which the total III contribution lowers the mass despite the repulsive $uds$ three-body term~\cite{OkaTakeuchi1991,ShinozakiOkaTakeuchi2004}.
The compact $H$-dibaryon is an important case in which the three-body repulsion is particularly large and can remove the deep binding obtained from the two-body attraction alone~\cite{TakeuchiOka1991,TakeuchiOka1992,IrieOkaYasui2018}.
At the baryon level, the KMT interaction also induces a repulsive short-range three-baryon contribution in hyperonic channels such as $NN\Lambda$ and $N\Lambda\Lambda$, while it vanishes for $NNN$~\cite{OhnishiKashiwaMorita2017}.
In multiquark systems containing compact $uds$ subclusters --- either a  $uds$ flavor singlet or antiquark-containing configurations such as $ud\bar{s}$ and $u\bar{d}\bar{s}$ --- the spectrum of possible stable configurations and their excitations inferred from the OGE color--spin interaction $V^{CS}$ alone may be incomplete. 
In such cases, instanton-induced two- and three-body interactions should be incorporated already at the operator level, as they can compete with $V^{CS}$ and, in certain systems, alter the channel ordering predicted by $V^{CS}$ alone; see, for example, Refs.~\cite{TakeuchiOka1991,IrieOkaYasui2018,OkaTakeuchi1991,ShinozakiOkaTakeuchi2004,OkaTakeuchi1989}. 
The relation between this KMT mechanism and the present TGE-inspired connected interaction, including the overlap with fitted two-body terms and the resulting need for a joint refit, is summarized in  Section~\ref{subsec:three_quark_definition}.

\subsection{Technical remarks on quark model spectroscopy}
\label{subsec:tech_remarks}
This subsection briefly collects the numerical tools used in the quark model spectroscopy: resonance extraction in an $L^2$ basis and the treatment of Jacobi-coordinate rearrangements in few-body calculations.

\subsubsection{Resonances in an $L^2$ basis: complex and real scaling}
\label{subsubsec:complex_real_scaling}
Near an open two-hadron threshold, a state may appear as a resonance rather than as a bound state.
A resonance satisfies an outgoing-wave boundary condition and is not square integrable in the usual sense.
An unmodified finite $L^2$ diagonalization instead replaces the continuum by a discrete set of basis-dependent pseudostates, so an individual positive-energy eigenvalue cannot by itself be identified with a resonance.
Complex scaling and real scaling provide two complementary ways of extracting resonance information from an $L^2$ basis~\cite{Moiseyev1998}.

A resonance corresponds to a pole of the analytically continued scattering amplitude at $E_{\rm pole}=M_R-i\Gamma_R/2$, as in Eq.~\eqref{eq:unquenched_complex_pole}.
In the complex scaling method (CSM), let $\mathbf k=-i\nabla_{\mathbf r}$ and introduce the Hermitian dilation generator and the associated non-unitary similarity transformation,
\begin{equation}
\hat G=\frac{1}{2}\left(\mathbf r\cdot\mathbf k+\mathbf k\cdot\mathbf r\right),\qquad
H(\theta)=U(\theta)HU^{-1}(\theta),\qquad
U(\theta)=e^{-\theta\hat G}.
\label{eq:complex_scaling_transformation}
\end{equation}
For a many-body system, $\hat G$ is summed over the independent relative coordinates to which the scaling is applied.
The transformation gives
\begin{equation}
U(\theta)\mathbf rU^{-1}(\theta)=e^{i\theta}\mathbf r,\qquad
U(\theta)\mathbf kU^{-1}(\theta)=e^{-i\theta}\mathbf k.
\label{eq:complex_scaling_coordinates}
\end{equation}
For dilation-analytic interactions, the continuum cut associated with each explicitly represented threshold is rotated by an angle $2\theta$ into the lower-half energy plane~\cite{Aguilar1971,Balslev1971}.
A resonance lying between the real axis and the relevant rotated cut then appears as an isolated eigenvalue of the non-Hermitian Hamiltonian $H(\theta)$.
In an exact calculation, the pole position is independent of $\theta$, while in a finite Gaussian basis it is identified from approximate stationarity under variations of $\theta$ and of the basis parameters.
The complex rotation regularizes the outgoing Gamow wave function and encodes the corresponding analytic continuation in the transformed Hamiltonian, so that a separate on-shell jump term need not be added when CSM is applied consistently to the full coupled-channel Hamiltonian.

A complementary $L^2$ approach is the real-scaling or stabilization method (RSM).
In a multicluster calculation, one enlarges the Gaussian ranges associated with the asymptotic intercluster coordinates, for example,
\begin{equation}
r_{n,\rm rel}(\alpha_{\rm sc})=\alpha_{\rm sc}r_{n,\rm rel}^{(0)},
\label{eq:real_scaling_ranges}
\end{equation}
while keeping the internal cluster scales fixed or varying them independently.
A finite basis then provides an effective infrared length $L(\alpha_{\rm sc})$, and discretized continuum levels typically move as $E-E_{\rm th}\propto L(\alpha_{\rm sc})^{-2}$.
A spatially localized bound or resonant component is much less sensitive to this asymptotic dilation and consequently produces a stabilization plateau or a sequence of avoided crossings with the moving continuum pseudostates.
The evolution of the level density can be related to the energy derivative of the scattering phase shift and to time-delay diagnostics based on the Krein--Friedel--Lloyd and Wigner--Smith relations~\cite{Newton1982,Krein1962,Friedel1958,Lloyd1967,Wigner1955,Smith1960}.
Practical procedures for extracting resonance parameters from stabilization spectra are discussed in Refs.~\cite{Hazi1970,Mandelshtam1993,Kukulin1989}.
Unlike CSM, however, RSM keeps the Hamiltonian and its finite-basis spectrum real and does not by itself impose an outgoing-wave boundary condition or select a Riemann sheet.
A plateau or avoided crossing is therefore a resonance diagnostic rather than, by itself, a direct solution of the complex pole equation, and broad, overlapping, or threshold-adjacent structures require a phase-shift, level-density, or time-delay analysis together with basis-convergence tests.

The Feshbach, RGM, and CSM formulations treat the same pole problem in different ways. In the Feshbach formulation, the continuum $\mathsf Q$ sector is eliminated and the pole is obtained from the zeros of $\mathsf D^{(\boldsymbol\eta)}(E)$ on the appropriate Riemann sheet. In a microscopic RGM calculation, the cluster channels are retained together with the exchange and norm kernels. In CSM, the outgoing-wave continuation is implemented by rotating the continuum cuts of the full Hamiltonian. If the same Hamiltonian, asymptotic channel space, and boundary conditions are treated exactly, the three formulations give the same $S$-matrix poles, although their numerical equations are different~\cite{Saito1977,FujiwaraEtAl2000LSRGM,LinChengHuangZhu2023}. If CSM is applied after the continuum has been eliminated, the energy dependence and branch structure of the self-energy must also be continued consistently. Scaling $H_{\mathsf P\mathsf P}$ alone is not sufficient.

This comparison requires the relevant continuum channels to be included in the Hamiltonian. In a fixed-particle-number multiquark system, compact configurations and color-singlet two-hadron rearrangements can belong to the same Hilbert space. A sufficiently complete multichannel CSM or RGM calculation can then describe a fall-apart resonance without introducing a separate phenomenological pair-creation vertex. For an ordinary valence $q\bar q$ meson or $qqq$ baryon, however, the physical two-hadron channels contain an additional $q\bar q$ pair and lie outside the valence space. These channels must first be included in an enlarged higher-Fock or hadronic channel space, or eliminated into the energy-dependent Hamiltonian described in Section~\ref{sec:unquenching}. Applying CSM or RSM only to the valence Hamiltonian cannot generate these missing channels.

In practice, a uniform complex rotation exposes only poles lying between the real axis and the rotated continuum cuts and within the Riemann-sheet sectors reached by the chosen contour.
Virtual states and some poles on more remote multichannel sheets may therefore require an improved contour deformation or the explicit sheet-by-sheet continuation used in Section~\ref{sec:unquenching}~\cite{ChenMengLinZhu2024}.

\subsubsection{Jacobi-coordinate rearrangements and angular-momentum recoupling}
In few-body calculations, the main bottleneck is often not the radial integration itself but the angular-momentum recoupling induced by transformations among different Jacobi coordinate sets.
One widely used approach is the hyperspherical-harmonics (HH) method, which handles this problem through hyperangular bases and Raynal--Revai transformations~\cite{Raynal1970,Das_book,Lin1995,Marcucci2020,Salom2017,Youping1987}. 
In the present work, we use the Gaussian expansion method (GEM) and avoid most explicit recoupling algebra by using infinitesimally shifted Gaussians (ISG)~\cite{hiyamaGEM}.

\paragraph{Gaussian expansion method (GEM):}

In GEM the few-body wave function is expanded in a finite set of Gaussian basis functions with geometric ranges, and the Schr\"odinger equation is solved by a Rayleigh--Ritz (Galerkin) variational diagonalization of the Hamiltonian matrix~\cite{hiyamaGEM}.  For three and more bodies the Jacobi coordinates are not unique: different rearrangement channels correspond to different pairings/clustering patterns and therefore to different Jacobi coordinates.  Accurate multi-Jacobi channel calculations thus require inter-channel transformations and angular-momentum recoupling between these Jacobi sets, which motivates using ISG to avoid explicit solid-harmonic and Racah algebra in the matrix elements.

\paragraph{Infinitesimally shifted Gaussians (ISG):}
The ISG method rewrites a solid-harmonic Gaussian as a finite superposition of shifted $S$-wave Gaussians. Each shifted Gaussian is $l=0$ with respect to its own center, so the angular dependence is carried by a fixed set of shift directions $\{\mathbf{D}_{lm,k}\}$ rather than explicit solid harmonics and Racah algebra. The key identity is the exact representation~\cite{hiyamaGEM,Suzuki1998}:
\begin{equation}\label{ISG_tech}
r^l e^{-\nu r^2}Y_{lm}(\Omega)=\lim_{\epsilon\to0}\Big(\frac{1}{\nu\epsilon}\Big)^l\sum_{k=1}^{k_\text{max}} C_{lm,k}e^{-\nu(\mathbf{r}-\epsilon \mathbf{D}_{lm,k})^2}.
\end{equation}
Here $\epsilon$ is an infinitesimal real scalar shift, $\{C_{lm,k}\}$ are dimensionless complex coefficients, and $\{\mathbf{D}_{lm,k}\}$ are dimensionless complex shift vectors. For $l>0$, one explicit construction is
\begin{align}
&C_{lm,k}  =\frac{l^l(-1)^{j+l-s-t-u+m(1-\text{sgn}(m))/2} (l+|m|)!}{2^{|m|+2l+2j} j! (|m|+j)! (l-|m|-2j)!}
\left[ \frac{(2l+1)(l-|m|)!}{4\pi(l+|m|)!} \right]^{1/2}
\binom{l-|m|-2j}{s} \binom{j+|m|}{t} \binom{j}{u}, \label{C_coeff}\\
&\mathbf{D}_{lm,k} = \frac{1}{l} \left[(2t-j-|m|) (\hat{\mathbf{e}}_x+i\;\text{sgn}(m)\hat{\mathbf{e}}_y)
+ (2u-j) (\hat{\mathbf{e}}_x-i\;\text{sgn}(m)\hat{\mathbf{e}}_y)
+ (2s-l+|m|+2j)\hat{\mathbf{e}}_z \right],\label{D_vector}
\end{align}
where $k=(j,s,t,u)$, each running from $(0,0,0,0)$ to $\left(\lfloor(l-\vert m\vert )/{2}\rfloor, l-\vert m\vert -2j, j+\vert m\vert , j\right)$.
For $l=0$, use a single unshifted term in Eq.~\eqref{ISG_tech}, with $C_{00,1}=Y_{00}=1/\sqrt{4\pi}$ and $\mathbf{D}_{00,1}=\mathbf{0}$.
Throughout we use associated Legendre functions $P_{lm}$ ($m>0$) without the Condon-Shortley phase $(-1)^m$:
\begin{equation}
P_{lm}(x)=\left(1-x^2\right)^{m/2} \frac{d^m}{dx^m}P_{l}(x).
\end{equation}
The ISG representation is not unique: many discrete sets $\{C_{lm,k},\mathbf D_{lm,k}\}$ lead to the same solid-harmonic Gaussian in the $\epsilon\to0$ limit. A convenient characterization is through the symmetric tensor
\begin{equation}
\mathcal M^{(n)}_{i_1\cdots i_n}(l,m)\equiv
\sum_k C_{lm,k}\,D_{lm,k,i_1}\cdots D_{lm,k,i_n}=0,
\qquad (n=0,1,\ldots,l-1),
\label{eq:ISG_moment_vanish}
\end{equation}
while the first nonvanishing $\mathcal{M}$ at rank $n=l$ reproduces the desired symmetric traceless tensor $T^{(lm)}$,
\begin{equation}\label{eq:T_exp}
\mathcal{M}_{i_1\cdots i_l}^{(l)}(l,m)=\frac{l!}{2^l}T^{(lm)}_{i_1\cdots i_l},
\qquad
r^lY_{lm}(\Omega)\equiv T^{(lm)}_{i_1\cdots i_l} r_{i_1}\cdots r_{i_l}.
\end{equation}
To see that these conditions imply Eq.~\eqref{ISG_tech}, expand
\begin{equation}
e^{-\nu(\mathbf r-\epsilon\mathbf D_k)^2} = e^{-\nu r^2} \sum_{p,q\ge0} \frac{(2\nu\epsilon)^p}{p!} \frac{(-\nu\epsilon^2)^q}{q!} (\mathbf r\cdot\mathbf D_k)^p(\mathbf D_k^2)^q .
\end{equation}
After summing over $k$, each contribution of total rank $n=p+2q$ is a contraction of $\mathcal M^{(n)}(l,m)$ with $p$ factors of $r_i$ and $q$ Kronecker deltas. Hence all terms with $n<l$ vanish by Eq.~\eqref{eq:ISG_moment_vanish}; the terms with $n=l$ and $q\ge1$ vanish because $T^{(lm)}$ is traceless; and the terms with $n>l$ are suppressed by positive powers of $\epsilon$. Therefore only the $p=l$, $q=0$ term survives, yielding
\begin{equation}
\lim_{\epsilon\to0} \Big(\frac{1}{\nu\epsilon}\Big)^l \sum_k C_{lm,k}e^{-\nu(\mathbf r-\epsilon \mathbf{D}_{lm,k})^2} = e^{-\nu r^2} \Big(\frac{1}{\nu\epsilon}\Big)^l \frac{(2\nu\epsilon)^l}{l!} \mathcal M^{(l)}_{i_1\cdots i_l}(l,m)\, r_{i_1}\cdots r_{i_l} = e^{-\nu r^2}r^lY_{lm}(\Omega),
\end{equation}
where Eq.~\eqref{eq:T_exp} was used in the last step. The explicit ISG construction in Eqs.~\eqref{C_coeff}--\eqref{D_vector} gives one realization of the ISG identity~\cite{hiyamaGEM,Suzuki1998,Yu2023}. However, it is not unique: any set $\{C_{lm,k},\mathbf D_{lm,k}\}$ satisfying Eqs.~\eqref{eq:ISG_moment_vanish} and~\eqref{eq:T_exp} yields the same $\epsilon\to0$ limit. A particularly useful alternative is obtained from a spherical cubature rule on $S^2$ that is exact for all polynomials of degree $\le 2l$. Let the weights and nodes $\{w_k,\hat{\mathbf n}_k\}_{k=1}^{N_l}$ denote such a rule,\footnote{Here $\{\hat{\mathbf n}_k,w_k\}_{k=1}^{N_l}$ denotes a chosen spherical cubature rule on $S^2$ that is exact for all polynomials of degree $\le 2l$, i.e.,~
\[ \sum_{k=1}^{N_l} w_k\,p(\hat{\mathbf n}_k) =
\int_{S^2} p(\hat{\mathbf n})\,d\Omega
\qquad (\deg p\le 2l). \]
The vectors $\hat{\mathbf n}_k$ are the cubature nodes (unit directions on the sphere), $w_k$ are the corresponding weights, and $N_l$ is the number of nodes in the chosen rule.} whose values are given in Refs.~\cite{Lebedev1976,LebedevLaikov1999,Delsarte1977} and codes are also available in Ref.~\cite{BurkardtSphereLebedev} (one must replace $w_k\to 4\pi \,w_k$ or equivalently absorb the factor $4\pi$ to use this reference code). 
Then one may choose
\begin{equation}
\mathbf D^{(\mathrm{cub})}_{lm,k}=\rho_l\,\hat{\mathbf n}_k, \qquad C^{(\mathrm{cub})}_{lm,k} = \frac{(2l+1)!!}{4\pi\,2^l\,\rho_l^{\,l}}\, w_k\,Y_{lm}(\hat{\mathbf n}_k),
\label{eq:ISG_cubature}
\end{equation}
with arbitrary nonzero scale $\rho_l$, using the same phase convention for $Y_{lm}$ as above. By polynomial exactness and spherical-harmonic orthogonality, this choice automatically satisfies Eqs.~\eqref{eq:ISG_moment_vanish} and~\eqref{eq:T_exp}. In this construction the direction set is real and independent of $m$, while all $m$-dependence resides in the coefficients $C_{lm,k}$; for Ref.~\cite{Delsarte1977}, one simply has $w_k=4\pi/N_l$. Thus the standard $k=(j,s,t,u)$ prescription can be understood as a convenient representative of a wider class of ISG constructions that satisfy the same tensor conditions. For complex spherical harmonics the coefficients $C_{lm,k}$ are generally complex even when the shift directions are real; if instead one adopts real tesseral harmonics, both $C_{lm,k}$ and $\mathbf D_{lm,k}$ may be taken real.

In practice, Eq.~\eqref{ISG_tech} replaces orbital recoupling/rotation algebra by sums of dot products among the fixed vectors $\mathbf{D}_{lm,k}$, enabling efficient mixed-channel matrix elements in GEM~\cite{hiyamaGEM,Suzuki1998,Varshalovich1988}. For any few-body basis whose coordinate-space wave functions can be written as finite linear combinations of Gaussians times polynomials (in the Jacobi coordinates), the evaluation of matrix elements is equivalent to a Wick-pairing problem: the continuous part is encoded in a Gaussian-type measure, while explicit spherical-harmonic, Racah manipulations are carried by discrete Wick contractions of the ISG direction vectors, replacing the usual Racah algebra. In what follows we refer the reader to Ref.~\cite{hiyamaGEM} for the detailed derivations and applicable examples. For direct application to a quark-model Hamiltonian, the computation-ready matrix-element formulas are collected in \ref{mat_calc}.
The GEM/ISG basis and the Yukawa-profile three-body matrix elements are presented explicitly in Sections~\ref{app:GEM_ref} and~\ref{GEM_3body_calc}, respectively.
These formulas show how the ISG method reduces explicit angular-momentum algebra to finite sums over fixed shift directions within Gaussian integrals.

\section{Meson and baryon spectrum}\label{sec:baryon_GEM}

In the preceding sections we reviewed and collected the theoretical ingredients of constituent quark models: the static color potential extracted from Wilson loops and the chromomagnetic hyperfine force. 
We now turn to a calculation in which these ideas are realized in a simple quark Hamiltonian and tested against meson and baryon data.
Our strategy is as follows. We first specify the two-body quark Hamiltonian that will serve as our baseline. 
We then determine the parameters of the quark model Hamiltonian by a high-precision fit to ground-state mesons using the Gaussian expansion method (GEM), and using the same parameters, compute the ground-state baryon spectrum across the light, strange, and charm sectors. 
The resulting pattern of mass deviations --- particularly the spin splittings and flavor-dependent trends --- quantifies the limitations of the two-body baseline.
It motivates the spatially contact three-quark interaction introduced later in this section and its finite-range extension in  Section~\ref{sec:short_range}.

\subsection{The quark model Hamiltonian}

Although quark models can be classified in various ways --- such as quenched vs unquenched, relativistic vs non-relativistic, or one-gluon-exchange based on color operators (OGE) vs Goldstone-boson-exchange based on flavor operators (GBE) --- here we will focus on the non-relativistic OGE quark models.

QCD contains eight massless color-octet gluons, which are confined and are seen only indirectly, for example through jets~\cite{FritzschGellMannLeutwyler1973}. Asymptotic freedom makes $\alpha_s$ small at large momentum transfer (e.g.,~Bjorken scaling in deep-inelastic scattering and the logarithmic scaling violations measured at LEP), so quarks behave as nearly free point particles in that regime~\cite{GrossWilczek1973,Politzer1973}. In contrast, the low-energy properties of the strong interactions cannot be derived using perturbative QCD. In the seminal work of De R\'ujula, Georgi, and Glashow~\cite{DeRujula:1975}, the original quark model Hamiltonian was constructed by combining two key QCD-inspired ingredients: long-range confinement motivated by Wilson's lattice QCD area law and short-range OGE. The non-relativistic constituent-quark model Hamiltonian\footnote{The overall factor $-3/4$ is chosen so that, in a color singlet channel, the effective $q\bar q$ potential reduces to the standard Cornell form, while the $qq$ interaction is one half of it ($V_{qq}= V_{q\bar q}/2$), as is customary in baryon spectroscopy.}  used in this work takes the form
\begin{equation}\label{hamiltonian}
H = \sum_{i}\Big( m_i + \frac{\mathbf{p}_i^2}{2m_i} \Big) + \sum_{i<j} \left( V_{ij}^C + V_{ij}^{CS} \right),
\end{equation}
where the center-of-mass kinetic energy $K_{\rm CM}$ is subtracted, and the actual calculations are carried out in Jacobi coordinates. The color-Coulomb and color-spin operator structures are motivated by the non-relativistic reduction of OGE.
The full central interaction used here also contains the linear confinement term and the additive constant discussed in Sections~\ref{VC_cornell} and~\ref{VC_constant}.
The color-Coulomb part follows schematically from
\begin{equation}
V(r)\sim\int \frac{d^3q}{(2\pi)^3}e^{i\mathbf{q}\cdot \mathbf{r}} \mathcal{M}(\mathbf{q}).
\end{equation}
From the temporal $\gamma^0\gamma^0$ component of $\mathcal{M}$, we obtain a color-Coulomb potential. Including a linear confinement term (see Section~\ref{VC_cornell}; with string tension $\sigma$) derived from the Wilson loop area law~\cite{Wilson1974} in lattice QCD and a constant shift $D$ (see Section~\ref{VC_constant}),
\begin{equation}
V_{ij}^C = -\frac{3}{4}F_i^c F_j^c\bigg(-\frac{\kappa}{r_{ij}} + \sigma {r_{ij}} - D\bigg), \label{2body_1}
\end{equation}
where $F_i^c= \lambda_i^c/2 \;(c=1, \ldots,8)$ denotes SU(3) color matrices for $i$th quark, and $r_{ij}$ is the relative distance between $i$ and $j$ quarks. 
Here $\kappa$ is a fitted effective Coulomb coefficient, conventionally written as $\kappa=4\alpha_s/3$.
The linear term represents confinement in the valence Hamiltonian.
For the color-spin interaction, we use the finite-range Gaussian form reviewed in  Section~\ref{VCS_intro},
\begin{align}
V_{ij}^{CS}= -\frac{3}{4}F_i^c F_j^c\frac{\mu_{ij}\mu_{ij}'}{m_i m_j c^4}\frac{\hbar c}{r_{ij}}{e^{-(\mu_{ij}r_{ij}/\hbar c)^2}} \boldsymbol{\sigma}_i\cdot \boldsymbol{\sigma}_j.\label{2body_2}
\end{align}
The mass-dependent range and normalization parameters $\mu_{ij}$ and $\mu_{ij}'$ are defined in Eq.~\eqref{eq:VCS_range_parameters}.
$\boldsymbol{\sigma}_i$ denotes the Pauli matrices of the $i$th quark acting on the spin space. 
As the quark masses increase, the bound-state wave functions become more localized and the spatial range of $V^{CS}$ decreases.
The resulting hyperfine splitting is determined by the competition between the larger short-distance overlap and the explicit $1/(m_i m_j)$ suppression.
The parameters $\mu_{ij}$ and $\mu'_{ij}$ describe this mass-dependent range and normalization.
In particular, the spatial interaction approaches a delta function as the quark mass increases (see Eq.~\eqref{approaches_to_delta_general}).

The central and color-spin interactions define our ground-state two-body baseline.
They reproduce the main spin pattern, but the common meson-fit parameter set leaves the large baryon residuals shown below.
Spin--orbit, tensor, and other higher-order terms are not included in this baseline.
If one attempts to treat a $1/r^3$-type sub-leading correction perturbatively, it diverges due to a failure of relative boundedness. Therefore, one must include a proper regularization step. This model successfully reproduces the ground-state masses of hadrons containing light, charm, and bottom quarks~\cite{park2019,noh2021}, and it has also been used to study possible compact exotic configurations~\cite{park2019,noh2021,park2017}. Although quark models with OGE work well at describing ground-state hadrons --- where short-range Coulomb and chromomagnetic interactions dominate --- they often face difficulties with excited states. At larger separations, coupling $\alpha_s$ grows, perturbative OGE (and its $V^{CS}$ term) loses validity, and non-perturbative flux-tube dynamics, string breaking, and other emergent forces must be invoked. Moreover, excited baryons often mix strongly with meson--baryon continua, a coupled-channel effect absent in simple potentials, and relativistic corrections become large. Without these long-range, channel-coupling, and dynamical sea-quark contributions, quark models cannot reliably predict highly excited spectra.

\subsection{From meson fit to baryon spectrum with two-body interactions: $u, d, s, c$ sectors}

From this point on, we determine the parameters of the two-body OGE quark model Hamiltonian by a fit to the observed meson spectrum, and using the same parameters, compute the ground-state baryon masses so as to isolate the impact of quark three-body forces. Unquenching (or coupled-channel) effects are taken to generate an approximately uniform downward self-energy shift and are therefore regarded as absorbed into the fitted parameters---most notably the overall constant $-D$ and constituent mass~\cite{Barnes2008,Hammer2016,Chen2018,Li2009,Kanwal2022,Ni2024}; this explains why quenched quark models typically require a larger $D$ (lower $-D$) than lattice, unquenched or continuum-model determinations (see Table~\ref{tab:cornell_lattice}). Moreover, higher-order terms that arise in a Pauli (Foldy-Wouthuysen) reduction --- such as spin--orbit and tensor interactions --- as well as long-range pion-exchange effects, which are negligible for ground states, are omitted in our baseline Hamiltonian~\cite{Glozman1996,Goldman2008}.

\subsubsection{Meson spectrum and model parameters}

\begin{table*}[!t]
\caption{\justifying
Ground-state meson and baryon masses obtained using the common two-body Hamiltonian.
The model parameters are fixed from the ground-state meson spectrum using 30 Gaussian basis functions.
The same parameters are subsequently applied to the baryon sector using 20 Gaussian ranges for each Jacobi coordinate in each of the three rearrangement channels, with the internal $SS$, $PP$, $DD$, and $FF$ components included within the Gaussian expansion method.
The RMSE is defined as
$\sigma=\sqrt{N^{-1}\sum_i(M_i-M_i^{\exp})^2}$.}
\label{tab:meson_baryon_two_body}

\centering

\textbf{(a) Ground-state meson spectrum}

\vspace{2pt}

\renewcommand{\pdev}[1]{\textcolor{devred}{\ensuremath{(#1)}}}
\renewcommand{\ndev}[1]{\textcolor{devgreen}{\ensuremath{(#1)}}}

\resizebox{\textwidth}{!}{
\begin{tblr}{
  width=\textwidth,
  colspec={Q[l,wd=5.5cm] *{8}{Q[c]}},
  row{1} = {bg=headgray, halign=c, valign=m},
  cell{1}{1} = {}{halign=l},
  hline{2} = {1-9}{0.3pt},
  hline{3} = {1-9}{0.3pt},
  rowsep = 1.5pt,
}
\toprule
Meson ($\mathcal{J}^{P(C)}$)
& $\eta_c\,(0^{-+})$
& $J/\psi\,(1^{--})$
& $D\,(0^-)$
& $D^*\,(1^-)$
& $\pi\,(0^{-+})$
& $\rho\,(1^{--})$
& $K\,(0^-)$
& $K^*\,(1^-)$ \\

Experimental mass (MeV)
& 2983.60
& 3096.90
& 1864.80
& 2010.30
& 139.57
& 775.11
& 493.68
& 891.66 \\

Calibrated mass ($\sigma=2.51$ MeV)
& 2985.10
& 3097.31
& 1868.18
& 2013.68
& 140.67
& 770.96
& 491.24
& 892.48 \\

Deviation ($M-M_{\rm exp}$ in MeV)
& \pdev{+1.50}
& \pdev{+0.41}
& \pdev{+3.38}
& \pdev{+3.38}
& \pdev{+1.10}
& \ndev{-4.15}
& \ndev{-2.44}
& \pdev{+0.82} \\
\bottomrule
\end{tblr}}

\vspace{10pt}

\textbf{(b) Ground-state baryon spectrum from meson-fit parameters}

\vspace{2pt}

\resizebox{\textwidth}{!}{
\begin{tblr}{
  width=\textwidth,
  colspec={Q[l,wd=4.8cm] *{9}{Q[c]}},
  row{1} = {bg=headgray, halign=c, valign=m},
  cell{1}{1} = {}{halign=l},
  hline{2} = {1-10}{0.3pt},
  hline{3} = {1-10}{0.3pt},
  rowsep = 1.5pt,
}
\toprule
Baryon ($\mathcal{J}^{P}$)
& $p\,(1/2^+)$
& $\Delta\,(3/2^+)$
& $\Sigma\,(1/2^+)$
& $\Lambda\,(1/2^+)$
& $\Sigma_c\,(1/2^+)$
& $\Lambda_c\,(1/2^+)$
& $\Xi_{cc}\,(1/2^+)$
& $\Sigma^*\,(3/2^+)$
& $\Sigma_c^*\,(3/2^+)$ \\

Experimental mass (MeV)
& 938.272
& 1232
& 1192.642
& 1115.683
& 2452.650
& 2286.460
& 3619.970
& 1383.800
& 2518.480 \\

Mass ($\sigma=65.04$ MeV)
& 1038.746
& 1338.864
& 1259.291
& 1171.919
& 2491.972
& 2317.914
& 3622.232
& 1450.640
& 2562.599 \\

Deviation ($M-M_{\rm exp}$ in MeV)
& \pdev{+100.474}
& \pdev{+106.864}
& \pdev{+66.649}
& \pdev{+56.236}
& \pdev{+39.322}
& \pdev{+31.454}
& \pdev{+2.262}
& \pdev{+66.840}
& \pdev{+44.119} \\
\bottomrule
\end{tblr}}

\end{table*}

To isolate the contribution of the three-quark interaction, we first determine the parameters of the two-body Hamiltonian in Eq.~\eqref{hamiltonian} exclusively from the meson sector.
The parameters are fitted to the ground-state meson masses by minimizing the root-mean-square error $\sigma=\sqrt{N^{-1}\sum_i(M_i-M_i^{\exp})^2}$.

For each meson, the relative wave function is expanded in 30 Gaussian basis functions,
\begin{equation}
\phi_{nlm}(\mathbf r)=N_{nl}\,r^l e^{-\nu_n r^2}Y_{lm}(\hat{\mathbf r}),\qquad \nu_n=\nu_0\varrho_M^{\,n-1},\qquad n=1,\ldots,30.
\label{eq:meson_gaussian_basis}
\end{equation}
Here $N_{nl}$ is a normalization constant, and the nonlinear scale $\nu_0$ is optimized for each flavor and spin channel.
The Gaussian widths are distributed geometrically with $\varrho_M$ chosen in the range 1.20--1.30.\footnote{For fixed orbital angular momentum $l$, the functions $r^l e^{-\nu_n r^2}$ efficiently represent smooth bound-state wave functions because the geometric progression $\nu_n=\nu_0\varrho_M^{\,n-1}$ samples the relevant length scales approximately logarithmically.} 
Values of $\varrho_M$ too close to unity generate strong linear dependence among neighboring basis functions and make the overlap matrix increasingly ill-conditioned.
Conversely, excessively large values of $\varrho_M$ undersample the intermediate length scales and may lead to underconverged variational energies.

For color-singlet $q\bar q$ mesons, the color matrix element is $\langle F_i^cF_j^c\rangle=-4/3$, so that the central interaction reduces to the conventional Cornell form.
The mass-dependent smearing parameters entering the color-spin interaction are those defined in Eq.~\eqref{eq:VCS_range_parameters}. The parameter $\mu_{ij}$ primarily controls the spatial range of the Gaussian regulator, while the product $\mu_{ij}\mu'_{ij}$ determines its overall normalization.
Allowing $\mu_{ij}$ and $\mu'_{ij}$ to have independent intercepts and mass dependences provides sufficient flexibility to reproduce the hyperfine splittings across the light, strange, and charm sectors.

Because the present meson calibration contains only a limited number of ground-state masses, the meson-only objective possesses an overfitted global minimum. 
At this minimum, the light-pair color-spin range $r_{qq}^{CS}\equiv{\hbar c}/{\mu_{qq}}$ is driven to a very large value, and the fitted color-spin interaction no longer retains the intended short-range character. We therefore adopt a different local minimum with a physically reasonable color-spin range. The physical scale used to select this solution is discussed below, together with the characteristic scales of the fitted parameters.
The parameter set used in the present calculation is
\begin{align}
\kappa&=78.22998~\mathrm{MeV\,fm},&
\sigma&=1144.73573~\mathrm{MeV/fm},&
D&=1098.22352~\mathrm{MeV},\nonumber\\
\alpha&=285.19763~\mathrm{MeV},&
\beta&=0.666766,&
\gamma&=167.28927~\mathrm{MeV},&
\delta&=0.150533,\nonumber\\
m_{u,d}&=329.55839~\mathrm{MeV},&
m_s&=592.42519~\mathrm{MeV},&
m_c&=1873.87408~\mathrm{MeV}.
\label{params}
\end{align}
In natural units, the Coulomb coefficient and string tension correspond to $\kappa=0.3964$ and $\sigma=0.2259~\mathrm{GeV}^2$, respectively.
The effective coupling associated with the central Coulomb term is therefore $\alpha_s^{(C)}=(3/4)\kappa=0.2973$.
The fitted string tension is somewhat larger than the commonly used Cornell value of approximately $0.18~\mathrm{GeV}^2$, but remains of the same characteristic QCD scale.
For a light $qq$ pair, the two smearing parameters and the associated spatial range are
\begin{equation}
\mu_{qq}=395.07~\mathrm{MeV},\qquad \mu'_{qq}=192.09~\mathrm{MeV},\qquad r_{qq}^{CS}=0.499~\mathrm{fm}.
\label{eq:qq_smearing_range}
\end{equation}
The light-sector regulator is therefore moderately broad for an interaction originating from a nominally short-range OGE contact term.
In the large reduced-mass limit, the ratio approaches $\mu'_{ij}/\mu_{ij}\to\delta/\beta=0.2258$, which corresponds within the present normalization convention to an effective color-spin coupling $\alpha_s^{(CS,\infty)}=(9/4)(\delta/\beta)=0.508$.
This effective hyperfine normalization is larger than the coupling $\alpha_s^{(C)}=0.2973$ inferred from the central Coulomb term.

This mismatch should not be interpreted as two independent determinations of the same bare QCD coupling.
The quantity $\alpha_s^{(CS,\infty)}$ is the formal contact-limit normalization obtained by extrapolating the globally fitted color-spin potential, rather than a direct extraction of the OGE coupling from asymptotically heavy states.
In particular, the instanton interpretation discussed below concerns the renormalization of the global fit by omitted light-sector physics and does not imply a direct instanton-induced interaction between genuinely heavy quarks.
As reviewed in  Section~\ref{subsec:instanton_3body}, the leading two-body instanton-induced interaction in a flavor-antisymmetric light-quark $S$-wave channel contains the same local color-spin structure, $\lambda_i^c\lambda_j^c\, \boldsymbol{\sigma}_i\cdot\boldsymbol{\sigma}_j$, as the OGE hyperfine interaction, although its complete operator also contains a flavor projector and spin-independent terms.
Low-lying light-hadron spectroscopy therefore constrains the combined short-range hyperfine strength coming from both the separate OGE and instanton-induced components.
Phenomenological OGE--III analyses typically assign $p_{\mathrm{III}}\simeq0.3$--0.4, corresponding to approximately 90--120~MeV of the $N-\Delta$ splitting, to the two-body instanton-induced interaction~\cite{Takeuchi1998,OkaTakeuchi1989}.
Since $V_{\mathrm{III}}^{(2)}$ is not included explicitly in Eq.~\eqref{2body_2}, part of its net light-sector contribution can be absorbed into the globally fitted parameters $\alpha$, $\beta$, $\gamma$, and $\delta$, together with relativistic, higher-order, and vertex corrections.
Because these parameters are determined from a simultaneous global fit rather than from asymptotically heavy states alone, the formal ratio $\delta/\beta$ can likewise be affected by light-sector renormalization through parameter correlations.
The numerical ratio $\alpha_s^{(CS,\infty)}/\alpha_s^{(C)}\simeq1.71$ is therefore qualitatively consistent with a sizable non-OGE contribution to the effective hyperfine interaction, including a possible two-body instanton-induced component.
It should not, however, be converted directly into $p_{\mathrm{III}}$, because the instanton-induced interaction is flavor selective and differs from OGE in both its projector and radial structure.
In the present ground-state calculation, we retain only an effective representation of the leading spin-spin contribution.
Noncentral instanton-induced terms have also been derived~\cite{Takeuchi1998}; what remains model dependent is their finite-range regularization and matching to the Gaussian regulator employed here.
The physical interpretation of the moderately broad light-sector smearing range and the effective hyperfine normalization is further examined in the $u,d,s,c,b$ compatibility benchmark of  Section~\ref{sec:udscb_benchmark}.

The ground-state meson masses alone also leave substantial correlations among $\kappa$, $\sigma$, $D$, the constituent masses, and the four smearing parameters.
The individual parameter values should consequently be regarded as one effective solution within the adopted model space rather than as unique determinations of QCD masses and couplings.
Panel~(a) of Table~\ref{tab:meson_baryon_two_body} presents the resulting meson spectrum.
The 30-Gaussian calculation reproduces the fitted ground-state mesons with an RMSE of $\sigma_M=2.51~\mathrm{MeV}$.
Because the meson Hamiltonian is central, the Hamiltonian and overlap matrices are block diagonal in the orbital angular momentum $l$.
The ground-state mesons included in the fit belong to the $l=0$ block, so they do not mix with higher orbital components.
Radially excited $nS$ states instead appear as higher eigenstates of the same $l=0$ generalized eigenvalue problem.

In Section~\ref{sec:meson_excitations}, we also test an extended fit that includes selected radial or orbital excitations.
Radial level spacings constrain the shape of the potential over a wider distance range than ground-state masses alone and should therefore reduce the flat parameter directions present in the current fit.
In particular, they provide additional constraints on the correlations among $\kappa$, $\sigma$, $D$, the constituent masses, and the smearing parameters.
We examine whether the enlarged fit favors a more physically natural parameter region, even if the ground-state RMSE increases moderately.
Since radially excited mesons are more sensitive to nearby thresholds and coupled-channel effects, we use only well-established states or assign appropriate theoretical uncertainties.

\subsubsection{Baryon spectrum}\label{baryon_spectrum_sec}

\begin{figure}[!b]
\centering
\includegraphics[width=0.9\linewidth]{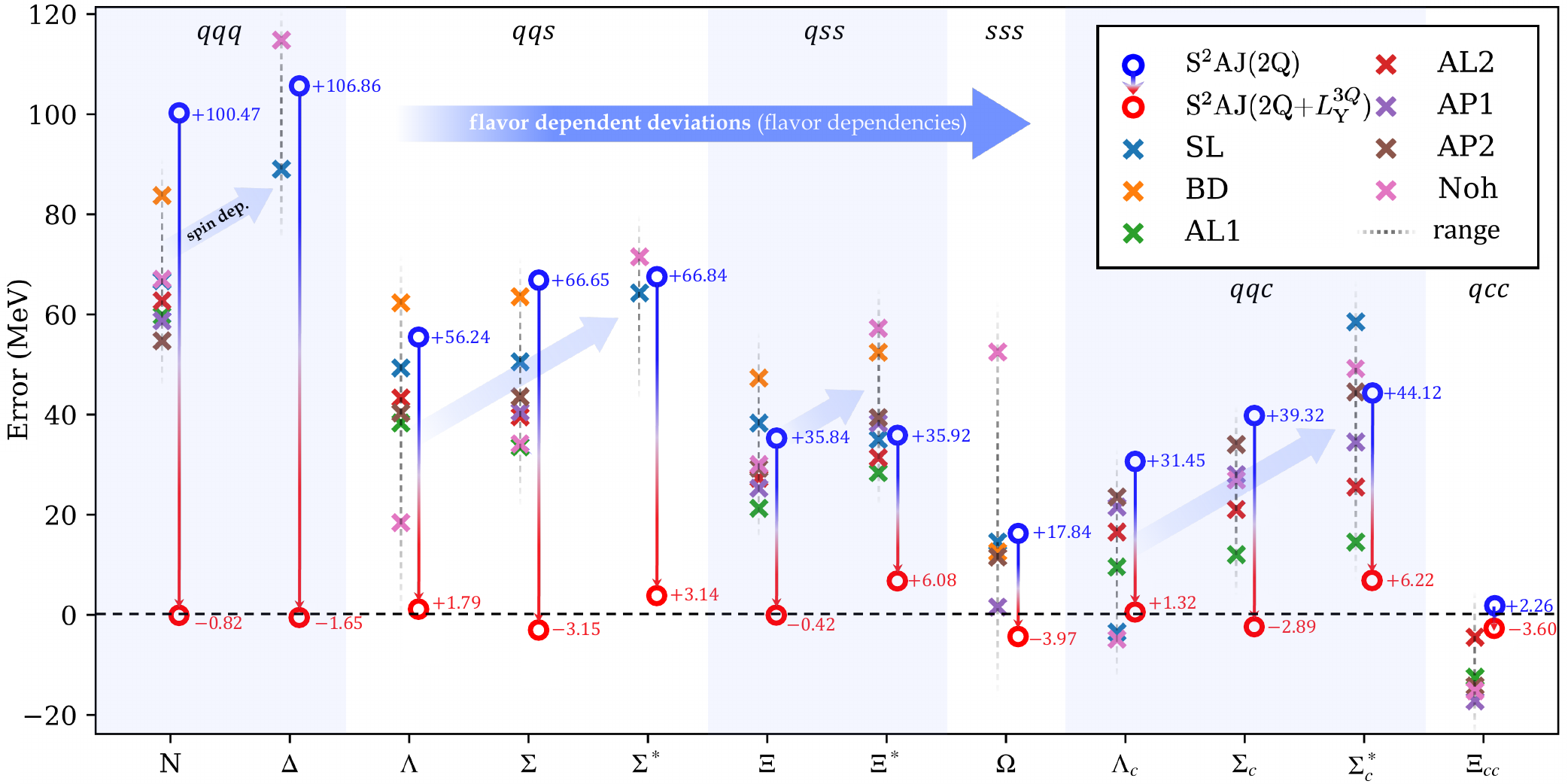}
\caption{\justifying Prediction errors of various quark models for the ground-state baryon spectrum. Each marker shows the difference between a model's predicted mass and the experimental value for a given baryon; models are distinguished by marker color as indicated in the legend. Baryons are grouped (and lightly shaded) by their quark-content sectors. Upward diagonal arrows annotate the systematic growth of spin-dependent errors in each sector, while the horizontal arrow at the top emphasizes the decrease in flavor-dependent deviations as the quark content becomes heavier. ``SL'' refers to Ref.~\cite{Stanley1980}; ``BD'' to Ref.~\cite{Bhaduri1981}; ``AL'' and ``AP'' to Refs.~\cite{Semay1994,SilvestreBrac1996}; and ``Noh'' to Ref.~\cite{noh2025inevitable} with $S$-wave estimation. All data points were either estimated for the ground state using one (or more) $S$-wave Gaussian bases or calculated by solving the Faddeev equations. ``$\mathrm{S}^{2}\mathrm{AJ}$(2Q)'' and ``$\mathrm{S}^{2}\mathrm{AJ}$(2Q+$L_{\mathrm Y}^{3Q}$)'' refer to predictions from our model using only two-body interactions in Eq.~\eqref{hamiltonian} and with the inclusion of the three-body force in Eq.~\eqref{eq:3bd_yuk_full}, respectively. The downward blue-to-red arrows show how adding the three-body force reduces the deviations.}
\label{fig:errors}
\end{figure}

Using the parameters determined exclusively from the meson sector in Eq.~\eqref{params}, we construct the baryon wave function as
\begin{align}
\Psi_{\mathcal{J}^{P}M}^{\rm baryon}
=\sum_{\mathcal{C}=a,b,c}\sum_{\alpha}\mathcal{B}_{\alpha}^{(\mathcal C)}\mathcal{P}_{B}\left\{\left[\left[\phi_{n_{\mathcal C}l_{\mathcal C}}(\mathbf r_{\mathcal C})\phi_{N_{\mathcal C}L_{\mathcal C}}(\mathbf R_{\mathcal C})\right]_{J}\otimes\chi_{S}\right]_{\mathcal{J}M}\otimes F\otimes\mathcal{P}_{\mathbf 1}C\right\},
\label{eq:baryon_wavefunction}
\end{align}
where $\alpha=\{n,N,l,L,J,S\}$ collectively denotes the spatial and spin quantum numbers.
Here $\chi_S$, $F$, and $C$ denote the spin, flavor, and color wave functions, respectively.
The orbital angular momenta $l_{\mathcal C}$ and $L_{\mathcal C}$ are coupled to the total orbital angular momentum $J$, which is subsequently coupled with the total quark spin $S$ to form the baryon angular momentum $\mathcal J$.
The parity of each spatial component is $P=(-1)^{l_{\mathcal C}+L_{\mathcal C}}$.
We restrict the three-quark basis to the color-singlet subspace using $\mathcal P_{\mathbf 1}=\ket{\mathbf 1}\bra{\mathbf 1}$.
The confining dynamics is supplied by the central interaction $V^C$ in Eq.~\eqref{2body_1}.
The operator $\mathcal{P}_{B}$ denotes the permutation projector appropriate to the flavor content of each baryon and ensures that the complete wave function satisfies the Pauli principle.
The explicit spin-flavor basis and the corresponding permutation symmetries are summarized in \ref{bases}.
The label $\mathcal C$ denotes the three possible Jacobi rearrangement channels.
For example, the $a$-channel coordinates are defined as
\begin{equation}
\mathbf r_a=\frac{\mathbf r_1-\mathbf r_2}{\sqrt{2}},\qquad \mathbf R_a=\sqrt{\frac{2}{3}}\left(\frac{m_1\mathbf r_1+m_2\mathbf r_2}{m_1+m_2}-\mathbf r_3\right).
\label{Jacobi_coord}
\end{equation}
The $b$ and $c$ coordinates are obtained through the cyclic permutations (123) and (132) of the particle labels, respectively.
The normalization factors in Eq.~\eqref{Jacobi_coord} are chosen such that the coordinate transformation has unit Jacobian in the convention adopted here.
These scaled Jacobi coordinates are used only to construct the spatial basis, while all interactions are evaluated using the physical pair separations $r_{ij}=\vert \mathbf r_i-\mathbf r_j\vert $.
For each rearrangement channel, the radial dependences are expanded as
\begin{equation}
\phi_{nlm}(\mathbf r)=N_{nl}r^le^{-\nu_n r^2}Y_{lm}(\hat{\mathbf r}),\qquad \phi_{NLM}(\mathbf R)=N_{NL}R^Le^{-\nu_N R^2}Y_{LM}(\hat{\mathbf R}).
\label{eq:baryon_gaussian_basis}
\end{equation}
The Gaussian widths are distributed geometrically according to
\begin{equation}
\nu_n=\frac{1}{r_n^2},\qquad r_n =r_{\min}\varrho_r^{\,n-1},\qquad n=1,\ldots,20,\qquad \nu_N=\frac{1}{R_N^2},\qquad R_N = R_{\min}\varrho_R^{\,N-1},\qquad N=1,\ldots,20.
\label{eq:baryon_gaussian_mesh}
\end{equation}
The endpoints and ratios of the two geometric meshes are optimized directly in the full nonorthogonal basis rather than being inferred from a single-Gaussian trial state.
The basis ranges are chosen to cover both the short-range correlations and the long-distance tails of the baryon wave functions.
Thus, each orbital channel uses a {20~\ensuremath{\times}~20} radial mesh in each of the three Jacobi rearrangement channels.

The coefficients $\mathcal B_{\alpha}^{(\mathcal C)}$ are determined by solving the generalized eigenvalue problem
\begin{equation}
\sum_{\alpha'}\left(H_{\alpha\alpha'}-E\,\mathcal O_{\alpha\alpha'}\right)\mathcal B_{\alpha'}=0,\qquad H_{\alpha\alpha'}=\braket{\Phi_\alpha|H|\Phi_{\alpha'}},\qquad \mathcal O_{\alpha\alpha'}=\braket{\Phi_\alpha|\Phi_{\alpha'}}.
\label{eq:baryon_generalized_eigenvalue}
\end{equation}
The three-body Schr\"odinger equation is solved using the Gaussian expansion method (GEM)~\cite{hiyamaGEM}.
Because basis states constructed in different Jacobi channels are not mutually orthogonal, the use of all three rearrangement channels leads naturally to the overlap matrix $\mathcal O$ in Eq.~\eqref{eq:baryon_generalized_eigenvalue}.
Numerically redundant directions associated with very small overlap eigenvalues are removed before the Hamiltonian is diagonalized.

Matrix elements between basis functions defined in different Jacobi coordinates become more involved when either $l_{\mathcal C}$ or $L_{\mathcal C}$ is nonzero.
They may be evaluated by transforming between Jacobi coordinates using angular-momentum recoupling algebra or by employing the infinitesimally shifted Gaussian representation.
We use the latter method, in which a Gaussian with nonzero orbital angular momentum is expressed as
\begin{equation}
r^le^{-\nu_n r^2}Y_{lm}(\hat{\mathbf r})=\lim_{\epsilon\rightarrow0}\left(\frac{1}{\nu_n\epsilon}\right)^l\sum_{k=1}^{k_{\max}}C_{lm,k}\exp\left[-\nu_n\left(\mathbf r-\epsilon\mathbf D_{lm,k}\right)^2\right].
\label{eq:isg_expansion}
\end{equation}
The coefficients $C_{lm,k}$ and shift vectors $\mathbf D_{lm,k}$ are defined in Eqs.~\eqref{C_coeff} and~\eqref{D_vector}, respectively.
This representation replaces matrix elements involving explicit solid harmonics by finite sums of elementary multidimensional Gaussian integrals and is particularly convenient for evaluating cross-channel matrix elements. 
For matched Gaussian meshes, Jacobi channels, symmetry projectors, and overlap cutoffs, the direct Cartesian-polynomial implementation of \ref{app:analytic_without_gem} reproduces the GEM/ISG overlap and two-body Hamiltonian matrices of \ref{app:GEM_ref}, as well as the three-quark matrices.

For the positive-parity ground states, we include the symmetry-allowed $J=0$ channels $(l_{\mathcal C},L_{\mathcal C})=(0,0)$, $(1,1)$, $(2,2)$, $(3,3)$, thereby incorporating the internal $SS$, $PP$, $DD$, and $FF$ basis components.
The odd-partial-wave components relevant to positive-parity baryons have $l_{\mathcal C}+L_{\mathcal C}$ even, as in the $PP$ and $FF$ channels.
Although these components carry nonzero angular momenta in the individual Jacobi coordinates, they may couple to total orbital angular momentum $J=0$ and therefore contribute to the positive-parity ground state.
All masses quoted in Tables~\ref{tab:meson_baryon_two_body} and~\ref{tab:combined_3bd_baryon} are obtained in this full $SS+PP+DD+FF$ variational space.
This hierarchy can be understood from the distinct permutation properties of the $PP$ and $DD$ couplings.

For $S=1/2$, we use the pair-spin basis
\begin{equation}\label{eq:spin_bases_rho_lambda}
\chi_{\rho}=\left|[(12)_{s_{12}=0}3]_{S=1/2}\right\rangle,\qquad
\chi_{\lambda}=\left|[(12)_{s_{12}=1}3]_{S=1/2}\right\rangle.
\end{equation}
Writing the radial part of the color-spin interaction as $V_{ij}^{CS}=v_{ij}^{CS}\,\boldsymbol{\sigma}_i\cdot\boldsymbol{\sigma}_j$, we decompose the two cross-pair terms as follows:
\begin{align}
v_{13}^{CS}\,\boldsymbol{\sigma}_1\cdot\boldsymbol{\sigma}_3
+v_{23}^{CS}\,\boldsymbol{\sigma}_2\cdot\boldsymbol{\sigma}_3
=&\frac{v_{13}^{CS}+v_{23}^{CS}}{2}
\left(\boldsymbol{\sigma}_1\cdot\boldsymbol{\sigma}_3
+\boldsymbol{\sigma}_2\cdot\boldsymbol{\sigma}_3\right)+\frac{v_{13}^{CS}-v_{23}^{CS}}{2}
\left(\boldsymbol{\sigma}_1\cdot\boldsymbol{\sigma}_3
-\boldsymbol{\sigma}_2\cdot\boldsymbol{\sigma}_3\right).
\end{align}
Both factors in the second term are odd under the exchange $1\leftrightarrow2$, so their product is permutation even and can connect a pair-even $SS$ component to a pair-odd $PP$ component.
The corresponding spin-recoupling matrix element is nonzero,
\begin{equation}
\left|
\left\langle\chi_{\rho}\left|
\boldsymbol{\sigma}_1\cdot\boldsymbol{\sigma}_3
-\boldsymbol{\sigma}_2\cdot\boldsymbol{\sigma}_3
\right|\chi_{\lambda}\right\rangle
\right|=2\sqrt{3}.
\end{equation}
Thus, the explicit $PP$ channel permits simultaneous changes of the spatial exchange symmetry and the pair-spin coupling $s_{12}=0\leftrightarrow1$.
For the $\Lambda$-type states, the dominant $SS$ component is $SS\chi_{\rho}$ and couples to a spin--triplet $PP\chi_{\lambda}$ component, whereas the corresponding roles are reversed for the $\Sigma$ states.
This reversal alone does not explain the different variational gains, because the off-diagonal spin-recoupling matrix element has the same magnitude in both directions.
The distinction follows from the diagonal cross-pair matrix elements,
\begin{equation}
\left\langle\chi_{\rho}\left|
\boldsymbol{\sigma}_1\cdot\boldsymbol{\sigma}_3
+
\boldsymbol{\sigma}_2\cdot\boldsymbol{\sigma}_3
\right|\chi_{\rho}\right\rangle=0,
\qquad
\left\langle\chi_{\lambda}\left|
\boldsymbol{\sigma}_1\cdot\boldsymbol{\sigma}_3
+
\boldsymbol{\sigma}_2\cdot\boldsymbol{\sigma}_3
\right|\chi_{\lambda}\right\rangle=-4.
\end{equation}
The dominant $\Lambda$-type $SS\chi_{\rho}$ component therefore contains no diagonal cross-pair hyperfine contribution, and the explicit $PP\chi_{\lambda}$ channel opens a correlation that is inaccessible within the pure $SS$ subspace.
For the $\Sigma$ states, the dominant $SS\chi_{\lambda}$ component already contains the corresponding cross-pair correlation, so the additional variational gain from mixing with $PP\chi_{\rho}$ is smaller.
The fully spin-aligned $S=3/2$ $\Sigma^{*}$ state has no independent spin--singlet pair component, which strongly suppresses this recoupling mechanism.

The $DD$ contribution has a different origin.
In a given Jacobi channel, a cross-pair central interaction depends on a linear combination of the two Jacobi coordinates and admits the multipole expansion
\begin{equation}\label{eq:cross_pair_multipole_expansion}
V\!\left(\left|a\mathbf r_{\mathcal C}+b\mathbf R_{\mathcal C}\right|\right)
=\sum_{\lambda=0}^{\infty}
V_{\lambda}(r_{\mathcal C},R_{\mathcal C})
P_{\lambda}\!\left(\cos\theta_{\mathcal C}\right),
\qquad
\cos\theta_{\mathcal C}
=\hat{\mathbf r}_{\mathcal C}\cdot\hat{\mathbf R}_{\mathcal C}.
\end{equation}
Since $P_{\lambda}(\cos\theta_{\mathcal C})$ is proportional to the scalar-coupled product $[Y_{\lambda}(\hat{\mathbf r}_{\mathcal C})\otimes Y_{\lambda}(\hat{\mathbf R}_{\mathcal C})]_{J=0}$, the $\lambda=1$, $\lambda=2$, and $\lambda=3$ multipoles generate $SS\leftrightarrow PP$, $SS\leftrightarrow DD$, and $SS\leftrightarrow FF$ couplings, respectively.
For two identical quarks, the odd multipoles cancel in the symmetric combination $V_{13}+V_{23}$, whereas the even multipoles survive.
Consequently, the central interaction retains a direct quadrupole $SS\leftrightarrow DD$ coupling, while the $SS\leftrightarrow PP$ and $SS\leftrightarrow FF$ couplings are more sensitive to exchange-odd structures and rearrangement-channel recoupling.
The present $DD$ component should therefore not be confused with the conventional tensor-induced $S$--$D$ mixing discussed after Eq.~\eqref{V_CS_cont}.
It is an internal $(l_{\mathcal C},L_{\mathcal C})=(2,2)$ correlation coupled to $J=0$, generated by the $\lambda=2$ central multipole even when no tensor force is included.

The labels $S$, $P$, $D$, and $F$ are defined with respect to a particular Jacobi rearrangement channel.
An $S$-wave Gaussian in one channel, when expressed in another Jacobi coordinate set, generally acquires a cross term of the form,
\begin{equation}
\exp\!\left[-a r_{\mathcal C}^{2}-b R_{\mathcal C}^{2}-c\,\mathbf r_{\mathcal C}\!\cdot\!\mathbf R_{\mathcal C}\right]
= e^{-a r_{\mathcal C}^{2}-b R_{\mathcal C}^{2}} \sum_{\lambda=0}^{\infty}(2\lambda+1)(-1)^{\lambda} \,i_{\lambda}\!\left(c r_{\mathcal C}R_{\mathcal C}\right) P_{\lambda}(\hat{\mathbf r}_{\mathcal C}\!\cdot\!\hat{\mathbf R}_{\mathcal C}),
\label{eq:jacobi_swave_partialwave_content}
\end{equation}
where $i_{\lambda}$ is the modified spherical Bessel of the first kind. 
Since each $P_{\lambda}(\hat{\mathbf r}_{\mathcal C}\!\cdot\!\hat{\mathbf R}_{\mathcal C})$ is proportional to the scalar-coupled spherical harmonic $[Y_{\lambda}(\hat{\mathbf r}_{\mathcal C})\otimes Y_{\lambda}(\hat{\mathbf R}_{\mathcal C})]_{00}$, the transformed Gaussian contains $(l_{\mathcal C},L_{\mathcal C})=(0,0),(1,1),(2,2),(3,3),\ldots$ components, corresponding to $SS$, $PP$, $DD$, $FF$, $\ldots$ in channel $\mathcal C$. Thus, a multi-rearrangement $S$-wave basis already incorporates part of the higher-partial-wave correlations implicitly when viewed in any single Jacobi channel, while explicit $PP$, $DD$, and $FF$ basis functions provide additional independent variational flexibility and quantify the residual improvement beyond these recoupled correlations.

The Gaussian meshes and basis dimensions are determined through direct convergence tests of the lowest generalized eigenvalues.
For each baryon, the short- and long-distance endpoints are varied while the number of Gaussian widths is increased until adding further basis functions changes the ground-state mass by less than 0.1~MeV.
Table~\ref{tab:meson_baryon_two_body}(b) presents the corresponding two-body baryon spectrum obtained without readjusting any parameter fixed in the meson sector.

Radial excitations are obtained as higher eigenstates within a fixed $\mathcal J^P$ block, while orbital excitations require basis components with nonzero $l_{\mathcal C}$ or $L_{\mathcal C}$.
The use of geometric Gaussian meshes together with all three Jacobi rearrangement channels provides an efficient representation of both radial and orbital structures~\cite{hiyama2012,varga1996,mitroy2013}.

\begin{figure}[!b]
\centering
\includegraphics[width=0.68\textwidth]{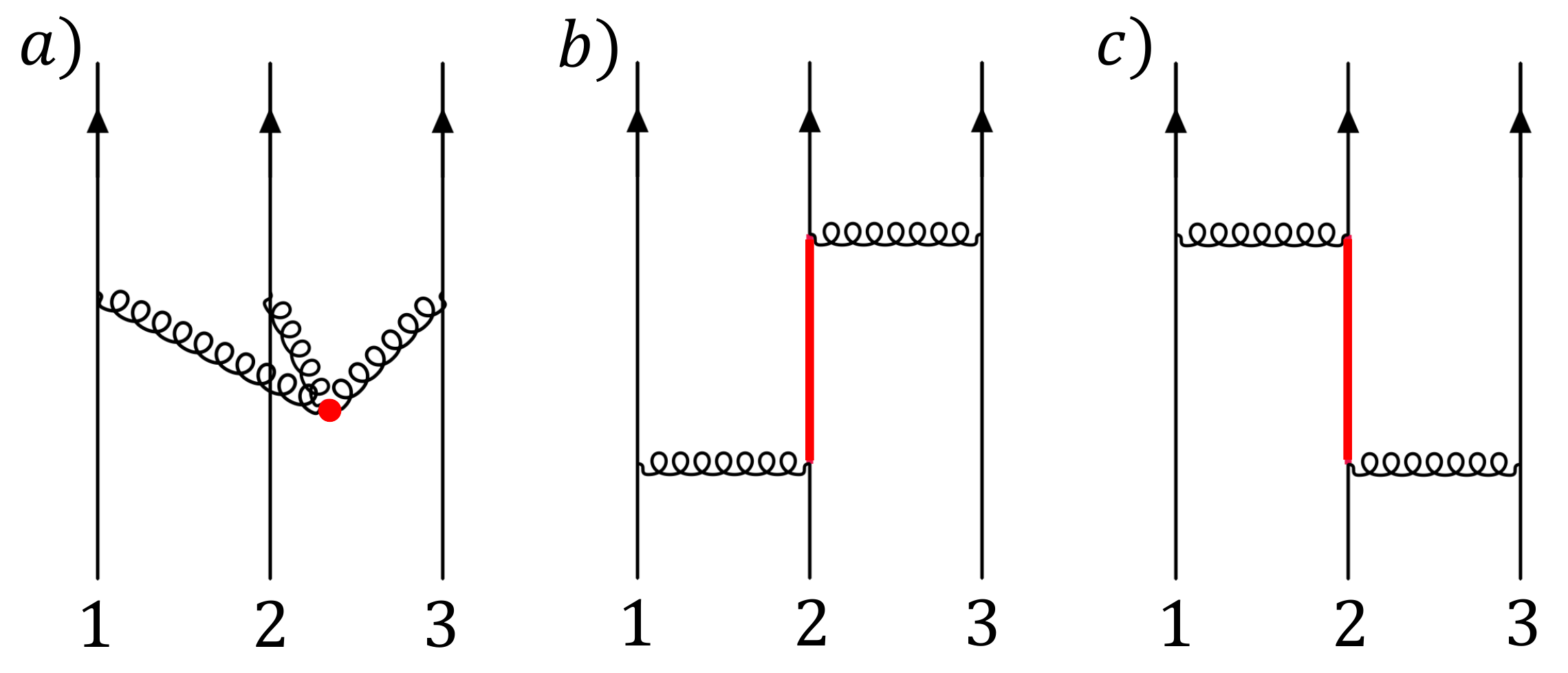}
\caption{\justifying Schematic quark three-body interaction topologies. Panel (a) represents a three-gluon interaction, whereas panels (b) and (c) show two-gluon-exchange (TGE) processes involving an intermediate configuration outside the retained valence $qqq$ space.}
\label{3body}
\end{figure}

\subsection{Color-spin structure of the three-body force}

Throughout this work, ``three-quark interaction'' means an explicit operator in the three-constituent sector after the chosen spectator-independent pair interactions have been fixed.
Its operational definition and representation dependence are discussed in  Section~\ref{subsec:three_quark_definition}.
We now translate the three-body diagrams of  Fig.~\ref{3body} into explicit color-spin operators. 
The red lines in (b) and (c) denote a generic intermediate configuration outside the retained valence $qqq$ space, which we schematically write as $q^\ast$. 
In this work, the static energy denominator is used only to organize the resulting color--spin operator basis. 
The relation to the Fujita--Miyazawa construction is discussed below after the model-space reduction has been specified.
For the topology where quarks 1 and 2 interact first and then 2 and 3 interact (Fig.~\ref{3body}b), the corresponding two-gluon-exchange (TGE) contribution can be summarized schematically as
\begin{equation}
  \mathcal M^{\mathrm{TGE}}_{23-12}
  \;\sim\;
  \bra{qqq}
V_{23}^{\mathrm{OGE}}\,
\frac{1}{M_2}\,
V_{12}^{\mathrm{OGE}}
  \ket{qqq},
  \label{eq:TGE_static_main}
\end{equation}
where $V_{ij}^{\mathrm{OGE}}$ is the OGE potential between quarks $i$ and $j$, and $M_2$ denotes the excitation-energy scale associated with the intermediate configuration carried by line 2.
\eqref{eq:TGE_static_main} is used to organize the color--spin operator basis. The unresolved transition and propagation effects are absorbed into the effective couplings $A$, $B$, and $C$ and, in  Section~\ref{sec:short_range}, into the spatial profile. After the intermediate sector is reduced into the $qqq$ sector and its induced one- and two-body contributions are assigned to the baseline Hamiltonian, the remaining connected contribution is represented by the operators constructed below.
In the pure color sector, a three-gluon $Y$-vertex attached to the three quark lines produces a color tensor proportional to $f^{abc}\lambda_1^a\lambda_2^b\lambda_3^c$, whose matrix element vanishes in a color-singlet baryon.

The inverse-mass factors depend on whether each insertion is central or chromomagnetic.
Up to a common factor $g^4$ and the corresponding color--spin tensors, the three operator classes scale schematically as
\begin{align}
L_{ij;k}^{C-C}\sim
\frac{g^4}{M_k}\,
\mathcal O_{C-C}, \qquad
L_{ij;k}^{C-S}
\sim \frac{g^4}{M_k} \left( \frac{1}{m_i m_k}
+ \frac{1}{m_j m_k} \right) \mathcal O_{C-S},
\qquad L_{ij;k}^{S-S}
\sim \frac{g^4}{m_i m_j m_k^2M_k}\, \mathcal O_{S-S}.
\label{eq:L3Q_scaling_main}
\end{align}
Here $i$ and $j$ denote the two external lines, while $k$ labels the line carrying the intermediate configuration.
Cyclic permutations and the two orderings of the pair insertions generate the operator basis used below.
The correspondence with the Fujita--Miyazawa mechanism is structural: in both cases, eliminating an intermediate excitation outside the retained three-particle space produces an effective three-particle operator.\footnote{In nuclear physics, the Fujita--Miyazawa three-nucleon ($3N$) force is obtained from time-ordered perturbation theory applied to two-pion exchange with an intermediate $\Delta$ excitation. Its operator structure may be written schematically as
\begin{equation*}
V^{2\pi}_{3N}\sim\sum_\text{cyc}\frac{(\boldsymbol{\sigma}_i\cdot \mathbf{q}_i)(\boldsymbol{\sigma}_k\cdot \mathbf{q}_k)}{(q_i^2+m_\pi^2)(q_k^2+m_\pi^2)} (\boldsymbol{\tau}_i \cdot \boldsymbol{\tau}_k)\frac{f_{\pi N\Delta}^2f_{\pi NN}^2}{E_\Delta - E_N}.
\end{equation*}
In the static approximation, the intermediate-$\Delta$ propagation is reduced to the excitation denominator $1/(E_\Delta-E_N)$.
The analogy used here is limited to the intermediate state topology: the $\Delta$ and $q^\ast$ play similar model-space roles, while the exchanged fields, transition vertices, quantum numbers, and spatial ranges are different.}

The products below are reduced using the SU(3) generator algebra
\begin{equation}
[F^a,F^b]=if^{abc}F^c,\qquad \{F^a,F^b\}=\frac13\delta^{ab}\mathbf{1}+d^{abc}F^c,\qquad F^a=\frac{\lambda^a}{2}.
\label{eq:SU3_generator_algebra}
\end{equation}
Using Eq.~\eqref{eq:SU3_generator_algebra}, the three-body color-color interaction arising from two $V^C$ insertions reads
\begin{align}
(\lambda_1^a \lambda_2^a)(\lambda_2^b \lambda_3^b)+ (\lambda_2^a \lambda_3^a)(\lambda_1^b \lambda_2^b) = \frac{4}{3}\lambda_1^a \lambda_3^a + 2d^{abc}\lambda_1^a \lambda_2^b \lambda_3^c.
\end{align}
Here $f$ and $d$ are the structure constants of SU(3). Including the spin factor $\boldsymbol{\sigma}\cdot\boldsymbol{\sigma}$ in each gluon exchange yields the three-body color-spin interaction (from the two-$V^{CS}$ exchange) as follows:
\begin{align}
&(\lambda_1^a \lambda_2^a \sigma_1^i \sigma_2^i)(\lambda_2^b \lambda_3^b \sigma_2^j \sigma_3^j)+ (\lambda_2^a \lambda_3^a \sigma_2^i \sigma_3^i)(\lambda_1^b \lambda_2^b \sigma_1^j \sigma_2^j)\nonumber\\&=-2f^{abc}\lambda_1^a \lambda_2^b \lambda_3^c \epsilon^{ijk}\sigma_1^i \sigma_2^j \sigma_3^k + \frac{4}{3}\lambda_1^a \lambda_3^a \sigma_1^i \sigma_3^i+ 2d^{abc}\lambda_1^a \lambda_2^b \lambda_3^c \sigma_1^i \sigma_3^i.
\end{align}
The three-body color-spin hybrid term (exchanging $V^{C}$ and $V^{CS}$) is
\begin{align}
&(\lambda_1^a \lambda_2^a \sigma_1^i \sigma_2^i)(\lambda_2^b \lambda_3^b)+ (\lambda_2^a \lambda_3^a)(\lambda_1^b \lambda_2^b \sigma_1^i \sigma_2^i) + (\lambda_1^a \lambda_2^a)(\lambda_2^b \lambda_3^b \sigma_2^i \sigma_3^i)+ (\lambda_2^a \lambda_3^a \sigma_2^i \sigma_3^i)(\lambda_1^b \lambda_2^b)\nonumber \\
&= \left(\frac{4}{3}\lambda_1^a \lambda_3^a + 2d^{abc}\lambda_1^a \lambda_2^b \lambda_3^c\right)\sigma_1^i \sigma_2^i + \left(\frac{4}{3}\lambda_1^a \lambda_3^a + 2d^{abc}\lambda_1^a \lambda_2^b \lambda_3^c\right)\sigma_2^i \sigma_3^i.
\end{align}
For the two insertions, we retain the central color and color-spin operator structures of Eqs.~\eqref{2body_1} and~\eqref{2body_2}.
Their effective strengths and spatial dependence are represented by the fitted couplings and profiles introduced below.
Gathering all possible excitations and including a factor of $M_i^{-1}$ for the propagator of the excited $i$th quark, we arrive at the following color-spin dependence. Here, the superscripts $C$ and $S$ label the color-color and color-spin interactions, respectively; the factor $\mathrm{sgn}(q)$ is \ensuremath{+}1 for quarks $q$ and \ensuremath{-}1 for antiquarks $\bar{q}$. The notation $\sum_{\mathrm{cyc}(l,i,j)}$ denotes the cyclic sum over the indices $(l,i,j)$, $(i,j,l)$, and $(j,l,i)$:
\begin{align}
&L^{C-C}_{123}
=\frac{4}{3}\sum_{\text{cyc}(l,i,j)} \frac{\lambda_i^c\lambda_j^c}{M_l}
+
2\,d^{abc}\,\lambda_1^a\lambda_2^b\lambda_3^c\,
\sum_{l=1}^3 \frac{\mathrm{sgn}(q_l)}{M_l},\label{eq:l12}\\
&L^{S-S}_{123}
=
\sum_{\text{cyc}(l,i,j)} \frac{4\,\boldsymbol{\sigma}_i\cdot\boldsymbol{\sigma}_j\,\lambda_i^c\lambda_j^c}{3\,m_l M_l(m_1 m_2 m_3)}
+
\frac{2\,d^{abc}\,\lambda_1^a\lambda_2^b\lambda_3^c}{m_1 m_2 m_3}
\sum_{\text{cyc}(l,i,j)} \frac{\mathrm{sgn}(q_l)}{m_l M_l}\,
\boldsymbol{\sigma}_i\cdot\boldsymbol{\sigma}_j-
\frac{2\,f^{abc}\,\lambda_1^a\lambda_2^b\lambda_3^c\,
\boldsymbol{\sigma}_1\cdot(\boldsymbol{\sigma}_2\times \boldsymbol{\sigma}_3)}{m_1 m_2 m_3}
\sum_{l=1}^3 \frac{1}{m_l M_l}
\nonumber\\&
L^{C-S}_{123}
=\frac{4}{3}\sum_{\text{cyc}(l,i,j)}
\frac{\lambda_i^c\lambda_j^c}{M_l}
\left(\frac{\boldsymbol{\sigma}_l\cdot\boldsymbol{\sigma}_i}{m_l m_i}
 +\frac{\boldsymbol{\sigma}_l\cdot\boldsymbol{\sigma}_j}{m_l m_j}\right)
+2\,d^{abc}\,\lambda_1^a\lambda_2^b\lambda_3^c
\sum_{\text{cyc}(l,i,j)}
\frac{\mathrm{sgn}(q_l)}{m_l M_l}
\left(\frac{\boldsymbol{\sigma}_l\cdot\boldsymbol{\sigma}_i}{m_i}
 +\frac{\boldsymbol{\sigma}_l\cdot\boldsymbol{\sigma}_j}{m_j}\right)
 .\nonumber
\end{align}
For an antiquark, $\bar F^a=-F^{aT}$. Applying Eq.~\eqref{eq:SU3_generator_algebra} in the conjugate representation gives $[\bar F^a,\bar F^b]=if^{abc}\bar F^c$ and $\{\bar F^a,\bar F^b\}=\delta^{ab}\mathbf{1}/3-d^{abc}\bar F^c$. The $f^{abc}$ structures therefore retain their sign, whereas a $d^{abc}$ structure changes sign when one of its legs is replaced by an antiquark.

The different inverse-mass factors follow from the number of chromomagnetic insertions.
They allow $L^{C-C}$, $L^{C-S}$, and $L^{S-S}$ to respond differently to the spin and flavor content of a baryon.
These operators, $L^{C-C}$, $L^{S-S}$, and $L^{C-S}$, are a natural three-body extension of the original De R\'ujula-Georgi-Glashow quark model~\cite{DeRujula:1975}.

\subsection{Color--spin three-quark forces and the
flavor-dependency puzzle}

\begin{table*}[!t]
\caption{\justifying
Ground-state baryon masses obtained using the meson-fitted two-body Hamiltonian without and with the spatial-profile-independent three-body interaction $L^{3Q}$ and its Gaussian- and Yukawa-profile extensions, $L_G^{3Q}$ and $L_Y^{3Q}$.
Values in parentheses denote deviations from the experimental masses.
The three-body contribution is defined as $M^{3Q}=M_{\rm with,3BD}-M_{\rm 2Q}$.
The determined couplings for the contact interaction are $A=-(24.4395~\mathrm{MeV})^2$, $B=(95.4733~\mathrm{MeV})^6$, and $C=-(48.0203~\mathrm{MeV})^4$.
Those for the Gaussian profile are $A=-(36.9003~\mathrm{MeV})^2$, $B=(102.744~\mathrm{MeV})^6$, and $C=-(62.462~\mathrm{MeV})^4$, while those for the Yukawa profile are $A=-(136.266~\mathrm{MeV})^2$, $B=(121.385~\mathrm{MeV})^6$, and $C=-(133.372~\mathrm{MeV})^4$.}
\label{tab:combined_3bd_baryon}
\renewcommand{\pdev}[1]{\textcolor{devred}{\ensuremath{(#1)}}}
\renewcommand{\ndev}[1]{\textcolor{devgreen}{\ensuremath{(#1)}}}

\centering
\small

\begin{tblr}{
  width=\textwidth,
  colspec={
Q[c,wd=1.15cm]
Q[c,wd=0.95cm]
Q[c,wd=1.35cm]
Q[c,wd=1.55cm]
Q[c,wd=1.55cm]
Q[c,wd=1.15cm]
Q[c,wd=1.55cm]
Q[c,wd=1.15cm]
Q[c,wd=1.55cm]
Q[c,wd=1.15cm]
  },
  row{1-2} = {bg=headgray, halign=c, valign=m},
  cell{1}{1} = {r=2}{},
  cell{1}{2} = {r=2}{},
  cell{1}{3} = {r=2}{},
  cell{1}{4} = {r=2}{},
  cell{1}{5} = {c=2}{},
  cell{1}{7} = {c=2}{},
  cell{1}{9} = {c=2}{},
  hline{2} = {5-10}{0.2pt},
  hline{3} = {1-10}{0.2pt},
  rowsep = 2pt,
}

\toprule
Baryon
& $\mathcal{J}^P$
& \shortstack[c]{$M_{\rm exp}$\\(MeV)}
& \shortstack[c]{Two-body\\($\sigma=65.04$)}
& \SetCell[c=2]{c}\shortstack[c]{Contact $L^{3Q}$\\($\sigma=18.87$)}
&
& \SetCell[c=2]{c}\shortstack[c]{Gaussian $L^{3Q}_{\rm G}$\\($\sigma=6.28$)}
&
& \SetCell[c=2]{c}\shortstack[c]{Yukawa $L^{3Q}_{\rm Y}$\\($\sigma=3.13$)}
& \\

&
&
&
&
Mass & $M^{3Q}$
& Mass & $M^{3Q}$
& Mass & $M^{3Q}$ \\

$p$ & $1/2^{+}$ & 938.272
& \massdev{1038.746}{\pdev{+100.474}}
& \massdev{965.782}{\pdev{+27.510}}
& \mthreeq{-72.964}{deltaBlue}
& \massdev{945.713}{\pdev{+7.441}}
& \mthreeq{-93.033}{gaussOrange}
& \massdev{937.449}{\ndev{-0.823}}
& \mthreeq{-101.297}{yukawaGreen} \\

$\Delta$ & $3/2^{+}$ & 1232
& \massdev{1338.864}{\pdev{+106.864}}
& \massdev{1257.170}{\pdev{+25.170}}
& \mthreeq{-81.694}{deltaBlue}
& \massdev{1240.358}{\pdev{+8.358}}
& \mthreeq{-98.506}{gaussOrange}
& \massdev{1230.352}{\ndev{-1.648}}
& \mthreeq{-108.512}{yukawaGreen} \\

$\Sigma$ & $1/2^{+}$ & 1192.642
& \massdev{1259.291}{\pdev{+66.649}}
& \massdev{1190.802}{\ndev{-1.840}}
& \mthreeq{-68.489}{deltaBlue}
& \massdev{1186.188}{\ndev{-6.454}}
& \mthreeq{-73.103}{gaussOrange}
& \massdev{1189.489}{\ndev{-3.153}}
& \mthreeq{-69.802}{yukawaGreen} \\

$\Lambda$ & $1/2^{+}$ & 1115.683
& \massdev{1171.919}{\pdev{+56.236}}
& \massdev{1117.274}{\pdev{+1.591}}
& \mthreeq{-54.645}{deltaBlue}
& \massdev{1113.396}{\ndev{-2.287}}
& \mthreeq{-58.523}{gaussOrange}
& \massdev{1117.473}{\pdev{+1.790}}
& \mthreeq{-54.446}{yukawaGreen} \\

$\Sigma_c$ & $1/2^{+}$ & 2452.650
& \massdev{2491.972}{\pdev{+39.322}}
& \massdev{2431.474}{\ndev{-21.176}}
& \mthreeq{-60.498}{deltaBlue}
& \massdev{2443.698}{\ndev{-8.952}}
& \mthreeq{-48.274}{gaussOrange}
& \massdev{2449.758}{\ndev{-2.892}}
& \mthreeq{-42.214}{yukawaGreen} \\

$\Lambda_c$ & $1/2^{+}$ & 2286.460
& \massdev{2317.914}{\pdev{+31.454}}
& \massdev{2274.460}{\ndev{-12}}
& \mthreeq{-43.454}{deltaBlue}
& \massdev{2282.061}{\ndev{-4.399}}
& \mthreeq{-35.853}{gaussOrange}
& \massdev{2287.781}{\pdev{+1.321}}
& \mthreeq{-30.133}{yukawaGreen} \\

$\Xi_{cc}$ & $1/2^{+}$ & 3619.970
& \massdev{3622.232}{\pdev{+2.262}}
& \massdev{3589.189}{\ndev{-30.781}}
& \mthreeq{-33.043}{deltaBlue}
& \massdev{3612.128}{\ndev{-7.842}}
& \mthreeq{-10.104}{gaussOrange}
& \massdev{3616.370}{\ndev{-3.600}}
& \mthreeq{-5.862}{yukawaGreen} \\

$\Sigma^{*}$ & $3/2^{+}$ & 1383.800
& \massdev{1450.640}{\pdev{+66.840}}
& \massdev{1380.424}{\ndev{-3.376}}
& \mthreeq{-70.216}{deltaBlue}
& \massdev{1380.290}{\ndev{-3.510}}
& \mthreeq{-70.350}{gaussOrange}
& \massdev{1386.937}{\pdev{+3.137}}
& \mthreeq{-63.703}{yukawaGreen} \\

$\Sigma_c^{*}$ & $3/2^{+}$ & 2518.480
& \massdev{2562.599}{\pdev{+44.119}}
& \massdev{2502.403}{\ndev{-16.077}}
& \mthreeq{-60.196}{deltaBlue}
& \massdev{2515.322}{\ndev{-3.158}}
& \mthreeq{-47.277}{gaussOrange}
& \massdev{2524.703}{\pdev{+6.223}}
& \mthreeq{-37.896}{yukawaGreen} \\

\bottomrule
\end{tblr}
\end{table*}

For an $n$-quark state, the $d$-term and $f$-term can be calculated as follows~\cite{Park:2022jpf}:
\begin{align}
\sum_{i<j<k}
d^{abc}\lambda_i^a\lambda_j^b\lambda_k^c
=\frac{4}{3}C_3 -\frac{10}{3}C_2 +\frac{80}{27}n,
\qquad \sum_{i<j<k}
f^{abc}\lambda_i^a\lambda_j^b\lambda_k^c =\frac{i}{2}\sum_{i<j<k}[\lambda_i^a\lambda_j^a,\,\lambda_j^b\lambda_k^b].
\end{align}
Here $C_2$ and $C_3$ denote the quadratic and cubic Casimir invariants of the total color representation.
For an irreducible SU(3) representation $\mathcal D(p,q)$, their eigenvalues are
\begin{equation}
\label{cas_1}
C_2(p,q) = \frac{1}{3}\left( p^2+q^2+pq+3p+3q \right),
\qquad C_3(p,q) = \frac{1}{18} (p-q)(2p+q+3)(p+2q+3).
\end{equation}
For a color-singlet baryon, the required color expectation values are
\begin{equation}
\left\langle\lambda_i^c\lambda_j^c\right\rangle_{\mathbf1_C}=-\frac83,\qquad
\left\langle d^{abc}\lambda_1^a\lambda_2^b\lambda_3^c\right\rangle_{\mathbf1_C}=\frac{80}{9},\qquad
\left\langle f^{abc}\lambda_1^a\lambda_2^b\lambda_3^c\right\rangle_{\mathbf1_C}=0.
\label{eq:qqq_color_singlet_factors}
\end{equation}
Using Eq.~\eqref{eq:qqq_color_singlet_factors}, the three-quark operators reduce to
\begin{align}
\langle L^{C-C}\rangle_\mathbf{1}&=\frac{128}{9}\left(\frac{1}{M_1}+\frac{1}{M_2}+\frac{1}{M_3}\right),\nonumber\\
\langle L^{S-S}\rangle_\mathbf{1}&=\frac{128}{9m_1m_2m_3}\left(\frac{\boldsymbol{\sigma}_2 \cdot \boldsymbol{\sigma}_3}{m_1M_1} + \frac{\boldsymbol{\sigma}_3 \cdot \boldsymbol{\sigma}_1}{m_2M_2} + \frac{\boldsymbol{\sigma}_1 \cdot \boldsymbol{\sigma}_2}{m_3M_3}\right), \nonumber\\
\langle L^{C-S}\rangle_\mathbf{1} &=\frac{128}{9} \left[ \frac{\boldsymbol{\sigma}_1\cdot\boldsymbol{\sigma}_2}{m_1 m_2}\left(\frac{1}{M_1}+\frac{1}{M_2}\right)+\frac{\boldsymbol{\sigma}_2\cdot\boldsymbol{\sigma}_3}{m_2 m_3}\left(\frac{1}{M_2}+\frac{1}{M_3}\right) + \frac{\boldsymbol{\sigma}_3\cdot\boldsymbol{\sigma}_1}{m_3 m_1}\left(\frac{1}{M_3}+\frac{1}{M_1}\right)\right].\label{organized_form}
\end{align}
The spatial-profile-independent three-quark operator is written as $L^{3Q}=A L^{C-C}+B L^{S-S}+C L^{C-S}$.
At this stage, no explicit spatial dependence is included.
We therefore use the term ``contact limit'' only as shorthand for a spatially unresolved three-body interaction.
Any common zero-range spatial matrix element is absorbed into the effective couplings $A$, $B$, and $C$, rather than being represented by an explicit coordinate-space potential.
Finite-range spatial profiles are introduced in  Section~\ref{sec:short_range}.
Using the two-body eigenstate $\boldsymbol{\mathcal{ B}}$ obtained in  Section~\ref{baryon_spectrum_sec}, we evaluate the three-quark interaction in first-order perturbation. The accuracy of this first-order treatment is tested independently for the Gaussian-profile interaction by rediagonalizing the retained $S$-wave Hamiltonian in \ref{subsec:gaussian_3q_perturbative_validation}. 
The quantity denoted by $M_i$ in Eqs.~\eqref{eq:l12} and~\eqref{organized_form} represents the excitation-energy denominator associated with the intermediate constituent-quark $q^{*}$ configuration, rather than the baryon mass.
We use $M_i=k\, m_i$ as a minimal phenomenological flavor-scaling prescription and absorb the common constant $k$ into $A$, $B$, and $C$.
Possible intermediate space realizations of $M_i$ are discussed in  Section~\ref{subsec:three_quark_definition}.
The couplings are determined by minimizing the baryon RMSE while keeping all two-body parameters fixed at their meson-fit values.
For the spatial-profile-independent interaction, the determined couplings are
\begin{equation}
A=-(24.4395~\mathrm{MeV})^2,\qquad B=(95.4733~\mathrm{MeV})^6,\qquad C=-(48.0203~\mathrm{MeV})^4.
\label{3bd_coeff_fitting}
\end{equation}
Only the nine ground-state baryons listed in Table~\ref{tab:combined_3bd_baryon} are included in this stage.

As shown in Table~\ref{tab:combined_3bd_baryon}, the spatial-profile-independent interaction reduces the baryon RMSE from 65.04 to 18.87~MeV.
The corresponding mean absolute error (MAE) decreases from 57.14 to 15.50~MeV, while the mean residual changes from \ensuremath{+}57.14 to $-3.44~\mathrm{MeV}$.
The principal effect of $L^{3Q}$ at this stage is therefore to remove the large common overestimate generated by transferring the meson-fitted two-body Hamiltonian to the baryon sector.

The color-spin $L^{3Q}$ interaction also reorganizes the relative residuals within fixed-flavor baryon multiplets.
We do not characterize this effect using only the $\Delta-N$, $\Sigma^{*}-\Sigma$, and $\Sigma_c^{*}-\Sigma_c$ splittings, because such a diagnostic omits the $\Lambda$--$\Sigma$ light-pair structure and the corresponding primed--unprimed structures in three-distinct-flavor baryons.
A broader spin structure RMSE, including both the calibration multiplets and the measured ground-state out-of-fit multiplets, is defined together with the out-of-fit test in Table~\ref{tab:primary_ground_state_holdout}.
We report the Pearson correlation between the constituent rest-mass sum $M_B^{\rm rest}=\sum_{i\in B}m_i$ and the mass residual $\varepsilon_B=M_B-M_B^{\exp}$ as a descriptive measure of the remaining flavor trend,
\begin{equation}
r_{M_{\rm rest},\varepsilon}=\frac{\sum_B(M_B^{\rm rest}-\overline{M^{\rm rest}})(\varepsilon_B-\bar\varepsilon)}{\sqrt{\sum_B(M_B^{\rm rest}-\overline{M^{\rm rest}})^2\sum_B(\varepsilon_B-\bar\varepsilon)^2}}.
\label{eq:flavor_correlation}
\end{equation}
For the present data, this coefficient changes from $r=-0.908$ in the two-body calculation to $r=-0.862$ after including $L^{3Q}$.
The contact-limit interaction therefore leaves a strong negative correlation, with lighter baryons remaining systematically higher relative to experiment than baryons containing charm quarks.
We use $r$ only as a descriptive measure of the residual trend because the three couplings are fitted to the same finite sample and the constituent rest-mass sum assumes only four distinct values for the baryons considered here.
Accordingly, the spatially contact interaction shows that the color-spin operator basis can substantially reduce the overall discrepancy in the baryon spectrum. 
However, this form alone does not provide the flavor selectivity required for a uniform description across flavor sectors. 
Since the couplings of the spatially contact form are determined phenomenologically from the same baryon data, the reduction of the in-sample residuals is expected by construction and should, at this stage, be interpreted primarily as evidence that the operator basis has sufficient flexibility to accommodate the dominant discrepancy, rather than as evidence that the contact form captures the underlying flavor-dependent dynamics.

\section{Finite-range effective three-quark interaction in the valence \(qqq\) representation}\label{sec:short_range}

In the previous section, the spatially contact three-quark interaction $L^{3Q}$ removed most of the common baryon mass offset but left a strong flavor-dependent residual pattern. This suggests that the remaining discrepancy comes from treating the interaction without an explicit spatial dependence. We therefore introduce short-range profiles whose physical range decreases as the constituent pair mass increases.

We tested exponential, Gaussian, and several other monotonic functions of the mass-weighted pair distances $x_{ij}=m_{ij}cr_{ij}/\hbar$. When their effective ranges are comparable, all of these profiles strongly reduce the baryon RMSE. In contrast, the same type of short-range profile with one common flavor-independent physical range does not remove the residual flavor pattern. We therefore find that the important ingredient is the mass scaling of the spatial range and not the detailed analytic shape of the profile.

We use the Yukawa-type form as the simplest representative of these profiles. Its spatial scale is fixed by the constituent reduced masses already determined from the meson calibration, and no additional continuous range parameter is introduced. We also calculate a Gaussian profile with the same effective range to examine the dependence on the profile shape.

\subsection{A minimal spatial ansatz for the three-body force and its contribution}

In the quark model Hamiltonian of Eq.~\eqref{hamiltonian}, the two-body interaction already incorporates short-distance behavior through the spatial dependence (non-perturbative effects are encoded). To go beyond the spatially unresolved interaction, the newly introduced three-body term must likewise respect QCD's asymptotic freedom at small separations. Asymptotic freedom causes the coupling $\alpha_s(Q^2)$ to decrease with increasing momentum transfer $Q$ (i.e.,~shorter distances), so that TGE naturally generates local three-body operators whose strength is confined to a short range, after unresolved intermediate configurations are eliminated. In non-perturbative QCD, the singular OGE contact term --- e.g.,~$V^{CS}_{\text{pQCD}}\propto\delta^{(3)}(\mathbf{r})F_i^cF_j^c\boldsymbol{\sigma}_i\cdot\boldsymbol{\sigma}_j$ --- is effectively smeared by gluon-propagator dressing and the freezing of $\alpha_s$, yielding a short but finite range set by the gluonic correlation length. By the same logic, the TGE three-body operator should inherit an analogous smearing. In the heavy-quark sector, although these higher-dimensional TGE operators carry formal suppressions of order $1/m_Q$ or higher, the bound-state wavefunction near the origin scales as $\psi(0)\sim(m_Q \alpha_s)^{3/2}$, and is dominated by its Coulombic core, $\psi(r)\sim e^{-r/r_0}$, where $r_0\sim1/(m_Q\alpha_s)$ is the QCD Bohr radius. For an unequal-mass pair, the corresponding relative-motion scale is controlled by the pair reduced mass $m_{ij}$, so this scaling generalizes schematically to $r_{0,ij}\propto1/(m_{ij}\alpha_s)$ and reduces to the heavy-heavy result when $m_i=m_j=m_Q$. As $m_{ij}$ increases, the corresponding pair scale contracts, further concentrating the short-distance wavefunction. To capture this intrinsic short-range behavior, we supplement the common algebraic operator with a minimal spatial operator applied uniformly to each of $L^{C-C}$, $L^{S-S}$, and $L^{C-S}$:
\begin{align}\label{eq:3bd_yuk_full}
L^{3Q}_{\text{Y}} &= \sum_{i<j<k} \left(AL^{C-C}_{ijk} + BL^{S-S}_{ijk} + CL^{C-S}_{ijk}\right)\times f_\text{Y}(r_{ij}, r_{jk}, r_{ki}),
\end{align}
where the explicit spatial operator adopted here is the minimal combination of $mr$ (we term it ``Yukawa-type'')
\begin{equation}\label{yukawa}
f_\text{Y}(r_{ij},r_{jk},r_{ki})=\exp\left[ -\frac{c}{\hbar}(m_{ij}r_{ij}+m_{jk}r_{jk} +m_{ki}r_{ki} )\right],
\end{equation}
where $m_{ij}$ is the reduced mass. With this minimal choice, both the two-body $V^{CS}$ and the TGE-induced three-body terms are regulated by the same type of short-range profile. For simplicity we keep $\alpha_s$ constant; the profile absorbs the dominant short-distance dressing, so that the scale dependence primarily enters through the reduced masses $m_{ij}$.
This profile has the same exponential form as the Coulomb-like heavy-quark wave function, and its range is of order the pair Compton wavelength $\hbar/m_{ij}c$. It is a simple way to use the short-distance scale of the two-body hyperfine interaction also in the three-body term.
The details of the matrix element calculation for $L_{\text{Y}}^{3Q}$ are provided in \ref{GEM_3body_calc}.

With the additional spatial dependence, we redetermine the couplings $A$, $B$, and $C$ by evaluating $L_{\mathrm Y}^{3Q}$ in first-order perturbation theory and minimizing the baryon RMSE.
The resulting couplings, which retain the nominal $\mathcal O(\alpha_s^2)$ counting of the TGE construction, are
\begin{equation}
A=-(136.266~\mathrm{MeV})^2,\qquad B=(121.385~\mathrm{MeV})^6,\qquad C=-(133.372~\mathrm{MeV})^4.
\label{3bd_short_coeff_fitting}
\end{equation}
No additional continuous range parameter is introduced relative to the spatial-profile-independent interaction.
The same three couplings are redetermined, while the spatial scales appearing in $f_{\mathrm Y}$ are fixed by the constituent-quark masses already determined from the meson spectrum.
The corresponding characteristic scales, $\vert A\vert ^{1/2}=136.266~\mathrm{MeV}$, $B^{1/6}=121.385~\mathrm{MeV}$, and $\vert C\vert ^{1/4}=133.372~\mathrm{MeV}$, are of the same hadronic order.

The resulting ground-state baryon masses are shown in Table~\ref{tab:combined_3bd_baryon} and  Fig.~\ref{fig:errors}.
The baryon RMSE decreases according to
\begin{equation}
\sigma_B=65.04~\mathrm{MeV}\;(\mathrm{two\mbox{-}body})\;\longrightarrow\;18.87~\mathrm{MeV}\;(L^{3Q})\;\longrightarrow\;3.13~\mathrm{MeV}\;(L_{\mathrm Y}^{3Q}).
\label{eq:yukawa_rmse_improvement}
\end{equation}
The Yukawa profile therefore improves the RMSE by a factor of approximately 20.8 relative to the direct two-body prediction and by a factor of approximately 6.0 relative to the spatial-profile-independent three-body interaction.
The spatial-profile-independent interaction already reduces the mean residual from \ensuremath{+}57.14 to $-3.44~\mathrm{MeV}$, showing that it primarily corrects the common baryon mass offset.
However, its RMSE remains 18.87~MeV because substantial state-dependent residuals remain.
After introducing the Yukawa profile, the MAE decreases to 2.73~MeV, the mean residual becomes $+0.04~\mathrm{MeV}$, and the largest absolute residual is reduced to 6.22~MeV.
So, the improvement is not only a common shift of the spectrum; the state-dependent residuals also become small.

The residual flavor dependence is characterized by the Pearson correlation between the constituent rest-mass sum $M_B^{\rm rest}=\sum_{i\in B}m_i$ and the mass residual $\varepsilon_B=M_B-M_B^{\exp}$.
The correlation changes as
\begin{equation}
r_{M_{\rm rest},\varepsilon}=-0.908\;(\mathrm{two\mbox{-}body})\;\longrightarrow\;-0.862\;(L^{3Q})\;\longrightarrow\;-0.136\;(L_{\mathrm Y}^{3Q}).
\label{eq:yukawa_flavor_correlation}
\end{equation}
Thus, the strong negative flavor trend of the two-body calculation remains largely unchanged by the spatial-profile-independent interaction but is almost removed once the mass-dependent short-range profile is included.\footnote{Because the same nine baryons are used to determine $A$, $B$, and $C$, and because $M_B^{\rm rest}$ assumes only four distinct flavor values in the present sample, we use $r$ only as a descriptive measure.} 

The same two-body parameter set reproduces the meson spectrum with $\sigma_M=2.51~\mathrm{MeV}$ but gives $\sigma_B=65.04~\mathrm{MeV}$ when transferred directly to baryons.
The Yukawa-profile three-quark interaction reduces the latter value to 3.13~MeV without readjusting any two-body parameter.
Within the nine-state calibration set, the algebraic color-spin three-body operators improve the relative spin placement and correct much of the common baryon centroid.
The fixed-parameter test below shows, however, that this improvement does not transfer uniformly in the spatially contact and common-orbital form; the mass-dependent finite-range profile is required to obtain the better simultaneous transfer of the absolute multiplet centroids and their internal spin structure.

\paragraph{Out-of-fit baryons and spin structure:}

As a fixed-parameter ground-state test, we next evaluate the baryons that were not used to determine the couplings $A$, $B$, and $C$.
The two-body and Yukawa-profile values are calculated with the same Hamiltonian and wave-function convention used in the primary fit.
For the spatially contact interaction, the shifts follow algebraically from Eq.~\eqref{organized_form} using the determined couplings of Eq.~\eqref{3bd_coeff_fitting}; no additional spatial matrix element or parameter refit is required.
For the $\Xi_c$ and $\Xi_c'$ states, we use the pure $s_{qs}=0$ and $s_{qs}=1$ pair-spin assignments, respectively, while the starred states use the aligned $S=3/2$ configuration.

\begin{table*}[!t]
\centering
\caption{\justifying
Out-of-fit ground-state baryon test for the present linear central plus color-spin two-body Hamiltonian, later denoted TH1 in Section~\ref{sec:excitation_spectra}.
None of the states in this table is used to determine the two-body parameters or the three-quark couplings.
Each calculated mass is followed by its signed deviation from the experimental reference mass.
The quoted out-of-fit RMSEs include the nine states with experimental entries.
For the $\Omega_{cc}^{+}$, we use the preliminary LHCb mass near 3727~MeV announced in 2026 \cite{LHCbOmegaCC2026}.
The $\Omega_{cc}^{*}$ and $\Omega_{ccc}$ entries are predictions and are excluded from the RMSE. All masses, signed deviations, three-quark contributions, and RMSE values are in MeV.}
\label{tab:primary_ground_state_holdout}
\scriptsize
\renewcommand{\pdev}[1]{\textcolor{devred}{\ensuremath{(#1)}}}
\renewcommand{\ndev}[1]{\textcolor{devgreen}{\ensuremath{(#1)}}}
\resizebox{\textwidth}{!}{%
\begin{tblr}{
  colspec={
Q[l,wd=0.8cm]
Q[c,wd=1.15cm]
Q[c,wd=1.45cm]
Q[c,wd=2.5cm]
Q[c,wd=1.3cm]
Q[c,wd=2.5cm]
Q[c,wd=1.3cm]
Q[c,wd=2.5cm]
  },
  row{1}={bg=headgray,halign=c,valign=m},
  cell{1}{1}={}{halign=l},
  hline{2}={1-8}{0.3pt},
  rowsep=1.45pt,
}
\toprule
State
& $\mathcal J^P$
& $M_{\rm exp}$
& 2BD ($\sigma_{\rm out}=14.31$)
& $M^{3Q}$
& 2BD $+$ contact 3Q ($\sigma_{\rm out}=32.89$)
& $M^{3Q,\mathrm Y}$
& 2BD $+$ Yukawa 3Q ($\sigma_{\rm out}=5.37$)
\\

$\Xi$
& $1/2^+$
& $1321$
& \massdevinline{1349.835}{\pdev{+28.835}}
& \textcolor{deltaBlue}{$-48.069$}
& \massdevinline{1301.767}{\ndev{-19.233}}
& \textcolor{yukawaGreen}{$-23.653$}
& \massdevinline{1326.182}{\pdev{+5.182}}
\\

$\Xi^*$
& $3/2^+$
& $1535$
& \massdevinline{1559.621}{\pdev{+24.621}}
& \textcolor{deltaBlue}{$-57.452$}
& \massdevinline{1502.169}{\ndev{-32.831}}
& \textcolor{yukawaGreen}{$-23.974$}
& \massdevinline{1535.647}{\pdev{+0.647}}
\\

$\Xi_c$
& $1/2^+$
& $2469$
& \massdevinline{2483.543}{\pdev{+14.543}}
& \textcolor{deltaBlue}{$-39.186$}
& \massdevinline{2444.356}{\ndev{-24.644}}
& \textcolor{yukawaGreen}{$-12.123$}
& \massdevinline{2471.419}{\pdev{+2.419}}
\\

$\Xi_c'$
& $1/2^+$
& $2579$
& \massdevinline{2583.399}{\pdev{+4.399}}
& \textcolor{deltaBlue}{$-46.003$}
& \massdevinline{2537.396}{\ndev{-41.604}}
& \textcolor{yukawaGreen}{$-14.080$}
& \massdevinline{2569.318}{\ndev{-9.682}}
\\

$\Xi_c^*$
& $3/2^+$
& $2646$
& \massdevinline{2657.502}{\pdev{+11.502}}
& \textcolor{deltaBlue}{$-46.703$}
& \massdevinline{2610.799}{\ndev{-35.201}}
& \textcolor{yukawaGreen}{$-12.233$}
& \massdevinline{2645.269}{\ndev{-0.731}}
\\

$\Omega$
& $3/2^+$
& $1672$
& \massdevinline{1677.312}{\pdev{+5.312}}
& \textcolor{deltaBlue}{$-44.756$}
& \massdevinline{1632.556}{\ndev{-39.444}}
& \textcolor{yukawaGreen}{$-9.799$}
& \massdevinline{1667.513}{\ndev{-4.487}}
\\

$\Omega_c$
& $1/2^+$
& $2695$
& \massdevinline{2692.684}{\ndev{-2.316}}
& \textcolor{deltaBlue}{$-33.515$}
& \massdevinline{2659.169}{\ndev{-35.831}}
& \textcolor{yukawaGreen}{$-5.763$}
& \massdevinline{2686.921}{\ndev{-8.079}}
\\

$\Omega_c^*$
& $3/2^+$
& $2766$
& \massdevinline{2765.384}{\ndev{-0.616}}
& \textcolor{deltaBlue}{$-34.143$}
& \massdevinline{2731.241}{\ndev{-34.759}}
& \textcolor{yukawaGreen}{$-4.853$}
& \massdevinline{2760.531}{\ndev{-5.469}}
\\

$\Omega_{cc}$
& $1/2^+$
& $\simeq3727$
& \massdevinline{3724.128}{\ndev{-2.872}}
& \textcolor{deltaBlue}{$-22.814$}
& \massdevinline{3701.313}{\ndev{-25.687}}
& \textcolor{yukawaGreen}{$-1.171$}
& \massdevinline{3722.957}{\ndev{-4.043}}
\\

$\Omega_{cc}^*$
& $3/2^+$
& --
& $3810.968$
& \textcolor{deltaBlue}{$-23.714$}
& $3787.255$
& \textcolor{yukawaGreen}{$-1.002$}
& $3809.967$
\\

$\Omega_{ccc}$ 
& $3/2^+$ 
& --
& $4797.099$ 
& \textcolor{deltaBlue}{$-13.667$} 
& $4783.432$ 
& \textcolor{yukawaGreen}{$-0.268$} 
& $4796.831$ \\

\bottomrule
\end{tblr}}
\end{table*}

The out-of-fit masses distinguish sharply between the spatially contact and finite-range interactions.
The spatially contact interaction lowers the out-of-fit states too uniformly and increases their absolute-mass RMSE ($\sigma_{\rm out}$) from 14.31 to 32.89~MeV.
The Yukawa profile instead gives $\sigma_{\rm out}=5.37~\mathrm{MeV}$ without any additional adjustment.
Thus, the mass-scaled finite-range interaction retains a substantial fixed-parameter advantage over both the two-body Hamiltonian and the spatially contact three-quark interaction.

To isolate the internal spin structure from a common flavor-dependent centroid, we group states having exactly the same constituent content:
\begin{equation}
\mathcal G_{\rm structure}
=
\bigl\{
\{N,\Delta\},
\{\Lambda,\Sigma,\Sigma^*\},
\{\Lambda_c,\Sigma_c,\Sigma_c^*\},
\{\Xi,\Xi^*\},
\{\Xi_c,\Xi_c',\Xi_c^*\},
\{\Omega_c,\Omega_c^*\}
\bigr\}.
\label{eq:spin_structure_groups}
\end{equation}
For each group $g$, let
\begin{equation}
\bar\varepsilon_g
=
\frac{1}{n_g}
\sum_{B\in g}\varepsilon_B,
\qquad
\varepsilon_B=M_B^{\rm th}-M_B^{\rm exp}.
\end{equation}
We define the spin structure RMSE by profiling out one common mass offset from each fixed-content group,
\begin{equation}
\sigma_{\rm structure}
=
\bigg[
\frac{1}{N_{\rm structure}}
\sum_{g\in\mathcal G_{\rm structure}}
\sum_{B\in g}
\left(
\varepsilon_B-\bar\varepsilon_g
\right)^2
\bigg]^{1/2},
\qquad
N_{\rm structure}
=
\sum_g(n_g-1)
=
9.
\label{eq:spin_structure_rmse}
\end{equation}
This construction measures the relative placement of the different internal spin configurations while remaining invariant under a common shift of every member of a fixed-flavor multiplet.
It includes the $\Lambda$--$\Sigma$ and $\Xi_c$--$\Xi_c'$ pair-spin structures that are absent from the previous three-splitting diagnostic.
The resulting values are
\begin{equation}
\sigma_{\rm structure}
=
5.17~\mathrm{MeV}\;(\mathrm{2BD})
\;\longrightarrow\;
5.75~\mathrm{MeV}\;(\text{spatially contact }3Q)
\;\longrightarrow\;
4.17~\mathrm{MeV}\;(\text{Yukawa }3Q).
\label{eq:spin_structure_rmse_results}
\end{equation}
The spatially contact interaction does not improve the aggregate centroid-subtracted spin structure of the calibration and out-of-fit multiplets; instead, it increases $\sigma_{\rm structure}$ relative to the two-body Hamiltonian.
The Yukawa profile gives the smallest value.
The contact interaction therefore acts mainly as a broad correction to the calibration-set baryon centroid and overcorrects several omitted multiplets, whereas the mass-scaled finite-range profile gives the better transfer of both the absolute masses and the relative spin placements.
Over the union of the nine calibration states and the nine measured out-of-fit states, the absolute-mass RMSE decreases monotonically,
\begin{equation}
\sigma_{B,18}
=
47.09~\mathrm{MeV}\;(\mathrm{2BD})
\;\longrightarrow\;
26.81~\mathrm{MeV}\;(\text{spatially contact }3Q)
\;\longrightarrow\;
4.39~\mathrm{MeV}\;(\text{Yukawa }3Q).
\label{eq:full_ground_state_rmse}
\end{equation}
The mass-scaled Yukawa profile gives both the best absolute-mass transfer and the smallest aggregate centroid-subtracted spin-structure error, whereas the spatially contact interaction mainly corrects the common calibration-set centroid and does not transfer the same relative spin accuracy to the omitted multiplets.
The single-state $\Omega$, $\Xi_{cc}$, and $\Omega_{cc}$ sectors contribute to the absolute-mass tests but not to $\sigma_{\rm structure}$ because no measured spin partner is available in the present data set.

\subsection{Generalization of the spatial operator}

The improvement obtained by introducing the Yukawa profile is not restricted to the particular exponential dependence chosen in Eq.~\eqref{yukawa}.
The essential requirement is that the three-body interaction be localized by dimensionless combinations of the pair masses and physical pair distances.
To examine this point, we replace $f_{\mathrm Y}$ by the mass-dependent Gaussian profile
\begin{equation}
f_{\mathrm G}(r_{ij},r_{jk},r_{ki})=\exp\left[-\frac{c^2}{8\hbar^2}\left(m_{ij}^2r_{ij}^2+m_{jk}^2r_{jk}^2+m_{ki}^2r_{ki}^2\right)\right].
\label{gaussian}
\end{equation}
Replacing $f_{\mathrm Y}$ by $f_{\mathrm G}$ in Eq.~\eqref{eq:3bd_yuk_full} defines the Gaussian-profile interaction $L_{\mathrm G}^{3Q}$.
As in the Yukawa case, the numerical coefficient in the exponent is fixed and no independent range parameter is introduced.

The Gaussian profile is particularly convenient within GEM because its product with the spatial basis remains a multidimensional Gaussian.
The corresponding matrix elements can therefore be reduced to finite sums of analytic Gaussian integrals, including contributions between different Jacobi rearrangement channels.
The couplings $A$, $B$, and $C$ are redetermined using the same nine ground-state baryons and the same two-body wave functions employed for the other profiles.
The resulting values are
\begin{equation}
A=-(36.9003~\mathrm{MeV})^2,\qquad B=(102.744~\mathrm{MeV})^6,\qquad C=-(62.462~\mathrm{MeV})^4.
\label{eq:3bd_gaussian_coeff_fitting}
\end{equation}
As shown in Table~\ref{tab:combined_3bd_baryon}, the Gaussian profile gives $\sigma_B=6.28~\mathrm{MeV}$, compared with 18.87~MeV for the spatial-profile-independent interaction and 3.13~MeV for the Yukawa profile.
Its MAE is 5.82~MeV, its mean residual is $-2.31~\mathrm{MeV}$, and its largest absolute residual is 8.95~MeV.
The Gaussian profile also suppresses the overall state-dependent and flavor-dependent residual pattern.
The common structure of the two spatial forms becomes transparent by introducing\footnote{The factor $1/8$ in Eq.~\eqref{gaussian} is fixed by requiring the Gaussian and Yukawa pair factors to have the same normalized spatial range, rather than treating their widths as independent fit parameters.
For a single pair, temporarily write the Gaussian factor as $\exp(-\xi_{\rm G}x_{ij}^{2})$ and normalize both $\exp(-x_{ij})$ and $\exp(-\xi_{\rm G}x_{ij}^{2})$ with the three-dimensional radial measure.
Their dimensionless second moments are $\langle x_{ij}^{2}\rangle_{\mathrm Y}=12$ and $\langle x_{ij}^{2}\rangle_{\mathrm G}
={3}/({2\xi_{\rm G}})$, so equality of the pairwise rms ranges uniquely gives $\xi_{\rm G}=1/8$.
Equivalently, after normalization, the two pair profiles have the same quadratic coefficient in the low-momentum expansion of their Fourier transforms, since that coefficient is proportional to $\langle r_{ij}^{2}\rangle$.
Thus, the coefficient $1/8$ defines a parameter-free equal-range comparison between $f_{\mathrm Y}$ and $f_{\mathrm G}$; it does not make the two profiles pointwise identical or equate their full three-body matrix elements, but leaves their different higher moments and tail behavior as the principal distinction.
Because $A$, $B$, and $C$ are refitted separately for the two profiles, differences in their overall normalization are absorbed into the corresponding effective couplings.}
\begin{equation}
x_{ij}\equiv\frac{m_{ij}cr_{ij}}{\hbar},\qquad f_{\mathrm Y}=\exp\Big[-\sum_{i<j}x_{ij}\Big],\qquad f_{\mathrm G}=\exp\Big[-\frac{1}{8}\sum_{i<j}x_{ij}^2\Big].
\label{eq:mass_scaled_profile_variables}
\end{equation}
Both profiles are permutation invariant, satisfy $f=1$ when the three quarks coincide, and suppress configurations as their mass-weighted spatial extent increases.
They therefore implement the same physical mechanism despite their different behavior away from the origin.
The numerical results indicate that the principal ingredient is the mass-scaled localization of the three-body force rather than the detailed analytic form of its tail.

This mechanism is seen directly in the distribution of the fitted three-body contributions.
For the spatial-profile-independent interaction, the corrections to the proton and $\Xi_{cc}$ masses are \ensuremath{-}72.964 and $-33.043~\mathrm{MeV}$, respectively.
After introducing the Gaussian profile, these corrections become \ensuremath{-}93.033 and $-10.104~\mathrm{MeV}$, while the Yukawa profile gives \ensuremath{-}101.297 and $-5.862~\mathrm{MeV}$.
After the couplings are refitted, the spatial profile moves the attraction to the light baryons and makes the correction for $\Xi_{cc}$ small. The spatial-profile-independent interaction does not have this flavor dependence.

\begin{table*}[!t]
\centering
\caption{\justifying
Sensitivity of the nine-state ground-state baryon fit to alternative mass-scaled spatial profiles.
Panel (a) summarizes the matched one-pair profiles, their characteristic shapes, and the resulting baryon RMSEs, while panel (b) gives the corresponding state-by-state masses and signed deviations.
All one-pair factors satisfy the common second-moment condition $\langle x^2\rangle_g=12$ of Eq. \eqref{eq:alternative_profile_second_moment}, and the three couplings $A$, $B$, and $C$ are refitted independently for each profile.}
\label{tab:alternative_spatial_profiles}
\scriptsize
\textbf{(a) Characteristics of alternative mass-scaled spatial profiles.}
\vspace{2pt}
\resizebox{\textwidth}{!}{%
\begin{tblr}{
  colspec={
    Q[l,wd=2.9cm]
    Q[l,wd=5.8cm]
    Q[l,wd=4.9cm]
    Q[c,wd=1.1cm]
    Q[c,wd=1.3cm]
    Q[c,wd=1.8cm]
  },
  row{1}={bg=headgray,halign=c,valign=m},
  cell{1}{1-3}={}{halign=l},
  hline{2}={1-6}{0.3pt},
  rowsep=1.8pt,
}
\toprule
Profile $(g)$
& Matched one-pair factor $g(x)$
& Characteristic shape
& $\langle x^2\rangle_g$
& $\langle x^4\rangle_g$
& $\sigma_B$ (MeV) \\

Yukawa
&
$\displaystyle
e^{-x}$
&
Exponential profile; reference shape for the matched-range comparison
&
$12$
&
$360.0$
&
$3.130$
\\

Algebraic, $n=50$
&
$\displaystyle
\left(1+{x}/{\sqrt{2070}}\right)^{-50}$
&
Long algebraic tail; close to an exponential over the relevant range
&
$12$
&
$393.9$
&
$3.601$
\\ 

Algebraic, $n=20$
&
$\displaystyle
\left(1+{x}/{\sqrt{240}}\right)^{-20}$
&
Algebraic tail broader than the $n=50$ case
&
$12$
&
$474.7$
&
$3.748$
\\

Hyperbolic secant
&
$\displaystyle
\operatorname{sech}\!\left(
\pi\sqrt{{5}/{48}}\,x
\right)$
&
Smooth core with an exponential asymptotic tail
&
$12$
&
$351.4$
&
$4.114$
\\

Relativistic bridge
&
$\displaystyle
\exp\!\left[
1-\sqrt{1+(a_{\mathrm R}x)^2}
\right]$
&
Gaussian-like core continuously connected to an exponential tail
&
$12$
&
$342.1$
&
$4.240$
\\ 

Stretched exponential, $p=3/4$
&
$\displaystyle
\exp\!\left[
-a_{3/4}x^{3/4}
\right],
\qquad
a_{3/4}
=
\left[{\Gamma(20/3)}/{72}
\right]^{3/8}$
&
Broader-than-exponential tail
&
$12$
&
$472.9$
&
$4.252$
\\ 

Stretched exponential, $p=5/4$
&
$\displaystyle
\exp\!\left[
-a_{5/4}x^{5/4}
\right],
\qquad
a_{5/4}
=
\left[
{2\Gamma(12/5)}
\right]^{-5/8}$
&
Intermediate between exponential and Gaussian damping
&
$12$
&
$305.8$
&
$4.573$
\\ 

Mat\'ern $3/2$
&
$\displaystyle
\left(
1+\sqrt{{3}/{2}}\,x
\right)
\exp\!\left(
-\sqrt{{3}/{2}}\,x
\right)$
&
Vanishing slope at the origin with an exponential tail
&
$12$
&
$320.0$
&
$4.664$
\\ 

Airy
&
$\displaystyle
{
\operatorname{Ai}(a_{\mathrm A}x)
}/{
\operatorname{Ai}(0)
}$
&
Confinement-like tail proportional asymptotically to
$x^{-1/4}e^{-{\rm const.}\,x^{3/2}}$
&
$12$
&
$282.7$
&
$5.222$
\\ 

Mat\'ern $5/2$
&
$\displaystyle
\left(
1+\sqrt{2}\,x+(2/3)x^2
\right)e^{-\sqrt{2}\,x}$
&
Smoother and flatter core than the Mat\'ern $3/2$ profile
&
$12$
&
$300.0$
&
$5.593$
\\ 

Integrated Gaussian
&
$\displaystyle
\operatorname{erfc}\!\left({x}/{\sqrt{10}}
\right)$
&
Gaussian asymptotic damping with a broader smooth shoulder
&
$12$
&
$257.1$
&
$6.317$
\\ 

\bottomrule
\end{tblr}}

\vspace{5pt}

\renewcommand{\massdev}[2]{%
  \shortstack[c]{%
    #1\\[-0.5pt]
    {\tiny #2}%
  }%
}
\scriptsize
\textbf{(b) State-by-state comparison of the nine-state fits.}
\vspace{2pt}
\renewcommand{\pdev}[1]{\textcolor{devred}{\ensuremath{(#1)}}}
\renewcommand{\ndev}[1]{\textcolor{devgreen}{\ensuremath{(#1)}}}
\resizebox{\textwidth}{!}{%
\begin{tblr}{
colspec={
Q[l,wd=0.5cm]
Q[c,wd=0.5cm]
Q[c,wd=1.1cm]
Q[c,wd=1.3cm]
Q[c,wd=1.3cm]
Q[c,wd=1.3cm]
Q[c,wd=1.3cm]
Q[c,wd=1.3cm]
Q[c,wd=1.3cm]
Q[c,wd=1.3cm]
Q[c,wd=1.3cm]
Q[c,wd=1.3cm]
Q[c,wd=1.3cm]
Q[c,wd=1.3cm]
},
cells={valign=m},
row{1}={bg=headgray,halign=c},
cell{1}{1}={}{halign=l},
hline{2}={1-14}{0.4pt},
rowsep=1.25pt,
}
\toprule
State
& $\mathcal J^P$
& \shortstack[c]{$M_{\rm exp}$\\(MeV)}
& \shortstack[c]{2BD\\$\sigma_B=65.04$}
& \shortstack[c]{Alg. $n=50$\\$\sigma_B=3.60$}
& \shortstack[c]{Alg. $n=20$\\$\sigma_B=3.75$}
& \shortstack[c]{$\operatorname{sech}$\\$\sigma_B=4.11$}
& \shortstack[c]{Bridge\\$\sigma_B=4.24$}
& \shortstack[c]{Str. $p=3/4$\\$\sigma_B=4.25$}
& \shortstack[c]{Str. $p=5/4$\\$\sigma_B=4.57$}
& \shortstack[c]{Mat\'ern $3/2$\\$\sigma_B=4.66$}
& \shortstack[c]{Airy\\$\sigma_B=5.22$}
& \shortstack[c]{Mat\'ern $5/2$\\$\sigma_B=5.59$}
& \shortstack[c]{$\operatorname{erfc}$\\$\sigma_B=6.32$} \\

$p$
& $1/2^+$
& 938.272
& \massdev{1038.746}{\pdev{+100.474}}
& \massdev{941.770}{\pdev{+3.498}}
& \massdev{940.935}{\pdev{+2.663}}
& \massdev{944.038}{\pdev{+5.766}}
& \massdev{944.314}{\pdev{+6.042}}
& \massdev{939.976}{\pdev{+1.704}}
& \massdev{944.922}{\pdev{+6.650}}
& \massdev{945.156}{\pdev{+6.884}}
& \massdev{946.066}{\pdev{+7.794}}
& \massdev{946.828}{\pdev{+8.556}}
& \massdev{947.845}{\pdev{+9.573}} \\

$\Delta$
& $3/2^+$
& 1232
& \massdev{1338.864}{\pdev{+106.864}}
& \massdev{1231.435}{\ndev{-0.565}}
& \massdev{1230.246}{\ndev{-1.754}}
& \massdev{1234.127}{\pdev{+2.127}}
& \massdev{1234.473}{\pdev{+2.473}}
& \massdev{1228.730}{\ndev{-3.270}}
& \massdev{1235.288}{\pdev{+3.288}}
& \massdev{1235.470}{\pdev{+3.470}}
& \massdev{1236.604}{\pdev{+4.604}}
& \massdev{1237.269}{\pdev{+5.269}}
& \massdev{1238.556}{\pdev{+6.556}} \\

$\Sigma$
& $1/2^+$
& 1192.642
& \massdev{1259.291}{\pdev{+66.649}}
& \massdev{1187.105}{\ndev{-5.537}}
& \massdev{1187.366}{\ndev{-5.276}}
& \massdev{1186.422}{\ndev{-6.220}}
& \massdev{1186.393}{\ndev{-6.249}}
& \massdev{1187.764}{\ndev{-4.878}}
& \massdev{1186.387}{\ndev{-6.255}}
& \massdev{1186.287}{\ndev{-6.355}}
& \massdev{1186.274}{\ndev{-6.368}}
& \massdev{1186.081}{\ndev{-6.561}}
& \massdev{1186.189}{\ndev{-6.453}} \\

$\Lambda$
& $1/2^+$
& 1115.683
& \massdev{1171.919}{\pdev{+56.236}}
& \massdev{1113.965}{\ndev{-1.718}}
& \massdev{1114.027}{\ndev{-1.656}}
& \massdev{1113.698}{\ndev{-1.985}}
& \massdev{1113.684}{\ndev{-1.999}}
& \massdev{1114.073}{\ndev{-1.610}}
& \massdev{1113.682}{\ndev{-2.001}}
& \massdev{1113.629}{\ndev{-2.054}}
& \massdev{1113.615}{\ndev{-2.068}}
& \massdev{1113.498}{\ndev{-2.185}}
& \massdev{1113.536}{\ndev{-2.147}} \\

$\Sigma_c$
& $1/2^+$
& 2452.65
& \massdev{2491.972}{\pdev{+39.322}}
& \massdev{2449.321}{\ndev{-3.329}}
& \massdev{2450.512}{\ndev{-2.138}}
& \massdev{2446.869}{\ndev{-5.781}}
& \massdev{2446.522}{\ndev{-6.128}}
& \massdev{2451.956}{\ndev{-0.694}}
& \massdev{2445.690}{\ndev{-6.960}}
& \massdev{2445.575}{\ndev{-7.075}}
& \massdev{2444.482}{\ndev{-8.168}}
& \massdev{2443.975}{\ndev{-8.675}}
& \massdev{2442.761}{\ndev{-9.889}} \\

$\Lambda_c$
& $1/2^+$
& 2286.46
& \massdev{2317.914}{\pdev{+31.454}}
& \massdev{2285.950}{\ndev{-0.510}}
& \massdev{2286.741}{\pdev{+0.281}}
& \massdev{2284.190}{\ndev{-2.270}}
& \massdev{2283.963}{\ndev{-2.497}}
& \massdev{2287.753}{\pdev{+1.293}}
& \massdev{2283.442}{\ndev{-3.018}}
& \massdev{2283.320}{\ndev{-3.140}}
& \massdev{2282.607}{\ndev{-3.853}}
& \massdev{2282.168}{\ndev{-4.292}}
& \massdev{2281.391}{\ndev{-5.069}} \\

$\Xi_{cc}$
& $1/2^+$
& 3619.97
& \massdev{3622.232}{\pdev{+2.262}}
& \massdev{3616.210}{\ndev{-3.760}}
& \massdev{3616.701}{\ndev{-3.269}}
& \massdev{3615.281}{\ndev{-4.689}}
& \massdev{3615.106}{\ndev{-4.864}}
& \massdev{3617.162}{\ndev{-2.808}}
& \massdev{3614.612}{\ndev{-5.358}}
& \massdev{3614.637}{\ndev{-5.333}}
& \massdev{3613.964}{\ndev{-6.006}}
& \massdev{3613.861}{\ndev{-6.109}}
& \massdev{3612.956}{\ndev{-7.014}} \\

$\Sigma^*$
& $3/2^+$
& 1383.8
& \massdev{1450.640}{\pdev{+66.840}}
& \massdev{1386.402}{\pdev{+2.602}}
& \massdev{1387.374}{\pdev{+3.574}}
& \massdev{1384.309}{\pdev{+0.509}}
& \massdev{1384.107}{\pdev{+0.307}}
& \massdev{1388.853}{\pdev{+5.053}}
& \massdev{1383.693}{\ndev{-0.107}}
& \massdev{1383.514}{\ndev{-0.286}}
& \massdev{1382.956}{\ndev{-0.844}}
& \massdev{1382.463}{\ndev{-1.337}}
& \massdev{1381.980}{\ndev{-1.820}} \\

$\Sigma_c^*$
& $3/2^+$
& 2518.48
& \massdev{2562.599}{\pdev{+44.119}}
& \massdev{2524.665}{\pdev{+6.185}}
& \massdev{2526.067}{\pdev{+7.587}}
& \massdev{2521.818}{\pdev{+3.338}}
& \massdev{2521.413}{\pdev{+2.933}}
& \massdev{2527.815}{\pdev{+9.335}}
& \massdev{2520.428}{\pdev{+1.948}}
& \massdev{2520.307}{\pdev{+1.827}}
& \massdev{2519.008}{\pdev{+0.528}}
& \massdev{2518.443}{\ndev{-0.037}}
& \massdev{2516.971}{\ndev{-1.509}} \\

\bottomrule
\end{tblr}%
}
\end{table*}

\begin{figure}[t]
\centering
\includegraphics[width=\textwidth]{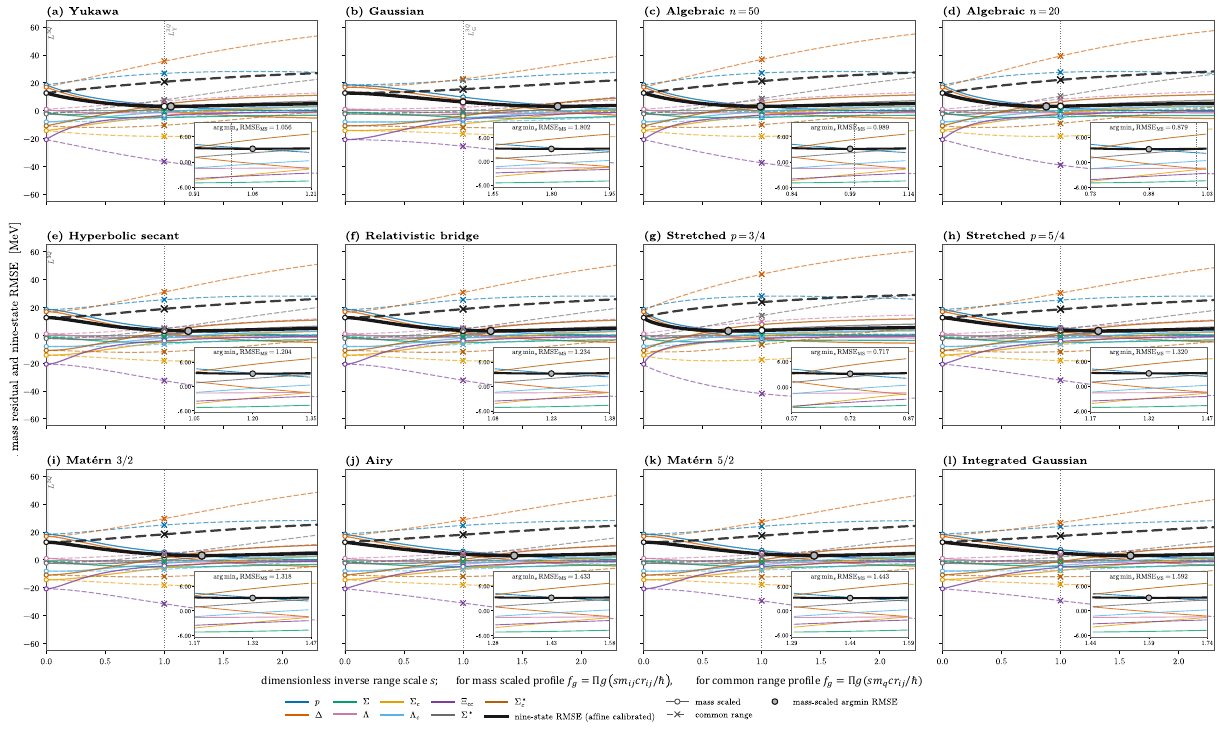}
\caption{Dependence of the nine-state baryon fit on the inverse range scale $s$ for the twelve one-pair profiles. Solid lines use the mass-scaled profile $f_g=\prod_{i<j}g(s\,m_{ij}c\,r_{ij}/\hbar)$ and dashed lines the common-range profile $f_g=\prod_{i<j}g(s\,m_q c\,r_{ij}/\hbar)$; the couplings $A$, $B$, and $C$ are refitted at every $s$. Colored lines show the state-by-state mass residuals and thick black lines the nine-state RMSE. At $s=0$ both forms reduce to the profile-independent interaction $L^{3Q}$, and the dotted vertical line marks the matched scale $s=1$ in Table~\ref{tab:alternative_spatial_profiles}. The common-range points of Table~\ref{tab:fixed_range_spatial_profiles}, with $r_0=1$~fm, correspond to $s=\hbar/(m_q c\,r_0)\simeq0.60$. Gray circles mark the RMSE minimum $s_*$ of the mass-scaled profile, enlarged in each inset.}
\label{fig:profile_scan}
\end{figure}

\paragraph{Sensitivity to the spatial-profile shape:}
We then test the more general profile dependence. In addition to the Yukawa and Gaussian forms discussed above, we consider algebraic, hyperbolic-secant, relativistic-bridge, stretched-exponential, Mat\'ern, Airy, and integrated-Gaussian profiles, spanning substantially different core and asymptotic behaviors. 
The couplings $A$, $B$, and $C$ were redetermined independently for each profile. 
We write the three-quark spatial factor as
\begin{equation}
f_g(r_{12},r_{23},r_{31})
=
\prod_{i<j}g(x_{ij}),
\qquad
x_{ij}
=
\frac{m_{ij}cr_{ij}}{\hbar},
\label{eq:alternative_profile_product}
\end{equation}
and match the effective range of every one-pair factor to that of $g_{\mathrm Y}(x)=e^{-x}$ through
\begin{equation}
\left\langle x^2\right\rangle_g \equiv {\displaystyle\int_0^\infty x^4g(x)\,dx}\bigg/ {\displaystyle\int_0^\infty x^2g(x)\,dx} =12=\left\langle x^2\right\rangle_{\mathrm Y}.
\label{eq:alternative_profile_second_moment}
\end{equation}
For the stretched-exponential and algebraic families, the exact matching factors are
\begin{equation}
a_p = \left[ \frac{\Gamma(5/p)}{12\,\Gamma(3/p)} \right]^{p/2},
\qquad
a_n = \frac{1}{\sqrt{(n-4)(n-5)}},  \qquad n>5.
\label{eq:alternative_profile_family_scales}
\end{equation}
The remaining scales used below are
\begin{align}
a_{\mathrm S} =\pi\sqrt{\frac{5}{48}},
\quad a_{\mathrm R} = \frac{1}{2} \sqrt{\frac{K_3(1)}{K_2(1)}},
\quad a_{\mathrm A} = 3^{2/3} \sqrt{ \frac{\Gamma(2/3)} {6\,\Gamma(1/3)} }
\label{eq:alternative_profile_exact_scales}
\end{align}
where $K_\nu$ is the modified Bessel function of the second kind and $\operatorname{Ai}$ is the Airy function.
The resulting mass-scaled profile comparison is summarized in Table~\ref{tab:alternative_spatial_profiles}.
Table~\ref{tab:alternative_spatial_profiles}(a) also lists the common second moment and the corresponding fourth moment of each matched one-pair factor. Since all profiles are constrained to the same value $\langle x^2\rangle_g=12$, the remaining shape dependence is reflected partly in $\langle x^4\rangle_g$: the better fits are generally obtained for profiles with larger fourth moments than the Gaussian-like forms, although the trend is not strictly monotonic.
As a direct control of whether a finite spatial range alone is sufficient, we repeat the same nine-state fits using the identical dimensionless functions $g$, but replace the mass-weighted coordinate $x_{ij}$ by $y_{ij}\equiv{r_{ij}}/{r_0}$ with $r_0=1~\mathrm{fm}$, where the same flavor-independent scale $r_0$ is assigned to every pair. The dimensionless profile shapes retain the matched condition $\langle y^2\rangle_g=12$, and the couplings $A$, $B$, and $C$ are again refitted independently for each profile. The corresponding spectra are shown in Table~\ref{tab:fixed_range_spatial_profiles}. 
These common-range profiles give $\sigma_B=22.39$--31.74~MeV, so none improves upon the spatially contact result $\sigma_B=18.87~\mathrm{MeV}$, whereas the corresponding mass-scaled profiles give $\sigma_B=3.60$--6.32~MeV. The state-by-state pattern is equally characteristic: the proton and $\Delta$ remain systematically overestimated, while the correction to $\Xi_{cc}$ becomes excessively attractive. The control calculation therefore shows that finite-range localization alone does not supply the required flavor selectivity.
Fig.~\ref{fig:profile_scan} extends the matched comparison to a continuous range scale $s$. For every shape, the mass-scaled RMSE decreases from its $s=0$ value to a shallow minimum at $s_*=0.72$--1.80. The common-range RMSE instead increases from the same $s=0$ value for every shape and remains above it for all $s>0$. The state-by-state residuals show where the common-range profiles fall short: as $s$ increases with a common range, the proton and $\Delta$ move upward and $\Xi_{cc}$ downward, whereas the nine mass-scaled residuals converge near $s_*$. The flavor selectivity therefore follows from the mass dependence of the range at every scale tested.
In the present valence $qqq$ representation, what matters is the mass scaling of the range, not the analytic form of the profile.

\begin{table*}[!t]
\centering
\caption{\justifying
Ground-state baryons with flavor-independent common-range spatial profiles.
The one-pair functions and column order are identical to those in panel~(b) of Table~\ref{tab:alternative_spatial_profiles}, but the mass-weighted variables $x_{ij}=m_{ij}cr_{ij}/\hbar$ are replaced by $y_{ij}=r_{ij}/r_0$, with the common scale fixed to $r_0=1~\mathrm{fm}$ for every flavor pair.
All dimensionless one-pair factors satisfy $\langle y^2\rangle_g=12$, and the complete three-quark profile is $\prod_{i<j}g(y_{ij})$.}
\label{tab:fixed_range_spatial_profiles}

\renewcommand{\massdev}[2]{%
  \shortstack[c]{%
    #1\\[-0.5pt]
    {\tiny #2}%
  }%
}
\renewcommand{\pdev}[1]{\textcolor{devred}{\ensuremath{(#1)}}}
\renewcommand{\ndev}[1]{\textcolor{devgreen}{\ensuremath{(#1)}}}
\scriptsize
\resizebox{\textwidth}{!}{%
\begin{tblr}{
colspec={
Q[l,wd=0.5cm]
Q[c,wd=0.5cm]
Q[c,wd=1.1cm]
Q[c,wd=1.3cm]
Q[c,wd=1.3cm]
Q[c,wd=1.3cm]
Q[c,wd=1.3cm]
Q[c,wd=1.3cm]
Q[c,wd=1.3cm]
Q[c,wd=1.3cm]
Q[c,wd=1.3cm]
Q[c,wd=1.3cm]
Q[c,wd=1.3cm]
Q[c,wd=1.3cm]
},
cells={valign=m},
row{1}={bg=headgray,halign=c},
cell{1}{1}={}{halign=l},
hline{2}={1-14}{0.4pt},
rowsep=1.25pt,
}
\toprule

State
& $\mathcal J^P$
& \shortstack[c]{$M_{\rm exp}$\\(MeV)}
& \shortstack[c]{2BD\\$\sigma_B=65.04$}
& \shortstack[c]{Alg. $n=50$\\$\sigma_B=27.75$}
& \shortstack[c]{Alg. $n=20$\\$\sigma_B=28.88$}
& \shortstack[c]{$\operatorname{sech}$\\$\sigma_B=23.95$}
& \shortstack[c]{Bridge\\$\sigma_B=23.92$}
& \shortstack[c]{Str. $p=3/4$\\$\sigma_B=31.74$}
& \shortstack[c]{Str. $p=5/4$\\$\sigma_B=24.32$}
& \shortstack[c]{Mat\'ern $3/2$\\$\sigma_B=23.59$}
& \shortstack[c]{Airy\\$\sigma_B=23.74$}
& \shortstack[c]{Mat\'ern $5/2$\\$\sigma_B=22.39$}
& \shortstack[c]{$\operatorname{erfc}$\\$\sigma_B=22.83$} \\

$p$
& $1/2^+$
& 938.272
& \massdev{1038.746}{\pdev{+100.474}}
& \massdev{976.060}{\pdev{+37.788}}
& \massdev{977.057}{\pdev{+38.785}}
& \massdev{972.236}{\pdev{+33.964}}
& \massdev{972.197}{\pdev{+33.925}}
& \massdev{979.159}{\pdev{+40.887}}
& \massdev{972.591}{\pdev{+34.319}}
& \massdev{971.799}{\pdev{+33.527}}
& \massdev{971.927}{\pdev{+33.655}}
& \massdev{970.377}{\pdev{+32.105}}
& \massdev{970.858}{\pdev{+32.586}} \\

$\Delta$
& $3/2^+$
& 1232
& \massdev{1338.864}{\pdev{+106.864}}
& \massdev{1276.234}{\pdev{+44.234}}
& \massdev{1278.952}{\pdev{+46.952}}
& \massdev{1267.856}{\pdev{+35.856}}
& \massdev{1267.769}{\pdev{+35.769}}
& \massdev{1285.957}{\pdev{+53.957}}
& \massdev{1268.479}{\pdev{+36.479}}
& \massdev{1266.994}{\pdev{+34.994}}
& \massdev{1267.184}{\pdev{+35.184}}
& \massdev{1264.443}{\pdev{+32.443}}
& \massdev{1265.218}{\pdev{+33.218}} \\

$\Sigma$
& $1/2^+$
& 1192.642
& \massdev{1259.291}{\pdev{+66.649}}
& \massdev{1191.229}{\ndev{-1.413}}
& \massdev{1191.350}{\ndev{-1.292}}
& \massdev{1190.951}{\ndev{-1.691}}
& \massdev{1190.948}{\ndev{-1.694}}
& \massdev{1191.722}{\ndev{-0.920}}
& \massdev{1190.959}{\ndev{-1.683}}
& \massdev{1190.928}{\ndev{-1.714}}
& \massdev{1190.924}{\ndev{-1.718}}
& \massdev{1190.880}{\ndev{-1.762}}
& \massdev{1190.882}{\ndev{-1.760}} \\

$\Lambda$
& $1/2^+$
& 1115.683
& \massdev{1171.919}{\pdev{+56.236}}
& \massdev{1123.070}{\pdev{+7.387}}
& \massdev{1123.984}{\pdev{+8.301}}
& \massdev{1120.054}{\pdev{+4.371}}
& \massdev{1120.053}{\pdev{+4.370}}
& \massdev{1126.526}{\pdev{+10.843}}
& \massdev{1120.436}{\pdev{+4.753}}
& \massdev{1119.866}{\pdev{+4.183}}
& \massdev{1120.067}{\pdev{+4.384}}
& \massdev{1119.069}{\pdev{+3.386}}
& \massdev{1119.470}{\pdev{+3.787}} \\

$\Sigma_c$
& $1/2^+$
& 2452.650
& \massdev{2491.972}{\pdev{+39.322}}
& \massdev{2425.466}{\ndev{-27.184}}
& \massdev{2425.086}{\ndev{-27.564}}
& \massdev{2427.345}{\ndev{-25.305}}
& \massdev{2427.371}{\ndev{-25.279}}
& \massdev{2424.703}{\ndev{-27.947}}
& \massdev{2427.173}{\ndev{-25.477}}
& \massdev{2427.611}{\ndev{-25.039}}
& \massdev{2427.559}{\ndev{-25.091}}
& \massdev{2428.468}{\ndev{-24.182}}
& \massdev{2428.192}{\ndev{-24.458}} \\

$\Lambda_c$
& $1/2^+$
& 2286.460
& \massdev{2317.914}{\pdev{+31.454}}
& \massdev{2276.392}{\ndev{-10.068}}
& \massdev{2277.005}{\ndev{-9.455}}
& \massdev{2274.763}{\ndev{-11.697}}
& \massdev{2274.772}{\ndev{-11.688}}
& \massdev{2279.037}{\ndev{-7.423}}
& \massdev{2274.977}{\ndev{-11.483}}
& \massdev{2274.728}{\ndev{-11.732}}
& \massdev{2274.854}{\ndev{-11.606}}
& \massdev{2274.499}{\ndev{-11.961}}
& \massdev{2274.670}{\ndev{-11.790}} \\

$\Xi_{cc}$
& $1/2^+$
& 3619.970
& \massdev{3622.232}{\pdev{+2.262}}
& \massdev{3572.013}{\ndev{-47.957}}
& \massdev{3569.759}{\ndev{-50.211}}
& \massdev{3580.103}{\ndev{-39.867}}
& \massdev{3580.112}{\ndev{-39.858}}
& \massdev{3563.936}{\ndev{-56.034}}
& \massdev{3579.044}{\ndev{-40.926}}
& \massdev{3580.667}{\ndev{-39.303}}
& \massdev{3580.110}{\ndev{-39.860}}
& \massdev{3583.038}{\ndev{-36.932}}
& \massdev{3581.876}{\ndev{-38.094}} \\

$\Sigma^*$
& $3/2^+$
& 1383.800
& \massdev{1450.640}{\pdev{+66.840}}
& \massdev{1389.289}{\pdev{+5.489}}
& \massdev{1390.936}{\pdev{+7.136}}
& \massdev{1384.495}{\pdev{+0.695}}
& \massdev{1384.478}{\pdev{+0.678}}
& \massdev{1395.660}{\pdev{+11.860}}
& \massdev{1384.970}{\pdev{+1.170}}
& \massdev{1384.153}{\pdev{+0.353}}
& \massdev{1384.371}{\pdev{+0.571}}
& \massdev{1382.966}{\ndev{-0.834}}
& \massdev{1383.465}{\ndev{-0.335}} \\

$\Sigma_c^*$
& $3/2^+$
& 2518.480
& \massdev{2562.599}{\pdev{+44.119}}
& \massdev{2500.629}{\ndev{-17.851}}
& \massdev{2501.001}{\ndev{-17.479}}
& \massdev{2500.178}{\ndev{-18.302}}
& \massdev{2500.205}{\ndev{-18.275}}
& \massdev{2502.795}{\ndev{-15.685}}
& \massdev{2500.296}{\ndev{-18.184}}
& \massdev{2500.311}{\ndev{-18.169}}
& \massdev{2500.413}{\ndev{-18.067}}
& \massdev{2500.595}{\ndev{-17.885}}
& \massdev{2500.617}{\ndev{-17.863}} \\

\bottomrule
\end{tblr}%
}
\end{table*}

\vspace{10pt}

The present calculation establishes this mechanism in the $u$, $d$, $s$, and $c$ ground-state baryons while retaining the two-body parameters fixed by the meson sector.
Its extension to the $b$ sector, its dependence on the meson-to-baryon fitting direction, and its robustness across alternative two-body Hamiltonians are examined separately in  Section~\ref{sec:udscb_benchmark}.

\subsection{Operational definition and representation dependence of the short-range three-quark interaction}
\label{subsec:three_quark_definition}
We now give an operational definition of the short-range three-quark interaction in the adopted constituent-quark representation, clarify the representation dependence of its separation from embedded pair dynamics, and relate its effective form to possible intermediate space realizations.

\paragraph{Definition in the adopted representation:}

An explicit three-body interaction can be identified only after the degrees of freedom and the one- and two-body parts of the Hamiltonian have been specified.
A two-body interaction $v_{ij}$ is embedded in the three-particle space as $v_{ij}\otimes I_k$: it acts on particles $i$ and $j$, while particle $k$ remains a spectator.
This prescription, referred to as spectator embedding, fixes the pairwise contribution that must be subtracted from the full three-particle Hamiltonian.
The remaining connected part is then defined as the explicit three-body interaction~\cite{Polyzou2010,CoesterPolyzou1982},
\begin{equation}
H_3=H_{\mathrm{1B}}+\sum_{i<j}v_{ij}\otimes I_k+V_{123}^{\mathrm{conn}},
\qquad k\neq i,j.
\label{eq:operational_cluster_decomposition}
\end{equation}
Here $V_{123}^{\mathrm{conn}}$ denotes the connected three-body remainder after the specified one-body term and spectator-embedded pair interactions have been subtracted.
For a local spectator-independent pair interaction, the spatial dependence has the additive form $W_{\mathrm{pair}}=w_{12}(r_{12})+w_{23}(r_{23})+w_{31}(r_{31})$. For example, the color-Cornell potential $V^C_{\rm pair}=V_{12}^C+V_{23}^C+V_{31}^C$ is a spectator-independent pair (two-body) interaction. These interactions satisfy $\partial^2W_{\mathrm{pair}}/(\partial r_{12}\partial r_{23})=0$ at interior points of the allowed triangle domain with $r_{31}$ held fixed.
However, the profiles in Eqs.~\eqref{yukawa} and~\eqref{gaussian} instead satisfy
\begin{align}
\frac{\partial^2f_{\mathrm Y}}{\partial r_{12}\partial r_{23}}
=\frac{c^2m_{12}m_{23}}{\hbar^2}f_{\mathrm Y}\neq0,\qquad\frac{\partial^2f_{\mathrm G}}{\partial r_{12}\partial r_{23}}
=\frac{c^4m_{12}^2m_{23}^2r_{12}r_{23}}{16\hbar^4}f_{\mathrm G}\neq0
\label{eq:profile_nonadditivity_test}
\end{align}

for generic configurations in the interior of the allowed triangle domain.
Thus, $L_{\mathrm Y}^{3Q}$ and $L_{\mathrm G}^{3Q}$ cannot be rewritten as sums of local spectator-independent pair functions in the present representation.
They also vanish in every coordinate-space $2+1$ limit because separating one quark makes the two distances connecting it to the remaining pair large.
Only the added short-range three-quark interaction vanishes in this limit.
The full Hamiltonian still contains the confining pair interactions and therefore does not approach a freely separated colored $qq$ subsystem and a quark.
Apart from the confining potential, the nonadditivity and $2+1$ falloff established above identify $L_{\mathrm Y}^{3Q}$ and $L_{\mathrm G}^{3Q}$ as connected three-body interactions within this representation.\footnote{Here, ``connected'' is used in the cluster-decomposition sense: $V_{123}^{\mathrm{conn}}$ is the part of the three-particle Hamiltonian that remains after the specified one-body term and spectator-embedded pair interactions have been removed and that cannot be represented by any one of the fixed terms $v_{ij}\otimes I_k$.}

\paragraph{Off-shell embedding and scattering equivalence:}
The representation ambiguity is most transparent in the standard scattering-theory setting, where a two-body subsystem has a physical asymptotic channel and its transition operator satisfies
\begin{equation}
t_{ij}(E)=v_{ij}+v_{ij}\frac{1}{E-h_{0,ij}+i0}t_{ij}(E).
\label{eq:operational_two_body_t}
\end{equation}
The matrix element $\langle \mathbf k'\lvert t_{ij}(E)\rvert  \mathbf k\rangle $ is on shell when $E=\hbar^2k^2/m_{ij}=\hbar^2k^{\prime 2}/m_{ij}$, while matrix elements not satisfying these kinematic conditions are off shell.
Thus, on- and off-shell refer to matrix elements of $t_{ij}(E)$ and not to two uniquely separable components of the potential.
In an ordinary three-body scattering problem without confinement, where the interacting pair can form a physical asymptotic subsystem,\footnote{Because a colored $qq$ pair is not a physical asymptotic channel of the present confining Hamiltonian, Eqs.~\eqref{eq:operational_two_body_t} and~\eqref{eq:operational_embedded_energy} do not define an observable on- or off-shell $qq$ amplitude in the present calculation; they are introduced only to illustrate the general representation freedom of embedded pair dynamics.}  the pair transition operator is embedded at the subsystem energy
\begin{equation}
E_{ij}=E_{\mathrm{int}}-\frac{\hbar^2q_k^2}{m_{k,(ij)}},
\qquad
m_{k,(ij)}=\frac{2m_k(m_i+m_j)}{m_i+m_j+m_k},
\label{eq:operational_embedded_energy}
\end{equation}
where the internal energy $E_{\mathrm{int}}$ excludes the constituent rest masses and $q_k$ is the spectator Jacobi momentum.
Integration over $q_k$ and the pair momenta then samples off-shell matrix elements of $t_{ij}$ in the corresponding Faddeev equation~\cite{PolyzouGlockle1990,Faddeev1961}.

Where physical asymptotic channels exist, equivalence requires a scattering-equivalent transformation $\mathcal U$, and the Hamiltonian and every operator used to calculate an observable must be transformed consistently,
\begin{equation}
H_3'=\mathcal U^\dagger H_3\mathcal U,
\qquad
\widehat{J}_X^{\mu\prime}=\mathcal U^\dagger\widehat{J}_X^\mu\mathcal U.
\label{eq:operational_unitary_transformation}
\end{equation}
Re-expanding $H_3'$ in particle clusters generally redistributes strength among the off-shell and nonlocal structure of the pair interactions, the explicit connected three-body term, and the short-distance wave function~\cite{PolyzouGlockle1990,Polyzou2010}.
A concrete constituent-quark-model example was given by Ref.~\cite{SengbuschPolyzou2004}, which constructed a scattering-equivalent model with the same meson spectrum and pion form factor as the original model, but with a pointlike constituent-quark impulse current in place of the original nontrivial current and a correspondingly transformed confining interaction and wave function. 
This example also illustrates the cost of simplifying one part of the dynamics: the transformed confining interaction acquires additional short-range nonlocal terms~\cite{SengbuschPolyzou2004}.
Scattering-equivalent transformations do not generally preserve the coordinate-space locality of a pair potential.
The transformed interaction can involve a kernel $v'_{ij}(\mathbf r,\mathbf r')$ with independent dependence on $\mathbf r$ and $\mathbf r'$, beyond the local potential ansatz adopted in this section.
Reducing the explicit connected three-body contribution can therefore require additional nonlocal pair structures, together with consistently transformed currents and induced many-body operators~\cite{Polyzou2010}.
This redistribution does not establish that the same few-body dynamics can be reproduced within the original local pair ansatz.
Masses and form factors therefore test a specified Hamiltonian--current representation and truncation rather than a unique off-shell pair potential or three-body potential~\cite{Polyzou2010,FurnstahlHammerTirfessa2001}.
The Polyzou--Gl\"ockle relation establishes this redistribution freedom and allows three- and higher-body interactions to be weakened in suitable representations, but it does not in general guarantee the existence of a representation with $V_{123}^{\mathrm{conn}}=0$ while scattering equivalence and the required cluster properties are both preserved~\cite{Polyzou2010}. 
This statement concerns equivalent representations of a specified effective few-body dynamics and should not be interpreted as implying that the present static valence-quark Hamiltonian is scattering equivalent to full QCD. 
An exact reduction of QCD to a restricted valence space would in general generate energy-dependent and nonlocal interactions, explicit channel dependence, induced many-body operators, and consistently transformed currents beyond the static local potential model adopted here.

\paragraph{Nuclear analogue, nonlocal pair interactions:}

A nuclear example of this redistribution is provided by the inside-nonlocal outside-Yukawa (INOY) nucleon--nucleon interactions of Refs.~\cite{Doleschall2000,Doleschall2003}.
These interactions describe the two-nucleon data with high precision and reproduce the ${}^3$H and ${}^3$He binding energies without an explicit three-nucleon force.
Their short-range nonlocal structure, however, is an off-shell freedom that two-nucleon observables do not fix, and its construction was guided by the requirement of the correct triton binding energy~\cite{Doleschall2000}.
The three-nucleon information is therefore not absent; it is carried by the off-shell structure of the pair interaction instead of an explicit connected term.
Explicit three-nucleon-force models are in a symmetric position, since their couplings are also fixed by three-nucleon data~\cite{Hammer2013,Epelbaum:2005pn}.
The representation-independent statement is that two-nucleon data alone do not determine three-nucleon observables, whereas whether the missing information appears as an explicit three-body operator depends on the chosen representation of the pair dynamics~\cite{PolyzouGlockle1990,Polyzou2010,Hammer2013}.
The INOY interactions are also not free of tension at higher particle number: they overbind ${}^4$He and give the lowest $n$--${}^3$H total cross section in the resonance region among the realistic potentials compared in Ref.~\cite{DeltuvaFonseca2007}, and nonlocality moves the nuclear-matter saturation point in an unfavorable direction~\cite{BaldoMaieron2005}.
The quark-model situation is analogous: meson masses constrain the color-singlet $q\bar q$ interaction, but they do not determine the color-antitriplet $qq$ interaction in baryons or its off-shell structure.
Baryon information must therefore enter either through an explicit connected three-quark term, as in the present work, or through a baryon-adjusted nonlocal pair interaction, and the statement that a three-quark interaction is required refers to the fixed pair representation defined in Eq.~\eqref{eq:operational_cluster_decomposition}.

\paragraph{A possible \(qqqg\) realization and effective matching:}

The discussion above shows that the present three-quark interaction is defined only after the valence space and the pair Hamiltonian have been fixed. The numerical values of $A$, $B$, and $C$ therefore depend on the representation, model space, spatial profile, and fitted data. These couplings may also contain unresolved gluonic, instanton, relativistic, and coupled-channel contributions. If any of these effects is included explicitly, the one-body, pair, and connected three-quark terms must be refitted together to avoid double counting.

One possible realization of the intermediate $q^\ast$ sector is a $qqqg$ configuration.
The intermediate excitation need not be interpreted as a new elementary quark level.
A gluonic excitation retained explicitly in an enlarged model space, beyond the gluonic dressing already absorbed into the constituent masses and pair interactions, may instead occur as part of a color-singlet $qqqg$ hybrid configuration: $\mathcal H_{qqq^\ast}\supset \mathcal H_{qqqg}$.
Since $\mathbf 3_C\otimes\mathbf 8_C = \mathbf 3_C\oplus\overline{\mathbf 6}_C\oplus\mathbf{15}_C$, the quark and this color-octet gluonic excitation contain a color-triplet recoupling component.
In a Casimir-scaled short-distance color interaction, this color-triplet component is the most attractive $qg$ channel, $\langle F_k^cF_g^c\rangle_{\mathbf 3_C}=-3/2$, where $F_k^c$ and $F_g^c$ denote the generators of the quark $k$ and the gluon in the adjoint representation.
One such color-singlet component of the $qqqg$ sector may be written as
\begin{align}
\ket{ [ (q_iq_j)_{\overline{\mathbf 3}_C} (q_k g_{\mathbf 8_C})_{\mathbf 3_C} ]_{\mathbf 1_C}}\ \longleftrightarrow\ \ket{ [ (q_iq_jq_k)_{\mathbf 8_C} g_{\mathbf 8_C} ]_{\mathbf 1_C}}.
\label{eq:qstar_qqqg_recoupling}
\end{align}
Coupling the spin-1 gluon to the spin-$1/2$ quark gives color-triplet $qg$ components with $j_{qg}^P=1/2^+$ and $3/2^+$.
This is consistent with lattice-QCD hybrid-baryon spectra containing positive-parity $\mathcal J=3/2$ states associated with a $j_g^P=1^+$ gluonic excitation~\cite{DudekEdwards2012HybridBaryons}.
The explicit $1/2\leftrightarrow3/2$ transition matrices and a minimal local $S$-wave reduction of the $j_{qg}^P=3/2^+$ component are given in \ref{app:qg_spin32}.
The displayed state represents one recoupling component of the full $qqqg$ sector and does not reduce its particle content.
The complete baryon construction includes the corresponding cyclic sum over the line $k$.
Hybrid baryons contain an explicit gluonic excitation and may therefore be viewed as having a $qqqg$ component, providing an example of the intermediate configuration considered here~\cite{DudekEdwards2012HybridBaryons}.

To express the relation to the effective interaction more precisely, let $\mathsf P$ project onto the retained $qqq$ valence space and let $\mathsf Q_g$ project onto the resolved $qqqg$ sector. 
Applying the Feshbach reduction in Eq.~\eqref{eq:unquenched_feshbach_equation} with $\mathsf Q\to\mathsf Q_g$ gives the induced contribution $\Delta H_{\mathsf P}^{(qqqg)}(E)\equiv\Sigma_{\mathsf Q_g}(E)$. 
In the valence three-quark space, its cluster decomposition takes the form
\begin{equation}
\Delta H_{\mathsf P}^{(qqqg)}(E)
=
\sum_i\Delta h_i(E)
+\sum_{i<j}\Delta v_{ij}(E)
+V_{123}^{\mathrm{conn}}(E).
\label{eq:qqqg_elimination_cluster_expansion}
\end{equation}
Thus, the same gluonic dynamics is represented by explicit $qqq\leftrightarrow qqqg$ transitions and propagation when the $qqqg$ sector is retained, but by constituent renormalizations, modified pair interactions, induced many-body operators, and a connected three-quark remainder after that sector is eliminated.

If the intermediate $\ket{qqq^\ast}$ state in Eq.~\eqref{eq:TGE_static_main} is interpreted as the recoupled $qqqg$ configuration in Eq.~\eqref{eq:qstar_qqqg_recoupling}, then Eq.~\eqref{eq:TGE_static_main} contains approximations at both the vertex and propagator levels.
At the vertex level, the full pair-labeled transition blocks $T_{\mathsf Q_g\mathsf P}^{(ij)}$ and $T_{\mathsf P\mathsf Q_g}^{(ij)}$ are projected onto transition operators carrying the same central-color and color-spin tensors as $V_{ij}^{C}$ and $V_{ij}^{CS}$ in Eqs.~\eqref{2body_1} and~\eqref{2body_2}. These transition operators are not the diagonal pair potentials themselves, because they connect different model-space sectors. Independent transition form factors, momentum and energy dependence, gluonic polarizations, and the separate spin couplings of the $qg$ subsystem are not treated explicitly.

At the propagator level, for an intermediate sector lying above the valence energy, the full resolvent $(E-H_{\mathsf Q_g\mathsf Q_g})^{-1}$ is replaced by its leading static form $-1/M_k$, with $M_k>0$. In the definitions of the effective operators, $M_k^{-1}$ denotes the magnitude of this excitation denominator, while its overall spectral sign is absorbed into the fitted couplings. Within the present truncation, only the components parallel to the retained operator basis are represented by $A$, $B$, $C$, and the adopted spatial profile; structures outside this basis are omitted.
Phenomenologically, only the composite operator obtained after the intermediate sector is eliminated is matched; the individual transition vertices and the hybrid propagator are not separately identified.
Accordingly, after projection onto the quark-like color-triplet component of the intermediate $qqqg$ sector and assignment of the induced one- and two-body terms to the baseline Hamiltonian, we make the schematic matching

\begin{align}
&\left[
T_{\mathsf P\mathsf Q_g}^{(jk)}
\frac{1}{E-H_{\mathsf Q_g\mathsf Q_g}}
T_{\mathsf Q_g\mathsf P}^{(ik)}
+
T_{\mathsf P\mathsf Q_g}^{(ik)}
\frac{1}{E-H_{\mathsf Q_g\mathsf Q_g}}
T_{\mathsf Q_g\mathsf P}^{(jk)}
\right]_{\mathrm{conn}} 
\xRightarrow[\;\text{onto the $qqq$ operator basis}]{\;\text{static projection }\;}
A\,L_{ij;k}^{C-C}
+ B\,L_{ij;k}^{S-S}
+ C\,L_{ij;k}^{C-S},
\label{eq:qstar_effective_operator_matching}
\end{align}
Here $L_{ij;k}^{C-C}$, $L_{ij;k}^{S-S}$, and $L_{ij;k}^{C-S}$ denote the terms associated with intermediate line $k$ in Eq.~\eqref{eq:l12}. 
Their unresolved coordinate dependence is represented by the adopted three-quark spatial profile.
The static scale $M_k^{-1}$ entering the definitions of $L_{ij;k}^{C-C}$, $L_{ij;k}^{S-S}$, and $L_{ij;k}^{C-S}$ represents the reduced excitation denominator associated with line $k$; it is not identified with the inverse physical mass of an isolated $(q_k g)_{\mathbf 3_C}$ constituent.
This projection retains the central-color and color-spin operator content generated by two insertions, $V^{C}\!\times V^{C}$, $V^{C}\!\times V^{CS}+V^{CS}\!\times V^{C}$, and $V^{CS}\!\times V^{CS}$ after their connected three-quark parts are isolated.
These combinations motivate $L^{C-C}$, $L^{C-S}$, and $L^{S-S}$, respectively.
It does not assume a common factorized microscopic vertex.
The present fit therefore treats $A$, $B$, and $C$ as independent effective couplings.
Only the components of these effects that project onto the retained static local operator basis are represented by the fitted couplings and spatial profile; energy-dependent, nonlocal, noncentral, and additional operator structures outside this basis are omitted.

For the nine calibrated ground-state baryons, the meson-constrained pair Hamiltonian gives $\sigma_B=65.04~\mathrm{MeV}$, the spatially unresolved (contact and common orbital) color-spin interaction $L^{3Q}$ gives 18.87~MeV, and the Yukawa- and Gaussian-profile interactions give 3.13~MeV and 6.28~MeV, respectively.
Because the same nine masses determine $A$, $B$, and $C$, these values demonstrate a large in-sample improvement and an economical organization of the residuals rather than a unique microscopic extraction.
The meson--baryon compatibility, cross-model, and fixed-parameter omitted-state tests in  Section~\ref{sec:udscb_benchmark} test the result more broadly. The excitation analysis in  Section~\ref{sec:excitation_spectra} shows where the static compact description is no longer sufficient.

The main result is that a small set of connected operators describes the short-distance contribution missing from the tested two-body Hamiltonians in the valence $qqq$ space.

\section{Meson--baryon compatibility benchmarks for the $u,d,s,c,b$ sectors across alternative quark models}
\label{sec:udscb_benchmark}
We now test the compatibility of the meson and baryon sectors within the adopted valence $q\bar q$ and $qqq$ representation, using the definition of the explicit three-quark interaction given above.
For this purpose, we include selected ground-state hadrons containing $u,d,s,c$, and $b$ quarks and repeat the calculation for different fitting prescriptions, kinetic-energy operators, and two-body interactions.
We then apply the same compatibility test to several widely used alternative quark models.

The flavor dependency appears through the systematic change of the mass deviation $\varepsilon_H=M_H^{\rm th}-M_H^{\exp}$ with the quark content of the hadron.
A meson-fitted two-body Hamiltonian typically gives positive deviations for light baryons, while the deviations decrease toward heavy-quark baryons.
When the pair parameters are refitted, the error in one sector may decrease while the error in the other sector increases.
In the $(\sigma_M,\sigma_B)$ plane, this behavior appears as a meson--baryon trade-off trajectory.
We use this trajectory to examine whether both sectors can be described with small errors by the two-body Hamiltonian, or whether the explicit three-quark interaction is still necessary.

For this purpose, we use the weighted-sum objective (loss) function
\begin{equation}
\label{eq:reciprocal_objective}
{\cal L}_{\lambda}
=
\lambda\,\frac{1}{N_M}\sum_{a=1}^{N_M}
\left(M_a^{\rm th}-M_a^{\rm exp}\right)^2
+
(1-\lambda)\,\frac{1}{N_B}\sum_{b=1}^{N_B}
\left(M_b^{\rm th}-M_b^{\rm exp}\right)^2 .
\end{equation}
Here $\lambda$ is the meson-sector weight.
Varying $\lambda$ from 0 to 1 connects the baryon-only and meson-only optimizations as we vary the relative importance of the meson and baryon sectors, thereby tracing what we term the ``meson--baryon trade-off trajectory''.

We work in the isospin-symmetric limit $m_u=m_d\equiv m_q$, where $q$ denotes a light $u$ or $d$ constituent quark.
Only established ground states are used in order to reduce sensitivity to coupled-channel and orbital-excitation effects.
The meson and baryon data sets are
\begin{align}
{\cal D}_M
={}&
\{\pi,\rho,K,K^{*},\phi,D,D^{*},D_s,D_s^{*},
\eta_c,J/\psi,B,B^{*},B_s,B_s^{*},B_c,\eta_b,\Upsilon\},
\nonumber\\
{\cal D}_B
={}&
\{N,\Lambda,\Sigma,\Delta,\Xi,\Sigma^{*},\Xi^{*},\Omega,
\Lambda_c,\Sigma_c,\Sigma_c^{*},\Omega_c,\Omega_c^{*},\Xi_{cc},
\Lambda_b,\Sigma_b,\Sigma_b^{*},\Omega_b\}.
\label{eq:reciprocal_datasets}
\end{align}
For the present benchmark, $N_M=N_B=18$.
The choice $\lambda=1/2$, hereafter called the equal-sector point, therefore assigns the same total weight to the meson and baryon sectors and the same average weight per state.

The selected baryons span the light, strange, charm, double-charm, and bottom sectors while restricting the sample to ground-state baryons whose spin-flavor configurations are treated consistently across all comparison models.
States containing three distinct constituent flavors, such as $\Xi_c$, $\Xi_c'$, $\Xi_b$, and $\Xi_b'$, are not included in this benchmark set because their $S=1/2$ spin diagonalization introduces an additional basis convention that is not common to all comparison models.

The limit $\lambda=1$ corresponds to a meson-only optimization of the nonlinear two-body parameters, whereas $\lambda=0$ corresponds to a baryon-only optimization.
Intermediate values give joint meson--baryon fits.
All allowed nonlinear two-body parameters, including the constituent-quark masses, are reoptimized at every value of $\lambda$.
Because the three-quark interaction does not contribute to mesons, its couplings $A$, $B$, and $C$ are determined from the baryon masses by linear least squares at every trial two-body parameter set, including the $\lambda=1$ endpoint.
The result at $\lambda=1$ then shows how much of the baryon residual left by the meson-calibrated two-body Hamiltonian is corrected by the three-quark interaction.

For the compatibility scan, all baryon calculations in this section use the same $S$-wave basis.
The quoted values should therefore be regarded as a uniform $S$-wave comparison.
The full $SS+PP+DD+FF$ GEM calculation in  Sections~\ref{sec:baryon_GEM} and~\ref{sec:short_range} is used for the principal quantitative result.

\subsection{Meson--baryon compatibility and trade-off trajectories}
\label{subsec:SAJ_udscb}
For compact notation in this compatibility benchmark, we denote the Hamiltonian defined in Eqs.~\eqref{hamiltonian}--\eqref{2body_2} as the $\mathrm{S}^{2}\mathrm{AJ}$ Hamiltonian, using the given-name initials of Su Houng Lee, Sungsik Noh, Aaron Park, and Jongheon Baek.
The benchmark contains 18 ground-state mesons and 18 ground-state baryons with $u,d,s,c$, and $b$ quarks.
At every value of $\lambda$, all nonlinear two-body parameters, including the constituent-quark masses, are reoptimized.\footnote{The same numerical basis is used consistently along each trade-off trajectory, with the final baryon masses evaluated using the converged {20~\ensuremath{\times}~20} Gaussian meshes.} 

Ground-state masses alone do not determine all components of the effective Hamiltonian independently.
In particular, the additive constant, constituent masses, confinement strength, Coulomb coefficient, and hyperfine regulator can compensate one another.
If all parameters are varied without restriction, the fit can move to a different branch by making the nominally short-range light-quark hyperfine interaction as broad as the hadron size.
Such a solution can reproduce the selected masses numerically, but it is not the short-range OGE-inspired Hamiltonian considered here.

We therefore use the same short-range benchmark domain for all $\mathrm{S}^{2}\mathrm{AJ}$ trade-off trajectories, including the original two-body, running-coupling, relativized, and three-quark-extended calculations.
This allows us to test whether each Hamiltonian describes both sectors on the same short-range branch.
We impose
\begin{equation}
\begin{aligned}
0.060&\leq\kappa\leq0.140~{\rm GeV\,fm},&
0.76&\leq\sigma\leq1.07~{\rm GeV/fm},&
0.45&\leq D\leq1.40~{\rm GeV},\\
0.50&\leq\alpha\leq0.90~{\rm GeV},&
0&\leq\beta\leq0.50,&
0.10&\leq Z_{\rm hf}\leq0.60.
\end{aligned}
\label{eq:SAJ_physical_bounds}
\end{equation}
together with
\begin{equation}
0.28\leq m_q\leq0.50~{\rm GeV},\qquad
0.48\leq m_s\leq0.75~{\rm GeV},\qquad
1.65\leq m_c\leq2.10~{\rm GeV},\qquad
5.00\leq m_b\leq5.55~{\rm GeV}.
\label{eq:SAJ_mass_bounds}
\end{equation}
Here the hyperfine scales are parametrized as $\mu_{ij}=\alpha+\beta m_i m_j/(m_i+m_j)$ and $\mu'_{ij}=Z_{\rm hf}\mu_{ij}$. 
The factor $Z_{\rm hf}$ independently controls the effective color-spin strength relative to the central Coulomb coupling, allowing $\alpha_s^{(CS)}/\alpha_s^{(C)}=3(\hbar c)Z_{\rm hf}/\kappa<1$ when part of the hyperfine splitting is generated by the two-body instanton-induced interaction rather than by OGE alone.\footnote{As discussed in  Section~\ref{subsec:instanton_3body}, OGE--III analyses typically attribute about 30\%--40\% of the $N-\Delta$ splitting to the instanton-induced contribution~\cite{Takeuchi1998,OkaTakeuchi1989}.} 
The lower bound on $\alpha$, together with $\beta\geq0$, directly constrains the range of the light-quark hyperfine regulator.
For a light $qq$ pair, $\mu_{qq} = \alpha+{\beta m_q}/{2}$ and $r_{qq}^{CS} ={\hbar c}/{\mu_{qq}}$, so that the imposed domain gives $r_{qq}^{CS}\lesssim0.395~{\rm fm}$, whose range is comparable to the size of a light hadron.
Compared with the original independent parametrization of $\alpha$, $\beta$, $\gamma $, and $\delta$, the condition $Z_{\rm hf}=\mu'_{ij}/\mu_{ij}$ reduces the regulator freedom from four parameters to three, with $\gamma=Z_{\rm hf}\alpha$ and $\delta=Z_{\rm hf}\beta$. 
Thus, $\beta$ controls the common mass dependence of two scales, while $Z_{\rm hf}$ fixes their relative normalization and prevents the range from being adjusted to excessively large or small values.
The interval for the string tension corresponds to $0.150\leq(\hbar c)\sigma\leq0.211~{\rm GeV}^2$, which covers the range commonly used in Cornell-type phenomenology around the canonical QCD string-tension scale.
The Coulomb interval corresponds to an effective central coupling $0.23\lesssim\alpha_s^{(C)} ={3\kappa}/{4\hbar c} \lesssim0.53$, while the relatively broad interval assigned to $D$ reflects the fact that the additive constant is convention and model dependent.
The constituent-mass intervals cover the usual phenomenological ranges of quark models.

For the two-body baseline, we consider the following three cases.
The first is the original $\mathrm{S}^{2}\mathrm{AJ}$ two-body Hamiltonian of Eqs.~\eqref{hamiltonian}--\eqref{2body_2}.
The second replaces the constant Coulomb coefficient by the reduced-mass running prescription~\cite{Vijande2005,Ebert2005,YangPingOrtegaSegovia2020}
\begin{equation}
\kappa_{ij}=\frac{4}{3}\alpha_s(Q_{ij})\hbar c,\qquad
\alpha_s(Q)=
\frac{\alpha_{\rm IR}}
{1+\dfrac{\beta_0\alpha_{\rm IR}}{4\pi}
\ln\left(1+\dfrac{Q^2}{Q_0^2}\right)},\qquad
Q_{ij}=m_{ij},
\label{eq:SAJ_running_benchmark}
\end{equation}
where $\beta_0=9$, $0.25\leq\alpha_{\rm IR}\leq0.80$, and $0.30\leq Q_0\leq3.00~{\rm GeV}$.\footnote{This infrared-finite form satisfies $\alpha_s(Q)\to\alpha_{\rm IR}$ as $Q\to0$, while for $Q\gg Q_0$ it recovers the usual one-loop behavior $\alpha_s(Q)\propto1/\ln(Q^2/\Lambda^2)$.} 
The third replaces only the kinetic-energy operator by the spinless-Salpeter form
\begin{equation}
K_M^{\rm rel}=\sqrt{{\bf p}^{\,2}+m_i^2}+\sqrt{{\bf p}^{\,2}+m_j^2},\qquad
K_B^{\rm rel}=\sum_{i=1}^{3}\sqrt{{\bf p}_i^{\,2}+m_i^2},
\label{eq:SAJ_relativized_benchmark}
\end{equation}
while retaining the same central and color-spin interactions.
The prescription given in Eq.~\eqref{eq:SAJ_relativized_benchmark} tests only the effect of relativistic kinematics.
It is not a fully relativized calculation because the vertices and spin-dependent operators are not momentum dressed.
To test the profile dependence of the three-quark interaction, we compare the contact, Yukawa, and Gaussian profiles.

\begin{figure*}[!t]
\centering
\includegraphics[width=\linewidth]{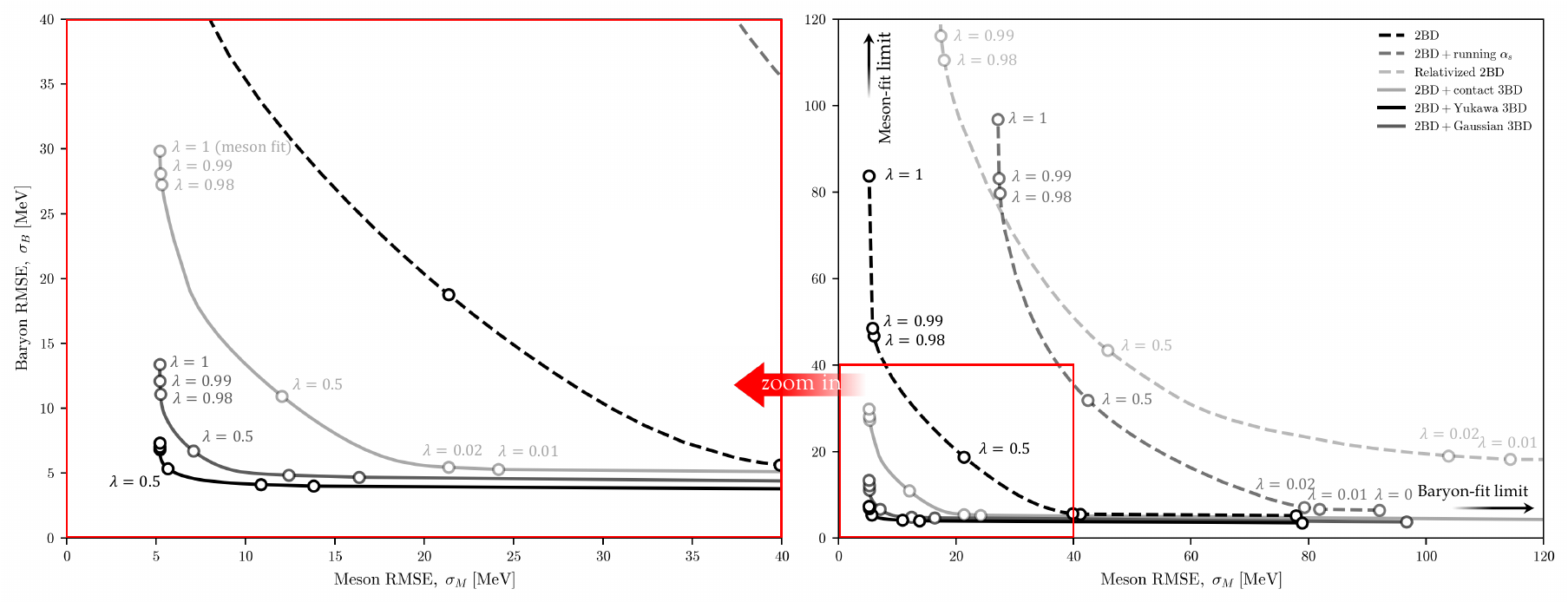}
\caption{\justifying
Meson--baryon RMSE trade-off trajectories in the $\mathrm{S}^{2}\mathrm{AJ}$ model for the $u,d,s,c,b$ benchmark under the common short-range bounds of Eqs.~\eqref{eq:SAJ_physical_bounds} and~\eqref{eq:SAJ_mass_bounds}.
The horizontal and vertical coordinates are the meson and baryon RMSEs, $\sigma_M$ and $\sigma_B$, respectively.
Each trajectory is obtained by varying the meson-sector weight $\lambda$ from the baryon-only endpoint $\lambda=0$ to the meson-only endpoint $\lambda=1$.
For the three-quark trajectories, the couplings $A$, $B$, and $C$ are profiled against the baryon masses at every point.
Hollow circles indicate $\lambda=0,0.01,0.02,0.50,0.98,0.99$, and 1.
The right panel shows the full trade-off trajectories, while the left panel enlarges the low-error region $0\leq\sigma_M,\sigma_B\leq40~\mathrm{MeV}$, which is populated mainly by the three-quark results.}
\label{fig:SAJ_reciprocal}
\end{figure*}

\begin{table*}[!t]
\centering
\caption{\justifying
Ground-state meson and baryon masses obtained at the equal-sector compatibility point $\lambda=1/2$ for the six $\mathrm{S}^2\mathrm{AJ}$ fitting schemes.
Each entry gives the calculated mass followed by the signed deviation $M-M_{\exp}$ in parentheses.
All values are given in MeV.
The overall RMSE, $\sigma=\sqrt{(\sigma_M^2+\sigma_B^2)/2}$, evaluated over the combined 18-meson and 18-baryon data set, is shown in each column heading.
All parameters are constrained by Eqs. \eqref{eq:SAJ_physical_bounds} and~\eqref{eq:SAJ_mass_bounds}.}
\label{tab:SAJ_lambda_half_spectrum}
\footnotesize
\renewcommand{\pdev}[1]{\textcolor{devred}{\ensuremath{(#1)}}}
\renewcommand{\ndev}[1]{\textcolor{devgreen}{\ensuremath{(#1)}}}

\resizebox{\textwidth}{!}{%
\begin{tblr}{
  colspec={
Q[l,wd=1.00cm]
Q[c,wd=1.00cm]
*{6}{Q[c,wd=2.15cm]}
  },
  row{1} = {bg=headgray, halign=c, valign=m},
  row{2,21} = {bg=headgray, halign=l, valign=m},
  cell{1}{1} = {}{halign=l},
  hline{2} = {1-8}{0.3pt},
  hline{3} = {1-8}{0.3pt},
  hline{21} = {1-8}{0.3pt},
  hline{22} = {1-8}{0.3pt},
  rowsep = 0.65pt,
}
\toprule
State
& \shortstack[c]{$M_{\rm exp}$\\(MeV)}
& \shortstack[c]{2BD\\($\sigma=20.10$)}
& \shortstack[c]{2BD[running $\alpha_s$]\\($\sigma=37.56$)}
& \shortstack[c]{2BD[relativized]\\($\sigma=44.66$)}
& \shortstack[c]{Contact 3BD\\($\sigma=11.50$)}
& \shortstack[c]{Gaussian 3BD\\($\sigma=6.90$)}
& \shortstack[c]{Yukawa 3BD\\($\sigma=5.51$)}
\\

\SetCell[c=8]{l}{\textit{Mesons}} & & & & & & & \\

$\pi$
& 139.57
& \massdevinline{108.41}{\ndev{-31.16}}
& \massdevinline{82.75}{\ndev{-56.82}}
& \massdevinline{99.47}{\ndev{-40.10}}
& \massdevinline{133.60}{\ndev{-5.97}}
& \massdevinline{138.91}{\ndev{-0.66}}
& \massdevinline{141.62}{\pdev{+2.05}}
\\

$\rho$
& 775.11
& \massdevinline{716.97}{\ndev{-58.14}}
& \massdevinline{709.06}{\ndev{-66.05}}
& \massdevinline{713.74}{\ndev{-61.37}}
& \massdevinline{747.44}{\ndev{-27.67}}
& \massdevinline{765.75}{\ndev{-9.36}}
& \massdevinline{770.09}{\ndev{-5.02}}
\\

$K$
& 497.61
& \massdevinline{476.25}{\ndev{-21.36}}
& \massdevinline{508.08}{\pdev{+10.47}}
& \massdevinline{480.85}{\ndev{-16.76}}
& \massdevinline{498.65}{\pdev{+1.04}}
& \massdevinline{496.89}{\ndev{-0.72}}
& \massdevinline{492.11}{\ndev{-5.50}}
\\

$K^*$
& 895.81
& \massdevinline{870.90}{\ndev{-24.91}}
& \massdevinline{868.71}{\ndev{-27.10}}
& \massdevinline{854.21}{\ndev{-41.60}}
& \massdevinline{894.78}{\ndev{-1.03}}
& \massdevinline{903.80}{\pdev{+7.99}}
& \massdevinline{903.82}{\pdev{+8.01}}
\\

$\phi$
& 1019.46
& \massdevinline{1010.39}{\ndev{-9.07}}
& \massdevinline{1002.40}{\ndev{-17.06}}
& \massdevinline{993.53}{\ndev{-25.93}}
& \massdevinline{1025.69}{\pdev{+6.23}}
& \massdevinline{1028.66}{\pdev{+9.20}}
& \massdevinline{1026.69}{\pdev{+7.23}}
\\

$D$
& 1864.84
& \massdevinline{1843.34}{\ndev{-21.50}}
& \massdevinline{1859.28}{\ndev{-5.56}}
& \massdevinline{1799.29}{\ndev{-65.55}}
& \massdevinline{1863.50}{\ndev{-1.34}}
& \massdevinline{1867.11}{\pdev{+2.27}}
& \massdevinline{1865.27}{\pdev{+0.43}}
\\

$D^*$
& 2010.26
& \massdevinline{1992.65}{\ndev{-17.61}}
& \massdevinline{1992.11}{\ndev{-18.15}}
& \massdevinline{1945.06}{\ndev{-65.20}}
& \massdevinline{2013.73}{\pdev{+3.47}}
& \massdevinline{2012.16}{\pdev{+1.90}}
& \massdevinline{2007.23}{\ndev{-3.03}}
\\

$D_s$
& 1968.30
& \massdevinline{1968.76}{\pdev{+0.46}}
& \massdevinline{1959.19}{\ndev{-9.11}}
& \massdevinline{1967.13}{\ndev{-1.17}}
& \massdevinline{1973.84}{\pdev{+5.54}}
& \massdevinline{1971.08}{\pdev{+2.78}}
& \massdevinline{1968.55}{\pdev{+0.25}}
\\

$D_s^*$
& 2112.10
& \massdevinline{2102.25}{\ndev{-9.85}}
& \massdevinline{2087.78}{\ndev{-24.32}}
& \massdevinline{2076.69}{\ndev{-35.41}}
& \massdevinline{2112.39}{\pdev{+0.29}}
& \massdevinline{2108.22}{\ndev{-3.88}}
& \massdevinline{2104.07}{\ndev{-8.03}}
\\

$\eta_c$
& 2984.10
& \massdevinline{3016.29}{\pdev{+32.19}}
& \massdevinline{2999.48}{\pdev{+15.38}}
& \massdevinline{3057.71}{\pdev{+73.61}}
& \massdevinline{3007.20}{\pdev{+23.10}}
& \massdevinline{2998.84}{\pdev{+14.74}}
& \massdevinline{2996.10}{\pdev{+12.00}}
\\

$J/\psi$
& 3096.90
& \massdevinline{3110.53}{\pdev{+13.63}}
& \massdevinline{3108.22}{\pdev{+11.32}}
& \massdevinline{3112.11}{\pdev{+15.21}}
& \massdevinline{3114.82}{\pdev{+17.92}}
& \massdevinline{3105.68}{\pdev{+8.78}}
& \massdevinline{3101.55}{\pdev{+4.65}}
\\

$B$
& 5279.72
& \massdevinline{5271.88}{\ndev{-7.84}}
& \massdevinline{5237.73}{\ndev{-41.99}}
& \massdevinline{5202.42}{\ndev{-77.30}}
& \massdevinline{5288.86}{\pdev{+9.14}}
& \massdevinline{5285.57}{\pdev{+5.85}}
& \massdevinline{5279.53}{\ndev{-0.19}}
\\

$B^*$
& 5325.20
& \massdevinline{5328.44}{\pdev{+3.24}}
& \massdevinline{5287.66}{\ndev{-37.54}}
& \massdevinline{5274.09}{\ndev{-51.11}}
& \massdevinline{5345.85}{\pdev{+20.65}}
& \massdevinline{5339.30}{\pdev{+14.10}}
& \massdevinline{5331.59}{\pdev{+6.39}}
\\

$B_s$
& 5366.77
& \massdevinline{5363.46}{\ndev{-3.31}}
& \massdevinline{5306.85}{\ndev{-59.92}}
& \massdevinline{5339.82}{\ndev{-26.95}}
& \massdevinline{5366.03}{\ndev{-0.74}}
& \massdevinline{5361.43}{\ndev{-5.34}}
& \massdevinline{5357.42}{\ndev{-9.35}}
\\

$B_s^*$
& 5415.40
& \massdevinline{5418.79}{\pdev{+3.39}}
& \massdevinline{5361.41}{\ndev{-53.99}}
& \massdevinline{5395.24}{\ndev{-20.16}}
& \massdevinline{5424.10}{\pdev{+8.70}}
& \massdevinline{5417.33}{\pdev{+1.93}}
& \massdevinline{5411.93}{\ndev{-3.47}}
\\

$B_c$
& 6274.47
& \massdevinline{6293.17}{\pdev{+18.70}}
& \massdevinline{6272.02}{\ndev{-2.45}}
& \massdevinline{6328.55}{\pdev{+54.08}}
& \massdevinline{6282.07}{\pdev{+7.60}}
& \massdevinline{6276.05}{\pdev{+1.58}}
& \massdevinline{6274.31}{\ndev{-0.16}}
\\

$\eta_b$
& 9398.70
& \massdevinline{9396.63}{\ndev{-2.07}}
& \massdevinline{9473.76}{\pdev{+75.06}}
& \massdevinline{9439.76}{\pdev{+41.06}}
& \massdevinline{9383.79}{\ndev{-14.91}}
& \massdevinline{9390.05}{\ndev{-8.65}}
& \massdevinline{9394.54}{\ndev{-4.16}}
\\

$\Upsilon$
& 9460.40
& \massdevinline{9447.38}{\ndev{-13.02}}
& \massdevinline{9544.23}{\pdev{+83.83}}
& \massdevinline{9456.78}{\ndev{-3.62}}
& \massdevinline{9458.82}{\ndev{-1.58}}
& \massdevinline{9461.14}{\pdev{+0.74}}
& \massdevinline{9462.12}{\pdev{+1.72}}
\\

\SetCell[c=8]{l}{\textit{Baryons}} & & & & & & & \\

$N$
& 938.27
& \massdevinline{986.61}{\pdev{+48.34}}
& \massdevinline{998.91}{\pdev{+60.64}}
& \massdevinline{1049.22}{\pdev{+110.95}}
& \massdevinline{956.31}{\pdev{+18.04}}
& \massdevinline{946.85}{\pdev{+8.58}}
& \massdevinline{937.61}{\ndev{-0.66}}
\\

$\Lambda$
& 1115.68
& \massdevinline{1155.37}{\pdev{+39.69}}
& \massdevinline{1174.05}{\pdev{+58.37}}
& \massdevinline{1203.91}{\pdev{+88.23}}
& \massdevinline{1126.26}{\pdev{+10.58}}
& \massdevinline{1116.84}{\pdev{+1.16}}
& \massdevinline{1108.49}{\ndev{-7.19}}
\\

$\Sigma$
& 1197.45
& \massdevinline{1213.46}{\pdev{+16.01}}
& \massdevinline{1244.76}{\pdev{+47.31}}
& \massdevinline{1240.81}{\pdev{+43.36}}
& \massdevinline{1200.92}{\pdev{+3.47}}
& \massdevinline{1197.51}{\pdev{+0.06}}
& \massdevinline{1202.01}{\pdev{+4.56}}
\\

$\Delta$
& 1232.00
& \massdevinline{1234.74}{\pdev{+2.74}}
& \massdevinline{1247.96}{\pdev{+15.96}}
& \massdevinline{1222.41}{\ndev{-9.59}}
& \massdevinline{1246.48}{\pdev{+14.48}}
& \massdevinline{1238.11}{\pdev{+6.11}}
& \massdevinline{1228.78}{\ndev{-3.22}}
\\

$\Xi$
& 1321.71
& \massdevinline{1348.87}{\pdev{+27.16}}
& \massdevinline{1375.39}{\pdev{+53.68}}
& \massdevinline{1378.51}{\pdev{+56.80}}
& \massdevinline{1330.46}{\pdev{+8.75}}
& \massdevinline{1327.57}{\pdev{+5.86}}
& \massdevinline{1328.39}{\pdev{+6.68}}
\\

$\Sigma^*$
& 1383.70
& \massdevinline{1386.02}{\pdev{+2.32}}
& \massdevinline{1401.24}{\pdev{+17.54}}
& \massdevinline{1366.85}{\ndev{-16.85}}
& \massdevinline{1392.95}{\pdev{+9.25}}
& \massdevinline{1389.24}{\pdev{+5.54}}
& \massdevinline{1389.52}{\pdev{+5.82}}
\\

$\Xi^*$
& 1535.00
& \massdevinline{1530.53}{\ndev{-4.47}}
& \massdevinline{1541.91}{\pdev{+6.91}}
& \massdevinline{1510.57}{\ndev{-24.43}}
& \massdevinline{1532.88}{\ndev{-2.12}}
& \massdevinline{1532.83}{\ndev{-2.17}}
& \massdevinline{1536.59}{\pdev{+1.59}}
\\

$\Omega$
& 1672.45
& \massdevinline{1669.13}{\ndev{-3.32}}
& \massdevinline{1672.29}{\ndev{-0.16}}
& \massdevinline{1653.74}{\ndev{-18.71}}
& \massdevinline{1667.61}{\ndev{-4.84}}
& \massdevinline{1669.27}{\ndev{-3.18}}
& \massdevinline{1672.97}{\pdev{+0.52}}
\\

$\Lambda_c$
& 2286.46
& \massdevinline{2296.22}{\pdev{+9.76}}
& \massdevinline{2311.94}{\pdev{+25.48}}
& \massdevinline{2316.17}{\pdev{+29.71}}
& \massdevinline{2271.56}{\ndev{-14.90}}
& \massdevinline{2274.51}{\ndev{-11.95}}
& \massdevinline{2282.22}{\ndev{-4.24}}
\\

$\Sigma_c$
& 2453.97
& \massdevinline{2432.22}{\ndev{-21.75}}
& \massdevinline{2453.72}{\ndev{-0.25}}
& \massdevinline{2417.14}{\ndev{-36.83}}
& \massdevinline{2433.62}{\ndev{-20.35}}
& \massdevinline{2437.65}{\ndev{-16.32}}
& \massdevinline{2445.59}{\ndev{-8.38}}
\\

$\Sigma_c^*$
& 2518.48
& \massdevinline{2503.69}{\ndev{-14.79}}
& \massdevinline{2516.89}{\ndev{-1.59}}
& \massdevinline{2472.91}{\ndev{-45.57}}
& \massdevinline{2510.28}{\ndev{-8.20}}
& \massdevinline{2510.55}{\ndev{-7.93}}
& \massdevinline{2514.53}{\ndev{-3.95}}
\\

$\Omega_c$
& 2695.20
& \massdevinline{2692.94}{\ndev{-2.26}}
& \massdevinline{2689.62}{\ndev{-5.58}}
& \massdevinline{2706.15}{\pdev{+10.95}}
& \massdevinline{2686.30}{\ndev{-8.90}}
& \massdevinline{2691.97}{\ndev{-3.23}}
& \massdevinline{2695.46}{\pdev{+0.26}}
\\

$\Omega_c^*$
& 2765.90
& \massdevinline{2758.28}{\ndev{-7.62}}
& \massdevinline{2750.50}{\ndev{-15.40}}
& \massdevinline{2751.21}{\ndev{-14.69}}
& \massdevinline{2755.74}{\ndev{-10.16}}
& \massdevinline{2758.51}{\ndev{-7.39}}
& \massdevinline{2758.93}{\ndev{-6.97}}
\\

$\Xi_{cc}$
& 3621.20
& \massdevinline{3612.39}{\ndev{-8.81}}
& \massdevinline{3623.52}{\pdev{+2.32}}
& \massdevinline{3622.75}{\pdev{+1.55}}
& \massdevinline{3606.19}{\ndev{-15.01}}
& \massdevinline{3619.91}{\ndev{-1.29}}
& \massdevinline{3621.98}{\pdev{+0.78}}
\\

$\Lambda_b$
& 5619.60
& \massdevinline{5639.64}{\pdev{+20.04}}
& \massdevinline{5613.08}{\ndev{-6.52}}
& \massdevinline{5661.50}{\pdev{+41.90}}
& \massdevinline{5613.62}{\ndev{-5.98}}
& \massdevinline{5619.37}{\ndev{-0.23}}
& \massdevinline{5629.49}{\pdev{+9.89}}
\\

$\Sigma_b$
& 5815.64
& \massdevinline{5809.25}{\ndev{-6.39}}
& \massdevinline{5784.20}{\ndev{-31.44}}
& \massdevinline{5802.20}{\ndev{-13.44}}
& \massdevinline{5811.27}{\ndev{-4.37}}
& \massdevinline{5812.72}{\ndev{-2.92}}
& \massdevinline{5816.02}{\pdev{+0.38}}
\\

$\Sigma_b^*$
& 5834.74
& \massdevinline{5837.34}{\pdev{+2.60}}
& \massdevinline{5808.86}{\ndev{-25.88}}
& \massdevinline{5827.58}{\ndev{-7.16}}
& \massdevinline{5841.11}{\pdev{+6.37}}
& \massdevinline{5840.67}{\pdev{+5.93}}
& \massdevinline{5842.37}{\pdev{+7.63}}
\\

$\Omega_b$
& 6046.10
& \massdevinline{6044.45}{\ndev{-1.65}}
& \massdevinline{5992.50}{\ndev{-53.60}}
& \massdevinline{6073.81}{\pdev{+27.71}}
& \massdevinline{6036.25}{\ndev{-9.85}}
& \massdevinline{6040.30}{\ndev{-5.80}}
& \massdevinline{6039.57}{\ndev{-6.53}}
\\

\bottomrule
\end{tblr}}
\end{table*}

At the equal-sector point $\lambda=1/2$, Table~\ref{tab:SAJ_lambda_half_spectrum} compares the complete ground-state meson and baryon spectra obtained with the six fitting schemes, with each calculated mass followed by its signed deviation from experiment.
Fig.~\ref{fig:SAJ_reciprocal} shows that when the three-quark interactions are not included, two-body Hamiltonian trajectories do not reach an RMSE below about $10~{\rm MeV}$.
At the equal-sector point, the six trajectories give
\begin{equation}
\begin{aligned}
(\sigma_M,\sigma_B)_{\lambda=1/2}
={}&(21.37,\,18.75)~{\rm MeV} && \text{for 2BD},\\
={}&(42.47,\,31.90)~{\rm MeV} && \text{for 2BD with running }\alpha_s,\\
={}&(45.86,\,43.42)~{\rm MeV} && \text{for kinematically relativized 2BD},\\
={}&(12.04,\,10.94)~{\rm MeV} && \text{with contact 3BD},\\
={}&(7.09,\,6.70)~{\rm MeV} && \text{with Gaussian 3BD},\\
={}&(5.66,\,5.36)~{\rm MeV} && \text{with Yukawa 3BD}.
\end{aligned}
\label{eq:SAJ_equal_weight_results}
\end{equation}
For the last three cases, the non-relativistic fixed-$\alpha_s$ two-body Hamiltonian is used as the baseline.
The contrast is even clearer at the meson-fit endpoint.
At $\lambda=1$, the baryon RMSEs are $\sigma_B=83.77$, 96.77, and $126.26~{\rm MeV}$ for the original, running-coupling, and relativized pairwise Hamiltonians, respectively, while the contact, Gaussian, and Yukawa three-body extensions reduce them to 29.85, 13.39, and $7.34~{\rm MeV}$.
The reduced-mass running prescription does not improve the meson--baryon compatibility.
The characteristic momentum scale of a bound state is generally determined by its inverse spatial size and wave function, and is therefore not fixed by the constituent reduced mass alone, particularly in heavy-light systems where the internal momentum remains largely controlled by the light degrees of freedom.
At $\lambda=1/2$, the fit drives $Q_0$ to its upper limit of $3.0~{\rm GeV}$, giving $\alpha_s^{qq}=0.399$, $\alpha_s^{cc}=0.365$, and $\alpha_s^{bb}=0.285$, and $Q_0$ remains saturated throughout the trade-off trajectory.

The residual pattern of the minimally relativized trajectory comes from changing only the kinetic-energy operator while keeping the non-relativistic central and hyperfine interactions, whose fitted parameters already contain part of the omitted relativistic dynamics.
At $\lambda=1/2$, the refit drives $m_q$ to its upper bound $0.50~{\rm GeV}$, $\alpha$ and $\beta$ to their lower bounds $0.50~{\rm GeV}$ and 0, respectively, and the hyperfine normalization to $Z_{\rm hf}=0.227$.
Since the light-pair color-spin strength scales approximately as $Z_{\rm hf}\mu_{qq}^{2}/m_q^{2}$, this parameter rearrangement substantially weakens the hyperfine attraction in the light baryons.
Consequently, the calculated $N-\Delta$ splitting is only $173~{\rm MeV}$, compared with the experimental value of approximately $294~{\rm MeV}$, leaving the nucleon and $\Lambda$ masses 111 and $88~{\rm MeV}$ above experiment, respectively.
The accompanying increase of the Coulomb strength and decrease of the string tension instead lower the spin-averaged heavy-light meson centroids, producing deviations of approximately $-65~{\rm MeV}$ and $-58~{\rm MeV}$ for the $D^{(*)}$ and $B^{(*)}$ centroids.
Heavy quarkonia exhibit a distinct failure: the nearly mass-independent regulator obtained from $\beta\simeq0$, together with the small $Z_{\rm hf}$, suppresses the $c\bar c$ and $b\bar b$ hyperfine splittings, so that the $\eta_c-J/\psi$ and $\eta_b-\Upsilon$ pairs cannot be reproduced even when their spin-averaged masses are partially compensated by the constituent masses.
This calculation shows that relativistic kinematics alone is not sufficient to remove the meson--baryon trade-off of the otherwise unchanged $\mathrm{S}^{2}\mathrm{AJ}$ Hamiltonian.
The corresponding momentum-dependent vertices and spin-dependent operators are also needed.

For every interior point $0<\lambda<1$, all three-body trajectories give smaller meson and baryon RMSEs than the original pairwise baseline.
At $\lambda=1$, they have the same meson-only optimum but retain substantially smaller baryon errors.
The spatially contact interaction already improves the two-body result, and the mass-dependent spatial profiles give a further reduction over nearly the entire trajectory.
The Gaussian profile gives a lower baryon RMSE than the contact form at 28 of the 29 sampled values of $\lambda$, with the baryon-only point $\lambda=0$ as the sole exception, and the Yukawa profile gives a lower baryon RMSE than the Gaussian form at all 29 sampled points.
For $0.1\leq\lambda\leq0.9$, the Yukawa trajectory remains within $5.25\leq\sigma_M\leq7.74~{\rm MeV}$ and $4.43\leq\sigma_B\leq6.24~{\rm MeV}$.
Thus, the improvement is obtained not only at the meson-only endpoint and the equal-sector point, but also over a broad range of meson-sector weights.

\begin{figure*}[!t]
\centering
\includegraphics[width=0.96\textwidth]{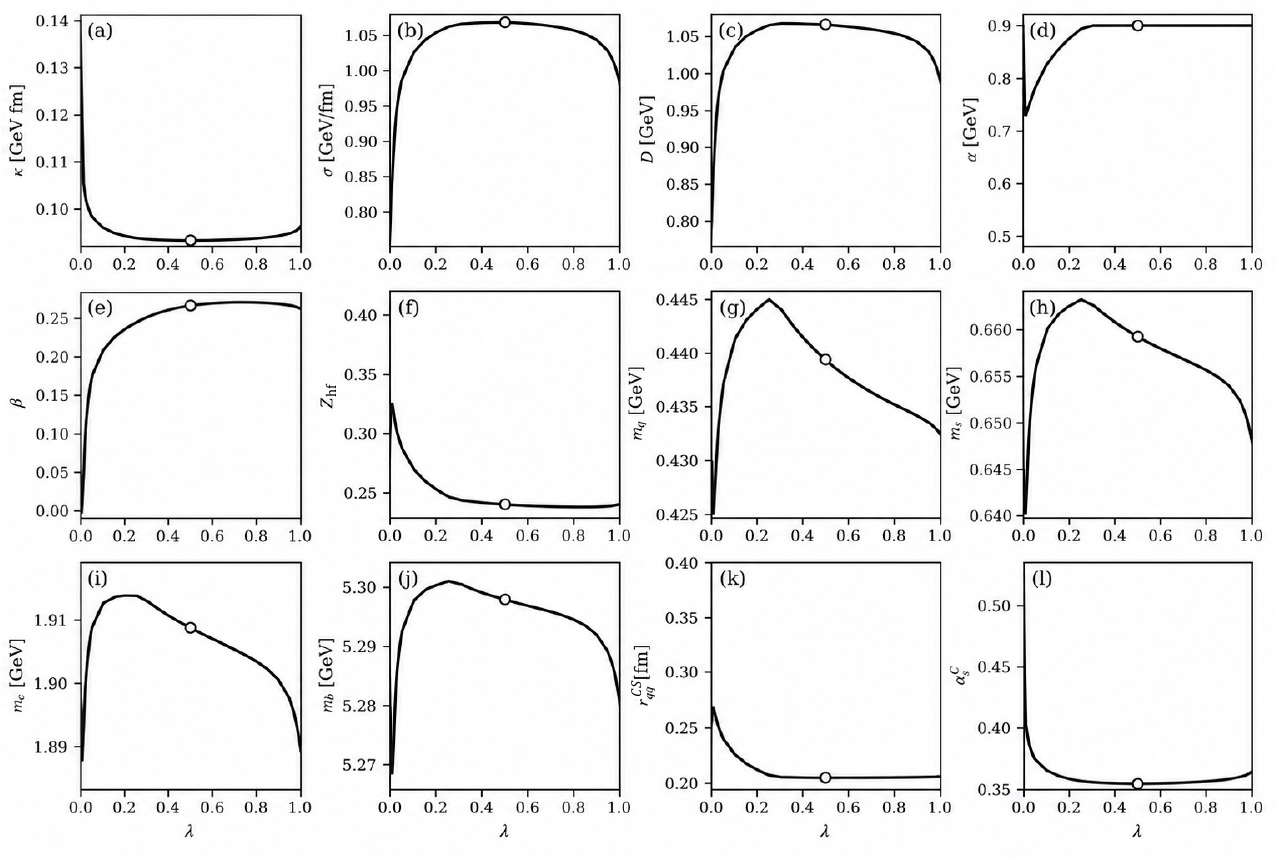}
\caption{\justifying Fitted nonlinear parameters and derived diagnostics along the Yukawa meson--baryon trade-off trajectory of the \(\mathrm{S^2AJ}\) model. Panels (a)--(j) show \(\kappa\), \(\sigma\), \(D\), \(\alpha\), \(\beta\), \(Z_{\rm hf}\), \(m_q\), \(m_s\), \(m_c\), and \(m_b\), respectively. Panels (k) and (l) show \(r_{qq}^{CS}=\hbar c/\mu_{qq}\) and \(\alpha_s^{(C)}=3\kappa/(4\hbar c)\). The hollow circle denotes the equal-sector point \(\lambda=1/2\).}
\label{fig:SAJ_yukawa_parameters}
\end{figure*}

We also examine how the fitted parameters change along the Yukawa trajectory.
Fig.~\ref{fig:SAJ_yukawa_parameters} shows the ten nonlinear fit parameters $\kappa$, $\sigma$, $D$, $\alpha$, $\beta$, $Z_{\rm hf}$, $m_q$, $m_s$, $m_c$, and $m_b$, together with the derived light-quark hyperfine range $r_{qq}^{CS}$ and effective central coupling $\alpha_s^{(C)}=3\kappa/(4\hbar c)$.
The largest parameter readjustment occurs between $\lambda=0$ and 0.01, where the meson block first enters the objective function.
Along the Yukawa trade-off trajectory, no nonlinear parameter is flagged on an imposed boundary for $0.01\leq\lambda\leq0.25$.
The hyperfine scale $\alpha$ reaches its upper bound at $\lambda=0.30$ and remains saturated through $\lambda=1$, while the string-tension parameter $\sigma$ is additionally at or near its upper bound for $0.35\leq\lambda\leq0.50$.
At $\lambda=1/2$, the fitted nonlinear two-body parameters are
\begin{align}
&\kappa=0.0934~{\rm GeV\,fm}, \quad \sigma=1.0694~{\rm GeV/fm}, \quad D=1.0681~{\rm GeV}, \quad \alpha=0.9000~{\rm GeV},\quad \beta=0.2685,\nonumber\\
&Z_{\rm hf}=0.2415, \quad m_q=0.4395~{\rm GeV},\quad m_s=0.6593~{\rm GeV},\quad m_c=1.9088~{\rm GeV},\quad m_b=5.2980~{\rm GeV}.  \label{eq:SAJ_yukawa_equal_weight_parameters}
\end{align}
The string tension, light-quark hyperfine range, and effective central coupling quoted below are derived from the fitted parameters:
\begin{equation}
(\hbar c)\sigma =0.2110~{\rm GeV}^2, \;\;\; \mu_{qq}=\alpha+\frac{\beta m_q}{2}=0.9590~{\rm GeV},\;\;\;
r_{qq}^{CS}=\frac{\hbar c}{\mu_{qq}}=0.2058~{\rm fm}, \;\;\;\; \alpha_s^{(C)}=\frac{3\kappa}{4\hbar c}=0.3550.
\label{eq:SAJ_yukawa_equal_weight_diagnostics}
\end{equation}
The solution at the equal-sector point is still in the intended constituent-quark and short-distance regime and has a particularly narrow light-quark hyperfine regulator.
The small errors are already obtained before these bounds become active: at $\lambda=0.25$, where no nonlinear parameter is flagged on a boundary, the Yukawa fit gives $(\sigma_M,\sigma_B)=(6.46,\,4.81)~{\rm MeV}$.

With all nonlinear two-body parameters refitted at each $\lambda$, the Yukawa three-quark interaction keeps \textit{both $\sigma_M$ and $\sigma_B$ below 8~MeV over $0.1\leq\lambda\leq0.9$}.
The numerical values of the fitted two-body parameters and the three-quark couplings change along the trajectory.

\begin{table}[!t]
\centering
\caption{\justifying
Masses and spin-component decomposition of the $\mathcal J^{P}=1/2^{+}$ $qsQ$ states obtained at the equal-sector compatibility point $\lambda=1/2$ of the Yukawa fit.
The signed mass deviation $\Delta M=M-M_{\exp}$ is shown in parentheses beneath each calculated mass.
Here, $\chi_{\rho}$ and $\chi_{\lambda}$ denote configurations in which the $qs$ pair is coupled to $s_{qs}=0$ and 1, respectively.
The effective admixture angle is defined state by state as $\theta_{\rm eff}=\tan^{-1}\!\sqrt{P_{\rm sub}/P_{\rm dom}}$, with $\Delta\theta_{\rm eff} =\theta_{\rm eff}^{2Q+3Q}-\theta_{\rm eff}^{2Q}$.}
\label{tab:S2AJ_XiQ_mixing}

\small
\renewcommand{\pdev}[1]{\textcolor{devred}{\ensuremath{(#1)}}}
\renewcommand{\ndev}[1]{\textcolor{devgreen}{\ensuremath{(#1)}}}

\begin{tblr}{
 width=\textwidth,
 colspec={
   Q[c,wd=1.2cm]
   Q[c,wd=1.3cm]
   Q[c,wd=1.1cm]
   X[1.8,c]
   X[1.8,c]
   Q[c,wd=1.5cm]
   X[0.78,c]
   X[0.78,c]
   X[0.95,c]
   X[0.95,c]
 },
 colsep=2.4pt,
 rowsep=2.2pt,
 row{1-2}={bg=headgray,halign=c,valign=m},
 row{3-Z}={halign=c,valign=m},
 hline{2}={7-10}{0.3pt},
 hline{3}={1-10}{0.3pt},
}
\toprule
\SetCell[r=2]{c,m} State
&
\SetCell[r=2]{c,m} Content
&
\SetCell[r=2]{c,m}
  \shortstack[c]{$M_{\rm exp}$\\$[\mathrm{MeV}]$}
&
\SetCell[r=2]{c,m}
  \shortstack[c]{$M^{2Q}$ $[\mathrm{MeV}]$\\$(\Delta M)$}
&
\SetCell[r=2]{c,m}
  \shortstack[c]{$M^{2Q+3Q}$ $[\mathrm{MeV}]$\\$(\Delta M)$}
&
\SetCell[r=2]{c,m}
  \shortstack[c]{Dominant\\spin}
&
\SetCell[c=2]{c}
  \shortstack[c]{Spin probabilities\\$(2Q+3Q)$}
&
&
\SetCell[c=2]{c}
  \shortstack[c]{Effective admixture\\angles}
&
\\
&
&
&
&
&
&
$P_{\rho}$ [\%]
&
$P_{\lambda}$ [\%]
&
$\theta_{\rm eff}^{2Q+3Q}$
&
$\Delta\theta_{\rm eff}$
\\

$\Xi_c$
& $qsc$
& 2470.50
& \mbox{\massdevinline{2517.39}{\pdev{+46.89}}}
& \mbox{\massdevinline{\textbf{2479.35}}{\pdev{+8.85}}}
& $\chi_\rho$
& 99.7162
& 0.2838
& $3.0546^\circ$
& $-0.0610^\circ$
\\

$\Xi_c^{\prime}$
& $qsc$
& 2578.30
& \mbox{\massdevinline{2597.38}{\pdev{+19.08}}}
& \mbox{\massdevinline{\textbf{2570.32}}{\ndev{-7.98}}}
& $\chi_\lambda$
& 0.3152
& 99.6848
& $3.2186^\circ$
& $-0.0526^\circ$
\\

$\Xi_b$
& $qsb$
& 5797.00
& \mbox{\massdevinline{5837.30}{\pdev{+40.30}}}
& \mbox{\massdevinline{\textbf{5807.98}}{\pdev{+10.98}}}
& $\chi_\rho$
& 99.9744
& 0.0256
& $0.9164^\circ$
& $-0.0197^\circ$
\\

$\Xi_b^{\prime}$
& $qsb$
& 5934.90
& \mbox{\massdevinline{5946.75}{\pdev{+11.85}}}
& \mbox{\massdevinline{\textbf{5928.76}}{\ndev{-6.14}}}
& $\chi_\lambda$
& 0.0341
& 99.9659
& $1.0589^\circ$
& $-0.0135^\circ$
\\

\bottomrule
\end{tblr}%

\end{table}

\paragraph{Coupled spin configurations and diagonalization in the $\boldsymbol{\Xi_Q}$ sector:}

The $\mathcal{J}^P=1/2^+$ $qsQ$ baryons require an explicit coupled-basis treatment of the two mixed-symmetry spin configurations
\begin{equation}
\chi_{\rho,Q} \equiv \big|\bigl[(qs)_{s_{qs}=0}Q\bigr]_{1/2}\big\rangle,\qquad  \chi_{\lambda,Q} \equiv \big|\bigl[(qs)_{s_{qs}=1}Q\bigr]_{1/2}\big\rangle,
\qquad Q=c,b.
\end{equation}
With the quark ordering $(q,s,Q)$, $\chi_\rho$ is antisymmetric and $\chi_\lambda$ is symmetric under interchange of the first two spin coordinates.
Because all three quark masses are different, the cross-pair hyperfine interactions connect the $\chi_\rho$ and $\chi_\lambda$ configurations, and the $\Xi_Q$ and $\Xi_Q'$ masses are therefore obtained by diagonalizing the full $(\chi_\rho,\chi_\lambda)$ coupled basis rather than by assigning either spin component a priori.
For each eigenstate, we define
\begin{equation}
P_\rho \equiv \frac{\langle\Psi|\mathcal{P}_{\chi_\rho}|\Psi\rangle}{\langle\Psi|\Psi\rangle},  \qquad P_\lambda \equiv \frac{\langle\Psi|\mathcal{P}_{\chi_\lambda}|\Psi\rangle}{\langle\Psi|\Psi\rangle}, \qquad \theta_{\mathrm{eff}} \equiv \tan^{-1}\!\sqrt{\frac{P_{\mathrm{sub}}}{P_{\mathrm{dom}}}},
\end{equation}
where $P_{\mathrm{dom}}$ and $P_{\mathrm{sub}}$ denote the larger and smaller of $P_\rho$ and $P_\lambda$, respectively.
As summarized in Table~\ref{tab:S2AJ_XiQ_mixing}, the subdominant spin-component probability is below 0.32\% in the charm sector and below 0.04\% in the bottom sector.
The resulting effective angles are approximately 3.1{\ensuremath{{}^{\circ}}}--3.2{\ensuremath{{}^{\circ}}} for the $\Xi_c$--$\Xi_c'$ pair and 0.9{\ensuremath{{}^{\circ}}}--1.1{\ensuremath{{}^{\circ}}} for the $\Xi_b$--$\Xi_b'$ pair.
The smaller bottom-sector angles show the expected suppression as the heavy-quark mass increases.
The small difference between the effective angles of the lower and upper eigenstates comes from their independently optimized radial amplitudes.
We therefore use these numbers as basis-dependent norm diagnostics and not as one exact two-state rotation angle.
The dependence of the $\Xi_c$--$\Xi_c'$ mixing angle on its definition, its comparison with lattice determinations, and the effect of virtual $qsQg$ configurations are analyzed in Ref.~\cite{KimNamBaek2026Xic}.
At each fitted parameter point, the three-quark interaction is evaluated perturbatively and does not directly change the two-body eigenvector.
The small values of $\Delta\theta_{\rm eff}$ arise from the reoptimization of the two-body parameters between the $2Q$ and $2Q+3Q$ fits, rather than from a direct first-order correction to the wave function.
The $\mathcal{J}^P=3/2^+$ states $\Xi_Q^{*}$ contain the totally symmetric spin configuration $\chi_S$ and therefore do not enter the $\chi_\rho$--$\chi_\lambda$ diagonalization.
The fixed-parameter masses are also improved after diagonalizing the Hamiltonian in the $\chi_\rho$--$\chi_\lambda$ basis.
Thus, the compatibility test can be extended to the $\Xi_Q$ and $\Xi_Q'$ states outside the main benchmark set.

\paragraph{Can chiral dynamics account for the phenomenological role of the TGE-inspired interaction?}

\begin{figure*}[!b]
\centering
\includegraphics[width=0.5\linewidth]{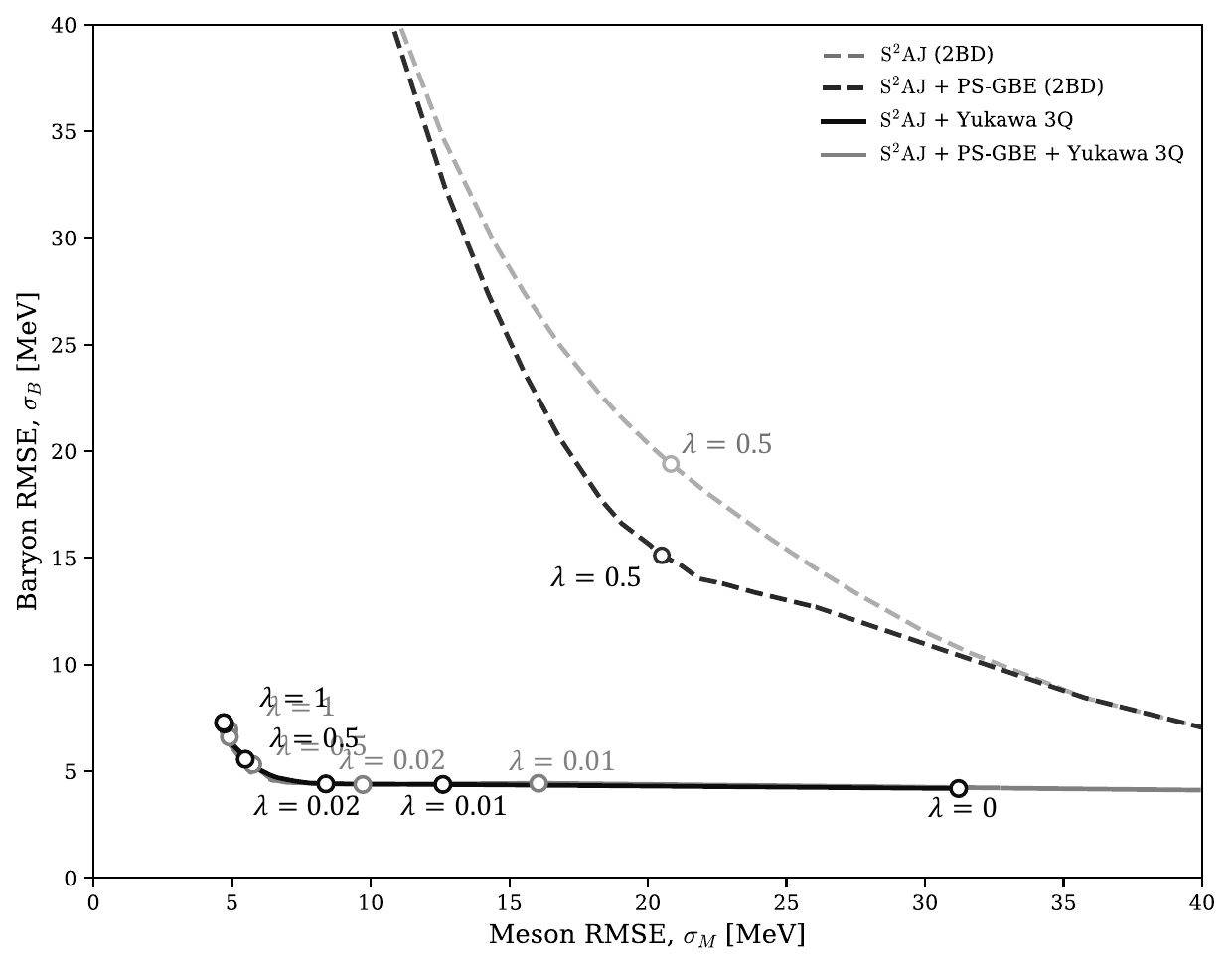}
\caption{\justifying Meson--baryon RMSE trade-off trajectories for the \(\mathrm{S^2AJ}\) pair Hamiltonian, its hybrid PS-GBE extension, and the corresponding Yukawa-profile three-quark calculations.}
\label{fig:chiral_vs_OGE}
\end{figure*}

\begin{table}[!t]
\centering
\caption{\justifying Equal-sector ($\lambda=1/2$) results of the chiral-dynamics benchmark. Here \(\sigma_{\rm eq}=[(\sigma_M^2+\sigma_B^2)/2]^{1/2}\), and \(s_\chi\) is the fitted overall PS-GBE strength.}
\label{tab:chiral_vs_TGE_equal_sector}
\begin{tblr}{
colspec={
Q[l,wd=6.3cm]
Q[c,wd=1.5cm]
Q[c,wd=2.0cm]
Q[c,wd=2.0cm]
Q[c,wd=2.0cm]
},
row{1}={bg=headgray,halign=c,valign=m},
cell{1}{1}={}{halign=l},
hline{2}={1-5}{0.3pt},
rowsep=1.5pt,
}
\toprule
Hamiltonian
& \(s_\chi\)
& \(\sigma_M\) (MeV)
& \(\sigma_B\) (MeV)
& \(\sigma_{\rm eq}\) (MeV) \\
\(\mathrm{S^2AJ}\) $+$ PS-GBE (2BD)
& 0.626
& 20.50
& 15.11
& 18.01 \\
\(\mathrm{S^2AJ}\) $+$ PS-GBE \(+\) Yukawa 3Q
& 0.070
& 6.25
& 5.50
& 5.89 \\
\bottomrule
\end{tblr}
\end{table}

As reviewed in  Section~\ref{chi_QM}, Goldstone-boson exchange gives a flavor-spin interaction which is different from the OGE color--spin interaction. It is therefore possible that part of the residual attributed to the connected three-quark interaction comes from the omitted chiral interaction. We test this possibility by adding the spin--spin (denoted by $SS$) parts of the regulated $\pi$, $K$, and $\eta$ exchange interactions in Eq.~\eqref{eq:GBE_general} to the $\mathrm{S}^{2}\mathrm{AJ}$ Hamiltonian. We do not replace the OGE interaction by a pure GBE Hamiltonian, because the present three-quark operators are constructed from the color-Coulomb and color--spin vertices of the OGE-based Hamiltonian. We instead use
\begin{equation}
H_{\mathrm{S^2AJ}+\chi}
=
H_{\mathrm{S^2AJ}}
+
s_\chi
\sum_{(ij)\in LL}
\sum_{\phi=\pi,K,\eta}
V_{ij}^{(\phi),SS},
\label{eq:SAJ_chiral_hybrid}
\end{equation}
where $LL$ denotes pairs composed only of $u,d,s$ quarks or antiquarks. No direct PS-GBE interaction is included for a pair containing $c$ or $b$. The tensor term has a vanishing diagonal matrix element in the common $S$-wave benchmark and is not included. We use the standard monopole-regulated parameters
\begin{equation}
\frac{g_{\rm ch}^{2}}{4\pi}=0.54,\qquad
(m_\pi,\Lambda_\pi)=(0.70,4.20)~{\rm fm}^{-1},\qquad
(m_K,\Lambda_K)=(2.51,5.20)~{\rm fm}^{-1},\qquad
(m_\eta,\Lambda_\eta)=(2.77,5.20)~{\rm fm}^{-1},
\label{eq:SAJ_chiral_reference_parameters}
\end{equation}
with $\theta_P=-15{\ensuremath{{}^{\circ}}}$. Only the common strength is fitted within the range $0.02\leq s_\chi\leq1.5$. We do not include an additional scalar exchange because it mainly changes the spin-independent central interaction and is strongly correlated with the central parameters already fitted in the $\mathrm{S}^{2}\mathrm{AJ}$ Hamiltonian. At every value of $\lambda$, all nonlinear $\mathrm{S}^{2}\mathrm{AJ}$ parameters and constituent masses are reoptimized within the bounds of Eqs.~\eqref{eq:SAJ_physical_bounds} and~\eqref{eq:SAJ_mass_bounds}. For the Yukawa branches, $A$, $B$, and $C$ are also refitted by linear variable projection. The PS-GBE and three-quark contributions are evaluated in first order using the same bounded $S$-wave eigenstates.

The meson--baryon trade-off trajectories are shown in  Fig.~\ref{fig:chiral_vs_OGE}, and the equal-sector ($\lambda=1/2$) results of the two hybrid Hamiltonians are summarized in Table~\ref{tab:chiral_vs_TGE_equal_sector}. At $\lambda=1/2$, PS-GBE changes $(\sigma_M,\sigma_B)$ from $(21.37,18.75)$ to $(20.50,15.11)~\mathrm{MeV}$, and the combined error decreases from 20.10 to 18.01~MeV. Thus, the chiral interaction removes part of the light-baryon residual, but the meson--baryon incompatibility remains. In comparison, the Yukawa-profile three-quark interaction gives an equal-sector error of 5.51~MeV. When both interactions are included, the fitted chiral strength decreases sharply from $s_\chi=0.626$ to 0.070. The intermediate-range PS-GBE interaction therefore cannot replace the short-range contribution identified by the compatibility benchmark, whereas the connected mass-scaled three-quark interaction drives both sectors into the low-error region. This strong decrease of $s_\chi$ shows that, in this benchmark, the missing contribution is described mainly by \textit{the short-range three-quark interaction and not by the omitted pseudoscalar pair exchange}.

\subsection{Model independence of the three-quark interactions}
\label{subsec:cross_model_udscb}
In  Section~\ref{subsec:SAJ_udscb}, we tested the fitting prescription, the Coulomb coupling, the kinetic-energy operator, and the explicit PS-GBE contribution within the $\mathrm{S}^{2}\mathrm{AJ}$ framework.
We now repeat the meson--baryon compatibility analysis for other OGE-based Hamiltonians.
For this purpose, we use the AL1, AL2, AP1, AP2, Bhaduri (BD), Stanley (SL), and Godfrey--Isgur/Capstick--Isgur-type (GI/CI-type) Hamiltonians~\cite{Godfrey1985,Semay1994,SilvestreBrac1996,CapstickIsgur:1986,Bhaduri1981,Stanley1980}.
These models include non-relativistic and semirelativistic kinematics, linear and noninteger-power confinement, Gaussian, exponential, and monopole-form-factor smearings of the short-distance interaction, and momentum-dressed relativized interactions.

We restrict the cross-model comparison below to OGE-based constituent-quark models containing both a color-Coulomb interaction and a chromomagnetic color-spin interaction. 
The operators $L^{C-C}$, $L^{S-S}$, and $L^{C-S}$ can then be added to every baseline with the same operator interpretation. 
We do not include a pure GBE Hamiltonian in this comparison. 
The models considered below are different realizations of the OGE-based color-Coulomb and color-spin interactions, whereas pure GBE provides a different flavor-spin description of the light-quark hyperfine interaction. The hybrid OGE+PS-GBE test was given separately in the preceding subsection.

The model labels below identify the functional forms of the two-body Hamiltonians rather than frozen historical parameter sets.
At each value of $\lambda$, the constituent masses and all allowed model-specific shape parameters are reoptimized using Eq.~\eqref{eq:reciprocal_objective}, separately for the common-offset condition $D_M=D_B\equiv D$ and for the separate-offset prescription in which $D_M$ and $D_B$ are independently profiled.

\paragraph{Kinematic conventions and additive offsets:}

For the non-relativistic baselines, the intrinsic kinetic operator is $K_{\rm NR}^{(N)} = \sum_{i=1}^{N} ( m_i+{\mathbf p_i^2}/{2m_i} ) -K_{\rm CM}$.
The semirelativistic SL- and GI/CI-type models instead use $K_{\rm rel}^{(N)} = \sum_{i=1}^{N} \sqrt{\mathbf p_i^2+m_i^2}$ in the hadron center-of-mass frame.

The zero-point convention for the parameter $D$ is treated separately from the state-dependent part of each interaction. 
The scheme and model dependence of the additive constant was reviewed in  Section~\ref{VC_constant}. 
The $\mathrm{S}^{2}\mathrm{AJ}$ trajectories in the preceding subsection retain a single common additive parameter.
For each alternative baseline, we use two prescriptions for the additive offset.

In the common-offset fit, a single color-additive pair constant is imposed,
$D_M=D_B\equiv D$, so that the total baryon offset is $3/2$ times the meson offset.
In the separate-offset fit, $D_M$ and $D_B$ are profiled independently, so the meson and baryon centroids can change separately without adding state-dependent freedom within either sector.
The common-offset calculation keeps the pairwise color-additive zero-point relation.
The separate-offset calculation removes the sector-wide centroid mismatch before testing the state-dependent baryon residuals.
We use both prescriptions to check how the result depends on the treatment of the additive offset.

\paragraph{The AL1, AL2, AP1, and AP2 interactions:}

The four AL/AP Hamiltonians share the non-relativistic kinetic operator and may be written in the common form
\begin{equation} 
H_X=K_{\rm NR}^{(N)}+\sum_{i<j}V_{ij}^{X},\qquad X\in\{\mathrm{AL1},\mathrm{AL2},\mathrm{AP1},\mathrm{AP2}\}. \label{eq:ALAP_hamiltonian_common} 
\end{equation}
The pair interaction is
\begin{equation} 
V_{ij}^{X}=-\frac{3}{4}F_i^cF_j^c\left[-\frac{\kappa_X}{r_{ij}}g_X(r_{ij})+\sigma_Xr_{ij}^{p_X}-D_{M/B}+\frac{2\pi\kappa_X^{\rm hf}}{3m_im_j}g_X(r_{ij})G_{ij}^{X}(r_{ij})\,\boldsymbol{\sigma}_i\cdot\boldsymbol{\sigma}_j\right]. \label{eq:ALAP_unified_our_notation} \end{equation}
Here and in the BD interaction below, $D_{M/B}$ denotes $D_M$ in the meson sector and $D_B$ in the baryon sector.
The common-offset calculation imposes $D_M=D_B\equiv D$, whereas the separate-offset calculation profiles the two constants independently.
The normalized Gaussian hyperfine profile is
\begin{equation} G_{ij}^{X}(r)=\frac{\exp[-r^2/(r_{0,ij}^{X})^2]}{\pi^{3/2}(r_{0,ij}^{X})^3},\qquad r_{0,ij}^{X}=r_{0,X}\left(\frac{m_{ij}}{1~{\rm GeV}}\right)^{-\zeta_X}. \label{eq:ALAP_smearing_our_notation} \end{equation}
For the AL1 and AL2 models, the power is fixed to $p_X=1$, whereas for the AP1 and AP2 models it is $p_X=2/3$.
The AL models therefore employ linear confinement, whereas the AP models use a softer $r^{2/3}$ confinement.
For $p_X=1$, $\sigma_X$ has the interpretation of a string tension.
For $p_X=2/3$, $\sigma_X$ denotes the corresponding power-law confinement coefficient.
The regulator function is $g_X(r)=1$ for AL1 and AP1.
For AL2 and AP2, the short-distance regulator is instead taken as $g_X(r)=1-e^{-r/r_{c,X}}$, where $r_{c,X}$ denotes the corresponding cutoff scale.
The AL2 and AP2 interactions additionally regularize the Coulomb and hyperfine terms at the origin through the common factor $1-e^{-r/r_{c,X}}$.

\paragraph{The pairwise BD interaction:}
The BD benchmark uses the non-relativistic kinetic energy and the pairwise part of the Bhaduri (BD) interaction,
\begin{equation} 
H_{\rm BD}=K_{\rm NR}^{(N)}-\frac{3}{4}\sum_{i<j} F_i^cF_j^c\bigg[-\frac{\kappa_{\rm BD}}{r_{ij}}+\sigma_{\rm BD}r_{ij}-D_{M/B}+\frac{\kappa_{\rm BD}}{m_im_j}\frac{e^{-r_{ij}/r_{0,\rm BD}}}{r_{0,\rm BD}^{\,2}r_{ij}}\,\boldsymbol{\sigma}_i\cdot\boldsymbol{\sigma}_j\bigg]. \label{eq:BD_potential_our_notation} \end{equation}
In contrast to the AL/AP family, the Coulomb and hyperfine strengths are tied to the same coefficient $\kappa_{\rm BD}$, and the exponential hyperfine range $r_{0,\rm BD}$ is flavor independent.
Only the pairwise BD interaction is retained in this benchmark.
Any additional non-pairwise baryon correction introduced in historical versions of the model is excluded before the present three-quark interaction is added.

\paragraph{The SL-type finite-size interaction:}
The SL benchmark uses the semirelativistic spinless-Salpeter kinetic energy and an orbital $S$-wave reduction of the pairwise finite-size interaction~\cite{Stanley1980}.
The Hamiltonian in the hadron center-of-mass frame is
\begin{align}
H_{\rm SL} =K_{\rm rel}^{(N)}
-\frac{3}{4} \sum_{i<j}^{N} F_i^cF_j^c \bigg[ -\kappa\,\hbar c\, \mathcal C_{ij}^{\rm Y}(r_{ij}) +\sigma r_{ij} -D_X +\frac{2\pi\kappa_{\rm hf}}{3m_i m_j} (\hbar c)^3 \delta_{ij}^{\rm Y}(r_{ij})\, \boldsymbol{\sigma}_i\cdot\boldsymbol{\sigma}_j \bigg].
\end{align}
Here $X=M,B$ labels the meson and baryon sectors, respectively.
The common-offset calculation imposes $D_M=D_B\equiv D$, whereas the separate-offset calculation profiles $D_M$ and $D_B$ independently.
The finite constituent size is introduced through one-quark monopole form factors
\begin{equation}
F_i(\mathbf q^{\,2})
= \frac{D_i^2}{\mathbf q^{\,2}+D_i^2}, \qquad D_i=\frac{1}{d_i},
\qquad d_i=d_n\left(\frac{m_q}{m_i}\right)^\eta .
\label{eq:SL_monopole_form_factor}
\end{equation}
Folding the point Coulomb potential and contact density with $F_iF_j$ gives, for $D_i\neq D_j$,
\begin{align}
\mathcal C_{ij}^{\rm Y}(r)
= \frac{1}{r} - \frac{ D_j^2 e^{-D_i r} - D_i^2 e^{-D_j r}}{ (D_j^2-D_i^2)r },\qquad 
\delta_{ij}^{\rm Y}(r)
= \frac{D_i^2D_j^2}{ 4\pi(D_j^2-D_i^2) }
\left( \frac{e^{-D_i r}}{r} - \frac{e^{-D_j r}}{r} \right).
\label{eq:SL_smeared_contact_unequal}
\end{align}
For equal form-factor scales $D_i=D_j=D$, the continuous limits are
\begin{align}
\mathcal C_{D}^{\rm Y}(r)
= \frac{1-e^{-Dr}}{r}
-\frac{D}{2}e^{-Dr},\qquad 
\delta_{D}^{\rm Y}(r) =\frac{D^3}{8\pi}e^{-Dr}.
\label{eq:SL_smeared_kernels_equal}
\end{align}
Thus the same constituent-size prescription regulates both the Coulomb and chromomagnetic interactions, rather than smearing only the contact term.

\paragraph{The GI/CI-type relativized interaction:}
To test whether the meson--baryon incompatibility is caused by the non-relativistic treatment of both the kinetic energy and the short-distance interaction, we use an $S$-wave projection of the Godfrey--Isgur meson and Capstick--Isgur baryon Hamiltonians with a common set of state-dependent interaction parameters~\cite{Godfrey1985,CapstickIsgur:1986}. We retain the relativized central and contact color--spin interactions required for the ground-state benchmark. Tensor and spin--orbit terms are not included.

The meson and baryon Hamiltonians are evaluated directly in their center-of-momentum frames, $\mathbf P_M=\mathbf0$ and $\mathbf P_B=\mathbf0$. For a baryon partition $(ij)k$, we use the physical pair momentum $\mathbf p_{ij}$ and spectator momentum $\mathbf q_k$,
\begin{align}
\mathbf p_{ij}&=\frac{m_j\mathbf p_i-m_i\mathbf p_j}{m_i+m_j},
&
\mathbf q_k&=\frac{m_k(\mathbf p_i+\mathbf p_j)-(m_i+m_j)\mathbf p_k}{m_i+m_j+m_k},
\nonumber\\
\mathbf p_i&=\mathbf p_{ij}+\frac{m_i}{m_i+m_j}\mathbf q_k,
&
\mathbf p_j&=-\mathbf p_{ij}+\frac{m_j}{m_i+m_j}\mathbf q_k,
&
\mathbf p_k&=-\mathbf q_k.
\label{eq:GICI_particle_momenta_from_jacobi}
\end{align}
The conjugate momenta of the scaled Jacobi coordinates are $\boldsymbol\pi_{\mathcal C}=\sqrt2\,\mathbf p_{ij}$ and $\boldsymbol\Pi_{\mathcal C}=\sqrt{3/2}\,\mathbf q_k$. The Gaussian basis is constructed with the scaled variables, whereas the pair interactions use the physical $r_{ij}$, $\mathbf p_{ij}$, and transfer momentum $\mathbf Q_{ij}=\mathbf p'_{ij}-\mathbf p_{ij}$.

The running coupling and the normalized flavor-dependent Gaussian smearing are written as
\begin{align}
\alpha_s(Q_{ij}^2)= \sum_{a=1}^{3}\alpha_a e^{-Q_{ij}^2/(4\gamma_a^2)},
\quad
\rho_{ij}^{\rm GI}(\mathbf x) =\frac{(\sigma_{ij}^{\rm sm})^3}{\pi^{3/2}}
e^{-(\sigma_{ij}^{\rm sm})^2x^2},\quad (\sigma_{ij}^{\rm sm})^2=\frac{\sigma_0^2}{2}\left[ 1+ \left(\frac{4m_im_j}{(m_i+m_j)^2}\right)^4\right]
+s_{\rm GI}^2m_{ij}^2,
\label{eq:GICI_smearing_scales}
\end{align}
and ${\tau_{a,ij}^{-2}} ={\gamma_a^{-2}} +{(\sigma_{ij}^{\rm sm})^{-2}}$.
Before relativistic dressing, the smeared color-singlet Coulomb and contact color--spin kernels are
\begin{align}
G_{ij}(\mathbf Q_{ij})
= -\frac{16\pi}{3Q_{ij}^2}
\sum_{a=1}^{3}
\alpha_a
e^{-Q_{ij}^2/(4\tau_{a,ij}^2)},\qquad
H_{ij}^{CS}(\mathbf Q_{ij})= \frac{8\pi\,\boldsymbol{\sigma}_i\cdot\boldsymbol{\sigma}_j}{9m_im_j}
\sum_{a=1}^{3}
\alpha_a
e^{-Q_{ij}^2/(4\tau_{a,ij}^2)}.
\label{eq:GICI_momentum_kernels}
\end{align}
The same density is used to smear the confinement term,
\begin{equation}
\overline r_{ij}
=
\int d^3\mathbf x\,
\rho_{ij}^{\rm GI}(\mathbf r_{ij}-\mathbf x)
|\mathbf x|.
\end{equation}
With $\varepsilon_i^{(ij)}(p)=\sqrt{p^2+m_i^2}$, the relativistic dressing factors are
\begin{align}
R_{ij}^{C}(p)
= \Bigg[ 1+ \frac{p^2} {\varepsilon_i^{(ij)}(p)\varepsilon_j^{(ij)}(p)} \Bigg]^{1/2+\epsilon_C},
\qquad
R_{ij}^{CS}(p) = \Bigg[ \frac{m_im_j} {\varepsilon_i^{(ij)}(p)\varepsilon_j^{(ij)}(p)} \Bigg]^{1/2+\epsilon_{CS}}.
\label{eq:GICI_relativistic_factors}
\end{align}
We define $\widetilde G_{ij}=R_{ij}^{C}G_{ij}R_{ij}^{C}$ and $\widetilde H_{ij}^{CS}=R_{ij}^{CS}H_{ij}^{CS}R_{ij}^{CS}$. These operators are nonlocal in the pair momentum but remain embedded two-body interactions in a baryon,
\begin{equation}
\langle\mathbf p'_{ij},\mathbf q'_k|
\widetilde V_{ij}
|\mathbf p_{ij},\mathbf q_k\rangle
=
(2\pi)^3
\delta^{(3)}(\mathbf q'_k-\mathbf q_k)
\langle\mathbf p'_{ij}|
\widetilde V_{ij}
|\mathbf p_{ij}\rangle,
\qquad
\widetilde V_{ij}
\in
\{\widetilde G_{ij},\widetilde H_{ij}^{CS}\}.
\label{eq:GICI_embedded_pair_kernels}
\end{equation}
Thus, the spectator momentum and spectator internal state are unchanged by the pair interaction.

The resulting $S$-wave meson and baryon Hamiltonians are
\begin{align}
H_{\rm GI/CI}^{M}\big|_{\mathbf P_M=0}
={}&
\sqrt{\mathbf p_{ij}^2+m_i^2}
+
\sqrt{\mathbf p_{ij}^2+m_j^2}
+
\sigma_{\rm GI}\overline r_{ij}
-D_M
+\widetilde G_{ij}
+\widetilde H_{ij}^{CS},
\label{eq:GICI_meson_hamiltonian}\\
H_{\rm GI/CI}^{B}\big|_{\mathbf P_B=0}
={}&
\sum_{\ell=1}^{3}
\sqrt{\mathbf p_\ell^2+m_\ell^2}
+
f_\Delta\sigma_{\rm GI}
\sum_{i<j}\overline r_{ij}
-\frac32D_B
+
\frac12
\sum_{i<j}
\left(
\widetilde G_{ij}
+
\widetilde H_{ij}^{CS}
\right),
\qquad
f_\Delta=0.5493.
\label{eq:GICI_baryon_hamiltonian}
\end{align}
The factor $1/2$ follows from the color-singlet $qq$ matrix element relative to the $q\bar q$ normalization, and the confinement term uses the Capstick--Isgur $\Delta$ ansatz. No additional center-of-mass kinetic term is subtracted because the Hamiltonians are defined directly on the $\mathbf P_M=\mathbf0$ and $\mathbf P_B=\mathbf0$ subspaces. We repeat the calculation with the common offset $D_M=D_B$ and with independently profiled $D_M$ and $D_B$, reoptimizing all nonlinear parameters for the two prescriptions.

The three running-coupling components are varied through
\begin{equation}
\alpha_a=\zeta_\alpha\alpha_a^{(0)},
\qquad
\gamma_a=\zeta_\gamma\gamma_a^{(0)},
\qquad
a=1,2,3.
\label{eq:GICI_template_scaling}
\end{equation}
The core trajectory fixes $\zeta_\gamma=1$, $\epsilon_C=0$, and $\epsilon_{CS}=-0.168$, and fits $\{\sigma_{\rm GI},\zeta_\alpha,\sigma_0,s_{\rm GI},m_q,m_s,m_c,m_b\}$. The extended trajectory also varies $\zeta_\gamma$, $\epsilon_C$, and $\epsilon_{CS}$.
The spinless-Salpeter kinetic matrix elements are evaluated directly in the Gaussian basis. For the nonlocal pair interactions, each dressing factor is represented by
\begin{equation}
R_{ij}^{X}(p)
\simeq
\sum_{n=1}^{N_R}
c_{n,ij}^{X}
e^{-d_np^2},
\qquad
X\in\{C,CS\},
\qquad
N_R=8.
\label{eq:GICI_numerical_dressing_expansion}
\end{equation}
The exponents $d_n$ are fixed and the coefficients $c_{n,ij}^{X}$ are recomputed for every flavor pair and trial parameter set over $0\leq p\leq15~{\rm GeV}$. This expansion is only a numerical representation of the dressing factors and is distinct from the three Gaussian components used for $\alpha_s(Q^2)$.

\paragraph{Three-quark interactions:} The three-quark interaction is evaluated in first-order perturbation theory using the two-body baryon eigenstate of the corresponding model, 
\begin{equation}
M_{B,X}^{3Q,(f)}
=
\frac{
\langle\Psi_{B,X}^{(2)}|
L_{X}^{3Q,(f)}
|\Psi_{B,X}^{(2)}\rangle
}{
\langle\Psi_{B,X}^{(2)}|\Psi_{B,X}^{(2)}\rangle
},
\qquad
f\in\{\mathrm{contact},\mathrm G,\mathrm Y\}.
\end{equation}
The three-body couplings $A_X^{(f)}$, $B_X^{(f)}$, and $C_X^{(f)}$ are refitted for each baseline, spatial profile, value of $\lambda$, and additive-offset prescription (whether $D_M=D_B$ or not).
The baryon eigenfunction $\Psi_{B,X}^{(2)}$ is obtained from the correspondingly reoptimized two-body Hamiltonian.
We do not assume that the fitted couplings are equal for the different schemes.
We instead test whether the same low-dimensional operator basis reduces the meson and baryon errors for the different two-body wave functions and for both the common- and separate-offset conventions.

The principal two-body ingredients of the seven alternative Hamiltonians and their equal-sector performance are summarized in Table~\ref{tab:cross_model_physics_summary}. This comparison tests whether the improvement from the three-quark interaction remains when the confinement, short-range regularization, kinematics, and additive-offset prescription are changed.

\paragraph{Kinematic regime and semirelativistic benchmarks.}

For the ground states obtained with the parameters of Eq.~\eqref{params}, the rms momentum-to-mass ratio $\xi_i\equiv\sqrt{\langle\mathbf p_i^2\rangle}/m_i$, evaluated in the hadron rest frame, is about 0.25--0.50 for charm quarks, 0.7--1.0 for strange quarks, and 1.1--1.8 for $u,d$ quarks.
Only the charm quarks are moderately non-relativistic; the strange-quark values approach unity and the light-quark values exceed it.
Because $\xi_i$ is defined through the non-relativistic kinetic operator, values above unity measure how far that operator is extrapolated beyond its domain of validity.

The effective-mass description discussed in  Section~\ref{sec:relativistic_quark_models} motivates absorbing the smooth spectral effects of omitted relativistic kinetic terms into $D$ and the fitted constituent masses (see Table~\ref{tab:cornell_lattice} and compare Eqs.~\eqref{params_mass_nonrel} and~\eqref{params_mass_rel}).
The semirelativistic baselines below test whether the improvement due to the three-quark interaction persists when relativistic kinematics is included explicitly.

\begin{figure*}[!b]
\centering
\includegraphics[width=\textwidth]{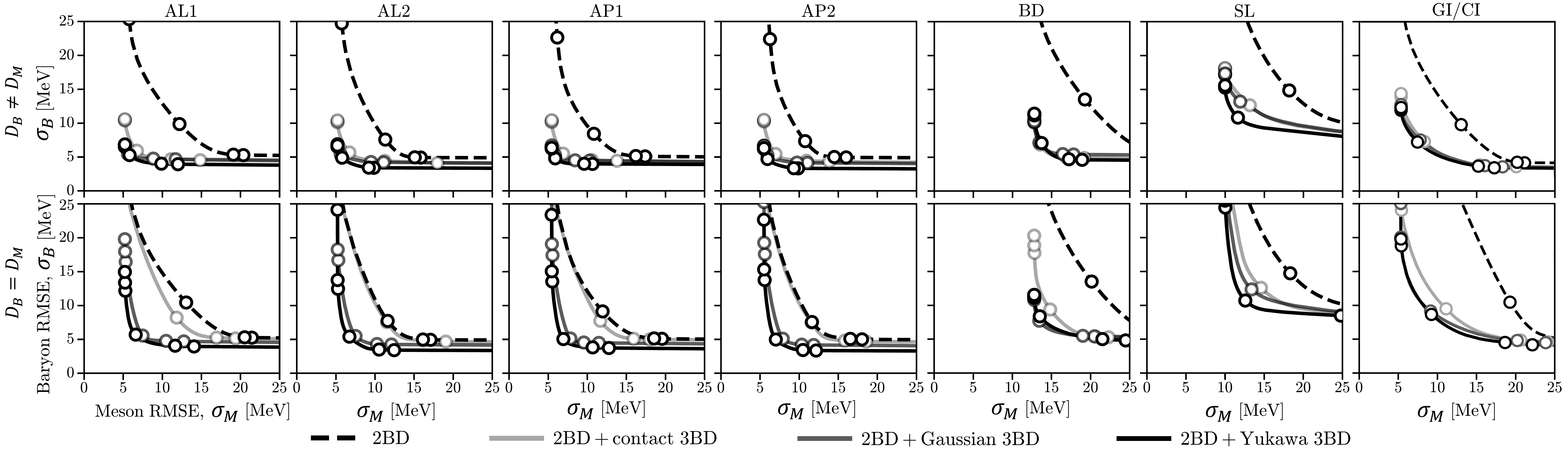}
\caption{\justifying
Meson--baryon trade-off trajectories for the AL1, AL2, AP1, AP2, BD, SL, and GI/CI models in the $u,d,s,c,b$ benchmark.
The upper row allows the meson and baryon constants $D_M$ and $D_B$ to vary independently, whereas the lower row imposes the common-$D$ condition $D_B=D_M$.
Open circles mark the sampled values of $\lambda=0$, 0.01, 0.02, 0.50, 0.98, 0.99, and 1 along each trajectory.
A common 0--25~MeV window is used for both axes to facilitate direct comparison among the models and the two $D$ prescriptions.}
\label{fig:reciprocal_other_models}
\end{figure*}

The SL and GI/CI baselines use square-root kinetic energies, with the parameters refitted within each model.
At the equal-sector point $\lambda=1/2$, the Yukawa three-quark interaction lowers $(\sigma_M,\sigma_B)$ from $(18.33,\,14.73)$ to $(11.62,\,10.84)$~MeV in SL and from $(26.86,\,12.99)$ to $(8.98,\,9.62)$~MeV in GI/CI (Table~\ref{tab:cross_model_physics_summary}).
The improvement therefore persists with square-root kinetic energies in both tested families 
(see Fig.~\ref{fig:reciprocal_other_models}).

\begin{table*}[!t]
\centering
\caption{\justifying
Physical content of the seven alternative OGE-based two-body Hamiltonians and their spectroscopic accuracy at the equal-sector point $\lambda=1/2$.
The 2BD column is obtained under the common color-additive offset constraint.}
\label{tab:cross_model_physics_summary}

\footnotesize

\begin{tblr}{
  width=\textwidth,
  colspec={
    Q[l,wd=1.20cm]
    X[6.2,l]
    Q[c,wd=2.75cm]
    Q[c,wd=3.10cm]
  },
  cells={valign=m},
  row{1}={bg=headgray,halign=c,valign=m},
  cell{1}{2}={}{halign=l},
  hline{2}={1-4}{0.3pt},
  rowsep=2.2pt,
}
\toprule
Model
&
Principal ingredients and physical content
&
\shortstack[c]{2BD\\$(\sigma_M,\sigma_B)$ [MeV]}
&
\shortstack[c]{2BD $+$ Yukawa 3BD\\$(\sigma_M,\sigma_B)$ [MeV]}
\\

AL1
&
non-relativistic kinematics, linear confinement, an unscreened Coulomb interaction, and reduced-mass-dependent Gaussian hyperfine smearing. This model represents a conventional Cornell-like OGE Hamiltonian with a flavor-dependent short-distance resolution~\cite{Semay1994,SilvestreBrac1996}.
&
$(13.05,\,10.44)$
&
$(5.82,\,5.28)$
\\  \hline

AL2
&
The AL1 dynamics supplemented by the common factor $g(r)=1-e^{-r/r_c}$, which regulates both the Coulomb and hyperfine interactions at the origin. It tests whether an explicit common ultraviolet softening of the pair interaction removes the meson--baryon residual~\cite{Semay1994,SilvestreBrac1996}.
&
$(11.62,\,7.73)$
&
$(5.79,\,4.90)$
\\ \hline

AP1
&
non-relativistic kinematics with the softer confinement law $r^{2/3}$, an unscreened Coulomb interaction, and reduced-mass-dependent Gaussian hyperfine smearing. Relative to AL1, it primarily tests sensitivity to the long-distance confinement profile~\cite{Semay1994,SilvestreBrac1996}.
&
$(11.96,\,9.13)$
&
$(5.90,\,4.81)$
\\ \hline

AP2
&
The $r^{2/3}$ confinement of AP1 combined with the common short-distance regulator $1-e^{-r/r_c}$. It simultaneously changes the long-distance confinement law and the ultraviolet regularization of the Coulomb and hyperfine interactions~\cite{Semay1994,SilvestreBrac1996}.
&
$(11.63,\,7.56)$
&
$(5.95,\,4.71)$
\\ \hline

BD
&
non-relativistic kinematics and linear confinement, with the Coulomb and exponential hyperfine strengths tied to the same coefficient $\kappa_{\rm BD}$ and with a flavor-independent hyperfine range. It is a comparatively constrained OGE realization; the historical non-pairwise baryon correction is excluded here~\cite{Bhaduri1981}.
&
$(20.14,\,13.51)$
&
$(13.72,\,7.09)$
\\ \hline

SL
&
Spinless-Salpeter kinematics and a finite constituent size represented by mass-dependent monopole form factors. The same finite-size prescription smears both the Coulomb and chromomagnetic interactions, testing semirelativistic motion together with constituent-size regularization~\cite{Stanley1980}.
&
$(18.33,\,14.73)$
&
$(11.62,\,10.84)$
\\ \hline

GI/CI
&
Spinless-Salpeter kinematics, a running coupling, mass-dependent Gaussian smearing, smeared confinement, and nonlocal momentum dressing of the Coulomb and color-spin vertices, together with the CI $\Delta$-ansatz for baryon confinement. It is the most complete relativized pairwise benchmark considered here~\cite{Godfrey1985,CapstickIsgur:1986}.
&
$(26.86,\,12.99)$
&
$(8.98,\,9.62)$
\\

\bottomrule
\end{tblr}
\end{table*}

\begin{table*}[!t]
\centering
\caption{\justifying
Ground-state meson and baryon spectra at the meson-optimized endpoint $\lambda=1$ for eight quark-model Hamiltonians.
Each calculated mass is followed by the signed deviation
$M-M_{\exp}$ in parentheses.
The three-quark interaction does not act in the meson sector.
For the baryon panels, the meson-optimized interaction parameters are retained; for AL1, AL2, AP1, AP2, BD, SL, and GI/CI, the baryonic constant $D_B$ is fitted independently of the mesonic value $D_M$ ($D_B\neq{D}_M$), whereas the $\mathrm{S}^2\mathrm{AJ}$ calculation retains the common choice $D_B=D_M\equiv{D}$.
All masses and RMSEs are in MeV.
The fits use either the bounds in Eq.~\eqref{eq:SAJ_physical_bounds} or model-specific parameter bounds chosen to enforce the same physically motivated domain.}
\label{tab:cross_model_lambda1_spectra}
\renewcommand{\pdev}[1]{\textcolor{devred}{\ensuremath{(#1)}}}
\renewcommand{\ndev}[1]{\textcolor{devgreen}{\ensuremath{(#1)}}}
\scriptsize

\textbf{(a) Ground-state meson spectrum}

\vspace{2pt}

\resizebox{\textwidth}{!}{%
\begin{tblr}{
  colspec={
    Q[l,wd=0.5cm]
    Q[c,wd=0.88cm]
    *{8}{Q[c,wd=2.05cm]}
  },
  row{1} = {bg=headgray, halign=c, valign=m},
  cell{1}{1} = {}{halign=l},
  hline{2} = {1-10}{0.3pt},
  rowsep = 0.7pt,
}
\toprule
State
& $M_{\rm exp}$
& \shortstack[c]{AL1\\($\sigma_M=5.21$)}
& \shortstack[c]{AL2\\($\sigma_M=5.21$)}
& \shortstack[c]{AP1\\($\sigma_M=5.44$)}
& \shortstack[c]{AP2\\($\sigma_M=5.49$)}
& \shortstack[c]{BD\\($\sigma_M=12.79$)}
& \shortstack[c]{SL\\($\sigma_M=9.98$)}
& \shortstack[c]{GI/CI\\($\sigma_M=5.33$)}
& \shortstack[c]{$\mathrm{S^2AJ}$\\($\sigma_M=5.21$)} \\

$\pi$
& 139.57
& \massdevinline{141.67}{\pdev{+2.10}}
& \massdevinline{141.60}{\pdev{+2.03}}
& \massdevinline{142.46}{\pdev{+2.89}}
& \massdevinline{142.65}{\pdev{+3.08}}
& \massdevinline{140.95}{\pdev{+1.38}}
& \massdevinline{141.52}{\pdev{+1.95}}
& \massdevinline{141.45}{\pdev{+1.88}}
& \massdevinline{140.69}{\pdev{+1.12}} \\

$\rho$
& 775.11
& \massdevinline{767.42}{\ndev{-7.69}}
& \massdevinline{767.26}{\ndev{-7.85}}
& \massdevinline{765.83}{\ndev{-9.28}}
& \massdevinline{765.60}{\ndev{-9.51}}
& \massdevinline{775.29}{\pdev{+0.18}}
& \massdevinline{782.99}{\pdev{+7.88}}
& \massdevinline{767.23}{\ndev{-7.88}}
& \massdevinline{769.96}{\ndev{-5.15}} \\

$K$
& 497.61
& \massdevinline{492.36}{\ndev{-5.25}}
& \massdevinline{492.62}{\ndev{-4.99}}
& \massdevinline{490.62}{\ndev{-6.99}}
& \massdevinline{490.17}{\ndev{-7.44}}
& \massdevinline{494.75}{\ndev{-2.86}}
& \massdevinline{495.73}{\ndev{-1.88}}
& \massdevinline{490.92}{\ndev{-6.69}}
& \massdevinline{494.54}{\ndev{-3.07}} \\

$K^*$
& 895.81
& \massdevinline{903.47}{\pdev{+7.66}}
& \massdevinline{903.50}{\pdev{+7.69}}
& \massdevinline{903.23}{\pdev{+7.42}}
& \massdevinline{903.04}{\pdev{+7.23}}
& \massdevinline{907.16}{\pdev{+11.35}}
& \massdevinline{903.88}{\pdev{+8.07}}
& \massdevinline{901.82}{\pdev{+6.01}}
& \massdevinline{903.23}{\pdev{+7.42}} \\

$\phi$
& 1019.46
& \massdevinline{1027.96}{\pdev{+8.50}}
& \massdevinline{1027.97}{\pdev{+8.51}}
& \massdevinline{1028.93}{\pdev{+9.47}}
& \massdevinline{1028.99}{\pdev{+9.53}}
& \massdevinline{1026.34}{\pdev{+6.88}}
& \massdevinline{1021.77}{\pdev{+2.31}}
& \massdevinline{1021.34}{\pdev{+1.88}}
& \massdevinline{1025.99}{\pdev{+6.53}} \\

$D$
& 1864.84
& \massdevinline{1869.69}{\pdev{+4.85}}
& \massdevinline{1870.55}{\pdev{+5.71}}
& \massdevinline{1871.91}{\pdev{+7.07}}
& \massdevinline{1872.18}{\pdev{+7.34}}
& \massdevinline{1848.53}{\ndev{-16.31}}
& \massdevinline{1849.65}{\ndev{-15.19}}
& \massdevinline{1869.09}{\pdev{+4.25}}
& \massdevinline{1865.81}{\pdev{+0.97}} \\

$D^*$
& 2010.26
& \massdevinline{2003.70}{\ndev{-6.56}}
& \massdevinline{2003.66}{\ndev{-6.60}}
& \massdevinline{2004.14}{\ndev{-6.12}}
& \massdevinline{2004.21}{\ndev{-6.05}}
& \massdevinline{2004.26}{\ndev{-6.00}}
& \massdevinline{1991.86}{\ndev{-18.40}}
& \massdevinline{2018.08}{\pdev{+7.82}}
& \massdevinline{2003.84}{\ndev{-6.42}} \\

$D_s$
& 1968.30
& \massdevinline{1967.87}{\ndev{-0.43}}
& \massdevinline{1966.89}{\ndev{-1.41}}
& \massdevinline{1968.24}{\ndev{-0.06}}
& \massdevinline{1968.73}{\pdev{+0.43}}
& \massdevinline{1970.01}{\pdev{+1.71}}
& \massdevinline{1971.24}{\pdev{+2.94}}
& \massdevinline{1977.29}{\pdev{+8.99}}
& \massdevinline{1968.65}{\pdev{+0.35}} \\

$D_s^*$
& 2112.10
& \massdevinline{2102.20}{\ndev{-9.90}}
& \massdevinline{2102.45}{\ndev{-9.65}}
& \massdevinline{2103.53}{\ndev{-8.57}}
& \massdevinline{2103.92}{\ndev{-8.18}}
& \massdevinline{2095.46}{\ndev{-16.64}}
& \massdevinline{2094.34}{\ndev{-17.76}}
& \massdevinline{2112.53}{\pdev{+0.43}}
& \massdevinline{2101.12}{\ndev{-10.98}} \\

$\eta_c$
& 2984.10
& \massdevinline{2988.52}{\pdev{+4.42}}
& \massdevinline{2987.16}{\pdev{+3.06}}
& \massdevinline{2986.88}{\pdev{+2.78}}
& \massdevinline{2986.77}{\pdev{+2.67}}
& \massdevinline{3011.69}{\pdev{+27.59}}
& \massdevinline{3004.99}{\pdev{+20.89}}
& \massdevinline{2980.93}{\ndev{-3.17}}
& \massdevinline{2992.58}{\pdev{+8.48}} \\

$J/\psi$
& 3096.90
& \massdevinline{3098.56}{\pdev{+1.66}}
& \massdevinline{3099.89}{\pdev{+2.99}}
& \massdevinline{3098.41}{\pdev{+1.51}}
& \massdevinline{3098.28}{\pdev{+1.38}}
& \massdevinline{3082.60}{\ndev{-14.30}}
& \massdevinline{3097.92}{\pdev{+1.02}}
& \massdevinline{3092.14}{\ndev{-4.76}}
& \massdevinline{3096.64}{\ndev{-0.26}} \\

$B$
& 5279.72
& \massdevinline{5283.71}{\pdev{+3.99}}
& \massdevinline{5284.01}{\pdev{+4.29}}
& \massdevinline{5284.00}{\pdev{+4.28}}
& \massdevinline{5283.94}{\pdev{+4.22}}
& \massdevinline{5278.98}{\ndev{-0.74}}
& \massdevinline{5276.11}{\ndev{-3.61}}
& \massdevinline{5275.10}{\ndev{-4.62}}
& \massdevinline{5282.35}{\pdev{+2.63}} \\

$B^*$
& 5325.20
& \massdevinline{5330.49}{\pdev{+5.29}}
& \massdevinline{5329.82}{\pdev{+4.62}}
& \massdevinline{5329.41}{\pdev{+4.21}}
& \massdevinline{5329.31}{\pdev{+4.11}}
& \massdevinline{5337.69}{\pdev{+12.49}}
& \massdevinline{5332.87}{\pdev{+7.67}}
& \massdevinline{5326.20}{\pdev{+1.00}}
& \massdevinline{5332.46}{\pdev{+7.26}} \\

$B_s$
& 5366.77
& \massdevinline{5361.27}{\ndev{-5.50}}
& \massdevinline{5361.35}{\ndev{-5.42}}
& \massdevinline{5361.67}{\ndev{-5.10}}
& \massdevinline{5361.84}{\ndev{-4.93}}
& \massdevinline{5360.74}{\ndev{-6.03}}
& \massdevinline{5365.57}{\ndev{-1.20}}
& \massdevinline{5363.34}{\ndev{-3.43}}
& \massdevinline{5360.78}{\ndev{-5.99}} \\

$B_s^*$
& 5415.40
& \massdevinline{5413.08}{\ndev{-2.32}}
& \massdevinline{5413.10}{\ndev{-2.30}}
& \massdevinline{5413.13}{\ndev{-2.27}}
& \massdevinline{5413.29}{\ndev{-2.11}}
& \massdevinline{5410.96}{\ndev{-4.44}}
& \massdevinline{5421.57}{\pdev{+6.17}}
& \massdevinline{5412.58}{\ndev{-2.82}}
& \massdevinline{5413.61}{\ndev{-1.79}} \\

$B_c$
& 6274.47
& \massdevinline{6274.33}{\ndev{-0.14}}
& \massdevinline{6274.35}{\ndev{-0.12}}
& \massdevinline{6273.13}{\ndev{-1.34}}
& \massdevinline{6272.22}{\ndev{-2.25}}
& \massdevinline{6287.76}{\pdev{+13.29}}
& \massdevinline{6282.46}{\pdev{+7.99}}
& \massdevinline{6266.28}{\ndev{-8.19}}
& \massdevinline{6274.44}{\ndev{-0.03}} \\

$\eta_b$
& 9398.70
& \massdevinline{9398.21}{\ndev{-0.49}}
& \massdevinline{9399.22}{\pdev{+0.52}}
& \massdevinline{9400.39}{\pdev{+1.69}}
& \massdevinline{9401.12}{\pdev{+2.42}}
& \massdevinline{9414.84}{\pdev{+16.14}}
& \massdevinline{9386.94}{\ndev{-11.76}}
& \massdevinline{9406.12}{\pdev{+7.42}}
& \massdevinline{9395.10}{\ndev{-3.60}} \\

$\Upsilon$
& 9460.40
& \massdevinline{9460.20}{\ndev{-0.20}}
& \massdevinline{9459.33}{\ndev{-1.07}}
& \massdevinline{9458.81}{\ndev{-1.59}}
& \massdevinline{9458.46}{\ndev{-1.94}}
& \massdevinline{9436.71}{\ndev{-23.69}}
& \massdevinline{9463.32}{\pdev{+2.92}}
& \massdevinline{9462.27}{\pdev{+1.87}}
& \massdevinline{9462.92}{\pdev{+2.52}} \\
\bottomrule
\end{tblr}}

\vspace{6pt}

\textbf{(b) Ground-state baryon spectrum}

\vspace{2pt}

\resizebox{\textwidth}{!}{%
\begin{tblr}{
  colspec={
    Q[l,wd=0.5cm]
    Q[c,wd=0.88cm]
    *{8}{Q[c,wd=2.05cm]}
  },
  row{1-2} = {bg=headgray, halign=c, valign=m},
  cell{1}{1} = {r=2}{halign=l, valign=m},
  cell{1}{2} = {r=2}{halign=c, valign=m},
  cell{1}{3} = {c=2}{halign=c},
  cell{1}{5} = {c=2}{halign=c},
  cell{1}{7} = {c=2}{halign=c},
  cell{1}{9} = {c=2}{halign=c},
  hline{2} = {3-10}{0.2pt},
  hline{3} = {1-10}{0.3pt},
  rowsep = 0.7pt,
}
\toprule
State
& $M_{\rm exp}$
& \SetCell[c=2]{c}{AL1}
& 
& \SetCell[c=2]{c}{AL2}
& 
& \SetCell[c=2]{c}{AP1}
& 
& \SetCell[c=2]{c}{AP2}
&  \\

& 
& \shortstack[c]{2BD\\($\sigma_B=50.02$)}
& \shortstack[c]{Yukawa\\($\sigma_B=6.86$)}
& \shortstack[c]{2BD\\($\sigma_B=39.95$)}
& \shortstack[c]{Yukawa\\($\sigma_B=6.71$)}
& \shortstack[c]{2BD\\($\sigma_B=43.49$)}
& \shortstack[c]{Yukawa\\($\sigma_B=6.35$)}
& \shortstack[c]{2BD\\($\sigma_B=45.01$)}
& \shortstack[c]{Yukawa\\($\sigma_B=6.30$)} \\

$N$
& 938.27
& \massdevinline{1058.64}{\pdev{+120.37}}
& \massdevinline{933.29}{\ndev{-4.98}}
& \massdevinline{1035.82}{\pdev{+97.55}}
& \massdevinline{934.73}{\ndev{-3.54}}
& \massdevinline{1045.69}{\pdev{+107.42}}
& \massdevinline{934.54}{\ndev{-3.73}}
& \massdevinline{1049.33}{\pdev{+111.06}}
& \massdevinline{934.46}{\ndev{-3.81}} \\

$\Lambda$
& 1115.68
& \massdevinline{1203.15}{\pdev{+87.47}}
& \massdevinline{1111.04}{\ndev{-4.64}}
& \massdevinline{1183.83}{\pdev{+68.15}}
& \massdevinline{1115.59}{\ndev{-0.09}}
& \massdevinline{1193.07}{\pdev{+77.39}}
& \massdevinline{1114.10}{\ndev{-1.58}}
& \massdevinline{1196.23}{\pdev{+80.55}}
& \massdevinline{1113.40}{\ndev{-2.28}} \\

$\Sigma$
& 1197.45
& \massdevinline{1245.91}{\pdev{+48.46}}
& \massdevinline{1206.24}{\pdev{+8.79}}
& \massdevinline{1237.10}{\pdev{+39.65}}
& \massdevinline{1202.75}{\pdev{+5.30}}
& \massdevinline{1240.14}{\pdev{+42.69}}
& \massdevinline{1203.89}{\pdev{+6.44}}
& \massdevinline{1241.34}{\pdev{+43.89}}
& \massdevinline{1204.36}{\pdev{+6.91}} \\

$\Delta$
& 1232.00
& \massdevinline{1255.26}{\pdev{+23.26}}
& \massdevinline{1226.92}{\ndev{-5.08}}
& \massdevinline{1258.40}{\pdev{+26.40}}
& \massdevinline{1226.52}{\ndev{-5.48}}
& \massdevinline{1247.49}{\pdev{+15.49}}
& \massdevinline{1228.19}{\ndev{-3.81}}
& \massdevinline{1245.55}{\pdev{+13.55}}
& \massdevinline{1228.70}{\ndev{-3.30}} \\

$\Xi$
& 1321.71
& \massdevinline{1365.32}{\pdev{+43.61}}
& \massdevinline{1337.54}{\pdev{+15.83}}
& \massdevinline{1355.30}{\pdev{+33.59}}
& \massdevinline{1338.61}{\pdev{+16.90}}
& \massdevinline{1360.74}{\pdev{+39.03}}
& \massdevinline{1338.03}{\pdev{+16.32}}
& \massdevinline{1362.45}{\pdev{+40.74}}
& \massdevinline{1337.71}{\pdev{+16.00}} \\

$\Sigma^*$
& 1383.70
& \massdevinline{1387.56}{\pdev{+3.86}}
& \massdevinline{1389.85}{\pdev{+6.15}}
& \massdevinline{1391.95}{\pdev{+8.25}}
& \massdevinline{1390.65}{\pdev{+6.95}}
& \massdevinline{1383.77}{\pdev{+0.07}}
& \massdevinline{1388.90}{\pdev{+5.20}}
& \massdevinline{1381.91}{\ndev{-1.79}}
& \massdevinline{1388.55}{\pdev{+4.85}} \\

$\Xi^*$
& 1535.00
& \massdevinline{1514.52}{\ndev{-20.48}}
& \massdevinline{1538.21}{\pdev{+3.21}}
& \massdevinline{1519.92}{\ndev{-15.08}}
& \massdevinline{1539.37}{\pdev{+4.37}}
& \massdevinline{1514.46}{\ndev{-20.54}}
& \massdevinline{1536.88}{\pdev{+1.88}}
& \massdevinline{1512.81}{\ndev{-22.19}}
& \massdevinline{1536.19}{\pdev{+1.19}} \\

$\Omega$
& 1672.45
& \massdevinline{1636.76}{\ndev{-35.69}}
& \massdevinline{1675.22}{\pdev{+2.77}}
& \massdevinline{1643.07}{\ndev{-29.38}}
& \massdevinline{1675.89}{\pdev{+3.44}}
& \massdevinline{1640.12}{\ndev{-32.33}}
& \massdevinline{1674.39}{\pdev{+1.94}}
& \massdevinline{1638.77}{\ndev{-33.68}}
& \massdevinline{1673.81}{\pdev{+1.36}} \\

$\Lambda_c$
& 2286.46
& \massdevinline{2316.48}{\pdev{+30.02}}
& \massdevinline{2279.08}{\ndev{-7.38}}
& \massdevinline{2302.43}{\pdev{+15.97}}
& \massdevinline{2277.04}{\ndev{-9.42}}
& \massdevinline{2310.90}{\pdev{+24.44}}
& \massdevinline{2277.93}{\ndev{-8.53}}
& \massdevinline{2313.67}{\pdev{+27.21}}
& \massdevinline{2278.32}{\ndev{-8.14}} \\

$\Sigma_c$
& 2453.97
& \massdevinline{2426.20}{\ndev{-27.77}}
& \massdevinline{2445.12}{\ndev{-8.85}}
& \massdevinline{2432.11}{\ndev{-21.86}}
& \massdevinline{2444.03}{\ndev{-9.94}}
& \massdevinline{2430.22}{\ndev{-23.75}}
& \massdevinline{2445.30}{\ndev{-8.67}}
& \massdevinline{2429.37}{\ndev{-24.60}}
& \massdevinline{2445.66}{\ndev{-8.31}} \\

$\Sigma_c^*$
& 2518.48
& \massdevinline{2481.64}{\ndev{-36.84}}
& \massdevinline{2509.19}{\ndev{-9.29}}
& \massdevinline{2489.32}{\ndev{-29.16}}
& \massdevinline{2509.22}{\ndev{-9.26}}
& \massdevinline{2484.10}{\ndev{-34.38}}
& \massdevinline{2508.05}{\ndev{-10.43}}
& \massdevinline{2482.47}{\ndev{-36.01}}
& \massdevinline{2507.83}{\ndev{-10.65}} \\

$\Omega_c$
& 2695.20
& \massdevinline{2650.83}{\ndev{-44.37}}
& \massdevinline{2696.71}{\pdev{+1.51}}
& \massdevinline{2657.59}{\ndev{-37.61}}
& \massdevinline{2695.83}{\pdev{+0.63}}
& \massdevinline{2658.90}{\ndev{-36.30}}
& \massdevinline{2698.04}{\pdev{+2.84}}
& \massdevinline{2658.73}{\ndev{-36.47}}
& \massdevinline{2698.62}{\pdev{+3.42}} \\

$\Omega_c^*$
& 2765.90
& \massdevinline{2706.19}{\ndev{-59.71}}
& \massdevinline{2756.66}{\ndev{-9.24}}
& \massdevinline{2715.81}{\ndev{-50.09}}
& \massdevinline{2757.58}{\ndev{-8.32}}
& \massdevinline{2714.95}{\ndev{-50.95}}
& \massdevinline{2758.20}{\ndev{-7.70}}
& \massdevinline{2714.03}{\ndev{-51.87}}
& \massdevinline{2758.23}{\ndev{-7.67}} \\

$\Xi_{cc}$
& 3621.20
& \massdevinline{3572.64}{\ndev{-48.56}}
& \massdevinline{3621.79}{\pdev{+0.59}}
& \massdevinline{3581.67}{\ndev{-39.53}}
& \massdevinline{3622.34}{\pdev{+1.14}}
& \massdevinline{3582.35}{\ndev{-38.85}}
& \massdevinline{3624.17}{\pdev{+2.97}}
& \massdevinline{3582.13}{\ndev{-39.07}}
& \massdevinline{3624.75}{\pdev{+3.55}} \\

$\Lambda_b$
& 5619.60
& \massdevinline{5648.62}{\pdev{+29.02}}
& \massdevinline{5626.85}{\pdev{+7.25}}
& \massdevinline{5636.27}{\pdev{+16.67}}
& \massdevinline{5621.92}{\pdev{+2.32}}
& \massdevinline{5643.43}{\pdev{+23.83}}
& \massdevinline{5622.59}{\pdev{+2.99}}
& \massdevinline{5645.79}{\pdev{+26.19}}
& \massdevinline{5623.10}{\pdev{+3.50}} \\

$\Sigma_b$
& 5815.64
& \massdevinline{5785.42}{\ndev{-30.22}}
& \massdevinline{5816.27}{\pdev{+0.63}}
& \massdevinline{5794.48}{\ndev{-21.16}}
& \massdevinline{5816.28}{\pdev{+0.64}}
& \massdevinline{5790.36}{\ndev{-25.28}}
& \massdevinline{5815.65}{\pdev{+0.01}}
& \massdevinline{5788.77}{\ndev{-26.87}}
& \massdevinline{5815.46}{\ndev{-0.18}} \\

$\Sigma_b^*$
& 5834.74
& \massdevinline{5806.26}{\ndev{-28.48}}
& \massdevinline{5839.72}{\pdev{+4.98}}
& \massdevinline{5815.33}{\ndev{-19.41}}
& \massdevinline{5839.65}{\pdev{+4.91}}
& \massdevinline{5810.12}{\ndev{-24.62}}
& \massdevinline{5838.11}{\pdev{+3.37}}
& \massdevinline{5808.31}{\ndev{-26.43}}
& \massdevinline{5837.74}{\pdev{+3.00}} \\

$\Omega_b$
& 6046.10
& \massdevinline{5992.14}{\ndev{-53.96}}
& \massdevinline{6043.84}{\ndev{-2.26}}
& \massdevinline{6003.15}{\ndev{-42.95}}
& \massdevinline{6045.55}{\ndev{-0.55}}
& \massdevinline{6002.76}{\ndev{-43.34}}
& \massdevinline{6046.58}{\pdev{+0.48}}
& \massdevinline{6001.90}{\ndev{-44.20}}
& \massdevinline{6046.69}{\pdev{+0.59}} \\
\bottomrule
\end{tblr}}

\vspace{5pt}

\resizebox{\textwidth}{!}{%
\begin{tblr}{
  colspec={
    Q[l,wd=0.5cm]
    Q[c,wd=0.88cm]
    *{8}{Q[c,wd=2.05cm]}
  },
  row{1-2} = {bg=headgray, halign=c, valign=m},
  cell{1}{1} = {r=2}{halign=l, valign=m},
  cell{1}{2} = {r=2}{halign=c, valign=m},
  cell{1}{3} = {c=2}{halign=c},
  cell{1}{5} = {c=2}{halign=c},
  cell{1}{7} = {c=2}{halign=c},
  cell{1}{9} = {c=2}{halign=c},
  hline{2} = {3-10}{0.2pt},
  hline{3} = {1-10}{0.3pt},
  rowsep = 0.7pt,
}
\toprule
State
& $M_{\rm exp}$
& \SetCell[c=2]{c}{BD}
& 
& \SetCell[c=2]{c}{SL}
& 
& \SetCell[c=2]{c}{GI/CI}
& 
& \SetCell[c=2]{c}{$\mathrm{S^2AJ}$}
&  \\

& 
& \shortstack[c]{2BD\\($\sigma_B=31.21$)}
& \shortstack[c]{Yukawa\\($\sigma_B=11.42$)}
& \shortstack[c]{2BD\\($\sigma_B=45.02$)}
& \shortstack[c]{Yukawa\\($\sigma_B=15.58$)}
& \shortstack[c]{2BD\\($\sigma_B=31.47$)}
& \shortstack[c]{Yukawa\\($\sigma_B=10.49$)}
& \shortstack[c]{2BD\\($\sigma_B=83.77$)}
& \shortstack[c]{Yukawa\\($\sigma_B=7.34$)} \\

$N$
& 938.27
& \massdevinline{986.37}{\pdev{+48.10}}
& \massdevinline{941.73}{\pdev{+3.46}}
& \massdevinline{1050.25}{\pdev{+111.98}}
& \massdevinline{935.03}{\ndev{-3.24}}
& \massdevinline{1007.95}{\pdev{+69.68}}
& \massdevinline{940.42}{\pdev{+2.15}}
& \massdevinline{1134.21}{\pdev{+195.94}}
& \massdevinline{939.56}{\pdev{+1.29}} \\

$\Lambda$
& 1115.68
& \massdevinline{1141.28}{\pdev{+25.60}}
& \massdevinline{1121.44}{\pdev{+5.76}}
& \massdevinline{1187.93}{\pdev{+72.25}}
& \massdevinline{1137.44}{\pdev{+21.76}}
& \massdevinline{1160.77}{\pdev{+45.09}}
& \massdevinline{1124.78}{\pdev{+9.10}}
& \massdevinline{1274.63}{\pdev{+158.95}}
& \massdevinline{1109.35}{\ndev{-6.33}} \\

$\Sigma$
& 1197.45
& \massdevinline{1214.91}{\pdev{+17.46}}
& \massdevinline{1188.38}{\ndev{-9.07}}
& \massdevinline{1250.30}{\pdev{+52.85}}
& \massdevinline{1204.89}{\pdev{+7.44}}
& \massdevinline{1227.95}{\pdev{+30.50}}
& \massdevinline{1197.87}{\pdev{+0.42}}
& \massdevinline{1311.57}{\pdev{+114.12}}
& \massdevinline{1203.76}{\pdev{+6.31}} \\

$\Delta$
& 1232.00
& \massdevinline{1296.47}{\pdev{+64.47}}
& \massdevinline{1221.28}{\ndev{-10.72}}
& \massdevinline{1272.17}{\pdev{+40.17}}
& \massdevinline{1233.91}{\pdev{+1.91}}
& \massdevinline{1277.44}{\pdev{+45.44}}
& \massdevinline{1229.71}{\ndev{-2.29}}
& \massdevinline{1321.52}{\pdev{+89.52}}
& \massdevinline{1234.39}{\pdev{+2.39}} \\

$\Xi$
& 1321.71
& \massdevinline{1329.45}{\pdev{+7.74}}
& \massdevinline{1328.94}{\pdev{+7.23}}
& \massdevinline{1360.00}{\pdev{+38.29}}
& \massdevinline{1358.35}{\pdev{+36.64}}
& \massdevinline{1345.06}{\pdev{+23.35}}
& \massdevinline{1340.36}{\pdev{+18.65}}
& \massdevinline{1429.98}{\pdev{+108.27}}
& \massdevinline{1328.38}{\pdev{+6.67}} \\

$\Sigma^*$
& 1383.70
& \massdevinline{1425.52}{\pdev{+41.82}}
& \massdevinline{1394.78}{\pdev{+11.08}}
& \massdevinline{1393.67}{\pdev{+9.97}}
& \massdevinline{1389.27}{\pdev{+5.57}}
& \massdevinline{1409.83}{\pdev{+26.13}}
& \massdevinline{1397.67}{\pdev{+13.97}}
& \massdevinline{1450.78}{\pdev{+67.08}}
& \massdevinline{1390.24}{\pdev{+6.54}} \\

$\Xi^*$
& 1535.00
& \massdevinline{1548.38}{\pdev{+13.38}}
& \massdevinline{1548.37}{\pdev{+13.37}}
& \massdevinline{1513.68}{\ndev{-21.32}}
& \massdevinline{1536.32}{\pdev{+1.32}}
& \massdevinline{1536.67}{\pdev{+1.67}}
& \massdevinline{1549.43}{\pdev{+14.43}}
& \massdevinline{1575.28}{\pdev{+40.28}}
& \massdevinline{1534.08}{\ndev{-0.92}} \\

$\Omega$
& 1672.45
& \massdevinline{1666.14}{\ndev{-6.31}}
& \massdevinline{1685.78}{\pdev{+13.33}}
& \massdevinline{1632.56}{\ndev{-39.89}}
& \massdevinline{1672.75}{\pdev{+0.30}}
& \massdevinline{1658.47}{\ndev{-13.98}}
& \massdevinline{1686.04}{\pdev{+13.59}}
& \massdevinline{1695.49}{\pdev{+23.04}}
& \massdevinline{1668.50}{\ndev{-3.95}} \\

$\Lambda_c$
& 2286.46
& \massdevinline{2264.60}{\ndev{-21.86}}
& \massdevinline{2271.94}{\ndev{-14.52}}
& \massdevinline{2283.43}{\ndev{-3.03}}
& \massdevinline{2265.47}{\ndev{-20.99}}
& \massdevinline{2284.15}{\ndev{-2.31}}
& \massdevinline{2276.65}{\ndev{-9.81}}
& \massdevinline{2385.53}{\pdev{+99.07}}
& \massdevinline{2278.95}{\ndev{-7.51}} \\

$\Sigma_c$
& 2453.97
& \massdevinline{2432.18}{\ndev{-21.79}}
& \massdevinline{2431.66}{\ndev{-22.31}}
& \massdevinline{2427.37}{\ndev{-26.60}}
& \massdevinline{2428.98}{\ndev{-24.99}}
& \massdevinline{2440.38}{\ndev{-13.59}}
& \massdevinline{2441.47}{\ndev{-12.50}}
& \massdevinline{2485.27}{\pdev{+31.30}}
& \massdevinline{2442.44}{\ndev{-11.53}} \\

$\Sigma_c^*$
& 2518.48
& \massdevinline{2516.27}{\ndev{-2.21}}
& \massdevinline{2517.65}{\ndev{-0.83}}
& \massdevinline{2481.85}{\ndev{-36.63}}
& \massdevinline{2495.84}{\ndev{-22.64}}
& \massdevinline{2506.26}{\ndev{-12.22}}
& \massdevinline{2513.52}{\ndev{-4.96}}
& \massdevinline{2543.17}{\pdev{+24.69}}
& \massdevinline{2509.10}{\ndev{-9.38}} \\

$\Omega_c$
& 2695.20
& \massdevinline{2660.61}{\ndev{-34.59}}
& \massdevinline{2692.46}{\ndev{-2.74}}
& \massdevinline{2658.80}{\ndev{-36.40}}
& \massdevinline{2705.51}{\pdev{+10.31}}
& \massdevinline{2668.57}{\ndev{-26.63}}
& \massdevinline{2701.67}{\pdev{+6.47}}
& \massdevinline{2707.97}{\pdev{+12.77}}
& \massdevinline{2689.91}{\ndev{-5.29}} \\

$\Omega_c^*$
& 2765.90
& \massdevinline{2729.35}{\ndev{-36.55}}
& \massdevinline{2762.57}{\ndev{-3.33}}
& \massdevinline{2705.75}{\ndev{-60.15}}
& \massdevinline{2755.25}{\ndev{-10.65}}
& \massdevinline{2730.46}{\ndev{-35.44}}
& \massdevinline{2765.86}{\ndev{-0.04}}
& \massdevinline{2763.94}{\ndev{-1.96}}
& \massdevinline{2751.17}{\ndev{-14.73}} \\

$\Xi_{cc}$
& 3621.20
& \massdevinline{3565.66}{\ndev{-55.54}}
& \massdevinline{3601.20}{\ndev{-20.00}}
& \massdevinline{3572.10}{\ndev{-49.10}}
& \massdevinline{3621.12}{\ndev{-0.08}}
& \massdevinline{3587.62}{\ndev{-33.58}}
& \massdevinline{3617.10}{\ndev{-4.10}}
& \massdevinline{3628.62}{\pdev{+7.42}}
& \massdevinline{3614.55}{\ndev{-6.65}} \\

$\Lambda_b$
& 5619.60
& \massdevinline{5606.80}{\ndev{-12.80}}
& \massdevinline{5621.91}{\pdev{+2.31}}
& \massdevinline{5621.81}{\pdev{+2.21}}
& \massdevinline{5601.45}{\ndev{-18.15}}
& \massdevinline{5606.16}{\ndev{-13.44}}
& \massdevinline{5605.19}{\ndev{-14.41}}
& \massdevinline{5717.86}{\pdev{+98.26}}
& \massdevinline{5629.91}{\pdev{+10.31}} \\

$\Sigma_b$
& 5815.64
& \massdevinline{5812.97}{\ndev{-2.67}}
& \massdevinline{5821.53}{\pdev{+5.89}}
& \massdevinline{5802.73}{\ndev{-12.91}}
& \massdevinline{5810.70}{\ndev{-4.94}}
& \massdevinline{5791.37}{\ndev{-24.27}}
& \massdevinline{5799.48}{\ndev{-16.16}}
& \massdevinline{5845.17}{\pdev{+29.53}}
& \massdevinline{5815.45}{\ndev{-0.19}} \\

$\Sigma_b^*$
& 5834.74
& \massdevinline{5845.41}{\pdev{+10.67}}
& \massdevinline{5854.90}{\pdev{+20.16}}
& \massdevinline{5824.94}{\ndev{-9.80}}
& \massdevinline{5838.12}{\pdev{+3.38}}
& \massdevinline{5815.25}{\ndev{-19.49}}
& \massdevinline{5825.61}{\ndev{-9.13}}
& \massdevinline{5867.81}{\pdev{+33.07}}
& \massdevinline{5840.64}{\pdev{+5.90}} \\

$\Omega_b$
& 6046.10
& \massdevinline{6011.16}{\ndev{-34.94}}
& \massdevinline{6047.02}{\pdev{+0.92}}
& \massdevinline{6014.19}{\ndev{-31.91}}
& \massdevinline{6063.14}{\pdev{+17.04}}
& \massdevinline{6002.16}{\ndev{-43.94}}
& \massdevinline{6038.10}{\ndev{-8.00}}
& \massdevinline{6048.54}{\pdev{+2.44}}
& \massdevinline{6037.68}{\ndev{-8.42}} \\
\bottomrule
\end{tblr}}

\end{table*}

\paragraph{Model comparison of the three-quark interactions:}

Because the three-quark interaction acts only in the baryon sector, the most direct cross-model test starts from the meson-optimized state-dependent pair interaction.
For the separate-offset results, only the sector-wide baryon constant $D_B$ is refitted, while all other meson-optimized pair parameters are retained.
We therefore compare the baryon RMSEs at the meson-optimized endpoint $\lambda=1$ for the two-body calculation and for the contact, Gaussian, and Yukawa three-quark extensions.
At this endpoint, the meson RMSEs range from 5.21 to $12.79~{\rm MeV}$ and are unchanged by the three-quark interaction, as expected because it does not contribute to mesons.

Table~\ref{tab:cross_model_lambda1_spectra} shows that the meson-calibrated two-body Hamiltonians still give baryon RMSEs between 31.21 and $83.77~{\rm MeV}$ when the three-quark interactions are not included.
For AL1, AL2, AP1, AP2, BD, SL, and GI/CI, the baryonic offset $D_B$ has already been fitted independently of $D_M$.
Thus, the remaining errors are not only due to a common meson--baryon centroid mismatch.
The alternative Hamiltonians contain different confinement laws, short-distance regulators, semirelativistic kinematics, constituent-size form factors, running couplings, mass-dependent smearing, and nonlocal momentum dressing.
These two-body ingredients change the magnitude of the residuals, but their systematic flavor dependencies remain.
On the other hand, including the Yukawa three-quark interaction substantially improves the baryon spectrum for all eight two-body Hamiltonians considered here.

\textit{None of the two-body mechanisms tested here} --- different confinement laws, short-distance regulators, running couplings, relativistic kinematics, nonlocal relativistic dressing, and explicit chiral pair dynamics --- removes the flavor-dependent mismatch between the meson and baryon sectors.
The reduction of the known two-body physics to a constituent-quark potential alone cannot explain the mismatch.
When the same TGE-inspired three-quark operator basis is added, the meson and baryon spectra become compatible \textit{for every two-body Hamiltonian tested here}, with an overall RMSE of about 5--12~MeV at $\lambda=1/2$ (Table~\ref{tab:cross_model_lambda_half_ABC_signs}).
Within the OGE-based models considered here, the need for the three-quark term therefore does not depend on the choice of the pair interaction.

The three-body couplings $A_X$, $B_X$, and $C_X$ are refitted for each two-body model, spatial profile, and offset prescription, so their numerical values are not universal (see Table~\ref{tab:cross_model_lambda_half_ABC_signs}).

\begin{table*}[!t]
\centering
\caption{\justifying
Sign patterns of the fitted three-body couplings $A$, $B$, and $C$ at the meson--baryon simultaneous fit $\lambda=1/2$ for the seven two-body Hamiltonians and the contact, Gaussian, and Yukawa three-body profiles.
Each entry gives  $(\operatorname{sgn}A,\operatorname{sgn}B,\operatorname{sgn}C)$, followed in parentheses by the overall RMSE $\sqrt{(\sigma_M^2+\sigma_B^2)/2}$ in MeV.}
\label{tab:cross_model_lambda_half_ABC_signs}

\centering 
(a) \(u,d,s,c\) subset couplings in Eq.~(\ref{eq:reciprocal_datasets})
\small
\resizebox{\textwidth}{!}{%
\begin{tblr}{
  width=\textwidth,
  colspec={
    Q[l,wd=1.20cm]
    *{6}{Q[c,wd=2.50cm]}
  },
  row{1-2} = {bg=headgray, halign=c, valign=m},
  cell{1}{1} = {r=2}{},
  cell{1}{2} = {c=3}{},
  cell{1}{5} = {c=3}{},
  hline{2} = {2-7}{0.2pt},
  hline{3} = {1-7}{0.3pt},
  rowsep = 2.0pt,
}
\toprule
Model
& Common $D\equiv D_M=D_B$ & &
& Separate $D_M,D_B$ & & \\
& Contact & Gaussian & Yukawa
& Contact & Gaussian & Yukawa \\

AL1
& $(-,+,-)\;(7.39)$
& $(-,+,-)\;(5.04)$
& $(-,-,+)\;(4.73)$
& $(-,+,-)\;(5.00)$
& $(-,+,-)\;(4.96)$
& $(-,+,-)\;(4.73)$ \\

AL2
& $(-,+,-)\;(7.41)$
& $(-,+,-)\;(5.06)$
& $(-,-,+)\;(4.76)$
& $(-,+,-)\;(5.02)$
& $(-,+,-)\;(4.98)$
& $(-,+,-)\;(4.76)$ \\

AP1
& $(-,+,-)\;(7.65)$
& $(-,+,-)\;(5.35)$
& $(-,-,+)\;(5.14)$
& $(-,+,-)\;(5.24)$
& $(-,+,-)\;(5.18)$
& $(-,+,-)\;(5.03)$ \\

AP2
& $(-,+,-)\;(7.66)$
& $(-,-,+)\;(5.36)$
& $(-,-,+)\;(5.15)$
& $(-,+,-)\;(5.26)$
& $(-,+,-)\;(5.19)$
& $(-,+,-)\;(5.04)$ \\

BD
& $(-,+,-)\;(8.23)$
& $(-,+,-)\;(8.01)$
& $(-,+,-)\;(7.92)$
& $(-,+,-)\;(8.08)$
& $(-,+,-)\;(7.98)$
& $(-,+,-)\;(7.75)$ \\

GI/CI
& $(-,+,-)\;(7.50)$
& $(-,+,-)\;(8.81)$
& $(-,+,-)\;(10.99)$
& $(-,+,-)\;(7.44)$
& $(-,+,-)\;(7.70)$
& $(-,+,-)\;(7.49)$ \\

SL
& $(-,+,-)\;(9.84)$
& $(-,+,-)\;(9.74)$
& $(-,+,-)\;(8.73)$
& $(-,+,-)\;(9.45)$
& $(-,+,-)\;(9.51)$
& $(-,+,-)\;(8.47)$ \\

\bottomrule
\end{tblr}}

\vspace{5pt}
\centering 
(b) Full $u,d,s,c,b$ set (all data) couplings in Eq.~(\ref{eq:reciprocal_datasets})
\small
\resizebox{\textwidth}{!}{%
\begin{tblr}{
  width=\textwidth,
  colspec={
    Q[l,wd=1.20cm]
    *{6}{Q[c,wd=2.50cm]}
  },
  row{1-2} = {bg=headgray, halign=c, valign=m},
  cell{1}{1} = {r=2}{},
  cell{1}{2} = {c=3}{},
  cell{1}{5} = {c=3}{},
  hline{2} = {2-7}{0.2pt},
  hline{3} = {1-7}{0.3pt},
  rowsep = 2.0pt,
}
\toprule
Model
& Common $D\equiv D_M=D_B$ & &
& Separate $D_M,D_B$ & & \\
& contact & Gaussian & Yukawa
& contact & Gaussian & Yukawa \\

AL1
& $(-,+,-)\;(10.18)$
& $(-,-,+)\;(6.59)$
& $(-,-,+)\;(6.17)$
& $(-,+,-)\;(6.36)$
& $(-,+,-)\;(5.52)$
& $(-,+,-)\;(5.56)$ \\

AL2
& $(-,+,-)\;(9.73)$
& $(-,-,+)\;(6.58)$
& $(-,-,+)\;(6.10)$
& $(-,+,-)\;(6.23)$
& $(-,+,-)\;(5.51)$
& $(-,+,-)\;(5.36)$ \\

AP1
& $(-,+,-)\;(9.90)$
& $(-,-,+)\;(6.63)$
& $(-,-,+)\;(6.07)$
& $(-,+,-)\;(6.25)$
& $(-,+,-)\;(5.58)$
& $(-,+,-)\;(5.38)$ \\

AP2
& $(-,+,-)\;(9.74)$
& $(-,-,+)\;(6.65)$
& $(-,-,+)\;(6.08)$
& $(-,+,-)\;(6.26)$
& $(-,+,-)\;(5.62)$
& $(-,+,-)\;(5.36)$ \\

BD
& $(-,+,-)\;(12.42)$
& $(-,+,-)\;(10.96)$
& $(-,+,-)\;(11.24)$
& $(-,+,-)\;(10.72)$
& $(-,+,-)\;(10.77)$
& $(-,+,-)\;(10.92)$ \\

GI/CI
& $(-,+,-)\;(10.93)$
& $(-,+,-)\;(11.68)$
& $(-,+,-)\;(11.70)$
& $(-,+,-)\;(9.38)$
& $(-,+,-)\;(9.78)$
& $(-,+,-)\;(9.31)$ \\

SL
& $(-,+,-)\;(13.62)$
& $(-,+,-)\;(12.88)$
& $(-,+,-)\;(11.68)$
& $(-,+,-)\;(12.91)$
& $(-,+,-)\;(12.58)$
& $(-,+,-)\;(11.24)$ \\
\bottomrule
\end{tblr}}
\end{table*}

\begin{figure*}[!b]
\centering
\includegraphics[width=\textwidth]{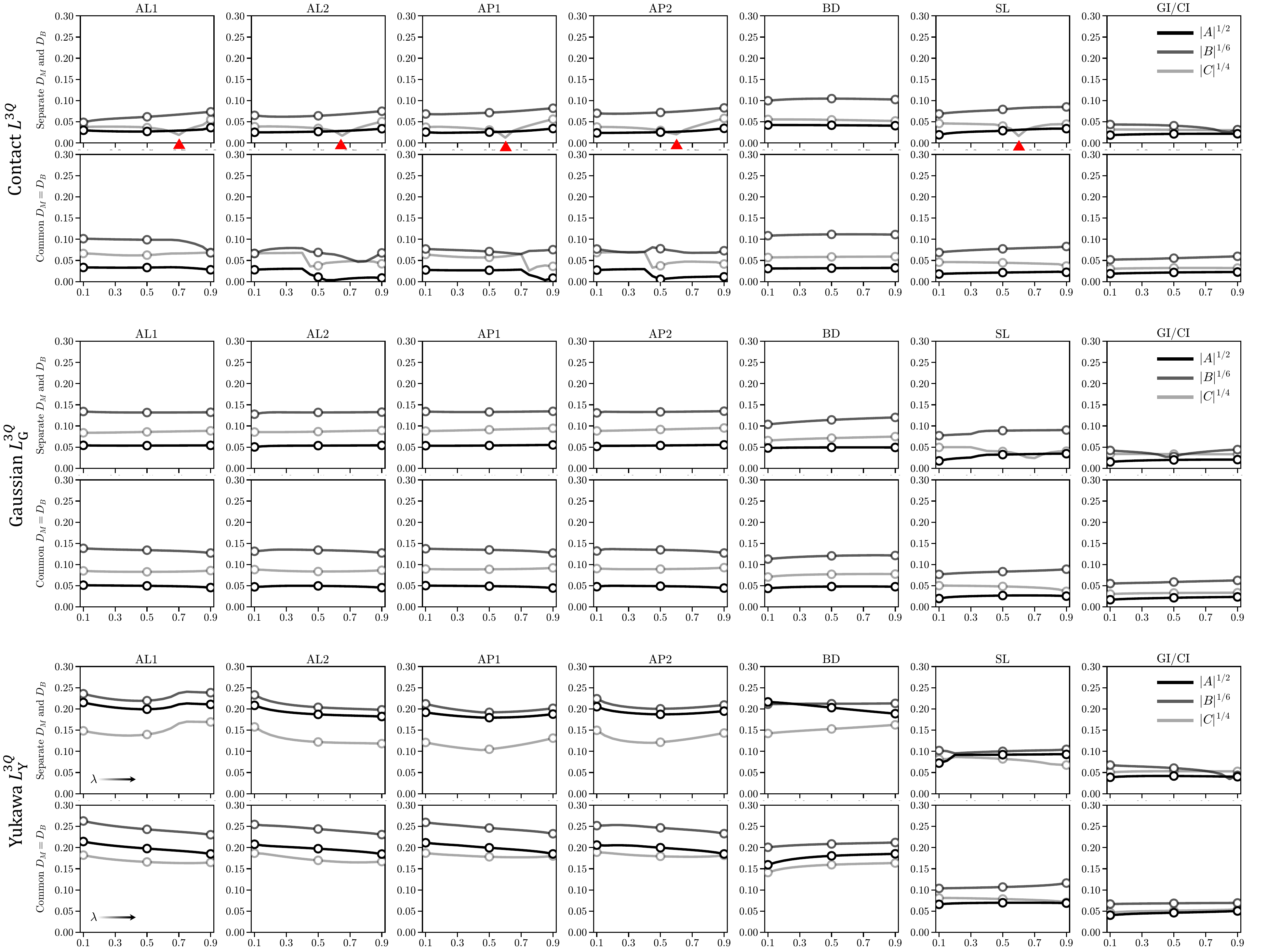}
\caption{\justifying
Stability of the fitted three-quark couplings along the meson--baryon trade-off trajectories.
For the contact (top), Gaussian (middle), and Yukawa (bottom) profiles, the dimension-matched coupling magnitudes $\vert A\vert ^{1/2}$, $\vert B\vert ^{1/6}$, and $\vert C\vert ^{1/4}$ are shown as functions of $\lambda$ for the AL1, AL2, AP1, AP2, BD, SL, and GI/CI two-body Hamiltonians.
Within each profile block, the upper and lower rows correspond, respectively, to fits with separate $D_M$ and $D_B$ and to fits with the common constraint $D_M=D_B$.
The open circles indicate the representative points at
$\lambda=0.1$, 0.5, and 0.9, while the curves connect the fitted solutions along the $\lambda$ scan.
Because only the coupling magnitudes are displayed, changes in their signs are indicated separately by the red triangles below the horizontal axes.}
\label{fig:ABC_stability}
\end{figure*}

\paragraph{Stability and signs of the three-quark couplings:}

The fitted values of $A$, $B$, and $C$ do not define a unique numerical minimum. Nearby local solutions with different sign assignments can appear when the flavor set, two-body Hamiltonian, spatial profile, or meson-sector weight $\lambda$ is changed. The cross-model survey nevertheless shows a clear tendency toward $A<0$, $B>0$, and $C<0$. The sign of $A$ is the most stable, while the remaining ambiguity lies mainly between the $(-,+,-)$ and $(-,-,+)$ branches. Most calculations favor $(-,+,-)$; when $(-,-,+)$ gives the lowest located minimum, the constrained $(-,+,-)$ solution usually remains nearby in RMSE.

The effect of the orbital truncation can be seen most clearly from the $\Lambda$--$\Sigma$ pair. As discussed after Eq.~\eqref{eq:spin_bases_rho_lambda}, the explicit $PP$ channel gives a larger variational lowering to the $\Lambda$-type state than to the $\Sigma$-type state. We now relate this result to the signs of $B$ and $C$.

Let quarks 1 and 2 be the equal-mass light pair, $m_1=m_2=m_q$, and let $m_3=m_s>m_q$. With $\sigma_{ij}\equiv\boldsymbol{\sigma}_i\cdot\boldsymbol{\sigma}_j$, the dominant $SS$ components satisfy
\begin{equation}
(\langle\sigma_{12}\rangle,\langle\sigma_{13}\rangle,\langle\sigma_{23}\rangle)_\Lambda=(-3,0,0),
\qquad
(\langle\sigma_{12}\rangle,\langle\sigma_{13}\rangle,\langle\sigma_{23}\rangle)_\Sigma=(1,-2,-2).
\label{eq:Lambda_Sigma_spin_expectations}
\end{equation}
Using $M_i=km_i$ in Eq.~\eqref{organized_form}, the common-orbital limit gives
\begin{align}
\Delta_{S-S}^{\Lambda\Sigma}
&\equiv
\langle L^{S-S}\rangle_\Lambda-
\langle L^{S-S}\rangle_\Sigma
=
\frac{512}{9k}
\frac{m_s^2-m_q^2}{m_q^4m_s^3}
>0,
\nonumber\\
\Delta_{C-S}^{\Lambda\Sigma}
&\equiv
\langle L^{C-S}\rangle_\Lambda-
\langle L^{C-S}\rangle_\Sigma
=
-\frac{512}{9k}
\frac{(2m_s+m_q)(m_s-m_q)}{m_q^3m_s^2}
<0.
\label{eq:Lambda_Sigma_operator_differences}
\end{align}
The $L^{C-C}$ contribution is common to the two states at this level because they have the same constituent content. It follows that
\begin{equation}
\delta M_\Lambda^{3Q}-\delta M_\Sigma^{3Q}
=
B\Delta_{S-S}^{\Lambda\Sigma}
+
C\Delta_{C-S}^{\Lambda\Sigma}.
\label{eq:Lambda_Sigma_branch_direction}
\end{equation}
Thus, the $(-,-,+)$ branch gives a more attractive three-quark shift to the $\Lambda$-type state, whereas the $(-,+,-)$ branch gives a more attractive shift to the $\Sigma$-type state.
To see the effect of the orbital space, define the $PP$ variational lowering by $\ell_B^{PP}=M_B^{2Q,SS}-M_B^{2Q,SS+PP}>0$ and the required three-quark correction by $d_B=M_B^{\exp}-M_B^{2Q}$. One then has
\begin{equation}
\left(d_\Lambda-d_\Sigma\right)_{SS+PP}
=
\left(d_\Lambda-d_\Sigma\right)_{SS}
+
\left(\ell_\Lambda^{PP}-\ell_\Sigma^{PP}\right).
\label{eq:Lambda_Sigma_PP_target_shift}
\end{equation}
Since $\ell_\Lambda^{PP}>\ell_\Sigma^{PP}$, restoring the $PP$ channel moves the required correction from the $(-,-,+)$ direction toward the $(-,+,-)$ direction. If the differential lowering carries Eq.~\eqref{eq:Lambda_Sigma_PP_target_shift} through zero, the preferred sign branch changes accordingly.

This explains why the restricted $S$-wave fits can support a nearby $(-,-,+)$ solution, while the full $SS+PP+DD+FF$ GEM calculation favors $(-,+,-)$. The argument fixes the direction of the orbital-space effect in the $\Lambda$--$\Sigma$ sector, but it does not establish mathematical uniqueness of the global fit because the finite-range spatial matrix elements and the remaining baryons also constrain $A$, $B$, and $C$. A separate matching interpretation of the remaining $(-,-,+)$ branch is considered below.

Fig.~\ref{fig:ABC_stability} also shows a difference between the contact and finite-range profiles.
For the contact interaction, some models show isolated sign changes as $\lambda$ is varied.
This reflects the larger sensitivity of a zero-range operator to the short-distance basis and to compensation among the fitted parameters.
By contrast, the Gaussian and Yukawa profiles generally keep their signs and give smooth trajectories for $\vert A\vert ^{1/2}$, $\vert B\vert ^{1/6}$, and $\vert C\vert ^{1/4}$ throughout the $\lambda$ scan.
The comparable dimension-matched scales obtained for the $u,d,s,c$ and $u,d,s,c,b$ data sets also show that the finite-range coupling scales depend only weakly on the fitted flavor set and on the meson-sector weight.
Thus, the finite-range profiles give a more stable spectral fit and smoother coupling scales across the flavor sets and fitting weights considered here.

After color-singlet projection, the $A L^{C-C}$ term acts mainly as a spin-independent, mass-dependent baryon shift. Its magnitude is therefore constrained primarily by the spin-averaged residual pattern across the different flavor sectors, and is correlated with the treatment of $D_M$ and $D_B$ and with the variational orbital space. The larger value of $\vert A\vert $ in the full GEM calculation can accordingly reflect the change of the spin-averaged residuals when the higher orbital correlations are restored.

We therefore adopt $(A,B,C)\sim(-,+,-)$ as \textit{the standard sign pattern} in this work. It gives the lowest or a comparable RMSE under the changes of the two-body model, flavor content, spatial profile, and meson-sector weight considered here, and it is recovered in the full GEM calculation. The fitted magnitudes and occasional sign changes remain scheme dependent.

\paragraph{Factorized interpretation of the alternative branch:}
The fit treats $A$, $B$, and $C$ as independent effective couplings. The alternative $(-,-,+)$ branch has a simple interpretation within the $qqqg$ realization of Section~\ref{subsec:three_quark_definition} if the reduced $qqq\leftrightarrow qqqg$ transition factorizes into real central-color and color-spin amplitudes with a common scalar propagator. With the excitation denominator and inverse-mass dependence already included in the definitions of $L_{ij;k}^{C-C}$, $L_{ij;k}^{S-S}$, and $L_{ij;k}^{C-S}$, let $g_{C,n}$ and $g_{CS,n}$ denote the remaining real reduced color and color-spin transition strengths for the intermediate configuration $n$. For intermediate configurations above the valence sector, the factorized contribution has the form
\begin{equation}
A=-\sum_n g_{C,n}^{\,2},
\qquad
B=-\sum_n g_{CS,n}^{\,2},
\qquad
C=-\sum_n g_{C,n}g_{CS,n}.
\label{eq:factorized_ABC_couplings}
\end{equation}
\eqref{eq:factorized_ABC_couplings} gives $A<0$ and $B<0$, while $C>0$ follows when the central-color and color-spin transition amplitudes have opposite relative signs. A single proportional intermediate component gives $AB=C^2$, up to the fixed normalization of the three operator classes. If several non-proportional components contribute, the Cauchy--Schwarz inequality gives $C^2<AB$. The fitted Yukawa solutions on the $(-,-,+)$ branch satisfy this inequality and are therefore consistent with a multi-component factorized reduction.

The $qqqg$ realization is not restricted to this scalar-factorized limit. In particular, the chromomagnetic $qg$ interpolator discussed in  Section~\ref{subsec:three_quark_definition} contains a $j_{qg}^{P}=3/2^+$ component. Its explicit transition matrices and a minimal local $S$-wave reduction are given in \ref{app:qg_spin32}. For this component, the propagation between two transition blocks is tensor valued,
\begin{equation}
\Gamma_{X,n}^{a\dagger}G_{ab,n}^{(3/2)}(E)\Gamma_{Y,n}^{b},
\qquad
X,Y\in\{C,CS\},
\label{eq:spin_three_half_qg_matching}
\end{equation}
rather than the scalar factorization $g_{X,n}\,G_n^{(1/2)}(E)\,g_{Y,n}$ underlying Eq.~\eqref{eq:factorized_ABC_couplings}. Spin projectors, momentum dependence, gluonic polarizations, and independent transition tensors can then generate additional structures before the result is projected onto $L^{C-C}$, $L^{S-S}$, and $L^{C-S}$. 
Since the $qqqg$ configuration appears only as a virtual intermediate state, its $qg$ substructure is sampled away from an on-shell limit, where the decomposition of the spin structure is not unique~\cite{KrebsEpelbaumMeissner2010Spin32}. 
The simple sign relations obtained from the scalar-factorized limit therefore need not hold for the general transition. 
If the intermediate component is represented by a relativistic vector-spinor field, this ambiguity appears explicitly in the off-shell lower-spin components, which can be shifted into local operators by field redefinitions without changing the on-shell $S$-matrix~\cite{KrebsEpelbaumMeissner2010Spin32,Pascalutsa2001Spin32}.

For an intermediate Hermitian sector lying above the valence state, the second-order correction is negative as a whole, as follows from $T_{\mathsf P\mathsf Q_g}(E-H_{\mathsf Q_g\mathsf Q_g})^{-1}T_{\mathsf Q_g\mathsf P}$.\footnote{Here $T_{\mathsf P\mathsf Q_g}$ denotes the full transition operator, of which the $\Gamma_{X,n}^{\ell}$ introduced above are resolved transition components.} 
This, however, does not require the fitted coefficients $A$, $B$, and $C$ to have the same negative sign.
Eliminating the $qqqg$ sector generates one-body, two-body, and connected three-body contributions.
After the one- and two-body pieces are absorbed into the baseline Hamiltonian, the remaining connected contribution is represented by the three operator structures used here.
The fitted values of $A$, $B$, and $C$ therefore need not inherit the overall sign of the full second-order correction.
The $(-,+,-)$ solution is thus not inconsistent with a $qqqg$ origin.
It instead shows that the simple scalar-factorized relation in Eq.~\eqref{eq:factorized_ABC_couplings} is too restrictive and that a more general nonfactorized transition structure is required.

We therefore distinguish the two results. The $(-,-,+)$ pattern is a possible near-factorized matching limit, whereas $(-,+,-)$ is the phenomenological sign pattern preferred by the spectrum. The fitted couplings should not be identified directly with a single factorized $qqq\leftrightarrow qqqg$ transition.

\section{Benchmark of radial and orbital excitations in the $u,d,s,c$ sectors}
\label{sec:excitation_spectra}
The preceding sections focused on ground-state mesons and baryons. In this region, the compact constituent description is most reliable, and the short-range three-quark interaction can be studied with less ambiguity. However, agreement with the ground states does not determine the long-distance confining law, relativistic kinematics, or the state-dependent interactions needed for the excited spectra. Radial excitations probe larger distances and nodal wave functions, orbital excitations test the tensor and spin--orbit interactions, and states near open thresholds may not be represented by a single valence eigenstate. We therefore use the excitation spectrum to examine the region where a static valence Hamiltonian remains useful.

We restrict the calculation to the $u,d,s,c$ sectors. Bottomonium and many bottom-hadron states are more compact and closer to the heavy-quark limit, where conventional potential models are already known to work particularly well~\cite{Godfrey1985,Barnes2005}. The lighter sectors give a more useful test because the same Hamiltonian must describe pseudo-Goldstone states, broad threshold-sensitive resonances, ordinary orbital multiplets, and relatively compact heavy-quark states.

We compare two Hamiltonians. TH1 is the linear-confinement Hamiltonian used in the preceding ground-state analysis. It contains the central and color-spin interactions but does not contain tensor or spin--orbit terms, and therefore gives unsplit multiplet centroids. TH2 replaces the linear confinement by an $r^{2/3}$ form and includes the symmetric spin--orbit, antisymmetric spin--orbit, and tensor interactions. In the meson sector, we fit TH2 to selected ground, radial, and fine-structure levels, while the broad threshold states and the higher spectrum are left as tests. In the baryon sector, no radial or orbital excitation is fitted. We fix each meson-calibrated two-body Hamiltonian and refit only the Yukawa-profile couplings $A$, $B$, and $C$ to the common ground-baryon calibration set.

We use the TH1--TH2 comparison as a diagnostic. The displacement of a complete TH2 multiplet relative to TH1 contains the effects of the changed confinement, the meson-sector refit, and the refitted three-quark couplings. The splittings inside a TH2 multiplet show more directly the effect of the added fine-structure interactions. If both the centroid and the internal splitting improve, the missing effect can be described within the extended valence Hamiltonian. If the discrepancy remains, relativistic dynamics, string breaking, chiral interactions, or explicit coupled channels are needed~\cite{BulavaEtAl2019StringBreaking,VijandeEtAl2004Screened}. For this reason, we discuss the results for each state and each flavor sector rather than using one global RMSE.

\subsection{Meson excitations}
\label{sec:meson_excitations}

\begin{figure*}[!t]
\centering
\includegraphics[width=\textwidth,height=0.95\textheight,keepaspectratio]{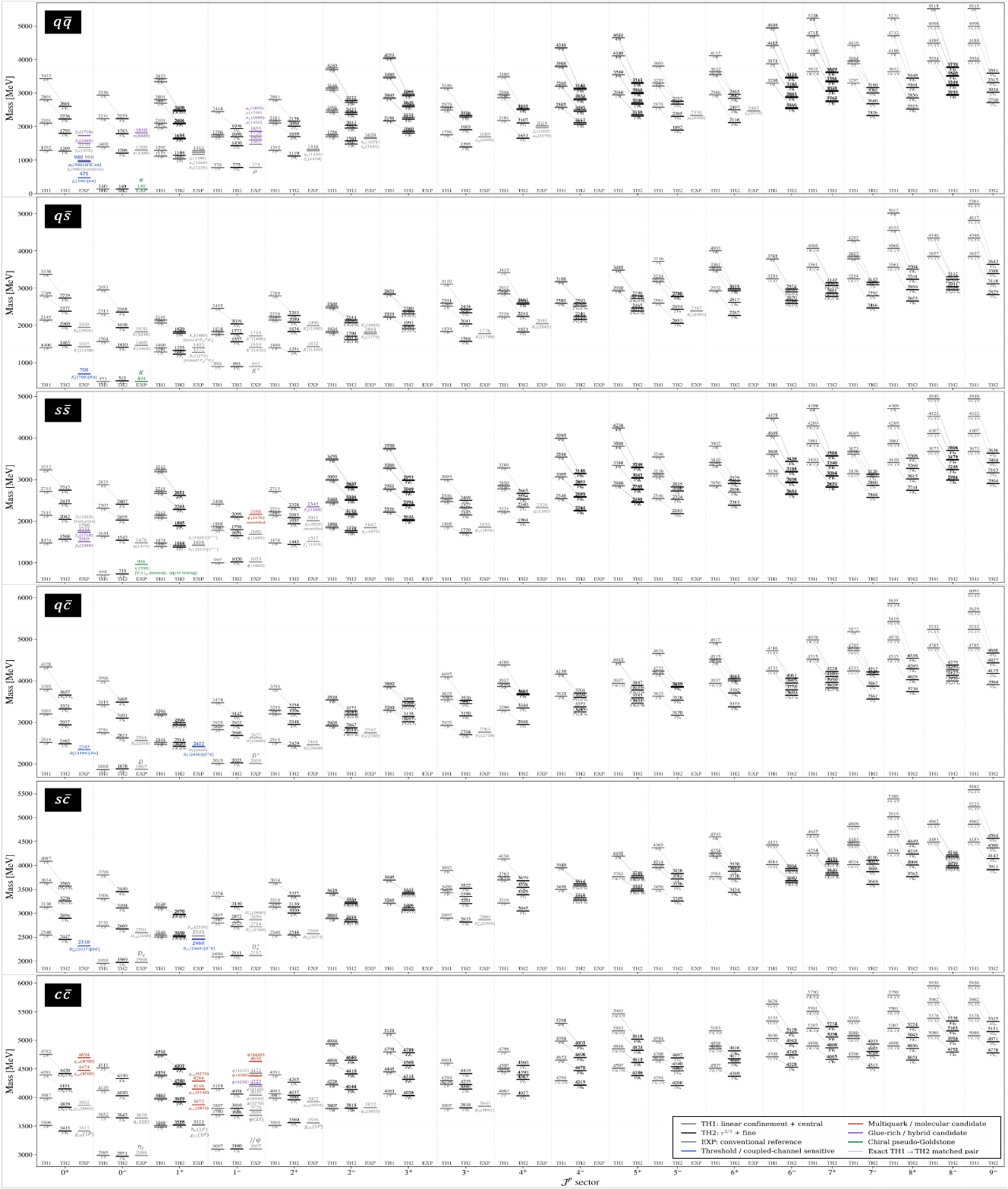}
\caption{\justifying (Enlarge online for details.) Meson spectra from the linear central calculation (TH1) and the $r^{2/3}$ fine-structure refit (TH2), compared with selected experimental levels.}
\label{fig:meson_excitation_spectra}
\end{figure*}

Since a meson contains only two valence constituents, the three-quark operators introduced in  Sections~\ref{sec:baryon_GEM} and~\ref{sec:short_range} do not act in this sector. The meson excitation spectrum therefore tests the two-body Hamiltonian directly. In particular, it tests whether the parameters fixed from the low-lying spectrum can reproduce the radial and orbital spacings of states that extend over larger distances.  Fig.~\ref{fig:meson_excitation_spectra} compares TH1 and TH2. TH1 is the original linear-confinement Hamiltonian used in the ground-state meson--baryon analysis. TH2 is an auxiliary meson-sector refit with an $r^{2/3}$ confining interaction and explicit symmetric spin--orbit, antisymmetric spin--orbit, and tensor interactions. We use TH2 only to study the excitation problem. It does not replace the parameter set of Eq.~\eqref{params} used for the ground-state baryon and three-quark-force results.

The TH1 and TH2 columns in  Fig.~\ref{fig:meson_excitation_spectra} have different meanings. TH1 gives the unsplit eigenvalues of the central Hamiltonian and is labeled by the $n^{2S+1}L$ centroids. TH2 is obtained by diagonalizing the full fixed-$\mathcal J^{P(C)}$ Hamiltonian and gives the individual $n^{2S+1}L_{\mathcal J}$-dominated states after spin--orbit, tensor, and configuration mixing. When several components mix, we use the dominant component as the label. The experimental masses and nominal assignments are taken from the Particle Data Group unless otherwise stated~\cite{PDG2024}. We use gray for comparatively conventional quark-model reference states, green for pseudoscalar states strongly affected by chiral or anomalous dynamics, blue for threshold- or coupled-channel-sensitive states, red for states with prominent multiquark or molecular interpretations, and purple for glue-rich or hybrid candidates. The colors only show why a state is or is not used as a clean valence-$q\bar q$ benchmark. They do not assign a unique Fock-space composition to the physical state.

\paragraph{TH1 central Hamiltonian with the linear-confinement baseline:}

TH1 uses Eqs.~\eqref{hamiltonian}--\eqref{2body_2} with the parameters in Eq.~\eqref{params}.  No excitation is used in determining this parameter set.  Radial excitations are the higher generalized eigenvalues of a fixed-$l$ block, and different orbital angular momenta are solved in independent blocks.  Because $V^{LS}$, $V^{ALS}$, and $V^T$ are absent, all members of a given $n^{2S+1}L$ multiplet are degenerate.  The physically observed $^1L_{\mathcal J}$--$^3L_{\mathcal J}$ mixing in unequal-mass mesons is also absent.  TH1 should therefore be compared with spin-weighted multiplet centroids or with states for which the omitted fine structure is known to be small, rather than with every individual resonance.

The low-lying charmonium spectrum illustrates why the linear Cornell-type interaction remains a useful baseline.  TH1 gives 3652 and 3700~MeV for the $2^1S_0$ and $2^3S_1$ levels, compared with $\eta_c(2S)$ and $\psi(2S)$ at approximately 3637 and 3686~MeV, and places the $1^3D$ centroid at 3807~MeV in the observed $1D$ charmonium region.  The $1^1P$ and $1^3P$ centroids, 3484 and 3509~MeV, likewise lie close to the $h_c$ and $\chi_{cJ}$ multiplets.  Thus, below and near the first open-charm thresholds, TH1 reproduces the gross radial and orbital scales at the several-tens-of-MeV level, in accord with the well-known success of Cornell-type potential models for heavy quarkonia~\cite{Godfrey1985,Barnes2005}.

The limitation becomes increasingly visible as the radial or orbital excitation is raised.  A linear potential has the semiclassical scaling $E_N\propto N^{2/3}$, and the TH1 levels rise rapidly at large $n$ and $l$.  This is particularly clear in the light and open-charm radial spectra.  For example, TH1 gives $M(2^1S_0)=1399.6~\mathrm{MeV}$ and $M(3^1S_0)=2231.1~\mathrm{MeV}$ in the $q\bar q$ sector, and $M(2^1S_0)=2751.2~\mathrm{MeV}$ in the $q\bar c$ sector.  The latter is more than 200~MeV above the nominal $D(2550)$ region.  The same upward tendency is visible in high-$l$ branches, such as the light $1^1G$ level at 2566.8~MeV and the charmonium $1L$ level near 5079.6~MeV.  Some low orbital states are still well reproduced. For example, the TH1 $s\bar s$ $1^3D$ centroid at 1868.1~MeV is close to the $\phi_3(1850)$. In most cases, the high-energy spectrum is too sparse and its levels lie too high compared with the spectrum expected after string breaking and continuum dressing. We therefore test a softer confining power together with explicit fine-structure interactions~\cite{BulavaEtAl2019StringBreaking,VijandeEtAl2004Screened}.

\paragraph{TH2 Hamiltonian with fine structure:}
The extended color-singlet meson Hamiltonian is
\begin{equation}
H_{\rm mes}^{\rm TH2}
=
 m_1+m_2+K_{\rm int}
+V_{12}^{C}
+V_{12}^{CS}
+V_{12}^{T}
+V_{12}^{LS}
+V_{12}^{ALS},
\label{eq:meson_power_hamiltonian}
\end{equation}
where the central potential is replaced by
\begin{equation}
V_{12}^{C}(r)
=-\frac{3}{4}F_1^cF_2^c
\left(-\frac{\kappa}{r}+\sigma_{2/3}r^{2/3}-D\right).
\label{eq:meson_power_central}
\end{equation}
For a color-singlet $q\bar q$ state, $\langle F_1^cF_2^c\rangle=-4/3$, so Eq.~\eqref{eq:meson_power_central} reduces to the radial potential in parentheses.  The color-spin interaction retains the finite-range Gaussian form and mass dependence of Eq.~\eqref{2body_2}. We keep the original parameters entering $\mu_{12}$ and $\mu'_{12}$ fixed and allow only an overall dimensionless normalization $c_{CS}$ to be refitted.  For a non-relativistic power-law potential $r^p$, the high-excitation growth is approximately $E_N\propto N^{2p/(p+2)}$. Therefore, fixing $p=2/3$ changes the asymptotic behavior from $N^{2/3}$ for linear confinement to approximately $N^{1/2}$. We use this form only as a phenomenological compression of the valence spectrum. It is not an explicit model of string breaking.
For the noncentral interactions, we use the scalar-confinement plus vector-Coulomb convention employed in relativized potential models~\cite{Godfrey1985,Barnes2005,Godfrey2016}.  After color-singlet projection, define
\begin{equation}
V_V(r)=-\frac{\kappa}{r},
\qquad
V_S(r)=\sigma_{2/3}r^{2/3}-D,
\end{equation}
and denote their regulated forms by $\widetilde V_V$ and $\widetilde V_S$.  The symmetric spin--orbit, antisymmetric spin--orbit, and tensor terms are
\begin{align}
V_{12}^{LS}(r)=&\;c_{LS}(\hbar c)^2
\left[
\frac{1}{4r}
\left(\frac{1}{m_1^2}+\frac{1}{m_2^2}\right)
\left(\widetilde V_V'(r)-\widetilde V_S'(r)\right)
+
\frac{\widetilde V_V'(r)}{m_1m_2r}
\right]
\mathbf L_{12}\cdot(\mathbf S_1+\mathbf S_2),
\nonumber\\
V_{12}^{ALS}(r)=&\;c_{ALS}(\hbar c)^2
\frac{1}{4r}
\left(\frac{1}{m_1^2}-\frac{1}{m_2^2}\right)
\left(\widetilde V_V'(r)-\widetilde V_S'(r)\right)
\mathbf L_{12}\cdot(\mathbf S_1-\mathbf S_2),
\nonumber\\
V_{12}^{T}(r)=&\;c_T\frac{(\hbar c)^2}{12m_1m_2}
\left(
\frac{\widetilde V_V'(r)}{r}-\widetilde V_V''(r)
\right)S_{12},
\label{eq:meson_power_spin_dependent}
\end{align}
where $S_{12} = 3(\boldsymbol\sigma_1\cdot\hat{\mathbf r}) (\boldsymbol\sigma_2\cdot\hat{\mathbf r}) - \boldsymbol\sigma_1\cdot\boldsymbol\sigma_2$.
The antisymmetric term vanishes for equal constituent masses.  In open-flavor mesons it mixes $^1l_{\mathcal J}$ and $^3l_{\mathcal J}$ components\footnote{For mesons $l$ denotes orbital angular momentum.} with $\mathcal J=l$, while the tensor force mixes triplet configurations with $l\leftrightarrow l\pm2$ at fixed $\mathcal J^P$.

The singular radial derivatives are regularized before constructing the spin-dependent operators. We convolve both the vector and scalar potentials with a flavor-dependent Gaussian,
\begin{align}
\widetilde V_X(\mathbf r)
&=
\int d^3\mathbf r'\,
\rho_{12}(\mathbf r-\mathbf r')V_X(r'),
\qquad X=V,S,
\nonumber\\
\rho_{12}(\mathbf r)
&=
\frac{(\sigma_{12}^{\rm sm})^3}{\pi^{3/2}}
\exp[-(\sigma_{12}^{\rm sm})^2r^2],\qquad
(\sigma_{12}^{\rm sm})^2=
(\sigma_0^{\rm sm})^2
\bigg[
\frac12+\frac12
\bigg(\frac{4m_1m_2}{(m_1+m_2)^2}\bigg)^4\,
\bigg]
+s_{\rm sm}^2
\left(\frac{2m_1m_2}{m_1+m_2}\right)^2.
\label{eq:meson_power_smearing}
\end{align}
In particular, $\widetilde V_V(r)=-\kappa\,\mathrm{erf}(\sigma_{12}^{\rm sm}r)/r$, and the derivatives in Eq.~\eqref{eq:meson_power_spin_dependent} are evaluated only after the convolution.  The confining interaction is taken to be purely scalar in the fine-structure sector, corresponding to a fixed vector-confinement fraction $f_V=0$.

We diagonalize the complete Hamiltonian in fixed $\mathcal J^{P(C)}$ blocks constructed from the $n^{2S+1}L_{\mathcal J}$ basis. Thus, the fine-structure interactions are not added as first-order shifts. Their diagonal and off-diagonal matrix elements are included in the same generalized eigenvalue problem as the central Hamiltonian. This treatment is important for the open-flavor $1^+$ states and for the tensor-coupled blocks. We label a calculated eigenstate by its dominant spectroscopic component and specify the lower or upper branch in the mixed $1^+$ sectors. In this way, the assignment is not changed only because two eigenvalues cross during the nonlinear fit.

\paragraph{Calibration set, numerical procedure, and fitted parameters:}

The TH2 refit uses 21 comparatively clean masses.  Eleven ground-state references, $\pi$, $\rho$, $K$, $K^{*}$, $D$, $D^{*}$, $D_s$, $D_s^{*}$, $\phi$, $\eta_c$, and $J/\psi$, constrain the constituent masses and the central and color-spin sectors.  The radial scale is constrained by $\eta_c(2S)$ and $\psi(2S)$.  The $\chi_{c0}$, $\chi_{c1}$, $h_c$, and $\chi_{c2}$ masses constrain the equal-mass $1P$ fine structure, and the narrow $D_1(2420)$, $D_2^{*}(2460)$, $D_{s1}(2536)$, and $D_{s2}^{*}(2573)$ states constrain the unequal-mass $P$-wave sector and the antisymmetric spin--orbit mixing.

The broad $D_0^{*}(2300)$ and $D_1(2430)$ and the subthreshold $D_{s0}^{*}(2317)$ and $D_{s1}(2460)$ are deliberately excluded from the calibration set.  The first two couple strongly to $D^{(*)}\pi$ in $S$ wave, while lattice calculations that include both $c\bar s$ and $DK$ or $D^{*}K$ interpolating fields find that the latter two are strongly affected by the nearby scattering channels~\cite{Godfrey2016,Hao2025,Hao2022,MohlerEtAl2013,LangEtAl2014}.  Their physical masses therefore do not provide clean constraints on a quenched local potential.  Similarly, the charmoniumlike $X$ and $Y$ states shown in the figure are not used in the fit.

To prevent sub-MeV experimental uncertainties from forcing the fit to reproduce a few states at the expense of the model as a whole, we assign sector-dependent model uncertainties.  The adopted values are 35~MeV for the light $\pi$ and $\rho$, 30~MeV for the strange and kaonic sectors, 20~MeV for the open-charm and charm-strange states, 10~MeV for the charmonium ground and $1P$ levels, and 12~MeV for the charmonium $2S$ levels.  The fit minimizes the corresponding weighted errors.  The optimization varies 12 parameters: $m_q$, $m_s$, $m_c$, $\kappa$, $\sigma_{2/3}$, $D$, $c_{CS}$, $\sigma_0^{\rm sm}$, $s_{\rm sm}$, $c_{LS}$, $c_{ALS}$, and $c_T$.  The confinement exponent, the vector-confinement fraction, and the four original color-spin range parameters are held fixed.
The fitting calculation uses 24 Gaussian widths over 0.03--10~fm, $l\leq4$, and 7 eigenstates per fixed-$\mathcal J^{P(C)}$ block. The final spectrum shown in  Fig.~\ref{fig:meson_excitation_spectra} is recomputed with 30 Gaussian widths over the same range, $l\leq8$, $\mathcal J=0,\ldots,9$, and 8 states per block. In the final calculation, eigenmodes of the overlap matrix with eigenvalues below 10\textsuperscript{\ensuremath{-}12} are removed to stabilize the generalized eigenvalue problem $Hc=E\mathcal O c$.

The resulting parameters are
\begin{align}
\kappa&=112.6077~\mathrm{MeV\,fm},
&\sigma_{2/3}&=1112.8578~\mathrm{MeV\,fm^{-2/3}},
&D&=1105.9836~\mathrm{MeV},
\nonumber\\
m_{u,d}&=318.0373~\mathrm{MeV},
&m_s&=603.3467~\mathrm{MeV},
&m_c&=1879.9559~\mathrm{MeV},
\nonumber\\
c_{CS}&=1.00617,
&\sigma_0^{\rm sm}&=2.05313~\mathrm{GeV},
&s_{\rm sm}&=2.80946,
\nonumber\\
c_{LS}&=1.00726,
&c_{ALS}&=0.44089,
&c_T&=0.03770.
\label{eq:meson_power_parameters}
\end{align}
The independent tensor normalization $c_T$ is only weakly constrained after the spin--orbit structure and open-flavor mixing are adjusted. For the 21 calibration masses, we obtain the raw RMSE $\sigma_M^{\rm TH2}=11.54~\mathrm{MeV}$. The four charmonium $1P$ states alone give an RMSE of approximately 4.18~MeV. The largest residuals are $-32.0~\mathrm{MeV}$ for $D_2^{*}(2460)$ and $-25.0~\mathrm{MeV}$ for $D_{s2}^{*}(2573)$. The TH1 value 2.51~MeV and the TH2 value 11.54~MeV are not obtained from the same data set and should not be compared directly. TH1 was fitted only to a small set of ground states, while TH2 was fitted simultaneously to ground states, radial spacings, fine structure, and unequal-mass mixing.

We keep three conventional levels as out-of-fit tests. Their calculated masses are
\begin{equation}
M[\psi_2(3823)]=3814.6~\mathrm{MeV},
\qquad
M[\psi_3(3842)]=3818.1~\mathrm{MeV},
\qquad
M[\phi_3(1850)]=1720.3~\mathrm{MeV}.
\label{eq:meson_power_holdouts}
\end{equation}
The first two residuals are \ensuremath{-}8.4 and $-24.6~\mathrm{MeV}$, showing that the charmonium $1D$ scale is maintained after the refit. The strangeonium result is approximately 134~MeV too low. Thus, the same $r^{2/3}$ softening does not work uniformly in all flavor sectors.

\subsubsection{The light sector and the pseudo-Goldstone states}
\label{subsubsec:light_meson_experiment}
The light sector exposes the principal limitation of the present TH2 refit.
TH1 already gives a reasonable centroid-level description of several low orbital references: the $1^1D$ and $1^3D$ centroids are 1694.5 and 1756.0~MeV, compared with the $\pi_2(1670)$ and $\rho_3(1690)$ regions.
TH2 overcompresses these states.
The $1^1D_2$ level falls to 1477.0~MeV and the $1^3D_3$-dominated level to 1393.3~MeV.
The triplet partners are spread over a broad interval extending to approximately 1696~MeV.
The light $1P$ spectrum is similarly distorted: the $1^1P_1$ level is near 1035~MeV, while the $1^3P_{0,1,2}$-dominated states lie near 1269, 1144, and 1119~MeV.
These values neither reproduce the observed multiplet centroid nor form a stable quark-model fine-structure pattern.

The $2S$ radial levels show a more mixed outcome.
The $2^1S_0$ mass changes from 1399.6 to 1206.5~MeV, moving from above to below the $\pi(1300)$, whereas the $2^3S_1$ level changes from 1688.9 to 1430.4~MeV and approaches the $\rho(1450)$.
Thus, the same compression that helps the vector radial state overshoots in the pseudoscalar and orbital channels.
A single flavor-independent power $p=2/3$ does not provide a uniformly improved light-meson spectrum.

\paragraph{The pseudo-Goldstone pion:}
The exceptionally small pion mass must be interpreted separately from the ordinary constituent-level spectrum.
In QCD, the pion is a pseudo-Goldstone boson of spontaneously broken chiral symmetry, and its mass is controlled by the axial Ward identity and the light current-quark masses rather than by a conventional balance of constituent rest masses and a static confining potential~\cite{Roberts1994,Gasser1982}.
Neither TH1 nor TH2 realizes this symmetry mechanism explicitly.
Reproducing the pion mass in the present fit means that the omitted chiral dynamics has been absorbed into the constituent masses, the additive constant, and the effective color-spin interaction; it does not mean that the Hamiltonian derives the Goldstone nature of the pion.
The green pion bar in Fig.~\ref{fig:meson_excitation_spectra} is therefore not placed on the same footing as an ordinary quark-model benchmark such as the charmonium $1P$ multiplet.

\paragraph{The broad $\boldsymbol{f_0(500)}$ pole:}
The $f_0(500)$ cannot be described as a narrow and isolated $1^3P_0$ eigenstate. It is a very broad pole with strong $\pi\pi$ rescattering and is determined from the analytic continuation of the isoscalar $S$-wave amplitude rather than from a Breit--Wigner mass that can be compared directly with a static Hamiltonian~\cite{CapriniColangeloLeutwyler2006,Pelaez2016Sigma}. A compact $q\bar q$ seed may be present in the scalar channel, but the physical pole is dominated by the long-distance two-pion dynamics. We therefore do not use the $f_0(500)$ as a clean calibration state of TH1 or TH2. In this case, the discrepancy with the calculated scalar eigenvalue mainly shows the missing chiral and continuum effects rather than an ordinary level-spacing error.

\paragraph{The $\boldsymbol{a_0(980)}$--$\boldsymbol{f_0(980)}$ threshold pair:}

The $a_0(980)$ and $f_0(980)$ have historically been interpreted in several competing ways, including compact $q\bar q$ states, tetraquarks, $K\bar K$ molecules, and poles generated or strongly reshaped by coupled $\pi\pi$, $\pi\eta$, and $K\bar K$ channels~\cite{Tornqvist1995ScalarNonet,MaianiEtAl2004LightScalars,DudekEtAl2016A0,BricenoEtAl2018Isoscalar}.
Their proximity to the $K\bar K$ threshold makes a description in terms of a single unmixed quark-model eigenvalue inadequate even when a compact seed is present.
For these states, we ask whether the pole contains a sizable two-constituent component in addition to the hadronic dressing.

Recent production measurements provide new evidence on this issue.
The CMS elliptic-flow analysis of $f_0(980)$ production in proton--lead collisions favors constituent-quark-number scaling with $n_q=2$ over $n_q=4$ by up to $7.7\sigma$, providing strong evidence for a dominant two-constituent production component~\cite{CMSF0980NCQ2025}.
The multiplicity dependence measured by ALICE is also better described when no significant strange or antistrange constituent content is assigned to the state~\cite{ALICEF0980Multiplicity2025}.
These measurements disfavor a simple predominantly $s\bar s$ assignment of the $f_0(980)$ and support the presence of a nonstrange $q\bar q$ seed. However, they do not show that the physical pole is a pure and undressed $q\bar q$ state. Hadronic coalescence calculations can reproduce important parts of the production and elliptic-flow data from a late-stage $K\bar K$ configuration~\cite{WangEtAlF0980Flow2025,ChaimongkonEtAlF0980Coalescence2026}. Therefore, constituent-number scaling alone does not determine the microscopic pole composition. We take these results to mean that the $f_0(980)$ contains a substantial compact two-constituent component, while its mass, width, and line shape are strongly modified by the nearby $\pi\pi$ and $K\bar K$ channels.

The scalar-nonet assignment proposed in Ref.~\cite{YasuiLeeLoSasaki2026} gives one possible realization of this picture. It identifies the $f_0(980)$ mainly with the nonstrange $q\bar q$ component and is consistent with the recent evidence against a predominantly hidden-strange assignment. The strange scalar and scalar-glueball assignments of this proposal are discussed below.

The remaining problem is the mass. A conventional static nonstrange $1^3P_0$ $q\bar q$ state is usually obtained near 1.2--$1.4~\mathrm{GeV}$, and both TH1 and TH2 place the corresponding bare state well above 980~MeV. Thus, the discrepancy is not removed by changing the confinement or by adding the standard tensor and spin--orbit interactions. If a sizable two-constituent seed is present, the low physical pole requires a large correction from chiral and coupled-channel dynamics or a more substantial change of the scalar interaction. The recent data support a compact nonstrange $q\bar q$ component, but the static quark model still gives a much higher bare mass. For this reason, we show the $f_0(980)$ as a threshold- and coupled-channel-sensitive but conventional $q\bar q$ state rather than as an exclusively molecular or tetraquark candidate in  Fig.~\ref{fig:meson_excitation_spectra}.

\paragraph{The spin-exotic $\boldsymbol{1^{-+}}$ states:}
A neutral $q\bar q$ meson has parity and charge conjugation $P=(-1)^{l+1}$ and $C=(-1)^{l+S}$.
The combination $\mathcal J^{PC}=1^{-+}$ is therefore forbidden for a single quark--antiquark pair.
The observed $\pi_1$ structures and the isoscalar $\eta_1(1855)$ consequently require additional gluonic or multiquark degrees of freedom and are natural hybrid candidates~\cite{Dudek2011Hybrid,BESIIIeta11855}.
They are shown in purple and should not be counted as failures of the TH1 or TH2 valence spectrum, because neither Hamiltonian contains an explicit excited-glue basis capable of producing spin-exotic quantum numbers.

\subsubsection{The ${q\bar s}$ and ${s\bar s}$ sectors}
\label{subsubsec:strange_meson_experiment}
In the kaon sector, TH2 preserves the $K^{*}(892)$ mass and lowers the radial levels relative to TH1.
The $2^1S_0$ level changes from 1564.2 to 1410.0~MeV, compared with the nominal $K(1460)$, while the $2^3S_1$ level changes from 1752.7 to 1557.0~MeV, compared with $K^{*}(1410)$.
The first radial pseudoscalar is therefore brought into the correct region, whereas the vector radial excitation remains high by more than 100~MeV.
The physical axial mesons require $^1P_1$--$^3P_1$ mixing.
TH1 places the unmixed centroids at 1284 and 1400~MeV, accidentally close to the two physical $K_1$ masses, while TH2 produces mixed branches near 1227 and 1325~MeV.
The latter result has the correct qualitative structure but is too low, demonstrating that agreement of the two TH1 centroids with $K_1(1270)$ and $K_1(1400)$ should not be mistaken for a dynamical description of their mixing~\cite{Godfrey1985,Vijande2005}.

The kaonic $D$-wave comparison gives the opposite trend from the radial sector.
TH1 places the $1^3D$ centroid at 1834.4~MeV, reasonably close to the $K_2(1770)$--$K_3^{*}(1780)$ region.
TH2 splits the triplet strongly: the $1^3D_3$-dominated branch falls to approximately 1568~MeV, whereas the $1^3D_1$ branch remains near 1772~MeV.
This spread is too large for a clean conventional assignment.
The result shows that the universal spin--orbit interaction fitted mainly to charmonium and narrow heavy-light states does not transfer reliably to the light--strange orbital spectrum.

The strangeonium vector radial level improves: $M(2^3S_1)$ changes from 1780.3~MeV in TH1 to 1650.9~MeV in TH2, close to the $\phi(1680)$.
The orbital scale, however, becomes too soft.
TH1 predicts the $1^3D$ centroid at 1868.1~MeV, nearly coincident with the $\phi_3(1850)$, while TH2 gives 1720.3~MeV for the $1^3D_3$ state.
The $1^1P_1$ and $1^3P_2$-dominated TH2 levels, 1374.6 and 1443.0~MeV, also lie below the $h_1(1415)$ and $f_2'(1525)$ references.
Consequently, the $r^{2/3}$ modification improves the first strangeonium radial spacing but worsens the best-established orbital benchmark.

\paragraph{The pseudo-Goldstone kaon and the anomaly-affected isoscalars:}

Like the pion, the kaon is a pseudo-Goldstone boson and cannot be interpreted as an ordinary deeply bound $1^1S_0$ level generated solely by a static constituent Hamiltonian.
The present fit reproduces its mass by absorbing the omitted chiral symmetry constraints into effective parameters.
Furthermore, the quark model does not contain contributions from the axial anomaly and quark--antiquark annihilation, making a direct comparison with the $\eta$--$\eta'$ system problematic.

\paragraph{The $\boldsymbol{K_0^*(700)}$ pole:}

The low-mass $K_0^{*}(700)$, historically denoted by $\kappa$, is the strange analogue of the broad light scalar problem.
It is established through the analytic structure of the $K\pi$ $S$-wave amplitude and is not naturally identified with the conventional $q\bar s$ $1^3P_0$ eigenvalue.
The latter is more plausibly associated with the substantially heavier $K_0^{*}(1430)$.
The $K_0^{*}(700)$ should therefore be treated as a strongly continuum-generated or continuum-dressed pole and shown as threshold sensitive rather than used to determine the local scalar confinement parameters~\cite{Pelaez2016Sigma}.

\paragraph{Scalar--isoscalar mixing and the glueball candidates:} The present calculation does not contain the quark--antiquark annihilation that mixes the nonstrange and strange isoscalar sectors, and it also does not contain an explicit glueball basis. Therefore, one cannot assign every observed $f_0$ state directly to a pure $q\bar q$ or $s\bar s$ state. Lattice QCD places the lightest scalar glueball in the same mass region as several isoscalar resonances, and phenomenological studies usually require mixing among nonstrange quarkonium, strange quarkonium, glueball, and meson--meson configurations. For this reason, the $f_0(1500)$ and $f_0(1710)$ have long been discussed as states containing sizable glueball components rather than as unmixed members of one conventional nonet~\cite{CloseKirk2000}. One possible assignment of this mixed spectrum was proposed in Ref.~\cite{YasuiLeeLoSasaki2026},
\begin{equation} 
f_0(980)\sim\frac{u\bar u+d\bar d}{\sqrt{2}},\qquad a_0(980)\sim q\bar q,\qquad K_0^*(1430)\sim q\bar s,\qquad f_0(1810)\;(\text{formerly }f_0(1770))\sim s\bar s, 
\end{equation}
where the leading scalar-glueball component is assigned to $f_0(1500)$. In this proposal, the $f_0(980)$ has a predominantly nonstrange two-constituent component and the heavier $f_0(1810)$ is mainly the strange scalar partner. This is a concrete and testable assignment, but it is not yet a unique classification because the distribution of the scalar-glueball strength among the physical poles remains model dependent. We therefore use the purple bars for states that require a coupled quarkonium--glueball calculation and do not compare them directly with the pure $s\bar s$ Hamiltonian.

\paragraph{The $\boldsymbol{\phi(2170)}$:}
The $\phi(2170)$ lies above the clean low-lying strangeonium spectrum and has no universally accepted assignment.
Interpretations include an excited $s\bar s$ vector, an $s\bar s g$ hybrid, a compact multiquark state, and a pole reshaped by channels such as $\phi f_0(980)$ and strange-meson pairs.
Its observed decay pattern has not produced a consensus that would justify using it as a unique $3^3S_1$ or $2^3D_1$ reference.
It should therefore remain outside the clean calibration set and be interpreted only after explicit decay and coupled-channel dynamics are included.

\subsubsection{Open charm and charm--strange mesons}
\label{subsubsec:open_charm_meson_experiment}
The clearest benefit of the softened confinement appears in the open-charm radial and $D$-wave sectors.
For $q\bar c$, the $2^1S_0$ level moves from 2751.2~MeV in TH1 to 2620.9~MeV in TH2, and the $2^3S_1$ level moves from 2825.3 to 2680.4~MeV.
The corresponding experimental references are near 2539 and 2628~MeV.
The remaining discrepancies are not negligible, but the overestimate is reduced by roughly one hundred MeV.
The $1^3D_3$ branch changes from the TH1 triplet centroid at 2925.3~MeV to 2703.5~MeV, close to the $D_3^{*}(2750)$ region.
In the open-charm sector, the compression brings these levels closer to the reference masses.

The narrow $P$-wave states provide the main fine-structure constraints.
TH2 gives 2419.9~MeV for the lower mixed $1^+$ branch and 2429.1~MeV for the $1^3P_2$-dominated state.
The former reproduces the fitted $D_1(2420)$, whereas the latter underestimates the $D_2^{*}(2460)$ by about 32~MeV.
The orthogonal $1^+$ branch is predicted near 2514~MeV.

The $s\bar c$ sector displays a similar but cleaner pattern.
TH2 gives 2722.5~MeV for the $2^3S_1$ level, close to the $D_{s1}^{*}(2700)$ region, and 2814.9~MeV for the $1^3D_3$ branch, within several tens of MeV of the $D_{s3}^{*}(2860)$.
The narrow $D_{s1}(2536)$ and $D_{s2}^{*}(2573)$ are obtained at 2529.8 and 2544.1~MeV, respectively.

The open-flavor $1^+$ eigenvectors also illustrate the role of the antisymmetric spin--orbit term.
The calculated states are mixtures of $^1P_1$ and $^3P_1$, and the fit constrains different branches in the $D$ and $D_s$ systems.
A comparison based only on the unmixed TH1 centroids can appear numerically favorable in some cases, but it does not reproduce the physical eigenstate structure.
TH2 supplies the necessary mixing mechanism, although the residual masses show that a universal local $V^{ALS}$ interaction cannot substitute for the missing continuum dynamics.

\paragraph{The broad $\boldsymbol{D_0^*(2300)}$ and $\boldsymbol{D_1(2430)}$:}

The broad $D_0^{*}(2300)$ and $D_1(2430)$ couple to $D\pi$ and $D^{*}\pi$ in $S$ wave and are consequently poor tests of a quenched local potential.
For the scalar channel, modern unitarized analyses find that the observed enhancement is more naturally described by two nearby poles with different flavor couplings than by one elementary Breit--Wigner resonance~\cite{AlbaladejoEtAl2017D0TwoPole}.
This picture explains why a single calculated $1^3P_0$ mass cannot be compared unambiguously with the nominal experimental label.
The broad axial state has the analogous difficulty because strong $D^{*}\pi$ dressing and mixing between the two $1^+$ configurations influence both its pole position and width.
These states should therefore be shown as coupled-channel sensitive and excluded from the clean fine-structure calibration.

\paragraph{The subthreshold $\boldsymbol{D_{s0}^*(2317)}$ and $\boldsymbol{D_{s1}(2460)}$:}

The TH2 $1^3P_0$-dominated and lower mixed $1^+$ levels remain at 2446.7 and 2507.0~MeV, well above the physical $D_{s0}^{*}(2317)$ and $D_{s1}(2460)$.
Here the bare $c\bar s$ mass and the physical pole position are clearly different.
Lattice calculations that include both $c\bar s$ and $DK$ or $D^{*}K$ interpolating fields obtain the two physical states as subthreshold levels in a Hilbert space containing compact and two-hadron components~\cite{MohlerEtAl2013,LangEtAl2014}.
These results do not mean that the compact $c\bar s$ component is absent; the strong $S$-wave coupling to the nearby threshold is needed to obtain the observed masses.
We therefore show these states in blue, because the missing part is the channel coupling, which cannot be obtained by changing only the local $LS$, $ALS$, or tensor strengths.

\subsubsection{Charmonium series}
\label{subsubsec:charmonium_experiment}
The $c\bar c$ sector provides the cleanest interpretation of TH2.
The $1S$ and $2S$ levels are below or only moderately affected by the open-charm thresholds, and the $1P$ quartet has well-established quantum numbers.
TH2 gives 3414.9, 3515.0, 3519.6, and 3560.4~MeV for the states dominated by $1^3P_0$, $1^3P_1$, $1^1P_1$, and $1^3P_2$, respectively.
These values reproduce the $\chi_{c0}$, $\chi_{c1}$, $h_c$, and $\chi_{c2}$ masses at the few-MeV level.
The $1D$ states are predicted at approximately 3810.3, 3810.8, 3814.6, and 3818.1~MeV for the $1^3D_1$, $1^1D_2$, $1^3D_2$, and $1^3D_3$-dominated branches.
The $\mathcal J=2$ and 3 states agree well with the out-of-fit levels, while the $1^3D_1$ branch remains about 35--40~MeV above the physical $\psi(3770)$.
This is precisely the member that couples directly to open-charm $D\bar D$ and is therefore expected to receive the largest continuum self-energy within the multiplet~\cite{Barnes2008,Li2009}.

\paragraph{The $\boldsymbol{\chi_{c1}(3872)}$:}

The $\chi_{c1}(3872)$ lies within experimental resolution of the neutral $D^0\bar D^{*0}$ threshold and cannot be described by a quenched $2^3P_1$ eigenvalue alone. Its near-threshold properties require a large long-distance $D\bar D^{*}$ component, while the prompt production and radiative transitions also indicate a compact $c\bar c$ component~\cite{Swanson2006,Li2009,Ferretti2013,BrambillaEtAl2020XYZ}. It is therefore usually described as a threshold pole in a coupled compact--molecular space rather than as a pure molecule or a pure charmonium state. We interpret the TH2 $2P$ bar only as a possible bare seed and not as the physical pole mass.

\paragraph{The vector charmonium and $\boldsymbol{Y}$ region:}

The $\psi(4415)$ is retained as a gray conventional reference, since it is commonly associated predominantly with a high radial charmonium excitation, usually the $4^3S_1$ level, although open-charm dressing and $S$--$D$ mixing may modify this simple assignment~\cite{Godfrey1985,Barnes2005}.
We do not identify the successive bare $c\bar c$ eigenvalues one-to-one with the $Y(4230)$ and $Y(4360)$. 
The colors indicate candidate assignments. We show the $Y(4230)$ in purple because the charmonium-hybrid interpretation has been widely discussed~\cite{BrambillaEtAl2020XYZ,KouPene2005}. 
We show the $Y(4360)$ in red because multiquark and hadronic-composite interpretations have also been proposed~\cite{BrambillaEtAl2020XYZ,KouPene2005}.
These assignments do not exclude a sizable conventional $c\bar c$ component. 
In particular, a mainly canonical $3^3D_1$ assignment of the $Y(4360)$ has also been suggested~\cite{DingZhuYan2008}. 
Since both states have channel-dependent line shapes, we do not include them in the quantitative TH2 calibration.

\paragraph{The $\boldsymbol{J/\psi\phi}$ structures:}
The $J/\psi\phi$ channel contains several structures, including the states denoted by $X(4140)$, $X(4274)$, $X(4500)$, and $X(4700)$, but their conventional assignments are not settled. Amplitude analyses find several contributions with different quantum numbers, and the proposed interpretations include excited charmonia, compact tetraquarks, hadronic molecules, and threshold cusps~\cite{BrambillaEtAl2020XYZ,LHCbJpsiPhi2021}. States with nonexotic quantum numbers may contain a compact $c\bar c$ component. However, the number of observed structures and their strong hidden-strangeness decay channel do not allow a direct one-to-one comparison with the pure TH2 $c\bar c$ spectrum. We therefore discuss them as anomalous states and do not include them in a global quark-model RMSE.

\subsubsection{High-energy compression and overall assessment}
\label{subsubsec:meson_high_energy_assessment}
The compression of the upper TH2 spectrum can be seen directly by comparing the corresponding branches.
Representative examples are
\begin{align}
q\bar q:\quad &2^1S_0:\ 1399.6\rightarrow1206.5~\mathrm{MeV}, &&3^1S_0:\ 2231.1\rightarrow1782.5~\mathrm{MeV},\nonumber\\
&1^1G_4:\ 2566.8\rightarrow2070.6~\mathrm{MeV}, &&1^1L_8:\ 3953.8\rightarrow2930.2~\mathrm{MeV},\nonumber\\
q\bar c:\quad &2^1S_0:\ 2751.2\rightarrow2620.9~\mathrm{MeV},\nonumber\\
c\bar c:\quad &1^3D_2:\ 3807.2\rightarrow3814.6~\mathrm{MeV}, &&1^1L_8:\ 5079.6\rightarrow4790.9~\mathrm{MeV}.
\label{eq:meson_power_compression}
\end{align}
The low-lying charmonium orbital scale changes only slightly, while states with high radial excitation or large $l$ are lowered progressively. This is the expected effect of the softer confinement. However, the same interaction lowers the light and strange orbital states too much. Thus, the present comparison does not show that the $r^{2/3}$ confinement is better than the linear potential for every flavor and angular momentum.

We find four main results. First, TH2 simultaneously describes the charmonium $2S$ spacing, the $1P$ fine structure, and the $1D$ out-of-fit states. Second, it improves several radial and high-orbital open-charm and charm--strange states, while the threshold-dominated $D_{s0}^{*}(2317)$ and $D_{s1}(2460)$ still cannot be described by a quenched local potential. Third, the strangeonium radial vector state improves, but the $\phi_3(1850)$ orbital result becomes worse. Fourth, the light and kaonic orbital spectra are overcompressed and have too large spin--orbit splittings. Therefore, the improvement is clear in several heavy and heavy-light states but is not common to all six flavor sectors.

We do not compute a single RMSE across all experimental bars in  Fig.~\ref{fig:meson_excitation_spectra}. The gray states do not all have equally certain assignments, and the green, blue, red, and purple states contain dynamics that are absent from a quenched two-constituent Hamiltonian. We use TH2 as an exploratory extension. It shows that a softer confining power can reduce the excessive growth of many high-energy levels and that a full fixed-$\mathcal J^{P(C)}$ calculation is needed for the fine structure and open-flavor mixing. However, the instability in the light sector and the flavor dependence of the orbital compression also show that changing only the confining exponent is not sufficient. Additional constraints, relativistic dynamics, and explicit coupled channels are needed for a stable excitation Hamiltonian.

\subsection{Baryon excitations}
\label{subsec:baryon_excitations}

\begin{table*}[!t]
\centering
\caption{Yukawa-profile three-quark couplings obtained after fixing the corresponding meson-fitted two-body Hamiltonian and refitting only the ground-baryon set  $\{N,\Delta,\Sigma,\Lambda,\Sigma^*,\Lambda_c,\Sigma_c,\Sigma_c^*,\Xi_{cc}\}$.}
\label{tab:baryon_th1_th2_3q_coefficients}
\renewcommand{\arraystretch}{1.15}
\small
\begin{tblr}{
  colspec={
    Q[l,wd=6.7cm]
    Q[c,wd=2.5cm]
    Q[c,wd=2.5cm]
    Q[c,wd=2.5cm]
  },
  row{1} = {bg=headgray, font=\bfseries, halign=c, valign=m},
  cell{1}{1} = {}{halign=l},
  hline{2} = {1-4}{0.3pt},
  rowsep = 2.0pt,
}
\toprule
Two-body baseline
& $A$
& $B$
& $C$
\\

\textbf{TH1} or \textbf{$\rm S^2AJ$} (linear central $+$ color-spin)
& $-\bigl(136.27~\mathrm{MeV}\bigr)^2$
& $+\bigl(121.39~\mathrm{MeV}\bigr)^6$
& $-\bigl(133.37~\mathrm{MeV}\bigr)^4$
\\

\textbf{TH2} ($r^{2/3}$ central $+$ fine structure)
& $-\bigl(140.51~\mathrm{MeV}\bigr)^2$
& $+\bigl(131.92~\mathrm{MeV}\bigr)^6$
& $-\bigl(118.06~\mathrm{MeV}\bigr)^4$
\\

\bottomrule
\end{tblr}
\end{table*}

We calculate the baryon excitations with the two meson-calibrated Hamiltonians without fitting any baryon excitation. For each branch, we redetermine only the Yukawa-profile couplings $A$, $B$, and $C$ from the common ground-baryon set $\{N,\Delta,\Sigma,\Lambda,\Sigma^{*},\Lambda_c,\Sigma_c,  \Sigma_c^{*},\Xi_{cc}\}$. No radial or orbital baryon state is used in this fit. The individual magnitudes of the couplings change from TH1 to TH2, but the sign pattern $A<0$, $B>0$, and $C<0$ remains the same, as shown in Table~\ref{tab:baryon_th1_th2_3q_coefficients}. Thus, the spectra below are out-of-sample tests at larger radii, nonzero orbital angular momenta, and different spin couplings.

TH1 contains the linear central and color-spin interactions together with the TH1 Yukawa-profile three-quark correction. Since $V^T$, $V^{LS}$, and $V^{ALS}$ are absent, a fixed TH1 $(n,J^P,S)$ eigenvalue is repeated for all allowed physical angular momenta $\mathcal J=\vert J-S\vert ,\ldots,J+S$. TH2 uses the meson-fitted $r^{2/3}$ interaction, includes the diagonal tensor, symmetric spin--orbit, and antisymmetric spin--orbit matrix elements, and uses the separately refitted TH2 Yukawa couplings. Therefore, the displacement of a complete multiplet from TH1 to TH2 contains the changes of the central interaction, the meson-sector parameters, and the three-quark expectation value. The splitting among the TH2 descendants of one TH1 parent shows more directly the effect of the fine-structure interactions. Since off-diagonal mixing between different configurations with the same $\mathcal J^P$ is not included, the TH2 states are fine-structure-resolved bare levels and not fully mixed resonance poles.

\begin{figure*}[!t]
\centering
\includegraphics[width=\textwidth,height=0.90\textheight,keepaspectratio]{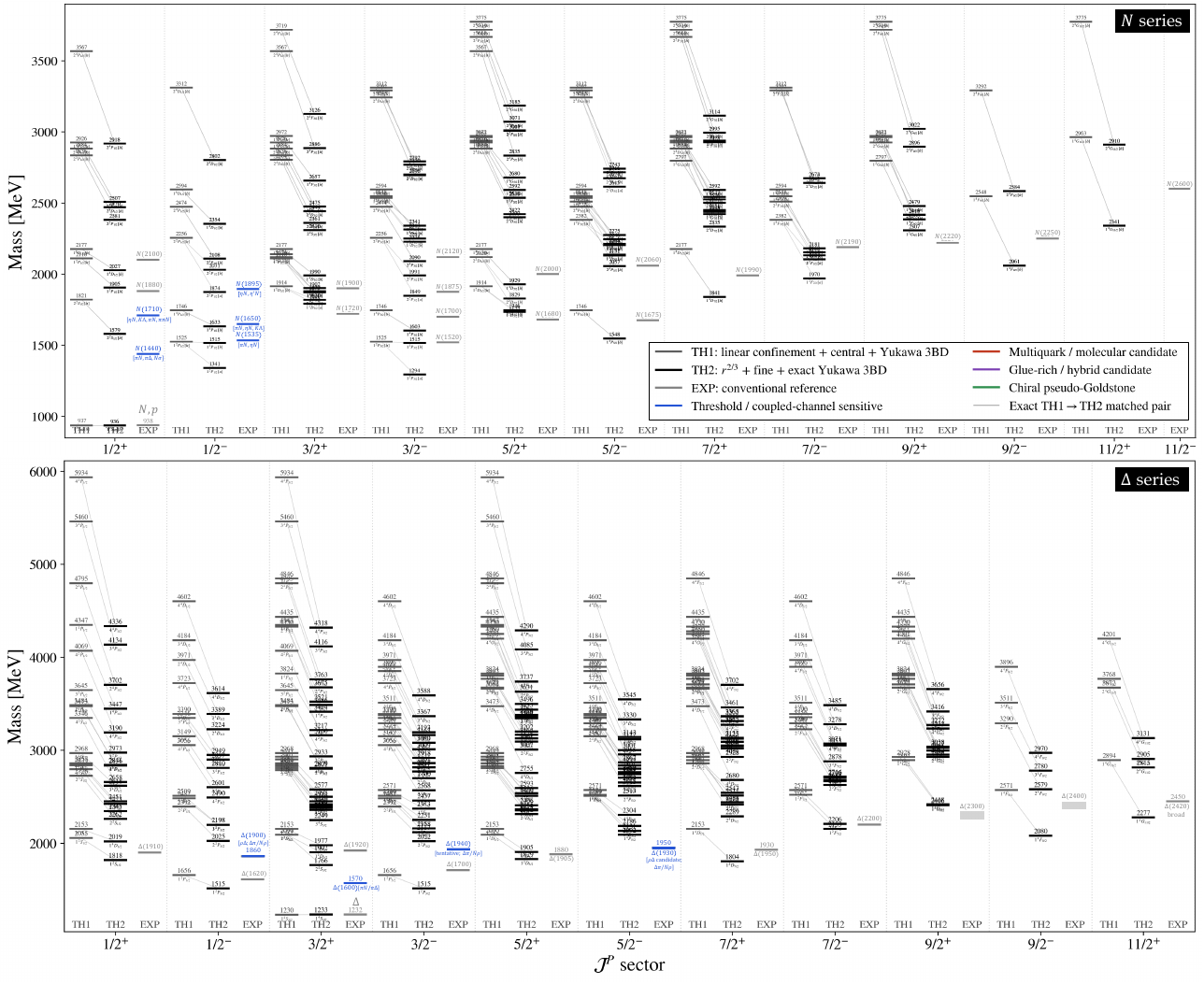}
\caption{\justifying (Enlarge online for details.) TH1, TH2, and reference $N$ and $\Delta$ spectra.}
\label{fig:N_Delta_spectra}
\end{figure*}

Figs.~\ref{fig:N_Delta_spectra}, \ref{fig:Lambda_Sigma_spectra}, \ref{fig:Xi_Omega_spectra}, \ref{fig:Lambda_c_Sigma_c_spectra}, \ref{fig:Xi_c_Omega_c_spectra}, and~\ref{fig:doubly_triply_charmed_spectra} show the baryon excitations in six flavor- and symmetry-related groups. These comparisons test the total-spin symmetry, light-pair symmetry, strange-quark-number dependence, light-diquark spin, flavor-$\overline{\mathbf 3}$--$\mathbf 6$ structure, and heavy-diquark systematics, respectively. The connecting lines only show which TH2 states originate from the same TH1 parent and do not denote physical transitions. In both calculations, the excited-baryon basis contains partial $S$-, $P$-, and $D$-wave components with $l_{\mathcal C},L_{\mathcal C}\leq2$. We use the same truncation for TH1 and TH2 because a larger basis greatly increases the computational cost of evaluating the TH2 fine-structure matrix elements and the Yukawa-profile three-quark integrals. Therefore, the highest angular-momentum states shown in the figures are not complete variational predictions. The Yukawa contribution also decreases rapidly with the excitation energy, so the upper spectrum mainly tests the two-body confinement and the spin-dependent interactions.

\subsubsection{The $N$ and ${\Delta}$ series}
\label{subsubsec:N_Delta_spectra}

The $N$ and $\Delta$ families contain the same nonstrange $qqq$ constituents but differ in isospin and spin-flavor permutation symmetry.
Their comparison therefore provides the most direct light-sector test of whether the same central and fine-structure interactions can describe distinct spin-flavor blocks.
Because both ground states belong to the baryon calibration set, their agreement with experiment mainly verifies the normalization of the two fitted branches.
TH1 and TH2 give 937.4 and 936.1~MeV for the nucleon and 1230.4 and 1232.6~MeV for the $\Delta$, respectively, compared with the experimental masses near 938 and $1232~\mathrm{MeV}$~\cite{PDG2026}.

The lowest nucleon negative-parity sector illustrates both the improvement and the instability of TH2.
Of the two TH1 $(J^P,S)=(1^-,1/2)$ roots at 1525.0~MeV, one becomes an almost degenerate TH2 pair at 1514.5 and 1514.6~MeV for $\mathcal J^P=1/2^-$ and $3/2^-$, whereas the other moves to 1341.1 and 1294.1~MeV.
The first pair has a realistic centroid for the $N(1535)$--$N(1520)$ region but does not reproduce the observed partner splitting, while the second has no established conventional counterpart.
The TH1 $(1^-,3/2)$ centroid at 1745.7~MeV is likewise lowered to TH2 branches at 1633.4, 1602.7, and 1548.2~MeV, giving a spin-weighted centroid near 1581~MeV.
TH2 therefore improves the lowest $S=1/2$ centroid but overcompresses the full $1P$ spectrum and produces too many negative-parity levels below $1.65~\mathrm{GeV}$.

The positive-parity nucleon spectrum shows a more systematic improvement.
The first radial excitation decreases from 1821.2 to 1579.3~MeV, and the lowest TH1 $(2^+,1/2)$ centroid at 1914.1~MeV becomes $3/2^+$ and $5/2^+$ branches at 1792.1 and 1735.1~MeV.
Their spin-weighted centroid is approximately 1758~MeV, and the predicted ordering agrees with that of the $N(1720)$--$N(1680)$ pair.
The higher TH1 $(2^+,3/2)$ centroid at 2176.7~MeV is resolved into TH2 branches at 2027.3, 1990.5, 1928.6, and 1840.5~MeV, which enter the $N(1880)$--$N(1900)$ region but remain too widely spread for a unique assignment.
The improvement is therefore clearest at the level of radial and orbital centroids rather than the complete physical fine structure~\cite{CapstickIsgur:1986,GlozmanPappPlessas1996,LoringMetschPetry2001,ZhongEtAl2024}.

For the $\Delta$ series, the lowest negative-parity multiplet behaves oppositely.
TH1 places its $(1^-,1/2)$ centroid at 1655.8~MeV, close to the $\Delta(1620)$--$\Delta(1700)$ region, whereas TH2 moves both descendants to approximately 1515~MeV with negligible splitting.
The modified central scale is therefore substantially less successful in this channel.
The positive-parity radial level is lowered from 2090.7 to 1765.8~MeV, while the TH1 $(2^+,3/2)$ centroid at 2153.3~MeV becomes TH2 branches at 2019.4, 1976.7, 1905.1, and 1804.3~MeV.
Their spin-weighted centroid near 1890~MeV is close to the experimental $\Delta(1910)$--$\Delta(1950)$ band, but the approximately 215~MeV internal spread is much too large and places the $7/2^+$ branch anomalously low.
The higher negative-parity TH2 branches at approximately 2.02--$2.20~\mathrm{GeV}$ improve upon their TH1 parents at 2.39--$2.51~\mathrm{GeV}$ but remain above the observed 1.90--$1.95~\mathrm{GeV}$ cluster.
The sectors with $\mathcal J\geq7/2$ are shown for continuity but are not used for detailed quantitative assessment because the basis does not contain complete $F$- and higher-orbital variational spaces.

\paragraph{The $\boldsymbol{N(1440)\,1/2^+}$ Roper resonance:}

TH2 removes most of the excessive TH1 radial excitation energy but still leaves the candidate bare level about 140~MeV above the nominal Roper mass.
Because the state has $(J^P,S)=(0^+,1/2)$, diagonal $LS$, $ALS$, and tensor terms cannot generate the remaining shift.
The residual discrepancy instead probes the central radial scale, relativistic and chiral flavor-spin dynamics, and continuum self-energies~\cite{SuzukiEtAl2010,GlozmanPappPlessas1996}.
Continuum-QCD, electrocoupling, and coupled-channel analyses consistently support a radial three-dressed-quark core accompanied by a substantial meson--baryon cloud rather than either a pure valence state or a pure molecule~\cite{SegoviaEtAl2015Roper,MokeevEtAl2023Electrocouplings,ZhouEtAl2026LeeFriedrichs}.
The TH2 value should therefore be interpreted as an improved bare-core scale whose physical pole requires additional dressing.

\paragraph{The $\boldsymbol{N(1535)\,1/2^-}$ and $\boldsymbol{N(1650)\,1/2^-}$ sector:}

The $N(1535)$ has important $\eta N$, $K\Lambda$, and $K\Sigma$ couplings, while valence calculations naturally supply nearby compact $1/2^-$ configurations for both resonances~\cite{KaiserSiegelWeise1995,GarzonOset2015}.
A unified coupled-channel treatment can reproduce the low negative-parity nucleons by dressing and mixing such bare $1P$ states through the $\pi N$, $\pi\Delta$, and $\eta N$ continua~\cite{ZhouEtAl2026LeeFriedrichs}.
Within the present spectrum, the TH2 pair near 1515~MeV provides a plausible lowest bare scale, whereas the additional 1341.1--1294.1~MeV pair is anomalously low.
Its approximately 215~MeV common displacement is far larger than its partner splitting and should be regarded as a pathology of the TH2 light-sector central branch rather than as evidence for undiscovered resonances.
A physical assignment therefore requires simultaneous mixing of the nearby quark configurations and continuum channels.

\paragraph{The $\boldsymbol{N(1675)\,5/2^-}$ electroexcitation anomaly:}
The principal anomaly of the $N(1675)$ is its sizeable proton electroexcitation despite the Moorhouse suppression of the direct three-quark transition.
The data consequently require an important non-valence contribution, naturally associated with meson--baryon dressing~\cite{AznauryanBurkert2015}.
TH1 provides a reasonable compact mass scale whereas the TH2 $5/2^-$ descendant is too low, but neither mass alone addresses this current-operator and wave-function test.

\paragraph{The positive-parity $\boldsymbol{N(1710)}$--$\boldsymbol{N(1880)}$ sector:}

The TH2 spectrum contains no isolated $1/2^+$ bare level naturally centered on the nominal $N(1710)$ mass.
A multichannel analysis instead obtains the $N(1710)$ close to the strange-channel thresholds with enhanced $\eta N$ and $K\Lambda$ residues, although an earlier fit of the same framework represented it through an explicit $s$-channel pole~\cite{RonchenEtAl2022KSigma}.
This change shows that the division between bare and dynamically generated components is parametrization dependent, so the state should be described as strongly channel sensitive rather than as an established pure molecule.
A TH2 branch near 1905~MeV can provide a compact candidate for the $N(1880)$ region, but the neighboring levels and omitted off-diagonal mixing prevent a unique radial or orbital assignment.

\paragraph{The $\boldsymbol{N(1875)\,3/2^-}$ and $\boldsymbol{N(1895)\,1/2^-}$ region:}

TH2 supplies plausible bare candidates at 1848.6 and 1873.8~MeV, but the two physical poles need not have the same dynamical origin.
The J\"ulich--Bonn analysis finds a possible very broad $N(1875)$ pole with strong $\pi\Delta$ coupling, although its existence and width remain uncertain~\cite{RonchenEtAl2022KSigma}.
The $N(1895)$ has more direct evidence for threshold sensitivity through the cusp at the opening of the $\eta'N$ channel and its substantial $\eta N$ and $\eta'N$ couplings~\cite{KashevarovEtAl2017EtaEtaPrime}.
The TH2 levels should therefore be viewed as candidate compact seeds whose pole positions and line shapes require the relevant channel dynamics.

\paragraph{The $\boldsymbol{\Delta(1620)\,1/2^-}$--$\boldsymbol{\Delta(1700)\,3/2^-}$ pair:}

These states are naturally associated with the lowest negative-parity $\Delta$ multiplet and do not require a predominantly exotic interpretation.
TH1 reproduces their gross centroid, whereas TH2 lowers both descendants to approximately 1515~MeV and generates almost no splitting.
Although $\pi N$, $\pi\Delta$, and $\rho N$ channels affect their poles, no established coupled-channel interpretation requires such a large common downward shift~\cite{ZhongEtAl2024,KamanoEtAl2009}.
So, the TH2 central interaction fails for this pair.

\paragraph{The $\boldsymbol{\Delta(1600)\,3/2^+}$:}

The $\Delta(1600)$ is the $\Delta$ analogue of the Roper problem.
TH2 substantially improves the radial bare scale but remains roughly 170--200~MeV above the nominal resonance region, and its $J=0$ parent cannot be lowered further by diagonal spin--orbit or tensor terms.
Electrocoupling, lattice, and Hamiltonian effective-field-theory studies support a compact radial core strongly dressed by $\pi N$ and $\pi\Delta$ rescattering~\cite{MokeevEtAl2023Electrocouplings,HockleyEtAl2024,HockleyEtAl2025}.
The J\"ulich--Bonn assignment also changes between a dynamically generated and a bare-pole description across fits, reinforcing that the separation of the two components is not unique~\cite{RonchenEtAl2022KSigma}.
TH2 should therefore be regarded as an improved bare scale rather than a complete description of the physical pole.

\paragraph{The negative-parity $\boldsymbol{\Delta(1900)}$--$\boldsymbol{\Delta(1940)}$ cluster:}
The approximate degeneracy of the $\Delta(1900)\,1/2^-$, $\Delta(1940)\,3/2^-$, and $\Delta(1930)\,5/2^-$ is not reproduced by either compact spectrum.
TH2 substantially lowers the corresponding TH1 parents but leaves the nearest branches too high and too widely separated.
A $\rho\Delta$-driven interpretation has been proposed in which the $\Delta(1930)$ is predominantly bound and the other two states contain important $\rho\Delta$ components~\cite{GonzalezOsetVijande2009}.
More recent multichannel analyses retain strong $\rho\Delta$, $\pi\Delta$, and strange-channel sensitivity but do not establish a unique dynamical assignment~\cite{RonchenEtAl2022KSigma}.
The residual excess is therefore a genuine limitation of the static valence Hamiltonian rather than merely a missing diagonal fine-structure correction.

\subsubsection{The ${\Lambda}$ and ${\Sigma}$ series}
\label{subsubsec:Lambda_Sigma_spectra}

\begin{figure*}[!t]
\centering
\includegraphics[width=\textwidth,height=0.90\textheight,keepaspectratio]{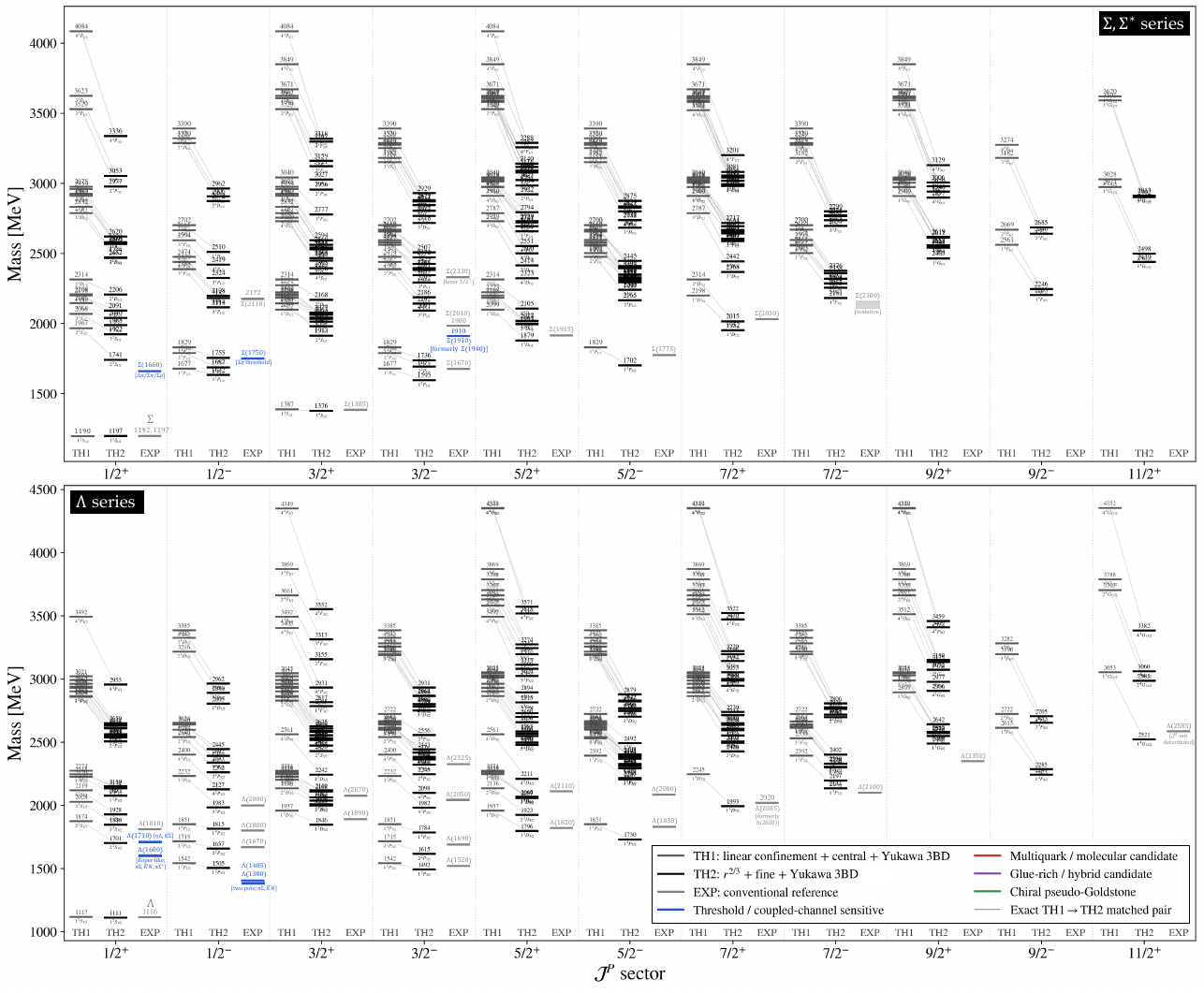}
\caption{\justifying (Enlarge online for details.) TH1, TH2, and reference $\Lambda$ and $\Sigma$ spectra.}
\label{fig:Lambda_Sigma_spectra}
\end{figure*}

The $\Lambda$ and $\Sigma$ series contain the same $uds$ quark content but differ in the symmetry and spin of the light pair.
The ground states remain accurately reproduced: TH1 and TH2 give 1117.5 and 1111.0~MeV for the $\Lambda$, and 1189.5 and 1197.2~MeV for the $\Sigma$.
The corresponding $\Sigma^{*}$ masses are 1386.9 and 1375.9~MeV.
The changes in the excitation spectrum are therefore not consequences of a failed ground-state transfer.

For the conventional $\Lambda$ negative-parity spectrum, the lowest TH1 $(1^-,1/2)$ centroid at 1542.5~MeV becomes TH2 branches at 1505.3 and 1491.9~MeV.
The next TH1 centroid at 1714.8~MeV moves to 1656.9 and 1615.1~MeV, while the $(1^-,3/2)$ centroid at 1850.9~MeV is resolved into 1814.8, 1784.1, and 1730.0~MeV.
TH2 consequently covers the $\Lambda(1520)$, $\Lambda(1670)$--$\Lambda(1690)$, and $\Lambda(1800)$--$\Lambda(1830)$ regions more compactly than TH1, although the diagonal ordering should not be interpreted without the omitted inter-configuration mixing~\cite{Loring2001,CapstickIsgur:1986,PDG2026}.

The $\Sigma$ negative-parity spectrum shows a similar but less uniform compression.
The first TH1 $(1^-,1/2)$ centroid at 1676.8~MeV becomes 1632.4 and 1595.1~MeV, while the next parent at 1788.3~MeV becomes 1687.3 and 1692.4~MeV.
The TH1 $(1^-,3/2)$ centroid at 1829.2~MeV is lowered to a TH2 triplet at 1755.3, 1735.9, and 1701.6~MeV.
The second doublet and the triplet overlap the $\Sigma(1670)$--$\Sigma(1775)$ region, whereas the lowest TH2 doublet is somewhat overcompressed.
The comparison favors neither branch uniformly: TH1 gives a cleaner lowest centroid, while TH2 produces a more realistic density across the full observed negative-parity band.

These branches overlap the $\Sigma(1915)$ and $\Sigma(2030)$ region substantially better than the TH1 centroids.

\paragraph{Recent experimental constraints on the higher $\boldsymbol{\Sigma}$ spectrum:}

An earlier BESIII partial-wave analysis of $\psi(3686)\to\Lambda\bar{\Sigma}^{0}\pi^{0}+\mathrm{c.c.}$ found significant $\Lambda(2325)$ and $\Sigma(1910)$ components, together with the lower $\Lambda^{*}$ and $\Sigma^{*}$ states displayed in  Fig.~\ref{fig:Lambda_Sigma_spectra}~\cite{BESIIILambdaSigmaPWA2025}.
A later BESIII analysis of $\psi(3686)\to\bar{p}K^{+}\Sigma^{0}+\mathrm{c.c.}$ requires the one-star $\Sigma(2010)$ and $\Sigma(2110)$ components and favors $\mathcal J^P=3/2^-$ and $1/2^-$, with fitted masses of $1980.1\pm10.0\pm18.1~\mathrm{MeV}$ and $2172.4\pm6.3\pm10.9~\mathrm{MeV}$, respectively~\cite{BESIIISigma2330_2026}.
The same analysis reports a new $\Sigma(2330)$ structure with $M=2334.7\pm7.9\pm16.0~\mathrm{MeV}$, $\Gamma=206.3\pm9.5\pm18.4~\mathrm{MeV}$, and a statistical significance of $11.9\sigma$, with $\mathcal J^P=3/2^-$ favored~\cite{BESIIISigma2330_2026}.
The $3/2^+$ and $5/2^-$ alternatives remain nearly competitive, so these higher experimental bars are retained as qualitative assignment-sensitive references and are not used for one-to-one TH1 or TH2 matching.

The positive-parity orbital levels benefit more clearly from TH2.
The lowest $\Lambda$ $(2^+,1/2)$ centroid moves from 1956.7~MeV to 1846.0 and 1796.4~MeV, close to the $\Lambda(1890)$ and $\Lambda(1820)$ pair.
The lowest $\Sigma$ $(2^+,1/2)$ centroid moves from 2099.5~MeV to 1980.7 and 1878.6~MeV, and the $(2^+,3/2)$ centroid at 2198.5~MeV becomes a quartet spanning 1952.2--2090.6~MeV.

\paragraph{The $\boldsymbol{\Lambda(1405)}$:}

Neither Hamiltonian produces the exceptionally low $\Lambda(1405)$.
The lowest TH2 $1/2^-$ branch remains near 1505~MeV, even though it lies below the corresponding TH1 centroid.
This failure is expected for a state whose line shape and pole structure are governed strongly by the coupled $\bar K N$ and $\pi\Sigma$ channels and are commonly described by a two-pole amplitude~\cite{Jido2003}.
The $\Lambda(1405)$ should therefore remain a threshold-sensitive reference rather than be counted as an ordinary three-quark level missed by roughly 100~MeV.

\paragraph{The Roper-like $\boldsymbol{\Lambda(1600)}$ and $\boldsymbol{\Sigma(1660)}$:}

The first $\Lambda$ radial state falls from 1873.8 to 1701.3~MeV, and the first $\Sigma$ radial state falls from 1966.7 to 1741.1~MeV.
TH2 removes approximately 170--225~MeV of the TH1 overestimate, but the levels remain about 100 and 80~MeV above the nominal $\Lambda(1600)$ and $\Sigma(1660)$ references.
Because these are predominantly $J=0$ radial states, the displacement is caused by the softened central interaction and the refit rather than by tensor or $LS$ splitting.
The residual offsets remain consistent with the known sensitivity of light and light--strange radial excitations to flavor-spin interactions, relativistic dynamics, and continuum dressing~\cite{Loring2001,CapstickIsgur:1986,Melde2008,FaustovGalkin2015}.

\subsubsection{The ${\Xi}$ and ${\Omega}$ series}
\label{subsubsec:Xi_Omega_spectra}

\begin{figure*}[!t]
\centering
\includegraphics[width=\textwidth,height=0.90\textheight,keepaspectratio]{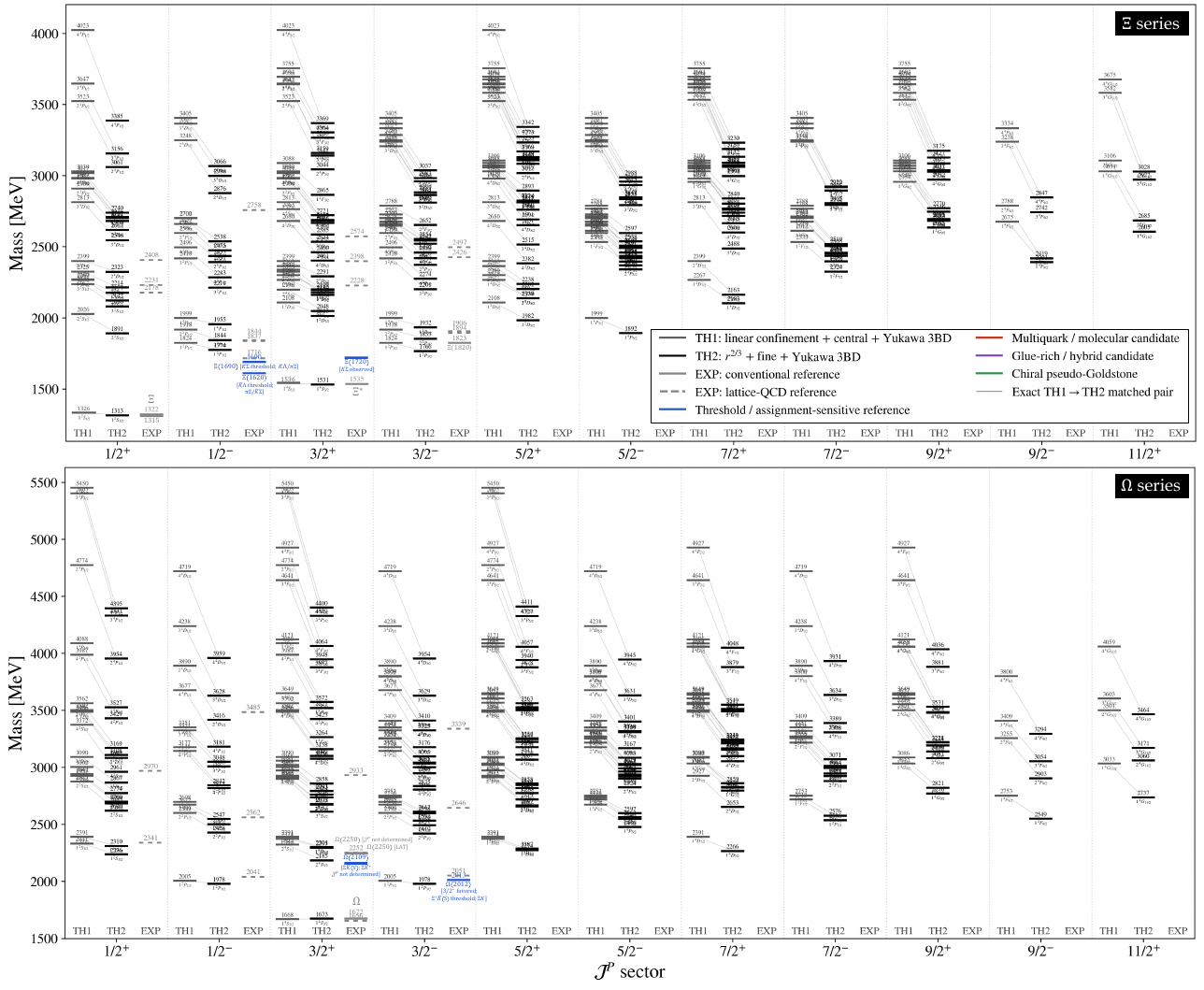}
\caption{\justifying (Enlarge online for details.) TH1, TH2, and reference $\Xi$ and $\Omega$ spectra.}
\label{fig:Xi_Omega_spectra}
\end{figure*}

The $\Xi$ and $\Omega$ spectra test how the excitation pattern changes as the number of strange quarks increases and the light-quark hyperfine interaction is suppressed.
Both ground-state multiplets remain at the expected mass scale under the TH1--TH2 replacement.
For the $\Xi$, TH1 and TH2 give 1326.2 and 1313.3~MeV for the $1/2^+$ ground state and 1535.6 and 1531.0~MeV for the $3/2^+$ partner.
Relative to the experimental masses near 1321 and 1535~MeV, TH1 is closer for both members of the ground-state multiplet.
For the $\Omega$, the corresponding ground-state masses are 1667.5 and 1672.9~MeV.

The lowest conventional $\Xi$ negative-parity branches remain substantially above the lightest reported $\Xi$ enhancements.
The first TH1 $(1^-,1/2)$ centroid at 1824.2~MeV becomes TH2 $1/2^-$ and $3/2^-$ states at 1774.1 and 1765.9~MeV.
A second parent at 1918.0~MeV becomes 1843.8 and 1852.7~MeV, while the $(1^-,3/2)$ centroid at 1998.6~MeV becomes a triplet spanning 1892.4--1954.9~MeV.
The TH1 centroid reproduces the $\Xi(1820)$ mass almost exactly, whereas the lowest TH2 $1/2^-$--$3/2^-$ pair lies about 50--60~MeV lower.
Both compact spectra nevertheless remain far above the $\Xi(1620)$ and $\Xi(1690)$ and therefore do not support identifying these two low-mass structures with the lowest local three-quark $1P$ multiplet.

The first $\Xi$ radial state moves from 2025.7 to 1890.8~MeV, and the lowest $(2^+,1/2)$ centroid moves from 2107.8~MeV to 2013.0 and 1981.9~MeV.
The $(2^+,3/2)$ centroid at 2266.8~MeV becomes a quartet spanning 2102.7--2177.0~MeV.
These values lie inside the broad and assignment-dependent region occupied by the $\Xi(1950)$, $\Xi(2030)$, and higher structures, so the mass comparison alone does not provide unique spectroscopic identifications.

The $\Omega$ series provides a clean mass-scale comparison for its lowest negative-parity parent, but not a unique state assignment.
TH1 places the $(1^-,1/2)$ centroid at 2005.5~MeV, while TH2 gives nearly degenerate $1/2^-$ and $3/2^-$ descendants at 1978.0~MeV.
The observed $\Omega(2012)$ is therefore reproduced within approximately 7~MeV by the TH1 compact scale and lies approximately 35~MeV above the TH2 doublet.
In the positive-parity sector, the symmetric radial $(0^+,3/2)$ state moves from 2321.8 to 2184.8~MeV, and the lowest symmetric $D$-wave quartet moves from the TH1 centroid 2390.5~MeV to 2265.8--2309.9~MeV.
For a radial assignment of the $\Omega(2250)$, the absolute TH1 and TH2 mass deviations are nearly equal, approximately 70 and 67~MeV, respectively.
By contrast, the lowest TH2 $D$-wave branches lie approximately 14--58~MeV above the PDG average and therefore populate the $\Omega(2250)$ region more closely.
Because the experimental $\Omega(2250)$ has no established quantum numbers, the radial and $D$-wave levels should be treated as alternative compact candidates rather than unique assignments~\cite{Loring2001,PDG2026,FaustovGalkin2015,HockleyOmegaLattice2025,LiuWangLuZhong2020}.

\paragraph{The $\boldsymbol{\Xi(1620)}$ and $\boldsymbol{\Xi(1690)}$:}

The failure of both TH1 and TH2 to produce compact negative-parity levels near 1.62--$1.69~\mathrm{GeV}$ is not an isolated peculiarity of the present calculation.
The $\Xi(1620)$ remains a two-star state with no established $\mathcal J^P$, whereas the recent BESIII partial-wave analysis determines $\mathcal J^P=1/2^-$ for the three-star $\Xi(1690)$; the $\Xi(1820)$ remains the cleanest established $3/2^-$  Ref.~\cite{PDG2026,BESIIIXi2024,LHCbXi2021}.
The extracted widths remain analysis dependent, so the improved spin-parity assignment does not remove the importance of coupled-channel dynamics.
Unitarized chiral calculations can generate the $\Xi(1620)$ and $\Xi(1690)$ dynamically from the coupled $\pi\Xi$, $\bar K\Lambda$, and $\bar K\Sigma$ channels~\cite{MiyaharaEtAl2017Xi,FeijooValcarceMagas2023}.
The blue classification consequently denotes strong channel sensitivity and does not imply a unique purely molecular assignment.

\paragraph{The post-PDG $\boldsymbol{\Xi(1720)}$:}
BESIII has reported a new $\Xi(1720)$ structure in the $K^-\Sigma^0$ channel with a mass of $1721.0\pm5.2_{\rm stat}\pm3.4_{\rm syst}~\mathrm{MeV}$ and a width of $31.3\pm18.3_{\rm stat}\pm15.4_{\rm syst}~\mathrm{MeV}$, with $\mathcal J^P=3/2^+$ favored~\cite{BESIIIXi1720_2026}.
The favored assignment places this state far below the first excited $3/2^+$ levels in both TH1 and TH2.
It is therefore displayed as an open-blue assignment-sensitive reference and excluded from quantitative state matching.
Its observation in the $K\Sigma$ channel does not by itself establish an $S$-wave threshold-generated interpretation, because a $3/2^+\to1/2^++0^-$ transition requires odd relative orbital angular momentum.

\paragraph{The $\boldsymbol{\Omega(2012)}$ and $\boldsymbol{\Omega(2109)}$:}

The TH1 mass of 2005.5~MeV shows that a compact $1P$ seed near the $\Omega(2012)$ mass is plausible, whereas the TH2 $1/2^-$ and $3/2^-$ descendants are nearly degenerate at 1978.0~MeV.
The experimental spin of the $\Omega(2012)$ remains undetermined, although lattice-QCD and Hamiltonian EFT studies favor a $3/2^-$ assignment~\cite{HockleyOmegaLattice2025,HanOmegaHEFT2025}.
Its proximity to the $\Xi(1530)\bar K$ threshold also indicates that the physical pole can receive a sizable coupled-channel contribution, so the TH1 result should be interpreted as a candidate bare-core scale rather than as a complete description of the state~\cite{LiuWangLuZhong2020,BelleOmega2012Decay2025,HuPing2022}.
BESIII has additionally reported an $\Omega(2109)^-$ structure, but its spin and parity are not known~\cite{BESIIIOmega2109_2025}.
If the $\Omega(2012)$ and $\Omega(2109)$ are identified with $3/2^-$ and $1/2^-$ poles, respectively, as suggested in a recent Hamiltonian effective-field-theory analysis, the nearly degenerate TH2 pair does not reproduce their approximately 96~MeV separation~\cite{HanOmegaHEFT2025}.
An alternative quark-model interpretation assigns the $\Omega(2109)$ to the $2S$, $3/2^+$ radial excitation~\cite{LuoLiuOmega2025}.
Under this assignment, the TH2 radial level at 2184.8~MeV is substantially closer to the observed mass than the TH1 value at 2321.8~MeV, although it remains approximately 76~MeV too high.
The two structures therefore provide useful tests of the compact spectrum, but their unique identification requires experimental spin assignments and an explicit coupled-channel treatment.

\subsubsection{The ${\Lambda_c}$ and ${\Sigma_c}$ series}
\label{subsubsec:Lambda_c_Sigma_c_spectra}

\begin{figure*}[!t]
\centering
\includegraphics[width=\textwidth,height=0.90\textheight,keepaspectratio]{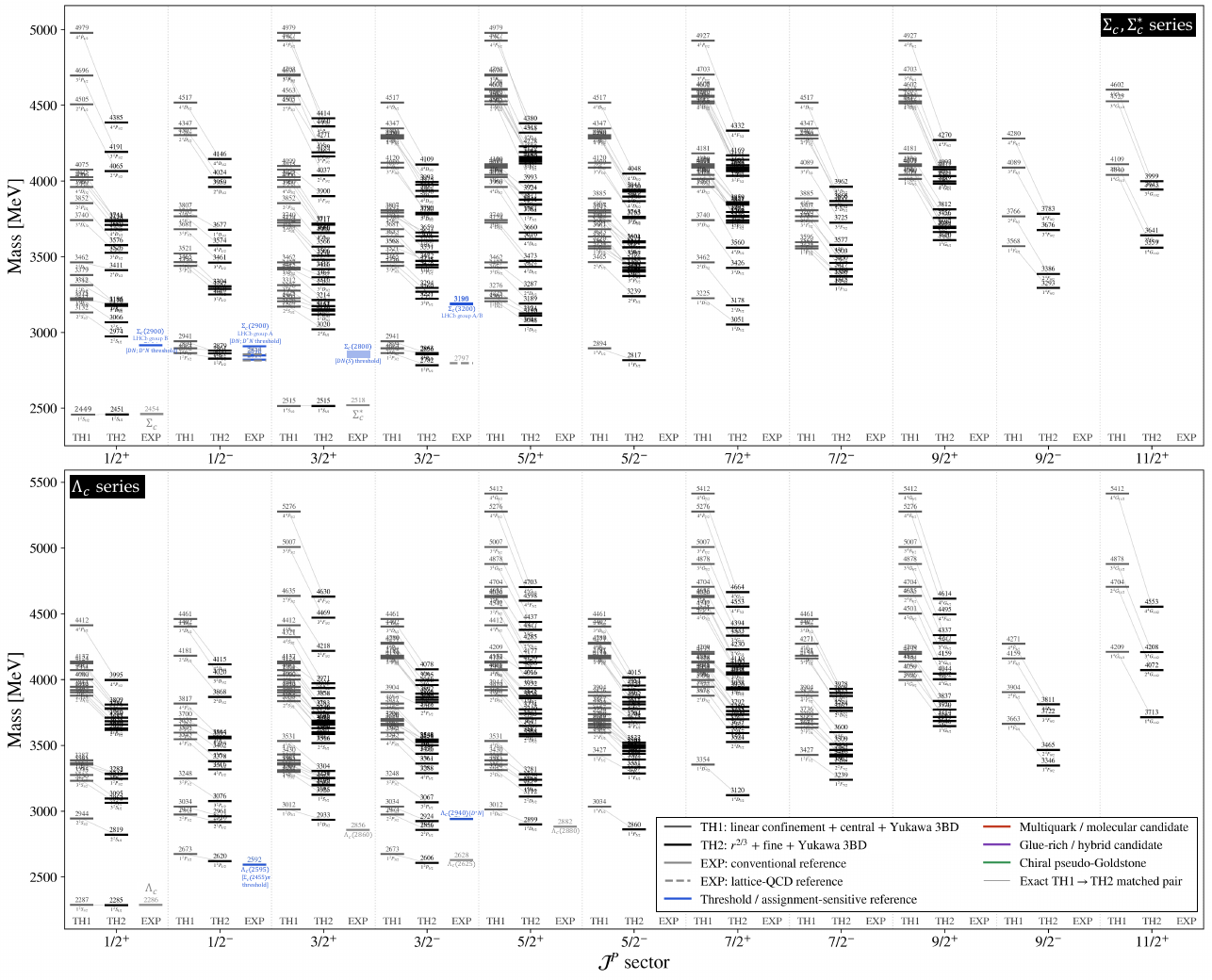}
\caption{\justifying (Enlarge online for details.) TH1, TH2, and reference $\Lambda_c$ and $\Sigma_c$ spectra.}
\label{fig:Lambda_c_Sigma_c_spectra}
\end{figure*}

The $\Lambda_c$ and $\Sigma_c$ both have quark content $udc$, with light-pair spins $s_\ell=0$ and $s_\ell=1$, respectively.
Both ground-state doublet structures remain well controlled by the calibration.
TH1 and TH2 give 2287.8 and 2284.8~MeV for the $\Lambda_c$, 2449.8 and 2450.7~MeV for the $\Sigma_c$, and 2524.7 and 2514.8~MeV for the $\Sigma_c^{*}$, respectively, compared with the corresponding experimental masses near 2286.5, 2454, and $2518.5~\mathrm{MeV}$~\cite{PDG2026}.
The excitation differences therefore probe the long-distance and spin-dependent dynamics rather than a failure of the charmed ground-state fit.

For the lowest $\Lambda_c$ negative-parity parent, TH1 gives a $(1^-,1/2)$ centroid at 2673.0~MeV.
TH2 resolves it into 2620.2~MeV for $\mathcal J^P=1/2^-$ and 2606.4~MeV for $3/2^-$, with a spin-weighted centroid of 2611.0~MeV.
The experimental $\Lambda_c(2595)$--$\Lambda_c(2625)$ centroid is approximately 2616~MeV.
TH2 therefore reduces the centroid error from roughly 57~MeV to about 5~MeV and lies within the 2.59--$2.64~\mathrm{GeV}$ range obtained in representative relativistic and non-relativistic calculations~\cite{CapstickIsgur:1986,Ebert2011,Yoshida2015,Roberts2008,ZhangXiaoZhong2025}.

The first $\Lambda_c$ radial state moves from 2943.5 to 2819.3~MeV, reducing its discrepancy from the $\Lambda_c(2765)$ region to roughly 50~MeV.
The lowest $(2^+,1/2)$ centroid moves from 3012.0~MeV to 2932.8 and 2899.2~MeV for the $3/2^+$ and $5/2^+$ branches.
Their spin-weighted centroid, 2912.6~MeV, remains about 41~MeV above the $\Lambda_c(2860)$--$\Lambda_c(2880)$ centroid. It is much closer to the observed centroid than the TH1 prediction is.
The higher negative-parity parents form a TH2 cluster between approximately 2.86 and $2.96~\mathrm{GeV}$, overlapping the $\Lambda_c(2910)$--$\Lambda_c(2940)$ region and conventional $2P$ bare-core predictions~\cite{PDG2026,ZhangXiaoZhong2025,BelleLambdaC2910}.

The $\Sigma_c$ spectrum is likewise compressed, but the updated experimental situation requires a solution-dependent comparison.
The first TH1 $(1^-,1/2)$ centroid at 2862.3~MeV becomes TH2 branches at 2827.0 and 2782.1~MeV.
The TH1 $(1^-,3/2)$ centroid at 2894.3~MeV becomes a triplet at 2879.5, 2856.2, and 2816.5~MeV, while the next TH1 parent at 2940.6~MeV produces a nearly degenerate TH2 doublet near 2862~MeV.
The recent LHCb amplitude analysis resolves the previously asymmetric enhancement in the 2.8--$2.9~\mathrm{GeV}$ region into two overlapping neutral structures, $\Sigma_c(2800)^0$ and $\Sigma_c(2900)^0$, and finds two statistically indistinguishable solution families~\cite{LHCbSigmaC2026}.
The best group-A solution assigns $\mathcal J^P=(3/2^+,1/2^-,3/2^-)$ to $\Sigma_c(2800)^0$, $\Sigma_c(2900)^0$, and $\Sigma_c(3200)^0$, with masses 2819, 2908, and 3186~MeV, whereas the best group-B solution assigns $(1/2^-,1/2^+,3/2^-)$ with masses 2848, 2914, and 3190~MeV, respectively~\cite{LHCbSigmaC2026}.
The low TH2 negative-parity band can therefore be compared at the gross-mass level with the group-B $1/2^-$ assignment of $\Sigma_c(2800)^0$ and the group-A $1/2^-$ assignment of $\Sigma_c(2900)^0$, but this comparison is conditional on the selected amplitude solution.
The first positive-parity radial states are lowered from 3131.6 and 3170.1~MeV to 2973.9 and 3020.4~MeV.
The TH2 $1/2^+$ radial branch lies approximately 60~MeV above the group-B $\Sigma_c(2900)^0$ solution, whereas the $3/2^+$ radial branch lies approximately 200~MeV above the group-A $\Sigma_c(2800)^0$ solution.
The lowest positive-parity $D$-wave branches span approximately 3.05--$3.18~\mathrm{GeV}$, but they should not be identified directly with $\Sigma_c(3200)^0$, for which both best LHCb solutions prefer $\mathcal J^P=3/2^-$.
Within those best solutions, the relevant compact comparison for $\Sigma_c(3200)^0$ is instead provided by the higher negative-parity TH2 spectrum.
Thus, TH2 supplies several candidate bare levels in the observed mass regions, but the present data do not permit a unique state-by-state mapping~\cite{Yoshida2015,LHCbSigmaC2026,BahtiyarEtAl2020,LiYuWangGu2024}.

\paragraph{The lowest $\boldsymbol{\Lambda_c(2595)}$--$\boldsymbol{\Lambda_c(2625)}$ doublet:}

TH2 improves the centroid and reverses the doublet ordering.
TH2 predicts $M(3/2^-)-M(1/2^-)=-13.9~\mathrm{MeV}$, whereas the experimental ordering is approximately $+35.9~\mathrm{MeV}$~\cite{PDG2026}.
The softened central interaction and the refitted three-quark contribution therefore correct the gross $1P$ scale, while the diagonal $LS$ and $ALS$ terms invert the doublet and underestimate its separation.
Since the calculation omits off-diagonal mixing among configurations with the same $\mathcal J^P$, the discrepancy should be assigned to the current fine-structure implementation rather than to the underlying centroid alone.

\paragraph{The $\boldsymbol{\Lambda_c(2765)}$ and the $\boldsymbol{1D}$ doublet:}

The TH2 radial mass is no longer a severe outlier, but a $\lambda$--$\rho$ mode decomposition and radial-node analysis are still required before identifying it uniquely with the $\Lambda_c(2765)$~\cite{ZhangXiaoZhong2025}.
The TH2 $1D$ centroid is likewise brought into the upper part of the common 2.84--$2.91~\mathrm{GeV}$ compact-model range~\cite{Ebert2011,Yoshida2015,Roberts2008,ZhangXiaoZhong2025}.
Its partner ordering is nevertheless reversed: TH2 gives $M(5/2^+)-M(3/2^+)=-33.6~\mathrm{MeV}$, whereas the experimental difference is approximately $+25.5~\mathrm{MeV}$.
The radial and orbital centroid corrections are therefore credible, but the individual fine-structure assignments require a full mixed-$\mathcal J^P$ diagonalization.

\paragraph{The threshold-adjacent $\boldsymbol{\Lambda_c(2940)}$:}

The dense TH2 $2P$ region precludes a unique assignment even before continuum dressing is included.
The $\Lambda_c(2940)$ lies close to the $D^{*}N$ threshold, and unquenched calculations find that the channel can lower a compact core by several tens of MeV and can alter the $1/2^-$--$3/2^-$ ordering~\cite{Luo2020,Zhang2023}.
The TH2 levels in this interval should therefore be interpreted as candidate bare configurations rather than physical pole positions.
Lattice calculations broadly support the expected low-lying positive- and negative-parity ordering but also show that a quantitative near-threshold assignment ultimately requires meson--baryon interpolators and a finite-volume scattering analysis~\cite{BahtiyarEtAl2020,BaliCollinsPerezRubio2015}.

\subsubsection{The ${\Xi_c}$, ${\Xi_c^{\prime}}$, and ${\Omega_c}$ series}
\label{subsubsec:Xi_c_Omega_c_spectra}

\begin{figure*}[!t]
\centering
\includegraphics[width=\textwidth,height=0.90\textheight,keepaspectratio]{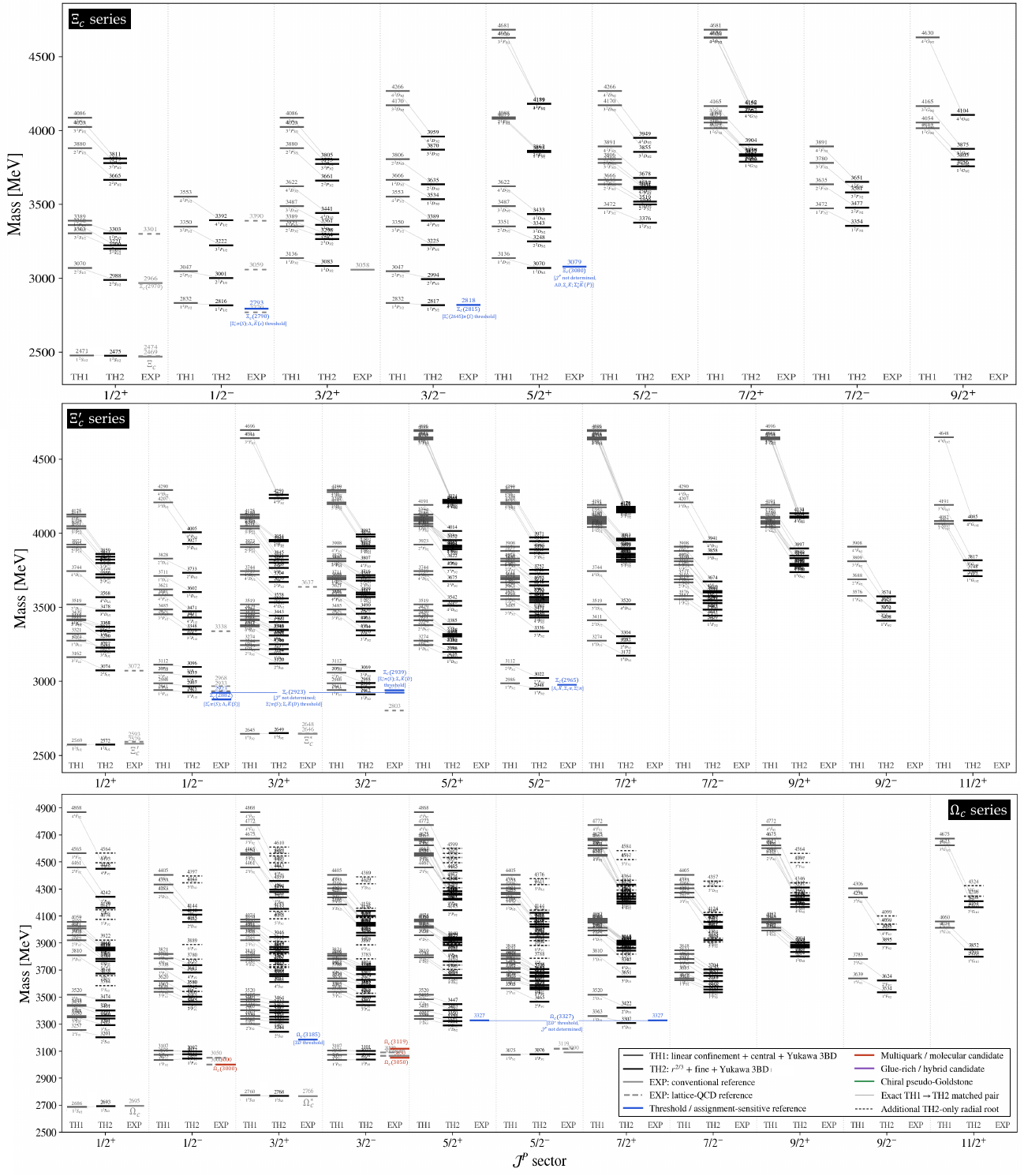}
\caption{\justifying (Enlarge online for details.) TH1, TH2, and reference $\Xi_c$, $\Xi_c^{\prime}$, and $\Omega_c$ spectra.}
\label{fig:Xi_c_Omega_c_spectra}
\end{figure*}

For the primed and unprimed $\Xi_c$ families, we adopt the pure pair-spin assignments of Eq.~\eqref{eq:spin_bases_rho_lambda}.
The unprimed series is identified with $\chi_\rho$, or $s_{qs}=0$, and the primed series with $\chi_\lambda$, or $s_{qs}=1$, while the $\Xi_c^{*}$ contains the aligned $S=3/2$ configuration.
Off-diagonal $\chi_\rho$--$\chi_\lambda$ mixing is omitted in this comparison.
Off-diagonal mixing among distinct orbital--spin parents with the same $\mathcal J^P$, including the two $\lambda$-mode sextet configurations with $\mathcal J^P=3/2^-$, is also omitted.
A complete treatment would therefore require a simultaneous diagonalization of all pair-spin and orbital--spin configurations in each common $\mathcal J^P$ block, so the present labels should be read as unmixed-basis labels rather than exact conserved quantum numbers.

The ground states are reproduced at the expected few-to-ten-MeV level.
For the unprimed $\Xi_c$, TH1 and TH2 give 2471.4 and 2475.0~MeV.
For the $\Xi_c^{\prime}$ they give 2569.3 and 2572.1~MeV, and for the $\Xi_c^{*}$ they give 2645.3 and 2649.3~MeV.
Relative to the experimental masses near 2469, 2579, and 2646~MeV, TH1 is closer for the unprimed $\Xi_c$ and the aligned $\Xi_c^{*}$, whereas TH2 remains closer for the $\Xi_c^{\prime}$ ground state.
For the $\Omega_c$ and $\Omega_c^{*}$ ground states, TH1 gives 2686.9 and 2760.5~MeV, while TH2 gives 2692.9 and 2768.2~MeV.
Both branches are already within about 8~MeV of the experimental masses near 2695 and 2766~MeV, with TH2 giving the smaller absolute deviations.

The lowest unprimed $\Xi_c$ negative-parity parent moves from the TH1 centroid 2832.4~MeV to TH2 branches at 2815.6 and 2817.4~MeV.
Its spin-weighted centroid is therefore close to the $\Xi_c(2790)$--$\Xi_c(2815)$ region.
The $\Xi_c^{\prime}$ parent at 2941.0~MeV becomes 2924.6 and 2912.5~MeV, while the $\Xi_c^{*}$ parent at 2985.7~MeV becomes a triplet at 2966.8, 2960.3, and 2948.5~MeV.
Together these sextet branches populate the same interval as the narrow $\Xi_c(2923)^0$, $\Xi_c(2939)^0$, and $\Xi_c(2965)^0$ structures.
The first radial levels are also compressed, from 3069.9 to 2987.9~MeV in the unprimed series, from 3162.5 to 3074.2~MeV in the primed series, and from 3213.5 to 3120.0~MeV in the spin-$3/2$ series.

The $\Omega_c$ negative-parity spectrum forms two compact TH2 groups.
Such clustering is partly expected from heavy-quark spin symmetry (HQSS): in the $m_Q\to\infty$ limit the heavy-quark spin decouples, and states with a common light-cloud angular momentum $j_\ell>0$ form approximately degenerate $\mathcal J=j_\ell\pm1/2$ doublets, while $j_\ell=0$ gives a singlet~\cite{Neubert1994,Yoshida2015}.
The TH1 $(1^-,1/2)$ centroid at 3034.6~MeV becomes 3043.7 and 3037.0~MeV, while the TH1 $(1^-,3/2)$ centroid at 3075.3~MeV becomes 3077.2, 3077.3, and 3076.4~MeV.
A further TH1 parent near 3106.6~MeV produces TH2 branches around 3097--3101~MeV.
The resulting 3.04--$3.10~\mathrm{GeV}$ band overlaps much of the observed narrow $\Omega_c$ spectrum, although it neither reaches the lowest $\Omega_c(3000)$ nor by itself identifies the highest established states.

\paragraph{The $\boldsymbol{\Xi_c(2790)}$--$\boldsymbol{\Xi_c(2815)}$ doublet:}

The TH2 centroid is successful, but the predicted splitting is only about 1.8~MeV.
Experiment separates the nominal $1/2^-$ and $3/2^-$ states by roughly 25~MeV.
TH2 consequently places the $3/2^-$ branch close to the $\Xi_c(2815)$ but leaves the $1/2^-$ branch approximately 25~MeV above the $\Xi_c(2790)$.
This is another case in which the gross orbital scale transfers well while the diagonal fine-structure is insufficient.

\paragraph{The $\boldsymbol{\Xi_c(2923)}$, $\boldsymbol{\Xi_c(2939)}$, and $\boldsymbol{\Xi_c(2965)}$ cluster:}

LHCb initially resolved three narrow neutral structures at 2923.04, 2938.55, and $2964.88~\mathrm{MeV}$~\cite{LHCbXiC2020}.
The charged $\Xi_c(2923)^+$ partner was subsequently observed, and evidence for a charged $\Xi_c(2939)^+$ contribution has also been reported.
The TH2 $\Xi_c^{\prime}$ and $\Xi_c^{*}$ branches naturally fill this interval, supporting a conventional sextet $1P$ interpretation at the level of the gross mass scale.
The quantum-number assignments are not unique, however, because decay analyses distribute the three observed peaks among different mixtures of $\mathcal J^P=1/2^-$, $3/2^-$, and $5/2^-$ states and sometimes identify the highest structure with a radial excitation~\cite{ChengCharmedBaryons2022,OrtizPacheco:2023}.
The present pure-$\chi_\lambda$ treatment therefore supports the existence of sufficient compact levels in the correct mass window but does not select a unique one-to-one correspondence.

\paragraph{The narrow excited $\boldsymbol{\Omega_c}$ states:}

LHCb established five narrow structures near 3000, 3050, 3066, 3090, and 3119~MeV and later confirmed them together with additional states near 3185 and $3327~\mathrm{MeV}$~\cite{LHCbOmegaC2017,LHCbOmegaC2023}.
The TH2 $1P$ band contains the expected number of low negative-parity descendants in the central part of this interval, but their near-degeneracies do not reproduce the full observed spacing.
Quark-model and heavy-quark-symmetry analyses frequently organize the five lower states as members of the $1P$ sextet, while detailed assignments differ and the upper states may contain radial or higher-orbital configurations~\cite{ChengCharmedBaryons2022,OrtizPacheco:2023}.
Coupled-channel calculations can also shift the lowest $\Omega_c$ negative-parity pole by order 100~MeV through $\Xi_c\bar K$ and $\Xi_c^{\prime}\bar K$ channels~\cite{Luo2021}.
The plotted TH2 bars should therefore be regarded as a compact-core inventory whose detailed mapping requires widths, decay modes, and channel mixing in addition to masses.

\subsubsection{The ${\Xi_{cc}}$, ${\Omega_{cc}}$, and ${\Omega_{ccc}}$ series}
\label{subsubsec:doubly_triply_charmed_spectra}

\begin{figure*}[!t]
\centering
\includegraphics[width=\textwidth,height=0.90\textheight,keepaspectratio]{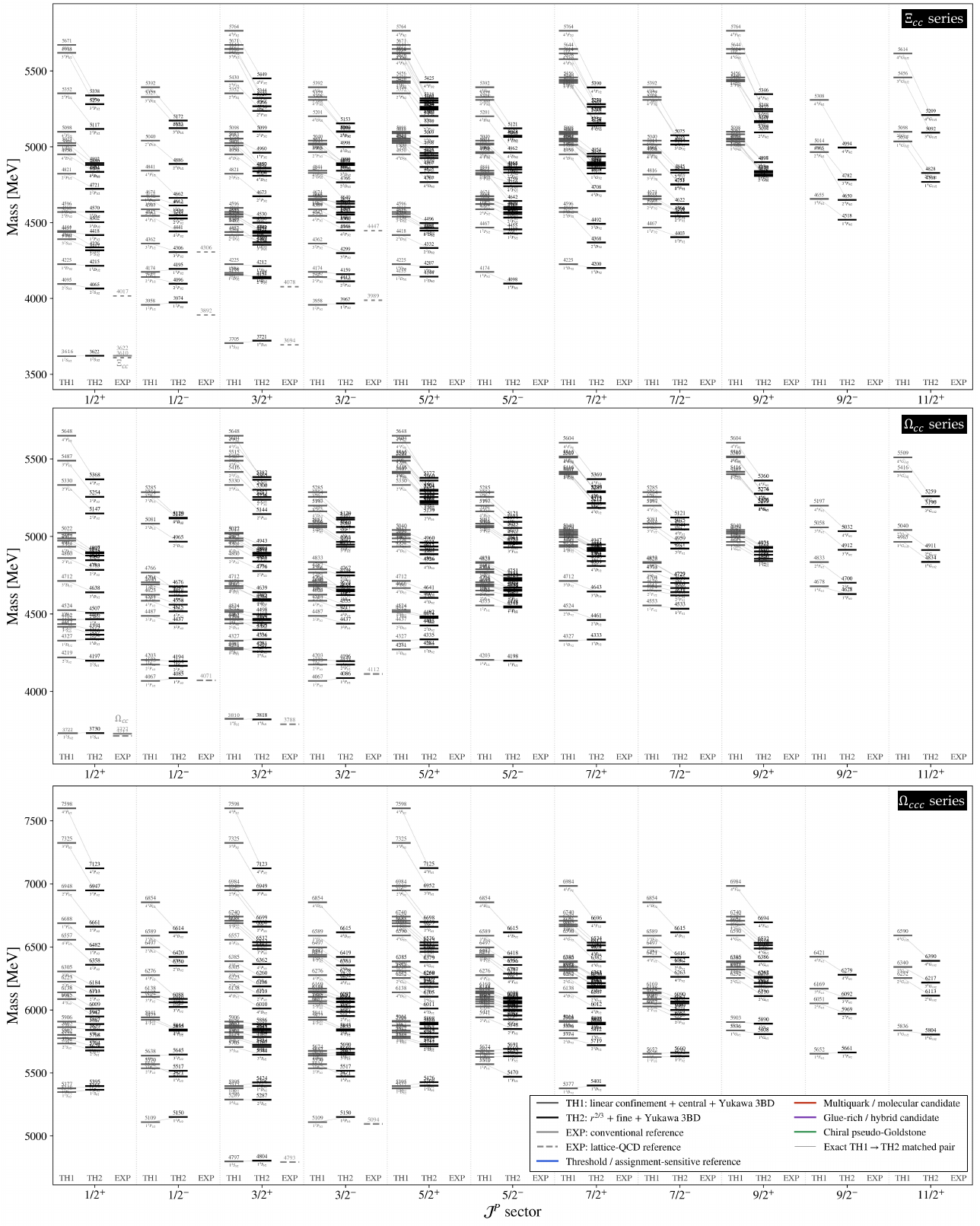}
\caption{\justifying (Enlarge online for details.) TH1, TH2, and reference $\Xi_{cc}$, $\Omega_{cc}$, and $\Omega_{ccc}$ spectra.}
\label{fig:doubly_triply_charmed_spectra}
\end{figure*}

The doubly and triply charmed spectra test heavy-diquark and heavy-quark-number systematics in a region where relativistic and chiral effects are reduced.
The TH1--TH2 displacements are markedly smaller than in the light Roper-like sectors, and the principal TH2 effect is generally a modest resolution of the multiplet fine structure.
This shows that the large changes in the light sector are not caused by the Yukawa three-quark interaction itself.

For the $\Xi_{cc}$, TH1 and TH2 give 3616.4 and 3622.5~MeV for the $1/2^+$ ground state and 3705.5 and 3721.1~MeV for the unobserved $3/2^+$ partner.
The lowest negative-parity centroid changes only from 3957.7~MeV to 3974.0 and 3967.4~MeV.
The first radial levels move from 4094.8 and 4168.1~MeV to 4064.9 and 4132.8~MeV, while the lowest $D$-wave branches shift by only several to a few tens of MeV.
These values remain within the ordinary spread of independent GEM, relativized, and three-quark calculations~\cite{Yoshida2015,LiYuWangEtAl2025,LuWangXiaoZhong2017}.

For the $\Omega_{cc}$, the $1/2^+$ ground state changes from 3723.0 to 3730.2~MeV, and the $3/2^+$ state from 3810.0 to 3818.4~MeV.
The lowest negative-parity centroid changes from 4066.6~MeV to 4084.9 and 4086.1~MeV.
The first radial levels move from 4219.2 and 4280.5~MeV to 4197.1 and 4255.7~MeV, and the lowest $D$-wave centroids remain close to 4.28--$4.34~\mathrm{GeV}$.
The ground and $1P$ scales are compatible with recent lattice and quark-model expectations~\cite{LiYuWangEtAl2025,LuWangXiaoZhong2017,MathurPadmanath2019,YuEtAl2023}.

The $\Omega_{ccc}$ spectrum is similarly stable in most positive-parity branches.
The ground state moves from 4796.8 to 4803.9~MeV, the first radial $3/2^+$ state from 5288.5 to 5286.6~MeV, and the lowest $D$-wave centroids from approximately 5.38--$5.40~\mathrm{GeV}$ to 5.40--$5.43~\mathrm{GeV}$.
The main exception is the lowest negative-parity parent, which moves upward from 5109.2 to approximately 5149.7~MeV.
The continuum-extrapolated lattice results place the ground and lowest negative-parity levels near 4793 and 5094~MeV, respectively, so TH1 is closer for both of these benchmarks while TH2 remains compatible with the broader theoretical spread~\cite{DhindsaEtAl2025,PadmanathEtAl2014,LiuLuZhong2020,FaustovGalkin2022}.

\paragraph{The observed $\boldsymbol{\Xi_{cc}}$ isodoublet:}

The TH1 and TH2 ground-state values both agree closely with the measured $\Xi_{cc}^{++}$ mass~\cite{LHCbXiCCMass2020}.
The 2026 observation of the $\Xi_{cc}^{+}$ at 3619.97~MeV confirms the expected near-degeneracy of the two isospin partners~\cite{LHCbXiCC2026}.
The present isospin-symmetric Hamiltonian assigns them one common mass and therefore cannot predict their few-MeV electromagnetic and $u$--$d$ mass splitting.
The agreement therefore tests only their common strong-interaction centroid.

\paragraph{The preliminary \(\boldsymbol{\Omega_{cc}^{+}}\) observation:}

LHCb announced evidence for the $\Omega_{cc}^{+}$ at the Beauty 2026 conference, with a preliminary mass peak near $3727~\mathrm{MeV}$~\cite{LHCbOmegaCC2026}.
If confirmed in a full publication, this state would complete the observed ground-state spin-$1/2$ doubly charmed flavor triplet.
TH2 lies about 3~MeV above the preliminary peak, while TH1 lies about 4~MeV below it.
The two branches are therefore comparably close, but the comparison should not be presented as a precision test before the full analysis and its systematic uncertainties are published.

\paragraph{The triply charmed spectrum:}

The $\Omega_{ccc}$ sector is almost free of light-diquark and pseudo-Goldstone complications, and the Yukawa-profile three-quark correction is extremely small.
Its agreement with lattice QCD therefore tests primarily the transferred two-body confinement and heavy-heavy dynamics.
The fact that TH1 slightly outperforms TH2 for the ground and lowest negative-parity lattice benchmarks, while the radial and $D$-wave spectra remain similar, shows that the $r^{2/3}$ softening is neither required nor decisively excluded in this compact sector.
Lattice results for the fine structure and future measurements of triply charmed baryons can be used to distinguish TH1 and TH2.

\subsection{Role of the short-range three-quark interaction in excited baryons}
\label{subsec:three_body_effect_excited_baryons}

The ground-state fit determines the strength of the Yukawa-profile three-quark interaction $L^{3Q}_{\rm Y}$. However, it does not show whether the same state dependence also works outside the calibration set. We therefore calculate its effect on the ground and excited baryons that can be compared with established experimental masses in a conventional valence-quark assignment. For each TH1 or TH2 state, we denote the mass without the three-quark interaction by $M_{\rm noY}$ and write the final mass as $M=M_{\rm noY}+\langle L^{3Q}_{\rm Y}\rangle$. We call the shift favorable when $M$ is closer to the experimental mass than $M_{\rm noY}$ and unfavorable when the absolute deviation becomes larger.

\begin{figure*}[!p]
\centering
\includegraphics[width=\textwidth]{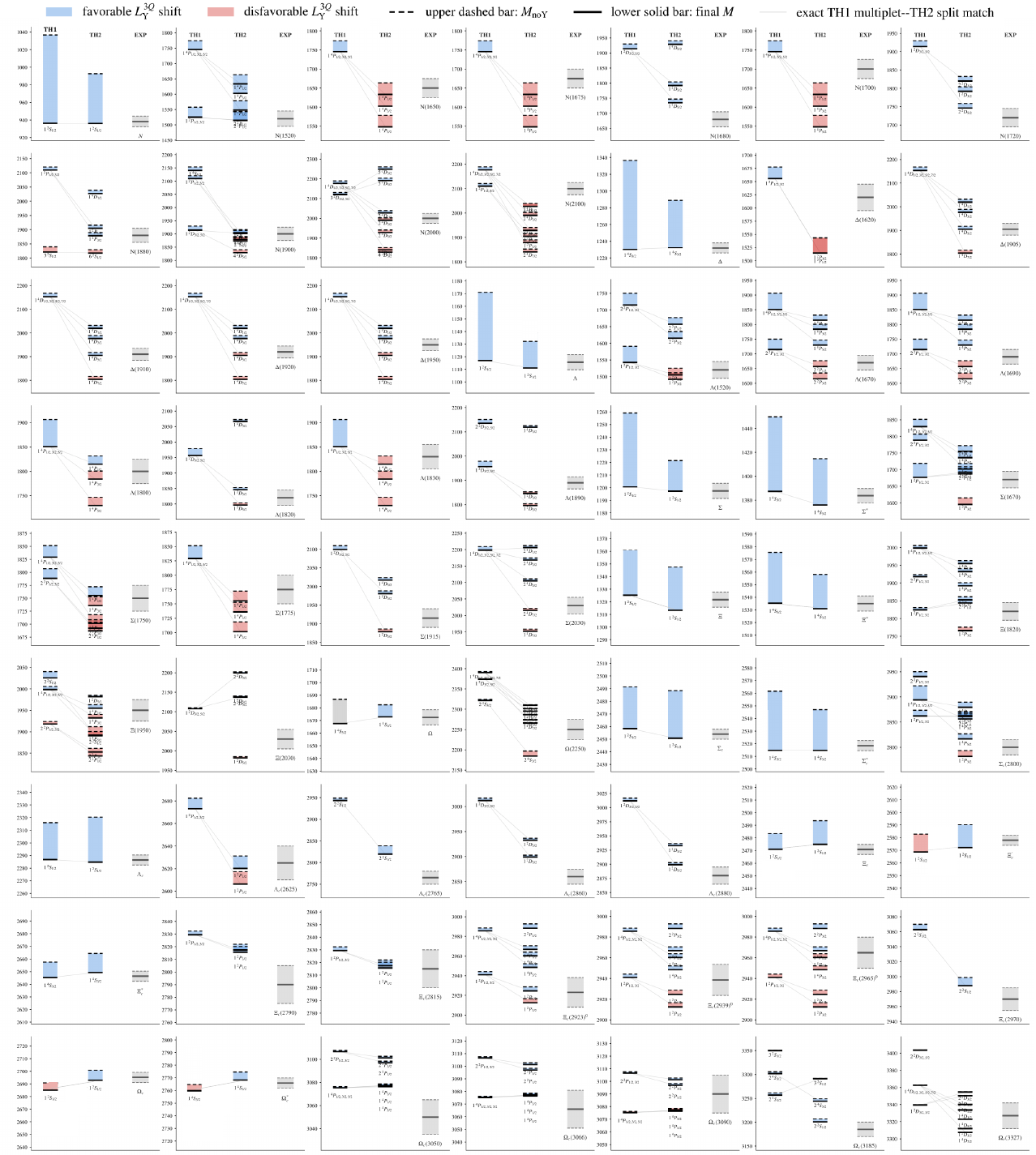}
\caption{(Enlarge online for details.) Effect of $L^{3Q}_{\rm Y}$ on the EXP-matched TH1 multiplets and their TH2 fine-structure partners.}
\label{fig:three_body_effect_excited_baryons}
\end{figure*}

Fig.~\ref{fig:three_body_effect_excited_baryons} presents this comparison separately for each retained experimental reference.
The upper dashed bar denotes $M_{\rm noY}$, the lower solid bar denotes the final mass $M$, and the intervening translucent rectangle represents the shift generated by $L^{3Q}_{\rm Y}$.
Blue rectangles indicate that the interaction moves the candidate closer to the experimental mass, whereas red rectangles indicate that it moves the candidate farther away.
The gray bar and surrounding dashed interval show the experimental central value and 
its reference window, respectively.

The TH1 and TH2 levels are connected only when they originate from the same radial, orbital, and spin configuration.
Because TH1 contains no spin--orbit, antisymmetric spin--orbit, or tensor interaction, a fixed $(n,L,S)$ parent remains degenerate over all allowed total angular momenta.
Its term symbol therefore lists every allowed $\mathcal J$, while the corresponding TH2 levels display the individual fine-structure partners generated by $V^{LS}$, $V^{ALS}$, and $V^T$.
Several experimental resonances also admit more than one plausible valence-quark assignment, so all retained candidate mappings are shown rather than imposing a unique identification.

The ground states $\{N,\Delta,\Sigma,\Lambda,\Sigma^{*},\Lambda_c,\Sigma_c,\Sigma_c^{*},\Xi_{cc}\}$ are used to fit the three-quark couplings and are not independent predictions. The excited states give the more relevant test because no additional parameter is adjusted. The correction is generally largest for compact light-quark states, remains sizable for several low-lying strange and singly charmed states, and becomes smaller for spatially extended or heavy-quark-dominated configurations. This is the expected behavior of the short-range, mass-weighted Yukawa profile of $L^{3Q}_{\rm Y}$.

For the EXP-matched states, $L^{3Q}_{\rm Y}$ tends to move the conventionally assigned baryon levels toward the experimental masses. If we exclude the highly excited TH1 states that are already strongly overestimated before the three-quark correction, approximately three quarters of the retained low-lying TH1 candidates and TH2 fine-structure partners receive a favorable shift. This number is not a state-by-state validation of the interaction. Several states move in the opposite direction, some experimental resonances have more than one possible assignment, and the fraction depends on which states are regarded as conventional three-quark configurations. Nevertheless, the overall tendency shows that the sign and approximate state dependence obtained from the ground-state fit can also be useful beyond the calibration set.

\subsection{Applicable regime of valence quark models}
\label{subsec:limitations_valence_models}
The results show that the relative performance of TH1 and TH2 depends more strongly on the excitation energy than on a simple ordering by flavor. TH1 gives the more stable description for many of the lowest light- and strange-quark orbital centroids. TH2 more often improves the radial and higher-orbital states and several charm and heavy-light branches by reducing the rapid increase of the linear spectrum. A similar behavior is found in conventional comparisons between linear and screened potentials, where the softening becomes more important in the upper radial and orbital spectrum~\cite{Godfrey1985,CapstickIsgur:1986,Li2009,VijandeEtAl2004Screened}.

The constituent valence-quark model describes the compact components of hadrons. It is not a complete description of every observed resonance. In its reliable region, it organizes the spectrum, identifies approximate spin-flavor multiplets, and gives bare mass scales and wave functions that can be used in more complete calculations. However, a discrete valence eigenvalue is not necessarily the pole position of a physical resonance, especially for a broad state or a state close to an open threshold. We discuss the limitations below to specify where the approximation can be used.

The reasons for these limitations were reviewed in the preceding sections. A constant constituent mass absorbs the momentum dependence of the dressed-quark mass, the wave-function renormalization, and part of the quark self-energy (see  Section~\ref{quasiparticle_pic}). A static local potential omits part of the momentum dependence, retardation, relativistic kinematics, and higher-order spin structure (see  Sections~\ref{VCS_intro} and~\ref{sec:pNRQCD_spin_dep}). Goldstone-boson exchange and instanton-induced interactions are important in the light sector (see  Section~\ref{chi_QM}), while hadronic loops generate mass shifts, continuum components, and energy-dependent widths (see  Section~\ref{sec:unquenching}). A bound-state eigenvalue and a resonance pole also require different treatments, as discussed in  Section~\ref{subsubsec:complex_real_scaling}. Finally, the separation into valence two-body, three-body, and intermediate-space contributions depends on the model space and representation (see  Section~\ref{subsec:three_quark_definition}).

The ground-state calculation already shows one limitation of the two-body Hamiltonian. Within the tested class of universal local pair interactions, parameters fitted to mesons do not reproduce the baryon spectrum, while a baryon-oriented fit worsens the meson spectrum. The compatibility scans also show that a reduced-mass-only running coupling or a spinless-Salpeter kinetic energy does not remove this problem when the other interactions are kept unchanged (see  Section~\ref{sec:udscb_benchmark}). The short-range three-quark interaction provides the required state- and flavor-dependent correction and greatly improves the compact ground-state spectrum (see  Section~\ref{sec:short_range}).

However, this result does not solve the excitation problem. The matrix element of the three-quark interaction decreases rapidly when the wave function becomes spatially extended. Therefore, the upper baryon spectrum tests mainly the transferred two-body confinement and the spin-dependent interactions rather than the short-range three-quark force (see  Section~\ref{subsec:baryon_excitations}).

The meson excitation spectrum demonstrates that no single modification of the central confining law is uniformly favored across all flavors and excitation scales.
The linear TH1 Hamiltonian gives a useful description of low-lying charmonium centroids but raises many high-radial and high-orbital levels too rapidly.
The softened $r^{2/3}$ confinement of TH2 corrects this tendency in charmonium and in several open-charm and charm-strange branches, while also permitting the observed fine structure to be resolved.
At the same time, the same modification overcompresses parts of the light and kaonic spectra and generates excessively large orbital splittings.
Quantitatively, the TH2 out-of-fit masses of the $\psi_2(3823)$ and $\psi_3(3842)$ remain within approximately 8 and 25~MeV of experiment, whereas the $\phi_3(1850)$ is underestimated by approximately 134~MeV.
The comparison in  Fig.~\ref{fig:meson_excitation_spectra} therefore does not select a universally superior confinement exponent.
Instead, it indicates that string breaking, relativistic dynamics, and state-dependent continuum renormalization cannot in general be represented by one flavor-independent change of the static power law (see Sections \ref{VC_cornell}, \ref{sec:pNRQCD_spin_dep}{, and} \ref{sec:unquenching})~\cite{BulavaEtAl2019StringBreaking,VijandeEtAl2004Screened}.

A second limitation concerns fine structure and configuration mixing.
TH1 contains only central and color-spin interactions, so states belonging to the same $n^{2S+1}L$ or baryonic $(n,J^P,S)$ parent are degenerate and should be interpreted as multiplet centroids rather than individual physical levels.
TH2 resolves this degeneracy, but in the present baryon calculation the tensor, symmetric spin--orbit, and antisymmetric spin--orbit interactions are included only through their diagonal expectation values.
The off-diagonal matrix elements that mix distinct orbital and spin configurations carrying the same $\mathcal J^P$ are not included.
The reversed ordering of the calculated $\Lambda_c$ $1D$ doublet and the too-small $\Xi_c(2790)$--$\Xi_c(2815)$ splitting directly show the limitation of this diagonal treatment.
The near-degenerate $\Omega_c$ groups are partly expected from heavy-quark spin symmetry, although their detailed spacing also requires configuration mixing.
The anomalously low nucleon pair at 1.29--$1.34~\mathrm{GeV}$ has a different origin: as discussed above, it mainly reflects the overcompressed TH2 light-sector central branch.
These examples show that a good multiplet centroid does not guarantee the correct ordering, mixing angle, or assignment of the individual states. For a precision calculation, all allowed orbital and pair-spin configurations with the same $\mathcal J^P$ must be diagonalized together, and the interaction vertices and noncentral operators must be relativized consistently (see  Sections~\ref{sec:OGE_color_spin} and~\ref{sec:pNRQCD_spin_dep})~\cite{Godfrey1985,CapstickRoberts2000,CapstickIsgur:1986}.

Near a threshold, one must distinguish a compact seed from the physical pole (see  Sections~\ref{sec:unquenching} and~\ref{subsubsec:complex_real_scaling}). A static Hermitian valence Hamiltonian gives real and discrete eigenvalues. An open-channel resonance instead has an energy-dependent complex pole, a width, and generally both compact and hadron--hadron components in its wave function.
The broad $D_0^{*}(2300)$ and $D_1(2430)$, the subthreshold $D_{s0}^{*}(2317)$ and $D_{s1}(2460)$, the $\Lambda(1405)$, the $\Xi(1620)$ and $\Xi(1690)$, the $\Omega(2012)$ and assignment-sensitive $\Omega(2109)$, and the threshold-adjacent $\Lambda_c(2940)$ illustrate different versions of this problem.
For such states, agreement between a valence eigenvalue and the measured mass may identify a plausible bare-core scale, but it neither determines the pole composition nor demonstrates that continuum dressing is small.
Conversely, a large mass discrepancy need not imply the absence of a compact seed, because an $S$-wave threshold can generate a large dispersive shift or even multiple nearby poles (see  Sections~\ref{sec:meson_excitations} and~\ref{subsec:baryon_excitations}).
Therefore, a calculated seed can be related to the measured state only after the relevant hadron--hadron interpolators or channels are included in a finite-volume or continuum scattering calculation~\cite{Luo2020,Zhang2023,Hyodo2012,Jido2003,MohlerEtAl2013,LangEtAl2014}.

The light-quark spectrum exposes an additional limitation of transferring intuition directly from heavy quarkonia.
Pseudo-Goldstone dynamics, flavor-spin interactions, and strong meson--baryon dressing are not small perturbations in many light states.
They affect the unusually low pseudoscalar masses, the ordering of radial and negative-parity baryons, and the Roper-like $N$ and $\Delta$ excitations (see  Sections~\ref{chi_QM} and~\ref{sec:unquenching}).
The comparison with lattice calculations based predominantly on three-quark interpolators is particularly instructive in the $\Delta(1600)$ channel, where a relatively high compact radial level can coexist with a lower physical pole generated through substantial $\pi N$ and $\pi\Delta$ rescattering~\cite{SuzukiEtAl2010,HockleyEtAl2024,HockleyEtAl2025}.
A constant constituent mass and a local smeared hyperfine interaction can absorb an average part of these effects in a low-lying calibration set, but the resulting parameters need not extrapolate uniformly to larger radii, nodal states, or different chiral representations.

There are also states that lie outside the valence Hilbert space altogether.
A neutral $q\bar q$ model cannot generate spin-exotic quantum numbers such as $\mathcal J^{PC}=1^{-+}$, and a pure $qqq$ model contains no explicit gluonic excitation from which hybrid baryons could be formed.
Glueballs and hybrids require explicit gluonic basis states, whereas compact multiquarks and hadronic molecules require enlarged color/Fock-space and continuum sectors rather than a further adjustment of the same valence potential (see  Sections~\ref{sec_V_CS_dom} and~\ref{sec:unquenching})~\cite{Dudek2011Hybrid,BESIIIeta11855,BrambillaEtAl2020XYZ}.
Even nonexotic quantum numbers do not guarantee a predominantly valence configuration, as illustrated by the threshold-sensitive scalar $a_0(980)$ and $f_0(980)$ and by the multiple structures observed in hidden-charm channels.
Such states should not be included indiscriminately in a global valence-model RMSE, because their discrepancy measures missing degrees of freedom as much as it measures the accuracy of the compact Hamiltonian.

Masses alone also do not determine all parameters uniquely. The ground-state masses constrain combinations of the constituent masses, additive constant, confinement, Coulomb attraction, and short-range hyperfine strengths, but they do not separate these contributions uniquely (see Sections \ref{quasiparticle_pic}, \ref{VC_cornell}, \ref{VC_constant}, and~\ref{VCS_intro}). TH1 and TH2 give one direct example: both reproduce the fitted ground states but predict very different upper spectra. Masses also give only limited information on the wave-function composition and do not determine uniquely the mixing angles, decay couplings, spatial sizes, or valence probabilities (see  Section~\ref{sec:unquenching}). A stronger test should therefore include strong and electromagnetic decay widths, branching fractions, transition form factors, static moments, and lattice finite-volume amplitudes together with the masses.
Electromagnetic radii and form factors require a current operator in addition to the mass Hamiltonian, and a pointlike one-body impulse current in a truncated valence space omits relativistic boosts, constituent-quark dressing, meson-cloud contributions, and interaction-induced exchange currents; agreement with masses therefore does not by itself validate these observables (see  Section~\ref{subsec:three_quark_definition})~\cite{DeSanctisSantopintoGiannini1998,WagnerBuchmannFaessler1998}.
Current constructions based on vector-meson dominance (VMD) or dispersion relations can efficiently incorporate part of this missing low-energy vector-channel physics, while covariant constituent-quark calculations can also provide a quantitative description of form factors when the current operator is constructed consistently~\cite{WagenbrunnEtAl2001,BelushkinHammerMeissner2007}.
Moreover, bare masses and valence probabilities depend on the chosen model space and transformation, so only consistently calculated observables are representation independent (see  Section~\ref{subsec:three_quark_definition})~\cite{Polyzou2010,FurnstahlHammerTirfessa2001}.

Electromagnetic observables remain valuable complementary tests of constituent-quark wave functions once a current operator is specified. \ref{app:em_observables} evaluates selected static moments and radiative transitions with a one-body impulse current. The reasonable $\Sigma^0\to\Lambda\gamma$ result and the underestimated $\Delta^+\to p\gamma$ and $\Sigma^{*0}\to\Lambda\gamma$ rates illustrate both the usefulness and the limitations of this truncation, while the sub-percent response of the allowed ground-state $M1$ quantities indicates only weak independent sensitivity to $L_{\rm Y}^{3Q}$.

Some limitations of the present figures are numerical and are not common to every valence quark model. We truncate the baryon orbital space at $l_{\mathcal C},L_{\mathcal C}\leq2$. Therefore, the highest angular-momentum states do not contain a complete $F$- or $G$-wave variational basis and may remain too high. The baryon spectra displayed here omit off-diagonal mixing between orbital configurations and, for $\Xi_c$ and $\Xi_c^{\prime}$, between the two pair-spin bases. The isospin-symmetric Hamiltonian gives one strong-interaction centroid and cannot reproduce the electromagnetic or $m_u-m_d$ splittings between the charged isospin partners. These are numerical restrictions rather than general failures of the valence approximation, but they limit the precision of the individual assignments in Figs.~\ref{fig:N_Delta_spectra}--\ref{fig:doubly_triply_charmed_spectra}.

The valence Hamiltonian therefore has different levels of applicability. Compact low-lying states, spin-averaged multiplet centroids, and heavy-quark-rich systems such as the $\Omega_{ccc}$ give the cleanest tests. For conventional but strongly mixed excitations, the valence calculation can still provide the compact-core states, but the complete relativistic and noncentral interactions must be included (see  Sections~\ref{sec:OGE_color_spin} and~\ref{sec:pNRQCD_spin_dep}). Near a threshold, the valence states must be coupled explicitly to the hadronic channels (see  Sections~\ref{sec:unquenching} and~\ref{subsubsec:complex_real_scaling}). Compact multiquarks and molecular states require enlarged color/Fock-space bases, while glueballs and hybrids additionally require explicit gluonic degrees of freedom (see  Sections~\ref{sec_V_CS_dom} and~\ref{sec:unquenching}). In this region of applicability, the short-range three-quark interaction gives an economical and quantitatively important correction to the compact baryon Hamiltonian. However, it does not replace the chiral, relativistic, long-distance, or continuum dynamics (see  Sections \ref{chi_QM}, \ref{VC_cornell}, \ref{sec:pNRQCD_spin_dep}, and~\ref{sec:unquenching}).

\section{Color and spin matrix elements of the two- and three-body operators}\label{sec:3bd_operators}

In this section, we provide a detailed compilation of color and spin matrix elements for two- and three-body operators in various multiquark configurations (baryons, tetraquarks, etc.). This both supports our calculations and serves as a reference for future studies.

\subsection{Casimir operators}

The quadratic Casimir $C_2$ essentially measures the color charge squared of a given representation, while the cubic Casimir $C_3$ for SU(3) distinguishes symmetric from antisymmetric charge combinations. For SU(3), a representation labeled by Dynkin indices $(p,q)$ has $C_2$ and $C_3$ given by Eq.~\eqref{cas_1}. For example, a quark (fundamental, $(1,0)$) has $C_2=4/3$, a diquark in $\bar{\mathbf{3}}$ has $C_2=4/3$ as well, etc. The two-body and three-body quark potentials are closely related to the Casimir operators of SU(3). In this section, we explore how to derive the quadratic and cubic Casimir operators of general SU($N$). We then present the matrix elements of the three-quark potentials for several cases of baryons, tetraquarks, dibaryons, and pentaquarks.

\paragraph{Quadratic Casimir operator of SU($\boldsymbol{N}$):}

For SU(3), the Casimir operators can be obtained from the highest weight in the rank space. However, this approach is difficult to apply to the general SU($N$) case. Therefore, we propose a new method based on Young tableaux. Let us consider the following irreducible representation of SU($N$).\\
\hspace*{2.5em}%
\begin{tabular}{|c|c|c|c|c|}
  \multicolumn{1}{c}{$\overbrace{\rule{.8cm}{0pt}}^{p_{N-1}}$} & \multicolumn{1}{c}{$\overbrace{\rule{.8cm}{0pt}}^{p_{N-2}}$} & \multicolumn{1}{c}{$\cdots$} & \multicolumn{1}{c}{$\overbrace{\rule{.8cm}{0pt}}^{p_2}$} & \multicolumn{1}{c}{$\overbrace{\rule{.8cm}{0pt}}^{p_1}$} \\
  \cline{1-5}
  $\cdots$ & $\cdots$ & $\cdots \cdots$ & $\cdots$ & $\cdots$ \\
  \cline{1-5}
  \quad \quad & \quad \quad & \quad \quad & \quad \quad \\
  \cline{1-4}
  \multicolumn{1}{c}{$\vdots$} \\
  \cline{1-2}
  \quad \quad & \quad \quad \\
  \cline{1-2}
  \quad \quad \\
  \cline{1-1}
\end{tabular}\\
Here, $p_i$ is the number of columns containing $i$ boxes. For SU($N$), we can represent the color operator using a permutation matrix.
\begin{align}\label{FF_rel}
F_i^a F_j^a = \frac{1}{2}(ij) -\frac{1}{2N}\mathbf{1}
\end{align}
where $F_i^a = \frac{1}{2}\lambda_i^a$ and $\mathbf{1}$ is the identity matrix.\footnote{This identity is a direct consequence of the SU($N$) Fierz relation
\begin{equation*}
F^c_{ij}F^c_{kl} =\frac{1}{2}\bigl(\delta_{il}\delta_{jk}-\frac{1}{N}\delta_{ij}\delta_{kl}\bigr)
\end{equation*}
At a more structural level it reflects Schur--Weyl duality: on the $n$-fold tensor product $(\mathbb{C}^N)^{\otimes n}$ the commuting actions of $\mathrm{SU}(N)$ and the permutation group $S_n$ generate each other's commutants. The operator $F_i^c F_j^c$ is then proportional to the transposition $(ij)$ in $S_n$ plus a multiple of the identity, which is precisely Eq.~\eqref{FF_rel}.}  For the system composed of $n$ fundamental representations,
\begin{align}
\sum_i^n F_i^c F_i^c = \frac{n}{2N}(N^2-1).
\end{align}
Since each irreducible representation is an eigenstate of the Casimir operator, we can compute the Casimir operator by considering only the diagonal components.
\begin{align}
C_2 &= \sum_{i,j} F_i^a F_j^a = \sum_{i}^n F_i^a F_i^a + 2\sum_{i<j}^n F_i^a F_j^a \\
&= n \times \frac{1}{2N}(N^2-1)  + 2 \left[ \binom{p_1+ \cdots + p_{N-1}}{2} + \binom{p_2 + \cdots + p_{N-1}}{2} + \cdots + \binom{p_{N-1}}{2} \right] \left(\frac{1}{2}-\frac{1}{2N}\right) \nonumber \\
& \quad \text{(only symmetric parts are considered: combinations in the same line)} \nonumber \\
& \quad +2 \bigg{[} (p_1+ \cdots + p_{N-1})(p_2 + \cdots + p_{N-1})\left(-\frac{1}{2(p_1 + \cdots + p_{N-1})} -\frac{1}{2N}\right) \nonumber\\
& \qquad + (p_1+ \cdots + p_{N-1})(p_3 + \cdots + p_{N-1})\left(-\frac{1}{2(p_1 + \cdots + p_{N-1})} -\frac{1}{2N}\right) \nonumber \\
& \qquad \qquad \qquad \vdots \nonumber \\
& \qquad + (p_1+ \cdots + p_{N-1})(p_{N-1})\left(-\frac{1}{2(p_1 + \cdots + p_{N-1})} -\frac{1}{2N}\right) \nonumber \\
& \qquad +(p_2+ \cdots + p_{N-1})(p_3 + \cdots + p_{N-1})\left(-\frac{1}{2(p_2 + \cdots + p_{N-1})} -\frac{1}{2N}\right) \nonumber \\
& \qquad +(p_2+ \cdots + p_{N-1})(p_4 + \cdots + p_{N-1})\left(-\frac{1}{2(p_2 + \cdots + p_{N-1})} -\frac{1}{2N}\right) \nonumber \\
& \qquad \qquad \qquad \vdots \nonumber \\
& \qquad +(p_2+ \cdots + p_{N-1})(p_{N-1})\left(-\frac{1}{2(p_2 + \cdots + p_{N-1})} -\frac{1}{2N}\right) \nonumber \\
& \qquad \qquad \qquad \vdots \nonumber \\
& \qquad + (p_{N-2} + p_{N-1})(p_{N-1})\left(-\frac{1}{2(p_{N-2}+p_{N-1})} -\frac{1}{2N}\right) \bigg{]} \\
&= \frac{n}{2N}(N^2-1) + \frac{1}{N}(N-1) \left[ \sum_{i=1}^{N-1} \binom{\sum_{j=1}^i p_{N-j}}{2} \right] -  \sum_{k=1}^{N-2} \left[ \sum_{j=k+1}^{N-1} (j-k)p_j \right] \left( 1+ \frac{1}{N}\sum_{i=k}^{N-1} p_i \right).
\end{align}
We represent the quadratic Casimir operators for SU(3), SU(4), SU(5), and SU(6) as follows.
\begin{align}\label{eq:C_2_casimir}
C_2^{\text{SU(3)}} &= p_1+\frac{p_1^2}{3}+p_2+\frac{1}{3} p_1 p_2+\frac{p_2^2}{3}, \\
C_2^{\text{SU(4)}} &= \frac{3 p_1}{2}+\frac{3 p_1^2}{8}+2 p_2+\frac{1}{2} p_1
   p_2+\frac{p_2^2}{2}+\frac{3 p_3}{2}+\frac{1}{4} p_1 p_3+\frac{1}{2} p_2
   p_3+\frac{3 p_3^2}{8}, \\
C_2^{\text{SU(5)}} &= 2 p_1+\frac{2 p_1^2}{5}+3 p_2+\frac{3}{5} p_1 p_2+\frac{3 p_2^2}{5}+3
   p_3+\frac{2}{5} p_1 p_3+\frac{4}{5} p_2 p_3+\frac{3 p_3^2}{5}+2
   p_4+\frac{1}{5} p_1 p_4+\frac{2}{5} p_2 p_4+\frac{3}{5} p_3 p_4 +\frac{2
   p_4^2}{5}, \\
C_2^{\text{SU(6)}} &= \frac{5 p_1}{2}+\frac{5 p_1^2}{12}+4 p_2+\frac{2}{3} p_1 p_2+\frac{2
   p_2^2}{3}+\frac{9 p_3}{2}+\frac{1}{2} p_1 p_3+p_2 p_3+\frac{3 p_3^2}{4}+4
   p_4+\frac{1}{3} p_1 p_4+\frac{2}{3} p_2 p_4+p_3 p_4 +\frac{2
   p_4^2}{3} \nonumber \\
   &\quad +\frac{5 p_5}{2}+\frac{1}{6} p_1 p_5+\frac{1}{3} p_2
   p_5+\frac{1}{2} p_3 p_5+\frac{2}{3} p_4 p_5+\frac{5 p_5^2}{12}.
\end{align}
For SU(3), substituting $(p_1,p_2)=(1,0)$ and $(1,1)$ into Eq.~\eqref{eq:C_2_casimir} gives $C_2(\mathbf{3})=4/3$ and $C_2(\mathbf{8})=3$, in agreement with the standard quadratic Casimirs of the fundamental and adjoint representations.

\paragraph{Cubic Casimir operator of SU($\boldsymbol{N}$):} Now we demonstrate how to compute the cubic Casimir operator using Young tableaux. The commutation and anticommutation relations of general SU($N$) are as follows:
\begin{align}
[F^a,F^b] = if^{abc}F^c, \qquad \{F^a, F^b \} = \frac{1}{N}\delta^{ab}\mathbf 1 + d^{abc}F^c,
\end{align}
where $a,b,c=1,2, \ldots, N^2-1$. The normalization condition takes the following form.
\begin{align}
\text{Tr}(F^a F^b) = \frac{1}{2}\delta^{ab}
\end{align}
Here, we decompose the cubic Casimir operator into three distinct parts.
\begin{align}
C_3 &= \sum_{i,j,k}d^{abc}F_i^a F_j^b F_k^c \nonumber\\
&= \sum_i d^{abc}F_i^a F_i^b F_i^c + 6\sum_{i<j}d^{abc}F_i^a F_i^b F_j^c + 6\sum_{i<j<k}d^{abc}F_i^a F_j^b F_k^c \quad (\because d^{abc} \text{\ is totally symmetric})
\end{align}
For the fundamental representation of SU($N$), the following identities hold.
\begin{align}
C_2 = -\frac{2i}{N}f^{abc}F^a F^b F^c =\frac{N^2-1}{2N}, \qquad C_3 = d^{abc}F^a F^b F^c = C_2\bigg(2C_2 -\frac{(N^2+2)}{2N}\bigg)
\end{align}
From the above formulae,
\begin{equation}
\sum_i d^{abc}F_i^a F_i^b F_i^c = n\Big(-\frac{5}{4}+\frac{1}{N^2}+\frac{N^2}{4}\Big),
\end{equation}
where $n$ is the total number of quarks. For the second term,
\begin{align}
& \{F_j^a,F_j^b \} = \frac{1}{N}\delta^{ab}+d^{abc}F_j^c \\
\Rightarrow \ & F_i^a F_i^b(F_j^a F_j^b + F_j^b F_j^a) = \frac{1}{N}F_i^a F_i^a + d^{abc}F_i^a F_i^b F_j^c \\
\Rightarrow \ & d^{abc}F_i^a F_i^b F_j^c = F_i^a F_i^b F_j^a F_j^b + F_i^a (F_i^b F_j^b)F_j^a -\frac{1}{N}F_i^a F_i^a
\end{align}
Note that $i \neq j$. Eq.~\eqref{FF_rel} reads
\begin{align}
d^{abc}F_i^a F_i^b F_j^c &= (F_i^a F_j^a) (F_i^b F_j^b) + F_i^a(F_i^b F_j^b)F_j^a -\frac{1}{N}F_i^a F_i^a  \nonumber\\
&= \bigg( \frac{1}{2}(ij)-\frac{1}{2N} \bigg)^2 + F_i^a \bigg( \frac{1}{2}(ij)-\frac{1}{2N} \bigg) F_j^a -\frac{1}{N} \times \frac{N^2-1}{2N} 
= \bigg(-\frac{2}{N}+\frac{N}{2}\bigg)F_i^a F_j^a \\
\Rightarrow \ \sum_{i<j} d^{abc}&F_i^a F_i^b F_j^c = \bigg(-\frac{2}{N}+\frac{N}{2}\bigg) \sum_{i<j} F_i^a F_j^a = \bigg(-\frac{1}{N}+\frac{N}{4}\bigg)C_2 +n\bigg(\frac{5}{8}-\frac{1}{2N^2}-\frac{N^2}{8} \bigg).
\end{align}
For SU($N$), the cubic Casimir becomes
\begin{align}
C_3 &= \frac{p_1}{2} + \frac{p_1^2}{2} + \frac{p_1^3}{9} -\frac{p_2}{2} +\frac{1}{6}p_1^2 p_2 -\frac{p_2^2}{2} -\frac{1}{6}p_1 p_2^2 - \frac{p_2^3}{9}\nonumber \\
&= n\Bigl(-\frac{5}{4}+\frac{1}{N^2}+\frac{N^2}{4}\Bigr) + 6\bigg[\Big(-\frac{1}{N}+\frac{N}{4}\Big)C_2 +n\Big(\frac{5}{8}-\frac{1}{2N^2}-\frac{N^2}{8} \Big) \bigg] + 6\sum_{i<j<k}d^{abc}F_i^a F_j^b F_k^c 
\end{align}
The formula for the third term is given in Ref.~\cite{Park:2022jpf}. Here, we represent the cubic Casimir operators for SU(3), SU(4), SU(5), and SU(6) as follows.
\begin{align}\label{eq:C_3_casimir}
C_3^{\text{SU(3)}} &= \frac{1}{18}(p_1-p_2)(p_1+2p_2+3)(2p_1+p_2+3), \\
C_3^{\text{SU(4)}} &= \frac{3}{2}p_1 + \frac{9}{8} p_1^2 + \frac{3}{16}p_1^3 + \frac{3}{4}p_1 p_2 + \frac{3}{8} p_1^2 p_2 -\frac{3}{2} p_3 + \frac{3}{16}p_1^2 p_3 -\frac{3}{4}p_2 p_3 -\frac{9}{8}p_3^2 -\frac{3}{16}p_1 p_3^2 -\frac{3}{8}p_2 p_3^2 -\frac{3}{16}p_3^3, \\
C_3^{\text{SU(5)}} &= 3 p_1+\frac{9 p_1^2}{5}+\frac{6 p_1^3}{25}+\frac{3 p_2}{2}+\frac{9}{5} p_1 p_2+\frac{27}{50} p_1^2 p_2+\frac{9 p_2^2}{10}+\frac{9}{50} p_1 p_2^2+\frac{3 p_2^3}{25}-\frac{3 p_3}{2}+\frac{3}{5} p_1 p_3+\frac{9}{25} p_1^2 p_3+\frac{6}{25} p_1 p_2 p_3 \nonumber \\
&\quad+\frac{6}{25} p_2^2 p_3-\frac{9p_3^2}{10}-\frac{3}{25} p_1 p_3^2-\frac{6}{25} p_2 p_3^2-\frac{3p_3^3}{25}-3 p_4+\frac{9}{50} p_1^2 p_4-\frac{3}{5} p_2 p_4+\frac{3}{25}p_1 p_2 p_4+\frac{3}{25} p_2^2 p_4-\frac{9}{5} p_3 p_4 \nonumber \\
&\quad-\frac{3}{25} p_1p_3 p_4-\frac{6}{25} p_2 p_3 p_4-\frac{9}{50} p_3^2 p_4-\frac{9p_4^2}{5}-\frac{9}{50} p_1 p_4^2-\frac{9}{25} p_2 p_4^2-\frac{27}{50} p_3 p_4^2-\frac{6 p_4^3}{25}, \\
C_3^{\text{SU(6)}} &= 5 p_1+\frac{5 p_1^2}{2}+\frac{5 p_1^3}{18}+4 p_2+3 p_1 p_2+\frac{2}{3} p_1^2 p_2+2 p_2^2+\frac{1}{3} p_1 p_2^2+\frac{2 p_2^3}{9}+\frac{3}{2} p_1 p_3+\frac{1}{2} p_1^2 p_3+\frac{3}{2} p_2 p_3+\frac{1}{2} p_1 p_2 p_3 \nonumber \\
&\quad+\frac{1}{2} p_2^2 p_3-4 p_4+\frac{1}{2} p_1 p_4+\frac{1}{3} p_1^2 p_4+\frac{1}{3} p_1 p_2 p_4+\frac{1}{3} p_2^2 p_4-\frac{3}{2} p_3 p_4-2 p_4^2-\frac{1}{6} p_1 p_4^2-\frac{1}{3} p_2 p_4^2-\frac{1}{2} p_3 p_4^2-\frac{2 p_4^3}{9} \nonumber \\
&\quad-5 p_5+\frac{1}{6} p_1^2 p_5-\frac{1}{2} p_2 p_5+\frac{1}{6} p_1 p_2 p_5+\frac{1}{6} p_2^2 p_5-\frac{3}{2} p_3 p_5 - 3 p_4 p_5-\frac{1}{6} p_1 p_4 p_5-\frac{1}{3} p_2 p_4 p_5-\frac{1}{2} p_3 p_4 p_5-\frac{1}{3} p_4^2 p_5 \nonumber \\
&\quad-\frac{5 p_5^2}{2}-\frac{1}{6} p_1 p_5^2-\frac{1}{3} p_2 p_5^2-\frac{1}{2} p_3 p_5^2-\frac{2}{3} p_4 p_5^2-\frac{5 p_5^3}{18}.
\end{align}
For instance, inserting $(p_1,p_2)=(1,0)$ (fundamental of SU(3)) into Eq.~\eqref{eq:C_3_casimir} gives $C_3(\mathbf{3})=10/9$, in agreement with the textbook result. Our explicit polynomials for SU(4) and SU(6) indeed vanish whenever $(p_1,p_2,\dots,p_{N-1})$ is equal to its conjugate, providing another nontrivial check on the Young-tableau construction.

\subsection{Baryons}

For baryons, since there is only one color-singlet state, $d^{abc}\lambda_i^a \lambda_j^b \lambda_k^c$ can be calculated easily: $d^{abc}\lambda_i^a \lambda_j^b \lambda_k^c=80/9$. Furthermore, as the color-color matrix $\lambda_i^c \lambda_j^c$ is determined by $-8/3$, the matrix elements of the three-quark potential can be computed solely from the spin-spin matrices corresponding to each symmetry. The results for various baryons are summarized below. We denote $L^{C-C}=\sum L_{ijk}^{C-C}$, $L^{S-S}=\sum L_{ijk}^{S-S}$, and $L^{C-S}=\sum L_{ijk}^{C-S}$.
\begin{align}
L^{C-C}&=\frac{128}{3m_u},\quad & L^{S-S}&=-\frac{128}{3m_u^5},\quad & L^{C-S}&=-\frac{256}{3m_u^3}\qquad &\text{for }p,\nonumber\\
L^{C-C}&=\frac{256}{9m_u}+\frac{128}{9m_s},\quad & L^{S-S}&=-\frac{128}{3m_u^2m_s^3},\quad & L^{C-S}&=-\frac{256}{3m_u^3}\qquad &\text{for }\Lambda,\nonumber\\
L^{C-C}&=\frac{256}{9m_u}+\frac{128}{9m_s},\quad & L^{S-S}&=-\frac{512}{9m_u^4m_s}+\frac{128}{9m_u^2m_s^3},\quad & L^{C-S}&=\frac{256}{9m_u^3}-\frac{512}{9m_u^2m_s}-\frac{512}{9m_um_s^2}\qquad &\text{for }\Sigma,\nonumber\\
L^{C-C}&=\frac{128}{9m_u}+\frac{256}{9m_s},\quad & L^{S-S}&=\frac{128}{9m_u^3m_s^2}-\frac{512}{9m_um_s^4},\quad & L^{C-S}&=-\frac{512}{9m_u^2m_s}-\frac{512}{9m_um_s^2}+\frac{256}{9m_s^3}\qquad &\text{for }\Xi,\nonumber\\
L^{C-C}&=\frac{128}{3m_u},\quad & L^{S-S}&=\frac{128}{3m_u^5},\quad & L^{C-S}&=\frac{256}{3m_u^3}\qquad &\text{for }\Delta,\nonumber\\
L^{C-C}&=\frac{256}{9m_u}+\frac{128}{9m_s},\quad & L^{S-S}&=\frac{256}{9m_u^4m_s}+\frac{128}{9m_u^2m_s^3},\quad & L^{C-S}&=\frac{256}{9m_u^3}+\frac{256}{9m_u^2m_s}+\frac{256}{9m_um_s^2}\qquad &\text{for }\Sigma^*,\nonumber\\
L^{C-C}&=\frac{128}{9m_u}+\frac{256}{9m_s},\quad & L^{S-S}&=\frac{128}{9m_u^3m_s^2}+\frac{256}{9m_um_s^4},\quad & L^{C-S}&=\frac{256}{9m_u^2m_s}+\frac{256}{9m_um_s^2}+\frac{256}{9m_s^3}\qquad &\text{for }\Xi^*.
\end{align}
As an illustration, consider the nucleon. Using the color-singlet factors in Eq.~\eqref{eq:qqq_color_singlet_factors}, the remaining distinction is the spin matrix element. For the spin-$1/2$ nucleon, $\sum_{i<j}\boldsymbol{\sigma}_i\cdot\boldsymbol{\sigma}_j=-3$, whereas for the $\Delta$ with $S=3/2$ the same sum gives \ensuremath{+}3.
Combining these with the mass factors reproduces, for example, the entries  $L^{C-C}=128/(3m_u)$ and $L^{S-S}=-128/(3m_u^5)$ for the proton, together with the opposite sign of $L^{S-S}$ for the $\Delta$, as listed above.

\subsection{Tetraquarks}

\paragraph{Three-body color matrix formula:}
A system that does not contain antiquarks, such as a baryon, enables the calculation of the three-quark potential using Casimir operators and permutation matrices of the color states. However, in the case of a system containing both quarks and antiquarks, the three-quark potential must be calculated using commutation or anticommutation relations. Let us first calculate $d^{abc}\lambda_i^a \lambda_j^b \overline{\lambda}_k^c$.
\begin{align}
&\overline{\lambda}_k^a \overline{\lambda}_k^b + \overline{\lambda}_k^b \overline{\lambda}_k^a = \frac{4}{3}\delta^{ab} -2d^{abc}\overline{\lambda}_k^c \\
\Rightarrow \ &\lambda_i^a \lambda_j^b \overline{\lambda}_k^a \overline{\lambda}_k^b + \lambda_i^a \lambda_j^b \overline{\lambda}_k^b \overline{\lambda}_k^a = \frac{4}{3}\lambda_i^a \lambda_j^a -2d^{abc}\lambda_i^a \lambda_j^b \overline{\lambda}_k^c \\
\Rightarrow \ &(\lambda_i^a \overline{\lambda}_k^a) (\lambda_j^b \overline{\lambda}_k^b) + (\lambda_j^a \overline{\lambda}_k^a) (\lambda_i^b \overline{\lambda}_k^b) = \frac{4}{3}\lambda_i^a \lambda_j^a -2d^{abc}\lambda_i^a \lambda_j^b \overline{\lambda}_k^c \\
\Rightarrow \ &d^{abc}\lambda_i^a \lambda_j^b \overline{\lambda}_k^c = -\frac{1}{2}(\lambda_i^a \overline{\lambda}_k^a) (\lambda_j^b \overline{\lambda}_k^b) -\frac{1}{2} (\lambda_j^a \overline{\lambda}_k^a) (\lambda_i^b \overline{\lambda}_k^b) + \frac{2}{3}\lambda_i^a \lambda_j^a
\end{align}
Similarly,
\begin{align}
d^{abc}\lambda_i^a \overline{\lambda}_j^b \overline{\lambda}_k^c = -\frac{1}{2}(\lambda_i^a \overline{\lambda}_k^a) (\overline{\lambda}_j^b \overline{\lambda}_k^b) -\frac{1}{2} (\overline{\lambda}_j^a \overline{\lambda}_k^a) (\lambda_i^b \overline{\lambda}_k^b) + \frac{2}{3}\lambda_i^a \overline{\lambda}_j^a
\end{align}
For $f^{abc}\lambda_i^a \lambda_j^b \overline{\lambda_k^c}$,
\begin{align}
&\overline{\lambda}_k^a \overline{\lambda}_k^b - \overline{\lambda}_k^b \overline{\lambda}_k^a =  2if^{abc}\overline{\lambda}_k^c \\
\Rightarrow \ &\lambda_i^a \lambda_j^b \overline{\lambda}_k^a \overline{\lambda}_k^b - \lambda_i^a \lambda_j^b \overline{\lambda}_k^b \overline{\lambda}_k^a = 2if^{abc}\lambda_i^a \lambda_j^b \overline{\lambda}_k^c \\
\Rightarrow \ &(\lambda_i^a \overline{\lambda}_k^a) (\lambda_j^b \overline{\lambda}_k^b) - (\lambda_j^a \overline{\lambda}_k^a) (\lambda_i^b \overline{\lambda}_k^b) = 2if^{abc}\lambda_i^a \lambda_j^b \overline{\lambda}_k^c \\
\Rightarrow \ &f^{abc}\lambda_i^a \lambda_j^b \overline{\lambda}_k^c = -\frac{i}{2}(\lambda_i^a \overline{\lambda}_k^a) (\lambda_j^b \overline{\lambda}_k^b) +\frac{i}{2} (\lambda_j^a \overline{\lambda}_k^a) (\lambda_i^b \overline{\lambda}_k^b) 
\end{align}
Similarly,
\begin{align}
f^{abc}\lambda_i^a \overline{\lambda}_j^b \overline{\lambda}_k^c = -\frac{i}{2}(\lambda_i^a \overline{\lambda}_k^a) (\overline{\lambda}_j^b \overline{\lambda}_k^b) +\frac{i}{2} (\overline{\lambda}_j^a \overline{\lambda}_k^a) (\lambda_i^b \overline{\lambda}_k^b) 
\end{align}
When the antiquark generators appear first, we can transform the above equations as follows.
\begin{align}
&d^{abc}\overline{\lambda}_i^a \lambda_j^b \lambda_k^c = -\frac{1}{2}( \overline{\lambda}_i^a \lambda_j^a) (\overline{\lambda}_i^b \lambda_k^b ) -\frac{1}{2} (\overline{\lambda}_i^a \lambda_k^a) (\overline{\lambda}_i^b \lambda_j^b) + \frac{2}{3}\lambda_j^a \lambda_k^a \\
&d^{abc}\overline{\lambda}_i^a \overline{\lambda}_j^b \lambda_k^c =  -\frac{1}{2} (\overline{\lambda}_i^a \overline{\lambda}_j^a) (\overline{\lambda}_i^b \lambda_k^b) -\frac{1}{2}(\overline{\lambda}_i^a \lambda_k^a) (\overline{\lambda}_i^b \overline{\lambda}_j^b)+ \frac{2}{3}\overline{\lambda}_j^a \lambda_k^a \\
&f^{abc}\overline{\lambda}_i^a \lambda_j^b \lambda_k^c = -\frac{i}{2}(\overline{\lambda}_i^a \lambda_j^a) (\overline{\lambda}_i^b \lambda_k^b) +\frac{i}{2} (\overline{\lambda}_i^a \lambda_k^a) (\overline{\lambda}_i^b \lambda_j^b) \\
&f^{abc}\overline{\lambda}_i^a \overline{\lambda}_j^b \lambda_k^c = -\frac{i}{2}(\overline{\lambda}_i^a \overline{\lambda}_j^a) (\overline{\lambda}_i^b \lambda_k^b) +\frac{i}{2} (\overline{\lambda}_i^a \lambda_k^a) (\overline{\lambda}_i^b \overline{\lambda}_j^b)
\end{align}

\paragraph{Color matrix elements:}

For tetraquarks $\overline{q}(1)\overline{q}(2)q(3)q(4)$, there are three different color basis representations: $\{\lvert \mathbf{3}_{12}\overline{\mathbf{3}}_{34}\rangle$, $\lvert \overline{\mathbf{6}}_{12}\mathbf{6}_{34}\rangle \}$, $\{\lvert \mathbf{1}_{13}\mathbf{1}_{24}\rangle $, $\lvert \mathbf{8}_{13}\mathbf{8}_{24}\rangle \}$, and $\{\lvert \mathbf{1}_{14}\mathbf{1}_{23}\rangle $, $\lvert \mathbf{8}_{14}\mathbf{8}_{23}\rangle \}$. We represent the two- and three-quark color matrices for antidiquark-diquark and singlet-octet representations. For the $\lvert \mathbf{3}_{12}\overline{\mathbf{3}}_{34}\rangle ,\;\lvert \overline{\mathbf{6}}_{12}\mathbf{6}_{34}\rangle $ color basis,
\begin{align}
&\bar\lambda_1^c\bar\lambda_2^c=\lambda_3^c\lambda_4^c=\begin{pmatrix}
-8/3 & 0 \\ 0 & 4/3
\end{pmatrix}, \quad \bar\lambda_1^c\lambda_3^c=\bar\lambda_2^c\lambda_4^c=\begin{pmatrix}
-4/3 & -2\sqrt{2}\\ -2\sqrt{2}&-10/3
\end{pmatrix}, \quad \bar\lambda_1^c\lambda_4^c=\bar\lambda_2^c\lambda_3^c=\begin{pmatrix}
-4/3 & 2\sqrt{2}\\ 2\sqrt{2}&-10/3
\end{pmatrix},\nonumber
\end{align}

\vspace{-0.5cm}

\begin{align}
f^{abc}\overline{\lambda}_1^a \overline{\lambda}_2^b \lambda_3^c &= f^{abc}\overline{\lambda}_1^a \lambda_3^b \lambda_4^c = \left(
\begin{tabular}{cc}
   $0$  & $-4i\sqrt{2}$ \\
   $4i\sqrt{2}$  & $0$
\end{tabular}\right), \
\quad f^{abc}\overline{\lambda}_1^a \overline{\lambda}_2^b \lambda_4^c = f^{abc}\overline{\lambda}_2^a \lambda_3^b \lambda_4^c =  \left(
\begin{tabular}{cc}
   $0$  & $4i\sqrt{2}$ \\
   $-4i\sqrt{2}$  & $0$
\end{tabular}\right), \nonumber\\
d^{abc}\overline{\lambda}_1^a \overline{\lambda}_2^b \lambda_3^c &= d^{abc}\overline{\lambda}_1^a \overline{\lambda}_2^b \lambda_4^c = \frac{1}{9}\left(
\begin{tabular}{cc}
   $-40$  & $0$ \\
   $0$  & $20$
\end{tabular}\right),   \quad d^{abc}\overline{\lambda}_1^a \lambda_3^b \lambda_4^c =  d^{abc}\overline{\lambda}_2^a \lambda_3^b \lambda_4^c = \frac{1}{9}\left(
\begin{tabular}{cc}
   $40$  & $0$ \\
   $0$  & $-20$
\end{tabular}\right).
\end{align}
For the $\lvert \mathbf{1}_{13}\mathbf{1}_{24}\rangle ,\;\lvert \mathbf{8}_{13}\mathbf{8}_{24}\rangle $ color basis,
\begin{align}
\bar\lambda_1^c\bar\lambda_2^c=\lambda_3^c\lambda_4^c= \begin{pmatrix}
   0 & 4\sqrt{2}/3 \\ 4\sqrt{2}/3 & -4/3
\end{pmatrix}, \quad \bar\lambda_1^c\lambda_3^c=\bar\lambda_2^c\lambda_4^c= \begin{pmatrix}
-16/3 & 0 \\0 & 2/3
\end{pmatrix}, \quad \bar\lambda_1^c\lambda_4^c=\bar\lambda_2^c\lambda_3^c= \begin{pmatrix}
0 & -4\sqrt{2}/3 \\ -4\sqrt2/3 &-14/3
\end{pmatrix}, \nonumber
\end{align}

\vspace{-0.5cm}

\begin{align}
f^{abc}\overline{\lambda}_1^a \overline{\lambda}_2^b \lambda_3^c &= f^{abc}\overline{\lambda}_1^a \lambda_3^b \lambda_4^c = \left(
\begin{tabular}{cc}
   $0$  & $-4i\sqrt{2}$ \\
   $4i\sqrt{2}$  & $0$
\end{tabular}\right), \quad f^{abc}\overline{\lambda}_1^a \overline{\lambda}_2^b \lambda_4^c = f^{abc}\overline{\lambda}_2^a \lambda_3^b \lambda_4^c = \left(
\begin{tabular}{cc}
   $0$  & $4i\sqrt{2}$ \\
   $-4i\sqrt{2}$  & $0$
\end{tabular}\right), \nonumber\\
d^{abc}\overline{\lambda}_1^a \overline{\lambda}_2^b \lambda_3^c &=  d^{abc}\overline{\lambda}_1^a \overline{\lambda}_2^b \lambda_4^c = \frac{1}{9} \left(
\begin{tabular}{cc}
   $0$  & $20\sqrt{2}$ \\
   $20\sqrt{2}$  & $-20$
\end{tabular}\right), \quad
d^{abc}\overline{\lambda}_1^a \lambda_3^b \lambda_4^c = d^{abc}\overline{\lambda}_2^a \lambda_3^b \lambda_4^c = \frac{1}{9}\left(
\begin{tabular}{cc}
   $0$  & $-20\sqrt{2}$ \\
   $-20\sqrt{2}$  & $20$
\end{tabular}\right).
\end{align}
The two-body color matrices in the $(\mathbf{3}_{12}\bar{\mathbf{3}}_{34},
\bar{\mathbf{6}}_{12}\mathbf{6}_{34})$ basis follow directly from the
projectors onto $\mathbf{3}$ and $\bar{\mathbf{6}}$ in the $\bar q\bar q$
subsystem: the operator $\bar\lambda_1^c\bar\lambda_2^c$ has eigenvalue
$-8/3$ on the $\mathbf{3}_{12}$ configuration and $+4/3$ on
$\bar{\mathbf{6}}_{12}$, reflecting the difference of quadratic Casimirs.
Similarly, the off-diagonal entries in $\bar\lambda_1^c\lambda_3^c$ arise
from Fierz rearrangements between the $(12)(34)$ and $(13)(24)$ couplings.

Using the above color matrices, we can calculate the three-quark potentials for tetraquarks. The three-quark potentials for $T_{cc}$ and $\chi_{c1}(3872)$ are summarized below. The color basis sets for the two states are $\lvert \mathbf{3}_{12}\overline{\mathbf{3}}_{34}\rangle ,\;\lvert \overline{\mathbf{6}}_{12}\mathbf{6}_{34}\rangle $ ($T_{cc}$) and $\lvert \mathbf{1}_{13}\mathbf{1}_{24}\rangle ,\;\lvert \mathbf{8}_{13}\mathbf{8}_{24}\rangle $ ($\chi_{c1}(3872)$), respectively. For $T_{cc}$: $\{\lvert \mathbf{3}_{12}\overline{\mathbf{3}}_{34}\rangle \otimes\lvert (\bar1\bar2)_{S=0}(34)_{S=1}\rangle ,\;\lvert \overline{\mathbf{6}}_{12}\mathbf{6}_{34}\rangle \otimes\lvert (\bar1\bar2)_{S=1}(34)_{S=0}\rangle \}$,
\begin{align}
\sum L^{C-C}_{ijk}&=\begin{pmatrix}
\frac{32}{9m_u}+\frac{32}{9m_c} & 0 \\
0 & -\frac{208}{9m_u}-\frac{208}{9m_c}
\end{pmatrix},\nonumber\\
\sum L^{S-S}_{ijk}&=\begin{pmatrix}
-\frac{224}{9m_u^3 m_c^2}+\frac{224}{3m_u^2 m_c^3}  & -\frac{224\sqrt{2}}{3m_u^4 m_c} +\frac{32\sqrt{2}}{m_u^3 m_c^2} - \frac{32\sqrt{2}}{m_u^2 m_c^3} + \frac{160\sqrt{2}}{3m_u m_c^4} \\
   -\frac{224\sqrt{2}}{3m_u^4 m_c} +\frac{32\sqrt{2}}{m_u^3 m_c^2} - \frac{32\sqrt{2}}{m_u^2 m_c^3} + \frac{160\sqrt{2}}{3m_u m_c^4}  & -\frac{112}{3m_u^3 m_c^2}+\frac{112}{9m_u^2 m_c^3}
\end{pmatrix},\nonumber\\
\sum L^{C-S}_{ijk}&=\begin{pmatrix}
-\frac{256}{3m_u^3}+\frac{256}{9m_c^3}  & \frac{32\sqrt{2}}{3m_u^2 m_c} +\frac{32\sqrt{2}}{3m_u m_c^2}  \\
   \frac{32\sqrt{2}}{3m_u^2 m_c} +\frac{32\sqrt{2}}{3m_u m_c^2}  & -\frac{320}{9m_u^3}+\frac{320}{3m_c^3}
\end{pmatrix}.
\end{align}
For $\chi_{c1}(3872)$: $\{\lvert \mathbf{1}_{13}\mathbf{1}_{24}\rangle ,\;\lvert \mathbf{8}_{13}\mathbf{8}_{24}\rangle \}\otimes\lvert ((\bar13)_{S=1}(\bar24)_{S=1})_{S=1}\rangle $,
\begin{align}
\sum L^{C-C}_{ijk}&=\begin{pmatrix}
-\frac{128}{9m_u}-\frac{128}{9m_c}  & -\frac{80\sqrt{2}}{9m_u} -\frac{80\sqrt{2}}{9m_c}  \\
   -\frac{80\sqrt{2}}{9m_u} -\frac{80\sqrt{2}}{9m_c}  & -\frac{16}{3m_u}-\frac{16}{3m_c}
\end{pmatrix},\nonumber\\
\sum L^{S-S}_{ijk}&=\begin{pmatrix}
-\frac{128}{9m_u^3 m_c^2}-\frac{128}{9m_u^2 m_c^3}  & -\frac{80\sqrt{2}}{9m_u^3 m_c^2} -\frac{80\sqrt{2}}{9m_u^2 m_c^3}  \\
   -\frac{80\sqrt{2}}{9m_u^3 m_c^2} -\frac{80\sqrt{2}}{9m_u^2 m_c^3}  & \frac{32}{3m_u^3 m_c^2}+\frac{32}{3m_u^2 m_c^3}+\frac{16}{m_u m_c^4}+\frac{16}{m_u^4 m_c}
\end{pmatrix},\nonumber\\
\sum L^{C-S}_{ijk}&=\begin{pmatrix}
\frac{256}{9m_u^2 m_c}+\frac{256}{9m_u m_c^2}  & \frac{160\sqrt{2}}{9m_u^2 m_c} +\frac{160\sqrt{2}}{9m_u m_c^2}  \\
   \frac{160\sqrt{2}}{9m_u^2 m_c} +\frac{160\sqrt{2}}{9m_u m_c^2} & -\frac{16}{m_u^3}-\frac{16}{3m_u^2 m_c} -\frac{16}{3m_u m_c^2}-\frac{16}{m_c^3}
\end{pmatrix}.
\end{align}
These operator matrices provide the algebraic input for future multiquark calculations. Numerical three-quark shifts are not quoted here because they require spatial wave functions and matrix elements recomputed with the final fitted Hamiltonian.

\subsection{Dibaryons}

We can also consider the three-quark potential in a dibaryon system. In this section, we represent the matrix elements of the three-quark potential using the flavor SU(3) symmetry-breaking parameter: $\delta = 1-{m_u}/{m_s}$. As can be seen from the following matrix elements, in a system composed solely of quarks, each flavor-spin state is an eigenstate of the three-quark potential in the flavor SU(3)-symmetric case. Among the various possible dibaryon states, we present here the matrix elements of the three-quark potential for the most attractive cases: the $H$-dibaryon and the $N\Omega$. For $uuddss\;(I=0,\,S=0)$: $F=\mathbf{1},\mathbf{27}$
\begin{align}
\sum L_{ijk}^{C-C}&=\frac{1}{m_u}\begin{pmatrix}
\frac{64}{3}-\frac{64  }{9}\delta & 0 \\
0 & \frac{64}{3}-\frac{64  }{9}\delta
\end{pmatrix},\nonumber\\
\sum L_{ijk}^{S-S}&=\frac{1}{m_u^5}\begin{pmatrix}
-320+\frac{1600  }{3}\delta-764 \delta ^2+\frac{1520 }{3}\delta ^3-\frac{400 }{3}\delta ^4 & -\frac{380 }{3 \sqrt{3}}\delta ^2+\frac{80 }{\sqrt{3}}\delta ^3-\frac{80}{3 \sqrt{3}}\delta ^4 \\
-\frac{380 }{3 \sqrt{3}}\delta ^2+\frac{80 }{\sqrt{3}}\delta ^3-\frac{80}{3 \sqrt{3}}\delta ^4 & -\frac{832}{3}-\frac{64 }{3}\delta +\frac{1580 }{9}\delta ^2-\frac{496}{3}\delta ^3+\frac{560 }{9}\delta ^4
\end{pmatrix},\nonumber\\
\sum L_{ijk}^{C-S}&=\frac{1}{m_u^3}\begin{pmatrix}
-256+256 \delta +\frac{32 }{3}\delta ^2-48 \delta ^3 & \frac{352 }{3 \sqrt{3}}\delta ^2-\frac{176}{3 \sqrt{3}}\delta ^3 \\
\frac{352 }{3 \sqrt{3}}\delta ^2-\frac{176 }{3 \sqrt{3}}\delta ^3 & 256-512 \delta +\frac{3104 }{9}\delta^2-\frac{784 }{9}\delta ^3
\end{pmatrix}.
\end{align}
For $uudsss\;(I=1/2,\;S=2)$: $F=\mathbf{27},\mathbf{8}$
\begin{align}
\sum L_{ijk}^{C-C}&=\frac{1}{m_u}\begin{pmatrix}
\frac{64}{3}-\frac{32  }{3}\delta & 0 \\
0 & \frac{64}{3}-\frac{32 }{3}\delta 
\end{pmatrix},\nonumber\\
\sum L_{ijk}^{S-S}&=\frac{1}{m_u^5}\begin{pmatrix}
\frac{64}{3}-\frac{7328  }{15}\delta+\frac{5088 }{5}\delta ^2-\frac{13264 }{15}\delta ^3+\frac{1136}{3}\delta ^4-\frac{304 }{5}\delta ^5 & \frac{2176 }{15}\delta -\frac{3728 }{15}\delta^2+\frac{2168 }{15}\delta ^3-\frac{56 }{3}\delta ^4-\frac{136 }{15}\delta ^5 \\
\frac{2176  }{15}\delta-\frac{3728 }{15}\delta ^2+\frac{2168 }{15}\delta ^3-\frac{56 }{3}\delta^4-\frac{136 }{15}\delta ^5 & -\frac{16}{3}-\frac{592  }{15}\delta+\frac{552 }{5}\delta^2-\frac{3116 }{15}\delta ^3+\frac{460 }{3}\delta ^4-\frac{236 }{5}\delta ^5
\end{pmatrix},\nonumber\\
\sum L_{ijk}^{C-S}&=\frac{1}{m_u^3}\begin{pmatrix}
\frac{640}{3}-\frac{3136 }{5}\delta +\frac{8512 }{15}\delta ^2-\frac{896 }{5}\delta ^3 & -\frac{128 }{5}\delta+\frac{512 }{5}\delta ^2-\frac{704 }{15}\delta ^3 \\
-\frac{128 }{5}\delta+\frac{512 }{5}\delta ^2-\frac{704 }{15}\delta ^3 &-\frac{320}{3}-\frac{544  }{5}\delta+\frac{3808 }{15}\delta ^2-\frac{544 }{5}\delta ^3
\end{pmatrix}.
\end{align}

The color-spin three-quark potential values, evaluated with the couplings of Eq.~\eqref{3bd_coeff_fitting} in the contact and common-orbital limit, are as follows (all matrix elements are given in MeV). For $uuddss\;(I=0,\,S=0)$,
\begin{equation}
\sum AL_{ijk}^{C-C}=\begin{pmatrix}
-32.946  & 0  \\
   0  & -32.946
\end{pmatrix}, \;\; \sum B L_{ijk}^{S-S}=\begin{pmatrix}
-37.927 & -2.135 \\
   -2.135  & -51.484
\end{pmatrix}, \;\; \sum CL_{ijk}^{C-S}=\begin{pmatrix}
21.467  & -1.542  \\
   -1.542 & -13.238
\end{pmatrix},
\end{equation}
and for $uudsss\;(I=1/2,\,S=2)$
\begin{equation}
\sum AL_{ijk}^{C-C}=\begin{pmatrix}
-30.086  & 0  \\
   0  & -30.086
\end{pmatrix}, \;\; \sum B L_{ijk}^{S-S}=\begin{pmatrix}
-11.437  & 5.296 \\
   5.296   & -2.752
\end{pmatrix}, \;\; \sum CL_{ijk}^{C-S}=\begin{pmatrix}
-4.621  & -0.698 \\
   -0.698 & 17.005
\end{pmatrix}.
\end{equation}
Collecting all the above,
\begin{equation}
L^{3Q}=\begin{pmatrix}
-49.405  & -3.677  \\
   -3.677 & -97.6682
\end{pmatrix} \quad \begin{matrix}
\text{for }uuddss \\(I=0,\,S=0)
\end{matrix} \;,\qquad L^{3Q}=\begin{pmatrix}
-46.144  & 4.597 \\
   4.597 & -15.833
\end{pmatrix}\quad \begin{matrix}
\text{for }uudsss \\(I=1/2,\,S=2)
\end{matrix}\;.
\end{equation}
The dominant flavor components of the $H$ dibaryon and the $N\Omega$ system are the singlet and the octet, respectively.
In the contact and common-orbital limit adopted here, the common orbital matrix element can be factored out, and the effect on dibaryon binding is determined by comparison with the corresponding two-baryon threshold.
Although the absolute three-quark matrix elements of the compact six-quark configurations are negative, the threshold baryons receive larger negative contributions.
The resulting threshold-subtracted shifts are therefore positive, implying that the three-quark interaction acts repulsively with respect to the binding of both the $H$ dibaryon and the $N\Omega$ system in this limit.

The original prediction framed the $H$ dibaryon as a compact six-quark object---an SU(3)-flavor singlet configuration bound by chromomagnetic (color-spin) attraction~\cite{Jaffe1977_2}. Subsequent studies, however, have shown that its nature is more subtle. In particular, lattice QCD in the flavor-SU(3)-symmetric limit supports a bound $H$ dibaryon~\cite{Inoue2011}, while phenomenological analyses based on hypernuclear constraints and effective descriptions disfavor a deeply bound compact state. Experimental searches at KEK and J-PARC have likewise not provided evidence for such a deeply bound six-quark configuration~\cite{Ahn1998,Ahn2018,Kim2019}. In this sense, whether the physical $H$ is realized as a compact state, a weakly bound near-threshold state, or not bound at all remains model dependent. If its dominant structure is spatially extended rather than compact, the effect of a short-range three-body force is expected to be correspondingly suppressed.

\subsection{Pentaquarks}

\begin{table}[!t]
\centering
\caption{\justifying
Matrix elements of the operator sum $\sum_{i<j<k} L^{X}_{ijk}$ in the chosen $\{\phi_\alpha\}$ basis for the $uudc\bar c$ system, shown separately for (a) $(I,S)=(1/2,3/2)$ and (b) $(1/2,1/2)$.  For each operator class $X=C\!-\!C$, $S\!-\!S$, and $C\!-\!S$, the table lists the coefficients of the expansion in the SU$(3)_F$-breaking parameter $\delta \equiv 1-m_u/m_c$, i.e.\ the coefficients of $1,\delta,\delta^2,\delta^3,\delta^4$ in each matrix element $\langle \phi_i \vert  \sum L^X \vert  \phi_j \rangle$.  Only independent entries with $i\le j$ are displayed.  The overall mass dimensions have been factored out as $m_u^{-1}$ for $C\!-\!C$, $m_u^{-5}$ for $S\!-\!S$, and $m_u^{-3}$ for $C\!-\!S$, so that the tabulated numbers are dimensionless coefficients.  For compactness, expressions of the form $\alpha/(\beta\sqrt{\gamma})$ are written as $\alpha/\beta\sqrt{\gamma}$.}
\label{tab:all_Lijk_twocol}

(a) $uudc\overline{c}\;(I=1/2,S=3/2)$

\resizebox{\columnwidth}{!}{%
\begin{tblr}{
  colspec={Q[c]|*{5}{Q[c]}|*{5}{Q[c]}|*{5}{Q[c]}},
  row{1-2} = {bg=headgray, halign=c, valign=m},
  hline{2} = {1-16}{0.3pt},
  hline{3} = {1-16}{0.3pt},
  vline{2,7,12} = {1-Z}{0.3pt},
  rowsep = 1.5pt,
}
\toprule
Type ($X$) & \SetCell[c=5]{c}{$C-C \;\;(\times m_u^{-1})$} & & & & & \SetCell[c=5]{c}{$S-S \;\;(\times m_u^{-5})$} & & & & & \SetCell[c=5]{c}{$C-S \;\;(\times m_u^{-3})$} & & & & \\
Component & $1$ & $\delta$ & $\delta^2$ & $\delta^3$ & $\delta^4$ & $1$ & $\delta$ & $\delta^2$ & $\delta^3$ & $\delta^4$ & $1$ & $\delta$ & $\delta^2$ & $\delta^3$ & $\delta^4$ \\

$\langle \phi_1| \sum L^X |\phi_1 \rangle $ & 0 & $+4/3$ & 0 & 0 & 0 & $-8/3$ & $-28$ & $+36$ & $-44/3$ & 0 & $+112/3$ & $-144$ & $+144$ & $-48$ & 0 \\
$\langle \phi_1| \sum L^X |\phi_2 \rangle $ & 0 & 0 & 0 & 0 & 0 & $+40\sqrt{5/3}$ & $-40\sqrt{5/3}$ & 0 & 0 & 0 & $+80\sqrt{5/3}$ & $-40\sqrt{15}$ & $+40\sqrt{5/3}$ & 0 & 0 \\
$\langle \phi_1| \sum L^X |\phi_3 \rangle $ & 0 & 0 & 0 & 0 & 0 & 0 & $+40\sqrt{2/3}$ & $-80\sqrt{2/3}$ & $+160\sqrt{2}/3\sqrt{3}$ & $-40\sqrt{2}/3\sqrt{3}$ & $-160\sqrt{2}/3\sqrt{3}$ & $+80\sqrt{2/3}$ & $-80\sqrt{2}/3\sqrt{3}$ & 0 & 0 \\
$\langle \phi_1| \sum L^X |\phi_4 \rangle $ & 0 & 0 & 0 & 0 & 0 & $+32\sqrt{2/3}$ & $+32\sqrt{6}$ & $-256\sqrt{2/3}$ & $+560\sqrt{2}/3\sqrt{3}$ & $-176\sqrt{2}/3\sqrt{3}$ & $-224\sqrt{2}/3\sqrt{3}$ & $+112\sqrt{2/3}$ & $-112\sqrt{2}/3\sqrt{3}$ & 0 & 0 \\
$\langle \phi_2| \sum L^X |\phi_2 \rangle $ & 0 & $+4/3$ & 0 & 0 & 0 & $+40/3$ & $-260/3$ & $+332/3$ & $-172/3$ & $32/3$ & $-400/9$ & $+224/3$ & $-512/9$ & $+16$ & 0 \\
$\langle \phi_2| \sum L^X |\phi_3 \rangle $ & 0 & 0 & 0 & 0 & 0 & $+64\sqrt{10}/3$ & $-88\sqrt{10}/3$ & $+16\sqrt{10}$ & $-32\sqrt{10}/3$ & $+8\sqrt{10}/3$ & $+32\sqrt{10}/9$ & $-16\sqrt{10}/3$ & $+16\sqrt{10}/9$ & 0 & 0 \\
$\langle \phi_2| \sum L^X |\phi_4 \rangle $ & 0 & 0 & 0 & 0 & 0 & 0 & 0 & 0 & 0 & 0 & $+160\sqrt{10}/9$ & $-80\sqrt{10}/3$ & $+80\sqrt{10}/9$ & 0 & 0 \\
$\langle \phi_3| \sum L^X |\phi_3 \rangle $ & 0 & $+4/3$ & 0 & 0 & 0 & $-16/3$ & $-188/3$ & $-4/3$ & $+12$ & $-16/3$ & $+32/9$ & $+104/3$ & $-392/9$ & $+16$ & 0 \\
$\langle \phi_3| \sum L^X |\phi_4 \rangle $ & 0 & 0 & 0 & 0 & 0 & $+160/3$ & $-160$ & $+160$ & $-160/3$ & 0 & $+320/9$ & $-160/3$ & $+160/9$ & 0 & 0 \\
$\langle \phi_4| \sum L^X |\phi_4 \rangle $ & 0 & $+64/3$ & 0 & 0 & 0 & $-128/3$ & $-64/3$ & $+128/3$ & $-64/3$ & 0 & $-1024/9$ & $+128/3$ & $-128/9$ & 0 & 0 \\
\bottomrule
\end{tblr}}
\vspace{6pt}

(b) $uudc\overline{c}\;(I=1/2,S=1/2)$

\resizebox{\columnwidth}{!}{%
\begin{tblr}{
  colspec={Q[c]|*{5}{Q[c]}|*{5}{Q[c]}|*{5}{Q[c]}},
  row{1-2} = {bg=headgray, halign=c, valign=m},
  hline{2} = {1-16}{0.3pt},
  hline{3} = {1-16}{0.3pt},
  vline{2,7,12} = {1-Z}{0.3pt},
  rowsep = 1.5pt,
}
\toprule
Type ($X$) & \SetCell[c=5]{c}{$C-C \;\;(\times m_u^{-1})$} & & & & & \SetCell[c=5]{c}{$S-S \;\;(\times m_u^{-5})$} & & & & & \SetCell[c=5]{c}{$C-S \;\;(\times m_u^{-3})$} & & & & \\
Component & $1$ & $\delta$ & $\delta^2$ & $\delta^3$ & $\delta^4$ & $1$ & $\delta$ & $\delta^2$ & $\delta^3$ & $\delta^4$ & $1$ & $\delta$ & $\delta^2$ & $\delta^3$ & $\delta^4$ \\

$\langle \phi_1| \sum L^X |\phi_1 \rangle $ & 0 & $+4/3$ & 0 & 0 & 0 & $+64/3$ & $-428/3$ & $+620/3$ & $-364/3$ & $+80/3$ & $-640/9$ & $+344/3$ & $-632/9$ & $+16$ & 0 \\
$\langle \phi_1| \sum L^X |\phi_2 \rangle $ & 0 & 0 & 0 & 0 & 0 & 0 & $-80/\sqrt{3}$ & $+160/\sqrt{3}$ & $-320/3\sqrt{3}$ & $+80/3\sqrt{3}$ & $+320/3\sqrt{3}$ & $-160/\sqrt{3}$ & $+160/3\sqrt{3}$ & 0 & 0 \\
$\langle \phi_1| \sum L^X |\phi_3 \rangle $ & 0 & 0 & 0 & 0 & 0 & $+128/3$ & $-176/3$ & $+32$ & $-64/3$ & $+16/3$ & $+64/9$ & $-32/3$ & $+32/9$ & 0 & 0 \\
$\langle \phi_1| \sum L^X |\phi_4 \rangle $ & 0 & 0 & 0 & 0 & 0 & $-640/\sqrt{3}$ & $+576\sqrt{3}$ & $-1984/\sqrt{3}$ & $+3488/3\sqrt{3}$ & $-800/3\sqrt{3}$ & $+448/3\sqrt{3}$ & $-224/\sqrt{3}$ & $+224/3\sqrt{3}$ & 0 & 0 \\
$\langle \phi_1| \sum L^X |\phi_5 \rangle $ & 0 & 0 & 0 & 0 & 0 & 0 & 0 & 0 & 0 & 0 & $+320/9$ & $-160/3$ & $+160/9$ & 0 & 0 \\
$\langle \phi_2| \sum L^X |\phi_2 \rangle $ & 0 & $+4/3$ & 0 & 0 & 0 & $-256/3$ & $+60$ & $-52$ & $+44/3$ & 0 & $+176/3$ & $-144$ & $+144$ & $-48$ & 0 \\
$\langle \phi_2| \sum L^X |\phi_3 \rangle $ & 0 & 0 & 0 & 0 & 0 & $-320/\sqrt{3}$ & $+560/\sqrt{3}$ & $-120\sqrt{3}$ & $+40\sqrt{3}$ & 0 & $+80/\sqrt{3}$ & $-40\sqrt{3}$ & $+40/\sqrt{3}$ & 0 & 0 \\
$\langle \phi_2| \sum L^X |\phi_4 \rangle $ & 0 & 0 & 0 & 0 & 0 & $+160/3$ & $-160$ & $+160$ & $-160/3$ & 0 & 0 & 0 & 0 & 0 & 0 \\
$\langle \phi_2| \sum L^X |\phi_5 \rangle $ & 0 & 0 & 0 & 0 & 0 & $+224/\sqrt{3}$ & $-224\sqrt{3}$ & $+800/\sqrt{3}$ & $-1408/3\sqrt{3}$ & $+352/3\sqrt{3}$ & $+448/3\sqrt{3}$ & $-224/\sqrt{3}$ & $+224/3\sqrt{3}$ & 0 & 0 \\
$\langle \phi_3| \sum L^X |\phi_3 \rangle $ & 0 & $+4/3$ & 0 & 0 & 0 & $-640/3$ & $+724/3$ & $-364/3$ & $+20$ & $+32/3$ & $-208/9$ & $+224/3$ & $-512/9$ & $+16$ & 0 \\
$\langle \phi_3| \sum L^X |\phi_4 \rangle $ & 0 & 0 & 0 & 0 & 0 & $-352/\sqrt{3}$ & $+416\sqrt{3}$ & $-1696/\sqrt{3}$ & $+3200/3\sqrt{3}$ & $-800/3\sqrt{3}$ & $+448/3\sqrt{3}$ & $-224/\sqrt{3}$ & $+224/3\sqrt{3}$ & 0 & 0 \\
$\langle \phi_3| \sum L^X |\phi_5 \rangle $ & 0 & 0 & 0 & 0 & 0 & $+160/3$ & $-160$ & $+160$ & $-160/3$ & 0 & $-640/9$ & $+320/3$ & $-320/9$ & 0 & 0 \\
$\langle \phi_4| \sum L^X |\phi_4 \rangle $ & 0 & $+64/3$ & 0 & 0 & 0 & $+128/3$ & $-192$ & $+128$ & $-64/3$ & 0 & $-256/3$ & 0 & 0 & 0 & 0 \\
$\langle \phi_4| \sum L^X |\phi_5 \rangle $ & 0 & 0 & 0 & 0 & 0 & 0 & 0 & 0 & 0 & 0 & 0 & 0 & 0 & 0 & 0 \\
$\langle \phi_5| \sum L^X |\phi_5 \rangle $ & 0 & $+64/3$ & 0 & 0 & 0 & $-128/3$ & $-64/3$ & $+128/3$ & $-64/3$ & 0 & $-256/9$ & $-256/3$ & $+256/9$ & 0 & 0 \\
\bottomrule
\end{tblr}}
\end{table}

We represent the matrix elements of three-quark potentials for $uudc\overline{c}\;(I=1/2,\;S=1/2,\;3/2)$ pentaquarks using the parameter $\delta = 1-{m_u}/{m_c}$. We construct the wave function of $q(1)q(2)q(3)c(4)\overline{c}(5)$ as follows. For $uudc\bar{c} \;(I=1/2,\;S=3/2)$
\begin{align}
\phi_1 &= [(123)_{I=\frac{1}{2}}4\overline{5}]\otimes [(123)_{C=\mathbf{8}} (4\overline{5})_{C=\mathbf{8}}] \otimes [(123)_{S=\frac{3}{2}}(4\overline{5})_{S=0}], \nonumber\\
\phi_2 &= [(123)_{I=\frac{1}{2}}4\overline{5}]\otimes [(123)_{C=\mathbf{8}} (4\overline{5})_{C=\mathbf{8}}] \otimes [(123)_{S=\frac{3}{2}}(4\overline{5})_{S=1}], \nonumber\\
\phi_3 &= [(123)_{I=\frac{1}{2}}4\overline{5}]\otimes [(123)_{C=\mathbf{8}} (4\overline{5})_{C=\mathbf{8}}] \otimes [(123)_{S=\frac{1}{2}}(4\overline{5})_{S=1}], \nonumber\\
\phi_4 &= [(123)_{I=\frac{1}{2}}4\overline{5}]\otimes [(123)_{C=\mathbf{1}} (4\overline{5})_{C=\mathbf{1}}] \otimes [(123)_{S=\frac{1}{2}}(4\overline{5})_{S=1}].
\end{align}
The matrix elements of the pair-color operators are diagonal in the present basis because the off-diagonal color couplings connect orthogonal spin--isospin components:
\begin{align}
&\lambda_i^c\lambda_j^c=\text{diag}\Big(-\frac{2}{3},-\frac{2}{3},-\frac{2}{3},-\frac{8}{3}\Big)\quad (i,j\in\{1,2,3\}), \qquad \lambda_4^c\bar \lambda_5^c=\text{diag}\Big(\frac{2}{3},\frac{2}{3},\frac{2}{3},-\frac{16}{3}\Big), \nonumber\\
&\lambda_i^c \lambda_{4}^c=\lambda_i^c\bar\lambda_5^c=\text{diag}(-2,-2,-2,0) \quad (i\in\{1,2,3\}).
\end{align}
For $uudc\bar{c} \;(I=1/2,\;S=1/2)$
\begin{align}
\phi_1 &= [(123)_{I=\frac{1}{2}}4\overline{5}]\otimes [(123)_{C=\mathbf{8}} (4\overline{5})_{C=\mathbf{8}}] \otimes [(123)_{S=\frac{3}{2}}(4\overline{5})_{S=1}], \nonumber\\
\phi_2 &= [(123)_{I=\frac{1}{2}}4\overline{5}]\otimes [(123)_{C=\mathbf{8}} (4\overline{5})_{C=\mathbf{8}}] \otimes [(123)_{S=\frac{1}{2}}(4\overline{5})_{S=0}], \nonumber\\
\phi_3 &= [(123)_{I=\frac{1}{2}}4\overline{5}]\otimes [(123)_{C=\mathbf{8}} (4\overline{5})_{C=\mathbf{8}}] \otimes [(123)_{S=\frac{1}{2}}(4\overline{5})_{S=1}], \nonumber\\
\phi_4 &= [(123)_{I=\frac{1}{2}}4\overline{5}]\otimes [(123)_{C=\mathbf{1}} (4\overline{5})_{C=\mathbf{1}}] \otimes [(123)_{S=\frac{1}{2}}(4\overline{5})_{S=0}], \nonumber\\
\phi_5 &= [(123)_{I=\frac{1}{2}}4\overline{5}]\otimes [(123)_{C=\mathbf{1}} (4\overline{5})_{C=\mathbf{1}}] \otimes [(123)_{S=\frac{1}{2}}(4\overline{5})_{S=1}].
\end{align}
The corresponding two-body matrices are:
\begin{align}
&\lambda_i^c\lambda_j^c=\text{diag}\Big(-\frac{2}{3},-\frac{2}{3},-\frac{2}{3},-\frac{8}{3},-\frac{8}{3}\Big)\quad (i,j\in\{1,2,3\}), \qquad \lambda_4^c\bar \lambda_5^c=\text{diag}\Big(\frac{2}{3},\frac{2}{3},\frac{2}{3},-\frac{16}{3},-\frac{16}{3}\Big), \nonumber\\
&\lambda_i^c \lambda_{4}^c=\lambda_i^c\bar\lambda_5^c=\text{diag}(-2,-2,-2,0,0) \quad (i\in\{1,2,3\}).
\end{align}

For $P_c(4380)$, the two-body color-spin term was evaluated in Ref.~\cite{park2017}.
Diagonalizing the color-spin term $\sum_{i<j}\lambda_i^c\lambda_j^c \boldsymbol{\sigma}_i\cdot\boldsymbol{\sigma}_j/(m_i m_j)$ in the basis of four orthonormal $[I\otimes C\otimes S]$ states gives a lowest eigenvalue of $-86.2~\mathrm{GeV}^{-2}$, which is essentially the same as the hyperfine contribution at the $pJ/\psi$ threshold.
After including the kinetic and confinement terms, the resulting mass is 4087.6~MeV, close to the uncorrelated $pJ/\psi$ threshold in the two-body model.
The wave function is dominated by the color-singlet $p\otimes J/\psi$ component, indicating no genuine compact binding.
It is therefore useful to determine whether the present three-quark interaction raises or lowers the pentaquark energy relative to the corresponding baryon--charmonium threshold.

The separate matrix elements of $\sum_{i<j<k}L^{C-C}_{ijk}$, $\sum_{i<j<k}L^{S-S}_{ijk}$, and $\sum_{i<j<k}L^{C-S}_{ijk}$ are listed in Table~\ref{tab:all_Lijk_twocol}.
Their numerical combination with the baryon-fitted couplings is given below.
The result refers to the contact and common-orbital limit.
\begin{equation}
L^{3Q}=
\begin{pmatrix}
 -2.297 & 0.183 & 1.040 & 3.269 \\
 0.183 & -2.088 & 1.851 & -0.864 \\
 1.040 & 1.851 & -14.147 & -0.490 \\
 3.269 & -0.864 & -0.490 & -27.169
\end{pmatrix}
~\mathrm{MeV}\qquad {\rm for}\; {uudc\bar c} \;(I,S)=(1/2,\,3/2),
\label{eq:pentaquark_L3Q_numeric_S32}
\end{equation}
and
\begin{equation}
L^{3Q}=
\begin{pmatrix}
 -1.950 & -1.471 & 1.171 & -5.333 & -0.546 \\
 -1.471 & -15.887 & -4.739 & 0.057 & -0.767 \\
 1.171 & -4.739 & -19.019 & -1.476 & 1.149 \\
 -5.333 & 0.057 & -1.476 & -27.091 & 0 \\
 -0.546 & -0.767 & 1.149 & 0 & -28.479
\end{pmatrix}
~\mathrm{MeV} \qquad {\rm for }\;{uudc\bar c}\;(I,S)=(1/2,\,1/2).
\label{eq:pentaquark_L3Q_numeric_S12}
\end{equation}
The color-singlet basis state $\phi_4$ in the $S=3/2$ sector corresponds to the $pJ/\psi$ configuration.
In the $S=1/2$ sector, $\phi_4$ and $\phi_5$ correspond to $p\eta_c$ and $pJ/\psi$, respectively.
The three-quark contribution to the proton in the same contact convention is $\langle L^{3Q}\rangle_p=-72.9639~\mathrm{MeV}$, whereas $\langle L^{3Q}\rangle_{\eta_c} =\langle L^{3Q}\rangle_{J/\psi}=0$.
Defining the threshold-subtracted contribution by $\Delta_{\mathrm{th}}^{3Q} \equiv \langle L^{3Q}\rangle_{uudc\bar c} - \left( \langle L^{3Q}\rangle_p + \langle L^{3Q}\rangle_{c\bar c} \right)$, we obtain
\begin{align}
\Delta_{\mathrm{th}}^{3Q} (pJ/\psi;S=3/2,\phi_4)
&=-27.1686-(-72.9639) = +45.7953~\mathrm{MeV}, \nonumber
\\
\Delta_{\mathrm{th}}^{3Q} (p\eta_c;S=1/2,\phi_4)
&= -27.0913-(-72.9639) = +45.8725~\mathrm{MeV}, \nonumber
\\
\Delta_{\mathrm{th}}^{3Q} (pJ/\psi;S=1/2,\phi_5)
&= -28.4794-(-72.9639) = +44.4845~\mathrm{MeV}.
\label{eq:pentaquark_threshold_shifts}
\end{align}
In the contact and common-orbital limit, the fitted three-quark interaction therefore acts repulsively with respect to the $p\eta_c$ and $pJ/\psi$ thresholds, raising these color-singlet pentaquark configurations relative to threshold by approximately 44--46~MeV.
These values refer to the indicated diagonal color-singlet components; the shift of a physically mixed pentaquark eigenstate requires the corresponding eigenvector of the complete two-body plus three-body Hamiltonian.

\section{Discussion and implications}\label{sec:discuss}

In the reduced valence Hamiltonian used here, we find that the pair interaction alone does not give the simultaneous meson--baryon compatibility obtained after the connected three-quark interaction is included. This result should be understood within the present valence-space formulation. We first fix the two-body Hamiltonian and its spectator embedding and then define the three-quark interaction as the remaining connected contribution. If the $qqq^\ast$ sector is retained explicitly, part of the same dynamics can instead be described by transitions to and propagation through the intermediate sector. The separation between pair and connected three-quark contributions is therefore representation dependent.

The conclusion also holds across the (u,d,s,c,b) compatibility scans, changes of the meson--baryon sector weight, common and separate additive constants, alternative OGE-based two-body Hamiltonians, and tests on baryons outside the principal calibration set.
To our knowledge, this provides the first internally consistent resolution of the flavor-dependency puzzle within a framework in which one meson-calibrated two-body Hamiltonian is transferred directly to the light, strange, and charm baryon sectors, followed only by an explicit short-range three-quark correction.

An analogy with nuclear three-body forces remains useful at the level of model-space reduction and many-body operator structure: after a lower-body Hamiltonian is fixed, eliminating intermediate configurations can induce connected many-body operators.
For compact multiquark systems, the same operator basis gives a useful starting point, but its contribution must be evaluated with the relevant wave functions and relative to the corresponding hadronic thresholds.
At the baryonic or dense-matter level, an additional matching calculation is required before the fitted quark-level interaction can be converted into hyperonic three-body forces or an equation-of-state contribution.
For example, systematic attempts have been made within nucleonic relativistic mean-field models to understand the maximum neutron-star mass~\cite{Nam:2025kca}, motivating a corresponding extension to hyperonic matter.
Hyperonic many-body interactions can substantially modify the onset of strangeness and the stiffness of the high-density equation of state, and several microscopic calculations have shown a strong sensitivity of neutron-star properties to hyperon---nucleon three-body forces~\cite{TolosFabbietti2020,Lonardoni2015HyperonPuzzle,Logoteta2019Hyperonic3BF,Gerstung2020Hyperon3BF,Vidana2025FemtoscopicNS}.

\begin{table}[!t]
\centering
\caption{\justifying 
Ground-state baryon masses across the $u$, $d$, $s$, and $c$ sectors, calculated with quark models or on the lattice. For every theoretical row, the signed deviations and RMSE are recomputed from the EXP row using only the states for which a mass is listed. The $\mathrm{S}^2\mathrm{AJ}$ results are obtained using the meson-calibrated two-body Hamiltonian together with the Yukawa-profile three-quark interaction, whereas the other quark-model calculations employ one or more of the strategies \textcircled{1}--\textcircled{5} discussed below. All masses, signed deviations, and RMSE values are in MeV.} \label{tab:previous_studies}
\resizebox{\textwidth}{!}{
\begin{tblr}{
  colspec={Q[l] *{16}{Q[c]}},
  row{1-2} = {bg=headgray, halign=c, valign=m},
  cell{1}{1} = {}{halign=l},
  cell{2}{1} = {}{halign=l},
  hline{3} = {1-17}{0.8pt},
  hline{5,7,9,11,13,15,17,19,21,23,25,27,29,31,33,35,37,39,41,43,45,47,49,51,53,55,57,59,61,63,65} = {1-17}{0.3pt},
  rowsep = 1.5pt,
}

\toprule
BARYONS
 & $N(1/2^+)$ & $\Delta(3/2^+)$ & $\Lambda(1/2^+)$ & $\Sigma(1/2^+)$ & $\Sigma^{*}(3/2^+)$ & $\Xi(1/2^+)$ & $\Xi^{*}(3/2^+)$ & $\Omega(3/2^+)$ & $\Lambda_{c}(1/2^+)$ & $\Sigma_{c}(1/2^+)$ & $\Sigma_{c}^{*}(3/2^+)$ & $\Xi_{c}(1/2^+)$ & $\Xi_{c}^{*}(3/2^+)$ & $\Omega_{c}(1/2^+)$ & $\Omega_{c}^{*}(3/2^+)$ & $\Xi_{cc}(1/2^+)$\\
EXP
 & 938.272 & 1232 & 1115.683 & 1192.642 & 1383.800 & 1315 & 1535 & 1672 & 2286.460 & 2452.650 & 2518.480 & 2469 & 2646 & 2695 & 2766 & 3619.970\\
$\mathrm{S^2AJ}$ (2Q+$L_{\mathrm Y}^{3Q}$)
& 937.449
& 1230.352
& 1117.473
& 1189.489
& 1386.937
&
&
&
& 2287.781
& 2449.758
& 2524.703
&
&
&
&
& 3616.370\\
ERR($\sigma=3.13$)
 & \textcolor{devgreen}{-0.82} & \textcolor{devgreen}{-1.65} & \textcolor{devred}{+1.79} & \textcolor{devgreen}{-3.15} & \textcolor{devred}{+3.14} &  &  &  & \textcolor{devred}{+1.32} & \textcolor{devgreen}{-2.89} & \textcolor{devred}{+6.22} &  &  &  &  & \textcolor{devgreen}{-3.60}\\
Ref.~\cite{Stanley1980} \textcircled{5}
 & 1005 & 1321 & 1165 & 1248 & 1448 & 1360 & 1570 & 1687 & 2283 & 2488 & 2577 &  &  &  &  &\\
ERR($\sigma=52.36$)
 & \textcolor{devred}{+66.73} & \textcolor{devred}{+89.00} & \textcolor{devred}{+49.32} & \textcolor{devred}{+55.36} & \textcolor{devred}{+64.20} & \textcolor{devred}{+45.00} & \textcolor{devred}{+35.00} & \textcolor{devred}{+15.00} & \textcolor{devgreen}{-3.46} & \textcolor{devred}{+35.35} & \textcolor{devred}{+58.52} &  &  &  &  & \\
Ref.~\cite{Stanley1980} \textcircled{5}
 & 939 & 1283 & 1108 & 1198 & 1416 & 1316 & 1543 & 1663 & 2249 & 2465 & 2536 &  &  &  &  &\\
ERR($\sigma=22.83$)
 & \textcolor{devred}{+0.73} & \textcolor{devred}{+51.00} & \textcolor{devgreen}{-7.68} & \textcolor{devred}{+5.36} & \textcolor{devred}{+32.20} & \textcolor{devred}{+1.00} & \textcolor{devred}{+8.00} & \textcolor{devgreen}{-9.00} & \textcolor{devgreen}{-37.46} & \textcolor{devred}{+12.35} & \textcolor{devred}{+17.52} &  &  &  &  & \\
Ref.~\cite{Richard1983} \textcircled{1},\textcircled{4}
 &  &  & 1111 & 1176 & 1392 & 1304 & 1538 & & & 2443 & 2542 &  &  &  &  &\\
ERR($\sigma=12.78$)
 &  &  & \textcolor{devgreen}{-4.68} & \textcolor{devgreen}{-16.64} & \textcolor{devred}{+8.20} & \textcolor{devgreen}{-11.00} & \textcolor{devred}{+3.00} &  &  & \textcolor{devgreen}{-9.65} & \textcolor{devred}{+23.52} &  &  &  &  & \\
Ref.~\cite{CapstickIsgur:1986} \textcircled{2}
 & 960 & 1230 & 1115 & 1190 & 1370 & 1305 & 1505 & 1635 & 2265 & 2440 & 2495 &  &  &  &  &\\
ERR($\sigma=19.57$)
 & \textcolor{devred}{+21.73} & \textcolor{devgreen}{-2.00} & \textcolor{devgreen}{-0.68} & \textcolor{devgreen}{-2.64} & \textcolor{devgreen}{-13.80} & \textcolor{devgreen}{-10.00} & \textcolor{devgreen}{-30.00} & \textcolor{devgreen}{-37.00} & \textcolor{devgreen}{-21.46} & \textcolor{devgreen}{-12.65} & \textcolor{devgreen}{-23.48} &  &  &  &  & \\
Ref.~\cite{Kalman1989} \textcircled{4}
 & 958 & 1215 & 1134 & 1279 & 1395 & 1317 & 1446 & 1620 & 2282 & 2515 & 2556 &  &  &  &  &\\
ERR($\sigma=47.23$)
 & \textcolor{devred}{+19.73} & \textcolor{devgreen}{-17.00} & \textcolor{devred}{+18.32} & \textcolor{devred}{+86.36} & \textcolor{devred}{+11.20} & \textcolor{devred}{+2.00} & \textcolor{devgreen}{-89.00} & \textcolor{devgreen}{-52.00} & \textcolor{devgreen}{-4.46} & \textcolor{devred}{+62.35} & \textcolor{devred}{+37.52} &  &  &  &  & \\
AL1~\cite{Semay1994} \textcircled{3}
 & 943 & 1235 & 1112 & 1187 & 1386 & 1329 & 1533 & 1679 & 2285 & 2471 & 2525 &  &  &  &  &\\
ERR($\sigma=8.05$)
 & \textcolor{devred}{+4.73} & \textcolor{devred}{+3.00} & \textcolor{devgreen}{-3.68} & \textcolor{devgreen}{-5.64} & \textcolor{devred}{+2.20} & \textcolor{devred}{+14.00} & \textcolor{devgreen}{-2.00} & \textcolor{devred}{+7.00} & \textcolor{devgreen}{-1.46} & \textcolor{devred}{+18.35} & \textcolor{devred}{+6.52} &  &  &  &  & \\
AP1~\cite{SilvestreBrac1996} \textcircled{3}
 & 932 & 1224 & 1105 & 1191 & 1375 & 1318 & 1532 & 1685 & 2290 & 2463 & 2541 & 2475 & 2645 & 2694 & 2770 &\\
ERR($\sigma=8.81$)
 & \textcolor{devgreen}{-6.27} & \textcolor{devgreen}{-8.00} & \textcolor{devgreen}{-10.68} & \textcolor{devgreen}{-1.64} & \textcolor{devgreen}{-8.80} & \textcolor{devred}{+3.00} & \textcolor{devgreen}{-3.00} & \textcolor{devred}{+13.00} & \textcolor{devred}{+3.54} & \textcolor{devred}{+10.35} & \textcolor{devred}{+22.52} & \textcolor{devred}{+6.00} & \textcolor{devgreen}{-1.00} & \textcolor{devgreen}{-1.00} & \textcolor{devred}{+4.00} & \\
AL2~\cite{Semay1994} \textcircled{3}
 & 936 & 1227 & 1110 & 1191 & 1380 & 1323 & 1528 & 1672 & 2283 & 2475 & 2534 &  &  &  &  &\\
ERR($\sigma=9.27$)
 & \textcolor{devgreen}{-2.27} & \textcolor{devgreen}{-5.00} & \textcolor{devgreen}{-5.68} & \textcolor{devgreen}{-1.64} & \textcolor{devgreen}{-3.80} & \textcolor{devred}{+8.00} & \textcolor{devgreen}{-7.00} & \textcolor{devgreen}{+0.00} & \textcolor{devgreen}{-3.46} & \textcolor{devred}{+22.35} & \textcolor{devred}{+15.52} &  &  &  &  & \\
AP2~\cite{SilvestreBrac1996} \textcircled{3}
 & 946 & 1243 & 1120 & 1205 & 1390 & 1340 & 1550 & 1690 & 2279 & 2482 & 2533 & 2520 & 2655 & 2711 & 2780 &\\
ERR($\sigma=19.71$)
 & \textcolor{devred}{+7.73} & \textcolor{devred}{+11.00} & \textcolor{devred}{+4.32} & \textcolor{devred}{+12.36} & \textcolor{devred}{+6.20} & \textcolor{devred}{+25.00} & \textcolor{devred}{+15.00} & \textcolor{devred}{+18.00} & \textcolor{devgreen}{-7.46} & \textcolor{devred}{+29.35} & \textcolor{devred}{+14.52} & \textcolor{devred}{+51.00} & \textcolor{devred}{+9.00} & \textcolor{devred}{+16.00} & \textcolor{devred}{+14.00} & \\
Ref.~\cite{Roncaglia1995} \textcircled{2},\textcircled{4}
 &  &  &  &  &  &  &  &  & 2285 & 2453 & 2520 & 2468 & 2650 & 2710 & 2770 & 3660\\
ERR($\sigma=15.27$)
 &  &  &  &  &  &  &  &  & \textcolor{devgreen}{-1.46} & \textcolor{devred}{+0.35} & \textcolor{devred}{+1.52} & \textcolor{devgreen}{-1.00} & \textcolor{devred}{+4.00} & \textcolor{devred}{+15.00} & \textcolor{devred}{+4.00} & \textcolor{devred}{+40.03}\\
Ref.~\cite{Loring2001} \textcircled{4},\textcircled{5}
 & 939 & 1261 & 1108 & 1190 & 1411 & 1310 & 1539 & 1636 &  &  &  &  &  &  &  &\\
ERR($\sigma=19.31$)
 & \textcolor{devred}{+0.73} & \textcolor{devred}{+29.00} & \textcolor{devgreen}{-7.68} & \textcolor{devgreen}{-2.64} & \textcolor{devred}{+27.20} & \textcolor{devgreen}{-5.00} & \textcolor{devred}{+4.00} & \textcolor{devgreen}{-36.00} &  &  &  &  &  &  &  & \\
Ref.~\cite{Loring2001} \textcircled{4},\textcircled{5}
 & 939 & 1231 & 1123 & 1188 & 1399 & 1316 & 1541 & 1656 &  &  &  &  &  &  &  &\\
ERR($\sigma=8.66$)
 & \textcolor{devred}{+0.73} & \textcolor{devgreen}{-1.00} & \textcolor{devred}{+7.32} & \textcolor{devgreen}{-4.64} & \textcolor{devred}{+15.20} & \textcolor{devred}{+1.00} & \textcolor{devred}{+6.00} & \textcolor{devgreen}{-16.00} &  &  &  &  &  &  &  & \\
Ref.~\cite{Ebert2005} \textcircled{4},\textcircled{5}
 &  &  &  &  &  &  &  &  & 2297 & 2439 & 2518 & 2481 & 2654 & 2698 & 2768 &\\
ERR($\sigma=8.61$)
 &  &  &  &  &  &  &  &  & \textcolor{devred}{+10.54} & \textcolor{devgreen}{-13.65} & \textcolor{devgreen}{-0.48} & \textcolor{devred}{+12.00} & \textcolor{devred}{+8.00} & \textcolor{devred}{+3.00} & \textcolor{devred}{+2.00} & \\
Ref.~\cite{Faessler2006} \textcircled{4}
 &  &  &  &  &  &  &  &  & 2286 & 2453 &  & 2472 &  & 2698 &  & 3519\\
ERR($\sigma=45.20$)
 &  &  &  &  &  &  &  &  & \textcolor{devgreen}{-0.46} & \textcolor{devred}{+0.35} &  & \textcolor{devred}{+3.00} &  & \textcolor{devred}{+3.00} &  & \textcolor{devgreen}{-100.97}\\
Ref.~\cite{Albertus2004} \textcircled{4},\textcircled{5}
 &  &  &  &  &  &  &  &  & 2270 & 2460 & 2440 & 2410 & 2550 & 2680 & 2660 &\\
ERR($\sigma=66.16$)
 &  &  &  &  &  &  &  &  & \textcolor{devgreen}{-16.46} & \textcolor{devred}{+7.35} & \textcolor{devgreen}{-78.48} & \textcolor{devgreen}{-59.00} & \textcolor{devgreen}{-96.00} & \textcolor{devgreen}{-15.00} & \textcolor{devgreen}{-106.00} & \\
Ref.~\cite{Albertus2004} \textcircled{4},\textcircled{5}
 &  &  &  &  &  &  &  &  & 2321 & 2494 &  & 2502 &  & 2707 &  &\\
ERR($\sigma=32.16$)
 &  &  &  &  &  &  &  &  & \textcolor{devred}{+34.54} & \textcolor{devred}{+41.35} &  & \textcolor{devred}{+33.00} &  & \textcolor{devred}{+12.00} &  & \\
Ref.~\cite{Albertus2004} \textcircled{4},\textcircled{5}
 &  &  &  &  &  &  &  &  & 2320 & 2498 & 2570 & 2506 & 2671 & 2713 & 2770 &\\
ERR($\sigma=34.16$)
 &  &  &  &  &  &  &  &  & \textcolor{devred}{+33.54} & \textcolor{devred}{+45.35} & \textcolor{devred}{+51.52} & \textcolor{devred}{+37.00} & \textcolor{devred}{+25.00} & \textcolor{devred}{+18.00} & \textcolor{devred}{+4.00} & \\
Ref.~\cite{Roberts2008} \textcircled{4}
 &  &  &  &  &  &  &  &  & 2268 & 2455 &  & 2466 &  & 2718 &  & 3676\\
ERR($\sigma=28.37$)
 &  &  &  &  &  &  &  &  & \textcolor{devgreen}{-18.46} & \textcolor{devred}{+2.35} &  & \textcolor{devgreen}{-3.00} &  & \textcolor{devred}{+23.00} &  & \textcolor{devred}{+56.03}\\
Ref.~\cite{Melde2008} \textcircled{4}
 & 939 & 1240 & 1136 & 1180 & 1389 & 1348 & 1528 &  &  &  &  &  &  &  &  &\\
ERR($\sigma=16.05$)
 & \textcolor{devred}{+0.73} & \textcolor{devred}{+8.00} & \textcolor{devred}{+20.32} & \textcolor{devgreen}{-12.64} & \textcolor{devred}{+5.20} & \textcolor{devred}{+33.00} & \textcolor{devgreen}{-7.00} &  &  &  &  &  &  &  &  & \\
Ref.~\cite{Melde2008} \textcircled{4}
 & 939 & 1231 & 1113 & 1213 & 1373 & 1346 & 1516 &  &  &  &  &  &  &  &  &\\
ERR($\sigma=16.31$)
 & \textcolor{devred}{+0.73} & \textcolor{devgreen}{-1.00} & \textcolor{devgreen}{-2.68} & \textcolor{devred}{+20.36} & \textcolor{devgreen}{-10.80} & \textcolor{devred}{+31.00} & \textcolor{devgreen}{-19.00} &  &  &  &  &  &  &  &  & \\
Ref.~\cite{Ebert2011} \textcircled{4},\textcircled{5}
 &  &  &  &  &  &  &  &  & 2286 & 2443 & 2519 & 2476 & 2649 & 2698 &  &\\
ERR($\sigma=5.17$)
 &  &  &  &  &  &  &  &  & \textcolor{devgreen}{-0.46} & \textcolor{devgreen}{-9.65} & \textcolor{devred}{+0.52} & \textcolor{devred}{+7.00} & \textcolor{devred}{+3.00} & \textcolor{devred}{+3.00} &  & \\
Ref.~\cite{Engel2013} \textbf{lattice}
 & 1000 & 1344 & 1149 & 1216 & 1471 & 1303 & 1553 & 1642 &  &  &  &  &  &  &  &\\
ERR($\sigma=58.08$)
 & \textcolor{devred}{+61.73} & \textcolor{devred}{+112.00} & \textcolor{devred}{+33.32} & \textcolor{devred}{+23.36} & \textcolor{devred}{+87.20} & \textcolor{devgreen}{-12.00} & \textcolor{devred}{+18.00} & \textcolor{devgreen}{-30.00} &  &  &  &  &  &  &  & \\
Ref.~\cite{Namekawa2013} \textbf{lattice}
 &  &  &  &  &  &  &  &  & 2333 & 2467 & 2538 & 2455 & 2674 & 2673 & 2738 & 3603\\
ERR($\sigma=25.71$)
 &  &  &  &  &  &  &  &  & \textcolor{devred}{+46.54} & \textcolor{devred}{+14.35} & \textcolor{devred}{+19.52} & \textcolor{devgreen}{-14.00} & \textcolor{devred}{+28.00} & \textcolor{devgreen}{-22.00} & \textcolor{devgreen}{-28.00} & \textcolor{devgreen}{-16.97}\\
Ref.~\cite{Santopinto2015} \textcircled{4},\textcircled{5}
 & 939 & 1247 & 1116 & 1211 & 1334 & 1317 & 1552 & 1672 &  &  &  &  &  &  &  &\\
ERR($\sigma=20.42$)
 & \textcolor{devred}{+0.73} & \textcolor{devred}{+15.00} & \textcolor{devred}{+0.32} & \textcolor{devred}{+18.36} & \textcolor{devgreen}{-49.80} & \textcolor{devred}{+2.00} & \textcolor{devred}{+17.00} & \textcolor{devgreen}{+0.00} &  &  &  &  &  &  &  & \\
Ref.~\cite{FaustovGalkin2015} \textcircled{4},\textcircled{5}
 &  &  & 1115 & 1187 & 1381 & 1330 & 1518 & 1678 &  &  &  &  &  &  &  &\\
ERR($\sigma=9.92$)
 &  &  & \textcolor{devgreen}{-0.68} & \textcolor{devgreen}{-5.64} & \textcolor{devgreen}{-2.80} & \textcolor{devred}{+15.00} & \textcolor{devgreen}{-17.00} & \textcolor{devred}{+6.00} &  &  &  &  &  &  &  & \\
Ref.~\cite{Yoshida2015} \textcircled{1},\textcircled{4}
 &  &  & 1116 & 1197 & 1391 & 1325 &  &  & 2285 & 2460 & 2523 &  &  & 2731 & 2779 & 3685\\
ERR($\sigma=24.38$)
 &  &  & \textcolor{devred}{+0.32} & \textcolor{devred}{+4.36} & \textcolor{devred}{+7.20} & \textcolor{devred}{+10.00} &  &  & \textcolor{devgreen}{-1.46} & \textcolor{devred}{+7.35} & \textcolor{devred}{+4.52} &  &  & \textcolor{devred}{+36.00} & \textcolor{devred}{+13.00} & \textcolor{devred}{+65.03}\\
Ref.~\cite{Garcilazo2007} \textcircled{1},\textcircled{4}
 &  &  &  &  &  &  &  &  & 2292 & 2448 & 2505 & 2496 & 2633 & 2701 & 2759 &\\
ERR($\sigma=13.19$)
 &  &  &  &  &  &  &  &  & \textcolor{devred}{+5.54} & \textcolor{devgreen}{-4.65} & \textcolor{devgreen}{-13.48} & \textcolor{devred}{+27.00} & \textcolor{devgreen}{-13.00} & \textcolor{devred}{+6.00} & \textcolor{devgreen}{-7.00} & \\
Ref.~\cite{wpark2023} \textcircled{2}
 & 936.67 & 1242.1 & 1111.21 & 1191.78 & 1395.7 & 1326.58 & 1540.28 &  & 2269.04 & 2438.13 & 2523.1 & 2471.6 & 2649.58 &  &  & 3620.81\\
ERR($\sigma=8.70$)
 & \textcolor{devgreen}{-1.60} & \textcolor{devred}{+10.10} & \textcolor{devgreen}{-4.47} & \textcolor{devgreen}{-0.86} & \textcolor{devred}{+11.90} & \textcolor{devred}{+11.58} & \textcolor{devred}{+5.28} &  & \textcolor{devgreen}{-17.42} & \textcolor{devgreen}{-14.52} & \textcolor{devred}{+4.62} & \textcolor{devred}{+2.60} & \textcolor{devred}{+3.58} &  &  & \textcolor{devred}{+0.84}\\
Ref.~\cite{wpark2024} \textcircled{2}
 & 941.58 & 1262.8 & 1105.1 & 1193.5 & 1400.8 & 1314.3 & 1530.4 &  & 2268.3 & 2448.2 & 2530.6 & 2465.4 & 2642.8 &  &  & 3623.8\\
ERR($\sigma=12.14$)
 & \textcolor{devred}{+3.31} & \textcolor{devred}{+30.80} & \textcolor{devgreen}{-10.58} & \textcolor{devred}{+0.86} & \textcolor{devred}{+17.00} & \textcolor{devgreen}{-0.70} & \textcolor{devgreen}{-4.60} &  & \textcolor{devgreen}{-18.16} & \textcolor{devgreen}{-4.45} & \textcolor{devred}{+12.12} & \textcolor{devgreen}{-3.60} & \textcolor{devgreen}{-3.20} &  &  & \textcolor{devred}{+3.83}\\
Ref.~\cite{Valcarce2008} \textcircled{4}
 &  &  &  &  &  &  &  &  & 2285 & 2435 & 2502 & 2471 & 2642 & 2699 & 2767 & 3579\\
ERR($\sigma=16.96$)
 &  &  &  &  &  &  &  &  & \textcolor{devgreen}{-1.46} & \textcolor{devgreen}{-17.65} & \textcolor{devgreen}{-16.48} & \textcolor{devred}{+2.00} & \textcolor{devgreen}{-4.00} & \textcolor{devred}{+4.00} & \textcolor{devred}{+1.00} & \textcolor{devgreen}{-40.97}\\
Ref.~\cite{HeHaradaZou2023} \textcircled{2},\textcircled{3}
 & 951.1 & 1208.9 & 1143 & 1192.5 & 1368.6 & 1322.4 & 1516.5 & 1647.5 & 2279.6 & 2476 & 2504.7 & 2455.2 & 2642.5 & 2729 & 2760.4 & 3653.9\\
ERR($\sigma=19.36$)
 & \textcolor{devred}{+12.83} & \textcolor{devgreen}{-23.10} & \textcolor{devred}{+27.32} & \textcolor{devgreen}{-0.14} & \textcolor{devgreen}{-15.20} & \textcolor{devred}{+7.40} & \textcolor{devgreen}{-18.50} & \textcolor{devgreen}{-24.50} & \textcolor{devgreen}{-6.86} & \textcolor{devred}{+23.35} & \textcolor{devgreen}{-13.78} & \textcolor{devgreen}{-13.80} & \textcolor{devgreen}{-3.50} & \textcolor{devred}{+34.00} & \textcolor{devgreen}{-5.60} & \textcolor{devred}{+33.93}\\
\bottomrule
\end{tblr}}
\end{table}

\begin{table}[!t]
\centering
\caption*{\justifying \textbf{Table~\ref{tab:previous_studies} (continued):} Results from the same reference correspond to different parameter sets or to calculations performed under different model assumptions and setups.}

\resizebox{\textwidth}{!}{
\begin{tblr}{
  colspec={Q[l] *{16}{Q[c]}},
  row{1-2} = {bg=headgray, halign=c, valign=m},
  cell{1}{1} = {}{halign=l},
  cell{2}{1} = {}{halign=l},
  hline{3} = {1-17}{0.8pt},
  hline{5,7,9,11,13,15,17,19,21,23,25,27,29,31,33,35,37,39,41,43,45,47,49,51,53,55,57,59,61,63,65} = {1-17}{0.3pt},
  rowsep = 1.5pt,
}
\toprule
BARYONS
 & $N(1/2^+)$ & $\Delta(3/2^+)$ & $\Lambda(1/2^+)$ & $\Sigma(1/2^+)$ & $\Sigma^{*}(3/2^+)$ & $\Xi(1/2^+)$ & $\Xi^{*}(3/2^+)$ & $\Omega(3/2^+)$ & $\Lambda_{c}(1/2^+)$ & $\Sigma_{c}(1/2^+)$ & $\Sigma_{c}^{*}(3/2^+)$ & $\Xi_{c}(1/2^+)$ & $\Xi_{c}^{*}(3/2^+)$ & $\Omega_{c}(1/2^+)$ & $\Omega_{c}^{*}(3/2^+)$ & $\Xi_{cc}(1/2^+)$\\
EXP
 & 938.272 & 1232 & 1115.683 & 1192.642 & 1383.800 & 1315 & 1535 & 1672 & 2286.460 & 2452.650 & 2518.480 & 2469 & 2646 & 2695 & 2766 & 3619.970\\
Ref.~\cite{Valcarce2005} \textcircled{4}
 & 939 & 1232 & 1122 & 1213 & 1382 & 1351 &  & 1650 &  &  &  &  &  &  &  &\\
ERR($\sigma=17.88$)
 & \textcolor{devred}{+0.73} & \textcolor{devgreen}{+0.00} & \textcolor{devred}{+6.32} & \textcolor{devred}{+20.36} & \textcolor{devgreen}{-1.80} & \textcolor{devred}{+36.00} &  & \textcolor{devgreen}{-22.00} &  &  &  &  &  &  &  & \\
Ref.~\cite{Yu2023} \textcircled{2}
 &  &  &  &  &  &  &  &  & 2288 & 2457 & 2532 &  &  & 2699 & 2762 &\\
ERR($\sigma=6.87$)
 &  &  &  &  &  &  &  &  & \textcolor{devred}{+1.54} & \textcolor{devred}{+4.35} & \textcolor{devred}{+13.52} &  &  & \textcolor{devred}{+4.00} & \textcolor{devgreen}{-4.00} & \\
Ref.~\cite{Bijker:2001} \textcircled{4}
 & 939 & 1246 & 1133 & 1170 & 1382 & 1334 & 1524 & 1670 &  &  &  &  &  &  &  &\\
ERR($\sigma=13.69$)
 & \textcolor{devred}{+0.73} & \textcolor{devred}{+14.00} & \textcolor{devred}{+17.32} & \textcolor{devgreen}{-22.64} & \textcolor{devgreen}{-1.80} & \textcolor{devred}{+19.00} & \textcolor{devgreen}{-11.00} & \textcolor{devgreen}{-2.00} &  &  &  &  &  &  &  & \\
Ref.~\cite{Jena2009} \textcircled{4}
 & 940 & 1232 & 1116 & 1190.27 & 1384.16 & 1318.31 & 1512.21 & 1688.44 &  &  &  &  &  &  &  &\\
ERR($\sigma=10.06$)
 & \textcolor{devred}{+1.73} & \textcolor{devgreen}{+0.00} & \textcolor{devred}{+0.32} & \textcolor{devgreen}{-2.37} & \textcolor{devred}{+0.36} & \textcolor{devred}{+3.31} & \textcolor{devgreen}{-22.79} & \textcolor{devred}{+16.44} &  &  &  &  &  &  &  & \\
Ref.~\cite{CutkoskyGeiger1993} \textcircled{4}
 &  &  &  &  &  &  &  &  & 2282.3 & 2453.8 & 2494 & 2473.3 & 2621 & 2716.3 & 2743 &\\
ERR($\sigma=17.90$)
 &  &  &  &  &  &  &  &  & \textcolor{devgreen}{-4.16} & \textcolor{devred}{+1.15} & \textcolor{devgreen}{-24.48} & \textcolor{devred}{+4.30} & \textcolor{devgreen}{-25.00} & \textcolor{devred}{+21.30} & \textcolor{devgreen}{-23.00} & \\
Ref.~\cite{OrtizPacheco:2023} \textcircled{4}
 &  &  &  &  &  &  &  &  & 2281 & 2455 & 2521 & 2474 & 2653 & 2733 & 2799 & 3619\\
ERR($\sigma=18.20$)
 &  &  &  &  &  &  &  &  & \textcolor{devgreen}{-5.46} & \textcolor{devred}{+2.35} & \textcolor{devred}{+2.52} & \textcolor{devred}{+5.00} & \textcolor{devred}{+7.00} & \textcolor{devred}{+38.00} & \textcolor{devred}{+33.00} & \textcolor{devgreen}{-0.97}\\
Ref.~\cite{Zhao2011} \textcircled{4},\textcircled{5} 
 &  &  &  &  &  &  &  &  & 2307.7 & 2473.7 & 2536.7 & 2541.9 & 2653.3 & 2692.5 & 2766.7 &\\
ERR($\sigma=30.71$)
 &  &  &  &  &  &  &  &  & \textcolor{devred}{+21.24} & \textcolor{devred}{+21.05} & \textcolor{devred}{+18.22} & \textcolor{devred}{+72.90} & \textcolor{devred}{+7.30} & \textcolor{devgreen}{-2.50} & \textcolor{devred}{+0.70} & \\
Ref.~\cite{Zhao2011} \textcircled{4},\textcircled{5} 
 &  &  &  &  &  &  &  &  & 2308.8 & 2457.9 & 2521.5 & 2535.9 & 2640.1 & 2684.1 & 2767.6 &\\
ERR($\sigma=27.17$)
 &  &  &  &  &  &  &  &  & \textcolor{devred}{+22.34} & \textcolor{devred}{+5.25} & \textcolor{devred}{+3.02} & \textcolor{devred}{+66.90} & \textcolor{devgreen}{-5.90} & \textcolor{devgreen}{-10.90} & \textcolor{devred}{+1.60} & \\
Ref.~\cite{Zhao2011} \textcircled{4},\textcircled{5} 
 &  &  &  &  &  &  &  &  & 2309.3 & 2452.3 & 2516.8 & 2533.8 & 2635.8 & 2681.3 & 2765 &\\
ERR($\sigma=26.77$)
 &  &  &  &  &  &  &  &  & \textcolor{devred}{+22.84} & \textcolor{devgreen}{-0.35} & \textcolor{devgreen}{-1.68} & \textcolor{devred}{+64.80} & \textcolor{devgreen}{-10.20} & \textcolor{devgreen}{-13.70} & \textcolor{devgreen}{-1.00} & \\
Ref.~\cite{Zhao2011} \textcircled{4},\textcircled{5} 
 &  &  &  &  &  &  &  &  & 2307.9 & 2468.6 & 2532 & 2539.8 & 2648.6 & 2688.5 & 2772.5 &\\
ERR($\sigma=29.28$)
 &  &  &  &  &  &  &  &  & \textcolor{devred}{+21.44} & \textcolor{devred}{+15.95} & \textcolor{devred}{+13.52} & \textcolor{devred}{+70.80} & \textcolor{devred}{+2.60} & \textcolor{devgreen}{-6.50} & \textcolor{devred}{+6.50} & \\
Ref.~\cite{Zhao2011} \textcircled{4},\textcircled{5} 
 &  &  &  &  &  &  &  &  & 2308.4 & 2462.6 & 2526.7 & 2537.6 & 2643.9 & 2685.6 & 2769.6 &\\
ERR($\sigma=27.93$)
 &  &  &  &  &  &  &  &  & \textcolor{devred}{+21.94} & \textcolor{devred}{+9.95} & \textcolor{devred}{+8.22} & \textcolor{devred}{+68.60} & \textcolor{devgreen}{-2.10} & \textcolor{devgreen}{-9.40} & \textcolor{devred}{+3.60} & \\
Ref.~\cite{Zhao2011} \textcircled{4},\textcircled{5} 
 &  &  &  &  &  &  &  &  & 2295.1 & 2452.8 & 2503.6 & 2525.4 & 2629.8 & 2672.8 & 2752.5 &\\
ERR($\sigma=25.11$)
 &  &  &  &  &  &  &  &  & \textcolor{devred}{+8.64} & \textcolor{devred}{+0.15} & \textcolor{devgreen}{-14.88} & \textcolor{devred}{+56.40} & \textcolor{devgreen}{-16.20} & \textcolor{devgreen}{-22.20} & \textcolor{devgreen}{-13.50} & \\
Ref.~\cite{Watanabe2008} \textcircled{4}
 & 939 & 1230 & 1166 & 1315 & 1435 & 1418 &  & 1673 &  &  &  &  &  &  &  &\\
ERR($\sigma=66.27$)
 & \textcolor{devred}{+0.73} & \textcolor{devgreen}{-2.00} & \textcolor{devred}{+50.32} & \textcolor{devred}{+122.36} & \textcolor{devred}{+51.20} & \textcolor{devred}{+103.00} &  & \textcolor{devred}{+1.00} &  &  &  &  &  &  &  & \\
Ref.~\cite{Watanabe2008} \textcircled{4}
 & 939 & 1232 & 1160 & 1264 & 1404 & 1401 &  & 1673 &  &  &  &  &  &  &  &\\
ERR($\sigma=46.08$)
 & \textcolor{devred}{+0.73} & \textcolor{devgreen}{+0.00} & \textcolor{devred}{+44.32} & \textcolor{devred}{+71.36} & \textcolor{devred}{+20.20} & \textcolor{devred}{+86.00} &  & \textcolor{devred}{+1.00} &  &  &  &  &  &  &  & \\
Ref.~\cite{Watanabe2008} \textcircled{4}
 & 939 & 1234 & 1135 & 1246 & 1396 & 1375 &  & 1673 &  &  &  &  &  &  &  &\\
ERR($\sigma=31.57$)
 & \textcolor{devred}{+0.73} & \textcolor{devred}{+2.00} & \textcolor{devred}{+19.32} & \textcolor{devred}{+53.36} & \textcolor{devred}{+12.20} & \textcolor{devred}{+60.00} &  & \textcolor{devred}{+1.00} &  &  &  &  &  &  &  & \\
Ref.~\cite{Watanabe2008} \textcircled{2},\textcircled{4}
 & 939 & 1235 & 1139 & 1244 & 1384 & 1374 &  & 1671 &  &  &  &  &  &  &  &\\
ERR($\sigma=30.88$)
 & \textcolor{devred}{+0.73} & \textcolor{devred}{+3.00} & \textcolor{devred}{+23.32} & \textcolor{devred}{+51.36} & \textcolor{devred}{+0.20} & \textcolor{devred}{+59.00} &  & \textcolor{devgreen}{-1.00} &  &  &  &  &  &  &  & \\
Ref.~\cite{Watanabe2008} \textcircled{2},\textcircled{4}
 & 939 & 1236 & 1120 & 1214 & 1382 & 1359 &  & 1670 &  &  &  &  &  &  &  &\\
ERR($\sigma=18.65$)
 & \textcolor{devred}{+0.73} & \textcolor{devred}{+4.00} & \textcolor{devred}{+4.32} & \textcolor{devred}{+21.36} & \textcolor{devgreen}{-1.80} & \textcolor{devred}{+44.00} &  & \textcolor{devgreen}{-2.00} &  &  &  &  &  &  &  & \\
Ref.~\cite{Yang2018} \textcircled{4}
 & 939 & 1232 & 1127 & 1263 & 1391 & 1386 & 1536 & 1663 & 2279 & 2454 & 2486 & 2522 & 2625 & 2704 & 2747 & 3580\\
ERR($\sigma=32.26$)
 & \textcolor{devred}{+0.73} & \textcolor{devgreen}{+0.00} & \textcolor{devred}{+11.32} & \textcolor{devred}{+70.36} & \textcolor{devred}{+7.20} & \textcolor{devred}{+71.00} & \textcolor{devred}{+1.00} & \textcolor{devgreen}{-9.00} & \textcolor{devgreen}{-7.46} & \textcolor{devred}{+1.35} & \textcolor{devgreen}{-32.48} & \textcolor{devred}{+53.00} & \textcolor{devgreen}{-21.00} & \textcolor{devred}{+9.00} & \textcolor{devgreen}{-19.00} & \textcolor{devgreen}{-39.97}\\
Ref.~\cite{Yang2018} \textcircled{4}
 & 939 & 1236 & 1128 & 1269 & 1398 & 1395 & 1549 & 1687 & 2246 & 2413 & 2459 & 2508 & 2613 & 2705 & 2759 & 3572\\
ERR($\sigma=39.39$)
 & \textcolor{devred}{+0.73} & \textcolor{devred}{+4.00} & \textcolor{devred}{+12.32} & \textcolor{devred}{+76.36} & \textcolor{devred}{+14.20} & \textcolor{devred}{+80.00} & \textcolor{devred}{+14.00} & \textcolor{devred}{+15.00} & \textcolor{devgreen}{-40.46} & \textcolor{devgreen}{-39.65} & \textcolor{devgreen}{-59.48} & \textcolor{devred}{+39.00} & \textcolor{devgreen}{-33.00} & \textcolor{devred}{+10.00} & \textcolor{devgreen}{-7.00} & \textcolor{devgreen}{-47.97}\\
Ref.~\cite{Yang2018} \textcircled{4}
 & 939 & 1233 & 1111 & 1254 & 1362 & 1373 & 1493 & 1631 & 2291 & 2467 & 2481 & 2521 & 2593 & 2693 & 2708 & 3540\\
ERR($\sigma=41.82$)
 & \textcolor{devred}{+0.73} & \textcolor{devred}{+1.00} & \textcolor{devgreen}{-4.68} & \textcolor{devred}{+61.36} & \textcolor{devgreen}{-21.80} & \textcolor{devred}{+58.00} & \textcolor{devgreen}{-42.00} & \textcolor{devgreen}{-41.00} & \textcolor{devred}{+4.54} & \textcolor{devred}{+14.35} & \textcolor{devgreen}{-37.48} & \textcolor{devred}{+52.00} & \textcolor{devgreen}{-53.00} & \textcolor{devgreen}{-2.00} & \textcolor{devgreen}{-58.00} & \textcolor{devgreen}{-79.97}\\
Ref.~\cite{Yang2018} \textcircled{4}
 & 936 & 1222 & 1103 & 1319 & 1384 & 1391 & 1535 & 1685 & 2286 & 2434 & 2449 & 2569 & 2615 & 2758 & 2777 & 3589\\
ERR($\sigma=52.06$)
 & \textcolor{devgreen}{-2.27} & \textcolor{devgreen}{-10.00} & \textcolor{devgreen}{-12.68} & \textcolor{devred}{+126.36} & \textcolor{devred}{+0.20} & \textcolor{devred}{+76.00} & \textcolor{devgreen}{+0.00} & \textcolor{devred}{+13.00} & \textcolor{devgreen}{-0.46} & \textcolor{devgreen}{-18.65} & \textcolor{devgreen}{-69.48} & \textcolor{devred}{+100.00} & \textcolor{devgreen}{-31.00} & \textcolor{devred}{+63.00} & \textcolor{devred}{+11.00} & \textcolor{devgreen}{-30.97}\\
Ref.~\cite{Yang2018} \textcircled{4}
 & 956 & 1224 & 1117 & 1333 & 1376 & 1399 & 1517 & 1655 & 2285 & 2428 & 2436 & 2559 & 2589 & 2724 & 2736 & 3544\\
ERR($\sigma=58.14$)
 & \textcolor{devred}{+17.73} & \textcolor{devgreen}{-8.00} & \textcolor{devred}{+1.32} & \textcolor{devred}{+140.36} & \textcolor{devgreen}{-7.80} & \textcolor{devred}{+84.00} & \textcolor{devgreen}{-18.00} & \textcolor{devgreen}{-17.00} & \textcolor{devgreen}{-1.46} & \textcolor{devgreen}{-24.65} & \textcolor{devgreen}{-82.48} & \textcolor{devred}{+90.00} & \textcolor{devgreen}{-57.00} & \textcolor{devred}{+29.00} & \textcolor{devgreen}{-30.00} & \textcolor{devgreen}{-75.97}\\
Ref.~\cite{Shah2016SCB} \textcircled{4}
 &  &  &  &  &  &  &  &  & 2286 & 2448 & 2506 & 2470 & 2635 & 2695 & 2748 &\\
ERR($\sigma=9.44$)
 &  &  &  &  &  &  &  &  & \textcolor{devgreen}{-0.46} & \textcolor{devgreen}{-4.65} & \textcolor{devgreen}{-12.48} & \textcolor{devred}{+1.00} & \textcolor{devgreen}{-11.00} & \textcolor{devgreen}{+0.00} & \textcolor{devgreen}{-18.00} & \\
Ref.~\cite{Shah2016SCB} \textcircled{4}
 &  &  &  &  &  &  &  &  & 2287 & 2454 & 2530 & 2471 & 2625 & 2695 & 2754 &\\
ERR($\sigma=10.17$)
 &  &  &  &  &  &  &  &  & \textcolor{devred}{+0.54} & \textcolor{devred}{+1.35} & \textcolor{devred}{+11.52} & \textcolor{devred}{+2.00} & \textcolor{devgreen}{-21.00} & \textcolor{devgreen}{+0.00} & \textcolor{devgreen}{-12.00} & \\
Ref.~\cite{Shah2016SCB} \textcircled{4}
 &  &  &  &  &  &  &  &  & 2286 & 2446 & 2506 & 2470 & 2616 & 2695 & 2742 &\\
ERR($\sigma=15.48$)
 &  &  &  &  &  &  &  &  & \textcolor{devgreen}{-0.46} & \textcolor{devgreen}{-6.65} & \textcolor{devgreen}{-12.48} & \textcolor{devred}{+1.00} & \textcolor{devgreen}{-30.00} & \textcolor{devgreen}{+0.00} & \textcolor{devgreen}{-24.00} & \\
Ref.~\cite{Shah2016SCB} \textcircled{4}
 &  &  &  &  &  &  &  &  & 2286 & 2454 & 2531 & 2471 & 2613 & 2695 & 2752 &\\
ERR($\sigma=14.38$)
 &  &  &  &  &  &  &  &  & \textcolor{devgreen}{-0.46} & \textcolor{devred}{+1.35} & \textcolor{devred}{+12.52} & \textcolor{devred}{+2.00} & \textcolor{devgreen}{-33.00} & \textcolor{devgreen}{+0.00} & \textcolor{devgreen}{-14.00} & \\
Ref.~\cite{Shah2016SCB} \textcircled{4}
 &  &  &  &  &  &  &  &  & 2286 & 2444 & 2509 & 2471 & 2635 & 2695 & 2748 &\\
ERR($\sigma=9.36$)
 &  &  &  &  &  &  &  &  & \textcolor{devgreen}{-0.46} & \textcolor{devgreen}{-8.65} & \textcolor{devgreen}{-9.48} & \textcolor{devred}{+2.00} & \textcolor{devgreen}{-11.00} & \textcolor{devgreen}{+0.00} & \textcolor{devgreen}{-18.00} & \\
Ref.~\cite{Shah2016SCB} \textcircled{4}
 &  &  &  &  &  &  &  &  & 2286 & 2454 & 2561 & 2470 & 2624 & 2695 & 2750 &\\
ERR($\sigma=19.09$)
 &  &  &  &  &  &  &  &  & \textcolor{devgreen}{-0.46} & \textcolor{devred}{+1.35} & \textcolor{devred}{+42.52} & \textcolor{devred}{+1.00} & \textcolor{devgreen}{-22.00} & \textcolor{devgreen}{+0.00} & \textcolor{devgreen}{-16.00} & \\
Ref.~\cite{Shah2016SCB} \textcircled{4}
 &  &  &  &  &  &  &  &  & 2286 & 2449 & 2506 & 2470 & 2625 & 2695 & 2740 &\\
ERR($\sigma=13.56$)
 &  &  &  &  &  &  &  &  & \textcolor{devgreen}{-0.46} & \textcolor{devgreen}{-3.65} & \textcolor{devgreen}{-12.48} & \textcolor{devred}{+1.00} & \textcolor{devgreen}{-21.00} & \textcolor{devgreen}{+0.00} & \textcolor{devgreen}{-26.00} & \\
Ref.~\cite{Shah2016SCB} \textcircled{4}
 &  &  &  &  &  &  &  &  & 2286 & 2454 & 2530 & 2470 & 2619 & 2695 & 2745 &\\
ERR($\sigma=13.66$)
 &  &  &  &  &  &  &  &  & \textcolor{devgreen}{-0.46} & \textcolor{devred}{+1.35} & \textcolor{devred}{+11.52} & \textcolor{devred}{+1.00} & \textcolor{devgreen}{-27.00} & \textcolor{devgreen}{+0.00} & \textcolor{devgreen}{-21.00} & \\
Ref.~\cite{Shah2016SCB} \textcircled{4}
 &  &  &  &  &  &  &  &  & 2286 & 2445 & 2509 & 2470 & 2628 & 2695 & 2751 &\\
ERR($\sigma=9.99$)
 &  &  &  &  &  &  &  &  & \textcolor{devgreen}{-0.46} & \textcolor{devgreen}{-7.65} & \textcolor{devgreen}{-9.48} & \textcolor{devred}{+1.00} & \textcolor{devgreen}{-18.00} & \textcolor{devgreen}{+0.00} & \textcolor{devgreen}{-15.00} & \\
Ref.~\cite{Shah2016SCB} \textcircled{4}
 &  &  &  &  &  &  &  &  & 2286 & 2454 & 2564 & 2473 & 2635 & 2695 & 2757 &\\
ERR($\sigma=18.10$)
 &  &  &  &  &  &  &  &  & \textcolor{devgreen}{-0.46} & \textcolor{devred}{+1.35} & \textcolor{devred}{+45.52} & \textcolor{devred}{+4.00} & \textcolor{devgreen}{-11.00} & \textcolor{devgreen}{+0.00} & \textcolor{devgreen}{-9.00} & \\
\bottomrule
\end{tblr}}

\end{table}

\paragraph{Three-body-extended quark models:}

We test whether the baryon spectrum can be reproduced while keeping the meson-calibrated two-body Hamiltonian fixed.
With the meson-calibrated two-body parameters kept unchanged, the direct baryon RMSE is 65.04~MeV.
The spatially contact interaction lowers it to 18.87~MeV, while the Gaussian and Yukawa profiles lower it to 6.28 and 3.13~MeV, respectively.
The finite-range interaction adds only the three couplings $A$, $B$, and $C$; its range is fixed by the constituent masses and introduces no additional continuous range parameter.

Table~\ref{tab:previous_studies} places this result beside representative quark-model and lattice calculations~\cite{Richard1983,Kalman1989,%
Roncaglia1995,Faessler2006,Albertus2004,Engel2013,Namekawa2013,Santopinto2015,Garcilazo2007,wpark2023,wpark2024,Valcarce2008,HeHaradaZou2023,Bijker:2001,Jena2009,CutkoskyGeiger1993,Zhao2011,Watanabe2008,Yang2018,Shah2016SCB}.
The tabulated model residuals and lattice uncertainties have different statistical meanings, so the table is used to compare numerical scales and fitting strategies rather than to assign a common statistical rank.
The important difference is the calibration protocol: the present baryon result follows from a meson-fixed pair Hamiltonian, whereas many small residuals in the literature are obtained after sector-specific or combined refits.
Table~\ref{tab:3q_force_comparison} compares the present two-body and two-body-plus-three-body results with corresponding predictions from earlier meson-calibrated quark models.

The first-order treatment of the three-quark interaction is also tested directly in \ref{subsec:gaussian_3q_perturbative_validation}.
In the common retained $S$-wave space, independent first-order and rediagonalized Gaussian fits change the dimension-matched coupling scales by at most 1.19\%, and the corresponding masses differ by at most 0.174~MeV.
The minimum two-body--three-body fidelity is 0.997913, showing that the required mass shifts are produced mainly by the diagonal three-quark matrix element rather than by a large distortion of the ground-state wave function.

The construction is not restricted to the $\mathrm{S}^{2}\mathrm{AJ}$ pair Hamiltonian.
The compatibility analysis of  Section~\ref{sec:udscb_benchmark} shows that the same three color--spin operator classes improve all tested OGE-based baselines, including AL1, AL2, AP1, AP2, BD, SL, and GI/CI.
Across these baselines, the same low-dimensional operator basis repeatedly improves the spectrum, while the determined coupling values vary with the underlying Hamiltonian and spatial profile.

\begin{table}[!t]
\centering
\caption{\justifying Comparison of two-body and two-body-plus-three-body predictions obtained from meson-fitted parameters. For each model, the RMSE is evaluated only over the states for which that model quotes a mass. The Silvestre--Brac three-body force \cite{SilvestreBrac1996} has the form $V_{3Q}=C/(m_1m_2m_3)$ and is denoted SB, while the model of Bhaduri \emph{et al.} \cite{Bhaduri1981} is denoted BD. All masses, signed deviations, and RMSE values are in MeV.}
\label{tab:3q_force_comparison}

\resizebox{\columnwidth}{!}{%
\begin{tblr}{
  colspec={Q[l]| *{12}{c}},
  row{1-2} = {bg=headgray, halign=c, valign=m},
  cell{1}{1} = {}{halign=l},
  cell{2}{1} = {}{halign=l},
  hline{1} = {1-13}{0.8pt},
  hline{3} = {1-13}{0.3pt},
  hline{7} = {1-13}{0.3pt},
  hline{11} = {1-13}{0.3pt},
  hline{15} = {1-13}{0.3pt},
  hline{19} = {1-13}{0.3pt},
  hline{23} = {1-13}{0.3pt},
  hline{27} = {1-13}{0.3pt},
  hline{31} = {1-13}{0.8pt},
  rowsep = 1.5pt
}
BARYONS& $N(1/2^+)$ & $\Delta(3/2^+)$ & $\Lambda(1/2^+)$ & $\Sigma(1/2^+)$ & $\Sigma^{*}(3/2^+)$ & $\Xi(1/2^+)$ & $\Omega(3/2^+)$ & $\Lambda_{c}(1/2^+)$ & $\Sigma_{c}(1/2^+)$ & $\Sigma_{c}^{*}(3/2^+)$ & $\Omega_{c}(1/2^+)$ & $\Xi_{cc}(1/2^+)$\\
EXP & 938.272 & 1232 & 1115.683 & 1192.642 & 1383.800 & 1315 & 1672 & 2286.460 & 2452.650 & 2518.480 & 2695 & 3619.970\\
S$^2$AJ (2Q) & 1038.746 & 1338.864 & 1171.919 & 1259.291 & 1450.640 &  &  & 2317.914 & 2491.972 & 2562.599 &  & 3622.232\\
ERR ($\sigma=65.04$)
 & $\textcolor{devred}{(+100.474)}$ & $\textcolor{devred}{(+106.864)}$ & $\textcolor{devred}{(+56.236)}$ & $\textcolor{devred}{(+66.649)}$ & $\textcolor{devred}{(+66.840)}$ &  &  & $\textcolor{devred}{(+31.454)}$ & $\textcolor{devred}{(+39.322)}$ & $\textcolor{devred}{(+44.119)}$ &  & $\textcolor{devred}{(+2.262)}$\\
S$^2$AJ (2Q+$L_{\mathrm Y}^{3Q}$) & 937.449 & 1230.352 & 1117.473 & 1189.489 & 1386.937 &  &  & 2287.781 & 2449.758 & 2524.703 &  & 3616.370\\
ERR ($\sigma=3.13$)
 & $\textcolor{devgreen}{(-0.823)}$ & $\textcolor{devgreen}{(-1.648)}$ & $\textcolor{devred}{(+1.790)}$ & $\textcolor{devgreen}{(-3.153)}$ & $\textcolor{devred}{(+3.137)}$ &  &  & $\textcolor{devred}{(+1.321)}$ & $\textcolor{devgreen}{(-2.892)}$ & $\textcolor{devred}{(+6.223)}$ &  & $\textcolor{devgreen}{(-3.600)}$\\
2Q(BD) & 1022   &  & 1178  & 1261  &   & 1369 & 1685 & 2325  & 2498  &  & 2707 & 3635\\
ERR ($\sigma=50.07$)
 & $\textcolor{devred}{(+83.73)}$ &  & $\textcolor{devred}{(+62.32)}$ & $\textcolor{devred}{(+68.36)}$ &  & $\textcolor{devred}{(+54.00)}$ & $\textcolor{devred}{(+13.00)}$ & $\textcolor{devred}{(+38.54)}$ & $\textcolor{devred}{(+45.35)}$ &  & $\textcolor{devred}{(+12.00)}$ & $\textcolor{devred}{(+15.03)}$\\
2Q(BD)+3Q(SB)  & 906&  & 1113  & 1196  &   & 1332 & 1664 & 2251  & 2424  &  & 2700 & 3621\\
ERR ($\sigma=19.77$)
 & $\textcolor{devgreen}{(-32.27)}$ &  & $\textcolor{devgreen}{(-2.68)}$ & $\textcolor{devred}{(+3.36)}$ &  & $\textcolor{devred}{(+17.00)}$ & $\textcolor{devgreen}{(-8.00)}$ & $\textcolor{devgreen}{(-35.46)}$ & $\textcolor{devgreen}{(-28.65)}$ &  & $\textcolor{devred}{(+5.00)}$ & $\textcolor{devred}{(+1.03)}$\\
2Q(AL1)& 998&  & 1154  & 1231  &   & 1343 & 1674 & 2296  & 2466  &  & 2678 & 3609\\
ERR ($\sigma=29.76$)
 & $\textcolor{devred}{(+59.73)}$ &  & $\textcolor{devred}{(+38.32)}$ & $\textcolor{devred}{(+38.36)}$ &  & $\textcolor{devred}{(+28.00)}$ & $\textcolor{devred}{(+2.00)}$ & $\textcolor{devred}{(+9.54)}$ & $\textcolor{devred}{(+13.35)}$ &  & $\textcolor{devgreen}{(-17.00)}$ & $\textcolor{devgreen}{(-10.97)}$\\
2Q(AL1)+3Q(SB) & 933&  & 1119  & 1196  &   & 1324 & 1663 & 2285  & 2455  &  & 2675 & 3607\\
ERR ($\sigma=9.36$)
 & $\textcolor{devgreen}{(-5.27)}$ &  & $\textcolor{devred}{(+3.32)}$ & $\textcolor{devred}{(+3.36)}$ &  & $\textcolor{devred}{(+9.00)}$ & $\textcolor{devgreen}{(-9.00)}$ & $\textcolor{devgreen}{(-1.46)}$ & $\textcolor{devred}{(+2.35)}$ &  & $\textcolor{devgreen}{(-20.00)}$ & $\textcolor{devgreen}{(-12.97)}$\\
2Q(AL2)& 1001   &  & 1159  & 1237  &   & 1349 & 1684 & 2303  & 2475  &  & 2686 & 3617\\
ERR ($\sigma=33.24$)
 & $\textcolor{devred}{(+62.73)}$ &  & $\textcolor{devred}{(+43.32)}$ & $\textcolor{devred}{(+44.36)}$ &  & $\textcolor{devred}{(+34.00)}$ & $\textcolor{devred}{(+12.00)}$ & $\textcolor{devred}{(+16.54)}$ & $\textcolor{devred}{(+22.35)}$ &  & $\textcolor{devgreen}{(-9.00)}$ & $\textcolor{devgreen}{(-2.97)}$\\
2Q(AL2)+3Q(SB) & 919&  & 1114  & 1192  &   & 1325 & 1671 & 2288  & 2460  &  & 2682 & 3614\\
ERR ($\sigma=9.05$)
 & $\textcolor{devgreen}{(-19.27)}$ &  & $\textcolor{devgreen}{(-1.68)}$ & $\textcolor{devgreen}{(-0.64)}$ &  & $\textcolor{devred}{(+10.00)}$ & $\textcolor{devgreen}{(-1.00)}$ & $\textcolor{devred}{(+1.54)}$ & $\textcolor{devred}{(+7.35)}$ &  & $\textcolor{devgreen}{(-13.00)}$ & $\textcolor{devgreen}{(-5.97)}$\\
2Q(AP1)& 997&  & 1156  & 1238  &   & 1347 & 1674 & 2308  & 2482  &  & 2678 & 3625\\
ERR ($\sigma=33.00$)
 & $\textcolor{devred}{(+58.73)}$ &  & $\textcolor{devred}{(+40.32)}$ & $\textcolor{devred}{(+45.36)}$ &  & $\textcolor{devred}{(+32.00)}$ & $\textcolor{devred}{(+2.00)}$ & $\textcolor{devred}{(+21.54)}$ & $\textcolor{devred}{(+29.35)}$ &  & $\textcolor{devgreen}{(-17.00)}$ & $\textcolor{devred}{(+5.03)}$\\
2Q(AP1)+3Q(SB) & 917&  & 1116  & 1198  &   & 1327 & 1664 & 2299  & 2473  &  & 2675 & 3623\\
ERR ($\sigma=13.62$)
 & $\textcolor{devgreen}{(-21.27)}$ &  & $\textcolor{devred}{(+0.32)}$ & $\textcolor{devred}{(+5.36)}$ &  & $\textcolor{devred}{(+12.00)}$ & $\textcolor{devgreen}{(-8.00)}$ & $\textcolor{devred}{(+12.54)}$ & $\textcolor{devred}{(+20.35)}$ &  & $\textcolor{devgreen}{(-20.00)}$ & $\textcolor{devred}{(+3.03)}$\\
2Q(AP2)& 993&  & 1156  & 1241  &   & 1351 & 1684 & 2310  & 2488  &  & 2681 & 3628\\
ERR ($\sigma=34.10$)
 & $\textcolor{devred}{(+54.73)}$ &  & $\textcolor{devred}{(+40.32)}$ & $\textcolor{devred}{(+48.36)}$ &  & $\textcolor{devred}{(+36.00)}$ & $\textcolor{devred}{(+12.00)}$ & $\textcolor{devred}{(+23.54)}$ & $\textcolor{devred}{(+35.35)}$ &  & $\textcolor{devgreen}{(-14.00)}$ & $\textcolor{devred}{(+8.03)}$\\
2Q(AP2)+3Q(SB) & 897&  & 1109  & 1194  &   & 1328 & 1673 & 2298  & 2476  &  & 2678 & 3626\\
ERR ($\sigma=18.02$)
 & $\textcolor{devgreen}{(-41.27)}$ &  & $\textcolor{devgreen}{(-6.68)}$ & $\textcolor{devred}{(+1.36)}$ &  & $\textcolor{devred}{(+13.00)}$ & $\textcolor{devred}{(+1.00)}$ & $\textcolor{devred}{(+11.54)}$ & $\textcolor{devred}{(+23.35)}$ &  & $\textcolor{devgreen}{(-17.00)}$ & $\textcolor{devred}{(+6.03)}$\\
2Q(Ref.~\cite{noh2025inevitable}) & 1005.30 & 1346.80 & 1134.10 & 1231.60 & 1455.20 &&& 2281.60 & 2480.90 & 2567.70 && 3606.30\\
ERR ($\sigma=55.83$)
 & $\textcolor{devred}{(+67.03)}$ & $\textcolor{devred}{(+114.80)}$ & $\textcolor{devred}{(+18.42)}$ & $\textcolor{devred}{(+38.96)}$ & $\textcolor{devred}{(+71.40)}$ &  &  & $\textcolor{devgreen}{(-4.86)}$ & $\textcolor{devred}{(+28.25)}$ & $\textcolor{devred}{(+49.22)}$ &  & $\textcolor{devgreen}{(-13.67)}$\\
2Q+3Q(Ref.~\cite{noh2025inevitable})  & 980.47  & 1272.10 & 1113.60 & 1196.50 & 1398.90 &&& 2266.70 & 2441.60 & 2522.90 && 3586.80\\
ERR ($\sigma=24.19$)
 & $\textcolor{devred}{(+42.20)}$ & $\textcolor{devred}{(+40.10)}$ & $\textcolor{devgreen}{(-2.08)}$ & $\textcolor{devred}{(+3.86)}$ & $\textcolor{devred}{(+15.10)}$ &  &  & $\textcolor{devgreen}{(-19.76)}$ & $\textcolor{devgreen}{(-11.05)}$ & $\textcolor{devred}{(+4.42)}$ &  & $\textcolor{devgreen}{(-33.17)}$\\

\end{tblr}}

\end{table}

\paragraph{Spin structure and the flavor-dependency puzzle:}

Several constituent-quark models reproduce large light-baryon spin gaps, most notably the $N$--$\Delta$ splitting, but retain systematic flavor-dependent residuals when the same interaction is applied across mesons and baryons.

The difficulty itself has a long history. Early studies explicitly explored whether dynamics constrained in the meson sector could be carried over to baryons through a common constituent-quark potential or through meson-to-baryon spectral relations and bounds~\cite{Bhaduri1981,Stanley1980,Richard1981MesonsBaryons,OnoSchoberl1982,BasdevantMartinRichard1990}. Later calculations made the sector tension more explicit. In the Semay--Silvestre-Brac program, interactions developed for mesons were subsequently tested in baryons, where direct transfer or simultaneous fitting did not give comparably satisfactory spectra without modifying the baryon Hamiltonian or refitting parameters~\cite{BrauSemaySilvestreBrac2002,SemaySilvestreBrac1997,BrauSemay1998,SemayBrauSilvestreBrac2001,SilvestreBracBrauSemay2003}; related potential-model analyses likewise noted difficulty in reproducing the meson and baryon sectors simultaneously~\cite{VargaGenoveseRichardSilvestreBrac1999}. Analogous cross-sector limitations appeared when baryon-motivated chiral interactions were extended to mesons and when a quarkonium-constrained confinement potential was transferred to heavy baryons~\cite{BlancoFernandezValcarce1999,Isgur2000Critique,JidoSakashita2016}. A related line of work emphasized the effective rather than universal character of fitted constituent quantities, while later calculations explicitly employed or extracted different constituent masses, parameter sets, or $qq$ and $q\bar q$ couplings in the two sectors~\cite{HuPing2022,CohenLipkin1980,BorkaJovanovicEtAl2010,KarlinerRosner2019Isospin,QinTanHuPing2021,HuPingPcs2022,HuEtAl2026XiC}. The difficulty common to these studies is the one tested here: transferring a two-body Hamiltonian fixed in the meson sector, unchanged, to baryons of different flavor content without sector-dependent prescriptions.

The representative quark models in Table~\ref{tab:previous_studies} use at least one of the following strategies:
{\textcircled{1}} separate meson and baryon parameter sets;
{\textcircled{2}} a combined fit that compromises between the two sectors;
{\textcircled{3}} an uncertainty-weighted objective dominated by very precisely measured states such as the nucleon;
{\textcircled{4}} calibration to a restricted flavor or state subset; or
{\textcircled{5}} direct transfer between sectors with the remaining spin- and flavor-dependent residuals accepted.
Among the surveyed two-body calculations, none gives the same direct meson-to-baryon transfer achieved here without one of these compensating procedures.
The compatibility scans of  Section~\ref{sec:udscb_benchmark} make this mismatch visible as a meson--baryon trade-off trajectory: refitting the pair interaction reduces the error in one sector mainly by moving the error to the other sector, while the finite-range three-quark interaction moves the trajectory into the region where both errors are small.

We use two complementary diagnostics for the residual pattern.
The spin-structure RMSE of Eq.~\eqref{eq:spin_structure_rmse} removes a common centroid from baryons with the same constituent content and tests the relative placement of their internal spin configurations.
For the two-body, spatially contact, and Yukawa-profile calculations, we obtain 5.17, 5.75, and 4.17~MeV, respectively.
The Yukawa-profile interaction therefore gives the smallest aggregate centroid-subtracted spin-structure error.
The spatially contact interaction improves the corresponding relative placement within the calibration set, but this in-sample improvement does not transfer to the omitted multiplets.
The remaining flavor trend is described by the Pearson correlation between the constituent rest-mass sum $M_B^{\rm rest}=\sum_{i\in B}m_i$ and the mass residual $\varepsilon_B=M_B^{\rm th}-M_B^{\exp}$.
For the two-body, spatially contact, and Yukawa-profile calculations in the nine-state calibration set, $r_{M_{\rm rest},\varepsilon} = -0.908$, \ensuremath{-}0.862, \ensuremath{-}0.136, respectively.
The decisive advantage of the Yukawa profile is its transfer outside the fit: the nine-state out-of-fit RMSE is 5.37~MeV, compared with 14.31~MeV for the two-body Hamiltonian and 32.89~MeV for the spatially contact interaction.
Thus, the spatially contact interaction mainly corrects the common calibration-set centroid, whereas the mass-scaled finite-range profile is required to obtain the better fixed-parameter transfer of both the absolute multiplet masses and their relative spin placements.

\paragraph{Exotics and dense matter:}

The effect of the three-quark interaction in a multiquark system is channel dependent and must be evaluated relative to the relevant hadronic threshold.
The contact and common-orbital estimates of  Section~\ref{sec:3bd_operators} illustrate this point.
For the dominant $H$-dibaryon and $N\Omega$ configurations, the absolute three-quark matrix elements are attractive, but the separated threshold baryons receive a larger attraction.
The resulting threshold-subtracted contribution is therefore repulsive.
The same contact analysis gives repulsive shifts of about 44--46~MeV for the indicated color-singlet $p\eta_c$ and $pJ/\psi$ components.
These results are operator diagnostics in a common-orbital limit; a physical multiquark mass requires the complete mixed wave function and the same Hamiltonian for both the compact state and its threshold.
For a molecular component, the short-range three-quark matrix element is suppressed by the separation of the hadronic clusters, whereas a compact component can receive a nonzero correction.
Near-threshold states therefore require a coupled compact--hadronic calculation.
For resonant fully heavy systems, the pole and its three-quark shift should be determined with a real- or complex-scaling calculation or an equivalent coupled-channel method; the pole width alone does not fix the size of the short-range correction.
Because the sign is channel dependent, the interaction can either reduce or increase near-threshold binding after the threshold shift is included consistently.

A simple contact estimate shows why a genuine three-body interaction can become more important as the quark density increases.
Let
\begin{align}
V_{2}^{\rm ct} =
G_2\sum_{i<j}
\mathcal O_{ij}^{(2)}
\delta^{(3)}(\mathbf r_i-\mathbf r_j), \qquad 
V_{3}^{\rm ct} =
G_3\sum_{i<j<k}
\mathcal O_{ijk}^{(3)}
\delta^{(3)}(\mathbf r_i-\mathbf r_j)
\delta^{(3)}(\mathbf r_i-\mathbf r_k),
\label{eq:dense_contact_scaling}
\end{align}
where $G_2$ and $G_3$ contain the corresponding coupling strengths and $\mathcal O^{(2,3)}$ denote dimensionless internal color--spin operators.
For $n$ quarks in the same Gaussian orbital $\phi(\mathbf r)=({\pi d^2})^{-3/4}e^{-r^2/2d^2}$, the two- and three-body contact matrix elements scale as $d^{-3}$ and $d^{-6}$, respectively.
The ratio therefore behaves schematically as
\begin{align}
\frac{\langle V_3^{\rm ct}\rangle}
     {\langle V_2^{\rm ct}\rangle} \sim
\frac{G_3}{G_2} \frac{\binom{n}{3}}{\binom{n}{2}} \frac{1}{d^3}
\frac{\overline{\mathcal O}_3}{\overline{\mathcal O}_2}
= \frac{G_3}{G_2} \frac{n-2}{3d^3} \frac{\overline{\mathcal O}_3}{\overline{\mathcal O}_2},
\label{eq:dense_three_to_two_ratio}
\end{align}
where $\overline{\mathcal O}_{2,3}$ are typical internal matrix elements in the state of interest.
The factor $G_3/G_2$ has the dimension of a volume, so Eq.~\eqref{eq:dense_three_to_two_ratio} is dimensionless.
In terms of the typical quark density $\rho_q\sim n/d^3$, the relative three-body contribution can thus grow approximately as $(G_3/G_2)\rho_q$, modulo the color--spin matrix elements.

This estimate is only a short-distance scaling argument.
The fitted interaction has finite Yukawa or Gaussian range, and its actual contribution depends on the three operator classes $L^{C-C}$, $L^{S-S}$, and $L^{C-S}$, their inverse-mass factors, and the color--spin composition of the many-quark state.
A repulsive matched contribution could stiffen an equation of state, whereas an attractive one could soften it.
The ground-state spectroscopy fit does not determine either outcome.
A quantitative statement about hyperonic matter, quark matter, or a phase transition therefore requires matching the quark-level operator to the relevant many-body degrees of freedom and solving the resulting matter problem.

\paragraph{Non-universal hadronization in high-energy collisions:}

The flavor-dependent pattern uncovered in our static quark-model spectra has an interesting qualitative link to high-energy hadron production.
It is now well established that baryon-to-meson ratios measured in hadronic and nuclear collisions are not universal when compared with the fragmentation baselines from $e^+e^-$ and $ep$ reactions.
Ratios such as $p/\pi$, $\Lambda/K$, and $\Lambda_c/D^0$ are significantly enhanced in high-multiplicity $pp$, $pA$, and $AA$ collisions relative to those baselines.

General-purpose Monte Carlo event generators tuned to describe baseline reactions tend to underpredict baryon production unless mechanisms such as color reconnection, which depend on the local density of colored partons, are included.
Several implementations of these effects exist, but they all introduce additional contributions associated with neighboring color sources~\cite{Altmann:2024icx}.
A commonly used approach for incorporating the surrounding partonic environment into hadronization is the coalescence model.
In this picture, the enhancement of the $p/\pi$ ratio can be understood from the fact that the hadron momentum is shared among its constituent partons.
Because the proton contains three quarks whereas the pion contains only two, the partons forming a proton carry smaller momenta for a given hadron momentum.
Proton production can therefore be enhanced by the larger thermal parton density at lower momenta~\cite{Greco:2003xt,Fries:2003vb}.

The ratio of charmed baryons to $D$ mesons provides a particularly direct probe of baryon formation involving a heavy quark, because charm quarks are produced early, their total number is approximately conserved during the subsequent evolution, and the ratio reflects the competition between heavy-meson and heavy-baryon formation.
The enhancement of $\Lambda_c/D$ beyond the fragmentation baseline is observed even in $pp$ collisions and can be described by coalescence calculations that include additional production in a dense partonic environment~\cite{Minissale:2020bif}.
On the other hand, existing implementations still underpredict the measured $\Xi_c^0/D^0$ ratio~\cite{Yun:2023kym,Altmann:2024kwx}.

These persistent discrepancies in charmed-baryon-to-$D$-meson ratios motivate testing whether an explicit connected three-quark interaction, which is not normally included in the pairwise Hamiltonians used to construct coalescence wave functions, provides an additional dynamical mechanism for charmed-baryon formation at finite local quark density.
Such an interaction could modify the compact baryon wave function and its Wigner distribution and thereby alter the coalescence probability.
A quantitative test requires calculating these wave functions with the same two-body-plus-three-body Hamiltonian and implementing them in a dynamical coalescence calculation.

\section{Summary and conclusions}

Section~\ref{sec:model_and_meson} reviewed the constituent-quark model as a QCD-motivated reduced description of low-energy hadrons. We first summarized how dressed constituent masses emerge from dynamical chiral symmetry breaking. We then discussed how the static and spin-dependent interactions are related to Wilson loops, one-gluon exchange (OGE), lattice QCD, and effective theories, and how chiral, instanton-induced, relativistic, and coupled-channel dynamics extend the conventional valence Hamiltonian. We also reviewed the GEM/ISG framework, model-space reduction, and resonance methods required for quantitative few-quark calculations. This review shows why constituent-quark Hamiltonians remain effective for compact low-lying hadrons. It also shows why their fitted masses, interactions, and operator decomposition depend on the adopted degrees of freedom and model space.

In  Section~\ref{sec:baryon_GEM}, we directly tested whether a pair Hamiltonian calibrated to mesons can describe baryons without sector-dependent readjustment. The selected ground-state mesons are reproduced with an RMSE of $\sigma_M=2.51~\mathrm{MeV}$. Applying the same two-body Hamiltonian to the nine principal ground-state baryons gives $\sigma_B=65.04~\mathrm{MeV}$ and a systematic residual pattern that decreases toward heavier flavor sectors. We then introduced the TGE-inspired operator classes $L^{C-C}$, $L^{S-S}$, and $L^{C-S}$ to represent a connected three-quark contribution after the pair Hamiltonian had been fixed. Their spatially contact form reduces the baryon RMSE to 18.87~MeV and removes most of the common baryon mass offset. However, the strong flavor-dependent residual pattern remains. These results show that the interaction has two distinct roles.
Within the calibration set, the color--spin operator basis corrects much of the common baryon centroid and improves the relative spin placement.
The fixed-parameter test shows, however, that this improvement does not transfer uniformly in the spatially contact form; the mass-scaled finite-range profile is needed for the better simultaneous transfer of the absolute multiplet centroids and their internal spin structure.

Section~\ref{sec:short_range} then examined whether the remaining discrepancy comes from the missing spatial localization of the connected interaction. 
The additional spatial dependence is characterized by an effective range taken to be inversely proportional to the reduced mass. This mass scaling is introduced phenomenologically to account for the more compact spatial distribution expected for systems containing heavier quarks.
The mass-scaled Yukawa profile reduces the baryon RMSE to 3.13~MeV, while the Gaussian profile with the same normalized pairwise range gives 6.28~MeV. In both cases, the spatial scales are fixed by the constituent pair masses already determined from the meson calibration, and no additional continuous range parameter is introduced. We also tested whether the result depends on the analytic form of the profile. The matched profile scan in Table~\ref{tab:alternative_spatial_profiles} includes algebraic, hyperbolic-secant, relativistic-bridge, stretched-exponential, Mat\'ern, Airy, and integrated-Gaussian (erfc) functions with substantially different core and asymptotic behavior. All tested mass-scaled profiles give $\sigma_B=3.60$--6.32~MeV. By contrast, when the same dimensionless functions are assigned a common flavor-independent physical range, the control calculations in Table~\ref{tab:fixed_range_spatial_profiles} give $\sigma_B=22.39$--31.74~MeV, and none improves upon the spatially contact result. Thus, finite-range localization alone is insufficient. The common feature of the successful profiles is the systematic shortening of the physical range with increasing constituent pair mass, rather than a special property of the Yukawa tail. The detailed profile shape determines the remaining few-MeV differences, but not the large improvement itself.

Section~\ref{subsec:three_quark_definition} defined the interaction operationally as the connected remainder after the one-body terms, the pair Hamiltonian, and its spectator embedding have been specified. In an enlarged $qqq\oplus qqqg$ or $qqq\oplus qqq^\ast$ space, part of the same short-distance dynamics may instead appear through explicit transitions and propagation. When this sector is reduced, its contribution is redistributed among constituent renormalizations, pair interactions, and connected many-body operators. Therefore, the fitted couplings $A$, $B$, and $C$ are effective, model-space-, representation-, and profile-dependent quantities rather than uniquely extracted microscopic QCD constants. The principal full-GEM solution favors the empirical sign pattern $(A,B,C)\sim(-,+,-)$. The nearby alternative branches found in the restricted scans are also consistent with this effective character. The independent rediagonalization test in \ref{subsec:gaussian_3q_perturbative_validation} confirms that the first-order treatment accurately reproduces the spectroscopic shifts in the retained ground-state space and changes the corresponding wave functions only weakly.

We next tested the interaction beyond the nine-state calibration.
With all parameters fixed, the two-body Hamiltonian gives an out-of-fit RMSE of 14.31~MeV, the spatially contact interaction gives 32.89~MeV, and the mass-scaled Yukawa interaction reduces the value to 5.37~MeV, as shown in Table~\ref{tab:primary_ground_state_holdout}.
Over the combined eighteen measured ground-state baryons, the corresponding RMSEs are 47.09, 26.81, and 4.39~MeV, respectively.

Section~\ref{sec:udscb_benchmark} extended the test to $u,d,s,c,b$ meson--baryon compatibility over the full range of fitting weights. Neither the tested reduced-mass running of the Coulomb coupling nor the minimal replacement of the kinetic energy by relativistic kinematics removes the pairwise meson--baryon trade-off if the three-body interactions are absent. The finite-range three-quark interaction instead moves the compatibility trajectory into the region where both sectors have small errors. The same qualitative improvement is found for the AL1, AL2, AP1, AP2, BD, SL, GI/CI, and $\mathrm{S}^{2}\mathrm{AJ}$ Hamiltonians, under both common- and separate-offset prescriptions. Thus, within the tested set, the result is not specific to the fitting scheme, the relativization prescription, the form of the confining potential, the regulator, or the selected model.

The Hamiltonians fitted to ground states give uneven results for radial and orbital excitations (Section~\ref{sec:excitation_spectra}). The diagnostic TH2 extension shows that a softer confining law together with spin--orbit, antisymmetric spin--orbit, and tensor interactions improves several conventional multiplets but does not remove all discrepancies. The constituent-quark description remains quantitatively most reliable for compact ground and low-lying conventional states. Highly excited, threshold-adjacent, strongly chiral, multiquark-dominated, or gluon-rich states require additional relativistic, long-distance, and coupled-channel dynamics.  Section~\ref{sec:3bd_operators} provides technical details for calculating the color-spin matrices of various multiquark states.
To calculate its physical contribution in such systems, one must evaluate their actual spatial wave functions and the corresponding hadronic thresholds. We discussed the implications for spectroscopy, multiquark systems, dense matter, and hadronization in  Section~\ref{sec:discuss}. 

Three-body interactions with a mass-dependent range provide a unified description of the low-lying meson and baryon masses. Within a fixed meson-calibrated pair representation, such a unified framework is required for a consistent quark-model description across hadron sectors, and it provides a natural basis for extending the same interaction to multiquark configurations.

\section*{Acknowledgments}
This work was supported by the National Research Foundation of Korea (NRF) under Grant No. RS-2023-NR077232. A.P. acknowledges support from the NRF under Grant No. RS-2025-23963552, and S.N. acknowledges support from the NRF under Grant No. RS-2026-25499570. K.C.H. acknowledges support from the 2025 Faculty Enhancement Program (FEP) grant of Prairie View A\&M University.
J.B. is grateful to Professor Makoto Oka of the RIKEN Nishina Center for valuable advice and for providing him with an opportunity to gain hands-on experience with instanton-induced interactions.
He also thanks his wife, Sangkyung Kim, for her unwavering support and encouragement.
The authors thank the anonymous reviewers for their careful reading and constructive comments, which substantially improved the manuscript, and the editors for handling the review process.

\appendix
\clearpage
\phantomsection
\section*{Notation and conventions}
\addcontentsline{toc}{section}{Notation and conventions}
This guide lists symbols used throughout the manuscript or in multiple sections.
Strictly local dummy indices and quantities defined and used within a single displayed derivation are omitted.
When a standard symbol is reused locally, its subsection, font, indices, or argument distinguish the meaning.
Boldface denotes vector-valued quantities or coefficient arrays, whereas auxiliary finite-dimensional matrices are generally set in sans serif type.

\begingroup
\footnotesize
\setstretch{1.0}
\setlength{\columnsep}{0.85cm}
\raggedcolumns
\begin{multicols}{2}

\notationgroup{General indices and conventions}
\begin{notationlist}
\item[$i,j,k,l,\ldots$] Constituent-particle labels unless another local role is stated; $i<j<k$ denotes an unordered constituent triplet.
\item[$a,b,c$] Jacobi rearrangement labels when used as $\mathcal C\in\{a,b,c\}$; Schwinger-channel labels in the Yukawa-profile appendix; adjoint SU$(3)_C$ indices when attached to $f^{abc}$, $d^{abc}$, or color generators.  In the GI/CI running-coupling parametrization, $a=1,2,3$ instead labels the three Gaussian components.
\item[$q$] A light constituent quark, $q\in\{u,d\}$, unless boldfaced as a momentum or used as a local summation variable.
\item[$Q$] A heavy constituent quark, normally $Q\in\{c,b\}$.  In the nonlinear chiral Lagrangian, $Q$ locally denotes the constituent-quark field rather than a heavy flavor; $\mathsf Q$ denotes a model-space projector.
\item[$\hbar,c$] Reduced Planck constant and speed of light; explicit factors are retained unless natural units are stated.
\item[$I$] Isospin quantum number or an identity operator when its operator role is explicit.  Projectors and block identities are written in calligraphic or sans serif form when needed.
\item[$\Re,\Im$] Real and imaginary parts.
\item[$\langle\cdots\rangle$] Matrix element or expectation value, with the relevant state or measure specified by the bra--ket or subscript.
\item[$\|\cdot\|_{\rm F}$] Frobenius norm of a finite-dimensional matrix.
\end{notationlist}

\notationgroup{QCD, chiral, and dressed-quark quantities}
\begin{notationlist}
\item[$D_{\mu\nu}(p)$] Gluon two-point function; distinct from the Cornell additive constant $D$ and the ISG shift vectors $\mathbf D_{lm,h}$.
\item[$d(p^2)$] Gluon dressing function in the Landau-gauge propagator.
\item[$S(p)$] Dressed-quark propagator; distinct from the total constituent spin $S$.
\item[$M(p^2)$] Momentum-dependent dressed-quark mass function; distinct from the chiral constituent scale $M$, hadron masses $M_H$, bare masses $M_i^{(0)}$, and TGE excitation-denominator scales $M_i$.
\item[$M$] Flavor-symmetric effective constituent-mass scale in the nonlinear chiral-quark Lagrangian, associated with the low-momentum limit of $M(p^2)$; in angular-momentum expressions, $M$ instead denotes a magnetic projection.
\item[$Z(p^2)$] Quark wave-function renormalization function.
\item[$A(p^2),B(p^2)$] Dirac-vector and scalar functions in $S^{-1}(p)=i\slashed p A+B$; distinct from the three-quark couplings $A$ and $B$.
\item[$\hat m$] Current-quark mass matrix in the chiral QCD Lagrangian.
\item[$U,U^{\gamma_5}$] Nonlinear Goldstone-field matrices in the chiral constituent-quark formulation.
\item[$\alpha_s$] Strong coupling; superscripts such as $(C)$, $(CS)$, and $(CS,\infty)$ identify effective central, hyperfine, and formal heavy-contact normalizations.
\item[$m_g$] Effective infrared gluon screening scale; where explicitly stated in the lattice-potential discussion, the same symbol follows the convention of the cited work and has a different local meaning.
\item[$f_\pi$] Pion decay constant.
\item[$\langle\bar q q\rangle$] Chiral quark condensate.
\end{notationlist}

\notationgroup{Coordinates, masses, and kinematics}
\begin{notationlist}
\item[$\mathbf r_i,\mathbf p_i$] Coordinate and canonical momentum of constituent $i$.
\item[$\mathbf r_{ij}$] Physical relative coordinate $\mathbf r_i-\mathbf r_j$; $r_{ij}=|\mathbf r_{ij}|$ and $\hat{\mathbf r}_{ij}=\mathbf r_{ij}/r_{ij}$.
\item[$\mathbf r_{\mathcal C}$] First scaled Jacobi coordinate in rearrangement channel $\mathcal C$.
\item[$\mathbf R_{\mathcal C}$] Second scaled Jacobi coordinate in rearrangement channel $\mathcal C$.
\item[$\mathbf X_{\mathcal C}$] Two-component array $(\mathbf r_{\mathcal C},\mathbf R_{\mathcal C})^T$ of three-dimensional Jacobi vectors.
\item[$U_{\mathcal C x}$] $2\times2$ Jacobi-coordinate transformation satisfying $\mathbf X_{\mathcal C}=U_{\mathcal C x}\mathbf X_x$ and $U_{x\mathcal C}=U_{\mathcal C x}^{-1}$.
\item[$\mathbf p_{ij}$] Physical pair-relative momentum used in the GI/CI-type benchmark.
\item[$\mathbf q_k$] Pair--spectator Jacobi momentum for the partition $(ij)k$; in the representation-dependence subsection, $q_k=|\mathbf q_k|$ enters the embedded pair energy.
\item[$\mathbf Q_{ij}$] Pair momentum transfer $\mathbf p'_{ij}-\mathbf p_{ij}$; distinct from the spectator momentum $\mathbf q_k$, the heavy-quark symbol $Q$, and the projector $\mathsf Q$.
\item[$\boldsymbol\pi_{\mathcal C},\boldsymbol\Pi_{\mathcal C}$] Momenta conjugate to the scaled Jacobi coordinates $\mathbf r_{\mathcal C}$ and $\mathbf R_{\mathcal C}$ in the GI/CI construction.
\item[$\mathbf P_M,\mathbf P_B$] Total meson and baryon momenta; the GI/CI Hamiltonians are evaluated in the corresponding center-of-momentum frames.
\item[$m_i$] Flavor-dependent constituent mass of particle $i$.
\item[$m_{ij}$] Reduced mass $2m_im_j/(m_i+m_j)$ used throughout the scaled-Jacobi convention.
\item[$m_{k,(ij)}$] Scaled reduced mass between spectator $k$ and the center of mass of pair $(ij)$ in the embedded three-body kinematics.
\item[$M_H,M_B$] Hadron and baryon masses, respectively.
\item[$M_i$] Effective excitation-denominator scale associated with the intermediate $q^\ast$ configuration in the TGE-inspired construction.  The prescription $M_i=k m_i$ is used, with the common factor $k$ absorbed into $A$, $B$, and $C$; $M_i$ is not the physical mass of an isolated $(q_i g)_{\mathbf 3_C}$ constituent.  The symbol $M_i^{(0)}$ instead denotes a bare valence-state mass in the coupled-channel discussion.
\item[$E_{ij},E_{\rm int}$] Embedded pair energy and total internal energy, excluding constituent rest masses, in the representation-dependence discussion.
\item[$K,K_{\rm int}$] Kinetic-energy operator and its internal, center-of-mass-subtracted form.
\item[$K_{\rm CM}$] Center-of-mass kinetic energy.
\item[$M^{-1}$] Inverse-mass metric in the GEM kinetic generating function of \ref{app:GEM_ref}.
\item[$\Lambda$] $2\times2$ kinetic coefficient matrix in the direct Cartesian calculation, defined from $Q_a\operatorname{diag}(m_i^{-1})Q_a^T$; distinct from $M^{-1}$ and from baryon symbols such as $\Lambda$.
\item[$Q_a$] $2\times3$ matrix mapping particle coordinates to the reference $a$-channel Jacobi coordinates in \ref{app:analytic_without_gem}.
\end{notationlist}

\notationgroup{Angular momentum and internal states}
\begin{notationlist}
\item[$l_{\mathcal C},L_{\mathcal C}$] Orbital angular momenta associated with $\mathbf r_{\mathcal C}$ and $\mathbf R_{\mathcal C}$.
\item[$J$] Total orbital angular momentum obtained by coupling $l_{\mathcal C}$ and $L_{\mathcal C}$.
\item[$S$] Total constituent spin.
\item[$\mathcal J$] Physical total angular momentum obtained by coupling $J$ and $S$.
\item[$M$] Magnetic projection of $\mathcal J$ or, where explicitly stated, of an orbital angular momentum; distinct from the chiral constituent scale $M$.
\item[$P$] Parity. For three-quark Jacobi-basis states, $P=(-1)^{l_{\mathcal C}+L_{\mathcal C}}$, whereas for a $q\bar q$ meson $P=(-1)^{l+1}$. The symbol is distinguished from the model-space projector $\mathsf P$.
\item[$n^{2S+1}L_{\mathcal J}$] Spectroscopic notation for radial, spin, orbital, and total angular-momentum quantum numbers; a label attached to a mixed TH2 eigenstate denotes its dominant component.
\item[$\mathcal C^{j_1j_2;J}_{m_1m_2;M}$] SU(2) Clebsch--Gordan coefficient; this indexed symbol is distinct from the Jacobi-channel label $\mathcal C$.
\item[$Y_{lm}$] Spherical harmonic with the phase convention specified in the ISG subsection.
\item[$\mathcal Y_{lm}(\mathbf z)$] Solid harmonic $z^lY_{lm}(\hat{\mathbf z})$.
\item[$P_{ij},P_{123},P_{132}$] Transposition and cyclic-permutation operators acting on particle labels and all attached degrees of freedom.
\item[{$[3],[21],[111]$}] Symmetric, mixed-symmetry, and antisymmetric irreducible representations of $S_3$.
\item[Subscripts $\lambda,\rho$] Young--Yamanouchi labels for the two components of $[21]$; $\lambda$ is symmetric and $\rho$ is antisymmetric under $P_{12}$.
\item[$\chi_S$] Total-spin wave function.
\item[$\chi_{[3]}$] Fully symmetric three-quark spin wave function with $S=3/2$.
\item[$\chi_\rho,\chi_\lambda,\chi_{\rho,Q},\chi_{\lambda,Q}$] $S=1/2$ pair-spin bases with the indicated pair coupled to spin 0 or 1, respectively.  The $Q$-indexed notation is used for the coupled $qsQ$ configurations with $Q=c,b$.
\item[$s_{12},s_{qs}$] Spin of the indicated constituent pair; $s_{qs}$ is used in the coupled-basis analysis of the $\Xi_Q$ and $\Xi_Q'$ states.
\item[$F_\lambda,F_\rho$] Flavor Young--Yamanouchi states in the $[21]$ representation.
\item[$\mathcal X_\lambda^N,\mathcal X_\rho^N$] Mixed-symmetry spin--flavor Young--Yamanouchi doublet of the nucleon-type $S=1/2$ basis defined in \ref{bases}.
\item[$\Phi_{[3]},\Phi_{[111]}$] Fully symmetric and fully antisymmetric orbital Gaussian combinations in \ref{bases}.
\item[$\Phi_\lambda,\Phi_\rho$] Mixed-symmetry orbital Gaussian combinations transforming as the $\lambda$ and $\rho$ components of $[21]$.
\item[$\mathscr O_\lambda^J,\mathscr O_\rho^J$] Generic pair-exchange-symmetric and pair-exchange-antisymmetric orbital representatives, respectively, selected from the reference-$a$ and combined-$b+c$ constructions in \ref{bases}; magnetic and radial labels are suppressed.
\item[$F,C$] Flavor and color wave functions.  The same letters are locally reused for the pair-creation form factor $F(\mathbf q)$, GEM generating functions, and the three-body coupling $C$.
\item[$\mathcal P_{\mathbf 1}$] Projector onto the color-singlet subspace.
\item[$\mathcal P_B$] Baryon permutation projector assembled from the relevant $S_3$ or pair-exchange projectors for the specified flavor content.
\item[$\mathcal P_{12}^{(\pm)}$] Pair-exchange projectors $(I\pm P_{12})/2$.
\item[$\mathcal P_{[3]},\mathcal P_{[21]},\mathcal P_{[111]}$] Projectors onto the three irreducible representations of $S_3$.
\end{notationlist}

\notationgroup{Color, flavor, and spin operators}
\begin{notationlist}
\item[$F_i^c,F_g^c$] Fundamental color generator $\lambda_i^c/2$ acting on constituent $i$ and adjoint color generator acting on the gluonic excitation, respectively.
\item[$\lambda_i^a$] Gell-Mann matrix acting in color space when $a$ is a color-adjoint index.
\item[$\lambda_i^F$] Flavor generator or exchanged-meson flavor operator in the GBE interaction; the unindexed $\lambda^a$ in $U$ and $U^{\gamma_5}$ likewise acts in flavor space.
\item[$\bar F_i^a,\bar\lambda_i^a$] Antiquark color generators, $\bar F^a=-F^{aT}$ and $\bar\lambda^a=-\lambda^{aT}$.
\item[$f^{abc},d^{abc}$] Antisymmetric and symmetric SU$(3)_C$ invariant tensors.
\item[$C_2(R),C_3(R)$] Quadratic and cubic Casimir invariants of representation $R$.  For an SU(3) representation with Dynkin indices $(p,q)$, their eigenvalues are denoted by $C_2(p,q)$ and $C_3(p,q)$.  Superscripts such as $(C)$, $(F)$, and $(CS)$ identify color, flavor, and color--spin groups when required; state labels such as $\ket{C_3}$ instead denote color-basis vectors.
\item[$\boldsymbol\sigma_i$] Pauli-spin vector of constituent $i$.
\item[$\mathbf S_i$] Spin operator $\boldsymbol\sigma_i/2$.
\item[$S_{ij}$] Two-body tensor operator $3(\hat{\mathbf r}_{ij}\!\cdot\!\boldsymbol\sigma_i)(\hat{\mathbf r}_{ij}\!\cdot\!\boldsymbol\sigma_j)-\boldsymbol\sigma_i\!\cdot\!\boldsymbol\sigma_j$.
\item[$\mathbf L_{ij}$] Pair orbital angular momentum $\mathbf r_{ij}\times(-i\nabla_{\mathbf r_{ij}})$.
\end{notationlist}

\notationgroup{Two-body Hamiltonian and potential parameters}
\begin{notationlist}
\item[$H$] Hamiltonian in the model space under discussion.
\item[$H_{\mathrm{1B}}$] Sum of one-body rest-mass and kinetic operators in the cluster decomposition.
\item[$V_{ij}^{C}$] Spin-independent color-electric central interaction.
\item[$V_{ij}^{CS}$] Color-spin hyperfine interaction.
\item[$V^{SS}$] Spin--spin interaction after the color factor has been separated according to $V^{CS}=(-3/4)F_i^cF_j^c\,V^{SS}$; it is distinct from the three-quark operator class $L^{S-S}$.
\item[$V_{ij}^{T}$] Tensor interaction.
\item[$V_{ij}^{LS},V_{ij}^{ALS}$] Symmetric and antisymmetric spin--orbit interactions.
\item[$V_V,V_S$] Vector and scalar radial components used in the spin-dependent reduction; tildes denote their regulated or smeared forms in TH2.
\item[$\kappa$] Cornell Coulomb coefficient; indexed forms such as $\kappa_{ij}$ denote a flavor-dependent or running prescription.
\item[$\sigma$] String tension or power-law confinement coefficient.  In tables and equations that explicitly define an error measure, an unsubscripted $\sigma$ may be locally reused for an RMSE.
\item[$D$] Additive constant in the central potential.
\item[$\mu_{ij},\mu'_{ij}$] Mass-dependent range and normalization scales of the color-spin regulator.
\item[$\alpha,\beta,\gamma,\delta$] Global fit parameters entering $\mu_{ij}$ and $\mu'_{ij}$.  Section~\ref{sec:3bd_operators} locally reuses $\delta$ for $1-m_u/m_s$ in the dibaryon expansion and for $1-m_u/m_c$ in the hidden-charm pentaquark expansion.  The derivations in \ref{app:GEM_ref} and \ref{app:analytic_without_gem} locally reuse $\alpha,\beta$ as bra/ket basis labels, while $\gamma$ also denotes the ${}^3P_0$ pair-creation strength and $\gamma_a$ denotes a GI/CI running-coupling scale.
\item[$Z_{\rm hf}$] Common hyperfine normalization ratio $\mu'_{ij}/\mu_{ij}$ used in the $\mathrm{S^2AJ}$ meson--baryon compatibility benchmark.
\item[$r_{qq}^{CS}$] Light-pair color-spin range $\hbar c/\mu_{qq}$, used in both the principal calibration and the compatibility benchmark.
\item[$D_M,D_B$] Sector-wide meson and baryon additive constants in the cross-model compatibility benchmark.  The common-offset prescription imposes $D_M=D_B\equiv D$, whereas the separate-offset prescription profiles $D_M$ and $D_B$ independently.
\item[$K_{\rm NR}^{(N)},K_{\rm rel}^{(N)}$] Nonrelativistic and semirelativistic $N$-constituent kinetic operators.
\end{notationlist}

\notationgroup{Chiral and instanton-induced interactions}
\begin{notationlist}
\item[$H_{\mathrm{S^2AJ}+\chi}$] Hybrid chiral benchmark Hamiltonian obtained by adding the regulated PS-GBE spin--spin interaction to the OGE-based $\mathrm{S^2AJ}$ Hamiltonian.
\item[$LL$] Set of constituent pairs composed only of $u,d,s$ quarks or antiquarks; used in the PS-GBE and instanton-induced interaction Hamiltonians and distinct from the orbital angular momentum $L_{\mathcal C}$.
\item[$\phi$] Exchanged pseudoscalar Goldstone boson, $\phi\in\{\pi,K,\eta\}$, in the chiral benchmark.
\item[$V_{ij}^{(\phi)},V_{ij}^{(\phi),SS}$] Regulated pseudoscalar Goldstone-boson-exchange interaction and its spin--spin component for exchanged boson $\phi$. Only $V_{ij}^{(\phi),SS}$ is retained in the common $S$-wave compatibility benchmark.
\item[$g_{\rm ch}$] Reference chiral quark--Goldstone-boson coupling entering the monopole-regulated PS-GBE interaction.
\item[$m_\phi,\Lambda_\phi$] Mass and monopole cutoff associated with exchanged pseudoscalar boson $\phi$.
\item[$\theta_P$] Pseudoscalar $\eta_8$--$\eta_0$ mixing angle.
\item[$s_\chi$] Dimensionless common strength multiplying the reference PS-GBE interaction in $H_{\mathrm{S^2AJ}+\chi}$; it is varied over $0.02\leq s_\chi\leq1.5$ in the compatibility benchmark.
\item[$G_D$] Coupling of the $N_f=3$ KMT determinant interaction.
\item[$V_{\rm III}^{(2)},V_{\rm III}^{(3)}$] Effective two-body and genuine three-body instanton-induced interactions after and before condensate contraction, respectively.
\item[$V_0,V_0^{(2)}(i,j)$] Three-body KMT coupling and the induced two-body coupling in the Takeuchi--Oka convention.
\item[$K_{\rm III}$] Parameter relating the current mass, constituent mass, and chiral condensate in the III coupling relation.
\item[$p_{\rm III}$] Phenomenological fraction of the light-pair hyperfine interaction assigned to the instanton-induced component.
\end{notationlist}

\notationgroup{Coupled channels and effective Hamiltonians}
\begin{notationlist}
\item[$H_0$] Bare or valence Hamiltonian before continuum dressing.
\item[$T$] Transition operator between valence and continuum sectors; $T_{{}^3P_0}$ denotes the pair-creation vertex.
\item[$\gamma$] Effective strength of the ${}^3P_0$ pair-creation operator; distinct from the color-spin regulator parameter $\gamma$ by its local context.
\item[$F(\mathbf q)$] Momentum-space form factor of the pair-creation vertex; distinct from the flavor wave function $F$ and the GEM generating function $F(\boldsymbol\epsilon;H)$.
\item[$M_A^{(0)},M_A$] Bare and physical masses of state $A$.
\item[$\mathcal M_{A\to BC}^{l,\mathcal J}(k)$] Partial-wave transition amplitude from valence state $A$ to channel $BC$.
\item[$\Sigma_A(E)$] Energy-dependent hadronic self-energy of a single retained valence state.
\item[$Z_A$] Valence wave-function renormalization factor for a stable bound state under the conditions stated in the unquenched formulation.
\item[$X_{A,BC},X_{A,\alpha}$] Bound-state continuum components associated with channel $BC$ or asymptotic channel $\alpha$ in the specified model-space representation.
\item[$\mathsf P,\mathsf Q$] Complementary retained-valence and explicit-continuum projectors in an orthogonal model-space representation; in an RGM coefficient equation the same labels identify basis blocks before norm-kernel orthogonalization.
\item[$H_{\mathsf P\mathsf P},H_{\mathsf Q\mathsf Q}$] Hamiltonian blocks in the retained-valence and explicit-continuum sectors.
\item[$T_{\mathsf P\mathsf Q},T_{\mathsf Q\mathsf P}$] Left and right transition blocks coupling the retained and continuum sectors.
\item[$R_{\mathsf Q}(E)$] Full continuum-space resolvent $(E-H_{\mathsf Q\mathsf Q}+i0)^{-1}$ on the branch specified in the surrounding discussion.
\item[$\Sigma_{\mathsf Q}(E)$] Continuum self-energy $T_{\mathsf P\mathsf Q}R_{\mathsf Q}(E)T_{\mathsf Q\mathsf P}$ in the retained sector.
\item[$H_{\mathsf Q\mathsf Q}^{(0)},V_{\mathsf Q\mathsf Q}$] Free continuum Hamiltonian and the direct elastic or channel-changing interaction in $H_{\mathsf Q\mathsf Q}=H_{\mathsf Q\mathsf Q}^{(0)}+V_{\mathsf Q\mathsf Q}$.
\item[$R_{\mathsf Q}^{(0)}(E)$] Free continuum resolvent $(E-H_{\mathsf Q\mathsf Q}^{(0)}+i0)^{-1}$.
\item[$R_{\mathsf Q}^{\rm ref}(E)$] Full interacting continuum resolvent evaluated on the chosen reference Riemann sheet.
\item[$\mathsf P_i,\mathsf Q_\alpha$] Projectors onto retained bare state $i$ and asymptotic continuum channel $\alpha$, with $\sum_i\mathsf P_i=\mathsf P$ and $\sum_\alpha\mathsf Q_\alpha=\mathsf Q$.
\item[$\mathcal O_{\mathsf P\mathsf Q},\mathcal O_{\mathsf Q\mathsf P},\mathcal O_{\mathsf Q\mathsf Q}$] RGM overlap or norm-kernel blocks in a nonorthogonal compact--continuum basis.
\item[$\mathcal W_{\mathsf P\mathsf Q},\mathcal W_{\mathsf Q\mathsf P}$] Generalized transition kernels $T-E\mathcal O$ entering a nonorthogonal coupled-channel equation.
\item[$\Sigma_{\mathsf Q}^{\rm RGM}(E)$] Schur-complement self-energy $\mathcal W_{\mathsf P\mathsf Q}(E)[E\mathcal O_{\mathsf Q\mathsf Q}-H_{\mathsf Q\mathsf Q}]^{-1}\mathcal W_{\mathsf Q\mathsf P}(E)$.
\item[$V_{\mathsf Q\mathsf Q}^{\rm RGM}(E)$] Energy-dependent RGM interaction $H_{\mathsf Q\mathsf Q}-H_{\mathsf Q\mathsf Q}^{(0)}-E(\mathcal O_{\mathsf Q\mathsf Q}-\mathbf 1_{\mathsf Q})$.
\item[$\Gamma_{\mathsf Q\mathsf P}(E),\Gamma_{\mathsf P\mathsf Q}(E)$] Right and left compact--continuum vertices dressed by continuum rescattering; superscript ``RGM'' identifies the corresponding nonorthogonal construction.
\item[$\boldsymbol\eta,\eta_\alpha$] Multichannel Riemann-sheet label and the sign specifying the momentum branch of channel $\alpha$.
\item[$k_\alpha(E)$] Analytically continued on-shell relative momentum in channel $\alpha$.
\item[$E_\alpha^{\rm th},\mu_\alpha$] Threshold energy and two-hadron reduced mass of channel $\alpha$ when a nonrelativistic channel dispersion is used.
\item[$\rho_\alpha(E)$] Two-hadron phase-space factor analytically continued to $k_\alpha(E)$; distinct from the GI smearing density $\rho_{ij}^{\rm GI}$ and the Yukawa Schwinger coefficients $\rho_x$.
\item[$\Sigma_{\mathsf Q}^{\rm ref}(E),\Sigma_{\mathsf Q}^{(\boldsymbol\eta)}(E)$] Continuum self-energy on the reference sheet and after analytic continuation to sheet $\boldsymbol\eta$.
\item[$J^\alpha(E)$] On-shell discontinuity, or Riemann-sheet-jump, contribution generated by crossing the cut of channel $\alpha$.
\item[$E_{\rm pole},M_R,\Gamma_R$] Complex resonance-pole energy and its parametrization $E_{\rm pole}=M_R-i\Gamma_R/2$.
\item[$\mathsf D^{(\boldsymbol\eta)}(E)$] Reduced inverse propagator in the retained sector on sheet $\boldsymbol\eta$.
\item[$\Psi_R,\Psi_L$] Independent right and left pole vectors of the analytically continued inverse propagator; for a Hermitian bound-state problem one may choose $\Psi_L=\Psi_R^*$.
\item[$\mathcal N$] Pole-residue normalization $\Psi_L^T\mathsf D^{(\boldsymbol\eta)\prime}(E_{\rm pole})\Psi_R$.
\item[$Z_{P,i}$] Contribution of retained bare state $i$ to the normalized pole residue.
\item[$X_\alpha$] Explicit continuum-propagation contribution associated with asymptotic channel $\alpha$ at a pole.
\item[$C_\alpha^{\rm jump}$] Contribution to the pole normalization from the derivative of the sheet-jump term of channel $\alpha$.
\item[$Y_{\rm metric}$] Explicit pole-normalization contribution generated by energy-dependent compact--continuum overlap kernels in a nonorthogonal representation.
\end{notationlist}

\notationgroup{Resonance extraction in an $L^2$ basis}
\begin{notationlist}
\item[$\hat G$] Hermitian dilation generator used in the complex-scaling transformation.
\item[$U(\theta),H(\theta)$] Non-unitary complex-scaling transformation and the transformed Hamiltonian.
\item[$\theta$] Complex-rotation angle in CSM; symbols such as $\theta^\mu$ in the Yukawa appendix instead denote coefficient arrays.
\item[$\alpha_{\rm sc}$] Real-scaling parameter multiplying asymptotic Gaussian ranges in the stabilization method.
\item[$L(\alpha_{\rm sc})$] Effective infrared length of the real-scaled finite basis.
\end{notationlist}

\notationgroup{Three-quark operators and representation dependence}
\begin{notationlist}
\item[$H_3$] Three-particle Hamiltonian in the fixed constituent-quark representation.
\item[$v_{ij}\otimes I_k$] Spectator-independent embedding of pair interaction $v_{ij}$ into the three-particle space.
\item[$V_{123}^{\mathrm{conn}}$] Connected three-particle remainder after the fixed embedded pair interactions have been subtracted in the adopted representation.
\item[$q^\ast$] Schematic label for a generic intermediate configuration outside the retained valence $qqq$ space.
\item[$\mathcal H_{qqq^\ast},\mathcal H_{qqqg}$] Generic nonvalence intermediate space and its illustrative color-singlet $qqqg$ subspace, with $\mathcal H_{qqq^\ast}\supset\mathcal H_{qqqg}$. The quark-like $(qg)_{\mathbf 3_C}$ component is one possible realization rather than a unique microscopic definition of $q^\ast$.
\item[$\mathsf Q_g$] Projector onto the resolved color-singlet $qqqg$ sector in the illustrative enlarged-space realization.
\item[$\Delta H_{\mathsf P}^{(qqqg)}(E)$] Effective valence-space contribution obtained by eliminating the resolved $qqqg$ sector, equal to $\Sigma_{\mathsf Q_g}(E)$ in the adopted Feshbach reduction.
\item[$\Delta h_i(E),\Delta v_{ij}(E)$] Induced one-body and spectator-embedded pair contributions in the cluster decomposition of $\Delta H_{\mathsf P}^{(qqqg)}(E)$.
\item[$T_{\mathsf P\mathsf Q_g}^{(ij)},T_{\mathsf Q_g\mathsf P}^{(ij)},H_{\mathsf Q_g\mathsf Q_g}$] Pair-labeled transition blocks between the retained $qqq$ and resolved $qqqg$ sectors, and the Hamiltonian in the resolved-gluon sector.
\item[$t_{ij}(E)$] Two-body transition operator used to distinguish on-shell and off-shell matrix elements in the representation-dependence discussion.
\item[$\mathcal U$] Scattering-equivalent unitary transformation; all Hamiltonians and external operators must be transformed consistently.
\item[$\widehat J_X^\mu$] External current operator of type $X$, transforming as $\mathcal U^\dagger\widehat J_X^\mu\mathcal U$.
\item[$L_{ij;k}^{C-C},L_{ij;k}^{S-S},L_{ij;k}^{C-S}$] Line-resolved central--central, color-spin--color-spin, and mixed central--color-spin contributions in which line $k$ carries the intermediate $q^\ast$ configuration.
\item[$L_{ijk}^{C-C},L_{ijk}^{S-S},L_{ijk}^{C-S}$] Fully assembled triplet operators after the cyclic placements and required insertion orderings are included.  Section~\ref{sec:3bd_operators} often writes $L^X=\sum_{i<j<k}L_{ijk}^X$ for $X\in\{C\!-\!C,S\!-\!S,C\!-\!S\}$.
\item[$L^{3Q}$] Spatially contact, spatial-profile-independent three-quark interaction $A L^{C-C}+B L^{S-S}+C L^{C-S}$; the common zero-range orbital matrix element is absorbed into $A$, $B$, and $C$.
\item[$L_{\rm Y}^{3Q},L_{\rm G}^{3Q}$] Yukawa- and Gaussian-profile finite-range connected three-quark interactions in the adopted local valence representation.
\item[$f_{\rm Y},f_{\rm G}$] Corresponding dimensionless spatial profiles.
\item[$x_{ij}$] Mass-scaled separation $m_{ij}cr_{ij}/\hbar$ in the three-quark profiles.
\item[$A,B,C$] Profile-, baseline-, and representation-dependent fitted couplings multiplying $L^{C-C}$, $L^{S-S}$, and $L^{C-S}$.  Their energy dimensions are $[A]=E^2$, $[B]=E^6$, and $[C]=E^4$; the corresponding dimension-matched scales are $|A|^{1/2}$, $|B|^{1/6}$, and $|C|^{1/4}$.
\item[$g_{C,n},g_{CS,n}$] Real reduced central-color and color-spin transition strengths of excluded configuration $n$ in the scalar-factorized interpretation of $A$, $B$, and $C$; the magnitude of the common scalar propagation weight is included in these quantities.
\item[$M_B^{3Q},M_{B,X}^{3Q,(f)}$] First-order three-quark mass contribution to baryon $B$; $X$ and $f$ specify the two-body baseline and the spatially contact, Gaussian, or Yukawa profile in the cross-model benchmark.
\end{notationlist}

\notationgroup{Illustrative $j_{qg}=3/2$ $qqqg$ realization}
\begin{notationlist}
\item[$j_g^P,j_{qg}^P,\ell_{qg}$] Spin--parity of the retained gluonic excitation, total angular momentum and parity of the color-triplet $qg$ subsystem, and their relative orbital angular momentum, respectively.
\item[$(\lambda_k^c)_{\rm tr}^{\dagger},(\lambda_k^c)_{\rm tr}$] Normalized color-transition tensors connecting the quark on line $k$ with the color-triplet $qg$ component and its reverse.
\item[$\mathsf S_{k,\ell}^{\dagger},\mathsf S_{k,\ell},P_{k,\ell\ell'}^{(3/2)}$] Off-diagonal rank-one transition tensors connecting the spin-$1/2$ quark subspace and the projected $j_{qg}=3/2$ component, together with the associated vector-spin projector $P_{k,\ell\ell'}^{(3/2)}=\mathsf S_{k,\ell}\mathsf S_{k,\ell'}^{\dagger}$.
\item[$\mathsf Q_{g,3/2},R_{\mathsf Q_{g,3/2}}(E)$] Projector onto the $j_{qg}^P=3/2^+$ component of the resolved $qqqg$ sector and the corresponding excluded-sector resolvent.
\item[$T_{\mathsf Q_{g,3/2}\mathsf P}^{(ik)},T_{\mathsf P\mathsf Q_{g,3/2}}^{(ik)}$] Minimal local color--spin transition blocks coupling the retained $qqq$ sector to the $j_{qg}^P=3/2^+$ excluded component through lines $i$ and $k$.
\item[$g_{CS}^{\rm tr},g_{CS}^{\rm red},u_{ik}^{(3/2)}$] Color--spin transition strength, its color-normalized reduced coupling, and the corresponding short-range radial transition function.
\item[$L_{ij;k}^{S-S,(3/2)}$] Line-resolved $S$--$S$-like connected operator generated by eliminating the minimal local $j_{qg}^P=3/2^+$ component.
\end{notationlist}

\notationgroup{Fit objectives and diagnostics}
\begin{notationlist}
\item[$\varepsilon_H$] Signed mass residual $M_H^{\rm th}-M_H^{\rm exp}$.
\item[$\sigma_M,\sigma_B$] Meson and baryon root-mean-square errors.
\item[$\sigma_{\rm eq}$] Equal-sector compatibility error $[(\sigma_M^2+\sigma_B^2)/2]^{1/2}$, corresponding to $\lambda=1/2$.
\item[$\sigma_{\rm out},\sigma_{B,18}$] Absolute-mass RMSE over the measured out-of-fit baryons and over the union of the nine calibration and nine measured out-of-fit baryons, respectively.
\item[$\sigma_{\rm structure}$] Centroid-subtracted RMSE of the measured internal spin structures within groups of baryons having identical constituent content, as defined in Eq.~(\ref{eq:spin_structure_rmse}).
\item[$\mathrm{MAE}$] Mean absolute mass error.
\item[$\lambda$] Meson-sector weight in the weighted-sum meson--baryon objective; distinguished from indexed Gell-Mann matrices $\lambda_i^a$ and local eigenvalue symbols.
\item[$\mathcal L_\lambda$] Weighted-sum meson--baryon objective used to trace the compatibility trade-off trajectory.
\item[$\mathcal D_M,\mathcal D_B$] Meson and baryon data sets used in the $u,d,s,c,b$ compatibility benchmark.
\item[$M_B^{\rm rest}$] Constituent rest-mass sum $M_B^{\rm rest}=\sum_{i\in B}m_i$ used in the baryon residual flavor-correlation diagnostic.
\item[$M^{2Q},M^{2Q+3Q}$] Calculated baryon masses obtained with the pair Hamiltonian and with the pair plus connected three-quark interaction, respectively.
\item[$r_{M_{\rm rest},\varepsilon}$] Pearson correlation between the constituent rest-mass sum $M_B^{\rm rest}$ and the mass residual.
\item[$P_\rho,P_\lambda$] Projected probabilities of the $\chi_\rho$ and $\chi_\lambda$ pair-spin components in the $qsQ$ baryons.
\item[$\theta_{\rm eff},\Delta\theta_{\rm eff}$] Effective admixture angle $\theta_{\rm eff}=\tan^{-1}\sqrt{P_{\rm sub}/P_{\rm dom}}$ and its change between the $2Q+3Q$ and $2Q$ fit points.
\end{notationlist}

\notationgroup{Alternative two-body benchmark notation}
\begin{notationlist}
\item[$X$] Two-body-model label in the cross-model compatibility benchmark, $X\in\{\mathrm{AL1},\mathrm{AL2},\mathrm{AP1},\mathrm{AP2},\mathrm{BD},\mathrm{SL},\mathrm{GI/CI}\}$.  The $\mathrm{S^2AJ}$ baseline is presented separately before this set.
\item[$Q_{ij}$] Scalar momentum scale in the reduced-mass running-coupling prescription, set to $m_{ij}$ there; distinct from the transfer momentum $\mathbf Q_{ij}$.
\item[$\alpha_{\rm IR},Q_0,\beta_0$] Infrared coupling, transition scale, and one-loop coefficient in the reduced-mass running-coupling benchmark.
\item[$p_X$] Confinement power of an AL/AP-type interaction; $p_X=1$ for AL and $p_X=2/3$ for AP.
\item[$g_X(r)$] Short-distance regulator multiplying the Coulomb and hyperfine terms in the AL2/AP2 variants.
\item[$G_{ij}^{X}(r),r_{0,ij}^{X}$] Normalized Gaussian hyperfine profile and its flavor-dependent range in an AL/AP-type model.
\item[$r_{c,X},\zeta_X$] AL2/AP2 short-distance cutoff and the mass-scaling exponent of $r_{0,ij}^{X}$.
\item[$d_i,D_i$] Constituent-size parameter and inverse monopole scale $D_i=1/d_i$ in the SL-type interaction.
\item[$d_n,\eta$] Reference nonstrange constituent size and its flavor-mass exponent in $d_i=d_n(m_q/m_i)^\eta$ for the SL benchmark; the indexed $d_n$ used in the GI/CI Gaussian expansion below is a separate local sequence.
\item[$\mathcal C_{ij}^{\rm Y}(r),\delta_{ij}^{\rm Y}(r)$] Monopole-folded Coulomb function and normalized contact density in the SL-type interaction.
\item[$\rho_{ij}^{\rm GI}(\mathbf x)$] Normalized coordinate-space Gaussian smearing density of the GI/CI-type interaction.
\item[$\sigma_{ij}^{\rm sm},\tau_{a,ij}$] Flavor-dependent GI/CI smearing scale and the combined running-coupling--smearing scale.
\item[$\alpha_a,\gamma_a$] Weight and momentum scale of Gaussian component $a=1,2,3$ in the GI/CI running coupling.
\item[$G_{ij},H_{ij}^{CS}$] Smeared GI/CI Coulomb and contact-hyperfine pair interactions before relativistic momentum dressing.
\item[$\overline r_{ij}$] Gaussian-smeared pair distance entering the GI/CI confinement term.
\item[$\varepsilon_i^{(ij)}(p)$] Relativistic pair-energy factor $\sqrt{p^2+m_i^2}$ used in the GI/CI vertex dressing; distinct from the hadron-mass residual $\varepsilon_H$.
\item[$R_{ij}^{C}(p),R_{ij}^{CS}(p)$] Relativistic dressing factors of the GI/CI Coulomb and color-spin interactions.
\item[$\widetilde G_{ij},\widetilde H_{ij}^{CS}$] Symmetrically momentum-dressed, nonlocal GI/CI Coulomb and color-spin operators.
\item[$\epsilon_C,\epsilon_{CS}$] Exponents controlling the GI/CI relativistic vertex factors; distinct from the ISG sources $\epsilon_i$ and mass residuals $\varepsilon_H$.
\item[$\zeta_\alpha,\zeta_\gamma$] Overall scaling parameters for the GI/CI running-coupling weights and momentum scales.
\item[$\sigma_{\rm GI},\sigma_0,s_{\rm GI}$] GI/CI confinement coefficient and the two parameters entering its flavor-dependent smearing scale.
\item[$f_\Delta$] Numerical factor multiplying the pairwise $\Delta$-ansatz confinement term in the GI/CI baryon Hamiltonian; unrelated to the three-quark spatial profiles.
\item[$N_R,c_{n,ij}^{X},d_n$] Rank and coefficients/exponents of the Gaussian expansion used to represent a GI/CI dressing factor $R_{ij}^{X}(p)$ numerically; this local exponent sequence is distinct from the SL reference size $d_n$.
\end{notationlist}

\notationgroup{Excitation-benchmark notation}
\begin{notationlist}
\item[TH1] Linear-confinement central baseline with the color-spin interaction but without tensor or spin--orbit terms.
\item[TH2] Softened $r^{2/3}$-confinement diagnostic Hamiltonian with symmetric spin--orbit, antisymmetric spin--orbit, and tensor interactions.  In the meson calculation these terms are included in the full fixed-$\mathcal J^{P(C)}$ diagonalization, whereas in the baryon benchmark only their diagonal expectation values are retained.
\item[$\sigma_{2/3}$] Coefficient of the $r^{2/3}$ confining potential in TH2.
\item[$c_{CS},c_{LS},c_{ALS},c_T$] Fitted normalizations of the TH2 color-spin, symmetric spin--orbit, antisymmetric spin--orbit, and tensor interactions.
\item[$\sigma_{12}^{\rm sm},\sigma_0^{\rm sm},s_{\rm sm}$] Flavor-dependent TH2 smearing scale and its two global parameters.
\item[$f_V$] Vector-confinement fraction used in the TH2 fine-structure convention; the adopted value is $f_V=0$.
\item[$M_{\rm noY}$] TH1 or TH2 baryon mass before adding the perturbative Yukawa-profile three-quark contribution.
\end{notationlist}

\notationgroup{GEM basis and generalized eigenvalue problem}
\begin{notationlist}
\item[$\phi_{nlm}(\mathbf r)$] Normalized Gaussian orbital $N_{nl}r^l e^{-\nu_n r^2}Y_{lm}(\hat{\mathbf r})$.
\item[$n,N$] Gaussian radial-basis indices for the first and second Jacobi coordinates.
\item[$N_{nl},N_{NL}$] Gaussian normalization factors.
\item[$\nu_n,\nu_N$] Gaussian widths associated with $\mathbf r$ and $\mathbf R$.
\item[$r_n,R_N$] Gaussian ranges, $\nu_n=r_n^{-2}$ and $\nu_N=R_N^{-2}$.
\item[$\varrho_M,\varrho_r,\varrho_R$] Geometric ratios for the meson mesh and the two baryon Jacobi meshes.
\item[$\alpha,\beta$] In \ref{app:GEM_ref} and \ref{app:analytic_without_gem}, collective labels of the bra and ket basis states, respectively.
\item[$\Phi_\alpha$] Gaussian basis state labeled by the collective quantum numbers $\alpha$.
\item[$\mathcal B_\alpha^{(\mathcal C)},\boldsymbol{\mathcal B}$] Variational coefficient in channel $\mathcal C$ and the assembled coefficient vector in the nonorthogonal basis.
\item[$H_{\alpha\beta}$] Hamiltonian matrix element between basis states $\alpha$ and $\beta$.
\item[$\mathcal O_{\alpha\beta}$] Nonorthogonal overlap matrix element; block-indexed $\mathcal O$ in the RGM discussion denotes a continuum norm kernel.
\item[$E$] Generalized eigenvalue in $H\boldsymbol{\mathcal B}=E\mathcal O\boldsymbol{\mathcal B}$.
\item[$U_{\mathcal O},o_n,\varepsilon_{\mathcal O}$] Overlap eigenvectors, eigenvalues, and relative cutoff used in canonical orthogonalization.
\item[$X_{\mathcal O},\mathbf y$] Canonical orthogonalizer and normalized coefficient vector in the retained orthonormal subspace.
\item[$\Delta E_{\mathcal O},\Delta\mathcal Q_{\mathcal O}$] Energy and operator-expectation changes used to test overlap-cutoff stability.
\end{notationlist}

\notationgroup{ISG and two-body generating functions}
\begin{notationlist}
\item[$\epsilon_i$] Infinitesimal ISG source variables, $i=1,\ldots,4$; distinct from $\varepsilon_H$ and the GI/CI exponents $\epsilon_C,\epsilon_{CS}$.
\item[$h_1,\ldots,h_4$] Discrete ISG indices for the four shifted orbital factors.
\item[$C_{lm,h}$] ISG coefficient associated with orbital quantum numbers $(l,m)$ and discrete index $h$; a generic local ISG index is sometimes denoted by $k$.
\item[$\mathbf D_{lm,h}$] Dimensionless complex ISG shift vector; the generic local notation $\mathbf D_{lm,k}$ is equivalent.
\item[$\mathbf D_1,\ldots,\mathbf D_4$] Shorthand for the four bra/ket ISG shift vectors.
\item[$\mathcal M^{(n)}_{i_1\cdots i_n}(l,m)$] Symmetric rank-$n$ ISG moment tensor; it vanishes for $n<l$, while its first nonzero rank reproduces the solid-harmonic tensor.
\item[$F(\boldsymbol\epsilon;H)$] Generating function whose specified source coefficient gives the matrix element of $H$.
\item[{$\mathrm{Co}[\boldsymbol\epsilon^{\mathbf q}]$}] Coefficient-extraction operator.
\item[$\mathbf q$] Orbital-degree tuple $(l_\alpha,L_\alpha,l_\beta,L_\beta)$ in the Bargmann--Fock discussion; not a constituent flavor or the spectator momentum $\mathbf q_k$.
\item[$\boldsymbol\epsilon$] Source tuple $(\epsilon_1,\epsilon_2,\epsilon_3,\epsilon_4)$.
\item[$\mathsf G_\alpha,\mathsf G_\beta$] $2\times2$ Gaussian exponent matrices of the bra and ket basis states; the direct Cartesian construction uses the same matrices in a common reference channel.
\item[$\mathbf D_\alpha,\mathbf D_\beta$] Two-component source-vector arrays for the bra and ket states.
\item[$Y_{\mathcal C\to\mathcal C'}$] Matrix $Y_{\mathcal C}$ transformed from Jacobi channel $\mathcal C$ to $\mathcal C'$; vector arrays use the analogous arrow notation.
\item[$\mathbf b_x$] Transformed two-component source array in Jacobi channel $x$.
\item[$\mathsf B_x$] Combined $2\times2$ Gaussian matrix in Jacobi channel $x$.
\item[$\mathsf W_K^{(1,2)}$] Symmetric $4\times4$ source-contraction matrices in the kinetic generating function.
\item[$\mathsf W_V^{(1,x)},\mathsf W_V^{(2,x)}$] Symmetric $4\times4$ source-contraction matrices for a two-body potential evaluated in channel $x$.
\item[$\mathsf E_1,\mathsf E_2$] Auxiliary $2\times2$ matrices entering $\mathsf W_V^{(1,x)}$ and $\mathsf W_V^{(2,x)}$.
\item[$y_x$] Schur-complement coefficient $\det\mathsf B_x/\mathsf B_{x,22}$.
\item[$\mathbf p$] Effective source vector $\mathbf b_{x,1}-(\mathsf B_{x,12}/\mathsf B_{x,22})\mathbf b_{x,2}$ in the radial potential integral; elsewhere bold $\mathbf p$ denotes a physical momentum.
\item[$u$] Nonnegative series index in the closed central and color-spin generating functions; unrelated to the light-quark flavor $u$.
\end{notationlist}

\notationgroup{Yukawa-profile GEM auxiliaries}
\begin{notationlist}
\item[$t_a,t_b,t_c$] Positive Schwinger parameters associated with the three pair distances.
\item[$\mathfrak m_x$] Pair-channel shorthand $(\mathfrak m_a,\mathfrak m_b,\mathfrak m_c)=(m_{12},m_{23},m_{31})$, used only in \ref{GEM_3body_calc}.
\item[$\rho_x$] Schwinger coefficient $c^2\mathfrak m_x^2/(2\hbar^2)$ for $x=a,b,c$; a local dummy index $i$ may replace $x$.  This symbol is distinct from $\rho_\alpha(E)$ and $\rho_{ij}^{\rm GI}$.
\item[$\mathcal A_a,\mathcal A_b,\mathcal A_c$] Positive-semidefinite $2\times2$ pair-coordinate matrices in a fixed reference Jacobi channel.
\item[$\mathsf B_{\rm Y}(t)$] Yukawa-profile Gaussian matrix $\mathsf B_\alpha+\sum_x t_x\mathcal A_x$.
\item[$\mathsf B_{\rm Y}^{(0,1)}(t)$] Commuting zeroth-order part and perturbative noncommuting remainder of $\mathsf B_{\rm Y}(t)$.
\item[$\mathsf W_{\rm Y}(t)$] Symmetric $4\times4$ source-contraction matrix in the Yukawa-profile generating function.
\item[$n_{pq}$] Wick-pairing multiplicity between source labels $p<q$.
\item[$\tau_{\rm p}$] Bookkeeping parameter for the perturbative decomposition of $\mathsf B_{\rm Y}$; set to unity after expansion.
\item[$\mathbf v_\pm$] Common zeroth-order mode vectors used to diagonalize the commuting part of the auxiliary matrices.
\item[$\Lambda_\pm(t)$] Eigenvalues of $\mathsf B_{\rm Y}^{(0)}(t)$ in the $\mathbf v_\pm$ modes.
\item[$\theta^\mu,\varphi^\nu$] Four-component coefficient arrays for one- and two-insertion master integrals; unrelated to the CSM angle $\theta$.
\item[$A(x),L_i(x),\Delta_s(x),\bar\rho$] Local scalar functions and the geometric mean $(\rho_a\rho_b\rho_c)^{1/3}$ used in the one-dimensional reduction of the Schwinger integrals.
\end{notationlist}

\notationgroup{Direct Cartesian cross-check}
\begin{notationlist}
\item[$\mathscr P_\alpha^{JM}$] Coupled Cartesian solid-harmonic polynomial of basis state $\alpha$.
\item[$T^{(lm)}_{i_1\cdots i_l}$] Symmetric traceless Cartesian tensor representing $\mathcal Y_{lm}$.
\item[$\mathsf B_{\alpha\beta}$] Combined Gaussian matrix $\mathsf G_\alpha+\mathsf G_\beta$.
\item[$\mathcal Z(\mathsf B)$] Six-dimensional Gaussian partition integral $\pi^3/(\det\mathsf B)^{3/2}$.
\item[{\(\mathscr G_{\alpha\beta}[\mathscr Q],\mathsf C_{\alpha\beta}\)}] Gaussian polynomial master integral and covariance matrix $\mathsf B_{\alpha\beta}^{-1}/2$.
\item[$\mathscr D_{\mu i}^{(\beta)}$] Gaussian derivative operator acting on the ket polynomial.
\item[$\mathscr K_\beta^{JM}$] Cartesian polynomial generated by the internal kinetic operator.
\item[$\mathbf w_{ij}$] Two-component coefficient vector expressing $\mathbf r_{ij}=\mathbf w_{ij}^T\mathbf X$.
\item[$\zeta_{ij}^{(\alpha\beta)}$] Component variance of pair coordinate $\mathbf r_{ij}$ under the combined Gaussian measure.
\item[$\mathbf c^{(ij)}$] Conditional-regression coefficient of $\mathbf X$ on $\mathbf r_{ij}$.
\item[$\boldsymbol\eta_\mu$] Conditional Gaussian fluctuation independent of $\mathbf r_{ij}$; distinct from the Riemann-sheet label $\boldsymbol\eta$.
\item[$\mathsf\Gamma^{(ij)}$] Conditional covariance matrix of the remaining Jacobi variables.
\item[$\mathcal I_{\alpha\beta}^{(ij)}(p,a)$] Radial Gaussian moment for a pair potential proportional to $r_{ij}^{p}e^{-a r_{ij}^2}$.
\item[$\mathcal R_q(\zeta,a)$] Elementary three-dimensional radial moment used in $\mathcal I_{\alpha\beta}^{(ij)}$.
\item[$\mathsf H_{\rm G},\mathsf B_{\alpha\beta}^{\rm G}$] Gaussian three-body profile matrix and the corresponding shifted exponent matrix.
\item[$\mathfrak A_{JP}^{(l_{\max})}$] Set of symmetry-allowed orbital channels $(l,L)$ in a fixed $(J,P)$ block.
\item[$\delta_{\mathcal O},\delta_H$] Relative matrix-level differences between Cartesian and ISG overlap/Hamiltonian matrices.
\item[$\eta_{\mathcal O},\eta_H$] Relative Hermiticity defects of the overlap and Hamiltonian matrices.
\end{notationlist}

\notationgroup{First-order and rediagonalization checks}
\begin{notationlist}
\item[$M_B^{(1)},M_B^{\rm diag}$] First-order and fully rediagonalized retained-space baryon masses in the Gaussian-profile validation.
\item[$\ket{\Psi_{B,n}^{(0)}},E_{B,n}^{(0)}$] Orthonormal two-body eigenstate and eigenvalue in a fixed baryon block.
\item[$\rho_r,\rho_R$] Normalized Jacobi radial densities obtained after tracing over internal spin--flavor degrees of freedom, with $\int_0^\infty\rho_r(r)\,dr=\int_0^\infty\rho_R(R)\,dR=1$.
\end{notationlist}

\notationgroup{Electromagnetic observables}
\begin{notationlist}
\item[$\mathcal R_B$] Retained set of low-lying orthonormal states in the same baryon flavor, parity, and angular-momentum block used in the first-order wave-function correction.
\item[$\ket{\delta\Psi_{B,n}^{\rm Y}},\ket{\Psi_{B,n}^{\rm Y}}$] First-order Yukawa-profile wave-function correction and the state retained through first order.
\item[$\widehat J_{\rm em}^{\mu}$] One-body electromagnetic current in the impulse approximation; $\widehat J_X^\mu$ denotes a generic external current in the representation-dependence discussion.
\item[$q_i^{\rm em}$] Constituent electric charge in units of $e$.
\item[$\widehat{\boldsymbol\mu}_{\rm em},\mu_N$] Magnetic-moment operator and nuclear magneton $e\hbar/(2m_pc)$.
\item[$\widehat{\mathbf d}_{E1}$] Center-of-mass-subtracted electric-dipole operator, defined without the overall factor $e$.
\item[$\mathbf R_{\rm CM}$] Constituent center-of-mass coordinate $\sum_i m_i\mathbf r_i/\sum_i m_i$.
\item[$\mu_{fi}$] Diagonal or transition magnetic moment between states $i$ and $f$; $f=i$ gives a static magnetic moment.
\item[$\mathcal S_{{\rm M1},fi},\mathcal S_{{\rm E1},fi}$] Magnetic- and electric-dipole strengths in the conventions of \ref{app:em_observables}; $\mathcal S_{{\rm M1},fi}$ is rotationally reduced, whereas $\mathcal S_{{\rm E1},fi}$ is defined from the adopted Cartesian $z$-component matrix element.
\item[$k_\gamma,E_f^\gamma$] Photon momentum and final-baryon recoil energy in the initial-baryon rest frame.
\item[$\alpha_{\rm em}$] Electromagnetic fine-structure constant.
\item[$\Gamma_{\rm M1},\Gamma_{\rm E1}$] Magnetic- and electric-dipole radiative widths.
\end{notationlist}

\notationgroup{Static three-quark confinement}
\begin{notationlist}
\item[$V_{3Q}$] Static three-quark potential containing the pairwise Coulomb contribution, the long-distance confining geometry, and an additive constant.
\item[$L_{\rm min}^{(Y)}$] Minimum total length of the $Y$-shaped flux-tube network joining three static quarks.
\item[$L_{\Delta}$] Pairwise $\Delta$ length $\tfrac12(r_{12}+r_{23}+r_{31})$, used as an approximation to the static confinement energy.
\item[$\kappa_{3Q},\sigma_{3Q},D_{3Q}$] Coulomb coefficient, string tension, and additive constant in the static three-quark $Y$-ansatz.
\end{notationlist}

\notationgroup{Frequently used abbreviations}
\begin{notationlist}
\item[QCD] Quantum chromodynamics.
\item[NRQCD/pNRQCD] Non-relativistic QCD/potential non-relativistic QCD.
\item[DCSB] Dynamical chiral symmetry breaking.
\item[EFT] Effective field theory.
\item[OGE/TGE] One-gluon exchange/two-gluon exchange.
\item[GEM/ISG] Gaussian expansion method/infinitesimally shifted Gaussian method.
\item[RGM] Resonating-group method.
\item[CSM/RSM] Complex-scaling method/real-scaling or stabilization method.
\item[GBE/PS-GBE/III] Goldstone-boson exchange/pseudoscalar Goldstone-boson exchange/instanton-induced interaction.
\item[KMT] Kobayashi--Maskawa--'t~Hooft determinant interaction.
\item[2BD/3BD] Two-body/three-body interaction in benchmark labels.
\item[2Q/3Q] Two-body-only/three-quark-extended calculation in fit and wave-function labels.
\item[3Q] Three-quark operator or contribution.
\item[GI/CI] Godfrey--Isgur/Capstick--Isgur.
\item[AL/AP/BD/SL] AL1--AL2/AP1--AP2/Bhaduri--Cohler--Nogami/Stanley comparison Hamiltonians.
\item[TH1/TH2] Linear central baseline/softened-confinement fine-structure diagnostic Hamiltonian in the excitation benchmark.
\item[RMSE/MAE] Root-mean-square error/mean absolute error.
\item[$\mathrm{S^2AJ}$] Baseline Hamiltonian of Eqs.~(\ref{hamiltonian})--(\ref{2body_2}) used in the meson--baryon compatibility benchmark.
\end{notationlist}

\end{multicols}
\endgroup

\section{Baryon basis}\label{bases}

\paragraph{Color and angular-momentum conventions:}
The normalized three-quark color-singlet state is fully antisymmetric and is written as
\begin{equation}
|C_{\mathbf 1}\rangle=\frac{1}{\sqrt{6}}\sum_{a,b,c=r,g,b}\epsilon_{abc}|a\rangle_1|b\rangle_2|c\rangle_3.
\end{equation}
The orbital, spin, and flavor parts must therefore combine into the symmetric representation for every exchange of identical quarks.
For a fixed set $\alpha=(n,N,l,L)$, we define the Gaussian orbital in rearrangement channel $\mathcal C$ by
\begin{equation}
\Phi_{\mathcal C,\alpha}^{JM}\equiv\left[\phi_{nl}(\mathbf r_{\mathcal C})\phi_{NL}(\mathbf R_{\mathcal C})\right]_{JM},\qquad \mathcal C=a,b,c.
\label{eq:A_channel_Gaussian}
\end{equation}
The orbital angular momenta satisfy $\vert l-L\vert \leq J\leq l+L$, and $J$ is coupled with the total spin $S$ according to $\vert J-S\vert \leq\mathcal J\leq J+S$.
All basis states displayed below have positive parity, so $l+L$ is even.
For $J=0$, this condition gives $L=l$, while for $J=2$ the allowed $L$ values satisfy the triangle condition and have the same parity as $l$.
These conditions are understood in every basis formula below.
The projector coefficients below are exact, while the displayed Gaussian representatives are used up to an overall nonzero constant because the physical normalization is determined by the full nonorthogonal overlap matrix.

\paragraph{$S_3$ projectors and Young--Yamanouchi states:}
The projectors onto the three irreducible representations of $S_3$ are
\begin{align}
\mathcal P_{[3]}&=\frac{1}{6}\left(I+P_{12}+P_{23}+P_{31}+P_{123}+P_{132}\right),\nonumber\\ \mathcal P_{[111]}&=\frac{1}{6}\left(I-P_{12}-P_{23}-P_{31}+P_{123}+P_{132}\right),\qquad \mathcal P_{[21]}=\frac{1}{3}\left(2I-P_{123}-P_{132}\right).
\label{eq:A_S3_projectors}
\end{align}
We use the Young--Yamanouchi convention
\begin{equation}
|[3]\rangle\leftrightarrow\vcenter{\hbox{\begin{ytableau}1&2&3\end{ytableau}}},\qquad |\lambda\rangle\leftrightarrow\vcenter{\hbox{\begin{ytableau}1&2\\3\end{ytableau}}},\qquad |\rho\rangle\leftrightarrow\vcenter{\hbox{\begin{ytableau}1&3\\2\end{ytableau}}},\qquad |[111]\rangle\leftrightarrow\vcenter{\hbox{\begin{ytableau}1\\2\\3\end{ytableau}}}.
\label{eq:A_Young_states}
\end{equation}
The $\lambda$ and $\rho$ states are symmetric and antisymmetric under $P_{12}$, respectively, and our relative phase is fixed by
\begin{equation}
P_{23}|\lambda\rangle=-\frac{1}{2}|\lambda\rangle+\frac{\sqrt{3}}{2}|\rho\rangle,\qquad P_{23}|\rho\rangle=\frac{\sqrt{3}}{2}|\lambda\rangle+\frac{1}{2}|\rho\rangle.
\label{eq:A_Young_recoupling}
\end{equation}

\paragraph{Orbital Young representations in the Gaussian basis:}

For fixed $(n,N,l,L,J,M)$, we abbreviate $\Phi_{\mathcal C,\alpha}^{JM}$ as $\Phi_{\mathcal C}$.
The fully symmetric and fully antisymmetric orbital states are obtained directly from Eq.~\eqref{eq:A_S3_projectors},

\begin{align}
\vcenter{\hbox{\begin{ytableau}1&2&3\end{ytableau}}}_{O}\quad&\longleftrightarrow\quad \Phi_{[3]}^{JM}\equiv\mathcal P_{[3]}\Phi_a=\frac{1}{3}\left(\Phi_a+\Phi_b+\Phi_c\right),\qquad l=0,2,4,\ldots,\label{eq:A_orbital_symmetric}\\
\vcenter{\hbox{\begin{ytableau}1\\2\\3\end{ytableau}}}_{O}\quad&\longleftrightarrow\quad \Phi_{[111]}^{JM}\equiv\mathcal P_{[111]}\Phi_a=\frac{1}{3}\left(\Phi_a+\Phi_b+\Phi_c\right),\qquad l=1,3,5,\ldots.\label{eq:A_orbital_antisymmetric}
\end{align}
The two $[21]$ components have the following Gaussian representatives for even $l$,
\begin{align}
\vcenter{\hbox{\begin{ytableau}1&3\\2\end{ytableau}}}_{O}\equiv\Phi_{\rho,e}^{JM}&=\frac{1}{\sqrt{2}}\left(\Phi_b-\Phi_c\right),\qquad l=0,2,4,\ldots,\label{mixed1}\\
\vcenter{\hbox{\begin{ytableau}1&2\\3\end{ytableau}}}_{O}\equiv\Phi_{\lambda,e}^{JM}&=\frac{1}{\sqrt{6}}\left(\Phi_b+\Phi_c-2\Phi_a\right),\qquad l=0,2,4,\ldots,\label{mixed2}
\end{align}
and for odd $l$,
\begin{align}
\vcenter{\hbox{\begin{ytableau}1&2\\3\end{ytableau}}}_{O}\equiv\Phi_{\lambda,o}^{JM}&=\frac{1}{\sqrt{2}}\left(\Phi_b-\Phi_c\right),\qquad l=1,3,5,\ldots,\label{mixed3}\\
\vcenter{\hbox{\begin{ytableau}1&3\\2\end{ytableau}}}_{O}\equiv\Phi_{\rho,o}^{JM}&=\frac{1}{\sqrt{6}}\left(2\Phi_a-\Phi_b-\Phi_c\right),\qquad l=1,3,5,\ldots.\label{mixed4}
\end{align}
Both pairs obey the same Young--Yamanouchi transformation in Eq.~\eqref{eq:A_Young_recoupling}.
In the following formulas, $\Phi_\lambda$ and $\Phi_\rho$ denote the appropriate even- or odd-$l$ realization.
Within one $[21]$ coupling, the $\lambda$ and $\rho$ components are always taken from the same parity family.

\paragraph{Pair-exchange basis:}

For a system with particles 1 and 2 identical, we use $\mathcal P_{12}^{(\pm)}=(I\pm P_{12})/2$.
The $a$-channel representatives are $\Phi_{a,\lambda}^{JM}\equiv\mathcal P_{12}^{(+)}\Phi_a=\Phi_a{\big\vert} _{l=0,2,4,\ldots}$, $\Phi_{a,\rho}^{JM}\equiv\mathcal P_{12}^{(-)}\Phi_a=\Phi_a{\big\vert} _{l=1,3,5,\ldots}$.
The corresponding $b+c$ representatives are
\begin{align}
\Phi_{bc,\lambda}^{JM}\equiv\mathcal P_{12}^{(+)}\Phi_b=\frac{1}{2}\left(\Phi_b+(-1)^l\Phi_c\right),\qquad  \Phi_{bc,\rho}^{JM}\equiv\mathcal P_{12}^{(-)}\Phi_b=\frac{1}{2}\left(\Phi_b-(-1)^l\Phi_c\right).\label{asym2}
\end{align}
We write $\mathscr O_\lambda^J\in\{\Phi_{a,\lambda}^{J},\Phi_{bc,\lambda}^{J}\}$ and $\mathscr O_\rho^J\in\{\Phi_{a,\rho}^{J},\Phi_{bc,\rho}^{J}\}$, where the magnetic and radial labels are suppressed.

\paragraph{Spin and flavor Young states:}
For total spin $S=1/2$, the two Young--Yamanouchi spin states are
\begin{equation}
\chi_\lambda\equiv\vcenter{\hbox{\begin{ytableau}\uparrow^1&\uparrow^2\\\downarrow^3\end{ytableau}}}_{\!S},\qquad \chi_\rho\equiv\vcenter{\hbox{\begin{ytableau}\uparrow^1&\uparrow^3\\\downarrow^2\end{ytableau}}}_{\!S},
\end{equation}
where $\chi_\lambda$ and $\chi_\rho$ have pair spin $s_{12}=1$ and 0, respectively.
The fully symmetric spin state is denoted by $\chi_{[3]}\equiv\vcenter{\hbox{\begin{ytableau}{}&{}&{}\end{ytableau}}}_{S=3/2}$.
For a proton, the two flavor states are

\begin{equation}
F_\lambda^N\equiv\vcenter{\hbox{\begin{ytableau}u^1&u^2\\d^3\end{ytableau}}}_{\!F},\qquad F_\rho^N\equiv\vcenter{\hbox{\begin{ytableau}u^1&u^3\\d^2\end{ytableau}}}_{\!F}.
\end{equation}
The neutron states follow by interchanging $u$ and $d$.
The three spin--flavor irreducible combinations for $S=1/2$ are
\begin{align}
(F\chi)_{[3]}^N&=\frac{1}{\sqrt{2}}\left(F_\lambda^N\chi_\lambda+F_\rho^N\chi_\rho\right),\qquad (F\chi)_{[111]}^N=\frac{1}{\sqrt{2}}\left(F_\lambda^N\chi_\rho-F_\rho^N\chi_\lambda\right),\label{eq:A_N_SF_sym_anti}\\
\mathcal X_\lambda^N&=\frac{1}{\sqrt{2}}\left(F_\lambda^N\chi_\lambda-F_\rho^N\chi_\rho\right),\qquad \mathcal X_\rho^N=-\frac{1}{\sqrt{2}}\left(F_\lambda^N\chi_\rho+F_\rho^N\chi_\lambda\right).\label{eq:A_N_SF_mixed}
\end{align}
The pair $(\mathcal X_\lambda^N,\mathcal X_\rho^N)$ transforms as the same $[21]$ Young--Yamanouchi doublet as $(\Phi_\lambda,\Phi_\rho)$.

\paragraph{Proton/neutron-type ($\mathcal J^P=1/2^+$) baryons:}
The even-$J$ sectors used here are $(J,S)=(0,1/2)$ and $(2,3/2)$.
For $(J,S)=(0,1/2)$, the three allowed permutation couplings are
\begin{align}
\Psi_{N,[3]}=\left[\Phi_{[3]}^{J=0}(F\chi)_{[3]}^N\right]_{\mathcal J=1/2}^{P=+},\;\;\; \Psi_{N,[21]}=\left[\frac{1}{\sqrt{2}}\left(\Phi_\lambda^{J=0}\mathcal X_\lambda^N+\Phi_\rho^{J=0}\mathcal X_\rho^N\right)\right]_{\mathcal J=1/2}^{P=+},\;\;\; \Psi_{N,[111]}=\left[\Phi_{[111]}^{J=0}(F\chi)_{[111]}^N\right]_{\mathcal J=1/2}^{P=+}.\label{eq:A_N_basis_anti}
\end{align}
For $(J,S)=(2,3/2)$, the orbital and flavor $[21]$ states are coupled to $[3]$,
\begin{equation}
\Psi_N^{(2,3/2)}=\left[\frac{1}{\sqrt{2}}\left(\Phi_\lambda^{J=2}F_\lambda^N+\Phi_\rho^{J=2}F_\rho^N\right)\chi_{[3]}\right]_{\mathcal J=1/2}^{P=+}.
\label{eq:A_N_basis_J2}
\end{equation}

\paragraph{$\Delta$-type ($I=3/2,\;\mathcal J^P=3/2^+$) baryons:}
The flavor state is fully symmetric and is denoted by $F_{[3]}^\Delta\leftrightarrow\vcenter{\hbox{\begin{ytableau}{}&{}&{}\end{ytableau}}}_{F}$.
The even-$J$ sectors used here are $(J,S)=(0,3/2)$, $(2,3/2)$, and $(2,1/2)$.
For $S=3/2$, the orbital state is fully symmetric,
\begin{equation}
\Psi_\Delta^{(J,3/2)}=\left[\Phi_{[3]}^{J}\chi_{[3]}\right]_{\mathcal J=3/2}^{P=+}F_{[3]}^\Delta,\qquad J=0,2.
\label{eq:A_Delta_basis_S32}
\end{equation}
For $(J,S)=(2,1/2)$, the orbital and spin doublets are coupled to $[3]$,
\begin{equation}
\Psi_\Delta^{(2,1/2)}=\left[\frac{1}{\sqrt{2}}\left(\Phi_\lambda^{J=2}\chi_\lambda+\Phi_\rho^{J=2}\chi_\rho\right)\right]_{\mathcal J=3/2}^{P=+}F_{[3]}^\Delta.
\label{eq:A_Delta_basis_S12}
\end{equation}
The same fully symmetric construction applies to $\Omega$-type states.

\paragraph{$\Sigma$-type ($I=1,\;\mathcal J^P=1/2^+$) baryons:}
The light-pair flavor state is symmetric and is written as $F_\lambda^\Sigma\equiv\vcenter{\hbox{\begin{ytableau}q^1&q^2\\s^3\end{ytableau}}}_{F}$.
The orbital--spin part is therefore symmetric under $P_{12}$.
For $(J,S)=(0,1/2)$ and $(2,3/2)$, the basis is
\begin{align}
\Psi_\Sigma^{(0,1/2)}\in\left\{\left[\mathscr O_\lambda^{J=0}\chi_\lambda\right]_{\mathcal J=1/2}^{P=+},\ \left[\mathscr O_\rho^{J=0}\chi_\rho\right]_{\mathcal J=1/2}^{P=+}\right\}F_\lambda^\Sigma, \qquad \Psi_\Sigma^{(2,3/2)}=\left[\mathscr O_\lambda^{J=2}\chi_{[3]}\right]_{\mathcal J=1/2}^{P=+}F_\lambda^\Sigma.\label{eq:A_Sigma_basis_J2}
\end{align}
The same formulas apply to $\Sigma_Q$ by replacing $s^3$ with $Q^3$.

\paragraph{$\Lambda$-type ($I=0,\;\mathcal J^P=1/2^+$) baryons:}
The light-pair flavor state is antisymmetric and is written as $F_\rho^\Lambda\equiv\vcenter{\hbox{\begin{ytableau}u^1&s^3\\d^2\end{ytableau}}}_{F}$.
The orbital--spin part is therefore antisymmetric under $P_{12}$.
For $(J,S)=(0,1/2)$ and $(2,3/2)$, the basis is
\begin{align}
\Psi_\Lambda^{(0,1/2)}\in\left\{\left[\mathscr O_\lambda^{J=0}\chi_\rho\right]_{\mathcal J=1/2}^{P=+},\ \left[\mathscr O_\rho^{J=0}\chi_\lambda\right]_{\mathcal J=1/2}^{P=+}\right\}F_\rho^\Lambda, \qquad 
\Psi_\Lambda^{(2,3/2)}=\left[\mathscr O_\rho^{J=2}\chi_{[3]}\right]_{\mathcal J=1/2}^{P=+}F_\rho^\Lambda.\label{eq:A_Lambda_basis_J2}
\end{align}
The same formulas apply to $\Lambda_Q$ by replacing $s^3$ with $Q^3$.

\paragraph{$\Sigma^*$-type ($I=1,\;\mathcal J^P=3/2^+$) baryons:}
The light-pair flavor state is again $F_\lambda^\Sigma$.
For the even-$J$ sectors $(J,S)=(0,3/2)$, $(2,1/2)$, and $(2,3/2)$, the basis is
\begin{align}
&\Psi_{\Sigma^*}^{(0,3/2)}=\left[\mathscr O_\lambda^{J=0}\chi_{[3]}\right]_{\mathcal J=3/2}^{P=+}F_\lambda^\Sigma, \qquad
\Psi_{\Sigma^*}^{(2,3/2)}=\left[\mathscr O_\lambda^{J=2}\chi_{[3]}\right]_{\mathcal J=3/2}^{P=+}F_\lambda^\Sigma.\label{eq:A_Sigmastar_basis_J2S32}, \nonumber\\
&\Psi_{\Sigma^*}^{(2,1/2)}\in\left\{\left[\mathscr O_\lambda^{J=2}\chi_\lambda\right]_{\mathcal J=3/2}^{P=+},\ \left[\mathscr O_\rho^{J=2}\chi_\rho\right]_{\mathcal J=3/2}^{P=+}\right\}F_\lambda^\Sigma.
\end{align}
The same formulas apply to $\Sigma_Q^{*}$ by replacing $s^3$ with $Q^3$.
A doubly heavy baryon is obtained by choosing the two identical heavy quarks as particles 1 and 2.

\paragraph{Block structure:}

For a central Hamiltonian, each $(J,P,S)$ sector is diagonalized independently.
Radial excitations are the higher generalized eigenstates in the same $(J,P,S)$ block, while different $J$ values define different orbital branches.
Tensor and spin--orbit interactions can mix different $(J,S)$ channels with the same physical $\mathcal J^P$.

\vspace{10pt}

\vspace{10pt}

\section{Matrix element calculations}\label{mat_calc}

The key methodological framework of the Gaussian expansion method (GEM) is summarized in the standard manual~\cite{hiyamaGEM}. In this appendix, we derive the explicit working formulae obtained by applying GEM to our quark model Hamiltonian, Eqs.~\eqref{hamiltonian}--\eqref{2body_2}, and discuss the computational considerations that are essential for stable and efficient numerical implementation. We then provide a supplementary perspective on the infinitesimally shifted Gaussian (ISG) method by recasting its full construction in the language of coherent states and the Bargmann-Fock representation. Finally, we derive the matrix elements of the Yukawa-type three-body spatial operator $L^{3Q}_{\text{Y}}$.

\subsection{Matrix elements with the Gaussian expansion method (GEM)}\label{app:GEM_ref}

Given the quantum numbers defining the basis functions, the Hamiltonian matrix elements are computed as follows.
The subscripts $\alpha$ and $\beta$ in $n_\alpha,\;N_\beta$, etc., denote the bra and ket states, respectively; likewise $\alpha$ and $\beta$ in $\mathbf r_\alpha,\;\mathbf R_\beta$, etc., may correspond to any of the coordinates $a,\;b$, or $c$.
$\mathcal C$ is the SU(2) Clebsch--Gordan (CG) coefficient, and $C$ and $\mathbf D$ are defined in Eqs.~\eqref{C_coeff} and~\eqref{D_vector}.
The Hamiltonian matrix element is then calculated as
\begin{align}\label{H_cal1}
&\bra{[[[\phi_{n_\alpha l_\alpha}(\mathbf r_\alpha)\phi_{N_\alpha L_\alpha}(\mathbf R_\alpha)]_{J_\alpha}\otimes(\text{Spin})_\alpha]_{\mathcal JM}\otimes(\text{Color})_\text{asym}}H\ket{[[[\phi_{n_\beta l_\beta}(\mathbf r_\beta)\phi_{N_\beta L_\beta}(\mathbf R_\beta)]_{J_\beta}\otimes(\text{Spin})_\beta]_{\mathcal JM}\otimes(\text{Color})_\text{asym}}\nonumber\\
&=N_{n_\alpha l_\alpha}N_{N_\alpha L_\alpha}N_{n_\beta l_\beta}N_{N_\beta L_\beta}
\sum_{M_{J_\alpha}M_{S_\alpha}}\mathcal C^{J_\alpha S_\alpha;\mathcal J}_{M_{J_\alpha}M_{S_\alpha};M}
\sum_{M_{J_\beta}M_{S_\beta}}\mathcal C^{J_\beta S_\beta;\mathcal J}_{M_{J_\beta}M_{S_\beta};M}
\sum_{M_{l_\alpha}M_{L_\alpha}}\mathcal C^{l_\alpha L_\alpha;J_\alpha}_{M_{l_\alpha}M_{L_\alpha};M_{J_\alpha}}
\sum_{M_{l_\beta}M_{L_\beta}}\mathcal C^{l_\beta L_\beta;J_\beta}_{M_{l_\beta}M_{L_\beta};M_{J_\beta}}\nonumber\\
&\;\times
\sum_{h_1}C_{l_\alpha M_{l_\alpha},h_1}
\sum_{h_2}C_{L_\alpha M_{L_\alpha},h_2}
\sum_{h_3}C_{l_\beta M_{l_\beta},h_3}
\sum_{h_4}C_{L_\beta M_{L_\beta},h_4}\nonumber\\
&\;\times
\lim_{\epsilon_1\to0}(\nu_{n_\alpha}\epsilon_1)^{-l_\alpha}
\lim_{\epsilon_2\to0}(\nu_{N_\alpha}\epsilon_2)^{-L_\alpha}
\lim_{\epsilon_3\to0}(\nu_{n_\beta}\epsilon_3)^{-l_\beta}
\lim_{\epsilon_4\to0}(\nu_{N_\beta}\epsilon_4)^{-L_\beta}
F_{n_\alpha,N_\alpha,n_\beta,N_\beta,h_1,h_2,h_3,h_4}^{l_\alpha,L_\alpha,l_\beta,L_\beta,S_\alpha,S_\beta}(\epsilon_1,\epsilon_2,\epsilon_3,\epsilon_4;H),
\end{align}
where $N_{nl}$ is the normalization factor and $F$ is the generating function
\begin{align}
&F_{n_\alpha,N_\alpha,n_\beta,N_\beta,h_1,h_2,h_3,h_4}^{l_\alpha,L_\alpha,l_\beta,L_\beta,S_\alpha,S_\beta}(\epsilon_1,\epsilon_2,\epsilon_3,\epsilon_4;H)
=(\text{Color part})\nonumber\\
&\;\times
\bra{e^{-\nu_{n_\alpha}(\mathbf r_\alpha-\epsilon_1\mathbf D_{l_\alpha M_{l_\alpha},h_1}^{*})^2-\nu_{N_\alpha}(\mathbf R_\alpha-\epsilon_2\mathbf D_{L_\alpha M_{L_\alpha},h_2}^{*})^2},S_\alpha}
H
\ket{e^{-\nu_{n_\beta}(\mathbf r_\beta-\epsilon_3\mathbf D_{l_\beta M_{l_\beta},h_3})^2-\nu_{N_\beta}(\mathbf R_\beta-\epsilon_4\mathbf D_{L_\beta M_{L_\beta},h_4})^2},S_\beta}.
\nonumber
\end{align}
This expression is exactly equivalent to Eq.~(162) of Ref.~\cite{hiyamaGEM}, with the color-spin degrees of freedom explicitly expanded.
Neglecting higher orders of the $\epsilon_i$, the last line of Eq.~\eqref{H_cal1} corresponds to the coefficient of $\epsilon_1^{l_\alpha}\epsilon_2^{L_\alpha}\epsilon_3^{l_\beta}\epsilon_4^{L_\beta}$ in the series expansion of the generating function $F$.
Terms of lower order than $\epsilon_1^{l_\alpha}\epsilon_2^{L_\alpha}\epsilon_3^{l_\beta}\epsilon_4^{L_\beta}$ all vanish when summed in $\sum_kC_{lm,k}(\cdots)$.
The most critical numerical issue in the above calculation is round-off error.
Integrating the bra-ket term within $F$ analytically and then performing numerical differentiation with respect to the $\epsilon_i$, or directly evaluating the numerical limit, tends to induce significant round-off errors.
For integer $l\geq1$, the same quantity can be written in three ways:
\begin{align}
(\mathbf{a} \cdot \mathbf{r})^l e^{-\nu r^2} \;=\;\lim_{\epsilon \to 0} \left( \frac{l}{4\epsilon\nu} \right)^l \left[ e^{-\frac{\nu}{l}(\mathbf{r}-\epsilon\mathbf{a})^2} - e^{-\frac{\nu}{l}(\mathbf{r}+\epsilon\mathbf{a})^2} \right]^l\;=\;\lim_{\epsilon \to 0} \left( \frac{l}{4\epsilon\nu} \right)^l \sum_{s=0}^l (-1)^{l-s}\binom{l}{s} e^{-\nu r^2+2\epsilon\nu\mathbf{a}\cdot\mathbf{r}({2s-l})/{l}}.
\end{align}
For $l=0$, use the unshifted Gaussian $e^{-\nu r^2}$.
Although the three expressions are mathematically equivalent, the third leads to significant round-off errors in numerical evaluation. In the $\sum_k C_{lm,k}\cdots$ loops, a binomial-type expansion appears, as in the third expression above. Moreover, this naive summation approach does not reduce computational complexity and consequently requires an enormous amount of computing time. In multi-loop calculations, for example, if the computational complexity is $\mathcal{O}(N_1N_2)$, it is preferable to separate it into $\mathcal{O}(N_1)+\mathcal{O}(N_2)$. The simplest approach is to factor out the dependence on $l$ and $L$ from $F_{\cdots}^{\cdots}(\epsilon_i)$ (i.e.,~by extracting coefficients of the $\epsilon$'s) so that it can be modularized with the $\sum_{k}C_{lm,k}(\cdots)$ loops and the $\sum_{M}\mathcal{C}_{\cdots}^{\cdots}(\cdots)$ loops. In other words, the loop structure in Eq.~\eqref{H_cal1} can be reorganized as follows (compare to Eq. (174) of Ref.~\cite{hiyamaGEM}):
\begin{align}
\bra{\text{left}\cdots}H\ket{\text{right}\cdots } &= \cdots \sum_{M(l,L)}\mathcal{C}\cdots\sum_{k(l,L)}C \cdots \frac{\partial^{\cdots} F_{l,L}(\epsilon, H)}{\partial \epsilon^{\cdots}}\nonumber\\
&=(\text{continuous parameter dependent}) \Big[\sum\cdots(\text{only discrete dependence})\Big]_\text{precomputed}. \nonumber
\end{align}
Terms depending on continuous parameters (such as the Gaussian parameter $\nu$) are computed outside the loops, thereby converting multiplicative computational complexity into additive complexity and dramatically reducing the overall computation time. This can be simply justified in the Bargmann-Fock representation, where $F_{\cdots }^{\cdots}(\epsilon;H)$ is interpreted as a coherent-state matrix element: projection onto harmonic oscillator (HO) basis states (via coefficient extraction in $\epsilon$) commutes with the finite sums over magnetic quantum numbers and ISG indices, so that all dependence on $(l,L,\nu)$ is collected into a single continuous factor multiplying purely discrete combinatorial coefficients. 

For notational convenience, we introduce the following arrays for $x=a,\,b,\,c$.
The transformation from the $x$ channel to the $\mathcal C$ channel is $U_{\mathcal C x}$, with $\mathbf X_{\mathcal C}=U_{\mathcal C x}\mathbf X_x$ and inverse $U_{x\mathcal C}=U_{\mathcal C x}^{-1}$.
\begin{align}
\mathbf X_{\alpha}=(\mathbf r_{\alpha},\mathbf R_{\alpha})^T,\qquad \mathbf X_{\beta}=(\mathbf r_{\beta},\mathbf R_{\beta})^T,\qquad \mathsf G_{\alpha}=\operatorname{diag}(\nu_{n_{\alpha}},\nu_{N_{\alpha}}),\qquad \mathsf G_{\beta}=\operatorname{diag}(\nu_{n_{\beta}},\nu_{N_{\beta}}),
\end{align}
where the subscripts $\alpha$ and $\beta$ denote the bra and ket basis states, respectively.
The shift-vector arrays are
\begin{align}\label{D_vecs_shorthand}
\mathbf D_1=\mathbf D_{l_{\alpha}M_{l_{\alpha}},h_1}^{*},\qquad
\mathbf D_2=\mathbf D_{L_{\alpha}M_{L_{\alpha}},h_2}^{*},\qquad
\mathbf D_3=\mathbf D_{l_{\beta}M_{l_{\beta}},h_3},\qquad
\mathbf D_4=\mathbf D_{L_{\beta}M_{L_{\beta}},h_4}.
\end{align}
We define the source arrays associated with the bra and ket states by
\begin{align}
\mathbf D_{\alpha}[\epsilon_1,\epsilon_2]=(\epsilon_1\mathbf D_1,\epsilon_2\mathbf D_2)^T,\qquad \mathbf D_{\beta}[\epsilon_3,\epsilon_4]=(\epsilon_3\mathbf D_3,\epsilon_4\mathbf D_4)^T.
\end{align}
Their coordinate-transformed arrays are given by
\begin{align}
\mathbf b_x=U_{\alpha x}^{T}\mathsf G_{\alpha}\mathbf D_{\alpha}+U_{\beta x}^{T}\mathsf G_{\beta}\mathbf D_{\beta},\qquad \mathsf B_x=U_{\alpha x}^{T}\mathsf G_{\alpha}U_{\alpha x}+U_{\beta x}^{T}\mathsf G_{\beta}U_{\beta x}.
\end{align}
For any matrix $Y_{\mathcal C}$ and vector array $\mathbf b_{\mathcal C}$ expressed in coordinates $\mathcal C$, we define $Y_{\mathcal C\to\mathcal C'}=U_{\mathcal C\mathcal C'}^{T}Y_{\mathcal C}U_{\mathcal C\mathcal C'}$, $\mathbf b_{\mathcal C\to\mathcal C'}=U_{\mathcal C\mathcal C'}^{T}\mathbf b_{\mathcal C}$.
With these definitions, we can explicitly compute $F(H)$.

\paragraph{From Racah to Wick algebra:}

For Eq.~\eqref{H_cal1}, it is convenient to reinterpret the ISG shifts as source variables of a multi-mode Bargmann-Fock space.
Let $\mathbf q=(q_1,q_2,q_3,q_4)=(l_\alpha,L_\alpha,l_\beta,L_\beta)$, $\boldsymbol\epsilon=(\epsilon_1,\epsilon_2,\epsilon_3,\epsilon_4)$, and define the coefficient-extraction operator
\begin{equation}
\mathrm{Co}[\boldsymbol\epsilon^{\mathbf q}]\,f(\boldsymbol\epsilon)\equiv\left.\prod_{a=1}^{4}\frac{1}{q_a!}\frac{\partial^{q_a}}{\partial\epsilon_a^{q_a}}f(\epsilon_1,\epsilon_2,\epsilon_3,\epsilon_4)\right|_{\epsilon_1=\cdots=\epsilon_4=0}.
\end{equation}
Then Eq.~\eqref{H_cal1} states that, up to the normalization factors and the finite CG and ISG sums written there explicitly, the orbital part of the matrix element is obtained from the coefficient of $\epsilon_1^{l_\alpha}\epsilon_2^{L_\alpha}\epsilon_3^{l_\beta}\epsilon_4^{L_\beta}$ in the generating function:
\begin{equation}
\bra{\alpha}H\ket{\beta}_{\rm orb}\sim\mathrm{Co}[\boldsymbol\epsilon^{\mathbf q}]\,F(\boldsymbol\epsilon;H).
\end{equation}
Since coefficient extraction is a linear operation, it commutes with all finite sums over magnetic quantum numbers and ISG indices.
This is the algebraic origin of the factorization used below: the dependence on continuous Gaussian parameters is collected, while the dependence on orbital quantum numbers is kept in purely discrete recoupling factors.
The Bargmann-Fock meaning of this operation is immediate.
Under the oscillator/coherent-state correspondence, a shifted Gaussian is the coordinate-space image of an unnormalized coherent state,
\begin{equation}
e^{-\nu(\mathbf x-\epsilon\mathbf D)^2}\longleftrightarrow e^{-\nu\epsilon^2\mathbf D^2}e^{\epsilon\mathbf D\cdot\mathbf a^\dagger}\ket{0}.
\end{equation}
Thus, the variables $\epsilon_i$ play the role of coherent-state labels, and extracting the coefficient of $\epsilon_i^{q_i}$ amounts to projecting onto the fixed oscillator degree required by the orbital quantum numbers.
Self-pairings are present algebraically through the factors $e^{-\nu\epsilon_i^2\mathbf D_i^2}$ and should therefore be kept in intermediate steps.
What matters is that, after the ISG projection, they do not modify the first nonvanishing orbital component.

For the operators treated in this Appendix, the analytically integrated generating function is an analytic function of the quadratic source invariants.  Before projection one may therefore expand it in the form
\begin{equation}
F(\boldsymbol\epsilon;H)
=
\sum_{\{s_a\}\ge 0}
\sum_{\{m_{ab}\}_{a<b}\ge 0}
\mathcal K^{(H)}_{\{s_a\},\{m_{ab}\}}
(\nu,U,\text{spin,color})
\prod_{a=1}^{4}
(\epsilon_a^2\,\mathbf D_a^2)^{s_a}
\prod_{1\le a<b\le 4}
(\epsilon_a\epsilon_b\,\mathbf D_a\cdot\mathbf D_b)^{m_{ab}}.
\end{equation}
Here $s_a$ counts self-pairings associated with the $a$th source variable $\epsilon_a\mathbf D_a$, while $m_{ab}$ counts pairings between the two distinct source variables $\epsilon_a\mathbf D_a$ and $\epsilon_b\mathbf D_b$.
After the ISG projection at fixed orbital rank, all surviving contributions can be reorganized into an effective cross-pairing expansion,
\begin{equation}
\mathrm{Co}[\boldsymbol\epsilon^{\mathbf q}]\,F(\boldsymbol\epsilon;H)\Big|_{\rm ISG\;proj.} = \sum_{\{m_{ab}\}_{a<b}} \widetilde{\mathcal K}^{(H)}_{\{m_{ab}\}}(\nu,U,\text{spin,color})\left[\prod_{a<b}(\mathbf D_a\cdot\mathbf D_b)^{m_{ab}}\right]\prod_{a=1}^{4}\delta_{\,q_a,\;\sum_{b\neq a}m_{ab}}. \label{eq:general_pairing_constraint}
\end{equation}
Eq.~\eqref{eq:general_pairing_constraint} is the precise discrete content of the Wick-type contraction algebra in the present problem.  The continuous dependence on Gaussian ranges, Jacobi transformations, and radial integrals is isolated in the coefficients $\widetilde{\mathcal K}^{(H)}$, whereas the orbital quantum numbers enter only through exact Kronecker constraints on the possible pairing numbers.  In particular, the explicit closed forms derived below for $F(K)$ and $F(V_{ij})$ are of precisely this type: the generating function is analytic in the quadratic source invariants, and the orbital matrix element is reduced, after coefficient extraction, to a finite sum over all possible (admissible) integer pairings.

This general structure becomes especially straightforward for the overlap, $H=1$.  Before restoring the full four-source problem, it is useful to isolate the simplest one-mode example, $G(\eta,\xi) \equiv \bra{0}\,e^{\eta\,\mathbf u\cdot\mathbf a}\,
e^{\xi\,\mathbf v\cdot\mathbf a^\dagger}\,\ket{0}$, where $\mathbf a$ and $\mathbf a^\dagger$ are the HO annihilation and creation operators of a given Jacobi mode.  Since $[\eta\,\mathbf u\cdot\mathbf a, \xi\,\mathbf v\cdot\mathbf a^\dagger]=\eta\,\xi\,\mathbf u\cdot\mathbf v$ is a $c$-number, the Baker-Campbell-Hausdorff identity gives $e^{\eta\,\mathbf u\cdot\mathbf a} e^{\xi\,\mathbf v\cdot\mathbf a^\dagger} = e^{\xi\,\mathbf v\cdot\mathbf a^\dagger} e^{\eta\,\mathbf u\cdot\mathbf a} e^{\eta\xi\,\mathbf u\cdot\mathbf v}$.  Using $\bra{0}\,e^{\xi\,\mathbf v\cdot\mathbf a^\dagger}=\bra{0}$ and $e^{\eta\,\mathbf u\cdot\mathbf a}\ket{0}=\ket{0}$, one obtains $G(\eta,\xi)=e^{\eta\xi\,\mathbf u\cdot\mathbf v}$. Coefficient extraction immediately yields $\mathrm{Co}[\eta^p\xi^q]\,G(\eta,\xi) = \delta_{pq}\, {(\mathbf u\cdot\mathbf v)^p}/{p!}$, or equivalently $\bra{0} (\mathbf u\cdot\mathbf a)^p (\mathbf v\cdot\mathbf a^\dagger)^q \ket{0} = \delta_{pq}\,p!\,(\mathbf u\cdot\mathbf v)^p$.
Thus the most elementary Bargmann-Fock selection rule is simply occupation-number matching: a nonvanishing matrix element requires the numbers of annihilation and creation operators to agree.  The explicit formulae derived next for $F(K)$ and $F(V_{ij})$ provide concrete realizations of Eq.~\eqref{eq:general_pairing_constraint} for the kinetic and two-body interaction terms used in the present quark model Hamiltonian in Eqs.~\eqref{2body_1} and~\eqref{2body_2}.

\paragraph{Closed form of two-body generating functions:}

The coherent-state viewpoint above can now be applied to obtain closed forms for the generating functions associated with two-body operators.  In practice, the Bargmann--Fock picture justifies computing once and for all the continuous Gaussian factors that depend on the HO ranges and Jacobi transformations, while keeping the dependence on orbital and spin quantum numbers in external Clebsch--Gordan and ISG summations.  The resulting factorization reduces the numerical complexity of the multi-loop structure in Eq.~\eqref{H_cal1} to a product of a universal continuous part and purely discrete combinatorial coefficients. 

First, consider the kinetic $F(K)$. After eliminating the CM motion, the three-body kinetic operator is given by
\begin{equation}\label{Kin_mat_def}
K = -\frac{\hbar^2}{2}(\nabla_{\mathbf{r}},\nabla_{\mathbf{R}})M^{-1}\binom{\nabla_{\mathbf{r}}}{\nabla_{\mathbf{R}}}
\end{equation}
where $M^{-1}$ contains the inverse reduced masses. For example, in our definition of Jacobi coordinates,
\begin{equation}\label{mass_mat_def}
M^{-1}=\text{diag}\left[\frac{m_1+m_2}{2m_1 m_2}, \frac{2(m_1+m_2+m_3)}{3(m_1+m_2)m_3} \right].
\end{equation}
The integral $F(K)$ in Eq.~\eqref{H_cal1} is then calculated as follows (neglecting higher order terms than $\mathcal{O}(\epsilon_i^{l_i}\cdots)$)
\begin{align}
{F(\{\epsilon\};K)}&= {2\hbar^2}\braket{S_{\alpha}|S_{\beta}}\left(\frac{\pi^2}{\det \mathsf B_{\alpha}}\right)^{3/2} \exp\bigg[ \sum_{i<j} (\mathsf W_K^{(1)})_{ij}\epsilon_i\epsilon_j\mathbf{D}_i \cdot \mathbf{D}_j \bigg] \nonumber
\\&\qquad \qquad \qquad \qquad \times\bigg[\frac{3}{2}  \text{Tr}\left(\mathsf B_{\alpha}^{-1} \mathsf G_{\alpha} M^{-1} \mathsf G_{\beta\to\alpha}\right) + \Big(\sum_{i<j} (\mathsf W_K^{(2)})_{ij}\epsilon_i\epsilon_j\mathbf{D}_i \cdot \mathbf{D}_j\Big)\bigg] 
\end{align}
where $\mathsf W_K^{(1)}$ and $\mathsf W_K^{(2)}$ are {4~\ensuremath{\times}~4} symmetric matrices
\begin{align}
\mathsf W_K^{(1)} &= 2\begin{bmatrix} \mathsf G_{\alpha} & 0 \\ 0 & \mathsf G_{\beta}U_{\beta\alpha}\end{bmatrix}\begin{bmatrix} \mathsf B_{\alpha}^{-1} & \mathsf B_{\alpha}^{-1} \\ \mathsf B_{\alpha}^{-1} & \mathsf B_{\alpha}^{-1}\end{bmatrix}\begin{bmatrix} \mathsf G_{\alpha} & 0 \\ 0 & U_{\beta\alpha}^{T}\mathsf G_{\beta}\end{bmatrix}, \nonumber\\
\mathsf W_K^{(2)} &=2\begin{bmatrix} \mathsf G_{\alpha} & 0 \\ 0 & \mathsf G_{\beta}U_{\beta\alpha}\end{bmatrix}
\begin{bmatrix}
-\mathsf B_{\alpha}^{-1}\mathsf G_{\beta\to\alpha}M^{-1}\mathsf G_{\beta\to\alpha}\mathsf B_{\alpha}^{-1} & \mathsf B_{\alpha}^{-1}\mathsf G_{\beta\to\alpha}M^{-1}\mathsf G_{\alpha}\mathsf B_{\alpha}^{-1} \\
\mathsf B_{\alpha}^{-1}\mathsf G_{\alpha}M^{-1}\mathsf G_{\beta\to\alpha}\mathsf B_{\alpha}^{-1} & -\mathsf B_{\alpha}^{-1}\mathsf G_{\alpha}M^{-1}\mathsf G_{\alpha}\mathsf B_{\alpha}^{-1}
\end{bmatrix}
\begin{bmatrix} \mathsf G_{\alpha} & 0 \\ 0 & U_{\beta\alpha}^{T}\mathsf G_{\beta}\end{bmatrix}.
\end{align}

Next, we compute the potential generating function $F(V_{ij})$; the case $V_{ij}=1$ trivially reduces to the overlap generating function. 
As before, we choose complex shift vectors $\mathbf D_i$ and set $\mathbf r_{ij}=\sqrt{2}\,\mathbf r_x$, transforming the overlapping $\alpha$ and $\beta$ coordinates into $x$.  
After the Gaussian integration over $\mathbf R_x$ one finds
\begin{align}
F_{n_{\alpha}, N_{\alpha}, n_{\beta}, N_{\beta}, h_1, h_2, h_3, h_4}^{l_{\alpha}, L_{\alpha}, l_{\beta}, L_{\beta}, S_{\alpha}, S_{\beta}} (\epsilon_1, \epsilon_2, \epsilon_3, \epsilon_4; V_{ij}(r_x)) &= \left(-\frac{2}{3}\right)_{\text{color}}\left(-\frac{3}{4}\right)
  \left(\frac{\pi}{\mathsf B_{x,22}}\right)^{3/2}
  \exp\left(\frac{\mathbf b_{x,2}^{\,2}}{\mathsf B_{x,22}}\right)\\
&\!\!\!\!\!\!\!\!\!\!\!\!\!\!\!\!\!\times
  \int d^3\mathbf r_x\;
  \bra{S_{\alpha}}V_{ij}(r_x)\ket{S_{\beta}}\,
  \exp\left[-\left(\frac{\det \mathsf B_x}{\mathsf B_{x,22}}\right)r_x^2
  + 2\left( \mathbf b_{x,1}
  - \frac{\mathsf B_{x,12}}{\mathsf B_{x,22}}\mathbf b_{x,2} \right)\cdot \mathbf r_x \right]. \nonumber
\end{align}
For notational convenience we introduced
\begin{equation}
  y_x \equiv \frac{\det \mathsf B_x}{\mathsf B_{x,22}},\qquad
  \mathbf p \equiv \mathbf b_{x,1} - \frac{\mathsf B_{x,12}}{\mathsf B_{x,22}}\mathbf b_{x,2},\qquad
  p^2 \equiv \mathbf p\cdot\mathbf p,
\end{equation}
so that the remaining angular integral has the closed form
\begin{equation}\label{eq:I_general}
  I[V] \equiv
  \int_{\mathbb R^3} d^3\mathbf r\;
  V(r)\,e^{-y_x r^2 + 2\mathbf p\cdot\mathbf r}= 4\pi \int_0^\infty dr\; r^2 V(r)\,e^{-y_x r^2}
 \frac{\sinh(2rp)}{2rp},
  \qquad p=\sqrt{\mathbf p\cdot\mathbf p},
\end{equation}
which is valid for any radial $V(r)$ for which the integral converges and for any $y_x$ with $\Re(y_x)>0$.  By analytic continuation this closed form, derived for real $\mathbf p$, extends to complex $\mathbf p\in\mathbb C^3$ with the same definition $p^2=\mathbf p\cdot\mathbf p$. We then only require the special cases
\begin{align}
&\int d^3\mathbf r\; r\,e^{-y_x r^2 + 2\mathbf p\cdot\mathbf r}
  = \left(\frac{\pi}{y_x}\right)^{3/2} e^{p^2/y_x}
 \left(\frac{1}{2}+\frac{p^2}{y_x}\right)
 \frac{1}{p}\,\mathrm{erf}\left(\frac{p}{\sqrt{y_x}}\right)
 + \frac{\pi}{y_x^2},
 \label{eq:I_r}\\
&\int d^3\mathbf r\; \frac{1}{r}\,e^{-y_x r^2 + 2\mathbf p\cdot\mathbf r}
  = \left(\frac{\pi^3}{y_x}\right)^{1/2} e^{p^2/y_x}
 \frac{1}{p}\,\mathrm{erf}\left(\frac{p}{\sqrt{y_x}}\right),
 \label{eq:I_inv_r}
\end{align}
and the Gaussian-smeared $V^{CS}$ case
\begin{equation}
V(r)\propto\frac{e^{-a r^2}}{r}
\quad\Rightarrow\quad
I[V] \;\text{is obtained from \eqref{eq:I_general} by the replacement}\; y_x\to y_x+a.
\label{eq:I_smeared}
\end{equation}
In our application to the Cornell-type central potential and the smeared contact interaction, $p^2$ is expressed as a quadratic form in the shift vectors,
\begin{equation}\label{eq:ext_ang}
  p^2
  = \sum_{m<n}(\mathsf W_V^{(2,x)})_{mn}\,\epsilon_m\epsilon_n\,
  \mathbf D_m\cdot\mathbf D_n,
\end{equation}
and Eqs.~\eqref{eq:I_r}--\eqref{eq:I_smeared} then yield directly the closed expressions for $F(V_{ij}^C(r_x))$ and $F(V_{ij}^{CS}(r_x))$ quoted below.
\begin{align}
&F_{n_{\alpha}, N_{\alpha}, n_{\beta}, N_{\beta}, h_1, h_2, h_3, h_4}^{l_{\alpha}, L_{\alpha}, l_{\beta}, L_{\beta}, S_{\alpha}, S_{\beta}} (\epsilon_1, \epsilon_2, \epsilon_3, \epsilon_4; V_{ij}^C (r_x)) = \left(-\frac{2}{3}\right)_\text{Color} \left(-\frac{3}{4}\right) \left(\frac{\pi}{\mathsf B_{x,22}}\right)^{3/2} \braket{S_{\alpha}|S_{\beta}} \\
&\times\Biggl[\frac{\sqrt{2}\pi\sigma}{y_x^2}\exp\bigg(\sum_{m<n}(\mathsf W_V^{(1,x)})_{mn}\epsilon_m \epsilon_n \mathbf{D}_m \cdot \mathbf{D}_n \bigg)+\exp\bigg\{\sum_{m<n} \bigg((\mathsf W_V^{(1,x)})_{mn}+\frac{(\mathsf W_V^{(2,x)})_{mn}}{y_x}\bigg)\epsilon_m \epsilon_n\mathbf{D}_m\cdot\mathbf{D}_n\bigg\}\nonumber\\
&\qquad\times\biggl[ (-D)\left(\frac{\pi}{y_x}\right)^{3/2}   + \sum_{u=0}^{\infty}\frac{(-1)^u}{(2u+1) u! y_x^u}\bigg( \sum_{m<n} (\mathsf W_V^{(2,x)})_{mn} \epsilon_m\epsilon_n \mathbf{D}_m \cdot \mathbf{D}_n \bigg)^u \nonumber\\
&\qquad\qquad\qquad\qquad\qquad\qquad\qquad\qquad\times \bigg\{\frac{\sqrt{2}\pi\sigma}{y_x^2}-\frac{\sqrt{2}\pi\kappa}{y_x}+\frac{2\sqrt{2}\pi\sigma}{y_x^3}\bigg( \sum_{m<n} (\mathsf W_V^{(2,x)})_{mn} \epsilon_m\epsilon_n \mathbf{D}_m \cdot \mathbf{D}_n \bigg)\bigg\}\biggr]\Biggr]. \nonumber\\
&F_{n_{\alpha}, N_{\alpha}, n_{\beta}, N_{\beta}, h_1, h_2, h_3, h_4}^{l_{\alpha}, L_{\alpha}, l_{\beta}, L_{\beta}, S_{\alpha}, S_{\beta}} (\epsilon_1, \epsilon_2, \epsilon_3, \epsilon_4; V_{ij}^{CS} (r_x))=\left(-\frac{2}{3}\right)_\text{Color} \left(-\frac{3}{4}\right) \left(\frac{\pi}{\mathsf B_{x,22}}\right)^{3/2} \\
&\quad \times \bra{S_{\alpha}} \boldsymbol{\sigma}_i \cdot \boldsymbol{\sigma}_j \ket{S_{\beta}}  \frac{\sqrt{2}(\hbar c) \mu_{ij}\mu_{ij}'\pi}{m_i m_j c^4(y_x+2(\mu_{ij}/\hbar c)^2)}\;\exp\bigg\{\sum_{m<n} \bigg((\mathsf W_V^{(1,x)})_{mn}+\frac{(\mathsf W_V^{(2,x)})_{mn}}{y_x+2(\mu_{ij}/\hbar c)^2}\bigg)\epsilon_m \epsilon_n\mathbf{D}_m\cdot\mathbf{D}_n\bigg\} \nonumber\\
&\quad\times \sum_{u=0}^{\infty}\frac{(-1)^u}{(2u+1) u! (y_x+2(\mu_{ij}/\hbar c)^2)^u}\bigg( \sum_{m<n} (\mathsf W_V^{(2,x)})_{mn} \epsilon_m\epsilon_n \mathbf{D}_m \cdot \mathbf{D}_n \bigg)^u. \nonumber
\end{align}
Here $\mathsf W_V^{(1,x)}$ and $\mathsf W_V^{(2,x)}$ are {4~\ensuremath{\times}~4} symmetric matrices
\begin{align}
\mathsf W_V^{(1,x)} &= \frac{2}{\mathsf B_{x,22}}\begin{bmatrix} \mathsf G_{\alpha}U_{\alpha x} & 0 \\ 0 & \mathsf G_{\beta}U_{\beta x}\end{bmatrix}\begin{bmatrix} \mathsf E_1 & \mathsf E_1\\ \mathsf E_1 & \mathsf E_1\end{bmatrix}\begin{bmatrix} U_{\alpha x}^{T}\mathsf G_{\alpha} & 0 \\ 0 & U_{\beta x}^{T}\mathsf G_{\beta}\end{bmatrix}, \qquad \mathsf E_1=\begin{bmatrix}0&0\\0&1\end{bmatrix},\nonumber\\
\mathsf W_V^{(2,x)} &= 2\begin{bmatrix} \mathsf G_{\alpha}U_{\alpha x} & 0 \\ 0 & \mathsf G_{\beta}U_{\beta x}\end{bmatrix}\begin{bmatrix} \mathsf E_2 & \mathsf E_2\\ \mathsf E_2 & \mathsf E_2\end{bmatrix}\begin{bmatrix} U_{\alpha x}^{T}\mathsf G_{\alpha} & 0 \\ 0 & U_{\beta x}^{T}\mathsf G_{\beta}\end{bmatrix}, \qquad \mathsf E_2=\begin{bmatrix}1&-\mathsf B_{x,12}/\mathsf B_{x,22}\\-\mathsf B_{x,12}/\mathsf B_{x,22}&(\mathsf B_{x,12}/\mathsf B_{x,22})^2\end{bmatrix}.
\end{align}
Once these generating functions are known in closed form, the Wick-type contraction algebra can be obtained directly by an algorithm that extracts the coefficient of $\epsilon_1^{l_{\alpha}}\epsilon_2^{L_{\alpha}}\epsilon_3^{l_{\beta}}\epsilon_4^{L_{\beta}}$ from $F({\epsilon_i};H)$.

\subsection{Three-body term in GEM}\label{GEM_3body_calc}

For the Gaussian spatial profile, the required expectation values and matrix elements can be easily obtained in closed form. 
In this appendix we therefore focus on the non-Gaussian three-body interaction $L^{3Q}_{\rm Y}=L^{3Q} f_{\rm Y}$ and describe an efficient procedure to evaluate its matrix elements within GEM.

In a set of test calculations, we found that the Yukawa-type exponential profile $f_{\rm Y}$ provides a systematically better phenomenological fit than the Gaussian profile $f_{\rm G}$ in the three-body sector.
The two profiles have different short-distance and asymptotic behaviors: the exponential profile has a nonzero slope at the origin and a longer tail, whereas the Gaussian profile has a vanishing slope at the origin and decreases faster at large distance.
The interaction used in the calculation has no independent range parameter and is written as $f_{\rm Y}=\exp[-(x_{12}+x_{23}+x_{31})]$, where $x_{ij}=m_{ij}cr_{ij}/\hbar$.
For one pair distance, the elementary identity
\begin{equation}
e^{-r/r_0}=\frac{1}{2\sqrt{\pi}\,r_0}\int_0^\infty dt\;t^{-3/2}\exp\left(-\frac{1}{4r_0^2t}\right)\exp(-tr^2)
\label{eq:exp_gaussian_mixture}
\end{equation}
shows that the exponential profile is an exact continuous Gaussian scale mixture.
In the present three-body profile, this identity is applied pair by pair with $r_0=\hbar/(m_{ij}c)$.
This representation is useful in GEM because the non-Gaussian profile is reduced to reusable Gaussian building blocks after the Schwinger parameters are introduced.

We exclude genuinely singular Yukawa-like potentials, such as $\exp[-(x_{12}+x_{23}+x_{31})]/(r_{12}r_{23}r_{31})$.
For regular Gaussian basis states, their first-order matrix elements are locally integrable, but the triple-coincidence singularity makes higher-order iterations and basis convergence sensitive to the ultraviolet treatment.
A consistent use of such a potential form would therefore require an explicit short-distance regulator and a corresponding refit of its coupling.
Finally, since the profile $f_{\rm Y}=\exp[-(x_{12}+x_{23}+x_{31})]$ does not in general reduce to a simple closed form in a Gaussian basis and direct six-dimensional numerical integration in GEM is computationally heavy, we present below an efficient auxiliary-parameter and coefficient-extraction procedure.
Since the three-body operator factorizes into color-spin and spatial parts, $L_{\rm Y}^{3Q}=L^{3Q}f_{\rm Y}$, the corresponding matrix element can be written as
\begin{align}
&F_{n_{\alpha}, N_{\alpha}, n_{\beta}, N_{\beta}, h_1, h_2, h_3, h_4}^{l_{\alpha}, L_{\alpha}, l_{\beta}, L_{\beta}, S_{\alpha}, S_{\beta}} \left[\epsilon_1, \epsilon_2, \epsilon_3, \epsilon_4; L^{3Q}_{\text{Y}}(r_{12},r_{23},r_{31})\right]=\bra{S_{\alpha},C_{\alpha}}L^{3Q}\ket{S_{\beta},C_{\beta}} \\ &\quad\quad\quad\quad\quad\quad\quad\quad\quad\quad\quad\quad\quad\quad\quad\quad\times \int d^3\mathbf{r}_{\alpha} d^3 \mathbf{R}_{\alpha} \exp\bigg[-\frac{c\sqrt{2}}{\hbar}(m_{12}r_a+m_{23}r_b+m_{31}r_c) -\mathbf{X}_{\alpha}^T \mathsf B_{\alpha} \mathbf{X}_{\alpha}+2\mathbf{b}_{\alpha}^T\mathbf{X}_{\alpha}\bigg]. \nonumber
\end{align}
Here, for example, $r_b$ satisfies $r_b=\sqrt{([U_{ba}\mathbf{X}_a]_1)^2} =\sqrt{(U_{ba})_{11}^2\mathbf{r}_a^2 +2(U_{ba})_{11}(U_{ba})_{12}\mathbf{r}_a\cdot\mathbf{R}_a +(U_{ba})_{12}^2\mathbf{R}_a^2}$. For compact notation in this appendix, we define 
\begin{equation} 
\mathfrak m_a\equiv m_{12}, \qquad \mathfrak m_b\equiv m_{23}, \qquad \mathfrak m_c\equiv m_{31}, \qquad \rho_x\equiv\frac{c^2\mathfrak m_x^2}{2\hbar^2}, \quad x=a,b,c. \label{eq:yukawa_pair_channel_masses} 
\end{equation}
Dummy indices such as $i$ and $j$ may subsequently run over the same set $\{a,b,c\}$. Using the Schwinger parametrization, \begin{equation} e^{-(\sqrt{2}c\mathfrak m_b/\hbar)r_b} = \frac{c\,\mathfrak m_b}{\hbar\sqrt{2\pi}} \int_0^{\infty}dt_b\,t_b^{-3/2} \exp\left( -t_b\mathbf{X}_{\alpha}^T\mathcal{A}_b\mathbf{X}_{\alpha} -\frac{\rho_b}{t_b} \right), \qquad \mathcal{A}_b= \begin{bmatrix} (U_{b\alpha})_{11}^2 & (U_{b\alpha})_{11}(U_{b\alpha})_{12} \\ (U_{b\alpha})_{11}(U_{b\alpha})_{12} & (U_{b\alpha})_{12}^2 \end{bmatrix},
\end{equation}
the integral in $F(L_{\rm Y}^{3Q})$ reads 
\begin{align} &\int d^3\mathbf r_{\alpha}\,d^3\mathbf R_{\alpha}\, \exp\bigg[ -\frac{c\sqrt{2}}{\hbar} \left( m_{12}r_a+m_{23}r_b+m_{31}r_c \right) -\mathbf X_{\alpha}^{T}\mathsf B_{\alpha}\mathbf X_{\alpha} +2\mathbf b_{\alpha}^{T}\mathbf X_{\alpha} \bigg] \nonumber\\ &= \left(\frac{c}{\hbar\sqrt{2\pi}}\right)^3 \prod_{x=a,b,c} \left[ \mathfrak m_x \int_0^\infty dt_x\, t_x^{-3/2} e^{-\rho_x/t_x} \right] \int d^3\mathbf r_{\alpha}\,d^3\mathbf R_{\alpha}\, \exp\bigg[ -\mathbf X_{\alpha}^{T} \bigg( \mathsf B_{\alpha} + \sum_{y=a,b,c}t_y\mathcal A_y \bigg) \mathbf X_{\alpha} + 2\mathbf b_{\alpha}^{T}\mathbf X_{\alpha} \bigg] \nonumber\\ &= \left(\frac{c}{\hbar\sqrt{2\pi}}\right)^3 \prod_{x=a,b,c} \left[ \mathfrak m_x \int_0^\infty dt_x\, t_x^{-3/2} e^{-\rho_x/t_x} \right] \left( \frac{\pi^2}{\det\mathsf B_{\rm Y}(t)} \right)^{3/2} \exp\left[ \mathbf b_{\alpha}^{T} \mathsf B_{\rm Y}(t)^{-1} \mathbf b_{\alpha} \right], 
\end{align} 
where $\mathsf B_{\rm Y}(t) = \mathsf B_{\alpha} + t_a\mathcal A_a + t_b\mathcal A_b + t_c\mathcal A_c$.
Since $\mathcal{A}_x$ ($x=a,b,c$) is a rank-one positive semi-definite pair-coordinate matrix, $F(L_{\rm Y}^{3Q})$ can be calculated as
\begin{align} &F_{n_{\alpha},N_{\alpha},n_{\beta},N_{\beta},h_1,h_2,h_3,h_4} ^{l_{\alpha},L_{\alpha},l_{\beta},L_{\beta},S_{\alpha},S_{\beta}} \left[ \epsilon_1,\epsilon_2,\epsilon_3,\epsilon_4; L_{\rm Y}^{3Q} \right] = \bra{S_{\alpha},C_{\alpha}} L^{3Q} \ket{S_{\beta},C_{\beta}} \left(\frac{c}{\hbar\sqrt{2\pi}}\right)^3 \nonumber\\ &\qquad\times \prod_{x=a,b,c} \left[ \mathfrak m_x \int_0^\infty dt_x\, t_x^{-3/2} e^{-\rho_x/t_x} \right] \left( \frac{\pi^2}{\det\mathsf B_{\rm Y}(t)} \right)^{3/2} \exp\bigg[ \sum_{i<j} \bigl(\mathsf W_{\rm Y}(t)\bigr)_{ij} \epsilon_i\epsilon_j\, \mathbf D_i\cdot\mathbf D_j \bigg]. \end{align}
where $\mathsf W_{\rm Y}(t)$ is the symmetric {4~\ensuremath{\times}~4} matrix
\begin{equation}
\mathsf W_{\rm Y}(t)=2\begin{bmatrix}
\mathsf G_{\alpha} & 0 \\ 0 & \mathsf G_{\beta}U_{\beta\alpha}
\end{bmatrix}\begin{bmatrix}
\mathsf B_{\rm Y}^{-1}(t) & \mathsf B_{\rm Y}^{-1}(t) \\ \mathsf B_{\rm Y}^{-1}(t) & \mathsf B_{\rm Y}^{-1}(t)
\end{bmatrix}\begin{bmatrix}
\mathsf G_{\alpha} & 0 \\ 0 & U_{\beta\alpha}^{T}\mathsf G_{\beta}
\end{bmatrix}.
\end{equation}
The coefficient of $\epsilon_1^{l_{\alpha}}\epsilon_2^{L_{\alpha}}\epsilon_3^{l_{\beta}}\epsilon_4^{L_{\beta}}$ in $F(L^{3Q}_{\rm Y})$ is
\begin{align}
&\mathrm{Co}[\epsilon_1^{l_{\alpha}}\epsilon_2^{L_{\alpha}}\epsilon_3^{l_{\beta}}\epsilon_4^{L_{\beta}}]F_{n_{\alpha},N_{\alpha},n_{\beta},N_{\beta},h_1,h_2,h_3,h_4}^{l_{\alpha},L_{\alpha},l_{\beta},L_{\beta},S_{\alpha},S_{\beta}}(\{\epsilon_i\};L^{3Q}_{\rm Y})
=\bra{S_{\alpha},C_{\alpha}}L^{3Q}\ket{S_{\beta},C_{\beta}}\left(\frac{c}{\hbar\sqrt{2\pi}}\right)^3\\
&\times\sum_{\substack{n_{12}+n_{13}+n_{14}=l_{\alpha}\\n_{12}+n_{23}+n_{24}=L_{\alpha}\\n_{13}+n_{23}+n_{34}=l_{\beta}\\n_{14}+n_{24}+n_{34}=L_{\beta}}}
\underbrace{ \left\{ \prod_{x=a,b,c} \left[ \mathfrak m_x \int_0^\infty dt_x\, t_x^{-3/2} e^{-\rho_x/t_x} \right] \right\} \left( \frac{\pi^2}{\det\mathsf B_{\rm Y}(t)} \right)^{3/2} \prod_{p<q} \left[ \bigl(\mathsf W_{\rm Y}(t)\bigr)_{pq} \right]^{n_{pq}} }_{\text{continuous parameters}} \times \underbrace{ \prod_{p<q} \frac{ (\mathbf D_p\cdot\mathbf D_q)^{n_{pq}} }{ n_{pq}! } }_{\text{discrete parameters}}.
\nonumber
\end{align}
The continuous $t$-integrals can be evaluated outside the outer Clebsch--Gordan or ISG recoupling (discrete parameters) loops, but in general they remain $n$-pattern dependent through $\mathsf W_{\rm Y}(t)$. To break the curse of dimensionality in the final $d$-dimensional (here, $d=3$) integral, one can replace the full tensor-product rule with a sparse grid~\cite{sparse_grids}, reducing the node count from $\mathcal{O}(N^d)$ to approximately $\mathcal{O}(N(\ln N)^{d-1})$. Pairing patterns with $n_{pq}=2$ first occur in quadratic source contractions required by rank-two orbital factors, including the $D$-wave sector, and they lead to a more complicated form of $[(\mathsf W_{\rm Y})_{pq}]^2$.
We first present a dimensional reduction for $n_{pq}=1$ and then extend it to $n_{pq}=2$; the case $n_{pq}=0$ follows by omitting the corresponding insertion.

For numerical stability, one can map the semi-infinite domain $s\in[0,\infty)$ to the unit interval with $s=x/(1-x)$, $x\in[0,1)$, i.e.,~
\begin{equation}
\int_0^\infty f(s)\,ds=\int_0^1f\Big(\frac{x}{1-x}\Big)\frac{dx}{(1-x)^2}.
\end{equation}

\paragraph{$S$- and $P$-wave contributions:} Our desired integral takes the form (temporarily generalizing $\mathsf S,\mathsf T\in \mathbb{R}^{2\times2}$)
\begin{align} I_{pq} = \prod_{x=a,b,c} \int_0^\infty dt_x\, t_x^{-3/2} e^{-\rho_x/t_x} \left( \frac{\pi^2}{\det\mathsf B_{\rm Y}(t)} \right)^{3/2} \left( \mathsf S\mathsf B_{\rm Y}^{-1}(t)\mathsf T \right)_{pq}. \end{align}
Since the integral does not admit a closed-form expression, we reduce its computational cost by separating a commuting part from the noncommuting remainder.
The matrix $\mathsf B_{\rm Y}(t)$ is written as
\begin{equation}\label{eq:M_perturbatively}
\mathsf B_{\rm Y}(t)=\mathsf B_{\rm Y}^{(0)}(t)+\tau_{\rm p}\mathsf B_{\rm Y}^{(1)}(t),\qquad \mathsf B_{\rm Y}^{(0)}(t)=\mathsf B_{\alpha}^{(0)}+\sum_{j=a,b,c}\mathcal A_j^{(0)}t_j,\qquad \mathsf B_{\rm Y}^{(1)}(t)=\left[\mathsf B_{\alpha}-\mathsf B_{\alpha}^{(0)}\right]+\sum_{j=a,b,c}\left[\mathcal A_j-\mathcal A_j^{(0)}\right]t_j.
\end{equation}
The orthonormal mode vectors $\mathbf v_\pm$ are chosen as a common basis that minimizes the total squared off-diagonal elements of $\mathsf B_\alpha$ and $\mathcal A_{a,b,c}$.
The commuting matrices are their diagonal projections,
\begin{equation}
\mathsf B_{\alpha}^{(0)}=\sum_{s=\pm}\mathbf v_s(\mathbf v_s^T\mathsf B_{\alpha}\mathbf v_s)\mathbf v_s^T,\qquad \mathcal A_j^{(0)}=\sum_{s=\pm}\mathbf v_s(\mathbf v_s^T\mathcal A_j\mathbf v_s)\mathbf v_s^T.
\end{equation}
The remainder $\mathsf B_{\rm Y}^{(1)}$ therefore contains the noncommuting off-diagonal parts in this common basis, and $\tau_{\rm p}$ is set to unity after the expansion.
We separate the determinant and inverse of $\mathsf B_{\rm Y}^{(0)}$ into the two modes,
\begin{equation}
\det\mathsf B_{\rm Y}^{(0)}(t)=\Lambda_+(t)\Lambda_-(t),\qquad [\mathsf B_{\rm Y}^{(0)}(t)]^{-1}=\frac{1}{\Lambda_+(t)}\mathbf v_+\mathbf v_+^T+\frac{1}{\Lambda_-(t)}\mathbf v_-\mathbf v_-^T.
\end{equation}
The inverse satisfies the resolvent identity
\begin{equation}
\mathsf B_{\rm Y}^{-1}=(\mathsf B_{\rm Y}^{(0)})^{-1}-(\mathsf B_{\rm Y}^{(0)})^{-1}\tau_{\rm p}\mathsf B_{\rm Y}^{(1)}\mathsf B_{\rm Y}^{-1}.
\end{equation}
Together with $\partial_{\tau_{\rm p}}\log\det\mathsf B_{\rm Y}=\mathrm{Tr}(\mathsf B_{\rm Y}^{-1}\partial_{\tau_{\rm p}}\mathsf B_{\rm Y})$, this gives
\begin{align}
\det \mathsf B_{\rm Y}&=\det \mathsf B_{\rm Y}^{(0)}\left[1+\tau_{\rm p}\,\mathrm{Tr}\left((\mathsf B_{\rm Y}^{(0)})^{-1}\mathsf B_{\rm Y}^{(1)}\right)+\frac{\tau_{\rm p}^2}{2}\left(\mathrm{Tr}\left((\mathsf B_{\rm Y}^{(0)})^{-1}\mathsf B_{\rm Y}^{(1)}\right)\right)^2-\frac{\tau_{\rm p}^2}{2}\mathrm{Tr}\left(\left((\mathsf B_{\rm Y}^{(0)})^{-1}\mathsf B_{\rm Y}^{(1)}\right)^2\right)+\mathcal O(\tau_{\rm p}^3)\right],\nonumber\\
\mathsf B_{\rm Y}^{-1}&=(\mathsf B_{\rm Y}^{(0)})^{-1}-\tau_{\rm p}(\mathsf B_{\rm Y}^{(0)})^{-1}\mathsf B_{\rm Y}^{(1)}(\mathsf B_{\rm Y}^{(0)})^{-1}+\tau_{\rm p}^2(\mathsf B_{\rm Y}^{(0)})^{-1}\mathsf B_{\rm Y}^{(1)}(\mathsf B_{\rm Y}^{(0)})^{-1}\mathsf B_{\rm Y}^{(1)}(\mathsf B_{\rm Y}^{(0)})^{-1}+\mathcal O(\tau_{\rm p}^3).
\label{order_expansion}
\end{align}
The expansion is used only when the largest absolute eigenvalue of $(\mathsf B_{\rm Y}^{(0)})^{-1/2}\mathsf B_{\rm Y}^{(1)}(\mathsf B_{\rm Y}^{(0)})^{-1/2}$ remains below unity over the numerically relevant Schwinger-parameter domain.
Its truncation error is checked against the full three-dimensional integration for representative basis pairs, and higher orders are included when the required accuracy is not reached.
The integral $I$ decomposes into a perturbative sum over $s=\pm$ modes ($\rho_x\equiv{c^2\mathfrak m_x^2}/{2\hbar^2}$):
\begin{align} 
&I^{(0)} = \pi^3 \sum_{s=\pm} (\mathsf S\mathbf v_s) (\mathbf v_s^T\mathsf T) I_s^{(0)}, \qquad I_s^{(0)} = \prod_{x=a,b,c} \int_0^\infty dt_x\, t_x^{-3/2} e^{-\rho_x/t_x} \Lambda_s(t)^{-5/2} \Lambda_{-s}(t)^{-3/2}, \nonumber\\ &I^{(1)} = \pi^3 \sum_{s=\pm} \left[ (\mathsf S\mathbf v_s) (\mathbf v_s^T\mathsf T) I_{ss}^{(1)} + (\mathsf S\mathbf v_s) (\mathbf v_{-s}^T\mathsf T) I_{s,-s}^{(1)} \right], \nonumber\\ &I_{s,-s}^{(1)} = - \prod_{x=a,b,c} \int_0^\infty dt_x\, t_x^{-3/2} e^{-\rho_x/t_x} \bigl[\mathsf B_{\rm Y}^{(1)}(t)\bigr]_{s,-s} \Lambda_s(t)^{-5/2} \Lambda_{-s}(t)^{-5/2}, \nonumber\\ &I_{ss}^{(1)} = - \prod_{x=a,b,c} \int_0^\infty dt_x\, t_x^{-3/2} e^{-\rho_x/t_x} \Bigg[ \frac{5}{2} \bigl[\mathsf B_{\rm Y}^{(1)}(t)\bigr]_{ss} \Lambda_s(t)^{-7/2} \Lambda_{-s}(t)^{-3/2} + \frac{3}{2} \bigl[\mathsf B_{\rm Y}^{(1)}(t)\bigr]_{-s,-s} \Lambda_s(t)^{-5/2} \Lambda_{-s}(t)^{-5/2} \Bigg].
\end{align}
Here we denote the eigenvalues of the matrices $\mathsf B_{\alpha}^{(0)}$, $\mathcal{A}_a^{(0)}$, $\mathcal{A}_b^{(0)}$, and $\mathcal{A}_c^{(0)}$ by $a^{(0)}_s$, $\lambda_{s,a}$ ($=1$ when the Jacobi channel of the $\alpha$ basis state is $a$), $\lambda_{s,b}$, and $\lambda_{s,c}$. Mode-$s$ components of $\mathsf B_{\rm Y}^{(1)}(t)$ are parametrized as $[\mathsf B_{\rm Y}^{(1)}(t)]_{ss}=\mathbf v_s^T\mathsf B_{\rm Y}(t)\mathbf v_s-\Lambda_s(t)=\delta a_s+\sum_j\delta\lambda_{s,j}t_j$ and $[\mathsf B_{\rm Y}^{(1)}(t)]_{s,-s} = \delta a_{s,-s} + \sum_j \delta\lambda_{s,-s,j} t_j$ (from now, we suppress the bookkeeping parameter, $\tau_{\rm p}\to1$). If we define a four-vector $t_\mu=(1,t_a,t_b,t_c)$ and a coefficient vector $\theta^\mu$, then $\theta^\mu=(\delta a_s,\delta\lambda_{s,a},\delta\lambda_{s,b},\delta\lambda_{s,c})$ for the $[\mathsf B_{\rm Y}^{(1)}(t)]_{ss}$ integration or $\theta^\mu=(\delta a_{s,-s},\delta\lambda_{s,-s,a},\delta\lambda_{s,-s,b},\delta\lambda_{s,-s,c})$ for the $[\mathsf B_{\rm Y}^{(1)}(t)]_{s,-s}$ integration, and either component can be written as $\theta^\mu t_\mu$. Greek indices represent four-vector components of $\mu=0,a,b,c$, whereas Latin indices $i,j, \ldots$ run only over the $a,b,c$ components. Let $\xi$ and $\zeta$ be negative half-integers, and $N_{\xi,\zeta}=-2\xi-2\zeta-1$. For notational convenience, we also define $A(x)=xa_s^{(0)}+(1-x)a_{-s}^{(0)}$, $L_i(x) = x\,\lambda_{s,i}+(1-x)\,\lambda_{-s,i}$, and $\Delta_s(x) = \sum_{i}\sqrt{\rho_i\,L_i(x)}$. Then all terms in $I_s^{(0)}$, $I_{ss}^{(1)}$, and $I_{s,-s}^{(1)}$ are organized with the following integrals ($\bar\rho=(\prod_i\rho_i)^{1/3}$):
\begin{align}
&I(\theta^\mu,\xi,\zeta)=\int_0^{\infty} \bigg(\prod_{i=a,b,c} dt_i \,t_i^{-3/2}\exp[-\rho_i/t_i]\bigg) (\theta^\mu t_\mu ) \Lambda_s(t)^{\xi}\Lambda_{-s}(t)^\zeta \label{formula1}\\
&=\frac{-(\partial/\partial w)|_{w=0}}{\Gamma(-\xi)\Gamma(-\zeta)}\int_0^\infty du\,\frac{e^{-ua_s^{(0)}}}{u^{\xi+1}}\int_0^{\infty}dv\,\frac{e^{-va_{-s}^{(0)}}}{v^{\zeta+1}}\int_0^\infty \bigg(\prod_{i} dt_i\,t_i^{-3/2}e^{-\rho_i/t_i-(u\lambda_{s,i}+v\lambda_{-s,i})t_i}\bigg)e^{-w\theta^\mu t_\mu}\label{formula2} \\
&=\left(\frac{\pi}{\bar\rho}\right)^{3/2}\frac{(-1)(\partial/\partial w)|_{w=0}}{\Gamma(-\xi)\Gamma(-\zeta)}e^{-w\theta^0}\int_0^\infty du\,u^{-\xi-1}e^{-ua_s^{(0)}}\int_0^{\infty}dv\,v^{-\zeta-1}e^{-va_{-s}^{(0)}}e^{-2\sum_i\sqrt{\rho_i(u\lambda_{s,i}+v\lambda_{-s,i}+w\theta^i)}} \label{formula3}\\
&=\left(\frac{\pi}{\bar\rho}\right)^{3/2}  \int_0^1 dx\,\frac{x^{-\xi -1}(1-x)^{-\zeta -1}}{\Gamma(-\xi)\Gamma(-\zeta)} \bigg[ \theta^0\,e^{\Delta_s(x)^2/A(x)} \sum_{k=0}^{N_{\xi,\zeta}} \binom{N_{\xi,\zeta}}{k} \frac{[-\Delta_s(x)]^{N_{\xi,\zeta}-k}} {A(x)^{-2\xi-2\zeta-(k+1)/2}} \Gamma\!\left(\frac{k+1}{2},\frac{\Delta_s(x)^2}{A(x)}\right)\nonumber \\  &\qquad +\sum_{i=a,b,c}\frac{\theta^i\sqrt{\rho_i}}{\sqrt{L_i(x)}}\,e^{\Delta_s(x)^2/A(x)}\sum_{k=0}^{N_{\xi,\zeta}-1} \binom{N_{\xi,\zeta}-1}{k} \frac{[-\Delta_s(x)]^{N_{\xi,\zeta}-1-k}} {A(x)^{-2\xi-2\zeta-(k+3)/2}} \Gamma\!\left(\frac{k+1}{2},\frac{\Delta_s(x)^2}{A(x)}\right) \bigg], \label{formula4}
\end{align}
From Eq.~\eqref{formula1} to Eq.~\eqref{formula2}, we apply the Schwinger parametrization, e.g.,~
\begin{align}
\Lambda_s(t)^{\xi}=\frac{1}{\Gamma(-\xi)}\int_0^{\infty} du\;u^{-\xi-1}\exp\left(-u\left[a_s^{(0)}+t_a\lambda_{s,a}+t_b\lambda_{s,b}+t_c\lambda_{s,c}\right]\right).
\end{align}
From Eq.~\eqref{formula2} to Eq.~\eqref{formula3}, we use the following results (3.471.9 of Ref.~\cite{Gradshteyn2007}):
\begin{equation}
\int_0^{\infty}x^{\nu-1}e^{-\beta/x-\gamma x}dx = 2\left(\frac{\beta}{\gamma}\right)^{\nu/2}K_\nu\left(2\sqrt{\beta\gamma}\right).
\end{equation}
From Eq.~\eqref{formula3} to Eq.~\eqref{formula4}, we change the variables as $u=qx$ and $v=q(1-x)$, then use the following polynomial $\times$ Gaussian integral:
\begin{align}
\int_0^\infty dq\,q^ne^{-aq-2b\sqrt{q}}=e^{b^2/a}\sum_{k=0}^{2n+1}\binom{2n+1}{k}\frac{(-b)^{2n+1-k}}{a^{2n+1+(1-k)/2}}\Gamma\!\left(\frac{k+1}{2},\frac{b^2}{a}\right),
\end{align}
where $n=0,1,2,\ldots$ and $\Gamma(s,z)$ is the upper incomplete gamma available in standard libraries. It can be easily proved with the substitution $q=p^2$. As a result, for example, one can compute $I^{(0)}_{s}$, $I^{(1)}_{s,-s}$, and $I^{(1)}_{ss}$ by calling
\begin{align}
I_s^{(0)}&=I\left((1,0,0,0),-\frac52,-\frac32\right), 
\qquad\qquad I^{(1)}_{s,-s}=-I\left((\delta a_{s,-s},\{\delta\lambda_{s,-s,j}\}),-\frac52,-\frac52\right)
\nonumber\\
I^{(1)}_{ss}&= -\frac{5}{2} I\left((\delta a_s,\{\delta\lambda_{s,j}\}),-\frac72,-\frac32\right)-\frac{3}{2}I\left((\delta a_{-s},\{\delta\lambda_{-s,j}\}),-\frac52,-\frac52\right). \nonumber
\end{align}
The first-order truncation is checked against the full three-dimensional integration for representative matrix elements. Accuracy can be further improved by including terms of order $\mathcal{O}(\tau_{\rm p}^2)$ or higher with an additional coefficient vector $\varphi^\nu$, i.e.,~
\begin{align}
&I(\theta^\mu,\varphi^\nu,\xi,\zeta)=\int_0^{\infty} \bigg(\prod_{i=a,b,c} dt_i \,t_i^{-3/2}e^{-\rho_i/t_i}\bigg)(\theta^\mu t_\mu )(\varphi^\nu t_\nu)\Lambda_s(t)^{\xi}\Lambda_{-s}(t)^\zeta \\
&=\left(\frac{\pi}{\bar\rho}\right)^{3/2}
\int_0^1 dx\,\frac{x^{-\xi-1}(1-x)^{-\zeta-1}}{\Gamma(-\xi)\Gamma(-\zeta)} \Bigg[
\theta^0\varphi^0\,e^{\Delta_s(x)^2/A(x)}
\sum_{k=0}^{N_{\xi,\zeta}}
\binom{N_{\xi,\zeta}}{k}
\frac{[-\Delta_s(x)]^{N_{\xi,\zeta}-k}}
 {A(x)^{-2\xi-2\zeta-(k+1)/2}}
\Gamma\!\left(\frac{k+1}{2},\frac{\Delta_s(x)^2}{A(x)}\right) \nonumber\\
&\quad+\Bigg(
\theta^0\sum_{j=a,b,c}\frac{\varphi^j\sqrt{\rho_j}}{\sqrt{L_j(x)}}
+\varphi^0\sum_{i=a,b,c}\frac{\theta^i\sqrt{\rho_i}}{\sqrt{L_i(x)}}
\Bigg)e^{\Delta_s(x)^2/A(x)}
\sum_{k=0}^{N_{\xi,\zeta}-1}
\binom{N_{\xi,\zeta}-1}{k}
\frac{[-\Delta_s(x)]^{N_{\xi,\zeta}-1-k}}
 {A(x)^{-2\xi-2\zeta-(k+3)/2}}
\Gamma\!\left(\frac{k+1}{2},\frac{\Delta_s(x)^2}{A(x)}\right)\nonumber \\
&\quad +\Bigg(
\sum_{i=a,b,c}\frac{\theta^i\sqrt{\rho_i}}{\sqrt{L_i(x)}}
\Bigg)
\Bigg(
\sum_{j=a,b,c}\frac{\varphi^j\sqrt{\rho_j}}{\sqrt{L_j(x)}}
\Bigg)e^{\Delta_s(x)^2/A(x)}
\sum_{k=0}^{N_{\xi,\zeta}-2}
\binom{N_{\xi,\zeta}-2}{k}
\frac{[-\Delta_s(x)]^{N_{\xi,\zeta}-2-k}}
 {A(x)^{-2\xi-2\zeta-(k+5)/2}}
\Gamma\!\left(\frac{k+1}{2},\frac{\Delta_s(x)^2}{A(x)}\right) \nonumber\\
&\quad + \Bigg(
\sum_{i=a,b,c}\frac{\theta^i \varphi^i \sqrt{\rho_i}}{2{L_i(x)}^{3/2}}
\Bigg)
e^{\Delta_s(x)^2/A(x)}
\sum_{k=0}^{N_{\xi,\zeta}-3}
\binom{N_{\xi,\zeta}-3}{k}
\frac{[-\Delta_s(x)]^{N_{\xi,\zeta}-3-k}}
 {A(x)^{-2\xi-2\zeta-(k+7)/2}}
\Gamma\!\left(\frac{k+1}{2},\frac{\Delta_s(x)^2}{A(x)}\right) 
\Bigg].
\end{align}
This two-insertion contribution is not, in general, expressible as a linear combination of the single master integrals $I(\theta^\mu,\xi,\zeta)$, because the same channel double differentiation generates the additional $L_i(x)^{-3/2}$ term. 

Higher partial wave contributions, including the $D$-wave ($n_{pq}=2$) sector, can be treated in a completely analogous manner by extending the auxiliary-parameter expansion to higher powers of the inverse-matrix elements.

\paragraph{Diagonalization and expectation values:}

Finally, we determine the baryon spectrum from the generalized eigenvalue problem $H\boldsymbol{\mathcal{ B}}=E\mathcal O\boldsymbol{\mathcal{ B}}$.
For the $S$- and $P$-wave sectors, the Yukawa-profile matrix elements reduce to one-dimensional integrals built from the master integral $I(\theta^\mu,\xi,\zeta)$.
Near-linearly dependent basis directions are removed by the canonical orthogonalization
\begin{align}
\mathcal O=U_{\mathcal O}\operatorname{diag}(o_1,o_2,\ldots)U_{\mathcal O}^{\dagger},\qquad \frac{o_n}{o_{\max}}>\varepsilon_{\mathcal O},\qquad
X_{\mathcal O}=U_{\mathcal O}^{\rm keep}\operatorname{diag}(o_n^{-1/2}),\qquad X_{\mathcal O}^{\dagger}\mathcal OX_{\mathcal O}=I.
\end{align}
The reduced eigenvalue problem is
\begin{equation}
\left(X_{\mathcal O}^{\dagger}HX_{\mathcal O}\right)\mathbf y=E\mathbf y,\qquad \boldsymbol{\mathcal B}=X_{\mathcal O}\mathbf y,\qquad \mathbf y^\dagger\mathbf y=\boldsymbol{\mathcal B}^{\dagger}\mathcal O\boldsymbol{\mathcal B}=1.
\end{equation}
For any operator $\mathcal Q$, its expectation value is
\begin{equation}
\langle\mathcal Q\rangle=\mathbf y^\dagger X_{\mathcal O}^{\dagger}\mathcal QX_{\mathcal O}\mathbf y.
\end{equation}
The cutoff stability is checked through
\begin{equation}
\Delta E_{\mathcal O}=\left|E(\varepsilon_{\mathcal O})-E(\varepsilon_{\mathcal O}/10)\right|,\qquad \Delta\mathcal Q_{\mathcal O}=\left|\langle\mathcal Q\rangle_{\varepsilon_{\mathcal O}}-\langle\mathcal Q\rangle_{\varepsilon_{\mathcal O}/10}\right|.
\end{equation}
The cutoff is accepted when $\Delta E_{\mathcal O}$ and the relevant $\Delta\mathcal Q_{\mathcal O}$ are below the required numerical accuracy.
Extended precision is used only when residual round-off sensitivity remains after the near-null overlap modes are removed.

\subsection{Analytic calculation without the GEM/ISG construction}
\label{app:analytic_without_gem}
As an independent cross-check of the GEM/ISG calculation described in Sections~\ref{app:GEM_ref} and~\ref{GEM_3body_calc}, we also evaluate the same finite-basis Hamiltonian using a direct Cartesian-polynomial representation of the solid harmonics.
This calculation does not employ the infinitesimal displacement of Eq.~\eqref{eq:isg_expansion}, the ISG coefficient sums, or a numerical angular integration.
The same Gaussian ranges, all three Jacobi rearrangement channels $\mathcal C=a,b,c$, the same spin-flavor permutation projectors, and the same generalized eigenvalue problem are retained.
Thus, the independence of the calculation lies in the matrix-element engine rather than in the variational trial space.
At a fixed basis truncation, the Cartesian and ISG formulations must therefore yield identical Hamiltonian and overlap matrices up to floating-point round-off.

\paragraph{Direct Cartesian representation of the coupled solid harmonics:}
We choose the $a$-channel Jacobi coordinates as a common reference and introduce the two-component array of three-dimensional vectors $\mathbf X\equiv (\mathbf r_a, \mathbf R_a)^T$.
The coordinates in an arbitrary rearrangement channel $\mathcal C$ are related to the reference coordinates by
\begin{equation}\label{eq:analytic_jacobi_transform}
\mathbf X_{\mathcal C}\equiv\begin{pmatrix}\mathbf r_{\mathcal C}\\ \mathbf R_{\mathcal C}\end{pmatrix}=U_{\mathcal C a}\mathbf X,\qquad |\det U_{\mathcal C a}|=1.
\end{equation}
For unequal constituent masses, $U_{\mathcal C a}$ is generally not orthogonal, although its determinant has unit magnitude in the scaled Jacobi convention of Eq.~\eqref{Jacobi_coord}.

For compact notation, define the solid harmonic $\mathcal Y_{lm}(\mathbf z)\equiv z^lY_{lm}(\hat{\mathbf z})=T^{(lm)}_{i_1\cdots i_l}z_{i_1}\cdots z_{i_l}$, where $T^{(lm)}$ is the symmetric traceless tensor introduced in Eq.~\eqref{eq:T_exp}, with the same spherical-harmonic phase convention used throughout this work.
For the basis label $\alpha=\{\mathcal C,n,N,l,L,J\}$, the angular and radial polynomial is written as
\begin{equation}\label{eq:analytic_coupled_polynomial}
\mathscr P_{\alpha}^{JM}(\mathbf X)=N_{nl}N_{NL}\sum_{m,m_L}\mathcal C^{lL;J}_{m\,m_L;M}\mathcal Y_{lm}\!\left([U_{\mathcal C a}\mathbf X]_1\right)\mathcal Y_{Lm_L}\!\left([U_{\mathcal C a}\mathbf X]_2\right).
\end{equation}
The corresponding spatial basis function is
\begin{equation}\label{eq:analytic_cartesian_gaussian_basis}
\Phi_{\alpha}^{JM}(\mathbf X)=\mathscr P_{\alpha}^{JM}(\mathbf X)\exp\left[-\sum_{\mu,\nu=1}^{2}\mathsf G_{\alpha,\mu\nu}\,\mathbf X_\mu\cdot\mathbf X_\nu\right], \qquad \mathsf G_\alpha=U_{\mathcal C a}^{T}\begin{pmatrix}\nu_n&0\\0&\nu_N\end{pmatrix}U_{\mathcal C a}.
\end{equation}
Eq.~\eqref{eq:analytic_cartesian_gaussian_basis} is algebraically identical to the basis in Eq.~\eqref{eq:baryon_gaussian_basis}, but all angular dependence is now represented by a finite Cartesian polynomial.
For $l,L\leq2$, the polynomial degree is at most four, so every matrix element reduces to a finite number of Gaussian moments.
This construction is closely related to the correlated-Gaussian formulation, but here the polynomial is generated directly from the same coupled solid harmonics used in the GEM basis~\cite{Suzuki1998,Varshalovich1988,varga1996}.

\paragraph{Gaussian moment algebra:}
For a bra basis $\alpha$ and ket basis $\beta$, define $\mathsf B_{\alpha\beta}\equiv\mathsf G_\alpha+\mathsf G_\beta$.
The elementary six-dimensional Gaussian integral is
\begin{equation}\label{eq:analytic_gaussian_partition}
\mathcal Z(\mathsf B_{\alpha\beta})\equiv\int d^3\mathbf r_a\,d^3\mathbf R_a\,\exp\left[-\sum_{\mu,\nu=1}^{2}\mathsf B_{\alpha\beta,\mu\nu}\mathbf X_\mu\cdot\mathbf X_\nu\right]=\frac{\pi^3}{(\det\mathsf B_{\alpha\beta})^{3/2}}.
\end{equation}
After dividing by $\mathcal Z$, the Cartesian components of $\mathbf X$ form a zero-mean Gaussian distribution with covariance
\begin{equation}\label{eq:analytic_gaussian_covariance}
\left\langle X_{\mu i}X_{\nu j}\right\rangle_{\mathsf B}=\frac{1}{2}(\mathsf B^{-1})_{\mu\nu}\delta_{ij},
\end{equation}
where $\mu,\nu=1,2$ label the two Jacobi vectors and $i,j=x,y,z$ label their Cartesian components.
All higher moments follow from Wick contraction,
\begin{equation}\label{eq:analytic_wick_contraction}
\left\langle X_{a_1}\cdots X_{a_{2n}}\right\rangle_{\mathsf B}=\sum_{\text{all pairings }\mathfrak p}\prod_{(r,s)\in\mathfrak p}\left\langle X_{a_r}X_{a_s}\right\rangle_{\mathsf B},\qquad \left\langle X_{a_1}\cdots X_{a_{2n+1}}\right\rangle_{\mathsf B}=0.
\end{equation}
For implementation and cross-checks, it is useful to record several low-order contractions explicitly.
We define the unit-polynomial Gaussian master integral
\begin{equation}
\mathscr G_{\alpha\beta}[\mathscr Q]
\equiv
\int d^3\mathbf r_a\,d^3\mathbf R_a\,
\mathscr Q(\mathbf r_a,\mathbf R_a)\,
\exp\left[-\sum_{\mu,\nu=1}^{2}
\mathsf B_{\alpha\beta,\mu\nu}
\mathbf X_\mu\cdot\mathbf X_\nu\right],
\qquad
\mathsf C_{\alpha\beta}
\equiv\frac{1}{2}\mathsf B_{\alpha\beta}^{-1},
\label{eq:analytic_low_order_master}
\end{equation}
and denote the entries of the covariance matrix by
$c_{\mu\nu}^{(\alpha\beta)}\equiv
(\mathsf C_{\alpha\beta})_{\mu\nu}$.
Then
\begin{align}
&\mathscr G_{\alpha\beta}[1]
=
\mathcal Z(\mathsf B_{\alpha\beta}), \qquad \qquad\qquad
\mathscr G_{\alpha\beta}
[\mathbf X_\mu\cdot\mathbf X_\nu]=
3\,\mathcal Z(\mathsf B_{\alpha\beta})
c_{\mu\nu}^{(\alpha\beta)},
\nonumber\\
&\mathscr G_{\alpha\beta}
[(\mathbf X_\mu\cdot\mathbf X_\nu)
 (\mathbf X_\rho\cdot\mathbf X_\sigma)]
=
3\,\mathcal Z(\mathsf B_{\alpha\beta})
\left[
3c_{\mu\nu}^{(\alpha\beta)}
 c_{\rho\sigma}^{(\alpha\beta)}
+c_{\mu\rho}^{(\alpha\beta)}
 c_{\nu\sigma}^{(\alpha\beta)}
+c_{\mu\sigma}^{(\alpha\beta)}
 c_{\nu\rho}^{(\alpha\beta)}
\right].
\label{eq:analytic_low_order_master_explicit}
\end{align}
For the positive-parity $J=0$ orbital block, the Cartesian scalar polynomials underlying the coupled $PP$ and $DD$ components may be chosen as
\begin{equation}
\mathscr Q_{P}
\equiv
\mathbf r_a\cdot\mathbf R_a,
\qquad
\mathscr Q_{D}
\equiv
3(\mathbf r_a\cdot\mathbf R_a)^2-r_a^2R_a^2.
\label{eq:analytic_J0_invariants}
\end{equation}
They are proportional to
$[\mathcal Y_1(\mathbf r_a)\otimes\mathcal Y_1(\mathbf R_a)]_{00}$
and
$[\mathcal Y_2(\mathbf r_a)\otimes\mathcal Y_2(\mathbf R_a)]_{00}$,
respectively.
The corresponding normalization and phase factors remain those contained in Eq.~\eqref{eq:analytic_coupled_polynomial}.
Writing
$c_{\mu\nu}\equiv c_{\mu\nu}^{(\alpha\beta)}$
within the following equation, the required scalar masters are
\begin{align}
&\mathscr G_{\alpha\beta}[\mathscr Q_P]
=
3\,\mathcal Z(\mathsf B_{\alpha\beta})\,c_{12},\qquad 
\mathscr G_{\alpha\beta}[\mathscr Q_P^2]=
3\,\mathcal Z(\mathsf B_{\alpha\beta})
\left(c_{11}c_{22}+4c_{12}^2\right),\qquad 
\mathscr G_{\alpha\beta}[\mathscr Q_D]
=
30\,\mathcal Z(\mathsf B_{\alpha\beta})\,c_{12}^2,
\nonumber\\
&\mathscr G_{\alpha\beta}[\mathscr Q_P\mathscr Q_D]=
30\,\mathcal Z(\mathsf B_{\alpha\beta})\,c_{12}
\left(2c_{11}c_{22}+5c_{12}^2\right),\qquad
\mathscr G_{\alpha\beta}[\mathscr Q_D^2]
=
60\,\mathcal Z(\mathsf B_{\alpha\beta})
\left(
3c_{11}^2c_{22}^2
+22c_{11}c_{22}c_{12}^2
+38c_{12}^4
\right).
\label{eq:analytic_J0_master_moments}
\end{align}
These expressions give the elementary Cartesian matrix-element building blocks for $SS\leftrightarrow PP$, $PP\leftrightarrow PP$, $SS\leftrightarrow DD$, $PP\leftrightarrow DD$, and $DD\leftrightarrow DD$ before multiplication by the basis normalization, Clebsch--Gordan, spin, flavor, and permutation factors.
For basis functions belonging to the same diagonal Jacobi channel, $c_{12}=0$, and the $SS\leftrightarrow PP$ and $SS\leftrightarrow DD$ scalar masters vanish as expected.
They become nonzero for cross-channel exponent matrices or for explicitly correlated Gaussian basis functions.
Consequently, the overlap matrix is obtained directly as
\begin{equation}\label{eq:analytic_overlap}
\mathcal O_{\alpha\beta}=\mathcal Z(\mathsf B_{\alpha\beta})\left\langle\mathscr P_{\alpha}^{JM*}(\mathbf X)\mathscr P_{\beta}^{JM}(\mathbf X)\right\rangle_{\mathsf B_{\alpha\beta}}.
\end{equation}
No limit in an auxiliary displacement parameter is required, and the cancellation among lower-order ISG terms is replaced by exact polynomial algebra.

\paragraph{Internal kinetic matrix element:}

Let $Q_a$ denote the {2~\ensuremath{\times}~3} matrix that defines the reference Jacobi coordinates from the particle coordinates, $(\mathbf r_a, \mathbf R_a)^T=Q_a(\mathbf r_1, \mathbf r_2, \mathbf r_3)^T$.
After subtracting the center-of-mass kinetic energy, the internal kinetic operator can be written as
\begin{equation}\label{eq:analytic_internal_kinetic}
K_{\rm int}=-\frac{\hbar^2}{2}\sum_{\mu,\nu=1}^{2}\Lambda_{\mu\nu}\nabla_{\mathbf X_\mu}\cdot\nabla_{\mathbf X_\nu},\qquad \Lambda=Q_a\,\operatorname{diag}\left(\frac{1}{m_1},\frac{1}{m_2},\frac{1}{m_3}\right)Q_a^T.
\end{equation}
For the ket polynomial, introduce the Gaussian derivative operator
\begin{equation}\label{eq:analytic_gaussian_derivative}
\mathscr D_{\mu i}^{(\beta)}\equiv\frac{\partial}{\partial X_{\mu i}}-2\sum_{\rho=1}^{2}\mathsf G_{\beta,\mu\rho}X_{\rho i}.
\end{equation}
It follows that
\begin{equation}\label{eq:analytic_kinetic_polynomial}
K_{\rm int}\Phi_\beta^{JM}(\mathbf X)=\mathscr K_\beta^{JM}(\mathbf X)\exp\left[-\sum_{\mu,\nu}\mathsf G_{\beta,\mu\nu}\mathbf X_\mu\cdot\mathbf X_\nu\right], \qquad \mathscr K_\beta^{JM}(\mathbf X)=-\frac{\hbar^2}{2}\sum_{\mu,\nu=1}^{2}\Lambda_{\mu\nu}\sum_{i=x,y,z}\mathscr D_{\mu i}^{(\beta)}\mathscr D_{\nu i}^{(\beta)}\mathscr P_\beta^{JM}(\mathbf X).
\end{equation}
Since $\mathscr K_\beta^{JM}$ is again a finite Cartesian polynomial, the kinetic-energy matrix element is
\begin{equation}\label{eq:analytic_kinetic_matrix}
K_{\alpha\beta}=\mathcal Z(\mathsf B_{\alpha\beta})\left\langle\mathscr P_{\alpha}^{JM*}(\mathbf X)\mathscr K_{\beta}^{JM}(\mathbf X)\right\rangle_{\mathsf B_{\alpha\beta}}.
\end{equation}
For the unit-polynomial $S$-wave case, $\mathscr P_\alpha=\mathscr P_\beta=1$, Eq.~\eqref{eq:analytic_kinetic_matrix} reduces to the closed form
\begin{equation}
K_{\alpha\beta}^{(SS,0)}
=
3\hbar^2
\mathcal Z(\mathsf B_{\alpha\beta})
\operatorname{Tr}\left[
\Lambda\,
\mathsf G_\alpha\,
\mathsf B_{\alpha\beta}^{-1}\,
\mathsf G_\beta
\right].
\label{eq:analytic_swave_kinetic_closed}
\end{equation}
The full normalized $SS$ matrix element is obtained by multiplying Eq.~\eqref{eq:analytic_swave_kinetic_closed} by the radial and spherical-harmonic normalization factors already contained in $\mathscr P_\alpha$ and $\mathscr P_\beta$.
The expression is manifestly equivalent under $\alpha\leftrightarrow\beta$ because $\Lambda$, $\mathsf G_\alpha$, $\mathsf G_\beta$, and $\mathsf B_{\alpha\beta}^{-1}$ are real symmetric matrices and the trace is cyclic.

\paragraph{Analytic pair-distance matrix elements:}
Every physical pair coordinate can be expressed in the reference Jacobi coordinates as $\mathbf r_{ij}=w_{ij,1}\mathbf r_a+w_{ij,2}\mathbf R_a\equiv\mathbf w_{ij}^{T}\mathbf X$.
Under the Gaussian measure associated with $\mathsf B_{\alpha\beta}$, the pair vector $\mathbf r_{ij}$ is an isotropic three-dimensional Gaussian variable with component variance $\zeta_{ij}^{(\alpha\beta)}=\mathbf w_{ij}^{T}\mathsf B_{\alpha\beta}^{-1}\mathbf w_{ij}/2$.
The remaining Jacobi variables may be decomposed into a part correlated with $\mathbf r_{ij}$ and an independent conditional fluctuation,
\begin{equation}\label{eq:analytic_conditional_decomposition}
\mathbf X_\mu=c_{\mu}^{(ij)}\mathbf r_{ij}+\boldsymbol{\eta}_\mu,\qquad \mathbf c^{(ij)}=\frac{\mathsf B_{\alpha\beta}^{-1}\mathbf w_{ij}}{\mathbf w_{ij}^{T}\mathsf B_{\alpha\beta}^{-1}\mathbf w_{ij}}.
\end{equation}
The conditional covariance is
\begin{equation}\label{eq:analytic_conditional_covariance}
\left\langle\eta_{\mu i}\eta_{\nu j}\right\rangle_{\rm cond}=\mathsf\Gamma_{\mu\nu}^{(ij)}\delta_{ij},\qquad \mathsf\Gamma^{(ij)}=\frac{1}{2}\left[\mathsf B_{\alpha\beta}^{-1}-\frac{\mathsf B_{\alpha\beta}^{-1}\mathbf w_{ij}\mathbf w_{ij}^{T}\mathsf B_{\alpha\beta}^{-1}}{\mathbf w_{ij}^{T}\mathsf B_{\alpha\beta}^{-1}\mathbf w_{ij}}\right].
\end{equation}
Substituting Eq.~\eqref{eq:analytic_conditional_decomposition} into $\mathscr P_\alpha^{JM*}\mathscr P_\beta^{JM}$ and contracting the conditional fluctuations with Eq.~\eqref{eq:analytic_conditional_covariance} reduces the angularly integrated polynomial to a finite radial expansion,
\begin{equation}\label{eq:analytic_radial_decomposition}
\left\langle\mathscr P_\alpha^{JM*}\mathscr P_\beta^{JM}\right\rangle_{\rm cond,\Omega_{ij}}=\sum_{n=0}^{n_{\max}}\mathcal C_{n,\alpha\beta}^{(ij)}r_{ij}^{2n}.
\end{equation}
The coefficients $\mathcal C_{n,\alpha\beta}^{(ij)}$ contain only finite binomial factors, the entries of $\mathbf c^{(ij)}$, and Wick contractions with $\mathsf\Gamma^{(ij)}$.

For any potential of the form $r_{ij}^{p}\exp(-a r_{ij}^{2})$, the remaining radial moment is analytic:
\begin{equation}\label{eq:analytic_radial_moment}
\mathcal R_q(\zeta,a)\equiv\left\langle r^q e^{-ar^2}\right\rangle_{\zeta}=(2\zeta)^{q/2}\frac{\Gamma\left(({q+3})/{2}\right)}{\Gamma\left({3}/{2}\right)}(1+2a\zeta)^{-(q+3)/2},\qquad q>-3.
\end{equation}
The low-order radial masters most frequently required in the present Hamiltonian are
\begin{align}
&\mathcal R_{-1}(\zeta,a)
=
\sqrt{\frac{2}{\pi\zeta}}\,
(1+2a\zeta)^{-1},\qquad 
\mathcal R_{0}(\zeta,a)
=
(1+2a\zeta)^{-3/2}, \qquad
\mathcal R_{1}(\zeta,a)
=
2\sqrt{\frac{2\zeta}{\pi}}\,
(1+2a\zeta)^{-2},
\nonumber\\
&\mathcal R_{2}(\zeta,a)
=
3\zeta\,
(1+2a\zeta)^{-5/2}, \qquad
\mathcal R_{4}(\zeta,a)
=
15\zeta^2\,
(1+2a\zeta)^{-7/2}.
\label{eq:analytic_radial_explicit}
\end{align}
All higher moments required by the polynomial partial-wave factors may be generated recursively from
\begin{equation}
\mathcal R_{q+2}(\zeta,a)
=
\frac{(q+3)\zeta}{1+2a\zeta}
\mathcal R_q(\zeta,a).
\label{eq:analytic_radial_recurrence}
\end{equation}
For the unit-polynomial $SS$ case, $n_{\rm max}=0$ and $\mathcal C_{0,\alpha\beta}^{(ij)}=1$.
Writing $\zeta\equiv\zeta_{ij}^{(\alpha\beta)}$, the basic pair matrix elements become
\begin{align}
&\left.
\mathcal I_{\alpha\beta}^{(ij)}(-1,0)
\right|_{SS}=
\mathcal Z(\mathsf B_{\alpha\beta})
\sqrt{\frac{2}{\pi\zeta}},
\qquad
\left.
\mathcal I_{\alpha\beta}^{(ij)}(1,0)
\right|_{SS}=
2\mathcal Z(\mathsf B_{\alpha\beta})
\sqrt{\frac{2\zeta}{\pi}},
\nonumber\\
&\left.
\mathcal I_{\alpha\beta}^{(ij)}(0,a)
\right|_{SS}
=
\mathcal Z(\mathsf B_{\alpha\beta})
(1+2a\zeta)^{-3/2},\qquad
\left.
\mathcal I_{\alpha\beta}^{(ij)}
\bigg(-1,\frac{\mu_{ij}^2}{(\hbar c)^2}\bigg)
\right|_{SS}
=
\mathcal Z(\mathsf B_{\alpha\beta})
\sqrt{\frac{2}{\pi\zeta}}
\bigg[
1+\frac{2\mu_{ij}^2\zeta}{(\hbar c)^2}
\bigg]^{-1}.
\label{eq:analytic_swave_pair_kernels}
\end{align}
The first, second, and fourth expressions are the explicit $SS$ masters for the Coulomb, linear-confining, and Gaussian-smeared color-spin interactions, respectively.
The complete pair-distance matrix element is therefore
\begin{equation}\label{eq:analytic_pair_matrix_master}
\mathcal I_{\alpha\beta}^{(ij)}(p,a)=\mathcal Z(\mathsf B_{\alpha\beta})\sum_{n=0}^{n_{\max}}\mathcal C_{n,\alpha\beta}^{(ij)}\mathcal R_{2n+p}\!\left(\zeta_{ij}^{(\alpha\beta)},a\right).
\end{equation}
The three radial structures needed for Eqs.~\eqref{2body_1} and~\eqref{2body_2} follow from
\begin{equation}\label{eq:analytic_required_radial_kernels}
\left\langle\Phi_\alpha\left|{r_{ij}}^{-1}\right|\Phi_\beta\right\rangle=\mathcal I_{\alpha\beta}^{(ij)}(-1,0),\quad \left\langle\Phi_\alpha\left|r_{ij}\right|\Phi_\beta\right\rangle=\mathcal I_{\alpha\beta}^{(ij)}(1,0),\quad \left\langle\Phi_\alpha\left|{e^{-(\mu_{ij}r_{ij}/\hbar c)^2}}/{r_{ij}}\right|\Phi_\beta\right\rangle=\mathcal I_{\alpha\beta}^{(ij)}\big(-1,{\mu_{ij}^{2}}/{(\hbar c)^2}\big).
\end{equation}
Thus, the Coulomb, linear-confining, and Gaussian-smeared color-spin matrix elements are obtained without a numerical radial quadrature or a fitted Gaussian representation of $1/r$.

\paragraph{Overall color-singlet two-body Hamiltonian:}
For a color-singlet baryon, $\langle F_i^cF_j^c\rangle=-2/3$, so that $-(3/4)\langle F_i^cF_j^c\rangle=1/2$.
The orbital part of the central color-type $V^C$ pair interaction and color-spin interaction $V^{CS}$ are therefore
\begin{equation}\label{eq:analytic_central_pair_matrix}
\mathcal V_{\alpha\beta}^{C,ij}=\frac{1}{2}\left[-\kappa\,\mathcal I_{\alpha\beta}^{(ij)}(-1,0)+\sigma\,\mathcal I_{\alpha\beta}^{(ij)}(1,0)-D\,\mathcal O_{\alpha\beta}\right], \qquad \mathcal V_{\alpha\beta}^{CS,ij}=\frac{1}{2}\frac{\mu_{ij}\mu_{ij}'}{m_im_jc^4}(\hbar c)\,\mathcal I_{\alpha\beta}^{(ij)}\bigg(-1,\frac{\mu_{ij}^{2}}{(\hbar c)^2}\bigg).
\end{equation}
Using the pair-spin basis of \ref{bases}, the full matrix in a fixed orbital-$J$, parity, and quark-spin-$S$ block is
\begin{align}\label{eq:analytic_full_hamiltonian_matrix}
H_{\alpha s,\beta s'}={}&\bigg[\bigg(\sum_{i=1}^{3}m_i\bigg)\mathcal O_{\alpha\beta}+K_{\alpha\beta}+\sum_{i<j}\mathcal V_{\alpha\beta}^{C,ij}\bigg]\delta_{ss'}+\sum_{i<j}\mathcal V_{\alpha\beta}^{CS,ij}\left\langle\chi_s\left|\boldsymbol{\sigma}_i\cdot\boldsymbol{\sigma}_j\right|\chi_{s'}\right\rangle.
\end{align}
For $S=1/2$, $s$ and $s'$ run over the pair-spin states $\chi_\rho$ and $\chi_\lambda$, whereas the $S=3/2$ space is one dimensional.
The off-diagonal $\chi_\rho$--$\chi_\lambda$ elements are retained explicitly and generate the pair-spin recoupling discussed in Section~\ref{baryon_spectrum_sec}.

\paragraph{Permutation projection and all Jacobi rearrangements:}

The direct basis is the union of the $a$, $b$, and $c$ rearrangement-channel bases.
All cross-channel matrix elements are evaluated after transforming their exponent matrices and Cartesian polynomials to the common reference coordinates through Eq.~\eqref{eq:analytic_jacobi_transform}.
No separate Racah transformation between the three rearrangement channels is required.
The same baryon permutation operator $\mathcal P_B$ used in Eq.~\eqref{eq:baryon_wavefunction} is then applied to the combined orbital-spin-flavor state.
For two-particle exchange symmetry one may use $\mathcal P_{12}^{(\pm)}=(I\pm P_{12})/2$.
The symmetric, mixed-symmetry, and antisymmetric $S_3$ sectors are selected by $\mathcal P_{[3]}$, $\mathcal P_{[21]}$, and $\mathcal P_{[111]}$, respectively, where
\begin{equation}
\mathcal P_{\Gamma}=\frac{d_{\Gamma}}{6}\sum_{\pi\in S_3}\chi_{\Gamma}^{*}(\pi)P_\pi,\qquad \Gamma\in\{[3],[21],[111]\}.
\end{equation}
Here $d_{\Gamma}$ and $\chi_{\Gamma}$ are the dimension and character of the corresponding irrep, and $P_\pi$ acts simultaneously on the orbital, spin, and flavor labels.
The projector $\mathcal P_{[21]}$ selects the full two-dimensional mixed-symmetry sector, while its $\lambda$ and $\rho$ Young--Yamanouchi components are defined in \ref{bases}.
Equivalently, after the flavor symmetry has been fixed, the corresponding projector can be applied directly to the orbital-spin space.

Because the multi-Jacobi basis is nonorthogonal and overcomplete, the overlap matrix is first canonically orthogonalized.
Writing $\mathcal O=U_{\mathcal O}\operatorname{diag}(o_1,o_2,\ldots)U_{\mathcal O}^{\dagger}$, we retain eigenmodes satisfying $o_n/o_{\rm max}>\varepsilon_{\mathcal O}$ and define
\begin{equation}\label{eq:analytic_canonical_orthogonalization}
X_{\mathcal O}=U_{\mathcal O}^{\rm keep}\operatorname{diag}(o_n^{-1/2}),\qquad X_{\mathcal O}^{\dagger}\mathcal O X_{\mathcal O}=I.
\end{equation}
The symmetry-projected Hamiltonian is subsequently diagonalized in this orthonormal subspace.
The calculation uses the same overlap cutoff as the GEM calculation, and stability is checked by varying $\varepsilon_{\mathcal O}$ over at least one order of magnitude.

\paragraph{Orbital and radial excitation blocks:}
For fixed orbital angular momentum $J$ and parity $P$, all channels satisfying
\begin{equation}\label{eq:analytic_allowed_channels}
\mathfrak A_{JP}^{(l_{\max})}=\left\{(l,L)\,\middle|\,0\leq l,L\leq l_{\max},\ |l-L|\leq J\leq l+L,\ (-1)^{l+L}=P\right\}
\end{equation}
are included simultaneously.
For the positive-parity $J=0$ orbital block and $l_{\rm max}=2$, this gives $\mathfrak A_{0+}^{(2)}=\{(0,0),(1,1),(2,2)\}$, so the $SS$, $PP$, and $DD$ components and all off-diagonal $SS\leftrightarrow PP$, $SS\leftrightarrow DD$, and $PP\leftrightarrow DD$ matrix elements are included in one diagonalization.
The $l_{\rm max}=2$ block displayed here is used as an analytic benchmark subset.
The production ground-state calculations reported in Tables~\ref{tab:meson_baryon_two_body} and~\ref{tab:combined_3bd_baryon} extend the same construction to $l_{\rm max}=3$ and include the $(l_{\mathcal C},L_{\mathcal C})=(3,3)$, or $FF$, channel.

The Hamiltonian of Eqs.~\eqref{hamiltonian}--\eqref{2body_2} contains neither tensor nor spin--orbit terms, and therefore $[H,\mathbf J^2]=[H,\mathbf S^2]=0$.
The calculation can consequently be organized in separate $(J,P,S)$ blocks.
The physical baryon angular momenta obtained by coupling $J$ and $S$ are $\mathcal J=\vert J-S\vert ,\vert J-S\vert +1,\ldots,J+S$.
They are degenerate within the present Hamiltonian.
Resolving the corresponding $\mathcal J^P$ fine structure requires the tensor, symmetric spin--orbit, and antisymmetric spin--orbit interactions discussed in  Section~\ref{VCS_intro}.

\paragraph{Explicit three-body spatial master integrals:}

For the $a$-channel reference coordinates of Eq.~\eqref{Jacobi_coord}, one convenient set of pair-coordinate vectors is
\begin{equation}
\mathbf w_{12}
=
\begin{pmatrix}
\sqrt{2}\\[1mm]0
\end{pmatrix},
\qquad
\mathbf w_{23}
=
\begin{pmatrix}
-\sqrt{2}\,m_1/(m_1+m_2)\\[1mm]
\sqrt{3/2}
\end{pmatrix},
\qquad
\mathbf w_{31}
=
\begin{pmatrix}
\sqrt{2}\,m_2/(m_1+m_2)\\[1mm]
\sqrt{3/2}
\end{pmatrix},
\label{eq:analytic_pair_vectors_a}
\end{equation}
such that
$r_{ij}=\vert \mathbf w_{ij}^{T}\mathbf X\vert $.
The overall sign of an individual $\mathbf w_{ij}$ is immaterial because only the magnitude of the corresponding pair vector enters the spatial profiles.
For the Gaussian three-body profile of Eq.~\eqref{gaussian}, define
\begin{equation}
\mathsf H_{\rm G}
\equiv
\frac{c^2}{8\hbar^2}
\sum_{{\rm cyc}(ij)}
m_{ij}^2\,
\mathbf w_{ij}\mathbf w_{ij}^{T},
\qquad
\mathsf B_{\alpha\beta}^{\rm G}
\equiv
\mathsf B_{\alpha\beta}+\mathsf H_{\rm G}.
\label{eq:analytic_threebody_gaussian_matrix}
\end{equation}
The complete Gaussian-profile spatial matrix element is then
\begin{equation}
\left\langle
\Phi_\alpha^{JM}
\middle|
f_{\rm G}
\middle|
\Phi_\beta^{JM}
\right\rangle
=
\mathcal Z(\mathsf B_{\alpha\beta}^{\rm G})
\left\langle
\mathscr P_\alpha^{JM*}(\mathbf X)
\mathscr P_\beta^{JM}(\mathbf X)
\right\rangle_{\mathsf B_{\alpha\beta}^{\rm G}}.
\label{eq:analytic_threebody_gaussian_master}
\end{equation}
Thus the Gaussian three-body profile requires no new integration formula: it is obtained from the overlap master by the replacement $\mathsf B_{\alpha\beta}\rightarrow \mathsf B_{\alpha\beta}^{\rm G}$.
The spatially contact interaction follows as the special case $\mathsf H_{\rm G}=0$.

For the Yukawa-type exponential profile, an elementary closed form is not available in general.
Nevertheless, the unit-polynomial $S$-wave matrix element admits an exact deterministic three-dimensional representation.
Writing $r=\vert \mathbf r_a\vert $, $R=\vert \mathbf R_a\vert $, and $z=\hat{\mathbf r}_a\cdot\hat{\mathbf R}_a$, define
\begin{equation}
r_{ij}(r,R,z)
=
\left[
w_{ij,1}^{\,2}r^2
+w_{ij,2}^{\,2}R^2
+2w_{ij,1}w_{ij,2}rRz
\right]^{1/2}.
\label{eq:analytic_pair_distance_rRz}
\end{equation}
The corresponding master integral is
\begin{align}
&\mathcal F_{\alpha\beta,{\rm Y}}^{(SS,0)}
\equiv
\int d^3\mathbf r_a\,d^3\mathbf R_a\,
\exp\bigg[
-\sum_{\mu,\nu=1}^{2}
\mathsf B_{\alpha\beta,\mu\nu}
\mathbf X_\mu\cdot\mathbf X_\nu
-\frac{c}{\hbar}
\sum_{{\rm cyc}(ij)}
m_{ij}r_{ij}
\bigg]
\label{eq:analytic_threebody_yukawa_direct}\\
&=
8\pi^2
\int_0^\infty dr\,r^2
\int_0^\infty dR\,R^2
\int_{-1}^{1}dz\,
\exp\bigg[
-\mathsf B_{\alpha\beta,11}r^2
-\mathsf B_{\alpha\beta,22}R^2
-2\mathsf B_{\alpha\beta,12}rRz
-\frac{c}{\hbar}
\sum_{{\rm cyc}(ij)}
m_{ij}r_{ij}(r,R,z)
\bigg].\nonumber
\end{align}
Eq.~\eqref{eq:analytic_threebody_yukawa_direct} provides a direct deterministic benchmark for the Schwinger-parameter representation of \ref{GEM_3body_calc}.
For nonzero Cartesian polynomials, each monomial may be reduced to the same $(r,R,z)$ integral with an additional finite polynomial in $r$, $R$, and $z$.

\paragraph{Cross-check criteria:}
When the Gaussian meshes, angular channels, permutation projectors, spin basis, and overlap cutoff are matched exactly, the direct Cartesian and ISG calculations represent the same finite-dimensional variational problem.
We quantify the matrix-level comparison by
\begin{equation}\label{eq:analytic_matrix_crosscheck}
\delta_{\mathcal O}=\frac{\|\mathcal O_{\rm Cart}-\mathcal O_{\rm ISG}\|_{\rm F}}{\|\mathcal O_{\rm ISG}\|_{\rm F}},\qquad \delta_H=\frac{\|H_{\rm Cart}-H_{\rm ISG}\|_{\rm F}}{\|H_{\rm ISG}\|_{\rm F}},
\end{equation}
and verify Hermiticity through
\begin{equation}\label{eq:analytic_hermiticity_check}
\eta_H=\frac{\|H-H^\dagger\|_{\rm F}}{\|H\|_{\rm F}},\qquad \eta_{\mathcal O}=\frac{\|\mathcal O-\mathcal O^\dagger\|_{\rm F}}{\|\mathcal O\|_{\rm F}}.
\end{equation}
The eigenvalues are additionally checked to be independent of the chosen magnetic projection and of the reference Jacobi channel.
The sequential enlargement of the variational space must satisfy
\begin{equation}\label{eq:analytic_variational_ordering}
E_{SS}\geq E_{SS+PP}\geq E_{SS+PP+DD},
\end{equation}
up to numerical precision.
Agreement at the matrix and eigenvalue levels provides a stronger validation than agreement of a small number of final masses, since it separately tests the Jacobi transformations, solid-harmonic phases, kinetic-energy metric, pair-distance matrix elements, spin recoupling, and permutation projection.

\begin{figure*}[!t]
\centering
\includegraphics[width=\textwidth]{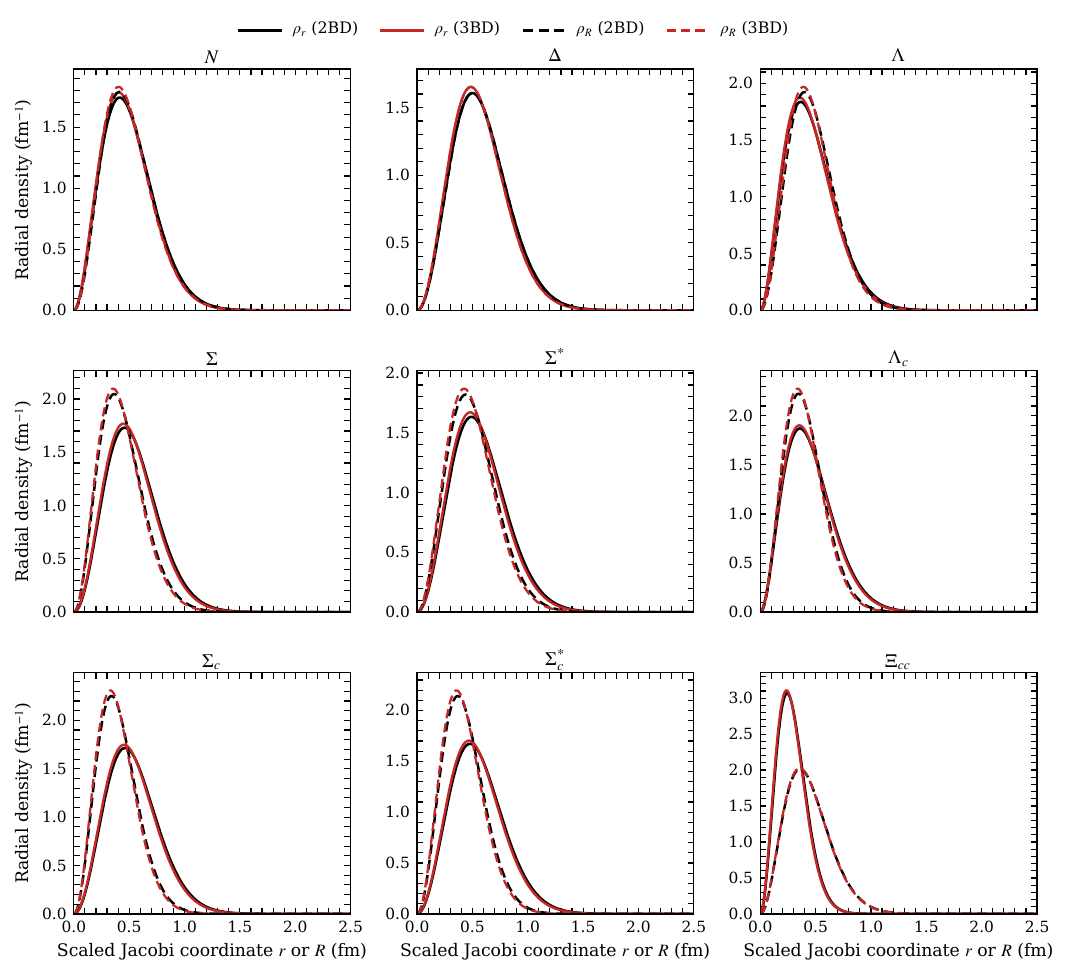}
\caption{\justifying
Normalized $S$-wave Jacobi radial densities for the nine calibration baryons.
Black curves are eigenstates of the meson-calibrated two-body Hamiltonian; red curves are obtained by full rediagonalization of $H_{2\rm BD}+L_{\rm G}^{3Q}$ with the independently refitted Gaussian couplings.
Solid and dashed curves denote $\rho_r$ and $\rho_R$, respectively.}
\label{fig:gaussian_3q_density_validation}
\end{figure*}

\subsection{Nonperturbative validation of the first-order three-quark treatment}
\label{subsec:gaussian_3q_perturbative_validation}
In the spectroscopy calculation above, the three-quark interaction is treated in first-order perturbation theory. We test this approximation by diagonalizing the full Gaussian-profile Hamiltonian in the same retained $S$-wave space. We use the Gaussian profile as a representative finite-range case because its matrix elements are analytic and its range follows the same mass-scaled convention as the Yukawa profile. The two-body parameters are fixed at the values in Eq.~\eqref{params}, and we use the same nine ground-state baryons as in Table~\ref{tab:combined_3bd_baryon}. The Gaussian couplings in Eq.~\eqref{eq:3bd_gaussian_coeff_fitting} were fitted in the larger $SS+PP+DD+FF$ space. We therefore refit the couplings separately in the present $S$-wave space for the first-order and fully diagonalized calculations.
The retained basis contains all three Jacobi rearrangements with {12~\ensuremath{\times}~12} geometric Gaussian ranges spanning 0.05--3.00~fm in each coordinate.
Overlap modes satisfying $o_n/o_{\rm max}\leq 10^{-10}$ are removed identically in every calculation.
Increasing the mesh to {14~\ensuremath{\times}~14} changes every mass reported below by at most
0.012~MeV.
At {14~\ensuremath{\times}~14}, enlarging the interval to 0.04--4.00~fm changes every mass by at most
0.005~MeV, while lowering the overlap cutoff to 10\textsuperscript{\ensuremath{-}11} leaves the {12~\ensuremath{\times}~12} results unchanged at the quoted precision.

The first-order mass and the fully rediagonalized retained-space mass are
\begin{align}
M_B^{(1)}(A,B,C)
=E_{B,0}^{(0)}+
\langle\Psi_{B,0}^{(0)}|L^{3Q}_{\rm G}(A,B,C)|\Psi_{B,0}^{(0)}\rangle, \qquad
M_B^{\rm diag}(A,B,C)
=\min\operatorname{eig}\!\left[H_{\rm 2BD}+L^{3Q}_{\rm G}(A,B,C)\right].
\end{align}
The largest relative change among the dimension-matched coupling scales $\vert A\vert ^{1/2}$, $\vert B\vert ^{1/6}$, and $\vert C\vert ^{1/4}$ is only 1.19\%, and the independently fitted masses differ from their first-order counterparts by at most 0.174~MeV.

At fixed first-order couplings, the largest magnitude of the nonperturbative remainder is 1.8~MeV; its sign is nonpositive for every state, as required by the variational principle.
The first-order mixing norm
\begin{equation}
    \left\|\sum_{n\ne0}|\Psi_{B,n}^{(0)}\rangle\frac{{\langle\Psi_{B,n}^{(0)}|L^{3Q}_{\rm G}|\Psi_{B,0}^{(0)}\rangle}}{E_{B,0}^{(0)}-E_{B,n}^{(0)}}\right\|
\end{equation}
does not exceed 0.048 in the nine-state set.
To test the wave functions directly, we trace over the internal spin-flavor degrees of freedom and define the normalized $a$-channel Jacobi densities
\begin{align}
\rho_r=r^2\sum_\xi\int d\Omega_{\mathbf r}\,d^3\mathbf R\,
|\Psi(\mathbf r,\mathbf R;\xi)|^2, \qquad
\rho_R=R^2\sum_\xi\int d\Omega_{\mathbf R}\,d^3\mathbf r\,
|\Psi(\mathbf r,\mathbf R;\xi)|^2,
\end{align}
with $\int_0^\infty\rho_r(r)dr=\int_0^\infty\rho_R(R)dR=1$.
Here $\mathbf r=(\mathbf r_1-\mathbf r_2)/\sqrt2$ and $\mathbf R=\sqrt{2/3}[(m_1\mathbf r_1+m_2\mathbf r_2)/(m_1+m_2)-\mathbf r_3]$.

Fig.~\ref{fig:gaussian_3q_density_validation} compares the two-body ground-state densities with the densities obtained after the full rediagonalization. The minimum two-body--three-body fidelity is 0.997913. The largest total-variation distances are 0.0255 for $\rho_r$ and 0.0255 for $\rho_R$, and the largest absolute relative change of the corresponding rms Jacobi radii is 2.732\%. These results show that the tens-of-MeV mass shifts mainly come from the diagonal matrix element of the Gaussian three-quark interaction. The higher-order energy correction and the change of the retained-space wave function are small.

\section{Transition through a \texorpdfstring{$j_{qg}^{P}=3/2^{+}$}{j(qg)=3/2+} component of the \texorpdfstring{$qqqg$}{qqqg} sector}
\label{app:qg_spin32}
In this appendix, we construct one explicit $j_{qg}^{P}=3/2^{+}$ component of the color-singlet $qqqg$ sector discussed in  Section~\ref{subsec:three_quark_definition}. We use the $qg$ state only as an internal color and angular-momentum component of the full color-singlet $qqqg$ configuration. It is not an independently observable colored state. We take a gluonic excitation with $j_g^P=1^+$ and couple it to the quark on line $k$ with $\ell_{qg}=0$. The resulting color-triplet $qg$ state contains both $j_{qg}^P=1/2^+$ and $3/2^+$ components. Lattice-QCD hybrid-baryon spectra give an example of positive-parity states that are compatible with such a $j_g^P=1^+$ excitation~\cite{DudekEdwards2012HybridBaryons}. We first define the normalized color-triplet state and the spin-transition projector. We then derive the local $S$-wave selection rule, eliminate the intermediate state in the static approximation, and project the result onto the color-singlet $qqq$ state. The result shows that the $j_{qg}^{P}=3/2^{+}$ component belongs to the $S$--$S$ operator class used in the main text. This construction is not a unique reduction of QCD and does not derive the other two operator classes.

\paragraph{Color-triplet $qg$ component:}
Using $F^c=\lambda^c/2$, a normalized color-triplet $qg$ state may be written as
\begin{equation}
\left|(qg)_{\mathbf 3_C};\xi,jm\right\rangle=\frac{\sqrt3}{2}\sum_{c,\eta,m_g,m_q}(F^c)_{\eta\xi}\,\mathcal C^{\frac12\,1;j}_{m_q\,m_g;m}\left|q_{\eta m_q}g_{c m_g}\right\rangle.
\label{eq:qg32_state_definition}
\end{equation}
The normalization factor follows from $\sum_cF^cF^c=(4/3)\mathbf 1$. In the adopted phase convention, $(\sqrt3/2)(F^c)_{\eta\xi}=\mathcal C[{\rm SU(3)}]^{\mathbf3\,\mathbf8;\mathbf3}_{\eta\,c;\xi}$ is the SU(3) Clebsch--Gordan coefficient for $\mathbf3_C\otimes\mathbf8_C\to\mathbf3_C$. In quark color indices, the same coupling is represented by $[(\lambda_k^c)_{\rm tr}^{\dagger}]_{\xi\eta}=(\sqrt3/4)(\lambda^c)_{\xi\eta}$, and therefore $(\lambda_k^d)_{\rm tr}(\lambda_k^c)_{\rm tr}^{\dagger}=(3/16)\lambda_k^d\lambda_k^c$. Two transition blocks consequently supply the common factor $3/16$, which is absorbed into the reduced transition strength used below.

\paragraph{Transition-spin matrices and the $j_{qg}=3/2$ projector:}

We order the spin-$1/2$ states as $\{\lvert 1/2,1/2\rangle ,\lvert 1/2,-1/2\rangle \}$ and the spin-$3/2$ states as $\{\lvert 3/2,3/2\rangle ,\lvert 3/2,1/2\rangle ,\lvert 3/2,-1/2\rangle ,\lvert 3/2,-3/2\rangle \}$. 
With the SU(2) Clebsch--Gordan convention adopted in this work, the spherical transition matrices are defined by $\langle 3/2,M\lvert \mathsf S_{k,m_g}^{\dagger}\rvert  1/2,m_q\rangle =\mathcal C^{\frac12\,1;\frac32}_{m_q\,m_g;M}$ for $m_g=+1,0,-1$. The rows and columns of the Cartesian matrices therefore follow the spin-$3/2$ and spin-$1/2$ orderings stated above. Using $V_{\pm1}=\mp(V_x\pm iV_y)/\sqrt2$ and $V_0=V_z$, one obtains
\begin{equation}
\mathsf S_{k,x}^{\dagger}=\begin{pmatrix}-1/\sqrt2&0\\0&-1/\sqrt6\\1/\sqrt6&0\\0&1/\sqrt2\end{pmatrix},\qquad \mathsf S_{k,y}^{\dagger}=i\begin{pmatrix}1/\sqrt2&0\\0&1/\sqrt6\\1/\sqrt6&0\\0&1/\sqrt2\end{pmatrix},\qquad \mathsf S_{k,z}^{\dagger}=\begin{pmatrix}0&0\\\sqrt{2/3}&0\\0&\sqrt{2/3}\\0&0\end{pmatrix}.
\label{eq:qg32_cartesian_transition_matrices}
\end{equation}
Their Hermitian conjugates $\mathsf S_{k,\ell}=(\mathsf S_{k,\ell}^{\dagger})^{\dagger}$ are {2~\ensuremath{\times}~4} matrices that return the $j_{qg}=3/2$ component to the spin-$1/2$ quark state. They are not the {4~\ensuremath{\times}~4} spin generators acting within the spin-$3/2$ representation. Direct multiplication gives
\begin{equation}
\mathsf S_{k,\ell}\mathsf S_{k,\ell'}^{\dagger}\equiv P_{k,\ell\ell'}^{(3/2)}=\delta_{\ell\ell'}-\frac13\sigma_{k,\ell}\sigma_{k,\ell'}=\frac23\delta_{\ell\ell'}-\frac{i}{3}\epsilon_{\ell\ell'n}\sigma_k^n.
\label{eq:qg32_transition_identity}
\end{equation}
The completeness relations are $\sum_{\ell=x,y,z}\mathsf S_{k,\ell}\mathsf S_{k,\ell}^{\dagger}=2\mathbf 1_{2\times2}$ and $\sum_{\ell=x,y,z}\mathsf S_{k,\ell}^{\dagger}\mathsf S_{k,\ell}=\mathbf 1_{4\times4}$.
The projector satisfies $\sigma_{k,\ell}P_{k,\ell\ell'}^{(3/2)}=P_{k,\ell\ell'}^{(3/2)}\sigma_{k,\ell'}=0$ and removes the $j_{qg}=1/2$ component from $1\otimes1/2=3/2\oplus1/2$. 

Eq.~\eqref{eq:qg32_transition_identity} also gives the local $S$-wave selection rule. Without an orbital derivative, the transition operator must be a positive-parity scalar. An operator acting only on line $k$ cannot connect spin $1/2$ and spin $3/2$. We therefore need a second spin vector. The simplest scalar is $\boldsymbol{\sigma}_i\cdot\boldsymbol{\mathsf{S}}_k^{\dagger}=\sum_{\ell}\sigma_i^{\ell}\mathsf S_{k,\ell}^{\dagger}$.
The explicit matrices then give
\begin{equation}
\big\langle\left[1/2_i\otimes3/2_{k^\ast}\right]_{s'_{ik}m'}\big|\boldsymbol{\sigma}_i\cdot\boldsymbol{\mathsf S}_k^{\dagger}\big|\left[1/2_i\otimes1/2_k\right]_{s_{ik}m}\big\rangle=\frac{2\sqrt6}{3}\,\delta_{s'_{ik},1}\delta_{s_{ik},1}\delta_{m'm}.
\label{eq:qg32_explicit_selection_matrix_element}
\end{equation}
One then finds that the local $S$-wave transition connects only
\begin{equation}
\left[(q_iq_k)_{s_{ik}=1}\right]_{m}\longleftrightarrow\left[q_i(q_kg)_{j_{qg}=3/2}\right]_{s'_{ik}=1,m}.
\label{eq:qg32_Swave_selection_rule}
\end{equation}
The matrix element vanishes for an initial spin--singlet pair and for the final $s'_{ik}=2$ state. Therefore, the local spin-independent central-color interaction $V_{ik}^{C}$ alone does not generate this channel. In the present local construction, a color-type transition requires additional derivative or orbital-momentum dependence.

\paragraph{Minimal transition block and static approximation:}
Accordingly, this appendix retains only the local color--spin transition allowed by the selection rule above. 
Let $\mathsf Q_{g,3/2}$ project onto the $j_{qg}^{P}=3/2^{+}$ subspace of the $qqqg$ sector represented by $\mathsf Q_g$. 
The corresponding transition block is
\begin{equation}
T_{\mathsf Q_{g,3/2}\mathsf P}^{(ik)}=\frac{g_{CS}^{\rm tr}}{m_i m_k}\,u_{ik}^{(3/2)}(r_{ik})\,\lambda_i^c(\lambda_k^c)_{\rm tr}^{\dagger}\,\boldsymbol{\sigma}_i\cdot\boldsymbol{\mathsf S}_k^{\dagger},\qquad T_{\mathsf P\mathsf Q_{g,3/2}}^{(ik)}=\left(T_{\mathsf Q_{g,3/2}\mathsf P}^{(ik)}\right)^{\dagger}.
\label{eq:qg32_quantum_transition_vertex}
\end{equation}
Here $g_{CS}^{\rm tr}$ is a reduced color--spin transition strength and $u_{ik}^{(3/2)}(r_{ik})$ is a short-range radial transition function.

Within this appendix, $M_k>0$ denotes the reference excitation scale of the $j_{qg}=3/2$ component on line $k$, in the same role as the quantity $M_k$ used in the main text. Write
\begin{equation}
H_{\mathsf Q_{g,3/2}\mathsf Q_{g,3/2}}=E_{\mathsf P}^{(0)}+M_k+\delta H_{g,3/2},\qquad E=E_{\mathsf P}^{(0)}+\delta E.
\label{eq:qg32_gap_decomposition}
\end{equation}
The corresponding resolvent is
\begin{align}
R_{\mathsf Q_{g,3/2}}(E)&\equiv\left(E-H_{\mathsf Q_{g,3/2}\mathsf Q_{g,3/2}}\right)^{-1}=-\frac{1}{M_k}+\frac{\delta H_{g,3/2}-\delta E}{M_k^2}+\mathcal O\!\left(\frac{(\delta H_{g,3/2}-\delta E)^2}{M_k^3}\right).
\label{eq:qg32_resolvent_expansion}
\end{align}
At leading order, we keep only $R_{\mathsf Q_{g,3/2}}(E)\longrightarrow-M_k^{-1}$. The spin-$3/2$ projector changes the spin operator in the second-order term, but it does not change the expansion parameter in Eq.~\eqref{eq:qg32_resolvent_expansion}. This approximation can be used when the recoil energy, the spread of the relevant hybrid levels, and the change of the external valence energy are small compared with $M_k$.

Lattice-QCD hybrid-baryon spectra suggest a gluonic excitation scale of order $1~\mathrm{GeV}$~\cite{DudekEdwards2012HybridBaryons}. This supports a separation between the low-lying valence states and hybrid excitations, but it does not determine the line-dependent denominator $M_k$. We therefore use $-M_k^{-1}$ as a leading phenomenological approximation. The higher terms give energy-, momentum-, and configuration-dependent corrections and may also generate nonlocal and noncentral three-quark operators.

\paragraph{Connected three-quark contribution:}
Within the static single-component approximation adopted here, terms in which the same pair creates and removes the excitation are assigned to the one- and two-body parts of the baseline Hamiltonian. We now apply the two insertion orderings on the left-hand side of Eq.~\eqref{eq:qstar_effective_operator_matching} to the $j_{qg}=3/2$ subspace, using the transition block in Eq.~\eqref{eq:qg32_quantum_transition_vertex} and the leading term of Eq.~\eqref{eq:qg32_resolvent_expansion}. For the first ordering, Eq.~\eqref{eq:qg32_transition_identity} gives
\begin{equation}
\sigma_j^{\ell}P_{k,\ell\ell'}^{(3/2)}\sigma_i^{\ell'}=\frac23\boldsymbol{\sigma}_i\cdot\boldsymbol{\sigma}_j+\frac{i}{3}\boldsymbol{\sigma}_i\cdot\left(\boldsymbol{\sigma}_j\times\boldsymbol{\sigma}_k\right).
\label{eq:qg32_first_order_spin_product}
\end{equation}
Interchanging $i$ and $j$ leaves the pair-spin term unchanged but reverses the sign of the triple-spin term. Defining the reduced coupling by $(g_{CS}^{\rm red})^2\equiv(3/16)(g_{CS}^{\rm tr})^2$, Eq.~\eqref{eq:SU3_generator_algebra} with $F^c=\lambda^c/2$ shows that reversing the order of the two generators leaves the $\delta^{cd}$ and $d^{cde}$ terms unchanged and reverses the sign of the $if^{cde}$ term. The corresponding sign changes in the spin and color factors determine which symmetric and antisymmetric color structures survive when the two insertion orderings are added.
Combining the two orderings gives the line-resolved $S$--$S$-like operator
\begin{align}
L_{ij;k}^{S-S,(3/2)}\equiv\frac{1}{m_i m_jm_k^2M_k}\Bigg[&\left(\frac89\lambda_i^c\lambda_j^c+\frac43d^{cde}\lambda_i^c\lambda_j^d\lambda_k^e\right)\boldsymbol{\sigma}_i\cdot\boldsymbol{\sigma}_j+\frac23f^{cde}\lambda_i^c\lambda_j^d\lambda_k^e\,\boldsymbol{\sigma}_i\cdot\left(\boldsymbol{\sigma}_j\times\boldsymbol{\sigma}_k\right)\Bigg].
\label{eq:qg32_color_spin_operator}
\end{align}
The minimal connected contribution is therefore
\begin{equation}
\Delta H_{\mathsf P,ij;k}^{(3/2)}=-(g_{CS}^{\rm red})^2u_{ik}^{(3/2)}(r_{ik})u_{jk}^{(3/2)}(r_{jk})L_{ij;k}^{S-S,(3/2)}.
\label{eq:qg32_effective_potential}
\end{equation}
We obtain the full contribution by summing over the three possible choices of the intermediate line $k$. In the phenomenological calculation, we represent the cyclic sum of $u_{ik}^{(3/2)}(r_{ik})u_{jk}^{(3/2)}(r_{jk})$ by $f_{\mathrm Y}(r_{ij},r_{jk},r_{ki})$ or $f_{\mathrm G}(r_{ij},r_{jk},r_{ki})$. We do not identify the cyclic radial structure pointwise with either profile. Their overall normalization is included in the fitted $S$--$S$ coupling $B$.
Using the color-singlet factors in Eq.~\eqref{eq:qqq_color_singlet_factors}, Eq.~\eqref{eq:qg32_color_spin_operator} gives $2/3$ of the corresponding line-$k$ contribution to $L_{123}^{S-S}$ in Eq.~\eqref{eq:l12}. Therefore, in the color-singlet $qqq$ state, the minimal local $j_{qg}=3/2$ transition gives the same $S$--$S$ operator class as the interaction used in the main text. The factor $2/3$ is absorbed into the fitted coupling $B$.

\section{Electromagnetic observables and radiative transitions}
\label{app:em_observables}
We calculate selected static magnetic moments and radiative transitions as an additional test of the baryon wave functions. We use the eigenstates of the two-body Hamiltonian and include their first-order wave-function corrections induced by the Yukawa-profile three-quark interaction $L_{\rm Y}^{3Q}$. Our purpose is to examine how the three-quark interaction changes the wave functions. We do not attempt to construct a complete electromagnetic current operator.

\paragraph{Basis and first-order three-quark correction:}
For the electromagnetic calculation, we use the TH1 central Hamiltonian but truncate the orbital space at $l_{\mathcal C},L_{\mathcal C}\leq2$.
For the positive-parity ground states, the retained orbital components are
\begin{equation}
(l_{\mathcal C},L_{\mathcal C};J)
=
(0,0;0),\ (1,1;0),\ (2,2;0),
\qquad
\mathcal C\in\{a,b,c\}.
\label{eq:app_em_ground_basis}
\end{equation}
The $PP$ and $DD$ components are therefore internal correlation components coupled to total orbital angular momentum $J=0$.
This $SS+PP+DD$ space is the $l_{\rm max}=2$ subset used in the direct Cartesian cross-check.
It is smaller than the production ground-state TH1 space, which additionally contains the $(l_{\mathcal C},L_{\mathcal C})=(3,3)$, or $FF$, component.
For the negative-parity central cores, we retain
\begin{equation}
(l_{\mathcal C},L_{\mathcal C};J)
=
(1,0;1),\ (0,1;1),\ (1,2;1),\ (2,1;1),
\qquad
\mathcal C\in\{a,b,c\}.
\label{eq:app_em_pwave_basis}
\end{equation}
The Gaussian ranges in both Jacobi coordinates satisfy $0.05~\mathrm{fm} \leq r_n,R_N \leq 3.0~\mathrm{fm}$.
Because TH1 contains neither tensor nor spin--orbit interactions, its negative-parity states are naturally organized in $(J,P,S)$ blocks.
In particular, the cores used below belong to $(J^P,S)=(1^-,1/2)$.
Coupling $J=1$ and $S=1/2$ gives the degenerate $\mathcal J^P=1/2^-$ and $3/2^-$ descendants in TH1; these cores should not be identified with fully mixed physical negative-parity states.

We first solve the generalized eigenvalue problem in Eq.~\eqref{eq:baryon_generalized_eigenvalue} and canonically orthogonalize the retained overlap space. We denote the resulting orthonormal two-body eigenstates by $\ket{\Psi_{B,n}^{(0)}}$ and their eigenvalues by $E_{B,n}^{(0)}$. We use $\mathcal R_B$ for the retained low-lying eigenstates in the same flavor, parity, and angular-momentum block. The first-order correction from $L_{\rm Y}^{3Q}$ is then
\begin{equation}
\ket{\delta\Psi_{B,n}^{\rm Y}}
=
\sum_{\substack{m\in\mathcal R_B\\m\neq n}}
\ket{\Psi_{B,m}^{(0)}}
\frac{
\bra{\Psi_{B,m}^{(0)}}
L_{\rm Y}^{3Q}
\ket{\Psi_{B,n}^{(0)}}
}{
E_{B,n}^{(0)}-E_{B,m}^{(0)}
}, \qquad \ket{\Psi_{B,n}^{\rm Y}}
=
\ket{\Psi_{B,n}^{(0)}}
+
\ket{\delta\Psi_{B,n}^{\rm Y}}.
\label{eq:app_em_state_correction}
\end{equation}
The correction in Eq.~\eqref{eq:app_em_state_correction} is orthogonal to the unperturbed state. Therefore, $\ket{\Psi_{B,n}^{\rm Y}}$ is normalized through first order.
For a current operator $\widehat J_X^\mu$ coupled to an external probe, and suppressing explicit baryon and radial labels on the initial and final states, the matrix element retained through first order is
\begin{align}
J_{X,fi}^{\mu,\rm Y}
={}&
\bra{\Psi_f^{(0)}}
\widehat J_X^\mu
\ket{\Psi_i^{(0)}}
+
\bra{\delta\Psi_f^{\rm Y}}
\widehat J_X^\mu
\ket{\Psi_i^{(0)}}
+
\bra{\Psi_f^{(0)}}
\widehat J_X^\mu
\ket{\delta\Psi_i^{\rm Y}} .
\label{eq:app_em_corrected_matrix_element}
\end{align}
Terms quadratic in $L_{\rm Y}^{3Q}$ are omitted.
As a numerical cross-check, we also diagonalize the finite retained-space matrix
\begin{equation}
\mathsf H_{B,mn}^{\rm ret}
=
E_{B,n}^{(0)}\delta_{mn}
+
\bra{\Psi_{B,m}^{(0)}}
L_{\rm Y}^{3Q}
\ket{\Psi_{B,n}^{(0)}}.
\label{eq:app_em_retained_hamiltonian}
\end{equation}
The resulting electromagnetic matrix elements are nearly identical to those obtained from the first-order construction above.

\paragraph{Electromagnetic operators:}

We calculate the radiative transitions in the one-body impulse approximation. The transition amplitude is proportional to $\epsilon_\mu^{*}(\mathbf k)\langle \Psi_f\lvert \widehat J_{\rm em}^{\mu}(\mathbf k)\rvert  \Psi_i\rangle $. The non-relativistic constituent-quark current contains the convection and spin--magnetization terms~\cite{OrtizPacheco:2023,KoniukIsgur1980,GarciaTecocoatziEtAl2025}. We use the leading dipole approximation. The electric amplitude is calculated with the center-of-mass-subtracted length-form $E1$ operator~\cite{PengLuoLiu2024}, and the magnetic amplitude is calculated with the spin $M1$ operator. We do not include the explicit finite-$k_\gamma$ convection-current matrix element. We denote the constituent charge numbers by $q_i^{\rm em}$ to distinguish them from the heavy-quark label $Q$ and the Cornell constant $D$.
The magnetic-moment operator
$\widehat{\boldsymbol{\mu}}_{\rm em}$ in units of the nuclear magneton $\mu_N$ and the reduced electric-dipole operator $\widehat{\mathbf d}_{E1}$ are
\begin{align}
\frac{\widehat{\boldsymbol\mu}_{\rm em}}{\mu_N}
&=
2\sum_{i=1}^{3}
q_i^{\rm em}\frac{m_p}{m_i}\mathbf S_i
=
\sum_{i=1}^{3}
q_i^{\rm em}\frac{m_p}{m_i}\boldsymbol{\sigma}_i,
\nonumber\\
\widehat{\mathbf d}_{E1}
&=
\sum_{i=1}^{3}
q_i^{\rm em}
\left(
\mathbf r_i-\mathbf R_{\rm CM}
\right),
\qquad
\mathbf R_{\rm CM}
=
\frac{\sum_i m_i\mathbf r_i}{\sum_i m_i},
\qquad
\mu_N=\frac{e\hbar}{2m_pc}.
\label{eq:app_em_operators}
\end{align}
Here $\mathbf S_i=\boldsymbol{\sigma}_i/2$, $q_u^{\rm em}=q_c^{\rm em}=2/3$, and $q_d^{\rm em}=q_s^{\rm em}=q_b^{\rm em}=-1/3$.
The operator $\widehat{\mathbf d}_{E1}$ is defined without the overall factor $e$ and has dimensions of length; the electromagnetic coupling is included explicitly through $\alpha_{\rm em}$ in the width formula.

For diagonal magnetic moments and $M1$ transition moments, we define
\begin{equation}
\frac{\mu_{fi}}{\mu_N}
=
\bra{\Psi_f;\mathcal J_f,M_f=1/2}
\frac{\widehat\mu_{{\rm em},z}}{\mu_N}
\ket{\Psi_i;\mathcal J_i,M_i=1/2},
\label{eq:app_em_transition_moment}
\end{equation}
where $\ket{\Psi_{i,f}}$ denotes either the unperturbed state $\ket{\Psi_{i,f}^{(0)}}$ or the first-order corrected state $\ket{\Psi_{i,f}^{\rm Y}}$.
For the $M1$ transitions considered here, we define the rotationally invariant strength by
\begin{equation}
\mathcal S_{{\rm M1},fi}
=
\frac{1}{2\mathcal J_i+1}
\left|
\left\langle
\mathcal J_f
\left\Vert
\frac{\widehat{\boldsymbol\mu}_{\rm em}}{\mu_N}
\right\Vert
\mathcal J_i
\right\rangle
\right|^2, \qquad \left\langle \mathcal J_f M_f \left| \frac{\widehat\mu_{{\rm em},q}}{\mu_N} \right| \mathcal J_i M_i \right\rangle
=
\frac{\mathcal C^{\mathcal J_i\,1;\mathcal J_f}_{M_i\,q;M_f}}{\sqrt{2\mathcal J_f+1}}
\left\langle \mathcal J_f \left\Vert \frac{\widehat{\boldsymbol\mu}_{\rm em}}{\mu_N} \right\Vert \mathcal J_i \right\rangle.
\end{equation}
Here we used the Wigner--Eckart convention~\cite{Varshalovich1988}.
For $\mathcal J_f=1/2$, $M_i=M_f=1/2$, and $q=0$, the relevant CG coefficients satisfy $\left\lvert  \mathcal C^{\mathcal J_i\,1;\mathcal J_f}_{M_i\,q;M_f}\right\rvert  ^2=1/3$ for both $\mathcal J_i=1/2$ and $\mathcal J_i=3/2$, so the squared coefficient multiplying the reduced matrix element is $1/6$ in either case.
Consequently,
\begin{equation*}
\mathcal S_{{\rm M1},fi}
=
3\left|\frac{\mu_{fi}}{\mu_N}\right|^2
\quad (\mathcal J_i=1/2),
\qquad
\mathcal S_{{\rm M1},fi}
=
\frac{3}{2}
\left|\frac{\mu_{fi}}{\mu_N}\right|^2
\quad (\mathcal J_i=3/2).
\end{equation*}
For the TH1 central-core $E1$ transitions with spectator spin $S=1/2$, we use $\mathcal S_{{\rm E1},fi}= \vert \bra{\Psi_f}\widehat d_{{\rm E1},z}\ket{\Psi_i}\vert ^2$.

The photon momentum and the final-baryon energy in the rest frame of the initial baryon are evaluated using experimental baryon masses, following the exact two-body recoil kinematics also used in constituent-quark radiative calculations~\cite{OrtizPacheco:2023,PengLuoLiu2024}:
\begin{equation}
k_\gamma = \frac{
\big(M_{B_i}^{\rm exp}\big)^2-\big(M_{B_f}^{\rm exp}\big)^2}{2M_{B_i}^{\rm exp}},
\qquad
E_f^\gamma = \frac{ \big(M_{B_i}^{\rm exp}\big)^2 + \big(M_{B_f}^{\rm exp}\big)^2 }{ 2M_{B_i}^{\rm exp} }.
\label{eq:app_em_kinematics}
\end{equation}
With energies and masses expressed in MeV and the $E1$ matrix element expressed in fm, the adopted Cartesian-component width convention is
\begin{align}
\Gamma_{\rm M1}(B_i\to B_f\gamma)
=
\frac{\alpha_{\rm em}}{3m_p^2}
\frac{E_f^\gamma}{M_{B_i}^{\rm exp}}
k_\gamma^3
\mathcal S_{{\rm M1},fi},
\qquad
\Gamma_{\rm E1}(B_i\to B_f\gamma)
=
\frac{4\alpha_{\rm em}}{3}
\frac{E_f^\gamma}{M_{B_i}^{\rm exp}}
\frac{k_\gamma^3}{(\hbar c)^2}
\mathcal S_{{\rm E1},fi}.
\label{eq:app_em_widths}
\end{align}
The explicit $k_\gamma^3$ factor is the leading dipole dependence generated by the two-body phase space and the dipole radiation amplitude.
At finite $k_\gamma$, the transition matrix element generally acquires additional momentum dependence through plane-wave factors and spatial form factors.
The numerical prefactors in Eq.~\eqref{eq:app_em_widths} follow from the Cartesian-component strength and state-normalization conventions defined above.
The same experimental masses, and therefore the same $k_\gamma$, are used for the unperturbed and $L_{\rm Y}^{3Q}$-corrected calculations.
Consequently, the difference between the two theory columns in Table~\ref{tab:app_em_summary} is generated solely by the change of the initial and final wave functions.

\begin{table*}[!t]
\centering
\caption{\justifying
Selected electromagnetic observables calculated with the eigenstates of the two-body Hamiltonian and after the first-order $L_{\rm Y}^{3Q}$ wave-function correction.
The experimental information is taken from the Particle Data Group, BESIII, and Belle \cite{PDG2026,BESIIILambdaRadiative2026,BelleXiCRadiative2020}.
States denoted by ``core'' are TH1 central configurations with $(J^P,S)=(1^-,1/2)$ and are not the fully mixed physical negative-parity states associated with the experimental labels.}
\label{tab:app_em_summary}
\small

\resizebox{\textwidth}{!}{%
\begin{tblr}{
  colspec={
    Q[l,wd=4.5cm]
    Q[c,wd=2.6cm]
    Q[c,wd=3.2cm]
    Q[c,wd=6.5cm]
  },
  row{1}={bg=headgray,halign=c,valign=m},
  cell{1}{1}={}{halign=l},
  hline{2}={1-4}{0.3pt},
  rowsep=1.5pt,
}
\toprule
Observable
& Two-body
& $L_{\rm Y}^{3Q}$ corrected
& Experimental or inferred value
\\

$\mu_p/\mu_N$
& $2.81331$
& $2.81346$
& $2.79285$
\\

$\mu_n/\mu_N$
& $-1.86429$
& $-1.86444$
& $-1.91304$
\\

$\Gamma(\Delta^+\to p\gamma)$ [keV]
& $396.47$
& $397.10$
& $630$--$780$ (PDG-derived range)
\\

$\Gamma(\Sigma^0\to\Lambda\gamma)$ [keV]
& $8.358$
& $8.380$
& $8.90(84)$ (from the radiative lifetime)
\\

$\mathrm{Br}(\Sigma^{*0}\to\Lambda\gamma)$ [$\%$]
& $0.5604$
& $0.5658$
& $1.25^{+0.13}_{-0.12}$
\\

$\Gamma[\Lambda(1520)_{\rm core}\to\Lambda\gamma]$ [keV]
& $116.65$
& $107.91$
& $149.3\pm26.4_{\rm stat}\pm31.0_{\rm syst}$ (inferred)
\\

$\Gamma[\Lambda(1520)_{\rm core}\to\Sigma^0\gamma]$ [keV]
& $21.73$
& $22.67$
& $46.8\pm6.6_{\rm stat}\pm9.3_{\rm syst}$
\\

$\Gamma_{\Lambda\gamma}/\Gamma_{\Sigma^0\gamma}$
for the $\Lambda(1520)$ core
& $5.367$
& $4.761$
& $3.19\pm0.34_{\rm stat}\pm0.19_{\rm syst}$
\\

$\Gamma[\Xi_c(2790)^0_{\rm core}\to\Xi_c^0\gamma]$ [keV]
& $225.69$
& $217.60$
& $\simeq800\pm320$ (Belle estimate)
\\

$\Gamma[\Xi_c(2815)^0_{\rm core}\to\Xi_c^0\gamma]$ [keV]
& $276.97$
& $267.05$
& $320\pm45{}^{+45}_{-80}$ (Belle estimate)
\\
\bottomrule
\end{tblr}}
\end{table*}

The $\Delta^+\to p\gamma$ interval is obtained from the PDG radiative branching-fraction and total-width ranges, whereas the $\Sigma^0\to\Lambda\gamma$ width is obtained from its measured lifetime.
The quoted $\Lambda(1520)\to\Lambda\gamma$ value is inferred from the BESIII $\Sigma^0\gamma$ width and the measured width ratio and is therefore not an independent experimental input.
Likewise, Belle directly measured branching-fraction ratios for the $\Xi_c(2790)$ and $\Xi_c(2815)$ transitions; the absolute widths in Table~\ref{tab:app_em_summary} require additional total-width and strong-decay inputs.
The theoretical $\Sigma^{*0}$ branching fraction is obtained using the experimental central total width, $\Gamma_{\rm tot}=44~\mathrm{MeV}$, without propagating its uncertainty.

Direct constituent-quark-model calculations of the $\Sigma^{*0}\to\Lambda\gamma$ transition have been reported in Refs.~\cite{YuChenDengZhu2006,WagnerBuchmannFaesslerRadiative1998}.
Meson-cloud corrections to the same class of $\Sigma^{*}$ radiative transitions have also been studied~\cite{RamalhoTsushima2013}.
The $\Lambda(1520)\to\Lambda\gamma$ and $\Lambda(1520)\to\Sigma^0\gamma$ channels have been calculated in several quark-model implementations, including investigations of multiplet mixing and nonvalence admixtures~\cite{YuChenDengZhu2006,DarewychHorbatschKoniuk1983,KaxirasMonizSoyeur1985}.
The corresponding $P$-wave-to-ground-state $\Xi_c(2790)$ and $\Xi_c(2815)$ radiative transitions have likewise been studied in constituent-quark calculations~\cite{OrtizPacheco:2023,GarciaTecocoatziEtAl2025,PengLuoLiu2024}.

The allowed ground-state $M1$ observables change by less than 1.0\% after including the first-order $L_{\rm Y}^{3Q}$ correction. The $\Sigma^0\to\Lambda\gamma$ width remains close to the experimental value. On the other hand, the calculated $\Delta^+\to p\gamma$ width and the $\Sigma^{*0}\to\Lambda\gamma$ branching fraction remain smaller than the experimental values. The $\Sigma^0\to\Lambda\gamma$ result shows that the normalization of the one-body magnetic operator is reasonable. However, its change due to $L_{\rm Y}^{3Q}$ is below one percent and gives little independent information on the three-quark interaction.

The selected $E1$ central-core quantities change by approximately 4\%--11\%. For the $\Lambda(1520)$ core, however, the $\Sigma^0\gamma$ width and the width ratio still change by approximately 16\%--20\% between the last two Gaussian-basis enlargements. The physical negative-parity states also require the omitted $S=3/2$ configurations, flavor-singlet--octet mixing, tensor and spin--orbit interactions, and coupled-channel dressing. The two neutral $\Xi_c$ widths are calculated from the same TH1 $(J^P,S)=(1^-,1/2)$ spatial core and differ mainly because of the experimental $k_\gamma^3$ factors. Therefore, they are not two independent predictions of the internal wave-function structure.

We do not include constituent-quark anomalous magnetic moments, finite-$k_\gamma$ corrections, relativistic boosts, meson-cloud contributions, exchange currents, or the interaction-induced many-body current associated with $L_{\rm Y}^{3Q}$~\cite{DeSanctisSantopintoGiannini1998,WagnerBuchmannFaessler1998}. This comparison gives an additional consistency check on the calculated baryon wave functions and the specified one-body electromagnetic current. It cannot independently determine the three-quark interaction.

\bibliographystyle{elsarticle-num}
\bibliography{ppnp_references}

\end{document}